\documentclass[11pt]{amsbook} 
\usepackage{amsfonts}
\usepackage{verbatim,amssymb,amsmath}

\usepackage{amscd}
\usepackage{color}

\makeindex %%%%%%

\begin{document}

\renewcommand{\thesection}{\thechapter.\arabic{section}}

\def\where{\quad\text{where}\quad}
\def\when{\quad\text{when}\quad}
\def\with{\quad\text{with}\;\;}
\def\for{\quad\text{for}\;\;}
\def\forany{\quad\text{for any}\;\;}
\def\foreach{\quad\text{for each}\;\;\;}
\def\et{\quad\text{and}\quad}
\def\Biint{{\mathrm{B}}\!\!-\!\!\!\!{\int{\!\!\!\!\!\!}\int}}
\def\VViint{{\mathrm{R}}\!{\mathrm{V}}\!{\mathrm{V}}\!\!-\!\!\!{\int{\!\!\!\!\!}\int}}
\def\Ber{\operatorname{Ber}} 

\def\coshsl{/\kern-0.6em{\mathrm{C}}}
\def\sinhsl{/\kern-0.54em{\mathrm{S}}}

\def\odotf{\odot} 
\def\odotb{\odot} 
\def\odots{\underset{\tiny{\mathrm s}}{\odot}}
\def\erf{\mathrm{erf}}
\numberwithin{equation}{section}
\newtheorem{definition}{Definition}[section] %{chapter}
\newtheorem{corollary}{Corollary}[section] %{chapter}
\newtheorem{proposition}{Proposition}[section] %{chapter}
\newtheorem{theorem}{Theorem}[section] %{chapter}
\newtheorem{lemma}{Lemma}[section] %{chapter}
\newtheorem{exercise}{Exercise}[section] %{chapter}
\newtheorem{axiom}{Axiom}[section] %{chapter}
\newtheorem{claim}{Claim}[section] %{chapter}
\newtheorem{remark}{Remark}[section] %{chapter}
\newtheorem{example}{Example}[section] %{chapter}
\newtheorem{comparison}{Comparison}[section] %{chapter}
\newtheorem{problem}{Problem}[section] %{chapter}
\newtheorem{question}{Question}[section] %{chapter}
\def\Hess{\operatorname{Hess}}
\def\proj{\operatorname{proj}}
\def\dist{\operatorname{dist}}
\def\ess.sup{{\operatornamewithlimits{ess.sup}}}
\def\Mat{\operatorname{Mat\,}}
\def\tr{\operatorname{tr\,}}
\def\sgn{\operatorname{sgn\,}}
\def\GL{\operatorname{GL}}
\def\arg{\operatorname{arg}}
\def\vol{\operatorname{vol}}
\def\Eta{\varUpsilon}
\def\sdet{\operatorname{sdet\,}}
\def\str{\operatorname{str\,}}
\def\dive{\operatorname{div}}
\def\Pf{\operatorname{Pf\,}}
\def\Ai{\operatorname{Ai\,}}
\def\trp{{}^{t\!}}
\def\ind{\operatorname{ind}}
\def\Ker{\operatorname{Ker}}

\def\CSS{{\mathcal C}_{\mathrm{SS}}}
\def\CSSC{{\mathcal C}_{\mathrm{SS,0}}}
\def\CSSE{{\mathcal C}_{\mathrm{SS,ev}}}
\def\SSS{{\mathcal S}_{\mathrm{SS}}}
\def\BSS{{\mathcal B}_{\mathrm{SS}}}
\def\DSS{{\mathcal D}_{\mathrm{SS}}}
\def\ESS{{\mathcal E}_{\mathrm{SS}}}
\def\COCSS{{\mathcal O}_{\mathrm{C,S\!S}}}
\def\ccsl{{/\kern-0.5em{\mathcal C}}}
\def\cbsl{{/\kern-0.65em{\mathcal B}}} 
\def\cssl{{/\kern-0.6em{\mathcal S}}} 
\def\cdsl{{/\kern-0.7em{\mathcal D}}} 
\def\cesl{{/\kern-0.58em{\mathcal E}}} 
\def\clsl{{/\kern-0.6em{\mathcal L}}} 
\def\chsl{{/\kern-0.65em{\mathcal H}}} 
\def\cpsl{{/\kern-0.6em{\mathcal P}}} 
\def\comsl{{/\kern-0.5em{\mathcal O}}_{M}}
\def\sinsl{\sinhsl} 
\def\cossl{\coshsl}

\def\unbu{\underline{u}}
\def\unbv{\underline{v}}
\def\unbw{\underline{w}}

\def\unbx{\underline{x}}
\def\unbq{\underline{q}}
\def\unbp{\underline{p}}
\def\unbxi{\underline{\xi}}
\def\unbtheta{\underline{\theta}}
\def\unbpi{\underline{\pi}}
\def\unbeta{\underline{\eta}}
\def\unbsigma{\underline{\sigma}}
\def\unby{\underline{y}}

\def\barx{\bar{x}}
\def\barxi{\bar{\xi}}
\def\barq{\bar{q}}
\def\barp{\bar{p}}
\def\bartheta{\bar{\theta}}
\def\barpi{\bar{\pi}}
\def\bargamma{\bar{\gamma}}

\def\bart{\bar{t}}
\def\unbt{\underline{t}}

\def\Hom{\operatorname{Hom}}

\def\clifpin{{\mathchar'26\mkern-9mu {\lambda}}}
\def\spin{\kbar}    
%{{{\scriptstyle /}\kern-0.43em{\slanted s}}} 
\def\kbar{{\mathchar'26\mkern-9mu k}}

\def\euc{{\mathbb R}}
\def\eucn{{\mathbb R}^n}
\def\eucd{{\mathbb R}^d}
\def\eucm{{\mathbb R}^m}
\def\eucN{{\mathbb R}^N}

\def\slim{\operatornamewithlimits{s-lim}}

\def\barX{\bar{X}}
\def\barXi{\bar{\Xi}}

%\inidata
\def\findata{{\barx},{\barxi},{\bartheta},{\barpi}}
\def\mixdata{{\barx},{\unbxi},{\bartheta},{\unbpi}}
\def\mixxdata{{\barx},{\bartheta},{\unbxi},{\unbpi}}

\def\frachi{\frac{\hbar}{i}}
\def\pdq{\frac{\partial}{\partial q}} %??
\def\sign{\operatorname{sign}}

\def\fC{\mathfrak{C}}
\def\fR{\mathfrak{R}}
\def\rev{\mathfrak{R}_{\mathrm{ev}}}
\def\rod{\mathfrak{R}_{\mathrm{od}}}
\def\fraccli{\frac{\clifpin}{i}}
\def\cev{\mathfrak{C}_{\mathrm{ev}}}
\def\cod{\mathfrak{C}_{\mathrm{od}}}
\def\supermn{{\fR}^{m|n}}
\def\supermo{{\fR}^{m|0}}
\def\superon{{\fR}^{0|n}}

\def\dx{\frac{d}{dx}}
\def\dt{\frac{d}{dt}}
\def\ds{\frac{d}{ds}}
\def\pdx{\frac{\partial}{\partial {x}}}
\def\pdt{\frac{\partial}{\partial {t}}}
\def\pdqi{\frac{\partial}{\partial q_i }}
\def\pdqj{\frac{\partial}{\partial q_j }}
\def\pdxij{\frac{\partial}{\partial \xi_j}}
\def\pdxj{\frac{\partial}{\partial x_j}}
\def\cbra(#1){\langle{#1}\rangle}
\def\pdtheta{\frac{\partial}{\partial \theta }}

\def\defeq{\overset{\mbox{def}}{=}}
\def\qed{\square} %{{\begin{flushright}{\square}\end{flushright}}}
\def\Hess{\roman {Hess}}   %{\operatorname{Hess}}
\def\arcsin{\operatorname{arcsin\,}}
\def\cos{\operatorname{cos\,}}
\def\arccos{\operatorname{arccos\,}}
\def\tan{\operatorname{tan\,}}
\def\arctan{\operatorname{arctan\,}}
\def\cotan{\operatorname{cotan\,}}
\def\arccot{\operatorname{arccot\,}}
\def\cosec{\operatorname{cosec\,}}
\def\arccosec{\operatorname{arccosec\,}}
\def\sec{\operatorname{sec\,}}
\def\arcsec{\operatorname{arcsec\,}}
\def\log{\operatorname{log\,}}
\def\diam{\operatorname{diam}}
\def\rank{\operatorname{rank}}
\def\exp{\operatorname{exp}}
\def\supp{\operatorname{supp\,}}

%\frontmatter %%%%%%

\begin{titlepage}
\vspace*{5cm}
\begin{center}
{\Huge{Lectures on Super Analysis}}\\
\vspace{2cm}
\hspace{0cm} {----- \huge{Why necessary and What's that?}}\\
\vspace{2cm}
\hspace{1.5cm} {\Large{Towards a new approach to a system of PDEs}}
\end{center}

\vspace{5cm}
\begin{center}
\hspace{1cm} {\large{e-Version1.5 \qquad \today}} %  2014.10.6, 10.10,}
\end{center}

\vspace{2cm}

\hspace{5cm}
{\huge{By Atsushi INOUE}}

\end{titlepage}

\frontmatter %%%%%%

\baselineskip 16truept 

\section*{Notice: Commencement of a class}
\begin{center}
{{{\color{red}Syllabus}}}\\[2ex] 
{\Large{\color{blue}{Analysis on superspace}}}\\[1ex]
{ {\color{blue}{----- \large a construction of non-commutative analysis}}}
\end{center}

\begin{center}
{{3 October 2008 -- 30 January 2009,\; 10.40-12.10,\;   H114B at TITECH, Tokyo}},\;\;\; A. Inoue
\end{center}

Roughly speaking, RA(=real analysis) means to study properties of (smooth) functions defined on real space, and CA(=complex analysis) stands for studying properties of (holomorphic) functions defined on spaces with complex structure.

On the other hand, we may extend the differentiable calculus to functions having definition domain in Banach space,
for example, S. Lang ``Differentiable Manifolds'' or J.A. Dieudonn\'e ``Treatise on Analysis''.
 But it is impossible in general to extend differentiable calculus to those defined on infinite dimensional Fr\'echet space, because the implicit function theorem doesn't hold on such generally given Fr\'echet space.

Then, if the ground ring (like $\euc$ or ${\mathbb{C}}$) is replaced by non-commutative one, what type of analysis we may develop under the condition
that  newly developed analysis should be applied to systems of PDE or RMT(=Random Matrix Theory).

In this lectures, we prepare as a ``ground ring", Fr\'echet-Grassmann  algebra  having countably many Grassmann generators and we define so-called superspace over such algebra.
On such superspace, we take a space of super-smooth functions as the main objects to study.

This procedure is necessary not only to associate a Hamilton flow for a given $2^d\times 2^d$ system of PDE
which supports to resolve  Feynman's murmur,
but also to make rigorous Efetov's result in RMT.

%\vspace{0.2cm}
{\small
\begin{enumerate}
\item {Feynman's path-integral representation of the solution for Schr\"odinger equation}
\item {Dirac and Weyl equations, Feynman's murmur} 
\item {Why is such new algebra necessary?Differential operator representations of $2^d\times 2^d$-matrices}
\item {Fr\'echet-Grassmann algebra $\mathfrak{R}$ and superspace $\mathfrak{R}^{m|n}$}
\item {Elementary linear algebra on superspace, super-determinant and super-trace, etc}
\item {Differential calculus on superspace; super smooth functions and implicit function theorem, etc}
\item {Integral calculus on superspace; integration by parts, change of variables under integral sign, etc}
\item {Fourier transformations on $\mathfrak{R}^{m|n}$ and its application}
\item {Path-integral representation of a fundamental solution of Weyl equation}
\end{enumerate}
}

Home Page:{\color{blue}{http://www.math.titech.ac.jp/〜inoue/SLDE2-08.html}}\\
\hspace{4cm}({\color{red} Closed after my retirement from TITech at 31th March 2009})

\begin{center}
{\color{red}To audience:}\quad {\color{blue} Please keep not only intellectual curioisity\\ 
{\hspace{5cm}}but also have patience to follow at least 3 lectures.}
\end{center}

\newpage

\section*{For what and why, this lecture note is written} 
I delivered 14 lectures, each 90 minutes,  for graduate students at Tokyo Institute of Technology,
started in Autumn 2008.

Since there is a special tendency based not only on Japanese culture but rather asian aesthetic feeling, students hesitate to make a question during class. 
Reasons about this tendency seems based on modesty and timidity and feeling in his or her noviciate, or more frankly speaking, they are afraid of making a stupid question  before friends which may probably exhibit their ignorance, which stems from their sense of guilty that they haven't been studied sufficiently enough, or bother those appreciation for lectures by stopping by questions.
In general, some one's any questions on  a lecture is very constructive that makes clear those which are not easy to understand for audience but also corrects miss understandings of speaker himself. 

Though ``Instantaneous response towards your uncomfortable feeling'' 
is embodied by very young peoples saying ``Why, Mamy?'', but it seems rather difficult to do so in student society. Even though, please make a question without hesitation in any time. 
Not only primitive stays near radical but also your slight doubt makes slow down the speed of speaker's explanation which gives some time to other students to help consider and to follow up.

To make easy to pose such questions, I prepare pre-lecture note for a week before my lecture and corrected version of it after lecture with answers to questions if possible and necessary, those are posted on my home-page.

This lecture note is translated and revised from them. Especially, I make big revision in Chapter 8 and Chapter 9.

At the delivering time of my lectures, I haven't yet clarified sufficiently the characterization of super-smooth functions but also the definition of integral on superspace which admits naturally the change of variables under integral sign.
Therefore, I change significantly these representations in this notes from those original lectures.
I owe much to my colleague Kazuo Masuda whose responses to my algebra related questions  help me very much not only to my understanding but also to correct some defects in other's publications with counter-examples.
Concerning the characterization of super-smoothness, we need to prepare Cauchy-Riemann equation for them which is deeply connected with the \underline{countably infinite Grassmann generators}. 

In his naive definition of integral, Berezin's formula of change of variables under integral sign holds only for integrand having a compact support. This point is ameliorated when we modify  the method of V.S. Vladimirov and I.V. Volovich~\cite{VV84} or A. Rogers~\cite{rog85-1} which relates to the different recognition of body part of ``super space'' from Berezin.

Leaving from syllabus, I give some application of super analysis to Random Matrix Theory (=RMT) and of SUSYQM=SUper SYmmetric Quantum Mechanics with Witten's index.

I gather some facts which are not explained fully during these lectures in the last chapter named ``Miscellaneous''. \\
(1) I give a precise proof of  Berezin's formula of change of variables under integral sign. I confess Rothstein's paper is not understandable to me, even a question letter to him without response and an explanation in Rogers' book is also outside my scope.\\
(2) Function spaces on superspace and Fourier transformation for supersmooth functions are breifly introduced.\\
(3) As a typical simplest example of application of superanalysis, I give another proof of Qi's result concerning an example of weakly hyperbolic equations, and\\
(4) I give also an example for a system version of Egorov's theorem which begins with Bernardi's question.\\
(5) I mentioned in the lecture that the problem posed by I.M. Gelfand at ICM congress address in 1954 concerning functional derivative equations related to QED or turbulence, which is more precisely explained there, and finally,\\
(6) in his famous paper, E. Witten introduced Supersymmetric Lagrangian, derived from his deformed $d_{\phi}*d_{\phi}
+d_{\phi}d_{\phi}^*$. Here, we derive ``the classical symbol of this operator'' as the supersymmetric extension of Riemannian metric $g_{ij}(q)dq^idq^j$, which is merely my poor interpretation from physicists calculation.\\
(7) As an example where we need the countably infinite number of Grassmann generators, we consider Weyl equation
with electro-magnetic external field. 
Besides whether it is phyisically meanigfull, we solve the Hamilton equation corresponding to this equation by degree of Grassmann generators.

Though, references are delivered at each time in lectures, but I gathered them at the last part.

Finally, this note may be unique since I confess several times ``their arguments are outside of my comprehension''
which are not so proud of but I do so.  Because not only I feel ashamed of my unmatureness of differential geometry and algebra's intuition, but also I hope these confessions encourage to those young mathematicians who try to do some thing new!

\tableofcontents

\mainmatter 

\baselineskip 16truept   

\chapter[A motivation]{A motivation of this lecture note}  %{What is motivation of this study?} 

\section{Necessity of the non-commutative analysis and its benefit }
\subsection{Another basic field?} % except $\euc$ or $\mathbb{C}$}
Is it truely necessary to introduce another ground ring  in analysis except for $\euc$ or $\mathbb{C}$? Why?

In the theory of linear PDEs(=Partial Differential Equations) of scalar type, 
the main problem is to reduce the non-commutativity inherited from the
so-called Heisenberg's uncertainty principle
\begin{equation}
\bigg[\frac\hbar{i}{\frac{\partial}{\partial q_j}}, q_k\bigg]
=\frac\hbar{i}\delta_{jk},\;\;{i.e.}\;\;
\frac\hbar{i}{\frac{\partial}{\partial q_j}}(q_ku(q))
-q_k\frac\hbar{i}{\frac{\partial}{\partial q_j}}u(q)=\frac\hbar{i}\delta_{jk}u(q)
\label{HR}
\end{equation}
to the one where commutative algebraic calculation is available.
This is done using Fourier transformations, that is, 
the non-commutativity caused by the Heisenberg's uncertainty principle,
is reduced to the commutative one with error terms
on the phase space by Fourier transformation, and then
this transformed one is analyzed there, and by the inverse Fourier transformation, 
it is transformed back to the original setting.
This procedure is done modulo error terms with suitable estimates. These ideas and various devices are unified as the theory of $\Psi$DO(=Pseudo Differential Operators) and FIO(=Fourier Integral Operators).

\underline{Whether this strategy is extendable also to a system of PDEs}, is our main concern.

Since there exists another non-commutativity stems from matrices coefficients for a system of PDEs, 
it seems difficult to treat it as similar as scaler cases.
But if we may diagonalize that system nicely then we may apply the standard method to its each component. 
Even if it is hard to diagonalize straightforwordly, then we impose certain conditions 
on the characteristic roots associated to that system
in order to assure that we may essentially reduce that system 
to the scalar pseudo differential operators.
But if this procedure fails, is there any detour?
Especially, if we need a ``Hamilton flow'' for the given system, how do we associate
that classical objects keeping matrix structure as it is? %without diagonalization

On the other hand, if the phenomenon is describable only
using a system of PDEs, it seems natural to adandone 
the idea of reducing it to the scalar case.
Of course, treating that system of PDEs, 
we need new idea to overcome the non-commutativity of matrices.

This difficulty is clearly claimed by Feynman where he asks what is the corresponding classical mechanics and action integral for Dirac equation. Moreover, he proposes to use quarternion to resolve this difficulty.

Here, {we} propose a new idea to overcome the non-commutativity of matrices.
This idea is essentially simple when we encounter $2^d\times 2^d$-matrices: Since those matrices are decomposed with elements in Clifford algebras and that algebras has the representation by differential operators on Grassmann algebras, {we} extend the ground ring to the one having Grassmann character. Developing analysis on this ground ring, {we} may apply the standard process which are used in the scalar PDE. This idea is based on F.A. Berezin and M.S. Marinov~\cite{BM77} of ``\underline{Treat bosons and fermions on equal}\\
\underline{footing}''.
Therefore, my answer to Feynman's proposal should be,

\begin{center}
`` Mr. Feynman, if you use Fr\'echet-Grassmann algebras\\ with countably infinite Grassmann generators
 instead of querternions, then it goes well!''
\end{center}

\subsection{Feynman's path-integral representation of the solution for Schr\"odinger equation}

More than seventy years ago, as a graduate student, R. Feynman~\cite{fey48} has a primitive question why
Schr\"odinger equation may be considered as the governing equation of quantum mechanics?
In other word, though Bohr's correspondence principle which is derived after many experiments and thoughts, that principle should be essential in Quantum Mechanics, 
but it seems difficult to derive it directly  from the Schr\"odinger equation itself. 

Mathematically, this question is interpreted as follows:
Let $u(t,q):\euc\times\eucm\to{\mathbb{C}}$ satisfy 
the initial value problem for the Schr\"odinger equation  
\begin{equation}
\left\{ 
\begin{aligned}
 & i\hbar
 {\frac{\partial}{\partial t}}
 u(t,q)=
 -{\frac{\hbar^2}{2}}\Delta u(t,q)
 +V(q)u(t,q),\\
 &u(0,q)={\underline{u}}(q).
\end{aligned}
\right.
\label{1022-10}
\end{equation}
How does the solution $u(t,q)$ depend on ${\hbar}$? Especially, can we deduce the Bohr's correspondence principle from this?

On the other hand, about fourty years before when I had been a student, main research subjects developing general theory of linear PDEs are ``existence, uniqueness and regularity'' of solutions for the given equation\footnote{Before advent of functional analytic approach to PDE, rather explicit solution is pursued at that time, therefore it is too hard to obtain a solution for a generally given PDE.}. 
Essential ingredients of these subjects is almost exhaustively studied and collected in L. H\"ormander's book~\cite{Hor83-85}\footnote{These books are not only so volumy to read through but also so difficult to find out problems for doctor thesis. Therefore, I recommend to use them as dictionary, but rather to look his doctor thesis~\cite{hor55} itself. Moreover, as he is a specialist to apply Hahn-Banach extension theorem, reconsider his procedure by using ``constructive extension theorem''?} 
and one of the recent problems is to pick up special properties from governing equation or to represent the solution as explicit as possible by using known objects (for example, R. Beals~\cite{bea98}).
From this point of view, \underline{to make clear the dependence of the solution of}\\
\underline{Schr\"odinger equation on Planck's constant $\hbar$} and to explain mathematically the appearance of Bohr's correspondence principle
is a good starting problem.

Therefore, we begin with retracing the heuristic procedure taken in Feynman's doctor thesis (see also, S.A. Albeverio and R.J. Hoegh-Krohn~\cite{AH-K76})
where he introduced his path-integral representation.

For the right-hand side of \eqref{1022-10}, we define the Hamiltonian operator on $C_0^{\infty}(\eucm)$ as
$$
 \hat{H}=-{\frac{\hbar^2}{2}}\Delta +V(\cdot)=\hat{H}_0+V,\quad 
  \hat{H}_0=\Delta=\sum_{j=1}^m{\frac{\partial^2}{\partial q_j^2}}.
$$
If above $ \hat{H}$ is essentially self-adjoint in $L^2(\eucm)$, applying
Stone's theorem, solution of \eqref{1022-10} is written by
$$
 u(t,q)=(e^{-i{\hbar}^{-1}t \hat{H}}{\underline{u}})(q).
$$
 Or generalizing a little, when and how the exponential function of a given operator $A$
 $$
 e^{tA}=\sum_{k=0}^{\infty}\frac{(tA)^k}{k!}\quad (t\in\euc^+,\;\mbox{or}\; t\in i{\euc})
 $$
 is well-defined? Guiding this problem, Hille-Yosida theory of semigroups is established.
 
{\small [Report problem 1-1]: Check what is the Stone's theorem. If the Hilbert space is finite-dimensional, what is the corresponding theorem in elementary linear algebra?
It is also preferable to check what is the theory of Hille-Yosida.}

On the other hand, Lie-Trotter-Kato's product formula says that if $\hat{H}=\hat{H}_0+V$, 
$ e^{-i{\hbar}^{-1}t \hat{H}} $ is given by
$$
 e^{- i{\hbar}^{-1}t \hat{H}} 
=\slim_{k\to\infty}
 \left( e^{- i{\hbar}^{-1}{\frac tk}V}
e^{- i{\hbar}^{-1}{\frac tk} \hat{H}_0}\right) ^k\quad\mbox{even if}\; [\hat{H}_0,V]\neq{0}.
$$

\begin{remark}
(i) In the above, if $[ \hat{H}_0,V]=0$, then $( \hat{H}_0+V)^k=\sum_{j=0}^k \binom{k}{j}  \hat{H}_0^jV^{k-j}$ and we have $e^{s( \hat{H}_0+V)}=e^{s \hat{H}_0}e^{sV}$, i.e. it isn't necessary to apply above product formula.\\
(ii) 
There doesn't exist the difference between strong and weak  convergence in finite-dimensional
vector spaces. Check the difference between the convergence  of operators in 
``strong'' or ``uniform'' sense in  infinite-dimensional Banach space.
\end{remark}

If the initial data ${\underline{u}}$ belongs to ${\mathcal S}(\eucm)$(=a space of Schwartz' rapidly decreasing functions), where
$$
\begin{aligned}
{\mathcal S}(\eucm)&=\{u\in C^{\infty}(\eucm:{\mathbb{C}})\;|\; p_{k,{\mathcal{S}}}(u)<{\infty},\quad \forall k\in{\mathbb{N}}\}\\
&\mbox{with}\;\;p_{k,{\mathcal{S}}}(u)=
\sup_{q\in\eucm, \ell+|\beta|\le k}\langle q\rangle^{\ell}|\partial_q^{\beta} u(q)|,\;\; \langle q\rangle=(1+|q|^2)^{1/2},
\end{aligned}
$$
since we know
$$
(e^{-i{\hbar}^{-1}t\hat{H}_0}{\underline{u}})({\barq})=
\left( 2\pi i{\hbar t}
\right)^{-m/2}\!\!\int_{\eucm}d{\unbq}\, e^{i{\hbar}^{-1}({\barq}-{\unbq})^2/(2t)}{\underline{u}}({\unbq}),
$$
we have
$$
\begin{aligned}
(e^{-i{\hbar}^{-1}t\hat{H}}{\underline{u}})(\barq)&\sim
(e^{-i{\hbar}^{-1}tV}(e^{-i{\hbar}^{-1}t\hat{H}_0}{\underline{u}}))(\barq)\\
&\sim
\left( 2\pi i{\hbar t}\right)^{-m/2}e^{-i{\hbar}^{-1}tV(\barq)}
%\\!
\int_{\eucm}d\unbq\, e^{i\hbar^{-1}(\barq-\unbq)^2/(2t)}{\underline{u}}(\unbq).
\end{aligned}
$$
Therefore, we get
{\allowdisplaybreaks
\begin{align*}
(e^{-i{\hbar}^{-1}s\hat{H}}&(e^{- i{\hbar}^{-1}t\hat{H}}{\underline{u}}))(\barq)
\sim
\left( 2\pi i{\hbar s}\right)^{-m/2}e^{- i{\hbar}^{-1}sV(\barq)}
\!\!\int_{\eucm}dq^{(1)}\, e^{i{\hbar}^{-1}(\barq-q^{(1)})^2/(2s)}
(e^{- i{\hbar}^{-1}t\hat{H}}{\underline{u}})(q^{(1)})\\
&\sim
\left( 2\pi i{\hbar}\right)^{-m}(ts)^{-m/2}
e^{- i{\hbar}^{-1}sV(\barq)}
\!\!\int_{\eucm}dq^{(1)}\, e^{i{\hbar}^{-1}(\barq-q^{(1)})^2/(2s)}\\
&\qquad\qquad\qquad\qquad\qquad\qquad\qquad\times
\bigg[\,e^{- i{\hbar}^{-1}tV(q^{(1)})}\!\!\int_{\eucm}d{\unbq}\, e^{i{\hbar}^{-1}(q^{(1)}-{\unbq})^2/(2t)}
{\underline{u}}(\unbq)\bigg]\\
&=\left( 2\pi i{\hbar}\right)^{-m}(ts)^{-m/2}\!\!\int_{\eucm}d{\unbq}\\
&\qquad\qquad\times
\bigg[\int_{\eucm}dq^{(1)}\,e^{- i{\hbar}^{-1}(sV({\barq})+tV(q^{(1)}))}
\, e^{i{\hbar}^{-1}({\barq}-q^{(1)})^2/(2s)+i{\hbar}^{-1}(q^{(1)}-{\unbq})^2/(2t)}\bigg]
{\underline{u}}({\unbq}).
\end{align*}}
Putting $t=s$ in the above, we have
$$
\begin{aligned}
- i{\hbar}^{-1}t(V({\barq})+V(q^{(1)}))+&i{\hbar}^{-1}[({\barq}-q^{(1)})^2+(q^{(1)}-{\unbq})^2]/(2t)\\
&=i{\hbar}^{-1}t\bigg[\frac{1}{2}\bigg(\frac{{\barq}-q^{(1)}}{t}\bigg)^2-V({\barq})+\frac{1}{2}\bigg(\frac{q^{(1)}-{\unbq}}{t}\bigg)^2-V(q^{(1)})\bigg].
\end{aligned}
$$
Repeating this procedures $k$-times and denoting $q^{(k)}={\barq},\,\, q^{(0)}={\unbq}$, we define
$$
 S_t(q^{(k)},\dots,q^{(0)})=\sum_{j=1}^k \left[
 {\frac 12}
\bigg({\frac{q^{(j)}-q^{(j-1)}}{t/k}}\bigg)^2-V(q^{(j)})\right]
 {\frac tk}
$$
and we get
$$
 \left( e^{- i{\hbar}^{-1}{\frac tk}V}
e^{- i{\hbar}^{-1}{\frac tk}\hat{H}_0}\right) ^k{\underline{u}}(\unbq)
\sim
\int d{\unbq}\, F_k(t,{\barq},q^{(k-1)},{\cdots},q^{(1)},{\unbq}){\underline{u}}(\barq).
$$
Here, we put
$$
F_k(t,{\barq},q^{(k-1)},{\cdots},q^{(1)},{\unbq})
=(2\pi i{\hbar}(t/k))^{-km/2}
\int\cdots\int dq^{(1)}\cdots
dq^{(k-1)}\, e^{i{\hbar}^{-1}
S_t({\barq},q^{(k-1)},\cdots,q^{(1)},{\unbq})}.
$$
Making $k\to{\infty}$ formally, we have
\begin{equation}
F(t,{\bar{q}},{\underline{q}})=
\slim_{k\to\infty}
(2\pi i{\hbar}(t/k))^{-km/2}
\int\cdots\int dq^{(1)}\cdots
dq^{(k-1)}\, e^{i{\hbar}^{-1}
S_t({\barq},q^{(k-1)},\cdots,q^{(1)},{\unbq})}.
\label{Fint}
\end{equation}
and 
$$
(e^{-i{\hbar}^{-1}t\hat{H}}{\underline{u}})({\barq})=
\int d{\unbq}\, F(t,{\barq},{\unbq}){\underline{u}}({\unbq}).
$$
\begin{comment}
Denoting $q^{(k)}={\barq},\,\, q^{(0)}={\unbq}$ and  putting
$$
 S_t(q^{(k)},\dots,q^{(0)})=\sum_{j=1}^k \left[
 {\frac 12}
\bigg({\frac{q^{(j)}-q^{(j-1)}}{t/k}}\bigg)^2-V(q^{(j)})\right]
 {\frac tk}
$$
and
\begin{equation}
F(t,{\barq},{\unbq})=
\slim_{k\to\infty}
(2\pi i{\hbar t})^{-km/2}
\int\cdots\int dq^{(1)}\cdots
dq^{(k-1)}\, e^{i{\hbar}^{-1}
S_t({\barq},q^{(k-1)},\cdots,q^{(1)},{\unbq})},
\label{Fint}
\end{equation}
we have
$$
(e^{-i{\hbar}^{-1}t\hat{H}}{\underline{u}})(\barq)\sim
\int d\unbq\, F(t,\barq,\unbq){\underline{u}}(\unbq).
$$
\end{comment}

{\small [Report Problem 1-2]: Show that the function space ${\mathcal S}(\eucm)$ forms a Fr\'echet space.}

{\bf Feynman's interpretation}: The set of ``paths'' is denoted by
$$
\begin{gathered}
C_{t,{\barq},{\unbq}}=\{\gamma(\cdot)\in AC([0,t]:\eucm)\,|\,
\gamma(0)={\unbq},\gamma(t)={\barq}\},\\
C_{t,loop}=\{\phi(\cdot)\in AC([0,t]:\eucm)\,|\,
\phi(0)=\phi(t)\},
\end{gathered}
$$
where AC stands for  absolute continuity.
In this case, for any $\gamma\in C_{t,{\barq},{\unbq}}$, we have
$$
C_{t,{\barq},{\unbq}}=\gamma+C_{t,loop}.
$$
For example, take as $\gamma$ the straight line combining ${\unbq}$ and ${\barq}$ such that 
$\gamma_{sl}=\gamma_{sl}(s)=(1-\frac{s}{t}){\unbq}+\frac{s}{t}{\barq}$. By connecting two paths and adjusting time scale, we may define the sum operation in $C_{t,loop}$ which makes it linear space.

We get a Lagrange function $L(\gamma,\dot{\gamma})$ from a Hamilton function $H(q,p)$ by Legendre transform with  a certain convexity;
$$
L(\gamma,\dot\gamma)
={\frac 12}{\dot\gamma}^2-V(\gamma)\in
C^\infty(T\eucm).
$$
For any path $\gamma\in C_{t,{\barq},{\unbq}}$, regarding 
$S_t(q^{(k)},\dots,q^{(0)})$ as a Riemann sum of an action function $S_t(\gamma)$, we get
$$
 S_{t}(\gamma)=\int_0^t d\tau \,
L(\gamma(\tau),\dot\gamma(\tau))
=\lim_{k\to\infty} S_t(q^{(k)},\cdots,q^{(0)}).
$$
Making $k\to\infty$, we ``construct" a limit of measures $dq^{(1)}\cdots dq^{(k-1)}$
$$
D_F\gamma=\prod_{0<\tau < t}d\gamma(\tau)
$$
which is regarded as ``the measure" on the path space $C_{t,{\barq},{\unbq}}$:
$$
F(t,{\barq},{\unbq})
=\int_{C_{t,{\barq},{\unbq}}}D_F\gamma \,e^{ i\hbar^{-1}\int_0^t d\tau\,
L(\gamma(\tau),\dot\gamma(\tau))}.
$$
Then, if we could apply the stationary phase method to this representation when $\hbar\to0$,
we got the main term which is obtained from the classical path $\gamma_c$, i.e.
$$
\delta \int_0^t d\tau\,
L(\gamma(\tau),\dot\gamma(\tau))=\frac{d}{d\epsilon}
 \int_0^t d\tau\,
L((\gamma_c+\epsilon\phi)(\tau),(\dot\gamma_c+\epsilon\dot\phi)(\tau))\big|_{\epsilon=0}
=0
\quad\for \forall \phi\in C_{t,loop}
$$
In this sense, 
\underline{Bohr's correspondence principle} is derived! (Probably, Feynman yelled with delight\\
``\underline{I did it!}'' ?).
\par
The \underline{obstruction of this beautiful expression}
is the claim that
``\underline{There doesn't exist a non-trivial}\\
\underline{Lebesgue-like measure  on any infinite-dimensional barreled locally convex vector space}''\footnote{Though to construct Lebesgue's integration theory, we are taught to prepare measure theory but is it truelly necessary to do so? For example, Berezin integral below works without measure.}.

{\small [Report Problem 1-3 (Campbell-Hausdorff's formula and its application)]:\\
(1) Search ``Campbell Hausdorff'' in Google and check what it is.\\
(2) Apply that formula to $e^{t{\mathbb{X}}}$ where
$$
{\mathbb{X}}=\begin{pmatrix}
0&\mu&1&0\\
-\mu&0&0&1\\
\Omega^2-\mu^2&0&0&\mu\\
0&\Omega^2-\mu^2&-\mu&0
\end{pmatrix}
$$
and get the concrete expression. Don't use the diagonalization procedure but apply 
Campbell-Hausdorff formula to the suitable decomposition of $X$. This matrix is derived from the 
Hamiltonian mechanics, 
for Lagrangian function $L$ below, which is called Bateman-model.
\begin{equation}
L(q,\dot{q})=\frac{1}{2}(\dot{q}_1^2+\dot{q}_2^2)+\mu(q_1\dot{q}_2-q_2\dot{q}_1)
+\frac{\Omega^2}{2}(q_1^2+q_2^2)\in C^{\infty}(T\euc^2:\euc).
\label{B-L}
\end{equation}
(3) Search also ``Lie-Trotter-Kato formula''.}

{\small [Report Problem 1-4]:  What is the meaning of AC function, what property it shares?}

\subsection{Non-existence of Feynman measure}
 To ``feel'' the reason why there doesn't exist Lebesgue-like measure (called Feynman measure), we give
 a simple theorem due to H.H. Kuo~\cite{Kuo75}. 
 Since that theorem is formulated in Hilbert space and the path space $C_{t,loop}$ is not
 Hilbert one, those who don't satisfy this explanation, consult the paper 
 by O.G. Smolyanov and S.V. Fomin~\cite{SF76}.

For the sake of those who forget teminology, we recall the following:
\begin{definition}[Complete $\sigma$-algebra]
For a given space $X$, a subset $\mathcal{B}$ of all subsets ${\mathcal{P}}^X$ satisfying
\begin{itemize}
\item $\emptyset\in{\mathcal{B}}$,
\item $A\in\mathcal{B} \Longrightarrow  A^c=X\setminus A\in\mathcal{B}$,
\item $A_n\in\mathcal{B}\Longrightarrow  \sum_{n=1}^{\infty}A_n\in\mathcal{B}$
\end{itemize}
is called \underline{complete $\sigma$-algebra}.
\end{definition}
\begin{definition}[measure]
A set function $\mu$ defined on a complete $\sigma$-algebra $\mathcal{B}$ of a space $X$ is called a \underline{measure} if it satisfies
\begin{itemize}
\item $0\le\mu(A)\le{\infty},\quad \mu(\emptyset)=0$,
\item $A_n\in\mathcal{B},\; A_j\cap A_k=\emptyset\;(j\neq k)\Longrightarrow 
\mu\big(\sum_{n=1}^{\infty}A_n\big)=\sum_{n=1}^{\infty}\mu(A_n)$.
\end{itemize}
\end{definition}
\begin{definition}[Borel-algebra]
A family $\mathcal{B}$ of sets  of a topological space $X$ is called
a \underline{Borel-algebra} and denoted $\mathcal{B}=\mathcal{B}(X)$ if it satisfies
\begin{itemize}
\item $A\in\mathcal{B} \Longrightarrow  A^c=X\setminus A\in\mathcal{B}$,
\item $A_n\in\mathcal{B} \Longrightarrow  \sum_{n=1}^{\infty}A_n \in\mathcal{B}$,
\item $\mathcal{O}(X)\subset\mathcal{B}$,
\item $\mathcal{B}$ is the minimum in ${\mathcal{P}}^X$ for the ordering by the set  inclusion.
\end{itemize}
\end{definition}
\begin{definition}
A Borel measure\footnote{measure defined on Borel algbra} satifying below is called Lebesgue-like:
\newline
(1) For any bounded Borel set, its measure is not only finite, but also positive if a set is not empty.
\newline
(2) That measure is translation invariant\footnote{assume the translation is defined on that topological space $X$ }.
\end{definition}
\begin{theorem} \label{Theorem 2.1}
There exists no non-trivial Lebesgue-like Borel measure on a inifinite dimensional separable Hilbert space.
\end{theorem}

{\it Proof}. Since $H$ is separable, there exists a countable orthonormal base $\{e_1,e_2,\cdots \}$\footnote{Hilbert-Schmidt's procedure of orthogonalization holds for countable number of bases}.

Assume that there exists a non-trivial Lebesgue-like Borel measure $\mu$ on ${\mathcal B}(H)$.
Define open sets as
$$
B_n=\{u\in H\,\big|\, \Vert u-e_n\Vert<\frac 12\} \quad \mbox{and}\quad
B=\{u\in H\,\big|\, \Vert u\Vert<2 \},
$$
then they satisfy 
$$
B_n\cap B_m=\emptyset \quad\text{and}\quad \cup_{n=1}^{\infty}B_n\subset B.
$$
Since the measure is Lebesgue-like, we have
$$
0<\mu(B_1)=\mu(B_2)=\cdots <\infty, \quad
\infty =\sum_{n=1}^\infty \mu(B_n) \le \mu(B) <\infty.
\quad\text{Contradiction!} \qquad\qquad\qed
$$ 

{\small  [Report Problem 1-5]:
 What occurs if bases has continuous cardinality? By the way, check whether there exist non-separable Hilbert space.
Check also the basis problem in general Banach space.
}

\begin{remark}
Recently, I recognized very radical idea from Hung Cheng, a Professor of Applied Mathematics in the theoretical physics group of MIT, he is a phisicist having job in math.department: He claimed in~\cite{che92},
\begin{quote}
The path integration approach is not only heuristic and non-rigorous; worse, it often leads to erroneous results when applied to non-Abelian gauge field.
\end{quote}
\end{remark}

\begin{remark}[Note added:2014.11]
Though \underline{full Feynman measure doesn't exist}\footnote{recall also, \underline{there doesn't exist `` functor'' called full quantization}, see R. Abraham and J.E. Marsden~\cite{AM80}}, the objects represented formally using path-integration should be carefully researched. How to make rigorous the partial differentiation in ``path-integral category'' is now under-construction by
Fujiwara~\cite{fuj13}, N. Kumano-go~\cite{n-Kum04}.Their trials are done in $(1+0)$-dimension, 
how to generalize it to $(1+d)$-dimensional case contains many interesting problem such as time-slicing should be replaced by something-like finite elements method corresponding to triangulation in topology, etc?.
\par
Not as integrand by measure, but something new, my concern is how one can prove that this object is a solution of Functional Differential Equations. See, the last chapter!
\end{remark}

\subsection{Resume of known procedures}
Assuming a certain convexity to apply Legendre transform, we have
$$
\begin{gathered}
\mbox{\fbox{Lagrange Mechanics}}{\longleftrightarrow}\mbox{\fbox{Hamilton Mechanics}}\\
L(\gamma,\dot\gamma)\overset{\mbox{Legendre transform}}{\longleftrightarrow}H(q,p).
%\quad\mbox{assuming a certain convexity}
\end{gathered}
$$

\fbox{\bf Classical Mechanics} 
$$
\mbox{\fbox{Hamilton equation}}\;
\left\{
\begin{aligned}
&\dot q=H_p(q,p),\\
&\dot p=-H_q(q,p),
\end{aligned}
\right.
\with
\begin{pmatrix}
q(0)\\
p(0)
\end{pmatrix}=
\begin{pmatrix}
\underline{q}\\
\underline{p}
\end{pmatrix},
$$
$$
\mbox{i.e.}\quad\dt\begin{pmatrix}
q\\
p
\end{pmatrix}
={\mathbb J}\begin{pmatrix}
H_q\\
H_p
\end{pmatrix}
\with
{\mathbb J}=\begin{pmatrix}
0&1\\
-1&0
\end{pmatrix}.
$$

$$
\mbox{\fbox{Liouville equation}}\;
\dot\phi=\{\phi, H\}=\sum_{j=1}^m
\bigg(\frac{\partial \phi}{\partial q_j}\frac{\partial H}{\partial p_j}
-\frac{\partial \phi}{\partial p_j}\frac{\partial H}{\partial q_j}\bigg)
\with \phi(0,q,p)=\underline{\phi}(q,p).
$$

\fbox{\bf Quantum Mechanics}
$$
\begin{CD}
\mbox{\fbox{Liouville equation}} @>\mbox{quantization}>>\mbox{\fbox{Heisenberg picture}}\\
@V{}V{}V  
@V{}V{}V\\
\mbox{\fbox{Hamilton equation}} @>\mbox{quantization}>>\mbox{\fbox{Schr\"odinger picture}}\\
\end{CD}
$$

\qquad $``L(\gamma,\dot{\gamma})$ or $H(q,p)\to \hat{H}=\hat{H}(q,-i{\hbar}\partial_q)''$

(S) A description of the movement of the state vector $u(t)$ w.r.t. time $t$:
$$
\mbox{\fbox{Schr\"odinger picture}}\quad
i\hbar\frac{\partial u(t)}{\partial t}=\hat{H}u(t)\with u(0)=\underline{u},
$$
 i.e. $u(t)=e^{-i\hbar^{-1}t\hat{H}}\underline{u}$.

(H) A description of the change of the kinetic operator $\hat{F}(t)$ w.r.t. time $t$:
$$
\mbox{\fbox{Heisenberg picture}}\quad
i\hbar\dt\hat{F}(t)=[\hat{F}(t), \hat{H}]\with \hat{F}(0)=\hat{\underline{F}}.
$$

(F) Path Integral method, clarifying Bohr's correspondence principle:

$$
\mbox{\fbox{Feynman picture}}\quad
u(t,q)=\int d\unbq\,E(t,0,q,{\unbq}){\unbu}(\unbq)
$$
with
$$
E(t,0,q,{\unbq})=\int_{C_{t,q,{\unbq}}}D_F\gamma\, 
\exp\big\{i\hbar^{-1}\int_0^t ds\,L(\gamma(s),\dot\gamma(s)) \big\}.
$$
Here,
$$
C_{t,q,{\unbq}}=\{\gamma\in C([0,t]:\euc^d)\;|\; \gamma(0)={\unbq},\; \gamma(t)=q\}
$$
and
$$
E(t,0,q,{\unbq})\sim D(t,0,q,{\unbq})^{1/2}e^{i\hbar^{-1}S(t,0,q,{\unbq})}\sim \delta_{q}(\unbq).
$$

\begin{problem}
Give a meaning to the symbolic representation
$$
\int D_F\gamma \,\exp{\{{i\hbar^{-1}\int_0^t 
L(\gamma(\tau),\dot\gamma(\tau))d\tau}\}}
$$
for a wider class of Lagrangian $L$.\\
(0) Concerning this question, D. Fujiwara~\cite{fuj79}
 gives a rigorous meaning without notorious measure when
the potential $V$ satisfies $|\partial_x^\alpha V(x)|\le C_\alpha$$\,(|\alpha|\ge 2)$.\\
%\newline
(i) For the Coulomb potential $V(q)=1/|q|$, i.e. hydrogen atom, because of the singularity, we have not yet established\footnote{Rather recently, I find a paper by C.Grosche~\cite{gro93} where he claims this problem is solved by path integral method. But from his explanation, seemingly, we don't have clearly the corresponding principle from their representation(2015.1.20)} the analogous result as Fujiwara. 
\begin{quotation}
(a) I propose to calculate this by replacing $1/|q|$ with $1/(|q|^2+\epsilon)^{1/2}$ for any $\epsilon>0$ and finally making $\epsilon\to 0$, or\\
(b) Use the fact that Schr\"odinger equation with $3$-dimensional Coulomb potential is obtained form $4$-dimensional harmonic oscillator(See, for example, N.E. Hurt~\cite{Hur83}).
\end{quotation}
%\newline
(ii) At least in dimension $1$, the essential selfadjointness of $-\Delta+|q|^4$ is proved by many methods (see, M. Reed and B. Simon vol I~\cite{RS72-79}). But we might not apply the procedure used by Fujiwara to construct a parametrix using classical quantities
(but, see, S. Albeverio and S. Mazzucchi~\cite{AM05}).\\
%\newline
(iii) How do we proceed when there exists many paths connecting points $q$ and $q'$ like
the dynamics on the circle or sphere (see, L. Schulman~\cite{schul68})? and\\
%\newline
(iv) When $|\partial_q^\alpha V(q)|\le C_\alpha$$\,(|\alpha|\ge 2)$, the above constructed parametrix converges in \underline{uniform operator}\\
\underline{norm}. On the other hand, Lie-Trotter-Kato product formula assures only for the \underline{strong convergence}. How can one express the reason for this difference?
In case of using polygonal line approximation for classical path to the harmonic oscillator, we get the strong but non-uniform convergence of parametrices.
One possibilty may to use non-standard analysis to check why there exists the difference of the convergence.
\end{problem}
\par
{\small [Report Problem 1-6]: What is the meaning of essential adjointness? Check~\cite{RS72-79}!}
\begin{problem}
Fujiwara adopted the Lagrangian formulation in his procedure, stressing without Fourier transform.
Does there exist the Hamiltonian object corresponding to this parametrix? (see for example, A. Intissar~\cite{int82} and A. Inoue~\cite{ino99}):
$$
\iint D_Fq\,D_Fp\,\exp{\{i\hbar^{-1}\int_0^t d\tau\,[\dot{q}(\tau)p(\tau)-H(q(\tau),p(\tau))\}}\, ?
$$
(Added August 2015): Concerning this problem, I find an interesting and important results by N. Kumano-go and D. Fujiwara~\cite{KF08}. They try to define ``path integral'' and corresponding calculation without ``measure'' and succeed it at least partially.
\end{problem}

\subsection{Feynman's murmur}
In p. 355 of their book~\cite{FH65},
Feynman wrote as follows (underlined by the author):
\begin{quote} 
$\cdots $ \underline{path integrals suffer grievously from a serious defect}.
They do not permit a discussion of spin operators or other such operators 
in a simple and lucid way.
They find their greatest use in systems for which coordinates and their
conjugate momenta are adequate.
Nevertheless, spin is a simple and vital part of real quantum-mechanical
systems.
It is a serious limitation that the half-integral spin of the electron
does not find a simple and ready representation.
\underline{It can be handled if the amplitudes and quantities are considered 
as quarternions}
instead of ordinary complex numbers,
but the lack of commutativity of such numbers is a serious complication.
\end{quote}

{\bf Main Problem}: \underline{How do we treat this murmur as a mathematical problem?}

Though for a given Schr\"odinger equation, we may associate a corresponding classical mechanics,
but how do we define  the classical mechanics corresponding to Dirac or Weyl equations?
In other word, since Schr\"odinger equations is obtained from Lagrangian or Hamiltonian function by quantization, can we define a Hamiltonian function from which we get Dirac equation after quantization?

One of my objects to this lecture note, is to answer this main problem affirmatively by preparing new tools and giving sketchy explanation. There exists at least two problems for this:\\
(1) How to define classical mechanics to Dirac or Weyl equations, or more generally for $2^d\times 2^d$ systems of LPDE?\\
(2) Like Dirac or Weyl equations who have only first order derivatives in space variables, it seems impossible, even if there exist Hamilton equations, to assign initial and final positions in configuration space as is done in Schr\"odinger equation. How to get rid of this?

Finally, my answer is affirmative, it is possible with not only using superspace formulation but also re-interpreting the method of characteristics by Hamilton flow and Fourier transformation.

\vspace{5mm}

\begin{center}

============   Mini Column 1:  Stationary phase method ============

\end{center}

Consider the integral with parameter
$$
I(\omega)=\int dq\,u(q) e^{i\omega \phi(q)}. 
$$
Study the asymptotic behavior of $I(\omega)$ when $\omega\to\infty$.  
Remembering Riemann-Lebesgue lemma, it seems natural to imagine the following  fundamental fact holds.
\begin{lemma} Let $\phi\in C^{\infty}(\euc^d:\euc)$ and $u\in C_0^{\infty}(\euc^d:\euc)$. Then,
$$
\phi'\neq{0}\quad\mbox{on}\quad \supp u \Longrightarrow I(\omega)=0(\omega^{-k})\;\when \omega\to{\infty}.
$$
\end{lemma}
This is a fundamental fact for the stationary phase method. Therefore, further study is to study the behavior when 
``$\phi'\neq{0}$ on $\supp u$''.
A typical answer for this is given
\begin{theorem}[Theorem 7.7.5, p.220 of H\"ormander I of ~\cite{Hor83-85}]
Let $K$ be a compact set of $\euc^d$, $X$ an open neighborhood of $K$ and $k$ a positive integer.
If $u\in C_0^{2k}(K)$, 
$\phi\in C^{3k+1}(X)$ and $\Im \phi\ge 0$ on $X$ and let there exists a point $q_0\in K$ such that $\Im \phi(q_0)=0$, $\phi'(q_0)=0$, $\det\phi''(q_0)\neq{0}$ and $\phi'\neq{0}$ on $K\setminus\{q_0\}$.
Then, we have
$$
\left| \int_{{\euc}^d} dq\, u(q)e^{i\omega \phi(q)}-e^{i\omega \phi(q_0)}(\det(\omega \phi''(q_0)/(2\pi i))^{-1/2}\sum_{j<k}\omega^{-j}L_j u\right|
\le C\omega^{-k}\sum_{|\alpha|\le 2k}\sup|D^{\alpha}u|.
$$
Here, $C$ is bounded when $\phi$ stays in a bounded set in $C^{3k+1}(X)$ and $|q-q_0|/\phi'(q)$ has a uniform bound.
With
$$
g_{q_0}(x)=\phi(q)-\phi(q_0)-\frac{\langle \phi''(q_0)^{-1}(q-q_0), q-q_0\rangle}{2}
$$
which vanishes of third order at $q_0$, we have
$$
L_ju=\sum_{\nu-\mu=j}\sum_{2\nu\ge 3\mu}i^{-j}2^{-\nu}\frac{\langle \phi''(q_0)D,D\rangle^{\nu}(g_{q_0}^{\mu} u)(q_0)}{\mu!\nu!}.
$$
This is a differential operator of order $2j$ acting on $u$ at $q_0$. The coefficients are rational homogeneous functions of degree $-j$ in $\phi''(q_0),\cdots, \phi^{(2j+2)}(q_0)$ with denominator $(\det \phi''(q_0))^{3j}$. In every term the total number of derivatives of $u$ and of $\phi''$ is at most $2j$.
\end{theorem}

\begin{remark}
\underline
{In mathematics society, it is regarded as $some one's$-conjecture if the statement}\\
\underline{ ``This is the main term'' goes without precise estimates of error terms}. But in papers of mathematical physics, seemingly there is not so many with estimating ``error terms''.
For example, the famous paper of E. Witten~\cite{witt82-1} doesn't have estimates of ``so-called small terms'' with precise calculation. Or more
frankly speaking, since there doesn't exist Feynman measure, the representation using such measure seems a castle in the air, though it shows us the goal or a dream as it is so.
Getting the main terms without error estimates, you may proceed very algebraically and geometrically, and you may have something-like solution, but it doesn't  mean its conclusion is true!  Even if you may have experiments based on that calculation and if you may claim the data is inside measurement error, how you may assure the theory is correct even in mathematical sense!
\end{remark}

\begin{center}  

  ======= End of Mini Column 1 =======

\end{center}

\subsection{Fujiwara's procedure}
 Since there doen't  exist the so-called Feynman measure which guarantees the beautiful path-integral expression, how do we represent the solution of the Schr\"odinger equation?

As the operator
$$
\hat{H}=\hat{H}(q,-i{\hbar}\partial_q)=-\frac{{\hbar}^2}{2}\Delta+V(q)
$$
is essentially self-adjoint on $L^2(\euc^m)$ under certain conditions on $V$,
there exists a solution $e^{i{\hbar}^{-1}t\hat{H}}{\underline{u}}$ (by Stone's theorem)  
of the initial value problem 
$$
i{\hbar}\frac{\partial u(t,q)}{\partial t}=\hat{H}u(t,q)\with u(0,q)={\unbu}(q).
$$
Moreover, by L. Schwartz's kernel theorem, we have a kernel $E(t,q,q')\in {\mathcal{D}}'(\euc\times\euc^m\times\euc^m)$ such that
$$
\langle e^{i{\hbar}^{-1}t\hat{H}}{\underline{u}},\varphi\rangle
=\langle E(t,q,q'){\unbu}(q'),\varphi(q)\rangle=\langle E(t,q,q'),{\unbu}(q')\varphi(q)\rangle
=\langle E(t,{\cdot},{\cdot}),{\unbu}\otimes\varphi\rangle.
$$
On the other hand, for the heat case $e^{t\hat{H}}{\underline{v}}$, the distributional kernel $H(t,q,q')$ has the representation by the ``classical quantities''?

\paragraph{Method of Fujiwara:}
About 30 years before, there doesn't exist a paper on the construction of a fundamental solution 
for the initial value problem of Schr\"odinger equation.
Fujiwara adopts the argument of Feynman modifying mathematically.
\par
(1) 
For given Lagrangian 
$L(\gamma,\dot{\gamma})=\frac{1}{2}|\dot{\gamma}|^2-V({\gamma})\in C^{\infty}(TM)$$(M=\euc^m)$,
by Legendre transform, we have the Hamilton function 
$\displaystyle{H(q,p)=\inf_{\dot{q}}[\dot{q}p-L(q,\dot{q})]\in C^{\infty}(T^*M)}$.
\par
(2) For the Hamilton function $H(q,p)=\frac{1}{2}|p|^2+V(q)$,
we construct a solution $S(t,q,{\unbq})$ of the Hamilton-Jacobi equation
$$
S_t(t,q,{\unbq})+H(q,S_q(t,q,{\unbq}))=0\with
\lim_{t\to0}tS(t,q,{\unbq})=\frac{1}{2}|q-{\unbq}|^2.
$$
\par
(3) For the action function $S(t,q,{\unbq})$ obtained above,
the amplitude function\footnote{\underline{How to recognize Feynman's idea of ``put equal weight for each path''}, I feel some difference between Fujiwara's idea and mine} defined by
$$
D(t,q,{\unbq})=\det\bigg(\frac{\partial^2 S(t,q,{\unbq})}{\partial q \partial{\unbq}}\bigg)\quad(\mbox{van Vleck determinant})
$$
satisfies the continuity equation
$$
D_t(t,q,{\unbq})+\partial_q(D(t,q,{\unbq})H_p(q,S_q(t,q,{\unbq})))=0\with
\lim_{t\to0}D(t,q,{\unbq})=1.
$$
\par
(4) Then we define the integral transformation
\begin{equation}
F(t){\underline{u}}(q)=(2\pi i{\hbar})^{-m/2}\int_{\eucm} d{\unbq}\, D(t,q,{\unbq})^{1/2}e^{i{\hbar}^{-1}S(t,q,{\unbq})}{\unbu}({\unbq}).
\label{parametrix}
\end{equation}

\begin{theorem}[{Theorem 2.2 of Fujiwara~\cite{fuj79}}] 
Assume $
\displaystyle{\sup_{q\in\eucm}|D^\alpha V(q)|\le C_\alpha}$$\,(|\alpha|\ge2)$.
Fix $0<T<\infty$ arbitrarily.
Put ${\mathbb H}=L^2(\eucm:{\mathbb C})$,
${\mathcal B}({\mathbb H})$=the set of bounded linear operators on ${\mathbb H}$.
\par
(i) $F(t)$ defines a bounded linear operator in ${\mathbb H}$
%That is, there exists a constant $C$ depending on $T$ such that 
%for any $ u\in L^2(\eucm:{\mathbb C})$ and $ t\in [-T,T]$,
$$
{\Vert} F(t) u {\Vert}\le C {\Vert} u {\Vert} \quad\text{by Cotlar's lemma}.
$$ 
\par
(ii) For any $ u\in L^2(\eucm:{\mathbb C})$, $t,\,s,\,t+s\in [-T,T]$, 
%it satisfies the following:
$$
\begin{gathered}
\lim_{t\to 0} {\Vert}F(t)u-u {\Vert}=0 ,\\
{i\hbar}\left.{\frac {\partial}{\partial t}}
 (F(t)u)(q)\right|_{t=0}
=\hat{H}(q,-i{\hbar}\partial_q)u(q),\\
{\Vert} F(t+s)-F(t)F(s) {\Vert}
\le
C({t^2}+{s^2}) .
\end{gathered}
$$
\par
(iii) Moreover, there exists a limit 
$\lim_{k\to\infty}( F({t/k}))^k=E(t)$
in ${\mathcal B}({\mathbb H})$,
i.e. in the operator norm of $L^2(\eucm:{\mathbb C})$, 
which satisfies 
the initial value problem below:
$$
\left\{
\begin{aligned}
&{i\hbar}{\frac {\partial}{\partial t}}
(E(t)u)(q)=\hat{H}(q,-i{\hbar}\partial_q)(E(t)u)(q), \\
&(E(0)u)(q)={\underline u}(q).
\end{aligned}
\right.
$$
\end{theorem}
\begin{remark}
The operator $F(t)$ is said to be a parametrix and $E(t)$ or its kernel is called the fundamental solution.
\end{remark}

{\bf {Outline of the proof}}:
In (2), for the construction of a solution of the Hamilton-Jacobi equation, he uses the Jacobi's method.

(a) For the given $H(q,p)$ and the initial data $({\unbq},{\unbp})$, there exists a unique  Hamilton flow
 $(q(s),p(s))=(q(s,{\unbq},{\unbp}), p(s,{\unbq},{\unbp}))$.
  
(b)  For the given time interval $t$ which is sufficiently small, and for any given terminal position ${\barq}$,
applying the implicit function theorem to
${\barq}=q(t,{\unbq},{\unbp})$, we get the unique ${\unbp}$ denoted by ${\unbp}=\xi(t,{\unbq},{\barq})$.

(c) Using this function, we put
 $$
S(t,{\barq},{\unbq})=S_0(t,{\unbq},{\unbp})\big|_{{\unbp}=\xi(t,{\unbq},{\barq})}.
$$
That is, there exists a unique path $\gamma_c$ in $C_{t,{\unbq},{\barq}}$
such that
$$
\inf_{\gamma\in C_{t,{\unbq},{\barq}}} S_t(\gamma)=S_t(\gamma_c)=
{S}(t,{\unbq},{\barq}) \with S_t(\gamma)=\int_0^t d\tau\,L(\gamma(\tau),\dot{\gamma}(\tau)).
$$
Moreover,
this function $S(t,{\barq},{\unbq})$ is a solution of the Hamiltonian-Jacobi equation.

\begin{remark}
By this construction, we have estimates for $\partial_t^{\ell}\partial_{\barq}^{\alpha}\partial_{\unbq}^{\beta}S(t,{\barq},{\unbq})$ with respect to $(t,{\barq},{\unbq})$.
\end{remark}

(3) is proved from (2) algebraically (see, Inoue and Maeda~\cite{IM85} or \cite{IM03} even on superspace).

(4) Since we have estimates of $S(t,{\barq},{\unbq})$ or $D(t,{\barq},{\unbq})$ w.r.t. $({\barq},{\unbq})$, we may prove the $L^2$-boundedness of the operator \eqref{parametrix} applying Cotlar's lemma. 
Since we take $D(t,{\barq},{\unbq})^{1/2}$ as the amplitude, the operator \eqref{parametrix} is considered as acting on the half-density bundle (or the intrinsic Hilbert space) ``$L^2(\euc^m:{\mathbb{C}})$''.
I regard this fact as corresponding to Copenhagen interpretation.

\par
(5)  Though above theorem is sufficient concerning the convergence of parametrix \eqref{parametrix}, but 
this convergence is not sufficient for the Feynman's expression.
Concerning this or the construction of the fundamental solution itself, there exists another paper
by Fujiwara~\cite{fuj80} which isn't discussed in this lecture because I haven't appreciated it fully\footnote{But a part of this problem is considered as the product formula for FIOp on superspace and is treated in Chapter 7 slightly}.

\begin{problem}\label{prob1.1.3}
In the above theorem,  the momentum energy is restricted on the flat Riemannian metric $\frac{1}{2}|p|^2$ on $\euc^m$.
Whether this procedure works for the Riemannian metric is calculated by physicist (see for example, B. DeWitt~\cite{deW84})  and he suggests the desired Laplace-Beltrami operator but with the term $R/12$ where $R$ is the scalar curvature of $g_{ij}(q)dq^idq^j$.
In general, to prove the $L^2$-boundedness of the FIO with suitable phase and amplitude of order $0$,
Fujiwara applied Cotlar' lemma which is formulated in flat space. Technically, we need new device to extend almost orthogonality in case the space is curved.
Therefore, it is an open problem to associate a quantum mechanics
for given Riemann metric $g_{ij}(q)$ on $\euc^m$ following Fujiwara's procedure.

\par
On the other hand, above procedure of Fujiwara was used also 
by Inoue and Maeda~\cite{IM85}
to explain mathematically the origin of the term $(1/12)R$, $R=$the scalar
curvature of the configuration manifold in the heat category,
which appeared when one wants 
to ``quantize with purely imaginary time" the Lagrangian on a curved manifold.
\end{problem}
\begin{problem} 
Feynman or Fujiwara used Lagrangian formulation. 
How do we connect the above procedure 
directly to the Hamiltonian without using Lagrangian?
\end{problem}
%Lastly, we need to remember the method of characteristics
%for the first order PDE and reformulate it for the future use.

\section{The first step towards Dirac and Weyl equations} %Elementary introduction to, primitive
\subsection{The origin of Dirac and Weyl equations}
Why and how does P. Dirac introduce, now so-called, Dirac equation? We modify the description of Nishijima~\cite{Nis73}.

Assume that energy $E$ and momentum $p$ of the free particle with mass $m$ satisfy the Einstein relation\footnote{Remember, for $p=0$, this gives the theoretical foundation of the possibility of atomic bomb!}
$$
E^2=c^2|p|^2+c^4m^2.
$$
Following the canonical quantization procedure of substitution
$$
p_j\longrightarrow \frac{\hbar}{i}\frac{\partial}{\partial q_j},\quad E\longrightarrow i\hbar\frac{\partial}{\partial t}
$$
which we did to get the Schr\"odinger equation, we have the Klein-Gordon equation
$$
{\hbar}^2\frac{\partial^2}{\partial t^2}u-c^2{\hbar}^2\Delta u+c^4m^2u=0.
$$
Unfortunately, the solution $u$ of this equation does not permit the Copenhagen interpretation, that is,
the quantity $\rho=|u|^2$ is not interpreted  as the probability density. In order to get rid of this inconvenience, it is claimed that it is necessary to have the first order derivative w.r.t. time in physics literature.

If this saying is accepted, the simplest prescription is to put
$$
E=\pm c\sqrt{|p|^2+c^2m^2}.
$$
But the right-hand side of above defines a $\Psi$DO, which doesn't have local property\footnote{If $P(q,p)=\sum_{|\alpha|\le N}a_{\alpha}(q)p^{\alpha}$, then roughly speaking, it is quantized as a PDE $P(q,-i\hbar\partial_q)$. Then, it satisfies $\supp P(q,-i\hbar\partial_q)u\subset \supp u$ for any $u\in C_0^{\infty}(\euc^m)$. This is the local property of PDE}.
This gives us a certain conflict if we insist on that the physical law(which is assured by experiments in laboratory system) should satisfy local property. Therefore, it is not so nice to accept such $\Psi$DO with symbol above, as the quantization of Einstein relation.

{\small
 [Report problem 2-1]: For a $\Psi$DO, it has pseudo-local property. Report on this subject.}
 
In order to have the equation which stems from Einstein relation and admits probabilistic interpretation,
we need to have
$$
\big(i\hbar\frac{\partial}{\partial t}-\hat{H}\big)\psi=0 
$$
which satisfies
$$
\big({\hbar}^2\frac{\partial^2}{\partial t^2}+\hat{H}^2\big)\psi=0.
$$
Assuming that this equation coincide with Klein-Gordon equation, we need that ``the symbol corresponding to the operator $\hat{H}$'' should satisfy
$$
H^2=c^2|p|^2+c^4m^2.
$$
Supposing that the state vector $\psi$ which satisfies the desired equation has multicomponents, then
we may have the option such that
$$
H=\sum_{j=1}^3\pmb{\alpha}_j\,p_j+mc^2\pmb{\beta}.
$$
Here, above appeared letters $\{\pmb{\alpha}_j,\,\pmb{\beta}\}$ satisfy
\begin{equation}
\pmb{\alpha}_j\,\pmb{\alpha}_k+\pmb{\alpha}_k\,\pmb{\alpha}_j=2\delta_{jk}{\mathbb{I}},\quad
\pmb{\alpha}_j\,\pmb{\beta}+\pmb{\beta}\,\pmb{\alpha}_j=0,\quad
\pmb{\beta}^2={\mathbb{I}}
\label{CLR}
\end{equation}

Dirac gave an example of $4\times 4$ matrices satisfying the relation \eqref{CLR}, which is now called Dirac matrices:
$$
\pmb{\alpha}_j=\begin{pmatrix}
0&\pmb{\sigma}_j\\
\pmb{\sigma}_j&0
\end{pmatrix}=\pmb\sigma_1\otimes \pmb\sigma_j
,\quad
\pmb{\beta}=\begin{pmatrix}
{\mathbb{I}}_2&0\\
0&-{\mathbb{I}}_2
\end{pmatrix}=\pmb\sigma_3\otimes{\mathbb{I}}_2.
$$
Here, Pauli matrices $\{\pmb\sigma_j\}_{j=1}^3$ are gievn by
\begin{equation}
\pmb\sigma_1=\begin{pmatrix}
0&1\\
1&0
\end{pmatrix},\quad
\pmb\sigma_2=\begin{pmatrix}
0&-i\\
i&0
\end{pmatrix},\quad
\pmb\sigma_3=\begin{pmatrix}
1&0\\
0&-1
\end{pmatrix}
\label{p-matrix}
\end{equation}
satisfying
\begin{equation}
\pmb\sigma_j\pmb\sigma_k+\pmb\sigma_k\pmb\sigma_k=2\delta_{jk}{\mathbb{I}},\quad
\pmb\sigma_1\pmb\sigma_2=i\pmb\sigma_3,\;\;
\pmb\sigma_2\pmb\sigma_3=i\pmb\sigma_1,\;\;
\pmb\sigma_3\pmb\sigma_1=i\pmb\sigma_2.
\label{CLR2}
\end{equation}
 
{\small
 [Report problem 2-2]: There are many representations satisfying \eqref{CLR}, named Majonara or chiral representation, etc.  Seek such representations as many as possible and check the relationship between them.
By the way of checking these, study also the Lorentz invariance.
If such relations are explained by unified manner, using the differential representation point of view, it will be good enough for master thesis, isn't it?}

\begin{problem}\label{DFrep}
For a given external electro-magnetic field, the IVP (=initial value problem) for the Dirac equation is given
as follows:
Find $\psi(t,q):{\mathbb R}\times{\mathbb R}^3\to {\mathbb C}^4$,
for the given initial data ${\underline{\psi}}(q)\in C_0^\infty({\mathbb R}^3:{\mathbb C}^4)$,
satisfying
\begin{equation}
\left\{
\begin{aligned}
&i\hbar \pdt \psi(t,q)={\mathbb H}(t)\psi(t,q),\\
&\psi({\unbt},q)={\underline{\psi}}(q). 
\end{aligned}\right.
\label{eD1.1}
\end{equation}
Here,
\begin{equation}
{\mathbb H}(t)
=c\,\sum_{k=1}^3{\pmb\alpha}_k
\bigg(\frac{\hbar}{i}\frac{\partial}{\partial q_k}-\frac{e}{c}  %\varepsilon
A_k(t,q)\bigg)+mc^2{\pmb\beta}+{e}A_0(t,q).
%+{\mathbb P}(t,q),
\label{eD1.1-1}
\end{equation}
Though it is well-known that this IVP has a solution, we want to know a ``good\footnote{representation implying Bohr's principle}'' parametrix or fundamental solution
as Feynman desired. More explicitly, show the mathematical proof for the phenomena called Zitterwebegung (see, Inoue~\cite{ino98-2} for free case).
\end{problem}

Seemingly H. Weyl had been at Dirac's talk as an audience, 
he proposed $2\times 2$-matrix representation \eqref{p-matrix} %with the relation \eqref{CLR2}
in stead of $4\times 4$-one when the mass $m=0$.
From this, he derived the initial value problem of the free Weyl equation:
For a ``vector'' $\psi(t,q)$, it satisfies
\begin{equation}
\left\{
\begin{aligned}
&i\hbar \pdt \psi(t,q)={\mathbb H}\psi(t,q), \quad
{\mathbb H}=-ic\,\hbar\sum_{j=1}^3\pmb\sigma_j\frac{\partial}{\partial q_j},\\ 
& \psi(0,q)=\underline{\psi}(q),
\end{aligned}
\right.
\with \psi(t,q)=\begin{pmatrix}
\psi_1(t,q)\\
\psi_2(t,q)
\end{pmatrix}.
\label{In-1.1}
\end{equation}
In spite of the beauty of this equation, it is not accepted in physicists society for a while,
because the ``parity'' is not preserved by this one. Its meaning is reconsidered after
Lee-Yang's theory and Wu's experiment in weak interaction, 
which shows that the parity is not necessarily preserved for certain spinning particles.

Since Neutrino has been considered as the particle with mass $0$, Weyl equation is believed to be the governing equation of Neutrino untill the recent experiment of Kamiokande which suggests that at least certain Neutrino
has non-zero mass.

{\small
 [Report Problem 2-3]: Search ``Weyl equation"  in  internet to check whether the usage of this equation
 in condensed matter physics, etc. Report  things what you appreciate interesting.}
 
\paragraph{Ordinary procedure:}
As a hint to get a result for Problem~\ref{DFrep}, we give a simple example.
Though the equation \eqref{In-1.1} is a system but with constant coefficients, applying Fourier transform,
we may have the solution rather by algebraic operation.
In fact, defining Fourier transform as
$$
\hat{u}(p)=(2\pi\hbar)^{-m/2}\int_{\euc^m}dq\,e^{-i{\hbar}^{-1}qp}u(q),\quad
u(q)=(2\pi\hbar)^{-m/2}\int_{\euc^m}dp\,e^{i{\hbar}^{-1}qp}\hat{u}(p),
$$
%for $m=3$ and 
and applying this to $q\in\euc^3$ of \eqref{In-1.1}, we get
\begin{equation}
i\hbar\pdt{\hat\psi}(t,p)=\hat{\mathbb H}{\hat\psi}(t,p)
\label{In-1.2}
\end{equation}
Here,
$$
\hat{\mathbb H}=c\sum_{j=1}^3\pmb\sigma_jp_j=c
\begin{pmatrix}
p_3&p_1-ip_2\\
p_1+ip_2&-p_3
\end{pmatrix}
\quad\text{and}\quad \hat{\mathbb H}^2=c^2|p|^2{\mathbb I}_2.
$$
From this, we have
\begin{proposition} %{Proposition 1.1}
For any $t\in\euc$ and ${\underline{\psi}}\in L^2(\euc^3:{\mathbb C}^2)$, we have
\begin{equation}
e^{-i\hbar^{-1}t{\mathbb H}}{\underline{\psi}}(q)=(2\pi\hbar)^{-3/2}\int_{\euc^3}dp\,
e^{i\hbar^{-1}qp}e^{-i\hbar^{-1}t\hat{\mathbb H}}\hat{\underline{\psi}}(p).
\label{In-1.3}
\end{equation}
If ${\underline{\psi}}\in {\mathcal S}(\euc^3:{\mathbb C}^2)$, then
\begin{equation}
{\mathbb E}(t,q)=
(2\pi\hbar)^{-3}\int_{\euc^3}dp\,e^{i\hbar^{-1}qp}
\big[\cos(c\hbar^{-1}t|p|){\mathbb I}_2
-i\frac{\sin(c\hbar^{-1}t|p|)}{c|p|}\hat{\mathbb H}\big]\in
{\mathcal S}'(\euc^3:{\mathbb C}^2)
\label{In-1.5}
\end{equation}
and
\begin{equation}
e^{-i\hbar^{-1}t{\mathbb H}}{\underline{\psi}}(q)
={\mathbb E}*{\underline{\psi}}(t,q)=
\int_{\euc^3}dq'\,
{\mathbb E}(t,q-q'){\underline{\psi}}(q').
\label{In-1.4}
\end{equation}
\end{proposition}

In spite of this, we may give another representation with action integral ${\mathcal{S}}$ and amplitude ${\mathcal{D}}^{1/2}$, which is proved in \cite{ino98-1} and will be explained in later chapter.

\begin{remark} Pauli said one day that ``There exists no classical counter-part
corresponding to quantum spinning particle", so I had seen somewhere but I can't remember where exactly.
Therefore, such saying didn't exist? Please give a look to the splendid book S. Tomonaga~\cite{Tom74}, written in Japanese. \\
%\par\noindent
In any way, it seems difficult to imagine the classical mechanics corresponding to the equation \eqref{In-1.1} from the formula \eqref{In-1.5}. This is the one reason why I denote
{Feynman's murmur} as Feynman's problem. 
\end{remark}

\begin{claim}\label{freeWeyl-par}
In spite of above, I claim that I may construct the classical mechanics corresponding to \eqref{In-1.1},
which yields a path-integral-like representation\footnote{Though I said on one hand that ``There doesn't exist path-integral''(more accurately, in path space, there doesn't exist Lebesgue-like Borel measure) but here I mention path-integral-like. Therefore, it seems better to find more suitable nomination for path-integral-like representation} in Theorem \ref{fW-pi-rep} of it!
%I will mention later that if configuration or phase space is not finite dimension, I propose to consider FDE(=Functional Derivative Equation) as PDE on infinite dimensional space), but I use the nomination ``path-integral-like''
\end{claim}

\subsection{The method of characteristics and Hamiltonian path-integral-like representation}
%, which resolve a problem caused by the spacial derivative being  of first order.
%Since Feynman's problem concerns with a system having only first order spacial derivative, 
%we can't assign start and end points in configuration space for  initial and final time, respectively,
%we need a device using
%momentum which is guaranteed with Fourier transformation.
Though Schr\"odinger equation has 2-times partial derivatives which guarantees to assign initial and final positions to the corresponding classical flow on configuration space,% w.r.t. the space variables
but there exists only 1-time partial derivatives w.r.t. the space variables in Dirac or Weyl equations, this is the very reason why we need Hamiltonian path-integral representation. We need to use phase space instead of configuration space.

Therefore, {we} want to give a simple example  exhibiting ``Hamiltonian path-integral-like representation'', which is a necessary device to resolve Feynman's problem.

We may solve the following equation readily:
\begin{equation}
\left\{
\begin{aligned}
&i\hbar\pdt u(t,q)=a\frachi{\pdq} u(t,q)+bqu(t,q),\\
&u(0,q)={\underline{u}}(q).
\end{aligned}\right.
\label{sqm}
\end{equation}
From the right-hand side of above, we get a Hamiltonian function
$$
H(q,p)=e^{-i\hbar^{-1}qp}\Big(a\frachi{\pdq}+bq\Big)e^{i\hbar^{-1}qp}\Big|_{\hbar=0}=ap+bq,
$$
then, the corresponding classical orbit is obtained easily
from the Hamilton equation
\begin{equation}
{\left\{
\begin{aligned}
&\dot q(t)=H_p=a,\\
&\dot p(t)=-H_q=-b
\end{aligned}\right.}
\with
{\begin{pmatrix}
q(0)\\
p(0)
\end{pmatrix}
=\begin{pmatrix}
{\unbq}\\
{\underline{p}}
\end{pmatrix}}
\label{scm}
\end{equation}
such as
\begin{equation}
q(s)={\unbq}+as,\quad p(s)={\underline{p}}-bs.
\label{scmq}
\end{equation}
Using these, by applying the method of characteristics, we get
$$
U(t,{\unbq})=
{\underline{u}}({\unbq})
e^{-i\hbar^{-1}(b{\overline{q}}t+2^{-1}ab{t^2})}.
$$
Using the inverse function ${\unbq}=y(t,\bar q)=\bar q-at$ of ${\overline{q}}=q(t,{\unbq})$,
the solution of \eqref{sqm} is given as
$$
u(t,{\overline{q}})=U(t,{\unbq})|_{{\unbq}=y(t,\bar q)}
={\underline{u}}({\overline{q}}-at)
e^{-i\hbar^{-1}(b{\overline{q}}t-2^{-1}ab{t^2})}.
$$

\begin{remark}  
 In the above procedure, the information from $p(t)$ is not used.
\end{remark}

{\small
 [Report problem 2-4]: Study the method of characteristics for the first order PDE.
Since from the information obtained from ODE(such as (non-linear) Hamilton equation), we get a solution of PDE(such as (linear) Louville equation), this is the core of the method of characteristics.
What is the linear Liouville equation corresponding to the non-linear field equation, for example,
the Hopf equation represented by functional derivatives is the Liouville equation corresponding to the
Navier-Stokes equation.}

\paragraph{\bf Another point of view from \underline{Hamiltonian path-integral-like method}:}
Put
$$
S_0(t,{\unbq},{\underline{p}})
=\int_0^t ds \,[\dot q(s)p(s)-H(q(s),p(s))]
=-b{\unbq}t-2^{-1}ab{t^2},
$$
$$
S(t,{\overline{q}},{\underline{p}})
=\bigg({\unbq}\,{\underline{p}}+S_0(t,{\unbq},{\underline{p}})\bigg)
\bigg|_{{\unbq}=y(t,{\overline{q}})}
={\overline{q}}{\underline{p}}-a{\underline{p}}t-b{\overline{q}}t+2^{-1}ab{t^2}.
$$
Then, the classical action
$S(t,{\bar{q}},{\underline{p}})$ 
satisfies the Hamilton-Jacobi equation.
$$
\left\{
\begin{aligned}
&\pdt S+H({\bar{q}},\partial_{\bar{q}}S)=0,\\
&S(0,{\bar{q}},{\underline{p}})={\bar{q}}\,{\underline{p}}.
\end{aligned}
\right.
$$
On the other hand, the van Vleck determinant
(though scalar in this case) is calculated as
$$
D(t,{\overline{q}},{\underline{p}})
={\frac{\partial^2S(t,{\overline{q}},{\underline{p}})}
{\partial{\overline{q}}\partial{\underline{p}}}}=1.
$$
This quantity satisfies the continuity equation:
$$
\left\{\begin{aligned}
&\pdt {D}+\frac12\partial_{\bar{q}}({D} H_p)=0
\quad\text{where
$ H_p=\frac{\partial H}{\partial {p}}
\big({\bar{q}},\frac{\partial S}{\partial \bar{q}}\big)$},\\
&{D}(0,{\bar{q}},{\underline{p}})=1.
\end{aligned}
\right.
$$

As an interpretation of Feynman's idea, 
we regard that {\bf the transition from classical to quantum}
is to study the following quantity or the one represented by this
 (be careful, the term ``quantization" is not so well-defined mathematically as functor, so ad-hoc):
$$
u(t,{\overline{q}})=(2\pi\hbar)^{-1/2}\int_{\euc} d{\underline{p}}\,
D^{1/2}(t,{\overline{q}},{\underline{p}})
e^{i\hbar^{-1}S(t,{\overline{q}},{\underline{p}})}
\hat{\underline{u}}({\underline{p}}).
$$
That is, in our case at hand, we should study the quantity defined by
$$
\begin{aligned}
u(t,{\overline{q}})
&=(2\pi\hbar)^{-1/2}\int_{\euc} d{\underline{p}}\,
e^{i\hbar^{-1}S(t,{\overline{q}},{\underline{p}})}
\hat{\underline{u}}({\underline{p}})\\
&=(2\pi\hbar)^{-1}\iint_{\euc^2} d{\underline{p}}d{\unbq}\,
e^{i\hbar^{-1}(S(t,{\overline{q}},{\underline{p}})-{\unbq}\,{\underline{p}})}
{\underline{u}}({\unbq})
\big(={\underline{u}}({\overline{q}}-at)
e^{i\hbar^{-1}(-b{\overline{q}}t+2^{-1}ab{t^2})}\big).
\end{aligned}
$$
Therefore, we may say that this second construction 
gives the explicit connection
between the solution \eqref{sqm} 
and the classical mechanics given by \eqref{scm}.
We feel the above expression ``good'' because there appears two classical
quantities $S$ and $D$ and also explicit dependence on $\hbar$.

\begin{claim}
Applying superanalysis,
we may extend the second argument above to a system of PDOs 
e.g.
quantum mechanical equations with spin such as Dirac, Weyl or Pauli equations,
(and if possible, any other $2^d\times 2^d$ system of PDOs),
after interpreting these equations
as those on superspaces.
\end{claim}

\subsection{Decomposition of $2\times 2$ matrix by Clifford algebra}
\paragraph{\bf How matrix does act on vectors?:}
Following matrices  form a special class in $2\times 2$ matrices.
$$
A=\begin{pmatrix}
a&-b\\
b&a
\end{pmatrix},\qquad a,b\in\euc.
$$
This set of matrices $\{A\}$ not only preserves their form under four rules of arithmetic but also is commutative each other.
Moreover, we identify this matrix $A$ with a complex number $\alpha=a+ib$.
If we regard a vector ${\bf z}=\begin{pmatrix}
x\\y\end{pmatrix}$ with a complex number $z=x+iy$, then the multiplication $\alpha$ to $z$ is considered as
$$
\begin{aligned}
(a+ib)(x+iy)
\sim\alpha{\bf z}
&\sim
\begin{pmatrix}
a&-b\\
b&a
\end{pmatrix}
\begin{pmatrix}
x\\y\end{pmatrix}
=\begin{pmatrix}
ax-by\\bx+ay\end{pmatrix}
\sim(ax-by)+i(bx+ay)=(a+ib)(x+iy)\\
&\sim
\begin{pmatrix}
ax-by&-(bx+ay)\\
bx+ay&ax-by
\end{pmatrix}
=\begin{pmatrix}
a&-b\\
b&a
\end{pmatrix}
\begin{pmatrix}
x&-y\\
y&x
\end{pmatrix}.
\end{aligned}
$$

Then, may we find another interpretation of making act $2\times 2$ matrix to a column vector? Since above mentioned interpretation gives you many stand points, is it possible this idea generalize?

Guided by the following theorem of C. Chevalley\footnote{Though I don't know how to prove this theorem itself, but I'm satisfied by constructing the differential operator representation of $2\times 2$-matrices using Pauli matrices. Oh, \underline{such a jerry-built attitude as a mathematician is allowed}?! Or I'm far from being a solid mathematician?}
below, we decompose a $2\times 2$-matrix. Here,
\begin{theorem}[C. Chevalley]
Any Clifford algebra has the representation on Grassmann algebra.
\end{theorem}

(I) For any $2\times 2$ matrix, we have
$$
\begin{aligned}
\begin{pmatrix}
a&c\\
d&b
\end{pmatrix}
&=\frac{a+b}{2}
\begin{pmatrix}
1&0\\
0&1
\end{pmatrix}
+\frac{a-b}{2}
\begin{pmatrix}
1&0\\
0&-1
\end{pmatrix}
+\frac{c+d}{2}
\begin{pmatrix}
0&1\\
1&0
\end{pmatrix}
+\frac{c-d}{2}
\begin{pmatrix}
0&1\\
-1&0
\end{pmatrix}\\
&=\frac{a+b}{2}{\mathbb{I}}_2+\frac{a-b}{2}\pmb{\sigma}_3+\frac{c+d}{2}\pmb{\sigma}_1
+i\frac{c-d}{2}\pmb{\sigma}_2.
\end{aligned}
$$
Here, $\{\pmb{\sigma}_j\}$ satisfies not only
\eqref{p-matrix} but also the relation \eqref{CLR2}.

This decomposition stands for that a set of all $2\times 2$ matrices is spanned by Pauli matrices
$\{\pmb{\sigma}_k\}$ having Clifford structure.

\par
(II-1) Now, preparing a letter $\theta$ satisfying $\theta^2=0$, we
identify Pauli matrices with differential operators acting on Grassmann algebra
$\Lambda=\{u(\theta)=u_0+u_1\theta\;|\; u_0,\; u_1\in{\mathbb{C}}\}$, i.e. for 
$$
u_0+u_1\theta\sim\begin{pmatrix}
u_0\\
u_1
\end{pmatrix},
$$
define the action as
$$
\theta u(\theta)=u_0\theta\sim\begin{pmatrix}
0\\
u_0
\end{pmatrix}
=\begin{pmatrix}
0&0\\
1&0
\end{pmatrix}
\begin{pmatrix}
u_0\\
u_1
\end{pmatrix},
\quad
 \frac{\partial}{\partial\theta}u(\theta)=u_1
 \sim\begin{pmatrix}
u_1\\
0
\end{pmatrix}
=\begin{pmatrix}
0&1\\
0&0
\end{pmatrix}
\begin{pmatrix}
u_0\\
u_1
\end{pmatrix}.
$$
Then, we have
$$
\begin{aligned}
&\bigg(\theta+\frac{\partial}{\partial\theta}\bigg)u(\theta)=u_0\theta+u_1\sim
\begin{pmatrix}
u_1\\
u_0
\end{pmatrix}
=
\begin{pmatrix}
0&1\\
1&0
\end{pmatrix}
\begin{pmatrix}
u_0\\
u_1
\end{pmatrix},\\
&\bigg(\theta-\frac{\partial}{\partial\theta}\bigg)u(\theta)=u_0\theta-u_1\sim
\begin{pmatrix}
-u_1\\
u_0
\end{pmatrix}
=
\begin{pmatrix}
0&-1\\
1&0
\end{pmatrix}
\begin{pmatrix}
u_0\\
u_1
\end{pmatrix},\\
&\bigg(1-2\theta\frac{\partial}{\partial\theta}\bigg)u(\theta)=u_0-u_1\theta\sim
\begin{pmatrix}
u_0\\
-u_1
\end{pmatrix}
=
\begin{pmatrix}
1&0\\
0&-1
\end{pmatrix}
\begin{pmatrix}
u_0\\
u_1
\end{pmatrix}.
\end{aligned}
$$
This means that Pauli matrices are represented as differential operators acting on $\Lambda$.
\newline
\underline{But such a representation is not unique !}

\par
(II-2) Here is another representation:
Preparing 2 letters $\theta_1,\;\theta_2$ satisfying $\theta_i\theta_j+\theta_j\theta_i=0$ for $i,j=1,2$,
we put
$$
\Lambda_{\mathrm{ev}}=\{u=u_0+u_1\theta_1\theta_2\;|\; u_0, u_1\in{\mathbb{C}}\},\quad
\Lambda_{\mathrm{od}}=\{v=v_1\theta_1+v_2\theta_2\;|\; v_1,v_2\in{\mathbb{C}}\},
$$
and define differential operators acting on $\Lambda_{\mathrm{ev}}$ as
$$
\begin{aligned}
&\sigma_1(\theta,\partial_{\theta})=\bigg(\theta_1\theta_2-\frac{\partial^2}{\partial\theta_1 \partial\theta_2}\bigg)u(\theta)
=u_0\theta_1\theta_2+u_1\sim
\begin{pmatrix}
u_1\\
u_0
\end{pmatrix}
=\begin{pmatrix}
0&1\\
1&0
\end{pmatrix}
\begin{pmatrix}
u_0\\
u_1
\end{pmatrix},\\
&-i\sigma_2(\theta,\partial_{\theta})=\bigg(\theta_1\theta_2+\frac{\partial^2}{\partial\theta_1 \partial\theta_2}\bigg)u(\theta)
=u_0\theta_1\theta_2-u_1\sim
\begin{pmatrix}
-u_1\\
u_0
\end{pmatrix}
=\begin{pmatrix}
0&-1\\
1&0
\end{pmatrix}
\begin{pmatrix}
u_0\\
u_1
\end{pmatrix},\\
&\sigma_3(\theta,\partial_{\theta})=\bigg(1-\theta_1\frac{\partial}{\partial \theta_1}-\theta_2\frac{\partial}{\partial \theta_2}\bigg)u(\theta)
=u_0-u_1\theta_1\theta_2\sim
\begin{pmatrix}
u_0\\
-u_1
\end{pmatrix}
=\begin{pmatrix}
1&0\\
0&-1
\end{pmatrix}
\begin{pmatrix}
u_0\\
u_1
\end{pmatrix}.
\end{aligned}
$$
\begin{remark}
Above defined differential operators
$\sigma_j(\theta,\partial_{\theta})$ annihilate $\Lambda_{\mathrm{od}}$.
Moreover, the symbols corresponding to them are ``even''.
\underline{This evenness of Hamiltonian function is crucial} to derive Hamilton flow corresponding to Weyl or Dirac equations.
\end{remark}

\chapter{Super number and Superspace}
To explain symbols $\theta$, $\theta_1$, $\theta_2$ appeared in previous lectures, we \underline{prepare} a set of countably infinite Grassmann generators. After introducing of these, {we} may consider the classical mechanics corresponding to PDE with spin,
which is rather easily solved.

\section{Super number}
\subsection{The Grassmann generators}
Preparing symbols
$\{\sigma_j\}_{j=1}^{\infty}$ which satisfy the Grassmann relation
\begin{equation}
\sigma_j \sigma_k+ \sigma_k \sigma_j=0,
\quad j,k=1,2,\cdots,
\label{GR}
\end{equation}
we \underline{put formally}
\begin{equation}
{\fC}=\{X=\sum_{{\mathbf{I}}\in\mathcal I} X_{\mathbf{I}}\sigma^{\mathbf{I}}\;\big|\; X_{\mathbf{I}}\in{\mathbb C}\}
\label{fC}
\end{equation}
and
$$
\left\{
{\begin{aligned}
&{\mathfrak  C}^{(0)}={\mathfrak  C}^{[0]}={\mathbb{C}},\\
&{\mathfrak  C}^{(j)}=\bigg\{X=\sum_{|{\mathbf{I}}|\leq j} X_{\mathbf{I}}\sigma^{\mathbf{I}}\bigg\} \et\\
&{\mathfrak  C}^{[j]}=\bigg\{X=\sum_{|{\mathbf{I}}|= j} X_{\mathbf{I}}\sigma^{\mathbf{I}}\bigg\}
={\mathfrak  C}^{(j)}/{\mathfrak  C}^{(j-1)},
\end{aligned}}
\right.
$$
where the index set is defined by
$$
\begin{gathered}
{\mathcal I}=\{{\mathbf{I}}=(i_k)\in \{0,1\}^{\mathbf N}\;\big|\;
|{\mathbf{I}}|=\sum_k i_k < \infty \},\\
\sigma^{\mathbf{I}}=\sigma_1^{i_1}\sigma_2^{i_2}\cdots,\quad {\mathbf{I}}=(i_1,i_2,\cdots ),\quad
\sigma^{\tilde 0}=1, \quad {\tilde 0}=(0,0,\cdots )\in\mathcal I.
\end{gathered}
$$

\begin{remark}
How do we construct symbols $\{\sigma_j\}_{j=1}^{\infty}$ satisfying the Grassmann relation?
What is the meaning of summation appeared above?
These will be soon explained.
\end{remark}

In today's lecture, we prove the following Proposition which guarantees that 
${\mathfrak  C}$ (or $\fR$, defined later) plays the alternative role of  ${\mathbb C}$ (or $\fR$) in analysis.
\begin{proposition}[A. Inoue and Y. Maeda~\cite{IM91}] \label{IM-sn}
$\fC$ forms an $\infty$-dimensional Fr\'echet-Grassmann algebra over ${\mathbb C}$,
that is, an associative, distributive and non-commutative ring with degree,
which is endowed with the Fr\'echet topology.
\end{proposition}

\begin{remark}
There exist some papers using $\fC$, for example,
S. Matsumoto and K. Kakazu~\cite{MK86}, Y. Choquet-Bruhat~\cite{C-B86}, P. Bryant~\cite{bry89}.
But, seemingly, they didn't try to construct ``elementary and real analysis'' on this ``ground ring'' $\fC$ (or $\fR$).
\end{remark}

\subsection{Sequence spaces and their topologies}
Following G. K\"othe~\cite{Kot69}, we introduce
the sequence spaces $\omega$ and $\phi$, (effm=except for finitely many)
\begin{equation}
\left\{
{\begin{aligned}
\phi & =\left\{{\mathfrak  x}=(x_k)=(x_1,x_2,\cdots,x_k,\cdots)\; \big| \; 
 \text{$x_k\in{\mathbb{C}}$ and $x_k=0$ effm $k$}
\right\},\\
\omega & =\left\{{\mathfrak  u}=(u_k)=(u_1,u_2,\cdots,u_k,\cdots)\; \big| \; 
 u_k\in{\mathbb{C}}\right\}.
\end{aligned}}
\right.
\label{EAI-1-5}
\end{equation} 
For any sequence space ${\mathcal{X}}$ containing $\phi$, 
we define the space ${\mathcal{X}}^\times$ by
$$
{\mathcal{X}}^\times=\left\{{\mathfrak  u}=(u_k)\; \big| \;  \sum_{k}|u_k||x_k|<\infty
\forany {\mathfrak  x}=(x_k)\in {\mathcal{X}}\right\},
$$
then, we get
$$
\phi^\times=\omega \et
 \omega^\times=\phi.
$$
We introduce the (normal) topology in ${\mathcal{X}}$ 
and ${\mathcal{X}}^\times$ by defining the seminorms
\begin{equation}
p_{{\mathfrak  u}}\,({\mathfrak  x})=\sum_{k}|u_k||x_k|=p_{{\mathfrak  x}}\,({{\mathfrak  u}})
\quad \text{for ${\mathfrak  x}\in {\mathcal{X}}$ and ${\mathfrak  u}\in {\mathcal{X}}^\times$.}
\label{EAI-1-6}
\end{equation}

Especially, ${\mathfrak  x}^{(n)}$ converges to $\mathfrak  x$ in $\phi$,
that is, $p_{\mathfrak  u}({\mathfrak  x}^{(n)}-{\mathfrak  x})\to 0$ as $n\to\infty$
for each ${\mathfrak  u}\in\omega$
if and only if for any $\epsilon>0$, 
there exist $L$ and $n_0$ such that
\begin{equation}
\left\{
\begin{aligned}
& \text{(i)}\quad
x^{(n)}_k=x_k=0 \for k>L \when  n \ge n_0, \et\\
& \text{(ii)}\quad
|x^{(n)}_k-x_k|<\epsilon  \for k\le L \when  n\ge n_0.
\end{aligned}
\right.
\label{EAI-1-7}
\end{equation}

Analogously, ${\mathfrak  u}^{(n)}$ converges to $\mathfrak  u$ in $\omega$, 
that is, $p_{\mathfrak  x}({\mathfrak  u}^{(n)}-{\mathfrak  u})\to 0$ as $n\to\infty$
for each ${\mathfrak  x}\in\phi$
if and only if for any $\epsilon>0$ and each $k$, 
there exists $n_0=n_0(\epsilon, k)$ such that
\begin{equation}
|u^{(n)}_k-u_k|<\epsilon  \when n\ge n_0.
\label{EAI-1-8}
\end{equation}
Clearly, $\omega$ forms a Fr\'echet space because the above topology  
in $\omega$ is equivalent to the one defined by countable seminorms:
$\{p_k({\mathfrak  u})\}_{k\in{\mathbb N}}$ where
$p_k({\mathfrak  u})=|u_k|$ for 
${\mathfrak  u}=(u_1,u_2,\cdots)=\sum_{j=1}^\infty u_j{\mathfrak  e}_j\in {\omega}$
with ${\mathfrak  e}_j=(\overbrace{0,\cdots,0,1}^{j},0,\cdots \,\,)\in\omega$.
%\end{quotation}

Now, we define the isomorphism (diadic-decomposition)
from $\mathcal{I}$ onto $\mathbb{N}$ defined by
\begin{equation}
r: {\mathcal{I}}\ni {\mathbf{I}}=(i_k) \to 
r({\mathbf{I}})=1+{\frac 12}\sum_{k=1}^\infty 2^k i_k\in{\mathbb{N}}
\where i_k=0\; \mbox{or}\; 1.
\label{EAI-1-9}
\end{equation}
Using $r({\mathbf{I}})$ in \eqref{EAI-1-9}, we define a map
$$
T: \,\, \sigma^{\mathbf{I}} \to {\mathfrak  e}_{r({\mathbf{I}})}  \for {\mathbf{I}}=(i_k)\in{\mathcal{I}}.
$$
Extending this map linearly, we put
\begin{equation}
T(X)=\sum_{} x_{r({\mathbf{I}})} {\mathfrak  e}_{r({\mathbf{I}})}\in \omega \for
 X=\sum_{|{\mathbf{I}}|\leq j} X_{\mathbf{I}}\sigma^{\mathbf{I}} \in{\fC}^{(j)}.
\label{EAI-1-10}
\end{equation}
More explicitly, we have the following first few terms:
$$
\sum_{} x_{r(I)} {\mathfrak  e}_{r(I)}
=(X_{(0,0,0,\cdots)},X_{(1,0,0,\cdots)},X_{(0,1,0,\cdots)},X_{(1,1,0,\cdots)},
X_{(0,0,1,\cdots)},X_{(1,0,1,\cdots)},X_{(0,1,1,\cdots)},\cdots).
$$

Then, since $T({\fC}^{[j]})$ and $T({\fC}^{[k]})$ 
are disjoint sets in $\omega$ if $j\neq k$, we have
\begin{equation}
\sum_{j=0}^\infty T({\fC}^{[j]})=\omega.
\label{EAI-1-11'}
\end{equation}

Therefore, it is reasonable to write as in \eqref{fC} and more precisely,
\begin{equation}
{\fC}=\oplus_{j=0}^\infty {\fC}^{[j]},\quad
\text{that is,}\quad
X=\sum_{j=0}^\infty X^{[j]}\with X^{[j]}=\sum_{|{\mathbf{I}}|=j}X_{\mathbf{I}}\sigma^{\mathbf{I}}.
\label{EAI-1-12}
\end{equation}
Here, $X^{[j]}$ is called the $j$-th degree component of $X\in {\fC}$.
By definition, we get
\begin{equation}
\left\{
\begin{aligned}
&{\fC}^{(j)}\subset{\fC}^{(k)}\for j\le k, \\
&{\fC}= \sum_{j=0}^{\infty}{\fC}^{[j]} \with
\cap_{j=0}^{\infty}{\fC}^{(j)} ={\mathbb{C}},
\end{aligned}
\right.
\label{EAI-1-13}
\end{equation}
\begin{equation}
{\fC}^{[j]} \cdot {\fC}^{[k]} 
\subset {\fC}^{[j+k]} \et
{\fC}^{(j)} \cdot {\fC}^{(k)} 
\subset {\fC}^{(j+k)}.
\label{EAI-1-14}
\end{equation}

\begin{remark} %{\it Remark}. 
The second relation with ${\fC}^{(*)}$ in \eqref{EAI-1-14} 
also holds for the Clifford algebras but
the first one with ${\fC}^{[\cdot]}$ is specific to the Grassmann algebras
satisfying \eqref{GR}.
Here, the Clifford relation for $\{{e}_j\}$ is defined by
\begin{equation}
{e}_i {e}_j+{e}_j{e}_i =2\delta_{ij}{\mathbb I} \forany i,j=1,2,\cdots .
\label{EAI-1-1-bis}
\end{equation}
Typical examples, though not countably many but finitely many elements, are the $2\times 2$-Pauli matrices 
${e}_j=\{\pmb{\sigma}_j\}_{j=1,2,3}$
and the $4\times 4$-Dirac matrices 
$\{{e}_j\}_{j=0,1,2,3}=\{\pmb{\beta},\pmb{\alpha}_j\}$.
\end{remark}

\subsection{Topology}
We introduce the weakest topology in $\fC$ which
makes the map $T$ continuous from $\fC$ to $\omega$,
that is,
$ X=\sum_{{\mathbf{I}}\in\mathcal{I}} X_{\mathbf{I}}\sigma^{\mathbf{I}}\to 0$ in ${\fC}$ if and only if
$ \proj_{\mathbf{I}}(X) \to 0$ for each  ${\mathbf{I}}\in{\mathcal{I}}$ with $ \proj_{\mathbf{I}}(X)=X_{\mathbf{I}}$;
it is equivalent to the metric $\dist(X,Y)=\dist(X-Y)$ defined by
\begin{equation}
\dist(X)=\sum_{{\mathbf{I}}\in{\mathcal{I}}} {\frac 1{2^{r({\mathbf{I}})}}}{\frac
{|\proj_{\mathbf{I}}(X)|}{1+|\proj_{\mathbf{I}}(X)|}} \forany
X\in {\fC}.
\label{EAI-1-15}
\end{equation}

For example, $X^{(\ell)}=f(\ell)\sigma_1\cdots\sigma_\ell\to 0$ in $\fC$
even if $f(\ell)\to\infty$ because $\dist(X^{(\ell)})\leq 2^{-2^\ell+1}$. 
\begin{comparison}
The sequence space $\omega$ is regarded as a formal power series ring of an interminate element $X$ (see, for example p.25 or p.91 of F. Treves~\cite{Tre67}). That is,
$$
{\mathbb{C}}[[X]]=\{u=u(X)=\sum_{n=0}^{\infty} u_n X^n\;|\; u_n\in{\mathbb{C}}\}
\cong
\{u=(u_0, u_1,{\cdots}, u_n,{\cdots})\;|\; u_n\in{\mathbb{C}}\}.
$$
Introducing ``standard'' algebraic operations and putting a fundamental neighbourhood system as
$$
V_{m,n}=\{u=u(X)=\sum_{p=0}^{\infty} u_p X^p\;|\; u_p\in{\mathbb{C}},\; |u_p|\le\frac{1}{m} \forany p\le n\},
$$
we may define a Fr\'echet topology on it.
\end{comparison}

\subsection{Algebraic operations -- addition and product}

For any $X,Y\in \fC$, we define
\begin{equation}
X+Y=\sum_{j=0}^\infty (X+Y)^{[j]}
\with (X+Y)^{[j]}=X^{[j]}+Y^{[j]} \for j\ge 0
\label{EAI-1-16}
\end{equation}
and
\begin{equation}
XY=\sum_{j=0}^\infty (XY)^{[j]}
\where
(XY)^{[j]}=\sum_{k=0}^jX^{[j-k]}Y^{[k]}=
\sum_{|{\mathbf{I}}|=j}(XY)_{\mathbf{I}}\sigma^{\mathbf{I}}.
\label{EAI-1-17}
\end{equation}
Here, 
$(XY)_{\mathbf{I}}=\sum_{{\mathbf{I}}={\mathbf{J}}+{\mathbf{K}}}(-1)^{\tau({\mathbf{I}};{\mathbf{J}},{\mathbf{K}})}X_{\mathbf{J}}Y_{\mathbf{K}}\in {\mathbb{C}}$ is well-defined 
because for any set ${\mathbf{I}}\in{\mathcal{I}}$, 
there exist only finitely many decompositions
by sets ${\mathbf{J}},\,{\mathbf{K}}$ satisfying ${\mathbf{I}}={\mathbf{J}}\dot{+}{\mathbf{K}}$ $(i.e. \;{\mathbf{I}}={\mathbf{J}}\cup {\mathbf{K}},\; {\mathbf{J}}\cap {\mathbf{K}}=\emptyset)$.
Here,
the indeces $\tau({\mathbf{I}};{\mathbf{J}},{\mathbf{K}})$, or more generally $\tau({\mathbf{I}};{\mathbf{J}}_1,\cdots,{\mathbf{J}}_k)$
 are defined by
\begin{equation}
(-1)^{\tau({\mathbf{I}};{\mathbf{J}}_1,\cdots,{\mathbf{J}}_k)}\sigma^{{\mathbf{J}}_1}\cdots \sigma^{{\mathbf{J}}_k}=\sigma^{\mathbf{I}}
\with {\mathbf{I}}={\mathbf{J}}_1\dot{+}{\mathbf{J}}_2\dot{+}{\cdots}\dot{+}{\mathbf{J}}_k.
\label{EA1-1.5}
\end{equation}
%when $I$ is decomposed by $I=J_1+\cdots+J_k$.
But for notational simplicity, we will use $(-1)^{\tau(*)}$ 
without specifying the decomposition if there occurs no confusion.
\begin{exercise}
Prove that for sets ${\mathbf{J}},\,{\mathbf{K}}$ satisfying ${\mathbf{I}}={\mathbf{J}}+{\mathbf{K}}$,
$$
(-1)^{|{\mathbf{J}}||{\mathbf{K}}|}(-1)^{\tau({\mathbf{I}};{\mathbf{J}},{\mathbf{K}})}=(-1)^{\tau({\mathbf{I}};{\mathbf{K}},{\mathbf{J}})}.
$$
\end{exercise}

Moreover, we get
\begin{lemma} \label{lemma EA-1.1}
The product defined by \eqref{EAI-1-17} is continuous from 
${\fC}\times{\fC} \to {\fC}$.
\end{lemma}

{\it Proof}.
It is simple by noting that there exist $2^{|{\mathbf{I}}|}$ elements
${\mathbf{J}}\in{\mathcal{I}}$ satisfying ${\mathbf{J}}\subset {\mathbf{I}}$ and that
$$
|(XY)_{\mathbf{I}}|\le \sum_{{\mathbf{I}}={\mathbf{J}}+{\mathbf{K}}}|X_{\mathbf{J}}||Y_{\mathbf{K}}|
\le 2^{r({\mathbf{I}})}(\max_{{\mathbf{J}}\subset {\mathbf{I}}}|X_{\mathbf{J}}|)(\max_{{\mathbf{K}}\subset {\mathbf{I}}}|Y_{\mathbf{K}}|) 
\forany X,Y\in{\fC}.  \qquad\quad\qed
$$

{\it Proof of Proposition \ref{IM-sn}}.
Clearly, we get
$$
\begin{cases}
 X(YZ)=(XY)Z  & (associativity),\\
 X(Y+Z)=XY+XZ & (distributivity).
\end{cases}
$$
Other properties have been proved. $\qquad\Box$ %\qed

\begin{remark} %{\it Remark 1.}
 We may consider that an element of $X\in \fC$
stands for the `state' such that the position labeled by $\sigma^{\mathbf{I}}$
is occupied by $X_{\mathbf{I}}\in {\mathbb{C}}$.
In other word, 
considering $\{\sigma_i\}$ as the countable indeterminate letters,
it seems reasonable to regard $\fC$ as the set of certain formal 
power series\footnote{with the special property that same letter appears only once in each monomials} 
with simple topology.
Therefore, it is permitted to reorder the terms freely under `summation sign'.
That is, the summation $\sum_{{\mathbf{I}}\in\mathcal{I}} X_{\mathbf{I}} {\mathfrak  e}_{r({\mathbf{I}})}$
is `unconditionally (though not absolutely) convergent' 
\footnote{diverting the terminology of the basis problem in the Banach spaces} 
and so is 
$\sum_{{\mathbf{I}}\in\mathcal{I}} X_{\mathbf{I}}\sigma^{\mathbf{I}}$.
 We use such a big space $\fC$ with rather weak topology because
this algebra is considered as the ambient space for reordering the places.
We feel such a big ambient space will be prefarable and tractable for our future use.
\end{remark}
\begin{remark} %{\it Remark 2.} 
(1) As $\{{\fC}^{(j)}\}$ forms a filter by \eqref{EAI-1-13} and \eqref{EAI-1-14}, 
it gives a 0-neighbourhood base of the linear topology of ${\fC}$ 
which is equivalent to the above one defined by \eqref{EAI-1-8}.
(See G. K\"{o}the~\cite{Kot69} for the linear topology of vector spaces.)
\newline
(2) We may introduce a stronger topology in ${\fC}$ 
called the topology by degree, that is,
$X^{(n)}{\overset{s}{\rightarrow}} X$ in ${\fC}$ means that
\par
(i) there exists $\ell\geq 0$ such that $X_{\mathbf{I}}^{(n)}=X_{\mathbf{I}}=0$ 
for any $n$ and $I$ when $|{\mathbf{I}}|>\ell$ 
and 
\par
(ii) $|X_{\mathbf{I}}^{(n)}-X_{\mathbf{I}}|\to 0$ as $n\to\infty$ when $|{\mathbf{I}}|\leq \ell$.
\end{remark}

\subsection{The supernumber}
The set $\fC$ defined by \eqref{fC} is called the {\it (complex) supernumber algebra} over $\mathbb{C}$ and 
any element $X$ of $ \fC$ is called {\it (complex) supernumber}.

{\bf Parity}:
We introduce the parity in $\fC$ by setting
\begin{equation}
p(X)=\begin{cases}
0 & \text{if $X=\sum_{{\mathbf{I}}\in\mathcal{I},|{\mathbf{I}}|={\mathrm{ev}}}X_{\mathbf{I}}\sigma^{\mathbf{I}}$},\\
1 & \text{if $X=\sum_{{\mathbf{I}}\in\mathcal{I},|{\mathbf{I}}|={\mathrm{od}}}X_{\mathbf{I}}\sigma^{\mathbf{I}}$}.
\end{cases}
\label{EAI-1-18}
\end{equation}
$X\in\fC$ is called homogeneous if it satisfies $p(X)=0$ or $=1$.
We put also
\begin{equation}
\left\{
\begin{aligned}
&{\cev}= \oplus_{j=0}^\infty {\fC}^{[2j]}
=\{X\in{\fC}\; | \;p(X)=0\},\\
&{\cod}= \oplus_{j=0}^\infty {\fC}^{[2j+1]}
=\{X\in{\fC}\; | \;p(X)=1\},\\
&{\fC}\cong {\cev}\oplus{\cod}\cong{\cev}\times{\cod}.
\end{aligned}
\right.
\label{EAI-1-19}
\end{equation}

Moreover,
it splits into its even and odd parts, 
called {\it (complex) even number} and {\it (complex) odd number}, respectively : 
\begin{equation}
X = X_{\mathrm{ev}} + X_{\mathrm{od}} = \sum_{|{\mathbf{I}}|={\mathrm{ev}}} X_{\mathbf{I}}\sigma^{\mathbf{I}}
+\sum_{|{\mathbf{J}}|=\mathrm{od}}X_{\mathbf{J}}\sigma^{\mathbf{J}} 
=\sum_{j={\mathrm{ev}}} X^{[j]}+\sum_{j={\mathrm{od}}} X^{[j]}.
\label{EAI-1-20}
\end{equation}
Using \eqref{EAI-1-20}, we decompose
\begin{equation}
X=X_{\mathrm B}+X_{\mathrm S}\quad \where
X_{\mathrm S}=\sum_{1\le j<\infty} X^{[j]}
\et
X_{\mathrm B}=X_{\tilde 0}=X^{[0]}
\label{EAI-1-21}
\end{equation}
and the number $X_{\mathrm B}$ is called {\it the body (part)} of $X$ and the
remainder $X_{\mathrm S }$ is called {\it the soul (part)}
of $X$, respectively.
We define the map $ \pi_{\mathrm B} $ from 
$\fC$ to $ \mathbb{C}$ by
$ \pi_{\mathrm B}(X)=X_{\mathrm B}$, called the body projection 
(or called the augmentation map). 

\begin{remark}[Important\footnote{This important fact is not mentioned at lecture time, why such bonehead!}]
$\fC$ does not form a field because $X^2=0$ for any $X\in\cod$.
But, it is easily proved that
\par 
(i) if $X$ satisfies $XY=0$ for any $Y\in\cod$, 
then $X=0$, and
\par 
(ii) the decomposition of $X$ with respect to degree 
in \eqref{EAI-1-12} is unique.
\newline
These properties
are shared only if the number of Grassmann generators is infinite.
For example, if the number of Grassmann generators is finite, say $n$,
then the number $\sigma_1\sigma_2{\cdots}\sigma_n$, which is not zero, is recognized $0$
for the multiplication of any odd number generated by $\{\sigma_j\}_{j=1}^n$.
\end{remark}

\begin{lemma}[the invertible elements]
Let $X\in{\mathfrak  C}$ with $X_{\mathrm B}\neq{0}$.
Then there exists a unique element $Y\in{\fC}$ such that
$XY=1=YX$. 
\end{lemma}

{\it Proof}. In fact, decomposing $X=X_{\mathrm B}+X_{\mathrm S}$ and
$Y=Y_{\mathrm B}+Y_{\mathrm S}$, we should have
$$
X_{\mathrm B}Y_{\mathrm B}=1,\quad
X_{\mathrm B}Y_{\mathrm S}+X_{\mathrm S}Y_{\mathrm B}+
X_{\mathrm S}Y_{\mathrm S}=0.
$$
Therefore, putting $X_{\mathrm S}=\sum_{|{\mathbf{I}}|>0}X_{\mathbf{I}}\sigma^{\mathbf{I}}$ 
and $Y_{\mathrm S}=\sum_{|{\mathbf{J}}|>0}Y_{\mathbf{J}}\sigma^{\mathbf{J}}$ and noting that $\sigma^{\mathbf{I}}\sigma^{\mathbf{J}}=(-1)^{\tau({\mathbf{K}};{\mathbf{I}},{\mathbf{J}})}\sigma^{\mathbf{K}}$ for ${\mathbf{K}}={\mathbf{I}}+{\mathbf{J}}$,
we have
$$
Y_{\mathrm B}=X_{\mathrm B}^{-1},\quad
Y_{\mathbf{K}}=-X_{\mathrm B}^{-1}\sum_{{\mathbf{K}}={\mathbf{I}}+{\mathbf{J}}}(-1)^{\tau({\mathbf{K}};{\mathbf{I}},{\mathbf{J}})}X_{\mathbf{I}}Y_{\mathbf{J}}.
$$
For example,
$$
\begin{gathered}
\mbox{for $|{\mathbf{K}}|=1$, then}\quad
Y_{\mathbf{K}}=-X_{\mathrm B}^{-1}X_{\mathbf{K}}Y_{\mathrm B},\cdots,\\
\mbox{for $|{\mathbf{K}}|=\ell$, then}\quad
Y_{\mathbf{K}}=-X_{\mathrm B}^{-1}\sum_{{\mathbf{K}}={\mathbf{I}}+{\mathbf{J}}}(-1)^{\tau({\mathbf{K}};{\mathbf{I}},{\mathbf{J}})}X_{\mathbf{I}}Y_{\mathbf{J}}.
\end{gathered}
$$
If $X_{\mathrm B}=0$, there exists no $Y$ satisfying $XY=1$ or $YX=1$. $\qquad\qquad\square$

Now, we define our {\it (real) supernumber algebra} 
by
\begin{equation}
{\fR}=\pi_{\mathrm B}^{-1}(\mathbb{R}) \cap {\fC}
=\left\{X=\sum_{{\mathbf{I}}\in\mathcal{I}} X_{\mathbf{I}}\sigma^{\mathbf{I}}\; | \; 
\,\, X_{\mathrm B}\in{\mathbb{R}}\et X_{\mathbf{I}}\in{\mathbb{C}} \for |{\mathbf{I}}|\ne 0 \right\}.
\label{EAI-1-26}
\end{equation}
Defining as same as before, we have
\begin{equation}
{\fR}={\rev}\oplus{\rod}, \quad
{\fR}=\oplus_{j=0}^\infty {\fR}^{[j]}.
\label{EAI-1-27}
\end{equation}
Analogous to $\fC$, we put
\begin{equation}
\left\{
\begin{aligned}
&{\fR}=\{X\in{\fC}\; | \;\pi_{\mathrm B}X\in{\mathbb{R}}\},\qquad
{\fR}^{[j]}={\fR}\cap {\fC}^{[j]},\\
&{\rev}={\fR}\cap{\cev},\quad {\rod}={\fR}\cap{\cod}={\cod},\\
&{\fR}\cong {\rev}\oplus{\rod}\cong{\rev}\times{\rod}.
\end{aligned}
\right.
\label{EAI-1-11}
\end{equation}
Here, we introduced the body (projection) map $\pi_{\mathrm B}$ by 
$\pi_{\mathrm B}X=\proj_{\mathrm B}(X)=X_{\tilde 0}=X_{\mathrm B}$.

${\fR}^{(j)}$ 
and other terminologies are analogously introduced.

\subsection{Conjugation}
We define the operation ``complex'' conjugation, denoted by $\overline{X}$ as follows:
Denoting the complex conjugation of $X_{\mathbf{I}}\in{\mathbb{C}}$ by $\overline{X_{\mathbf{I}}}$ and
defining $\overline{\sigma^{\mathbf{I}}}=\sigma_n^{i_n}\cdots\sigma_1^{i_1}$
for ${\mathbf{I}}=(i_1,\cdots,i_n)$, we put
\begin{equation}
\overline{X}=\sum_{{\mathbf{I}}\in{\mathcal{I}}}\overline{X_{\mathbf{I}}}\overline{\sigma^{\mathbf{I}}}
=\sum_{{\mathbf{I}}\in{\mathcal{I}}}(-1)^{\frac{|{\mathbf{I}}|(|{\mathbf{I}}|-1)}{2}}\overline{X_{\mathbf{I}}}\sigma^{\mathbf{I}}.
\label{2.1.25}
\end{equation}
Then, 
\begin{lemma} 
For $X, Y\in{\fC}$ and $\lambda\in{\mathbb{C}}$, we have
\begin{equation}
\overline{(\overline{X})}=X,\quad \overline{XY}=\overline{Y}\,\overline{X},\quad \overline{\lambda X}=\bar\lambda\overline{X}.
\label{conj}\end{equation}
\end{lemma}
\par{\it Proof}.
To prove the second equality, we remark $\displaystyle{
\overline{\sigma^{\mathbf{I}}\sigma^{\mathbf{J}}}
=\overline{\sigma^{\mathbf{J}}}\,\overline{\sigma^{\mathbf{I}}}}$. 
In fact,
$$
\begin{gathered}
\overline{\sigma^{\mathbf{I}}\sigma^{\mathbf{J}}}=\overline{(-1)^{\tau({\mathbf{K}};{\mathbf{I}},{\mathbf{J}})}\sigma^{\mathbf{K}}}
=(-1)^{\tau({\mathbf{K}};{\mathbf{I}},{\mathbf{J}})}(-1)^{\frac{|{\mathbf{K}}|(|{\mathbf{K}}|-1)}{2}}\sigma^{\mathbf{K}},\\
\overline{\sigma^{\mathbf{J}}}\,\overline{\sigma^{\mathbf{I}}}
=(-1)^{\frac{|{\mathbf{I}}|(|{\mathbf{I}}|-1)}{2}}(-1)^{\frac{|{\mathbf{J}}|(|{\mathbf{J}}|-1)}{2}}{\sigma^{\mathbf{J}}}{\sigma^{\mathbf{I}}}
=(-1)^{\frac{|{\mathbf{I}}|(|{\mathbf{I}}|-1)}{2}}(-1)^{\frac{|{\mathbf{J}}|(|{\mathbf{J}}|-1)}{2}}(-1)^{\tau({\mathbf{K}};{\mathbf{J}},{\mathbf{I}})}\sigma^{\mathbf{K}},\\
(-1)^{|{\mathbf{I}}||{\mathbf{J}}|}(-1)^{\tau({\mathbf{K}};{\mathbf{I}},{\mathbf{J}})}=(-1)^{\tau({\mathbf{K}};{\mathbf{J}},{\mathbf{I}})}.
%(-1)^{|{\mathbf{J}}||{\mathbf{K}}|}(-1)^{\tau({\mathbf{I}};J,K)}=(-1)^{\tau({\mathbf{I}};K,J)}.
\end{gathered}
$$
Therefore, we get the desired result. $\qquad\qed$

Moreover,  we have, if ${\mathbf{K}}={\mathbf{I}}+{\mathbf{J}}$,
$$
(-1)^{\tau({\mathbf{K}};{\mathbf{I}},{\mathbf{J}})}(-1)^{\frac{|{\mathbf{K}}|(|{\mathbf{K}}|-1)}{2}}
=(-1)^{\frac{|{\mathbf{I}}|(|{\mathbf{I}}|-1)}{2}}(-1)^{\frac{|{\mathbf{J}}|(|{\mathbf{J}}|-1)}{2}}(-1)^{\tau({\mathbf{K}};{\mathbf{J}},{\mathbf{I}})}.
$$

\begin{remark} %{\it Remark 1. } 
We may introduce ``real" as $\overline{X}=X$ for $X\in{\fC}$, 
or from purely aethetical point of view,
the set of ``reals" may be defined by
$$
{\fR}^{\mathbb{R}}=\{X=\sum_{{\mathbf{I}}\in{\mathcal{I}}}X_{\mathbf{I}}\sigma^{\mathbf{I}}\; | \; X_{\mathbf{I}}\in\euc\},
$$
but we don't use this ``real" in the sequel.
Because the analysis is really done for
the body part and the soul part is used not only for reordering the places but also ``imaginary'',
therefore, we imagine that the set 
$$
{\mathcal R}_F=
\left\{ x=\sum_{{\mathbf{I}}\in{\mathcal{I}}} 
x_{\mathbf{I}}\sigma^{\mathbf{I}}\; \big| \; x_{\mathrm B}\in{\mathbb{R}} \et
x_{\mathbf{I}}\in F \right\}
$$ 
would be more natural as our ``supernumber algebra''.
Here, $F$ should be an associative algebra such that we may define
seminorms analogously as before.
This point of view will be discussed if necessity occurs.
\end{remark}
\begin{remark} %{\it Remark 2. }  
There is another possible way of defining the conjugation:
By Hahn-Banach extension theorem, we may define ${\bar\sigma}_j$ as a linear mapping from
${\fC}$ to $\mathbb{C}$ 
such that $\langle {\bar\sigma}_j,\sigma_k \rangle =\delta_{jk}$, and
by this, we may introduce the duality $\langle \cdot,\cdot \rangle$
between ${\fC}$ and $\bar{\fC}$ which is
the Grassmann algebra generated by $\{{\bar\sigma}_j\}$,
%Then, the algebraic dual space of our ${\fC}$ is 
and whose Fr\'echet topology is compatible with the duality above.
In this case, putting
$\overline{\sigma^I}={\bar\sigma}_n^{i_n}\cdots{\bar\sigma}_1^{i_1}$
for $I=(i_1,\cdots,i_n)$ and
$$
X^*=\sum_{{\mathbf{I}}\in{\mathcal{I}}}\overline{X_{\mathbf{I}}}\overline{\sigma^{\mathbf{I}}}
=\sum_{{\mathbf{I}}\in{\mathcal{I}}}(-1)^{\frac{|{\mathbf{I}}|(|{\mathbf{I}}|-1)}{2}}\overline{X_{\mathbf{I}}}{\bar\sigma}^{\mathbf{I}},
$$
we have also \eqref{conj}.
\end{remark}

\section{Superspace}
\begin{definition}\label{Definition 1.3}
The {super Euclidean space} or (real) superspace $\supermn$ of dimension $m|n$ 
is defined by 
\begin{equation}
\begin{gathered}
\supermn={\fR}_{\mathrm{ev}}^m \times {\fR}_{\mathrm{od}}^n \ni 
X={}^{t\!}({}^{t\!}x,{}^{t\!}\theta),\\
\where
x={}^{t\!}(x_1,\cdots, x_m) 
\et\theta={}^{t\!}(\theta_1,\cdots,\theta_n) \with
x_j\in{\rev},\, \theta_s\in{\rod}. 
\end{gathered}
\label{1.12}
\end{equation}
\end{definition}

Notation: In the following,
we abbreviate the symbol `transposed' ${}^{t\!}(x_1,\cdots, x_m)$ 
and denote $x=(x_1,\cdots, x_m)$, etc.
unless there occurs confusion.

The topology of $\supermn$ is induced from 
the metric defined by
$\dist_{m|n}(X,Y)=\dist_{m|n}(X-Y)$ for $X,Y\in\supermn$, where we put 
\begin{equation}
\dist_{m|n}(X)=
\sum_{j=1}^m
\bigg(
\sum_{{\mathbf{I}}\in\mathcal{I}}{\frac 1{2^{r({\mathbf{I}})}}}
{\frac{|\proj_{\mathbf{I}}(x_j)|}{1+|\proj_{\mathbf{I}}(x_j)|}}
\bigg)
+\sum_{s=1}^n
\bigg( 　\sum_{{\mathbf{I}}\in\mathcal{I}}
{\frac 1{2^{r({\mathbf{I}})}}}
{\frac{|\proj_{\mathbf{I}}(\theta_s)|}{1+|\proj_{\mathbf{I}}(\theta_s)|}}
\bigg).
\label{1.29}
\end{equation}
Clearly, $\dist_{1|1}(X)=\dist(X)$ for 
$X\in{\fR}^{1|1}\cong{\fR}\subset{\fC}$.
Analogously, 
the complex superspace of dimension $m|n$ is defined by 
\begin{equation}
{\fC}^{m|n}= {\fC}_{\mathrm{ev}}^m \times {\fC}_{\mathrm{od}}^n .
\label{1.30}
\end{equation}

We generalize the body map $\pi_{\mathrm B}$ from $\supermn$ 
or $\supermo$ to $\eucm$ by
$\pi_{\mathrm B}X=\pi_{\mathrm B}x
=(\pi_{\mathrm B}x_1,\cdots, \pi_{\mathrm B}x_m)\in\eucm$
for $X=(x,\theta)\in\supermn$.
The (complex) superspace ${\fC}^{m|n}$ is
defined analogously. 

{\bf Dual superspace}. \ \ % Euclidean 
We denote the superspace $\supermn$ by ${\fR}_X^{m|n}$ 
whose point is presented by 
$X=(x,\theta)=(x_1,\cdots,x_m,\theta_1,\cdots,\theta_n)$.
We prepare another superspace ${\fR}_{\Xi}^{m|n}$ 
whose point is denoted by 
$\Xi=(\xi,\pi)=(\xi_1,\cdots,\xi_m,\pi_1,\cdots,\pi_n)$,
such that they are ``dual" each other by
\begin{equation}
\langle X|\Xi \rangle_{m|n} =\sum_{j=1}^m\langle x_j|\xi_j\rangle 
+ \sum_{k=1}^n \langle \theta_k|\pi_k\rangle
 \in {\fR}_{\mathrm{ev}}.
\label{dual}\end{equation}
Or, for any $\hbar\in{\euc}^{\times}$ and ${\spin}\in{\mathbb{C}}^{\times}$, we may put
\begin{equation}
\langle X|\Xi \rangle_{\hbar|\spin} ={\hbar}^{-1}\sum_{j=1}^m\langle x_j|\xi_j\rangle 
+ {\spin}^{-1}\sum_{k=1}^n \langle \theta_k|\pi_k\rangle
 \in {\fR}_{\mathrm{ev}}.
\label{dual-bis}
\end{equation}
We abbreviate above $\langle\cdot|\cdot\rangle_{m|n}$ or $\langle\cdot|\cdot\rangle_{\hbar|\spin}$
by $\langle\cdot|\cdot\rangle$ unless there occurs confusion. 

\section{Rogers' construction of a countably infinite Grassmann generators}\label{rogers-bg}
{We} borrow her construction in A. Rogers~\cite{rog80}.
Denote by ${\mathcal M}_L$ the set of integer sequences given by
$$
{\mathcal M}_L=\{\mu \;|\; \mu=(\mu_1,\mu_2,\cdots,\mu_k ),
\, 1\le \mu_1<\mu_2<\cdots<\mu_k \le L\}
\et{\mathcal M}_\infty=\cup_{L=1}^\infty {\mathcal M}_L.
$$
We regard $\emptyset\in {\mathcal M}_L$ and for any $j\in{\mathbb N}$, 
we put $(j)\in {\mathcal M}_\infty$.
For each $r\in{\mathbb N}$, we may correspond 
a member $\mu \in {\mathcal M}_\infty$
by using
\begin{equation}
r={\frac12}(2^{\mu_1}+2^{\mu_2}+\cdots+2^{\mu_k}).
\label{EA1.1.r}
\end{equation}
Conversely, for each $\mu \in {\mathcal M}_\infty$, we define $e_\mu$ as
$e_\mu=({\overbrace{0,\cdots,0,1}^r},0,\cdots \ )$
where $r$ and $\mu$ are related by (\ref{EA1.1.r}).
Then,
$w=\sum_{\mu}w_\mu e_\mu$.
Now, we introduce the multiplication by
\begin{equation}
\left\{
\begin{aligned}
&e_{\mu} e_{\emptyset}=
e_{\emptyset}e_{\mu} =e_{\mu} \for \mu\in{\mathcal M}_\infty, \\
&e_{(i)} e_{(j)}=-e_{(j)} e_{(i)} \for i,j\in {\mathbb N},\\
&e_\mu=e_{(\mu_1)}e_{(\mu_2)}\cdots e_{(\mu_k)}\where \mu=(\mu_1,\mu_2,\cdots,\mu_k).
\end{aligned}
\right.
\label{Rog-Gras}
\end{equation}
That is, we identify
$$
\omega\ni w=(w_1,w_2,w_3,w_4,\cdots)=\sum_{j=1}w_je_{(j)}
\longleftrightarrow
(w_{(1)},w_{(2)},w_{(1,2)},w_{(3)},\cdots)=\sum_{\mu}w_\mu e_\mu
$$
where
$$
\begin{gathered}
e_{(j)}\leftrightarrow\sigma_j,\;
e_{(1)}e_{(2)}=e_{(1,2)}\leftrightarrow 
\sigma_1\sigma_2=\sigma^I,\; I_{(1,2)}=(1,1,0,\cdots),\\
e_\mu=e_{(\mu_1)}e_{(\mu_2)}\cdots e_{(\mu_k)}\leftrightarrow
\sigma_{\mu_1}\sigma_{\mu_2}\cdots \sigma_{\mu_k}=\sigma^I,\;
I_{\mu}=(\underbrace{\overbrace{0,\cdots,0,1}^{\mu_1},0,\cdots,0,1}_{\mu_k},
0,\cdots).
\end{gathered}
$$
Defining $\sigma_j=e_{(j)}$, we have a countably infinite Grassmann algebra by \eqref{Rog-Gras}.

In stead of the sequence space $\omega$, Rogers uses $\ell^1$ to construct 
the real Banach-Grassmann algebra, which is the set of absolutely convergent sequences 
$$
\Vert X \Vert= \sum_{{\mathbf{I}}\in\mathcal{I}} |X_{\mathbf{I}}| <\infty 
\quad\mbox{for 
$X=\sum_{{\mathbf{I}}\in\mathcal{I}} X_{\mathbf{I}}\sigma^{\mathbf{I}}$ with $X_{\mathbf{I}}\in {\mathbb{R}}$,
such that
$\Vert XY \Vert \leq \Vert X \Vert \Vert Y \Vert$}.
$$
\begin{proposition}[Roger]
$\ell^1$ with the above multiplication forms a Banach-Grassmann algebra with
countably infinite generators.
\end{proposition}

\begin{remark}
There are many papers treating super manifolds which are based on ground ring with Banach-Grassmann structure (for example, V.S. Vladimirov and I.V. Volovich~\cite{VV83}, etc).
This phenomenon is rather reasonable because inverse or implicit function theorems hold in Banach space as same as Euclidian case, but not so in general, in Fr\'echet space. But the condition $\sum_{{\mathbf{I}}\in{\mathcal{I}}}|X_{\mathbf{I}}|<\infty$ is too difficult to check in our concrete problem. This has the similarity to indeterminate coefficients method to solve Cauchy-Kovalevsky theorem and to check its convergence by majorant test.
\end{remark}
\begin{remark}
Concerning  inverse or implicit function theorems in certain Fr\'echet  space, see R. Hamilton~\cite{ham82} or J.T. Schwartz~\cite{Sch-jt69} for Nash's implicit function theorem.
\end{remark}

\chapter{Linear algebra on the superspace}
In this chapter, we quote results from
 F.A. Berezin~\cite{Ber87}, B.S. deWitt~\cite{deW84} and D.A. Leites~\cite{lei80} with modifications if necessary.

\begin{remark} {Almost all papers prefixed ``super'', treated the case of finite number of odd variables also with the finite number of Grassmann generators, or rather, they don't distinguish odd variables and Grassmann generators.
But in any way, after slight modification if necessary, algebraic operations not affected with the topology
is borrowed from these papers.}
\end{remark}

\section{Matrix algebras on the superspace}
\subsection{Super matrices}
\begin{definition}
A rectangular array $M$,  whose cells are indexed by pairs consisting of
a row number and a column number, is called a supermatrix
and denoted by $M\in\Mat((m|n) \times (r|s):{\fC})$,
if it satisfies the following:
\begin{enumerate}
\item
A $(m+n) \times (r+s)$ matrix $M$ 
is decomposed blockwisely as
$M=
\begin{pmatrix}
A&C\\ 
D&B 
\end{pmatrix}$
where $A$, $B$, $C$ and $D$ are $m\times r$, $n\times s$, $m\times s$ and 
$n\times r$ matrices with elements in ${\fC}$, respectively.
\item
One of the following conditions is satisfied: Either
\begin{itemize}
\item
$p(M)=0$, that is, $p(A_{jk})=0=p(B_{uv})$ and $p(C_{jv})=1=p(D_{uk})$ or
\item 
$p(M)=1$, that is, $p(A_{jk})=1=p(B_{uv})$ and $p(C_{jv})=0=p(D_{uk})$. 
\end{itemize}
\end{enumerate}
We call $M$ is even 
denoted by $\Mat_{\mathrm{ev}}((m|n) \times (r|s):\fC)$
(resp. odd denoted by $\Mat_{\mathrm{od}}((m|n) \times (r|s):\fC)$) 
if $p(M)=0$ (resp. $p(M)=1$). Therefore, we have
$$
\Mat((m|n) \times (r|s):{\fC})=\Mat_{\mathrm{ev}}((m|n) \times (r|s):\fC)
\oplus\Mat_{\mathrm{od}}((m|n) \times (r|s):\fC).
$$
Moreover, we may decompose $M$ as $M=M_{\mathrm B}+M_{\mathrm S}$ where
$$
M_{\mathrm B}=
\begin{cases}
\begin{pmatrix}
A_{\mathrm B}&0\\ 
0&B_{\mathrm B} 
\end{pmatrix}
&\when p(M)=0,\\
{}&\\
\begin{pmatrix}
0&C_{\mathrm B}\\ 
D_{\mathrm B}&0
\end{pmatrix}
&\when p(M)=1.
\end{cases}
$$
\end{definition}

The summation of two matrices in $\Mat_{\mathrm{ev}}((m|n) \times (r|s):\fC)$
or in $\Mat_{\mathrm{od}}((m|n) \times (r|s):\fC)$ is defined as usual,
but the sum of $\Mat_{\mathrm{ev}}((m|n) \times (r|s):\fC)$ 
and $\Mat_{\mathrm{od}}((m|n) \times (r|s):\fC)$ is not defined 
except at least one of them being zero matrix.

It is clear that if $M$ is the $(m+n) \times (r+s)$ matrix and $N$ is the 
$(r+s)\times (p+q)$ matrix, 
then we may define the product $MN$ and its parity $p(MN)$ as
$$
(MN)_{ij}=\sum_kM_{ik}N_{kj},\quad
p(MN)=p(M)+p(N)\mod 2.
$$
Moreover, we define
$\Mat[m|n:{\fC}]$ as the algebra of $(m+n)\times (m+n)$ supermatrices.

\subsection{Matrices as Linear Transformations}
By definition of matrix operation to vector, we have
$$
\begin{aligned}
\Mat_{\mathrm{ev}}((m|n)\times(r|s):{\fC})\ni M=\begin{pmatrix}
A&C\\ 
D&B 
\end{pmatrix}:
{\fR}^{r|s}\to\supermn,\\
\Mat_{\mathrm{od}}((m|n)\times(r|s):{\fC})\ni M=\begin{pmatrix}
A&C\\ 
D&B 
\end{pmatrix}:
{\fR}^{r|s}\to{\fR}_{\mathrm{od}}^m\times{\fR}_{\mathrm{ev}}^n,
\end{aligned}
$$
$$
\Mat_{\mathrm{od}}((n|m)\times(m|n):{\fC})\ni\Lambda_{n,m}=
\begin{pmatrix}
0&{\mathbb  I}_n\\
{\mathbb  I}_m&0
\end{pmatrix}:{\fR}_{\mathrm{od}}^m\times{\fR}_{\mathrm{ev}}^n\to
{\fR}_{\mathrm{ev}}^n\times{\fR}_{\mathrm{od}}^m={\fR}^{n|m}.
$$

For elements $X=(x,\theta)=(x_1,\cdots,x_m,\theta_1,\cdots,\theta_n)$ and $\Xi=(\xi,\pi)=(\xi_1,\cdots,\xi_m,\pi_1,\cdots,\pi_n)$ in ${\fR}^{m|n}$, we define $\overline{X}$ and $\langle X|\Xi \rangle_{m|n}$ as in \eqref{2.1.25} and \eqref{dual}, respectively.

If we introduce the duality between $\supermn$ as in \eqref{dual},
we may define the transposed operator as
$$
\langle MX|\Xi\rangle_{m|n}=\langle  X|\,{}^t\!M\,\Xi\rangle_{r|s}
\forany M\in\Mat_{\mathrm{ev}}((m|n)\times(r|s):{\fC}),
$$
for $X=(x,\theta)\in{\fR}^{r|s}$ and $\Xi=(\xi,\omega)\in\supermn$.
More precisely, we have
$$
{}^t\!M=
{}^t\!\begin{pmatrix}
A&C\\ 
D&B 
\end{pmatrix}
=\begin{pmatrix}
{}^t\!A&{}^t\!D\\ 
-{}^t\!C&{}^t\!B
\end{pmatrix}\et {}^{tttt}M=M.
$$
Analogously, 
defining the duality between ${\fC}^{m|n}_Z$ and ${\fC}^{m|n}_\Eta$ for $Z=(z,\theta)\in{\fC}^{r|s}$, $\Eta=(w,\rho)\in{\fC}^{m|n}$ by
$$
\langle Z|\Eta\rangle_{m|n}
=\sum_{j=1}^m \overline{z_j}w_j
+\sum_{k=1}^n \overline{\theta_k}{\rho}_k,
\quad\mbox{or}\quad
=\sum_{j=1}^m  {z}_j\overline{w_j}
+\sum_{k=1}^n {\theta}_k\overline{\rho_k},
$$
we denote the conjugate (or adjoint) matrix of $A$ by $A^*={}^t\!\bar{A}=\overline{{}^t\!A}$ etc.
Then, we may introduce $M^*$, the conjugate (or adjoint) of matrix $M$, by
$$
\langle MZ|\Eta\rangle_{m|n}=\langle Z|M^*\Eta\rangle_{r|s}.
$$
Therefore, we have
$$
M^*=
\begin{pmatrix}
A&C\\ 
D&B 
\end{pmatrix}^*
=\begin{pmatrix}
A^*&-D^*\\ 
-C^*&B^*
\end{pmatrix}\et M^{**}=M.
$$

\begin{lemma} 
For $M\in\Mat((m|n)\times(r|s):{\fC})$
and $N\in \Mat((r|s)\times(p|q):{\fC})$, we have
$$
\begin{gathered}
(MN)^t=N^t\,M^t,\quad
(MN)^*=N^*M^*,\quad % \overline{MN}=\bar M\bar N,\\
(M^t)^t=\Lambda M\Lambda,
%\quad (\bar M)^t=M^*,
\where %=\Lambda\overline{{}^TM}\Lambda
\Lambda=\begin{pmatrix}
{\mathbb  I}_m&0\\ 
0&-{\mathbb  I}_n 
\end{pmatrix}.
\end{gathered}
$$
\end{lemma}

If $M\in \Mat[m|n:{\fC}]$ is even, denoted by 
$M\in \Mat_{\mathrm{ev}}[m|n:{\fC}]$, then $M$ acts on $\supermn$ linearly.
Denoting this by $T_M$, we call it super linear transformation on $\supermn$
and $M$ is called the representative matrix of $T_M$.

\begin{proposition}
Let $M\in \Mat_{\mathrm{ev}}[m|n:{\fC}]$ and assume 
$\det M_{\mathrm B}\neq 0$. %, that is, the body $M_B$ of $M$ is non-singular.
Then, for given $Y\in \supermn$,
\begin{equation}
T_M X=Y
\label{LE}
\end{equation}
has the unique solution $X\in\supermn$, which is denoted by $X=M^{-1}Y$.
\end{proposition}

{\it Proof}.
Since $M_{\mathrm B}$ has the inverse matrix $M_{\mathrm B}^{-1}$, 
\eqref{LE} is reduced to
$$
X+N_{\mathrm S} X=Y',\quad Y'=M_{\mathrm B}^{-1} Y
$$
where $N_{\mathrm S}=M_{\mathrm B}^{-1} M_{\mathrm S}$.
Remark that $N_{\mathrm S} X^{[j]}\in \sum_{k\ge j+1}^\infty {{\fC}^{[k]}}$ 
for $j\ge 0$.
Decomposing by degree, we get
$$
X^{[j]}=Y'^{[j]}- (N_{\mathrm S} X^{(j-1)})^{[j]} \for j=1,2,\dots.
%\sum_{j=0}^\infty 
$$
As $X^{(0)}=X^{[0]}=Y'^{[0]}$, we get 
$X^{[j]}$ from $X^{(j-1)}$ for $j\ge 1$ by induction. $\qquad\qed$

\begin{exercise}
How about $M\in\Mat_{\mathrm{od}}((m|n)\times(n|m):{\fC})$ ?
\end{exercise}

\begin{definition}
$M\in \Mat_{\mathrm{ev}}[m|n:{\fC}]$ is called invertible or non-singular
if $M_{\mathrm B}$ is invertible, 
i.e. $\det A_{\mathrm B}{\cdot} \det B_{\mathrm B} \neq 0$,
and denoted by $M\in \GL_{\mathrm{ev}}[m|n:{\fC}]$.
% if $p(M)=0$ or
%$\det C_{\mathrm B} \det D_{\mathrm B} \neq 0$ if $p(M)=1$.
\end{definition}

\section{Supertrace, superdeterminant}
\subsection{Supertrace}
\begin{lemma}\label{oddtr}
Let $V$, $W$ be two rectangular matrices with odd elements, 
$m\times n$, $n\times m$, respectively.
We have
\newline
(1) $\displaystyle{\tr (VW)^k=-\tr (WV)^k}$ for any  $k=1,2,\cdots$.
\newline
(2) $\displaystyle{\det({\mathbb  I}_m+VW)=\det({\mathbb  I}_n+WV)^{-1}}$.
\end{lemma}
\par{\it Proof. } Let $V=(v_{ij})$, $W=(w_{jk})$ with $v_{ij},w_{jk}\in\cod$.
$$
\begin{aligned}
\tr (VW)^k&=\sum v_{ij_1}w_{j_1j_2}v_{j_2j_3}\cdots 
v_{j_{k-1}j_\ell}w_{j_k i}\\
&=-\sum w_{j_1j_2}v_{j_2j_3}\cdots 
v_{j_{\ell-1}j_k} w_{j_k i}v_{ij_1}=-\tr (WV)^k.
\end{aligned}
$$
Using this, we have $\tr((WV)^{\ell-1}WV)=-\tr(V(WV)^{\ell-1}W)$
which yields$$
\begin{aligned}
\log\det({\mathbb  I}_n+WV)&=\tr\log({\mathbb  I}_n+WV)
=\sum_{\ell}\frac{(-1)^{\ell+1}}{\ell}\tr((WV)^{\ell-1}WV)\\
&=\sum_{\ell}\frac{(-1)^{\ell+1}}{\ell}\big[-\tr(V(WV)^{\ell-1}W)\big]
=-\sum_{\ell}\frac{(-1)^{\ell+1}}{\ell}\tr(VW)^\ell\\
&=-\log\det({\mathbb  I}_m+VW). \qquad \qed
\end{aligned}
$$
\begin{comparison} %\par{\it Comparison. }
If $A=(a_{ij})\in\Mat(m\times n:\cev)$, $B=(b_{jk})\in\Mat(n\times m:\cev)$, 
then we have
\newline
(1) $\displaystyle{\tr (AB)^k=\tr (BA)^k}$,
\newline
(2) $\displaystyle{\det({\mathbb  I}_m+AB)=\det({\mathbb  I}_n+BA)}$.
\end{comparison}

\begin{definition}
Let $M=
\begin{bmatrix}
A&C\\
D&B
\end{bmatrix} \in \Mat[m|n:\fC]$. 
We define the supertrace of $M$ by
$$
\str M %=\sum_{i} (-1)^{(p(M)+1)p_{row}(i)}M_{ii}
= \tr A - (-1)^{p(M)} \tr B.
$$
\end{definition}
 
Using Lemma \ref{oddtr}, we get readily
\begin{proposition}
(a) Let $M,N \in \Mat[m|n:{\fC}]$ such that $p(M)+p(N)\equiv 0 \bmod 2$. 
Then, we have
$$
\str (M+N)= \str M + \str N.
$$
(b) $M$ is a matrix of size $(m+n)\times (r+s)$ and
 $N$ is a matrix of size $(r+s)\times (m+n)$. Then,
$$
\str (MN)= (-1)^{p(M)p(N)} \str (NM).
$$
\end{proposition}

\subsection{Super determinant}

For even supermatrix, we put
\begin{definition}
Let $M$ be a supermatrix. %\in \!\Mat[m|n:{\fC}]$ be given. 
When $\det B_{\mathrm B}\neq 0$, we put
$$
\sdet M=\det (A-CB^{-1}D){\cdot}(\det B)^{-1}
$$
and call it superdeterminant or Berezinian of $M$.
\end{definition}

\begin{corollary}
When $\det B_{\mathrm B}\neq 0$ and $\sdet M\neq 0$, 
then $\det A_{\mathrm B}\neq 0$.
\end{corollary}
\par
\begin{exercise}
Prove the above corollary.
\end{exercise}

For reader's sake, we recall  the definition
in commutative setting.
\begin{definition}
 Let $B=(B_{jk})$ be $(\ell\times \ell)$-matrix with elements in ${\cev}$, 
denoted by, $B\in\Mat[\ell:\cev]$.
As ${\cev}$ is a commutative ring, we may define $\det B$ as usual:
$$
\det B=\sum_{\rho\in\wp_\ell} \sgn (\rho) B_{1\,\rho(1)}\cdots B_{\ell\,\rho(\ell)}.
$$
\end{definition}
Then, analogously as ordinary case, we have, 
\begin{equation}
\det(AB)=\det A{\cdot}\det B,\quad \det(\exp A)=\exp(\tr A) \for A,B\in\Mat[\ell:\cev].
\end{equation}
Moreover, for block matrix case, we have
\begin{comparison} 
Let 
$$
A=\begin{pmatrix}
A_{11}&A_{12}\\
A_{21}&A_{22}
\end{pmatrix},\quad
M=\begin{pmatrix}
{\mathbb  I}_m&0\\
-A_{22}^{-1}A_{21}&{\mathbb  I}_n
\end{pmatrix},
$$
be block matrices of even elements. Then, we have
\begin{equation}
\det A=\det (AM)=\det
\begin{pmatrix}
A_{11}-A_{12}A_{22}^{-1}A_{21}&A_{12}\\
0&A_{22}
\end{pmatrix}
=\det(A_{11}-A_{12}A_{22}^{-1}A_{21}){\cdot}\det A_{22}.
\label{block-det-evn}
\end{equation}
\end{comparison}
\par
\begin{remark} 
It seems meaningful to cite here the result of F.J. Dyson~\cite{dys72},
\begin{quotation}
{\small
\begin{theorem}[Dyson]
Let $R$ be a ring with a unit element and without divisors of zero.
Assume that on the matrix ring $A$ with $n>1$, 
a mapping $D$ exists satisfying the following axioms:
\par
Axiom 1. For any $a\in A$, $D(a)=0$ if and only if 
there is a non-zero $w\in W$ with
$aw=0$. Here, $W$ is the set of single-column matrices with elements in $R$.
\par 
Axiom 2. $D(a)D(b)=D(ab)$.
\par 
Axiom 3. Let the elements of $a$ be $a_{ij}$ $i,j=1,\cdots,n$,
and similarly for $b$ and $c$.
If for some row-index $k$ we have
$$
\begin{cases}
a_{ij}=b_{ij}=c_{ij}, &\quad i\neq k\\
a_{ij}+b_{ij}=c_{ij}, &\quad i=k,
\end{cases}
$$
then
$$
D(a)+D(b)=D(c).
$$
Then, $R$ is commutative.
\end{theorem}}
\end{quotation}
This theorem states that 
if the elements of matrix are taken from non-commutative algebra,
then it is impossible to define the determinant 
having above three properties.
But, he claims a certain `determinant' is defined for some class of
matrices with elements in `quarternion' requiring 
only one or two properties above 
(By the way, Moore's point of view, 
is reconsidered significantly in that paper).
%\newline
In fact, we may define ``superdeterminant" for ``supermatrix" as above
which staisfies the properties below.
\end{remark}

 Now, we continue to study the properties of super-determinant defined in the previous lecture.
 
 Following decomposition of a even supermatrix $M$ will be useful:
\begin{equation}
\begin{aligned}
	{\begin{bmatrix}
	  A&C\\ 
	  D&B 
	  \end{bmatrix}}
&=
  \begin{bmatrix}
     {{\mathbb  I}_m}&{CB^{-1}}\\ 
     0&{{\mathbb  I}_n} 
     \end{bmatrix}\,
  \begin{bmatrix}
     {A-CB^{-1}D}&0\\ 
     0&B 
     \end{bmatrix}\,
  \begin{bmatrix}
     {{\mathbb  I}_m}&0\\ 
     {B^{-1}D}&{{\mathbb  I}_n} 
     \end{bmatrix} \quad\text{if $\det B_{\mathrm B}\neq{0}$},\\
&=
  \begin{bmatrix}
     {{\mathbb  I}_m}&0\\ 
     {DA^{-1}}&{{\mathbb  I}_n} 
     \end{bmatrix}\,
  \begin{bmatrix}
     {A}&0\\ 
     0&{B-DA^{-1}C} 
     \end{bmatrix}\,
  \begin{bmatrix}
     {{\mathbb  I}_m}&{A^{-1}C}\\ 
     0&{{\mathbb  I}_n} 
     \end{bmatrix} \quad\text{if $\det A_{\mathrm B}\neq{0}$}.
\end{aligned}
\end{equation}

Moreover, we have
\begin{equation}
\begin{aligned}
&\begin{bmatrix}
	{\tilde A}&{\tilde C}\\ 
	{\tilde D}&{\tilde B} 
	\end{bmatrix}
 =
\begin{bmatrix}
	{A}&{C}\\ 
	D&{B} 
	\end{bmatrix}^{-1}
=\begin{bmatrix}
(A-CB^{-1}D)^{-1}& -A^{-1}C(B-DA^{-1}C)^{-1}\\
-B^{-1}D(A-CB^{-1}D)^{-1}&(B-DA^{-1}C)^{-1}
\end{bmatrix}\nonumber\\
& \qquad =
	\begin{bmatrix}
{({\mathbb  I}_m-A^{-1}CB^{-1}D)^{-1}A^{-1}}
&{-({\mathbb  I}_m-A^{-1}CB^{-1}D)^{-1}A^{-1}CB^{-1}}\\ 
{-({\mathbb  I}_n-B^{-1}DA^{-1}C)^{-1}B^{-1}DA^{-1}}
&{({\mathbb  I}_n-B^{-1}DA^{-1}C)^{-1}B^{-1}} 
	\end{bmatrix}\\
& \qquad =
	\begin{bmatrix}
{A^{-1}({\mathbb  I}_m-CB^{-1}DA^{-1})^{-1}}
&{-A^{-1}CB^{-1}({\mathbb  I}_n-DA^{-1}CB^{-1})^{-1}}\\ 
{-B^{-1}DA^{-1}({\mathbb  I}_m-CB^{-1}DA^{-1})^{-1}}
&{B^{-1}({\mathbb  I}_n-DA^{-1}CB^{-1})^{-1}} 
	\end{bmatrix}
\end{aligned}
\end{equation}

\begin{equation}
\begin{aligned}
\sdet
	\begin{bmatrix}
	{A}&{C}\\ 
	D&{B} 
	\end{bmatrix}
& =(\det A)(\det B)^{-1}
	\det ({\mathbb  I}_m-A^{-1}CB^{-1}D)\\
&  =(\det A)(\det B)^{-1}
	\det ({\mathbb  I}_m-CB^{-1}DA^{-1})=(\det {\tilde A})^{-1}(\det B)^{-1}\\
& =(\det A)(\det B)^{-1}
	\det ({\mathbb  I}_n-B^{-1}DA^{-1}C)\\
& =(\det A)(\det B)^{-1}
	\det ({\mathbb  I}_n-DA^{-1}CB^{-1})=(\det A)(\det {\tilde B}).
\end{aligned}
\end{equation}
As we have the following
\begin{equation}
	\begin{bmatrix}
	{A}&{C}\\ 
	D&{B} 
	\end{bmatrix}
	\begin{bmatrix}
	{A^{-1}}&{0}\\ 
	0&{B^{-1}} 
	\end{bmatrix}
	\begin{bmatrix}
	{{\mathbb  I}_m}&{0}\\ 
	0&{-{\mathbb  I}_n} 
	\end{bmatrix}
	\begin{bmatrix}
	{A}&{C}\\ 
	D&{B} 
	\end{bmatrix}
	\begin{bmatrix}
	{{\mathbb  I}_m}&{0}\\ 
	0&{-{\mathbb  I}_n} 
	\end{bmatrix}
=
	\begin{bmatrix}
	{A-CB^{-1}D}&{0}\\ 
	0&{B-DA^{-1}C} 
	\end{bmatrix},
\end{equation}
we guarantee the invertibility of matrices appeared above.

\begin{lemma}
(1) Let $L\in\Mat_{\mathrm {ev}}[\ell:\cev]$ 
such that the product of
any two entries of it is zero. Then
$$
({\mathbb  I}_\ell+L)^{-1}={\mathbb  I}_\ell-L,\quad \det({\mathbb  I}_\ell+L)=1+\tr L.
$$
(2) Let $M\in\Mat_{\mathrm {ev}}[m|n:{\fC}]$
such that the product of
any two entries of it is zero. Then
$$
\sdet({\mathbb  I}_{m+n}+M)=1+\str M.
$$
\end{lemma}
\par{\it Proof. }
(1) Remarking
$$
({\mathbb  I}_\ell+L)^{-1}={\mathbb  I}_\ell-L+L^2-L^3+\cdots
\et
\det(e^L)=e^{\tr L},
$$
we get the result readily.
\newline
(2) For $M=\begin{bmatrix}
	A&C\\ 
	D&B 
	\end{bmatrix}$,
satisfying $C({\mathbb  I}_n+B)^{-1}D=0$ and $\tr A\tr B=0$ guaranteed by 
the product of any two entries of $M$ being zero,
$$
\begin{aligned}
\sdet({\mathbb  I}_{m+n}+M)&=\det({\mathbb  I}_m+A-C({\mathbb  I}_n+B)^{-1}D){\cdot}
\det({\mathbb  I}_n+B)^{-1}\\
&=\det({\mathbb  I}_m+A){\cdot}\det({\mathbb  I}_n-B)
=1+\tr A-\tr B=1+\str M.  \qquad\qed
\end{aligned}
$$

\begin{corollary}
When $\det B_{\mathrm B}\neq 0$ and $\sdet M\neq 0$, 
then $\det A_{\mathrm B}\neq 0$.
\end{corollary}
\par
\begin{exercise}
Prove the above corollary.
\end{exercise}

\begin{theorem}
Let $M,N \in \Mat[m|n:{\fC}]$.
\par
(1)
If $M$ is invertible, then we have $\sdet M\neq 0$. 
Moreover, if $A$ is nonsingular, then 
\begin{equation}
(\sdet M)^{-1}=(\det A)^{-1} {\cdot}\det (B-DA^{-1}C).
\label{inv-sdet}
\end{equation}
\par
(2) Multiplicativity of $\sdet$:
\begin{equation}
\sdet (MN)=\sdet M{\cdot} \sdet N.
\end{equation}
\par
(3)
$\str$ and $\sdet$ are matrix invariants.
That is, if $N$ is invertible, then
\begin{equation}
\str M=(-1)^{p(M)+p(N)}\str (NMN^{-1}), \quad
\sdet M= \sdet (NMN^{-1}).
\label{strsdet}\end{equation}
%\end{enumerate}
\end{theorem}

{\it Proof} (due to Leites \cite{lei80}).
(1) By
\begin{equation}
{\begin{bmatrix}
	  A&C\\ 
	  D&B 
	  \end{bmatrix}}
=
  \begin{bmatrix}
     {{\mathbb I}_m}&0\\ 
     {DA^{-1}}&{{\mathbb I}_n} 
     \end{bmatrix}\,
  \begin{bmatrix}
     {A}&0\\ 
     0&{B-DA^{-1}C} 
     \end{bmatrix}\,
  \begin{bmatrix}
     {{\mathbb I}_m}&{A^{-1}C}\\ 
     0&{{\mathbb I}_n} 
     \end{bmatrix} \quad\text{if $\det A_{\mathrm B}\neq{0}$},
\end{equation}
we have readily by definition,
$\sdet M=\det A(\det(B-DA^{-1}C))^{-1}$, which yields
\eqref{inv-sdet}.
\par
(2) [Step 1]: Let ${\mathcal   G}_+$, ${\mathcal   G}_0$ and ${\mathcal   G}_-$ 
be subgroups of $\GL[m|n:{\fC}]$, 
given by
$$
{\mathcal   G}_+=\bigg\{
\begin{bmatrix}
{\mathbb I}_m&C\\
0&{\mathbb I}_n\end{bmatrix}\bigg\},\quad
{\mathcal   G}_0=\bigg\{
\begin{bmatrix}
A&0\\
0&B\end{bmatrix}\bigg\},\quad
{\mathcal   G}_-=\bigg\{
\begin{bmatrix}
{\mathbb I}_m&0\\
D&{\mathbb I}_n\end{bmatrix}\bigg\}.
$$
Then, we have, $M=M_+M_0M_-$ with
$M_+\in {\mathcal   G}_+$, $M_0\in {\mathcal   G}_0$ and $M_-\in {\mathcal   G}_-$.
i.e., for any $M\in\GL[m|n:{\fC}]$,
\begin{equation}
M=\begin{bmatrix}
A&C\\
D&B\end{bmatrix}=
\begin{bmatrix}
     {{\mathbb I}_m}&{CB^{-1}}\\ 
     0&{{\mathbb I}_n} 
     \end{bmatrix}\,
  \begin{bmatrix}
     {A-CB^{-1}D}&0\\ 
     0&B 
     \end{bmatrix}\,
  \begin{bmatrix}
     {{\mathbb I}_m}&0\\ 
     {B^{-1}D}&{{\mathbb I}_n} 
     \end{bmatrix} \quad\text{if $\det B_{\mathrm B}\neq{0}$}.
\end{equation}
Remarking that
$$
\begin{bmatrix}
     {{\mathbb I}_m}&C\\ 
     0&{{\mathbb I}_n} 
     \end{bmatrix}\times
\begin{bmatrix}
     {{\mathbb I}_m}&C'\\ 
     0&{{\mathbb I}_n} 
     \end{bmatrix}
=\begin{bmatrix}
     {{\mathbb I}_m}&C+C'\\ 
     0&{{\mathbb I}_n} 
     \end{bmatrix},
$$
we introduce the notion of elemantary matrices having the form
$$
\begin{bmatrix}
     {{\mathbb I}_m}&E\\ 
     0&{{\mathbb I}_n} 
     \end{bmatrix}
$$
where $E$ has only one non-zero entry.

[Step 2]: We claim $\sdet(MN)=\sdet M{\cdot}\sdet N$ whenever 
$M\in {\mathcal   G}_+$ or $M\in {\mathcal   G}_0$,
and similarly, whenever $N\in {\mathcal   G}_0$ or $N\in {\mathcal   G}_-$.
For example, when
$$
M=\begin{bmatrix}
     {{\mathbb I}_m}&C'\\ 
     0&{{\mathbb I}_n} 
     \end{bmatrix}\in {\mathcal   G}_+\quad
N=\begin{bmatrix}
     A&C\\ 
     D&B
     \end{bmatrix},
$$
we have
$$
\begin{aligned}
\sdet(MN)&=\sdet
\begin{bmatrix}
     {{\mathbb I}_m}&C'\\ 
     0&{{\mathbb I}_n} 
     \end{bmatrix}
\begin{bmatrix}
     A&C\\ 
     D&B
     \end{bmatrix}
=\sdet\begin{bmatrix}
     A+C'D&C+C'B\\ 
     D&B
     \end{bmatrix}\\
&=\det(A+C'D-(C+C'B)B^{-1}D){\cdot}(\det D)^{-1}
=\det(A-CB^{-1}D){\cdot}(\det D)^{-1}\\
&=\sdet M{\cdot}\sdet N.
\end{aligned}
$$
\begin{exercise}
Check other cases analogously.
\end{exercise}

[Step 3]: We claim that $\sdet(MN)=\sdet M{\cdot}\sdet N$ for any elementary matrix $N$
$$
N=\begin{bmatrix}
     {{\mathbb I}_m}&E\\ 
     0&{{\mathbb I}_n} 
     \end{bmatrix}\in {\mathcal   G}_+.
$$
Since we have
$$
\begin{aligned}
&\sdet(MN)=\sdet(M_+(M_0M_-N))=\sdet M_+{\cdot}\sdet(M_0(M_-N)))=\sdet M_0{\cdot}\sdet(M_-N),\\
&\sdet M{\cdot}\sdet N=\sdet M_0{\cdot}\sdet M_-{\cdot}\sdet N,
\end{aligned}
$$
by Step 1 and Step 2, we need to prove
$$
\sdet(M_-N)=\sdet M_-{\cdot}\sdet N=1
$$
when $N$ is an elementary matrix.
By definition,
$$
\sdet
\begin{bmatrix}
     {{\mathbb I}_m}&0\\ 
     D&{{\mathbb I}_n} 
     \end{bmatrix}
\begin{bmatrix}
     {{\mathbb I}_m}&E\\ 
     0&{{\mathbb I}_n} 
     \end{bmatrix}
=\sdet
\begin{bmatrix}
     {{\mathbb I}_m}&E\\ 
     D&{{\mathbb I}_n+DE} 
     \end{bmatrix}
=\det(1-E(1+DE)^{-1}D){\cdot}\det(1+DE)^{-1}.
$$
As $E$ has only one non-zero entry, the product of any two of
the matrices $E$, $DE$, $E(1+DE)^{-1}D$ is zero.
Applying Lemma, we get, by $(1+DE)^{-1}=1-DE$ and $E\cdot DE=0$,
$$
\sdet(M_-N)=\det(1-DE){\cdot}(\det(1+DE))^{-1}=(1-\tr DE)(1+\tr DE)^{-1}.
$$
As $\tr DE=-\tr ED$, we have
$$
\sdet(M_-N)=1=\sdet M_-{\cdot}\sdet N.
$$

[Step 4]: Put
$$
{\mathcal   G}=\bigg\{N\in\GL[m|n:{\fR}]\;|\;\sdet(MN)=\sdet M{\cdot}\sdet N
\forany M\in\GL[m|n:{\fR}]
\bigg\}.
$$
For $N_1, N_2\in {\mathcal   G}$, we have
\begin{equation}
\begin{aligned}
\sdet(M\cdot N_1N_2)&=\sdet((MN_1)N_2)=\sdet(MN_1){\cdot}\sdet N_2\\
&=\sdet M{\cdot}\sdet N_1{\cdot}\sdet N_2=\sdet M{\cdot}\sdet (N_1N_2),
\end{aligned}\label{comp}
\end{equation}
which implies ${\mathcal   G}$ froms a group. By Steps 2 and 3, 
${\mathcal   G}$ contains ${\mathcal   G}_-$ and ${\mathcal   G}_0$ 
and all elementary matrices $N\in {\mathcal   G}_+$.
By Step1, $\GL[m|n:{\fC}]$ is generated by these matrices,
we have ${\mathcal   G}=\GL[m|n:{\fC}]$, that is,
$\sdet(MN)=\sdet M{\cdot}\sdet N$.  
\par
(3)
Let $N,M$ be given.
Then, using \eqref{comp}, we get
$$
\str NMN^{-1}=(-1)^{p(N)p(MN^{-1})}\str MN^{-1}N=(-1)^{p(N)+p(M)}\str M,
$$
since $p(MN^{-1})=p(M)+p(N^{-1})\mod 2$ and $0=p(NN^{-1})=p(N)+p(N^{-1})\mod 2$,
we have
$p(N)p(MN^{-1})=p(N)+p(M)\mod 2$.
\par
Using \eqref{comp}, we have $\sdet (MN)=\sdet (NM)$ which implies
$\sdet (NMN^{-1})=\sdet (N^{-1}NM)=\sdet M$.   $\qquad\qed$

\begin{theorem}[Liouville's theorem: Theorem 3.5 of ~\cite{Ber87}]
Let $M(t)\in\Mat[m|n:{\fC}]$ with a real parameter $t$.
Let $X(t)\in \Mat[m|n:{\fC}]$ satisfy 
\begin{equation}
\dt X(t)=M(t)X(t),\quad X(0)={\mathbb I}_{m+n}.
\label{L1}
\end{equation}
Then $X(t)\in \GL[m|n:{\fC}]$, and
\begin{equation}
\sdet X(t)=\exp\{\int_0^t ds\str M(s)\}.
\label{L2}
\end{equation}
\end{theorem}
\par{\it Proof} (with slight modification of Berezin's proof in \cite{Ber87}).
Let $\tilde X(t)$ be a solution of
$$
\dt {\tilde X}(t)=-{\tilde X}(t)M(t),\quad {\tilde X}(0)={\mathbb I}_{m+n}.
$$
Then, since
$$
\dt({\tilde X}(t)X(t))=0\with {\tilde X}(0)X(0)={\mathbb I}_{m+n},
$$
we have
${\tilde X}(t)X(t)={\mathbb I}_{m+n}$ which implies $X(t)\in\GL[m|n:{\fC}]$.
\par
Let 
$$
M(t)=\begin{bmatrix}
A(t)&C(t)\\
D(t)&B(t)
\end{bmatrix},\quad
X(t)=\begin{bmatrix}
X_{11}(t)&X_{12}(t)\\
X_{21}(t)&X_{22}(t)
\end{bmatrix}.
$$
Then, we put
$Y(t)=X_{11}(t)-X_{12}(t)X_{22}^{-1}(t)X_{21}(t)$ and 
$Z=X_{22}^{-1}(t)$.
Differentiating $X_{22}^{-1}X_{22}={\mathbb{I}}_n$ w.r.t. $t$ and substituting
$\dot{X_{22}}=DA_{12}+BX_{22}$ which is obtained from \eqref{L1}, we have, 
$$
\dt Z=-Z(DX_{12}X_{22}^{-1}+B).
$$
Analogously calculating, we get
$$
\dt Y=(A-X_{12}X_{22}^{-1}D)Y.
$$
As all elements appeared in the above equations are even, 
we may apply the classical Liouville theorem to have
$$
\dt\det Y=\tr(A-X_{12}X_{22}^{-1}D)\det Y,\quad
\dt \det Z=-\tr(DX_{12}X_{22}^{-1}+B)\det Z.
$$
Putting $V=X_{12}X_{22}^{-1}$ and $W=D$ in Lemma~\ref{oddtr}, we get
$\tr(A-X_{12}X_{22}^{-1}D)=\tr(A+DX_{12}X_{22}^{-1})$,
therefore, recalling the definition of super-determinant, we have
$$
\dt\sdet X=\dt(\det Y{\cdot}\det Z)=\tr(A-B){\cdot}\det Y{\cdot}\det Z=\str M{\cdot}\sdet X
\with
\sdet X(0)=1.
$$
This yields the desired result after integrating w.r.t. $t$.  $\qquad\qed$

\begin{corollary}
For $M,N\in\Mat_{\mathrm{ev}}[m|n:{\fC}]$ we have
\begin{align}
&\sdet (MN)=\sdet M{\cdot}\sdet N,\nonumber\\
&\exp(\str M)=\sdet(\exp M).
\end{align}
\end{corollary}
\par{\it Proof. }
(1) Put $X(t)=(1-t){\mathbb I}_{m+n}+tM$ and $Y(t)=(1-t){\mathbb I}_{m+n}+tN$.
As $X(t)$ and $Y(t)$ are differentiable in $t$ and invertible except at most
one $t$, we my define
$$
A(t)=\frac{d X(t)}{dt}X(t)^{-1},\quad
B(t)=\frac{d X(t)}{dt}Y(t)^{-1}.
$$
Then
$$
\dt(X(t)Y(t))=(A(t)+B_1(t))X(t)Y(t) 
\where B_1(t)=X(t)B(t)X(t)^{-1}.
$$
Applying above theorem, we have
$$
\begin{aligned}
\sdet(MN)=\sdet(X(1)Y(1))&=\exp\{\int_0^1 ds\str(A(t)+B_1(t))\}
=\exp\{\int_0^1 ds(\str A(t)+\str B(t))\}\\
&=\sdet X(1){\cdot}\sdet Y(1)=\sdet M{\cdot}\sdet N.
\end{aligned}
$$
(2) Putting $M(t)=M$, $X(t)=e^{tM}$ and $t=1$ in theorem above, 
we get the desired result.  $\quad\qed$%  \qquad

\begin{comparison}[cited from ``Encyclopaedia of Mathematics'' ed. M. Hazewinkel]
\par
Liouville-Ostrogradski formula (or Liouville formula) :
A relation that connects the Wronskian of a system of solutions and the coefficients of an ordinary linear differential equation.

Let $x_1(t),{\cdots}, x_n(t)$  be an arbitrary system of  solutions of a homogeneous system of  linear first-order equations
	\begin{equation}
	x'=A(t)x,\qquad x\in\euc^n
	\label{LO-1}
	\end{equation}
with an operator $A(t)$ that is continuous on an interval $I$, and let
$$
W(x_1(t),{\cdots}, x_n(t))=W(t)
$$
be the Wronskian of this system of solutions. The Liouville-Ostrogradski formula has the form
	\begin{equation}
	\dt W(t)=A(t){\cdot}\tr A(t),\qquad t\in I
	\label{LO-2}
	\end{equation}
or, equivalently,
	\begin{equation}
W(x_1(t),{\cdots}, x_n(t))=
W(x_1({\unbt}),{\cdots}, x_n({\unbt})){\cdot} \exp\{\int_{\unbt}^t ds\,\tr A(s)\},\quad t, {\unbt}\in I.
	\label{LO-3}
	\end{equation}
Here, $\tr A(t)$ is the trace of the operator $A(t)$. The Liouville-Ostrogradski formula can be written by means of the Cauchy operator  $X(t,{\unbt})$ of the system \eqref{LO-1} as follows:
	\begin{equation}
	\det X(t,{\unbt})=\exp\{\int_{\unbt}^t ds\,\tr A(s)\},\quad t, {\unbt}\in I.
	\label{LO-4}
	\end{equation}
The geometrical meaning of  \eqref{LO-4}  (or  \eqref{LO-3} ) is that as a result of the transformation $X(t,{\unbt}):\euc^n\to\euc^n$  the oriented volume of any body is increased by a factor $\exp\{\int_{\unbt}^t ds\,\tr A(s)\}$.
\end{comparison} 

\section{An example of diagonalization}
\begin{definition}
A supermatrix $M=\begin{pmatrix}
A&C\\
D&B
\end{pmatrix}\in\Mat[m|n:{\fC}]$ is called generic 
if all eigenvalues of $M_{\mathrm B}$ as $\Mat[m+n:{\mathbb  C}]$ are
different each others.
\end{definition}
\begin{theorem}[Berezin]\label{diagonal-generic}%\cite{bere87}
Let $M\in\Mat[m|n:{\fC}]$ be generic. Then, there exists a matrix
$X\in\GL[m|n:{\fC}]$ such that $E=XMX^{-1}$ is diagonal.
\end{theorem}
\par{\it Proof. }
Decomposing the equality $EX=XM$ with respect to the degree, we have
\begin{equation}
(EX)^{[k]}=\sum_{j=0}^k E^{[j]}X^{[k-j]}
=\sum_{j=0}^k X^{[j]}M^{[k-j]}=(XM)^{[k]}.
\label{b-generic}
\end{equation}
From this, we want to construct $X^{[k]}$ and $E^{[k]}$:
For $k=0$, we have
\begin{equation}
E^{[0]}X^{[0]}=X^{[0]}M^{[0]}.
\label{b-generic0}
\end{equation}
By the assumption, there exist $X^{[0]}_{11}$, 
$E^{[0]}_{11}=\mbox{diagonal matrix with\;}(\lambda_1^{[0]},\cdots,\lambda_m^{[0]})$
and $X^{[0]}_{22}$,$E^{[0]}_{22}=\mbox{diagonal matrix with\;}(\lambda_{m+1}^{[0]},\cdots,\lambda_{m+n}^{[0]})$
such that 
$$
X^{[0]}_{11}A_{\mathrm B}=E^{[0]}_{11}X^{[0]}_{11}\et
X^{[0]}_{22}B_{\mathrm B}=E^{[0]}_{22}X^{[0]}_{22}.
$$
Defining
$$
X^{[0]}=\begin{pmatrix}
X^{[0]}_{11}&0\\
0&X^{[0]}_{22}
\end{pmatrix},\;
E^{[0]}=\begin{pmatrix}
E^{[0]}_{11}&0\\
0&E^{[0]}_{22}
\end{pmatrix},
$$
we have the desired one satisfying \eqref{b-generic0}.
\par
Assume that there exist $X^{[j]}$ and $E^{[j]}$ for $0\le j\le k-1$
satisfying \eqref{b-generic}.
Multiplying $(X^{[0]}){}^{-1}$ from the right to \eqref{b-generic} for $k$, 
we have
\begin{equation}
E^{[0]}X^{[k]}(X^{[0]}){}^{-1}-X^{[k]}(X^{[0]}){}^{-1}E^{[0]}+E^{[k]}=K^{[k]}
\label{b-generic2}\end{equation}
where
$$
K^{[k]}=(\sum_{j=0}^{k-1}X^{[j]}M^{[k-j]})(X^{[0]}){}^{-1}-
(\sum_{j=1}^{k-1} E^{[j]}X^{[k-j]})(X^{[0]}){}^{-1}.
$$
By inductive assumption, the matrix $K^{[k]}$ is known 
and belongs to $\Mat[m|n:{\fC}]$.
From \eqref{b-generic2}, we have
\begin{equation}
(\lambda_i^{[0]}-\lambda_j^{[0]})(X^{[k]}(X^{[0]}){}^{-1})_{ij}
+\lambda_i^{[k]}\delta_{ij}=(K^{[k]})_{ij}.
\label{b-generick}
\end{equation}
This equation is uniquely solvable 
since $\lambda_i^{[0]}\neq\lambda_j^{[0]}$ and
$$
\begin{cases}
\lambda_i^{[k]}=(K^{[k]})_{ii},&{}\\
(X^{[k]}(X^{[0]}){}^{-1})_{ij}
=\frac{(K^{[k]})_{ij}}{\lambda_i^{[0]}-\lambda_j^{[0]}},
&\for i\neq j.
\end{cases}
$$
Therefore, we define $X^{[j]}$ and $E_{[j]}$ for any $j\ge0$.  
Since $X^{[0]}$ is invertible, $X\in \GL[m|n:{\fC}]$.
This implies $X$ and $E$ are defined as desired.  $\qquad\qed$

\begin{problem}
Find a condition for a supermatrix $M$ being diagonalizable? Is ``generic'' condition in Theorem~\ref{diagonal-generic} necessary?
\end{problem}

\subsection{A simple example}
Let
$$
Q=\begin{pmatrix}
x_1&\theta_1\\
\theta_2&ix_2
\end{pmatrix}\with x_1,\,x_2\in\rev,\; \theta_1,\,\theta_2\in\rod,
$$
which maps ${\fR}^{1|1}$ to ${\fR}^{1|1}$ 
or $\rod\times{i}\rev$ to $\rod\times{i}\rev$.
This supermatrix appears in Efetov's calculation in Random Matrix Theory (see for example, K.B. Efetov~\cite{efe83},
A. Inoue and Y. Nomura~\cite{IN00}).

\subsection{Invertibility of $Q$}
Find $Y$ for a given $V$ such that
$$
QY=V\with Y=\begin{pmatrix}
y_1\\
\omega_2\end{pmatrix},\; V=\begin{pmatrix}
v_1\\
\rho_2\end{pmatrix}\in{\fR}^{1|1},
$$
$$
x_1y_1+\theta_1\omega_2=v_1,\; \theta_2y_1+ix_2\omega_2=\rho_2.
$$
If $(x_1x_2)_{\mathrm B}\neq{0}$, we have readily
$$
y_1=\frac{ix_2v_1-\theta_1\rho_2}{D_-},\;
\omega_2=\frac{x_1\rho_2-\theta_2v_1}{D_+}
\with
D_{\pm}=ix_1x_2{\pm}\theta_1\theta_2.
%,\; D_+=ix_1x_2+\theta_1\theta_2.
$$
Analogously, for
$$
\tilde Y=\begin{pmatrix}
\omega_1\\iy_2
\end{pmatrix}\in \rod\times{i}\rev,\; \tilde V=\begin{pmatrix}
\rho_1\\
v_2\end{pmatrix}\in\rod\times\rev,
$$
satisfying $Q\tilde Y=\tilde V$, we have
$$
\omega_1=\frac{ix_2\rho_1-\theta_1v_2}{D_-},\;
iy_2=\frac{x_1v_2-\theta_2\rho_1}{D_+}.
$$

To relate the above quantity with the $\sdet Q$, we proceed as follows:
Let
$$
Y=\begin{pmatrix}
y_1&\omega_1\\
\omega_2&iy_2
\end{pmatrix}\with QY=YQ={I}_2.
$$
Then, from $QY={I}_2$, we have
$$
\begin{matrix}
x_1y_1+\theta_1\omega_2=1,& x_1\omega_1+iy_2\theta_1=0,\\
\theta_2y_1+ix_2\omega_2=0,&\theta_2\omega_1-x_2y_2=1.
\end{matrix}
$$
Therefore, we have
$$
Y=\begin{pmatrix}
\frac{ix_2}{D_-}
&-\frac{\theta_1}{D_-}\\
-\frac{\theta_2}{D_+}
&\frac{x_1}{D_+}
\end{pmatrix}
=(\sdet Q)^{-1}
\begin{pmatrix}
\frac{1}{ix_2}
&\frac{\theta_1}{x_2^2}\\
\frac{\theta_2}{x_2^2}
&-\frac{x_1x_2+2i\theta_1\theta_2}{x_2^3}
\end{pmatrix},
$$
which yields $YQ=I_2$ also.
Here, we used
$$
\sdet Q=\det(x_1-\theta_1(ix_2)^{-1}\theta_2){\cdot}(\det(ix_2))^{-1}
=\frac{ix_1x_2-\theta_1\theta_2}{(ix_2)^2},\;
(\sdet Q)^{-1}=\frac{ix_1x_2+\theta_1\theta_2}{x_1^2}.
$$
Therefore,
$$
\begin{pmatrix}
y_1\\
\omega_2\end{pmatrix}
=\frac{D_+}{x_1^2}\begin{pmatrix}
\frac{1}{ix_2}&\frac{-\theta_1}{(ix_2)^2}\\
\frac{-\theta_2}{(ix_2)^2}&\frac{ix_1x_2-2\theta_1\theta_2}{(ix_2)^3}
\end{pmatrix}
\begin{pmatrix}
v_1\\
\rho_2
\end{pmatrix}
=\frac{D_+}{x_1^2}\begin{pmatrix}
\frac{ix_2v_1-\theta_1\rho_2}{(ix_2)^2}\\
\frac{-ix_2\theta_2v_1+(ix_1x_2-2\theta_1\theta_2)\rho_2}{(ix_2)^3}
\end{pmatrix},
$$
$$
\begin{pmatrix}
\omega_1\\iy_2
\end{pmatrix}=\frac{D_+}{x_1^2}\begin{pmatrix}
\frac{1}{ix_2}&\frac{-\theta_1}{(ix_2)^2}\\
\frac{-\theta_2}{(ix_2)^2}&\frac{ix_1x_2-2\theta_1\theta_2}{(ix_2)^3}
\end{pmatrix}
\begin{pmatrix}
\rho_1\\
v_2
\end{pmatrix}
=\frac{D_+}{x_1^2}\begin{pmatrix}
\frac{ix_2\rho_1-i\theta_1v_2}{(ix_2)^2}\\
\frac{-ix_2\theta_2\rho_1+(ix_1x_2-2\theta_1\theta_2)v_2}{(ix_2)^3}
\end{pmatrix}.
$$

\subsection{Eigenvalues of $Q$}
Let
$$
QU=\lambda U\with U=\begin{pmatrix}
u\\
\omega\end{pmatrix},\quad u\in\rev,\;\omega\in\rod,\;\lambda\in\rev.
$$
Then,
$$
(x_1-\lambda)u+\theta_1\omega=0,\; \theta_2u+(ix_2-\lambda)\omega=0.
$$
Putting
$$
D_+(\lambda)=(x_1-\lambda)(ix_2-\lambda)+\theta_1\theta_2,\;
D_-(\lambda)=(x_1-\lambda)(ix_2-\lambda)-\theta_1\theta_2,
$$
we have
$$
D_-(\lambda)u=0,\;
D_+(\lambda)\omega=0.
$$
To guarantee the existence of $u_{\mathrm B}\neq{0}$ satisfying above,
we take $\lambda$ satisfying
$$
D_-(\lambda)=\lambda^2-(x_1+ix_2)\lambda+ix_1x_2-\theta_1\theta_2=0.
$$
This yields
$$
\lambda=x_1+\frac{\theta_1\theta_2}{x_1-ix_2}\quad\mbox{(or
$\lambda=ix_2-\frac{\theta_1\theta_2}{x_1-ix_2}$, but $\pi_{\mathrm{B}}(\lambda)\not\in\euc$)}
$$
and
$$
U=\begin{pmatrix}
1\\
\frac{\theta_2}{x_1-ix_2}
\end{pmatrix},\quad
QU=(x_1+\frac{\theta_1\theta_2}{x_1-ix_2})U.
$$
Analogously, we seek $\tilde\lambda\in{i}\rev$, $\tilde{U}\in\rod\times\rev$
satisfying $Q\tilde{U}=\tilde\lambda\tilde{U}$ which is given
$$
\tilde U=\begin{pmatrix}
\frac{-\theta_1}{x_1-ix_2}\\
1
\end{pmatrix},\quad
Q\tilde U=(ix_2+\frac{\theta_1\theta_2}{x_1-ix_2})\tilde U.
$$
Therefore,
$$
Q\begin{pmatrix}
1&-\frac{\theta_1}{x_1-ix_2}\\
\frac{\theta_2}{x_1-ix_2}&1
\end{pmatrix}
=
\begin{pmatrix}
1&-\frac{\theta_1}{x_1-ix_2}\\
\frac{\theta_2}{x_1-ix_2}&1
\end{pmatrix}
\begin{pmatrix}
x_1+\frac{\theta_1\theta_2}{x_1-ix_2}&0\\
0&ix_2+\frac{\theta_1\theta_2}{x_1-ix_2}
\end{pmatrix}.
$$

\subsection{Diagonalization of $Q$}
We may diagonalize the matrix $Q$ by using the change of variables
\begin{equation}
\varphi^{-1}(x,\theta)=(y,\omega)=\left\{
{\begin{aligned}
&y_1=x_1+\frac{\theta_1\theta_2}{x_1-ix_2},\;
y_2=x_2-\frac{i\theta_1\theta_2}{x_1-ix_2},\\
&\omega_1=\frac{\theta_1}{x_1-ix_2},\;
\omega_2=-\frac{\theta_2}{x_1-ix_2},
\end{aligned}}
\right.
\label{LAC2}
\end{equation}
or
\begin{equation}
\varphi(y,\omega)=(x,\theta)=\left\{
{\begin{aligned}
&x_1=y_1+{\omega_1\omega_2}(y_1-iy_2),\;
x_2=y_2-{i\omega_1\omega_2}(y_1-iy_2),\\
&\theta_1=\omega_1(y_1-iy_2),\;
\theta_2=-\omega_2(y_1-iy_2),
\end{aligned}}
\right.
\label{LAC3}
\end{equation}
such that
\begin{equation}
GQG^{-1}=\begin{pmatrix}
y_1&0\\
0&iy_2
\end{pmatrix},\quad
GQ^2G^{-1}=\begin{pmatrix}
y_1^2&0\\
0&-y_2^2
\end{pmatrix}
\end{equation}
where
$$
G=\begin{pmatrix}
1+2^{-1}\omega_1\omega_2&\omega_1\\
\omega_2&1-2^{-1}\omega_1\omega_2
\end{pmatrix},\quad
G^{-1}=\begin{pmatrix}
1+2^{-1}\omega_1\omega_2&-\omega_1\\
-\omega_2&1-2^{-1}\omega_1\omega_2
\end{pmatrix}.
$$
It is clear that 
$$
\begin{gathered}
\str Q=x_1-ix_2=y_1-iy_2=\str GQG^{-1},\et\\
\str Q^2=x_1^2+x_2^2+2\theta_1\theta_2=y_1^2+y_2^2=\str(GQG^{-1})^2.
\end{gathered}
$$

\begin{center}

========  Mini Column 2:  Tensor and exterior algebras, interior product ========

 \end{center}

As a characteristic feature of mathematical thought, it some times happens that to generalize that situation makes it easier to understand. Though I have a tendency to feel bothered and sleepy following lengthy algebraic procedure, but I try to collect some terminology from T. Yokonuma~\cite{Yok77}(in Japanese).

\paragraph{Tensor algebras:}
Let $V$ be a $d$-dimensional vector space with inner product $(\cdot,\cdot)$.
Put $V^*$ is a set of linear operators on $V$, then $V^*$ and $V$ is dual each other by ${}_{V^*}\langle\cdot,\cdot\rangle_{V}$.

\begin{theorem}
Let $k$ be a field with characteristic 0, and let $V_1,V_2,{\cdots},V_n$ be finite dimensional linear spaces on $k$.
Then, we have a unique pair $(U_0, \iota_n)$ satisfying following properties $(\otimes)_1$, $(\otimes)_2$, here $U_0$ is a linear space on $k$ and $n$-times linear map $\iota_n\in{\mathcal{L}}(V_1,V_2,{\cdots},V_n:U_0)$.
\par
$(\otimes)_1$  $U_0$ is generated by image of $\iota_n$, $\iota_n(V_1\times V_2\times{\cdots}\times V_n:U_0)$.
\par
$(\otimes)_2$  For any $\Phi\in{\mathcal{L}}(V_1,V_2,{\cdots},V_n:U)$, there exists a linear map $F:U_0\to U$ such that $\Phi=F {\circ} \iota_n$.
\begin{comment}
$$
\xymatrix{
{V_1\times V_2\times{\cdots}\times V_n}  \ar@{->}[rd]_-\Phi  \ar[r]^-\iota & B \ar@{.>}[d]^{F}\\
&D}
$$
\end{comment}
%%
\end{theorem}
\begin{definition}
$(U_0,\iota_n)$ defined in the above theorem is called tensor product of $V_1,V_2,{\cdots},V_n$, and denoted by
$$
U_0=V_1\otimes V_2\otimes{\cdots}\otimes V_n,\quad
\iota_n(v_1,v_2,{\cdots}.v_n)=v_1\otimes v_2\otimes{\cdots}\otimes v_n\;\;(v_i\in V_i).
$$
\end{definition}
\begin{definition}
For matrices $A=(\alpha_{ij}),\; B=(\beta_{ij})$, we define a matrix
$$
\begin{pmatrix}
\alpha_{11}B&\alpha_{12}B&{\cdots}&\alpha_{1n}B\\
\alpha_{21}B&{\cdots}&{\cdots}&\alpha_{2n}B\\
{\vdots}&{}&{}&{\vdots}\\
\alpha_{m1}B&\alpha_{m2}B&{\cdots}&\alpha_{mn}B
\end{pmatrix},
$$
which is called the tensor (or Kronecker) product of $A$ and $B$ and denoted by $A\otimes B$. If $A$ is $(m,n)$-matrix and $B$ is $(m',n')$-matrix, then $A\otimes B$ is $(mm',nn')$-matrix.
\end{definition}

Let $V$ be a linear space over $k$ and $V^*$ be the dual of $V$. Then,
$$
{T_q}^{p}(V)=\overbrace{V\otimes {\cdots}\otimes V}^{\text{$p$-times}}\otimes\overbrace{V^*\otimes{\cdots}\otimes V^*}^{\text{$q$-times}}
$$
is denoted  by $(p,q)$-tensor space.  We put ${T_0}^{0}(V)=k$, ${T_0}^p(V)=T^p(V)$,
${T_q}^0(V)=T_q(V)$ and ${T_0}^{0}(V)=T^0(V)=T_0(V)$.

\begin{remark}
Here, we use identification $V\otimes V^*\equiv V^*\otimes V$, 
$V\otimes V^*\otimes V\otimes V\otimes V^*
\equiv  V^*\otimes V^*\otimes V\otimes V\otimes V$, etc.
\end{remark}

$(p,0)$- or $(0,q)$- tensors are called $p$-th order contravariant  or $q$-th order covariant tensor, respectively.
An element in ${T_0}^1(V)=V$ is called contravariant vector, 
one in ${T_1}^0(V)=V^*$ is called covariant vector and one in ${T_0}^{0}(V)=k$ is scalar.

From $\Hom(V,V)\equiv V^*\otimes V\equiv {T_1}^1(V)$, linear transformation  on $V$ is regarded as $(1,1)$-tensor.
Since ${\mathcal{L}}(V,V:k)\equiv V^*\otimes V^*={T_2}^0(V)$, bilinear form onり$V$ is $2$-th covariant tensor.

\begin{proposition}[contraction] Take integers $p>0$, $q>0$ and consider a tensor space ${T_q}^p(V)$.
For any integers $r$, $s$ satisfying $1\le r\le p$ and $1\le s\le q$, there exists a unique linear map
${c_s}^r:{T_q}^p(V)\to {T_{q-1}}^{p-1}(V)$ satisfying the following:
For any $v_i\in V$, $\varphi_j\in V^*$,
$$
{c_s}^r(v_1\otimes{\cdots}\otimes v_p\otimes\varphi_1\otimes{\cdots}\otimes\varphi_q)
=\varphi_s(v_r)v_1\otimes{\cdots}\otimes\check{v}_r\otimes{\cdots}\otimes v_p
\otimes\varphi_1\otimes{\cdots}\otimes\check{\varphi}_s\otimes{\cdots}\otimes \varphi_q.
$$
\end{proposition}

{\bf Remark.} In the above, $\check{v}_r$ or $\check{\varphi}_s$ stands for deleting that component, respectively.
This map ${c_s}^r$ is called contraction w.r.t $r$-th contravariant index and $s$-th covariant index.

The permutation group with $p$ letters, $\{1,2,{\cdots},p\}$, is denoted by $\mathfrak{S}_p$
and each $\sigma\in\mathfrak{S}_p$ has the signature $\sgn\sigma=\pm$.
\begin{proposition}
(1) For each $\sigma\in\mathfrak{S}_p$, there exists uniquely a linear transformation $P_{\sigma}$ of $T^p(V)$ satisfying
$$
P_{\sigma}(v_1\otimes{\cdots}\otimes v_p)=v_{\sigma^{-1}(1)}\otimes{\cdots}\otimes v_{\sigma^{-1}(p)}
\quad (v_i\in V).
$$
(2) For $\sigma, \tau, 1 \in\mathfrak{S}_p$, we have
$$
P_{\sigma}P_{\tau}=P_{\sigma\tau}, P_1=I.
$$
\end{proposition}

\begin{definition}
An element $t\in T^p(V)$ is called a symmetric tensor when it satisfies for any $\sigma\in\mathfrak{S}_p$. All such elements is denoted by $S^p(V)$.
In case $P_{\sigma}(t)=(\sgn\sigma) t$ for any $\sigma\in\mathfrak{S}_p$, it is called alternating (=anti-symmmetric?)  tensor whose set is denoted by $A^p(V)$.
\par
For $t\in S^p(V)$, $t'\in S^q(V)$, we define product of them as
$t{\cdot}t'={\mathcal{S}}_{p+q}(t\otimes t')$.
\par
For $t\in A^p(V)$, $t'\in A^q(V)$, we define exterior product of them as
$t{\wedge}t'={\mathcal{A}}_{p+q}(t\otimes t')$.
\end{definition}

\begin{definition}
We put 
$$
{\mathcal{S}}_p=\frac{1}{p!}\sum_{\sigma\in \mathfrak{S}_p} P_{\sigma},\;\;
{\mathcal{A}}_p=\frac{1}{p!}\sum_{\sigma\in \mathfrak{S}_p} (\sgn\sigma)P_{\sigma}.
$$
\end{definition}
Unless there occurs confusion, we simply denote them as ${\mathcal{S}}_p={\mathcal{S}}$, ${\mathcal{A}}_p={\mathcal{A}}$.

\begin{definition}
In infinite direct sum
$$
T(V)=\bigoplus_{p=0}^{\infty} T^p(V),
$$
is called tensor algebra,  if we introduce addition and product as follows:
 $$
t=\sum_{j=0}^{\infty}t_j, t'=\sum_{j=0}^{\infty}t'_j \in T(V),\;\; t_j, t'_j\in T^j(V),\;\;\alpha\in k
\Longrightarrow
\begin{cases}
t+t'=\sum_{j=0}^{\infty}(t_j+t'_j),\\
\alpha t=\sum_{j=0}^{\infty}\alpha t_j,\\
t\otimes t'=\sum_{p=0}^{\infty}\big(\sum_{r+s=p}t_r\otimes t'_s\big).
\end{cases}
$$
\end{definition}

Analogously, we put 
\begin{definition}
Introducing product ${\cdot}$ in
$$
S(V)=\bigoplus_{p=0}^{\infty} S^p(V)
$$
as
$$
t\in S^p(V),\; t'\in S^q(V)\Longrightarrow t{\cdot} t'={\mathcal{S}}_{p+q}(t\otimes t'),
$$
we have a symmetric algebra $S(V)$ on $V$.
\end{definition}

\paragraph{\bf Exterior algebra:}
\begin{definition}
Remarking $A^p(V)=0$ for  $p>n$, we have
$$
A(V)=\bigoplus_{p=0}^{\infty} A^p(V)=\bigoplus_{p=0}^{n} A^p(V).
$$
We define the exterior product $\wedge$ as
$$
t\wedge t'=\sum_{p,q}t_p\wedge t'_q=\sum_{k=0}^{n}{\mathcal{A}}_k(\sum_{p+q=k}t_p\otimes t'_q).
$$
$A(V)$ is called the exterior algebra on $V$.
\end{definition}
\begin{lemma}
(1) $v_i\in V\Longrightarrow {\mathcal{A}}(v_1\otimes{\cdots}\otimes v_p)=v_1\wedge{\cdots}\wedge v_p$,\\
(2) $\sigma\in{\mathfrak{S}}_p\Longrightarrow v_{\sigma^{-1}1}\wedge{\cdots}\wedge v_{\sigma^{-1}p}
=\sign\sigma (v_1\wedge{\cdots}\wedge v_p)$.\\
(3) $t\in A^p(V), t'\in A^q(V)\Longrightarrow t'\wedge t=(-1)^{pq}t\wedge t'$.
\end{lemma}

The $p$-th order covariant tensor space $T_p(V)=T^p(V^*)$ with inner (or scalar) product
$$
\begin{aligned}
\varphi_1\otimes{\cdots}\otimes\varphi_p\in T_p(V),\;
& v_1\otimes{\cdots}\otimes v_p\in T^p(V)\\
&\longrightarrow \langle v_1\otimes{\cdots}\otimes v_p, \varphi_1\otimes{\cdots}\otimes\varphi_p\rangle
=\varphi_1(v_1){\cdots} \varphi_p(v_p)\\
&\qquad\qquad\qquad\qquad(v_i\in V, \varphi_j\in V^*)
\end{aligned}
$$
is regarded as the dual of $T^p(V)$.

A bilinear form $\langle{\cdot}|{\cdot}\rangle_p$ on $A^p(V)\times A^p(V^*)$ is defined as
$$
\langle z|\xi \rangle_p=p!\langle z,\xi \rangle\;\;(z\in A^p(V),\; \xi\in A^p(V^*)).
$$
Then,
\begin{proposition}
(1) For $z=v_1\wedge{\cdots}\wedge v_p\;(v_i\in V)$ and
$\xi=\varphi_1\wedge{\cdots}\wedge \varphi_p\;(\varphi_i\in V^*)$, we have
$$
\langle z|\xi \rangle_p=\det(\varphi_i(v_j)).
$$
(2) $A^p(V)$ and $A^p(V^*)$ are dual each other by the scalar product $\langle{\cdot}|{\cdot}\rangle_p$.\\
(3) For $z=\sum_{p=0}^n z_p\in A(V)\;(z_p\in A^p(V))$ and $\xi=\sum_{p=0}^n \xi_p\in A(V^*)\;(\xi_p\in A^p(V^*))$, we define the scalar product
$$
\langle z|\xi\rangle=\sum_{p=0}^n \langle z_p|\xi_p\rangle_p
$$
then. $A(V)$ and $A(V^*)$ form dual spaces each other.
\end{proposition}

\begin{definition}
For $\xi\in A(V^*)$, we define a linear transformation $\delta(\xi)$ on $A(V^*)$ as
$$
\delta(\xi)\zeta=\xi\wedge\zeta\;\;(\zeta\in A(V^*))
$$
which is called (left)exterior multiplication.
Transposed map of this $\delta(\xi)$ is denoted by $\partial(\xi)$ and called interior product (or multiplication) by $\xi$:$$
\langle \partial(\xi)z|\zeta\rangle=\langle z|\delta(\xi)\zeta\rangle=\langle z|\xi\wedge\zeta\rangle.
$$
\end{definition}

{\bf Remark:} Above defined operations are denoted also by $\delta(\xi){\cdot}=\xi\wedge{\cdot}$, $\partial(\xi){\cdot}=\xi\rfloor{\cdot}$.

\begin{center}  

======= End of Mini Column 2 =======

\end{center}

\chapter{Elementary differential calculus on superspace}
On real Euclidian space $\eucm$, to begin with, we consider a real-valued, continuous and smooth function.
On the other hand if we work on complex space ${\mathbb{C}}^m$ with a complex valued function, it seems natural to develop complex analytic functions. From these, what is a natural candidate for a function on superspace $\supermn$.
This chapter and the next one are rewritten rather significantly from the original lectures.

\section{G\^ateaux or Fr\'echet differentiability on Banach spaces}
For the future use, we prepare the following lemma:
\begin{lemma}[see, Lemma 2.1.14 of Berger~\cite{Berg77}]
Let $X_1, X_2, Y$ be Banach spaces.
The Banach spaces $L(X_1,X_2:Y )$ and $L(X_1:L(X_2:Y ))$ are identical up to a linear isometry.
\end{lemma}
\begin{definition} %\par{\it Remark.}
Let $(X, \Vert\cdot\Vert_X)$ and $(Y, \Vert\cdot\Vert_Y)$ be two Banach spaces.\\
(i) A function $\Phi:X\to Y$ is called \underline{G\^ateaux(or G-) differentiable at $x\in X$ 
in the direction $h\in X$} if there exists an element
$\Phi'_G(x;h)\in Y$ such that %, linear with respect to $h$
$$
\Vert\Phi(x+th)-\Phi(x)-t\Phi'_G(x;h)\Vert_Y\to 0\when t\to0,\;\mbox{i.e.}\;
\frac{d}{dt}\Phi(x+th)\big|_{t=0}=\Phi'_G(x;h).
$$
$\Phi'_G(x;h)$ is also denoted by $\Phi'_G(x)(h)$, $d_G\Phi(x;h)$ or $(d_G\Phi(x))(h)$.
The second order G\^ateaux-derivatives $d_G^{(2)}\Phi(x;h)$ at $x\in X$ in the direction $h=(h_1,h_2)\in X^2$ is defined by
$$
\begin{aligned}
d_G^{\prime\prime}\Phi(x;h)=d_G^{(2)}\Phi(x;h)&=d_G(d_G\Phi(x;h_1);h_2)\\
&=\frac{d}{dt}d_G\Phi(x+th_2;h_1)\big|_{t=0}
=\frac{\partial^2}{\partial t_1\partial t_2}\Phi(x+t_1h_1+t_2h_2)\big|_{t_1=t_2=0}.
\end{aligned}
$$
Analogously, we may define N-th G\^ateaux-derivatives $d_G^{(N)}\Phi(x;h)$ (or $\Phi_G^{(N)}(x;h)$) with $h=(h_1,{\cdots},h_N)\in X^N$.
If this $d_G^N\Phi(x;h_1,{\cdots},h_{N})$ exists, then it is symmetric w.r.t. $(h_1,{\cdots},h_N)$.\\
(ii)  $\Phi:X\to Y$ is called \underline{Fr\'echet(or F-) differentiable at $x\in X$} 
if there exist a bounded linear operator 
$\Phi'_F(x):X\to Y$ and an element $\tau(x,h)\in Y$ such that
$$
\Phi(x+h)-\Phi(x)-\Phi'_F(x)h=\tau(x,h) 
\with \Vert \tau(x,h)\Vert_Y=o(\Vert h\Vert_X).
$$
It is clear that if $\Phi'_F(x)$ (or $d_F\Phi(x)$) exists, then $\Phi'_G(x)$ exists also and $\Phi'_G(x)=\Phi'_F(x)$.
The second order Fr\'echet-derivatives $\Phi_F^{\prime\prime}(x;h)$ at $x\in X$ is defined
if $\Phi'_F:X\to L(X:Y)$ is differentiable at $x\in X$ in the Fr\'echet sense.
In this case, $\Phi_F^{\prime\prime}\in L(X:L(X:Y))\cong L_2(X:Y)=L(X,X:Y)$.
It is denoted by $\Phi\in C^2(U:Y)$ if (a) $\Phi$ is twice Fr\'echet differentiable, and (b) $\Phi_F^{\prime\prime}(x):U\to L(X,X:Y)$ is continuous.
We define analogously N-th  Fr\'echet differivative $\Phi_F^{(N)}$ and a class of N-times Fr\'echet differentiable functions $C^N(U:Y)$. That is, 
$\Phi\in C_F^N(X:Y)$ means for each $x\in U\subset X$, $\Phi$ is $N$-times Fr\'echet- differentiable and
$\Phi_F^{(N)}(x)$ is  a continuous map from $U$ to $L(\overbrace{X\times{\cdots}\times X}^{N}:Y)=L_N(X:Y)$  w.r.t. $x$ \end{definition}
\begin{theorem}[see, Theorem 2.1.13 of Berger~{\cite{Berg77}}]
If $\Phi: X\to Y$ be Fr\'echet-differentiable at $x$, it is G\^ateaux-differentiable at $x$.
Conversely, if the G\^ateaux derivative of $\Phi$ at $x$, $d_G\Phi(x,h)$, is linear in $h$ and is continuous  in $x$ as a map from $X\to L(X:Y)$, then $\Phi$ is  Fr\'echet-differentiable at $x$. In either case, we have
$\Phi_G'(x)y=\Phi'_F(x,y)$.
\end{theorem}
\begin{theorem}[see, Theorem 2.1.27 of Berger~{\cite{Berg77}}]
If $\Phi: X\to Y$ be N-times Fr\'echet-differentiable in a neighbourhood $U$ at $x$
and $\Phi_F^{(N)}(x)(h^{(1)},{\cdots}, h^{(N)})$ denotes the N-th Fr\'echet derivative, 
the $\Phi$ is N-times G\^ateaux-differentiable and
$$
d_G^{(N)}\Phi(x;h^{(1)},{\cdots}, h^{(N)})= \Phi_F^{(N)}(x)(h^{(1)},{\cdots}, h^{(N)}).
$$
Conversely, if the N-th G\^ateaux derivative $d_G^{(N)}\Phi(x;h^{(1)},{\cdots}, h^{(N)})$ of $\Phi$ exists in a neighbourhood of $U$ of $x$, $d_G^{(N)}\Phi(x;h^{(1)},{\cdots}, h^{(N)})\in L_N(X:Y)$, and as a function of $x$, $d_G^{(N)}\Phi(x;h^{(1)},{\cdots}, h^{(N)})$ is continuous from $U$ to $L_N(X:Y)$,
then $\Phi$ is N-times Fr\'echet-differentiable  and the two derivatives are equal at x.
\end{theorem}

\begin{problem}
How does one extend these notion of differentiability to those on functions on ${\fR}^{m|n}$?
\end{problem}

\section{G\^ateaux or Fr\'echet differentiable functions on Fr\'echet spaces}\label{GFonF}
In this section, I borrow representations in R. Hamilton's paper~\cite{ham82} which I overooked  when lecture had been prepared.

\subsection{G\^ateaux-differentiability}
\begin{definition}[G\^ateaux-derivative, -differential and -differentiability]
(i) Let ${X}$, ${Y}$ be Fr\'echet spaces with countable seminorms $\{p_m\}$, $\{q_n\}$, respectively.
Let $U$ be an open subset of ${X}$. For a function $f:U{\to} {Y}$,
we say that $f$ is {1-time G\^ateaux (or $G$-)differentiable} at $x\in U$ in the direction $y{\in} X$ 
if there exists the following limit in ${Y}$:
$$
\lim_{t\to0}\frac{f(x+ty)-f(x)}{t}=\frac{df(x+ty)}{dt}\bigg|_{t=0}=d_Gf(x;y)=d_Gf(x)\{y\}=d_Gf(x)y=f'_G(x)y,
$$
i.e., for given $x\in U$ and $y\in {X}$  there exists an element $d_Gf(x;y)\in Y$ such that
for any $n\in{\mathbb{N}}$, we have
$$
q_n({f(x+ty)-f(x)}-{t}d_Gf(x;y))=o(t).
$$
We call this $d_Gf(x;y)$ the $G$-differential of $f$ at $x$ in the direction $y$ and denoted as above,
and $d_Gf(x)$ or $f_G'(x)$ are called  the {$G$-derivative}.
Moreover,  $f$ is said to be $G$-differentiable in $U$ and denoted by $f\in C_G^{1-}(U:{Y})$ if $f$ has
the $G$-differential $d_Gf(x;y)$ for every $x\in U$ and any direction $y\in {X}$. 
A map $f:U\to {Y}$ is said to be 1-time continuously $G$-differentiable on $U$, denoted by $f\in C_G^1(U:{Y})$, if $f$ has $G$-derivative in $U$ and
if $d_Gf: U\times {X}\ni(x,y)\to d_Gf(x;y)\in{Y}$ is jointly continuous.\\
(ii) If ${X}$, ${Y}$ are Banach spaces with norms $\Vert{\cdot}\Vert_X$, 
$\Vert{\cdot}\Vert_Y$, respectively,
then, $f$ has $G$-differential
$df(x;y)\in{Y}$ at $x\in U$ in the direction $y\in {X}$ {if and only if} 
$$
\Vert f(x+ty)-f(x)- t d_Gf(x;y)\Vert_Y=o(|t|)\;\;\mbox{as $t\to0$}.
$$
Moreover,  $f\in C_G^1(U:Y)$ {if and only if} $f$ is $G$-differentiable at $x$ 
and $d_Gf$ is continuous from $U\ni x$ to $d_Gf(x)\in{\mathbf{L}}(X:Y)$.
\end{definition}
\begin{proposition}[see, pp.76-77 of \cite{ham82}]
Let ${X}$, ${Y}$ be Fr\'echet spaces and let $U$ be an open subset of ${X}$.
If $f\in C_G^1(U:{Y})$, then $d_Gf(x;y)$ is linear in $y$.
\end{proposition}

\begin{remark}[see, p.70 of \cite{ham82}]
It should be remarked that even if $X, Y, Z$ are Banach spaces and $U\subset X$, 
there exists the difference between 
$$
``L:U\times Y\to Z\;\mbox{is continuous}"\et
``L: U\to {\mathbf{L}}(Y:Z)\;\mbox{is continuous}".
$$
\end{remark}
\begin{definition}[Higher order derivatives, see, p.80 of \cite{ham82}]
Let ${X}$, ${Y}$ be Fr\'echet spaces.\\
(i) If the following limit exists, we put
$$
d^2_Gf(x)\{y,z\}=d^2_Gf(x;y,z)=\lim_{t\to0}\frac{d_Gf(x+tz;y)-d_Gf(x;y)}{t}.
$$
Moreover, $f$ is said to be  $C_G^2(U:{Y})$ if $d_Gf$ is $C_G^1(U\times {X}:{Y})$, which happens {if and only if} $d^2_Gf$ exists and is continuous, that is,
$d^2_Gf$ is jointly continuous from $U\times{X}\times{X}\to {Y}$.\\
(ii) Analogously, we define
$$ 
d^n_Gf:U\times \overbrace{{X}\times{\cdots}\times{X}}^n\ni (x,y_1,{\cdots}, y_n)
\to d^n_Gf(x)\{y_1,{\cdots},y_n\}=d^n_Gf(x;y_1,{\cdots},y_n)\in {Y}.
$$
%%%%%
$$
\begin{aligned}
\frac{\partial^N}{\partial t_1{\cdots}\partial t_N}\Phi(x+\sum_{j=1}^Nt_jh_j)\bigg|_{t_1={\cdots}=t_N=0}
&=\frac{d}{dt_N}d_G^{N-1}\Phi(x+t_Nh_N;h_1,{\cdots},h_{N-1})\bigg|_{t_N=0}\\
&=d_G(d_G^{N-1}\Phi(x;h_1,{\cdots},h_{N-1});h_N)\\
&=d_G^N\Phi(x;h_1,{\cdots},h_{N})=\Phi_G^{(N)}(x;h_1,{\cdots},h_{N}).
\end{aligned}
$$
%%%%%
$f$ is  said to be $C_G^n(U:{Y})$ {if and only if} $d^n_Gf$ exists and is continuous. We put
$C_G^{\infty}(U:{Y})=\cap _{n=0}^{\infty}C_G^n(U:{Y})$.
\end{definition}
\begin{definition}[Many variables case]
(i) Let ${X}_1$, ${X}_2$, ${Y}$ be Fr\'echet spaces.
For  $x=(x_1,x_2)\in X_1\times X_2$ and $z=(z_1,z_2)\in X_1\times X_2$, we put
$$
\begin{gathered}
\partial_{x_1} f(x)\{z_1\}=f_{x_1}(x;z_1)=f_{x_1}(x)z_1=\lim_{t\to0}\frac{f(x_1+t z_1,x_2)-f(x_1,x_2)}{t},\\
\partial_{x_2} f(x)\{z_2\}=f_{x_2}(x;z_2)=f_{x_2}(x)z_2=\lim_{t\to0}\frac{f(x_1,x_2+t z_2)-f(x_1,x_2)}{t}.
\end{gathered}
$$
They are called partial derivatives.
We define the total $G$-derivative as
$$
d_Gf(x)\{z\}=f'_{x}(x;z)=\lim_{t\to0}\frac{f(x_1+tz_1,x_2+tz_2)-f(x_1,x_2)}{t}.
$$
For $f:X\to Y$ with $X=\prod_{i=1}^n X_i$, we define $\partial_{x_j}f(x)$ and $d_Gf(x)$ for $x=(x_1,{\cdots},x_n)$, analogously.\\
(ii) If ${X}_1$, ${X}_2$, ${Y}$ are Banach spaces, we may define analogously the above notion.
\end{definition}

\begin{proposition}
Let $\{X_i\}_{i=1}^{n}$, $Y$ be Fr\'echet spaces and let $U$ be an open set in $X=\prod_{i=1}^{n}X_i$. \\
(a)  $f\in C_G^{1}(U:{Y})$, i.e. $d_Gf(x)\{y\}$ exists and is continuous, 
if and only if $\partial_{x_j}f(x)\{{\cdot}\}$ exist and are continuous, and we have, for $x=(x_i)_{i=1}^{n},\; y=(y_i)_{i=1}^{n}\in X$,
\begin{equation}
d_Gf(x;y)=d_Gf(x)\{y\}=\sum_{i=1}^{n}f_{x_i}(x;y_i)=\sum_{i=1}^{n}f_{x_i}(x)\{y_i\}.
\label{total}
\end{equation}
(b)[Taylor's formula]  Moreover, if $f\in C_G^{p}(U:{Y})$, we have
\begin{equation}
\begin{gathered}
f(x+y)=\sum_{k=0}^p\frac{1}{k!} d_G^kf(x)\{\overbrace{y,{\cdots},y}^k\}+R_pf(x,y)\\
\with
\lim_{t\to0}t^{-p}R_pf(x,ty)=0 \for y\in{X},
\end{gathered}
\label{tay-f}
\end{equation}
where
$$
R_{p}f(x,y)=\int_0^1\frac{(1-s)^{p-1}}{(p-1)!}\frac{d^{p}}{ds^{p}}f(x+sy) ds.
$$
\end{proposition}
\par{\it Proof}.
\eqref{total} is proved in Theorem 3.4.3 of \cite{ham82} for $N=2$.
\eqref{tay-f} is given, for example, in p.101 of Keller~\cite{Kel74}, et al. $\qquad\qed$

\subsection{Fr\'echet-differentiability}
\begin{definition}[see, Definition 1.8. of L. Schwartz~\cite{Sch-jt69}]
(i) Let ${X}$, ${Y}$ be Fr\'echet spaces, and let $U$ be an open subset of ${X}$. 
A function $\varphi:U\to {Y}$ is said to be
{horizontal (or tangential) at $0$} {if and only if} for each neighbourhood $V$ of $0$ in
${Y}$ there exists a neighbourhood $U'$ of $0$ in ${X}$, and a
function $o(t):(-1,1)\to{\euc}$ such that
\begin{equation}
\varphi(tU')\subset o(t)V \with \lim_{t\to0}\frac{o(t)}{t}=0,
\label{FD11}
\end{equation}
i.e. for any seminorm $q_n$ on ${Y}$ and $\epsilon>0$, there exists a seminorm $p_m$ on $E$ and $\delta>0$ such that 
\begin{equation}
q_n(\varphi(tx))\le \epsilon t \for
p_m(x)<1,\;\;|t|\le \delta
\label{FD12}
\end{equation}
From \eqref{FD12}, putting
$V=\{z\in {Y}\;|\; q_n(z)<1\},\; U'=\{x\in {X}\;|\; p_m(x)<1\}$,
we may recover \eqref{FD11}.\\
(ii) For given Banach spaces $(X,\Vert{\cdot}\Vert_X)$ 
and $(Y,\Vert{\cdot}\Vert_Y)$, ``horizontal" implies
$$
\Vert\varphi(x)\Vert_Y\le\Vert x\Vert_X\psi(x)\with \psi:X\to\mathbb{R},\;\; \lim_{x\to0}\psi(x)=0
\quad \mbox{i.e. $\Vert\varphi(x)\Vert_Y=o(\Vert x\Vert_X)$ as $\Vert x\Vert_X\to0$.}
$$
\end{definition}
\begin{definition}[Fr\'echet differentiability] 
(i)(Definition 1.9. of \cite{Sch-jt69}) Let ${X}$, ${Y}$ be Fr\'echet spaces with $U$ being an open subset of ${X}$. 
We say that $f$ has a Fr\'echet (or is $F$-)derivative (or $f$ is $F$-differentiable) at $x\in U$,  
{if} there exists a continuous
linear map $A=A_x:{X}\to {Y}$ such that $\varphi(x;y)$ is horizontal w.r.t $y$ at $0$, where $\varphi(x;y)$ is defined by
$$
\varphi(x;y)=f(x+y)-f(x)-A_xy.
$$
We call $A=A_x$ the $F$-derivative of $f$ at $x$, and we denote
$A_xy$ as $d_Ff(x;y)$. 
Moreover, we denote $f\in C_F^1(U:Y)$ if $f$ is $F$-differentiable and 
$d_F f : U{\times} X{\ni} (x,y){\to} d_F f(x;y){\in}Y$ is jointly continuous. \\
(ii) For Banach spaces, $f$ is $F$-differentiable at $x$ {if} there exists a continuous
linear map $A=A_x:{X}\to {Y}$ satisfying
$$
\Vert f(x+y)-f(x)-Ay\Vert_Y=o(\Vert y\Vert_X)\;\;\mbox{as $\Vert y\Vert_X\to0$}.
$$
Moreover, $f\in C_F^1(U:Y)$ if $f$ is $F$-differentiable and $X\ni x\to A_x\in {\mathbf{L}}(X:Y)$ is continuous.
\end{definition}
\begin{remark}
If $f$ is $F$-differentiable, then it is also $G$-differentiable. Moreover, 
$$
f'_G(x;y)=d_Gf(x;y)=d_Ff(x;y)=f'_F(x;y).
$$
\end{remark}
\begin{definition}[Higher order derivatives]
(i) Let ${X}$, ${Y}$ be Fr\'echet spaces with $U$ being an open subset of ${X}$. 
A $F$-differentiable function $f:U\to Y$ is twice $F$-differentiable at $x\in U$ 
if  $d_Ff:U\times X\ni (x,y)\to d_Ff(x;y)\in Y$ is $F$-differentiable at $x\in X$. That is, the function
$$
\psi(x;y,z)=d_Ff(x+z;y)-d_Ff(x;y)-d_F^2f(x)\{y,z\},
$$
is horizontal w.r.t. $z$ at $0$.\\
(ii) (p.72 of \cite{Berg77}) Let $X$, $Y$ be Banach spaces.
A $F$-differentiable function $f:U\to Y$ is twice $F$-differentiable at $x\in U$ 
if $f_F':X\to {\mathbf{L}}(X:Y)$ is $F$-differentiable at $x\in X$ and
$f_F''(x)$, the derivative of $f_F'(x)$, belongs to ${\mathbf{L}}(X: {\mathbf{L}}(X:Y))= {\mathbf{L}}(X\times X:Y)$.
$f\in C^2(U:Y)$ if (a) $f$ is twicely $F$-differentiable for each $x\in U$ 
and (b) $f_F''(x):U\to  {\mathbf{L}}(X\times X:Y)$ is continuous.\\
(iii) Analogously  $N$-times $F$-differentiability is defined.
\end{definition}
\begin{definition}[Many variables case]
(i) Let $U=\prod_{i=1}^N U_i$ with each $U_i$ being an open subset of Fr\'echet spaces $X_i$. 
For $x=(x_1,{\cdots}, x_N)\in U$ with $x_i\in U_i$ and $h_i\in X_i$ s.t. $x_i+h_i\in U_i$, if there exists $F_i(x;h_i)\in Y$ such that
$$
\varphi_i(x;h_i)=f(x_1,{\cdots}, x_{i-1},x_i+h,x_{i+1},{\cdots} x_N)-f(x_1,{\cdots}, x_{i-1},x_i,x_{i+1},{\cdots} x_N)-F_i(x;h_i)
$$
is horizontal w.r.t. $h_i$. We denote $F_i(x;h_i)$ as $\partial_{x_i}f(x)h_i$, the partial derivative of $f$ w.r.t. $x_i$.\\
(ii)(p.69 of \cite{Berg77})
 In case $X_i$ are Banach spaces,
the partial derivative of $f$ w.r.t. $x_i$, $\partial_{x_i}f(x)$, is defined by
$$
f(x_1,{\cdots}, x_{i-1},x_i+h_i,x_{i+1},{\cdots} x_N)-f(x_1,{\cdots}, x_{i-1},x_i,x_{i+1},{\cdots} x_N)=\partial_{x_i}f(x)h_i+o(\Vert h_i\Vert).
$$
More generally, for each $x$ if there exists a continuous linear map $d_F f:X\ni h\to d_Ff(x;h)\in Y$ such that
$$
\Vert f(x+h)-f(x)-d_Ff(x;h)\Vert_Y=o(\Vert h\Vert_X) \for h\in X.
$$
We denote also $d_Ff(x;h)=f'_F(x)\{h\}$ with $f'_F(x)\in  {\mathbf{L}}(X:Y)$.
Moreover, there exist operators $\partial_{x_i}f(x)\in  {\mathbf{L}}(X_i:Y)$ such that
\begin{equation}
f'_F(x;h)=f'_F(x)\{h\}=\sum_{i=1}^N\partial_{x_i}f(x)\{h_i\}=\sum_{i=1}^N\partial_{x_i}f(x;h_i)\with h=(h_1,{\cdots}, h_N).
\label{FD11-1}
\end{equation}
\end{definition}

\section{Functions on superspace}

\subsection{Grassmann continuation}
Let $\phi(q)$ be a $\fC$-valued function on an open set $\Omega\subset\euc^m$,
that is,
$$
\phi(q)=\sum_{{\mathbf{I}}\in{\mathcal I}}\phi_{\mathbf{I}}(q)\sigma^{\mathbf{I}}\with 
\phi_{\mathbf{I}}:\Omega\ni q\to\phi_{\mathbf{I}}(q)\in{\mathbb C}.
$$
By the definition of the topology of $\fC$, we have
$$
\lim_{q\to q_0}\phi(q)=\sum_{{\mathbf{I}}\in{\mathcal I}}
\big(\lim_{q\to q_0}\phi_{\mathbf{I}}(q)\big)\sigma^{\mathbf{I}}.
$$
The differentiation and integration of such $\phi(q)$ are defined by
$$
\frac{\partial}{\partial q_j}\phi(q)=\sum_{{\mathbf{I}}\in{\mathcal I}}
\frac{\partial}{\partial q_j}\phi_{\mathbf{I}}(q)\sigma^{\mathbf{I}} \et
\int_\Omega dq\,\phi(q)=\sum_{{\mathbf{I}}\in{\mathcal I}}
\bigg(\int_\Omega dq\,\phi_{\mathbf{I}}(q)\bigg)\sigma^{\mathbf{I}}.
$$
We say $\phi\in C^\infty(\Omega:\fC)$ 
if $\phi_{\mathbf{I}}\in C^\infty(\Omega:{\mathbb C})$ for each ${\mathbf{I}}\in{\mathcal I}$.
\begin{remark} %\par{\it Remark. }
If we use Banach-Grassmann algebra instead of Fr\'echet-Grassmann algebra,
we need to check whether $\sum_{{\mathbf{I}}\in{\mathcal I}}|\phi_{\mathbf{I}}(q)|<\infty$, etc.,
which seems cumbersome or rather impossible to check for applying it to concrete problems.
\end{remark}

\begin{lemma}\label{lem3-1.1}
Let $\phi(t)$ and $\Phi(t)$ be continuous $\fC$-valued functions on an interval 
$[a,b]\subset\euc$. Then,
\par
(1) $\int_a^b dt\,\phi(t)$ exists,
\par
(2) if $\Phi'(t)=\phi(t)$ on $[a,b]$, then
$\displaystyle{
\int_a^b dt\,\phi(t)=\Phi(b)-\Phi(a)},
$
\par
(3) if $\lambda\in\fC$ is a constant, then
$$
\int_a^b dt\,(\phi(t)\cdot\lambda)=\bigg(\int_a^b dt\,\phi(t)\bigg)\cdot\lambda
\et
\int_a^b dt\,(\lambda\cdot\phi(t))=\lambda\cdot\int_a^b dt\,\phi(t).
$$
\end{lemma}

Moreover, we may generalize above lemma for
a $\fC$-valued function $\phi(q)$ on an open set 
$\Omega\subset\euc^m$.

\begin{definition}\label{def3-1.1}
A set ${\mathfrak{U}}_{\mathrm{ev}}\subset{{\fR}^{m|0}}={\fR}_{\mathrm{ev}}^m$ 
is called an \underline{even superdomain} 
if ${{U}}=\pi_{\mathrm B}({\mathfrak{U}}_{\mathrm{ev}})
\subset {\mathbb R}^m$ is open and connected and
$\pi_{\mathrm B}^{-1}(\pi_{\mathrm B}({\mathfrak{U}}_{\mathrm{ev}}))={\mathfrak{U}}_{\mathrm{ev}}$.
When ${\mathfrak{U}} \subset \supermn$ is represented by
${\mathfrak{U}}={\mathfrak{U}}_{\mathrm{ev}} \times {\fR}_{\mathrm{od}}^n$ with a even superdomain 
${\mathfrak{U}}_{\mathrm{ev}}\subset{{\fR}^{m|0}}$,
${\mathfrak{U}}$ is called a \underline{superdomain} in ${\fR}^{m|n}$.
\end{definition}

\begin{proposition}\label{prop3-1.2}
Let  ${\mathfrak{U}}_{\mathrm{ev}} \subset{{\fR}^{m|0}} $ be a even superdomain.
Assume that $f$ is a smooth function from
$\euc^m\supset {{U}}=\pi_{\mathrm B}({\mathfrak{U}}_{\mathrm{ev}})$ 
into $\fC$, denoted simply by
$f\in C^\infty({{U}}:{\fC})$.
That is, we have the expression 
\begin{equation}
f(q)=\sum_{{\mathbf{J}}\in{\mathcal I}} f_{\mathbf{J}}(q)\,\sigma^{\mathbf{J}} \with 
f_{\mathbf{J}}(q)\in C^\infty({{U}}:{\mathbb C})
\quad \text{for each ${\mathbf{J}}\in{\mathcal I}$}.
\label{II-3-1}
\end{equation}
Then, we may define a mapping
$\tilde f$ of
${\mathfrak{U}}_{\mathrm{ev}}$ into $ {\fC} $, 
called the Grassmann continuation of $f$, by
\begin{equation}
{\tilde f}(x) = \sum_{ |\alpha|\ge 0 } { 1 \over {\alpha ! }}
\partial_q^\alpha f(x_{\mathrm B} )\,{x_{\mathrm S}^\alpha}
\where
\partial_q^\alpha f(x_{\mathrm B})
=\sum_{\mathbf{J}} \partial_q^\alpha f_{\mathbf{J}}(x_{\mathrm B})\,\sigma^{\mathbf{J}} .
\label{II-3-2}
\end{equation}
Here, we put $ x = ( x_1, \cdots , x_m ), \, x=x_{\mathrm B} +x_{\mathrm S} $
with 
$ x_{\mathrm B} = ( x_{1,{\mathrm B}},\cdots,x_{m,{\mathrm B}})
=(q_1,\cdots,q_m )=q \in {{U}},$
$ x_{\mathrm S} = ( x_{1,{\mathrm S}},\cdots,x_{m,{\mathrm S}} )$
and $ x^{\alpha} = x_1^{\alpha_1} \cdots x_m^{\alpha_m}$.
\end{proposition}
\par{\it Proof}. 
[Since circulation of our paper A. Inoue and Y. Maeda~\cite{IM91} is so-limited, I repeat here the proof whose main point  is to check whether this mapping
\eqref{II-3-2} is well-defined or not. Therefore,
by using the degree argument, we need to define ${\tilde f}^{[k]}$,
the $k$-th degree component of $\tilde f$.]

Denoting by $x_{1,{\mathrm S}}^{[k_1]}$, 
the $k_1$-th degree component of $x_{1,{\mathrm S}}$,
we get
$$
(x_{1,{\mathrm S}}^{\alpha_1 })^{[k_1]}
=\sum
(x_{1,{\mathrm S}}^{[r_1]})^{p_{1,1}}
\cdots \,
(x_{1,{\mathrm S}}^{[{r_\ell}]})^{p_{1,\ell}}.
$$
Here, the summation is taken for all partitions of
an integer $\alpha_1$ into
$\alpha_1=p_{1,1}+\cdots+p_{1,\ell}$ satisfying
$ \sum_{i=1}^{\ell} r_i p_{1,i} =k_1,\, r_i\ge0 $. 
Using these notations, we put
\begin{equation}
{\tilde f}^{[k]}(x) = \sum\limits_{\scriptstyle |\alpha|\le k,\;
k_0+k_1+ \cdots + k_m=k
\atop\scriptstyle k_1,\cdots,k_m\;\text{are even}}
\frac{1} {\alpha ! }
( \partial_q^\alpha f)^{[k_0]}(x_{\mathrm B})\,
(x_{1,{\mathrm S}}^{\alpha_1})^{[k_1]} \cdots 
(x_{m,{\mathrm S}}^{\alpha_m})^{[k_m]}
\label{II-3-3}
\end{equation}
where
$$
(\partial_q^\alpha f)^{[k_0]}(x_{\mathrm B})
=\sum_{|{\mathbf{J}}|=k_0}\partial_q^\alpha f_{\mathbf{J}}(x_{\mathrm B})\,\sigma^{\mathbf{J}}.
$$
Or more precisely, we have
$$
\begin{aligned}
{\tilde f}^{[0]}(x)& = f^{[0]}(x_{\mathrm B} ),\\
{\tilde f}^{[1]}(x) & = f^{[1]}(x_{\mathrm B} ),\\
{\tilde f}^{[2]}(x) &= f^{[2]}(x_{\mathrm B})
+\sum_{j=1}^m(\partial_{q_j} f)^{[0]}(x_{\mathrm B})(x_{j,{\mathrm S}})^{[2]},\\
{\tilde f}^{[3]}(x) & = f^{[3]}(x_{\mathrm B})
+\sum_{j=1}^m(\partial_{q_j} f)^{[1]}(x_{\mathrm B})(x_{j,{\mathrm S}})^{[2]},\\
{\tilde f}^{[4]}(x)  &= f^{[4]}(x_{\mathrm B})
+\sum_{j=1}^m(\partial_{q_j} f)^{[2]}(x_{\mathrm B})(x_{j,{\mathrm S}})^{[2]}\\
& \quad\quad
+{\frac 12}\sum_{j=1}^m( \partial_{q_j}^2 f)^{[0]}(x_{\mathrm B})
(x_{j,{\mathrm S}}^2)^{[4]}
+\sum_{j\neq k}( \partial_{q_j\,q_k}^2 f)^{[0]}(x_{\mathrm B})
(x_{j,{\mathrm S}})^{[2]}(x_{k,{\mathrm S}})^{[2]},\quad etc.\\
\end{aligned}
$$
Since ${\tilde f}^{[j]}(x)\ne {\tilde f}^{[k]}(x)\,\,(j\ne k)$ in $\fC$,
we may take the sum $\sum_{j=0}^\infty {\tilde f}^{[j]}(x)\in
\fC=\oplus_{k=0}^\infty{\fC}^{[k]}$, 
which is denoted by ${\tilde f}(x)$.
Therefore, rearranging the above ``summation'', we get rather
the ``familiar'' expression as in \eqref{II-3-2}. $\qquad\qed$

\begin{remark}
Concerning the summation in \eqref{II-3-3}, summation w.r.t. $\alpha$ is clearly finite, but that in $(x_{1,{\mathrm S}}^{\alpha_1})^{[k_1]}$ w.r.t. ${\bf J}\in{\mathcal{I}}$ is infinite for $|{\bf J}|=k_1$.
\end{remark}

\begin{corollary}\label{cor3-1.3}
If $f$ and $\tilde f$ be given as above, 
then
\par (i) $\tilde f$ is continuous and 
\par (ii) $\tilde f(x)=0$ in ${\mathfrak{U}}_{\mathrm{ev}}$ 
implies $f(x_{\mathrm B})=0$ in ${{U}}$.
\par
Moreover, if we define the partial derivative of $\tilde f$ in the $j$-direction by
\begin{equation}
\partial_{x_j} \tilde f (x)
=\left.{\frac d{dt}}\tilde f(x+te_{(j)})\right|_{t=0}
\where
e_{(j)}=(\overbrace{0,\cdots,0,1}^{j},0,\cdots,0)\in{{{\fR}^{m|0}}},
\label{II-3-4}
\end{equation}
then we get
\begin{equation}
\partial_{x_j} \tilde f (x)={\widetilde{\partial_{q_j} f}}(x)
\for j=1,\cdots,m.
\label{II-3-5}
\end{equation}
\end{corollary}
\par
{\it Proof}.
Let $y_j=y_{j,\mathrm B}+y_{j,\mathrm S} \in\rev$. 
For $y_{(j)}=y_je_{(j)}=y_{j,\mathrm B}e_{(j)}+y_{j,\mathrm S}e_{(j)}
=y_{(j),\mathrm B}+y_{(j),\mathrm S}\in{{{\fR}^{m|0}}}$,
as
$$
{\frac d{dt}}\tilde f(x+ty_{(j)})={\frac d{dt}}
\left\{
\sum_\alpha {\frac 1{\alpha!}}
\left(
\sum_{\mathbf{J}} \partial_q^\alpha f_{\mathbf{J}}(x_{\mathrm B}+ty_{(j),\mathrm B})\sigma^{\mathbf{J}}
\right)
(x_{\mathrm S}+ty_{(j),\mathrm S})^\alpha \right\},
$$
we get easily, 
$$
\begin{aligned}
{\frac d{dt}}\tilde f(x+ty_{(j)})\Big|_{t=0}
&=y_{(j),\mathrm B}\sum_\alpha{\frac 1{\alpha!}}
\left(
\sum_{\mathbf{J}} \partial_q^\alpha f_{\mathbf{J}}(x_{\mathrm B})\sigma^{\mathbf{J}}
\right)
x_{\mathrm S}^\alpha
+y_{(j),\mathrm S}\sum_{\check{\alpha}} {\frac 1{{\check{\alpha}}!}}
\left(
\sum_{\mathbf{J}} \partial_q^{\check{\alpha}}\partial_{q_j} f_{\mathbf{J}}(x_{\mathrm B})\sigma^{\mathbf{J}}
\right)x_{\mathrm S}^{\check{\alpha}}\\
&=y_j\sum_\alpha {\frac 1{\alpha!}} 
\partial_q^\alpha \partial_{q_j}f(x_{\mathrm B}) \,x_{\mathrm S}^\alpha 
=y_j{\widetilde{\partial_{q_j}f}}(x).
\end{aligned}
$$
Here, ${\check{\alpha}}=(\alpha_1,\cdots,\alpha_j,\alpha_j-1,\alpha_{j+1},\cdots,\alpha_m)$.
Putting $y_j=y_{j,\mathrm B}+y_{j,\mathrm S}=1$ in the above, we have \eqref{II-3-5}. $\qquad\qed$

\begin{remark} %{\it Remark}.
(i) By the same argument as above, we get, for $y=(y_1,\cdots,y_m)\in{{\fR}^{m|0}}$,
\begin{equation}
\left.{\frac d{dt}}\tilde f(x+ty)\right|_{t=0}=
\sum_{j=1}^m y_j 
\sum_\alpha {\frac 1{\alpha!}} \partial_q^\alpha \partial_{q_j}f(x_{\mathrm B})
\,x_{\mathrm S}^\alpha=\sum_{j=1}^m y_j\partial_{x_j} \tilde f (x).
\label{II-3-6}
\end{equation}
(ii) Unless there occurs confusion, we denote $\tilde{f}$ simply by $f$.
\end{remark}

\subsection{Supersmooth functions and their derivatives}
How to define the continuity and differentiability of functions from $\supermn$ to $\fC$?
\begin{problem}
Since $\fR$, $\fC$ and $\supermn$ are Fr\'echet spaces, we may define $G$- or $F$-differentiable functions as before.
By the way, how to take into account the ring structure of Fr\'echet-Grassmann algebra in the definition of total differentiability?
\end{problem}

In order to answer this problem, we introduce ``the desired or tractable form of functions on $\supermn$'' and called them as ``\underline{supersmooth} (or called \underline{superfield} by physicist)''. In the next section, we study their properties and we characterize them.

\begin{definition}\label{def3-1.2}
\par
(1) Let $ {\mathfrak{U}}_{\mathrm{ev}}  \subset {{\fR}^{m|0}}$ be a even super domain.
A mapping $F$ from ${\mathfrak{U}}_{\mathrm{ev}}$ to $ {\fC}$ is called supersmooth if there exists a smooth mapping
$f$ from ${{U}}=\pi_{\mathrm B}({\mathfrak{U}}_{\mathrm{ev}})$ to $ {\fC}$ such that
$F=\tilde{f}$. We denote the set of supersmooth functions on $ {\mathfrak{U}}_{\mathrm{ev}} $ as ${\CSS}({\mathfrak{U}}_{\mathrm{ev}}:{\fC} )$.
\par
(2) Let ${\mathfrak{U}}$ be a superdomain in $\supermn$.
A mapping $f$ from ${\mathfrak{U}}$ to $\fC$ is called supersmooth if it is decomposed as
\begin{equation}
f(x,\theta) = \sum_{ |a|\le n }\theta^a f_a (x) .
\label{II-3-7}
\end{equation}
Here, $ a=(a_1, \cdots a_n) \in \{0,1\} ^n $,
$\theta^a = \theta_1^{a_1} \cdots \, \theta_n^{a_n} $
 and $ f_a (x) \in {\CSS} ({\mathfrak{U}}_{\mathrm{ev}}:{\fC})$.
Without mentioning it,   we assume always that $f_a(x)\in\cev (\mbox{or}\;  \in \cod)$ for all $a$,  and call them as
 even (or odd) supersmooth functions denoted by $ {\CSS}({\mathfrak{U}}:{\fC})$. Moreover,
 $$
\ccsl_{\mathrm{SS}}=\{f(x,\theta)\in{\CSS}({\mathfrak{U}}:{\fC})\,|\, f_a(x_{\mathrm{B}})\in{\mathbb C}\}.
$$
Therefore,  \underline{if $f\in \ccsl_{\mathrm{SS}}$, $f_a(x)$ may be put any side of $\theta^a$}.
\par
(3) Let $ f \in  {\CSS} ({\mathfrak{U}}:{\fC})$. We put
\begin{equation}
\left \{
\begin{aligned}
F_j(X)&= \sum_{|a|\le n} \theta^a\partial_{x_j} f_a(x) \for  j=1,2, \cdots ,m, \\
F_{s+m} (X)&=
\sum_{|a| \le n} (-1)^{l(a)} \theta_1^{a_1}\cdots
\theta_{s}^{{a_s}-1}
\cdots
\theta_n^{a_n} f_a(x) \for s = 1,2, \cdots ,n
\end{aligned}
\right .
\label{II-3-8}
\end{equation}
with $l(a)= \sum_{j=1}^{s-1} a_j $ and
$\theta_{s}^{-1}=0$. In this case, 
$F_{\kappa}(X)$ is the partial derivative of $f$ at $ X = (x, \theta)=(X_{\mu})$ w.r.t. $X_{\kappa}$
\begin{equation}
\begin{cases}
F_j (X) = {\displaystyle{{\partial}\over{\partial x_j}}} f(x, \theta )
= \partial_{x_j} f(x, \theta )
=f_{x_j}(x, \theta ) \for j=1,2, \cdots ,m,\\
F_{m+s}(X) = {\displaystyle{{\partial}\over{\partial\theta_s}}}f(x,\theta )
= \partial_{\theta_s} f(x, \theta )
=f_{\theta_s}(x, \theta )  \for s = 1,2, \cdots ,n \\
\end{cases}
\label{II-3-9}
\end{equation}
\begin{equation}
F_{\kappa}(X)={\partial}_{X_{\kappa}}f(X)=f_{X_{\kappa}}(X) 
\for \kappa=1,\cdots,m+n.
\label{II-3-10}
\end{equation}
\end{definition}

\begin{remark} 
(1) In this lecture, we use the left odd derivatives.
This naming stemms from putting most left  the variable w.r.t. which we differentiate. There are some authors (see, for example V.S. Vladimirov and I.V. Volovich~\cite{VV83})
who give the name right derivative to this.
\par
Put
$$
 {\CSS}^{(r)} ({\mathfrak{U}}:{\fC})=\{f(x,\theta)= \sum_{ |a|\le n } f_a (x)\theta^a\,|\, f_a(x)\in {\CSS} ({\mathfrak{U}}_{\mathrm{ev}}:{\fC})\}.
$$
For $ f \in  {\CSS}^{(r)} ({\mathfrak{U}}:{\fC})$ with $ j=1,2, \cdots ,m $ and $ s = 1,2, \cdots ,n$, 
we note here the right-derivatives:
$$ 
\left \{
\begin{aligned} 
F^{(r)}_j(X) &= \sum_{|a|\le n} \partial_{x_j} f_a(x) \theta^a,\\
F^{(r)}_{s+m} (X) &=
\sum_{|a| \le n} (-1)^{r(a)} f_a(x)
\theta_1^{a_1}\cdots
\theta_{s}^{{a_s}-1}
\cdots
\theta_n^{a_n}
\end{aligned}
\right .
$$ 
We put here $r(a)= \sum_{j=s+1}^{n} a_j $.
$F^{(r)}_{\kappa}(X)$is called the (right) partial $\kappa$-derivative w.r.t. $ X_{\kappa} $ at $ X = (x, \theta ) $
denoted by
$$
F^{(r)}_j (X) = {{\partial}\over{\partial x_j}} f(x, \theta ) 
= \partial_{x_j} f(x, \theta ),\quad
F^{(r)}_{m+s}(X) = f(x, \theta )
\overset{\leftarrow}{{\partial} \over { \partial\theta_s }}
= f(x, \theta )
\overset{\leftarrow}{{\partial}_{\theta_s}}.
$$
\newline
(2) Since we use a countably infinite Grassmann generators,  the decomposition 
\eqref{II-3-8} is unique. In fact, if $ \sum_a \theta^a f_a(x) \equiv 0 $ on ${\mathfrak{U}}$, then $ f_a(x) \equiv 0$.
(see, p~322 in Vladimirov and Volovich~\cite{VV83}.) 
\newline
(3) The higher derivatives are defined analogously.
For a multiindex $\alpha=(\alpha_1,\cdots,\alpha_m)\in({\mathbb N}\cup\{0\})^m$ and
$a=(a_1,\cdots,a_n)\in\{0,1\}^n$, we put
$$
\partial_x^\alpha = \partial_{x_1}^{\alpha_1} 
\cdots \partial_{x_m}^{\alpha_m} \et
\partial_{\theta}^a = \partial_{\theta_1}^{a_1} \cdots 
\partial_{\theta_n}^{a_n}.
$$
\end{remark}

Assume that for $X=(x,\theta),Y=(y,\omega)\in \supermn$, we have $X+tY\in {\mathfrak{U}}$ (for any $t\in[0,1]$).
Repeating the proof  used in the proof of Corollary \ref{cor3-1.3}, for
$ f\in{\CSS}({\mathfrak{U}}:{\fC})$, the following holds:
\begin{equation}
\left.{\frac d{dt}}f(X+tY)\right|_{t=0}=
\sum_{j=1}^{m}
y_j{\frac {\partial}{\partial x_j}}f(X)
+\sum_{s=1}^{m}
\omega_s{\frac {\partial}{\partial \theta_s}}f(X)
\label{II-3-11}
\end{equation}
\begin{definition} 
A function $f$ from the super domain ${\mathfrak{U}}\subset\supermn$ to $\fC$, is called
G-differentiable at $X=(x,\theta)$  if
$$
f(x+y,\theta+\omega)-f(x,\theta)=\sum (y_iF_i+\omega_sF_s)+
\sum(y_iR_i+\omega_s R_s).
$$
Here,
$$
d(R_i,0)\to0,\quad d(R_s,0)\to0,\quad d_{m|n}((y,\omega),0)\to0.
$$
\end{definition}
\subsubsection{Taylor's Theorem}
For $ f\in{\CSS}(U:{\fC})$, we have
\begin{equation}
\left.{\frac d{dt}}f(X+tY)\right|_{t=0}=
\sum_{j=1}^{m}
y_j{\frac {\partial}{\partial x_j}}f(X)
+\sum_{s=1}^{m}
\omega_s{\frac {\partial}{\partial \theta_s}}f(X).
\label{II-3-11-b}
\end{equation}
From this, we define
\begin{definition} For a supersmooth function $f$, we define its differential $df$ as %${\CSS}$
$$
df(X)=d_Xf(X)=\sum_{\kappa=1}^{m+n} 
dX_\kappa\frac{\partial f(X)}{\partial X_\kappa},
$$
or
$$
df(x,\theta)=
\sum_{j=1}^m dx_j\frac{\partial f(x,\theta)}{\partial x_j}
+\sum_{s=1}^n d\theta_s\frac{\partial f(x,\theta)}{\partial \theta_s}.
$$
\end{definition}

From the before mentioned Definition \ref{def3-1.2}, we have
%\ref{def3-1.2}
\begin{proposition}\label{prop3-3.13}
Let ${\mathfrak{U}}$ be a superdomain in $\supermn$.
For any $f,g \in {\CSS} ({\mathfrak{U}}:{\fC})$, the product
$fg$ belongs to ${\CSS} ({\mathfrak{U}}:{\fC})$ and their differentials
$ {d_X} f(X)$ and ${d_X} g(X)$ are continuous linear maps from
$\supermn$ to ${\fC}^{m+n}$.
\par
Moreover, 
%\begin{enumerate}
\par (1) %\item
For any homogeneous elements $\lambda,\, \mu \in {\fC}$,
\begin{equation}
{d_X} ( \lambda f + \mu g)(X) =
(-1)^{p(\lambda)p(X)} \lambda\, {d_X} f(X) 
+ (-1)^{p(\mu)p(X)} \mu\, {d_X} g(X).
\label{3-21}
\end{equation}
\par (2) %\item
(Leibnitz' formula)
\begin{equation}
\partial_{X_{\kappa}} [f(X)g(X)] 
= 
(\partial_{X_{\kappa}} f(X)) g(X)+ 
(-1)^{p( X_{\kappa} ) p(f(X))}
f(X) (\partial_{X_{\kappa}}g(X)).
\label{3-22}
\end{equation}
%\end{enumerate}
\end{proposition}
\par{\it Proof}. \eqref{3-21} is trivial.
For $f, g\in{\CSS}({\mathfrak{U}}:{\fC})$, we have
\begin{equation}
\begin{aligned}
\left.{\frac d{dt}}f(X+tY)g(X+tY)\right|_{t=0}=&
\left(\sum_{j=1}^{m}
y_j{\frac {\partial}{\partial x_j}}f(X)
+\sum_{s=1}^{m}
\omega_s{\frac {\partial}{\partial \theta_s}}f(X)\right)g(X)\\
&+%(-1)^{p( X_{\kappa} ) p(f(X))}
f(X)\left(\sum_{j=1}^{m}
y_j{\frac {\partial}{\partial x_j}}g(X)
+\sum_{s=1}^{m}
\omega_s{\frac {\partial}{\partial \theta_s}}g(X)\right).
\end{aligned}
\label{II-3-11-a}
\end{equation}
Therefore, we get the desired result.  $\qquad\qed$

\subsection{Characterization of supersmooth functions}
In previous lecture, we introduce abruptly a class $\CSS({\mathfrak{U}}:{\fC})$ of functions on super domain ${\mathfrak{U}}\subset\supermn$.
But such introduction is reasonable or it is stable under rather ordinary operations? Or how may we characterize it?
\begin{quotation}
{\small
Though there exists multiplication in ${\mathbb C}$ but not in $\euc^2$. 
How the ring structure of the definition domain affects the total-differentiability of functions?
How do we characterize such functions?
 
(a) We decompose a function $f(z)$ from ${\mathbb C}$ to ${\mathbb C}$ as
$$
{\mathbb C}\ni z=x+iy\longrightarrow f(z)=u(x,y)+iv(x,y),\quad u(x,y)=\Re{f(z)}\in{\mathbb R},\;
v(x,y)=\Im{f(z)}\in{\mathbb R}
$$
For $z=x+iy$ and $z_0=x_0+iy_0$, since $|z|=\sqrt{x^2+y^2}$, we have
$$
\begin{aligned}
|f(z)-f(z_0)|&=|u(x,y)+iv(x,y)-(u(x_0,y_0)+iv(x_0,y_0))|\\
&=\sqrt{(u(x,y)-u(x_0,y_0))^2+(v(x,y)-v(x_0,y_0))^2}
\end{aligned}
$$
Therefore, if $f(z)$ is continuous at $z=z_0$,
$u(x,y), v(x,y)$ are continuous at $(x_0,y_0)$ as real-valued functions with 2 real variables.

(b) A function $f(z)$ from ${\mathbb C}$ to ${\mathbb C}$ is called total differentiable at $z=z_0$ if there exists a number $\gamma\in{\mathbb C}\cong L({\mathbb C}:{\mathbb C})$ such that it satisfies
\begin{equation}
|f(z_0+w)-f(z_0)-\gamma w|=o(|w|)\qquad (|w|\to0).
\label{c-bibun}
\end{equation}
This number $\gamma=\alpha+i\beta$
$(\alpha, \beta\in{\mathbb R})$ is denoted by $f'(z_0)$. We check a little bit more precisely. putting
$w=h+ik$ $(h,k\in{\mathbb R})$, then
%\newline
$\gamma w=(\alpha+i\beta)(h+ik)=(h\alpha-k\beta)+i(k\alpha+h\beta)$, we have
$$
\begin{aligned}
|f(z_0+w)-&f(z_0)-\gamma w|\\
&=|u(x_0+h,y_0+k)-u(x_0,y_0)-(h\alpha-k\beta)\\
&\qquad\qquad
+i(v(x_0+h,y_0+k)-v(x_0,y_0)-(k\alpha+h\beta))|\\
&=\big([u(x_0+h,y_0+k)-u(x_0,y_0)-(h\alpha-k\beta)]^2\\
&\qquad\qquad
+[v(x_0+h,y_0+k)-v(x_0,y_0)-(k\alpha+h\beta)]^2
\big)^{1/2}
\end{aligned}
$$
Therefore, when $(h,k)\to0$(i.e. $\sqrt{h^2+k^2}\to0$),
\begin{equation}
\begin{gathered}
|u(x_0+h,y_0+k)-u(x_0,y_0)-(h\alpha-k\beta)|=o(\sqrt{h^2+k^2}),\\
|v(x_0+h,y_0+k)-v(x_0,y_0)-(k\alpha+h\beta)|=o(\sqrt{h^2+k^2}).
\end{gathered}
\label{CR1}
\end{equation}
From the first equation above, putting $h=0$ and $k\to0$, we get $\beta=-u_y(x_0,y_0)$, and
putting $k=0$ and $h\to0$ then $\alpha=u_x(x_0,y_0)$.
From the second one above, we have $\alpha=v_y(x_0,y_0)$ and $\beta=v_x(x_0,y_0)$.
Therefore, we get a system of PDE
\begin{equation}
u_x=v_y,\quad u_y=-v_x
\label{CR}
\end{equation}
called Cauchy-Riemann equation.
If real valued functions $u$, $v$ with two real variables satisfy Cauchy-Riemann equation, then they belong to $C^{\infty}$, moreover, $u(x,y)+iv(x,y)$ is shown as a convergent power series\footnote{in general, this is proved by applying Cauchy's integral representation} in $z=x+iy$ which is written $f(z)$,
and called analytic.
Without confusion, we write
$$
\frac{\partial}{\partial z}f(z)=f'(z)=\alpha+i\beta=u_x-iu_y=v_y+iv_x.
$$

(c) Using above notation, we consider a map $\Phi$ from $\euc^2$ to $\euc^2$
$$
\Phi:\euc^2\ni \binom{x}{y}\to\binom{u(x,y)}{v(x,y)}\in\euc^2.
$$
Denoting $\binom{x_0}{y_0}$ by $({x_0},{y_0})$, $\Phi$ is said to be totally differentiable at $({x_0},{y_0})$
if there exists  $\Phi'_F({x_0},{y_0})\in L(\euc^2:\euc^2)$ such that
$$
\Vert\Phi({x_0+h},{y_0+k})-\Phi({x_0},{y_0})-\Phi'_F({x_0},{y_0})\binom{h}{k}\Vert=o(\Vert\binom{h}{k}\Vert)
$$
Representing $\Phi'_F(x,y)$ as
\begin{equation}
\Phi'_F(x,y)=\begin{pmatrix}
u_x(x,y)&u_y(x,y)\\
v_x(x,y)&v_y(x,y)
\end{pmatrix},
\label{mult}
\end{equation}
we have
$$
\begin{aligned}
&\big[(u(x_0+h,y_0+k)-u(x_0,y_0)-(hu_x(x_0,y_0)+ku_y(x_0,y_0)))^2\\
&\qquad+(v(x+h,y+k)-v(x,y)-(hv_x(x_0,y_0)+kv_y(x_0,y_0)))^2\big]^{1/2}=o(\sqrt{h^2+k^2}),
\end{aligned}
$$
i.e.
\begin{equation}
\begin{gathered}
|u(x_0+h,y_0+k)-u(x_0,y_0)-(hu_x(x_0,y_0)+ku_y(x_0,y_0))|=o(\sqrt{h^2+k^2}),\\
|v(x_0+h,y_0+k)-v(x_0,y_0)-(hv_x(x_0,y_0)+kv_y(x_0,y_0))|=o(\sqrt{h^2+k^2}).
\end{gathered}
\label{CR2}
\end{equation}

Identifying $\euc^2$ as ${\mathbb{C}}$, we seek a condition that $\Phi'_F({x_0},{y_0})\in L(\euc^2:\euc^2)$ is regarded as a multiplication in ${\mathbb{C}}$.
When an element $a+ib\in{\mathbb{C}}$ acts as a multiplication operator, then a linear operator in $\euc^2$, that is,
a matrix is identified with 
$
a+ib\sim\begin{pmatrix}
a&-b\\
b&a\end{pmatrix}.
$
Since
$$
\Phi'_F(x,y)=\begin{pmatrix}
u_x(x,y)&u_y(x,y)\\
v_x(x,y)&v_y(x,y)
\end{pmatrix},
$$
we have $u_x=v_y$, $u_y=-v_x$. In another presentation, 
 $\Phi'_F({x_0},{y_0})\in L(\euc^2:\euc^2)$ is not only ${\mathbb{R}}$ but also ${\mathbb{C}}$-linear, that is, for any $a, b\in\euc$,
$$
\begin{pmatrix}
a&-b\\
b&a\end{pmatrix}\Phi'_F({x_0},{y_0})=
\Phi'_F({x_0},{y_0})\begin{pmatrix}
a&-b\\
b&a\end{pmatrix}
$$
hold. Here, $b\neq{0}$ is essential.

(d) Generalizing above to a map $\phi$ from ${\mathbb{C}}^m$ to ${\mathbb{C}}$,
$$
\phi:{\mathbb{C}}^m\ni z={}^t(z_1,{\cdots},z_m)\to \phi(z)\in{\mathbb{C}}.
$$
This is totally differentiable if  there exists $\phi'_F({z})\in L({\mathbb{C}}^m:{\mathbb{C}})$ at ${z}$ such that
$$
\Vert\phi({z}+w)-\phi({z})-\phi'_F({z})w\Vert=o(\Vert w\Vert).
$$
By same calculation, $\phi'_F(z)$ is
\begin{equation}
\phi'_F(z)=\left(\frac{\partial \phi(z)}{\partial z_1},{\cdots},\frac{\partial \phi(z)}{\partial z_m}\right)
\label{mult-m}
\end{equation}
and each component of $\phi$ is analytic w.r.t. each variable. From Hartogs' theorem, $\phi$ is holomorphic having a convergent power series expansion.

(e)  A map $\Phi$ from ${\mathbb{C}}^m$ to ${\mathbb{C}}^n$ is totally differentiable at $z$ if there exists $\Phi'_F({z})\in L({\mathbb{C}}^m:{\mathbb{C}}^n)$
$$
\Phi'_F({z})=\begin{pmatrix}
\frac{\partial \Phi_1(z)}{\partial z_1}&{\cdots}&\frac{\partial \Phi_1(z)}{\partial z_m}\\
{\vdots}&{\cdots}&{\vdots}\\
\frac{\partial \Phi_n(z)}{\partial z_1}&{\cdots}&\frac{\partial \Phi_n(z)}{\partial z_m}
\end{pmatrix}
$$
such that
$$
\Vert\Phi({z}+w)-\Phi({z})-\Phi'_F({z})w\Vert=o(\Vert w\Vert).
$$
}
\end{quotation}

\begin{problem}
Does there exist Cauchy-Riemann equation corresponding to supersmooth functions?
\end{problem}

\section{Super differentiable functions on ${\fR}^{m|n}$}
\subsection{Superdifferentiability of functions on ${\fR}^{m|n}$}
\begin{definition}[see, A. Yagi~\cite{yag88}]\label{yagi247}
Let $f$ be a $\fC$-valued function on a superdomain ${\mathfrak{U}}\subset  {\fR}^{m|n}$.
Then, a function $f$ is said to be super $C_G^1$-differentiable, denoted by  $f\in {\mathfrak{G}} _{S\!D}^1({\mathfrak{U}}:\fC)$ or simply $f\in {\mathfrak{G}} _{S\!D}^1$ {if} there exist  $\fC$-valued continuous functions $F_A$ $(1\le A\le m+n)$ on ${\mathfrak{U}}$
such that
\begin{equation}
\dt f(X+tH)\bigg|_{t=0}=f_G'(X,H)=\sum_{A=1}^{m+n}H_A F_A(X)
\label{sg1}
\end{equation}
or
$$
f(X+tH)-f(X)-tf_G'(X,H)\to0\;\;\mbox{in $\fC$, when $H\to0$ in $\fC$},
$$
for each $X\in {\mathfrak{U}}$ and $H\in {\fR}^{m|n}$ where $ f(X+tH)$ is considered as a $\fC$-valued function w.r.t. $t\in\euc$.
We denote $F_A(X)$ by $f_{X_A}(X)$. 
Moreover, for $r\ge 2$, $f$ is said to be in ${\mathfrak{G}} _{S\!D}^r$ if  $F_A$ are ${\mathfrak{G}} _{S\!D}^{r-1}$.
$f$ is said to be ${\mathfrak{G}} _{S\!D}^{\infty}$ or superdifferentiable if $f$ is ${\mathfrak{G}} _{S\!D}^r$ for all $r\ge 1$.
\end{definition}

\begin{definition}
Let $f$ be a $\fC$-valued function on a superdomain ${\mathfrak{U}}\subset  {\fR}^{m|n}$.
 A function $f$ is said to be  super $C_F^1$-differentiable, denoted by  $f\in {\mathfrak{F}} _{S\!D}^1({\mathfrak{U}}:\fC)$ or simply $f\in {\mathfrak{F}} _{S\!D}^1$ {if} there exist  $\fC$-valued continuous functions $F_A$ $(1\le A\le m+n)$ on ${\mathfrak{U}}$
and functions $\rho_A:{\mathfrak{U}}\times\supermn\to {\mathfrak{C}}$ 
such that 
\begin{equation}
\begin{aligned}
& (a)\; f(X+H)-f(X)=\sum_{{j}=1}^{m+n}H_AF_A(X)+\sum_{A=1}^{m+n}H_A \rho_A(X;H)
\for X\in {\mathfrak{R}}^{m|n},\\
& (b)\; \rho_A(X,H)\to 0 \;\;\mbox{in}\;\; {\mathfrak{C}} \when H\to 0 \;\;\mbox{in}\;\; {\mathfrak{R}}^{m|n},
\end{aligned}
\label{sf1}
\end{equation}
for each $X\in {\mathfrak{U}}$ and $X+H\in {\mathfrak{U}}$.
$f$ is said to be super  $C_F^2$-differentiable,
when $F_A \in {\mathfrak{F}} _{S\!D}^1(\supermn:\fC)$ $(1\le A\le m+n)$. Analogously, we may define
super $C_F^r$-differentiablity and we say it {superdifferentiable} 
if it is super $C_F^{\infty}$-differentiable, denoted by ${\mathfrak{F}} _{S\!D}^{\infty}$.
\end{definition}
\begin{question}\label{quest1}
Does there exist the difference between ${\mathfrak{G}} _{S\!D}^1$ and ${\mathfrak{F}} _{S\!D}^1$, or between ${\mathfrak{G}} _{S\!D}^{\infty}$ and ${\mathfrak{F}} _{S\!D}^{\infty}$?
\end{question}
\begin{remark}
Let ${\mathfrak{V}}$ be an open set ${\fC}^{m|n}$.
When $f:{\mathfrak{V}}\to{\mathfrak{C}}$ is in ${\mathfrak{F}} _{S\!D}^{\infty}$, 
$f$ is also said to be {superanalytic}.
\end{remark}

\subsection{Remarks on Grassmann continuation}
From Taylor's expansion formula~\eqref{tay-f} mentioned before in general Fr\'echet space, we get
\begin{lemma}\label{XXX}
For $f(q)\in C^\infty({\euc}^m)$, its Grassmann continuation ${\tilde f}$ has the following Taylor's expansion formula:
For any $N$, there exists $\tilde\tau_N(x,y)\in{\fC}$ such that
\begin{equation}
{\tilde f}(x+y)
=\sum_{|\alpha|=0}^N\frac{1}{\alpha!}
\partial_x^{\alpha}{\tilde f}(x)y^\alpha+\tilde\tau_N(f;x,y).
\label{taylor-exp}
\end{equation}
Here,
$$
\tilde\tau_N(f;x,y)=\sum_{|\alpha|=N+1} y^\alpha\int_0^1 dt\frac{1}{N!}(1-t)^N
\partial_x^\alpha {\tilde f}(x+ty).
$$
\end{lemma}
\par{\it Proof:} Putting $q=x_{\mathrm{B}}$ and $q'=y_{\mathrm{B}}$ into Taylor's expansion formula~\eqref{tay-f},
we have
$$
{f}(q+q')
=\sum_{|\alpha|=0}^N\frac{1}{\alpha!}
\partial_q^{\alpha}{f}(q)q^{\prime\alpha}
+\sum_{|\alpha|=N+1} q^{\prime\alpha}\int_0^1 dt\frac{1}{N!}(1-t)^N
\partial_q^\alpha {f}(q+tq').
$$
Taking Grassmann continuation of both sides, and remarking $\widetilde{\partial_q^{\alpha}f}(x)=\partial_x^{\alpha} f(x),\; \widetilde{q^{\prime\alpha}}=y^{\alpha}$, we get the desired equality \eqref{taylor-exp}. $\qquad\qed$

\begin{corollary}\label{11.21.cor5.3}
For $f(q)\in C^\infty({\mathbb{R}}^m)$,
${\tilde f}$ is super F-differentiable.
\end{corollary}

\par{\it Proof:} We prove the case $m=1$. From Lemma, we have
$$
\tilde{f}(x+y)-\tilde{f}(x)=y{\tilde{f}}'(x)+y\tau(x,y),\quad
\tau(x,y)=y\int_0^1dt(1-t){\tilde f}''(x+ty)
$$
and because when $y\to0$ in ${\rev}$, then $\tau(x,y)\to0$ in ${\fC}$. $\qquad\qed$
\begin{exercise}
Prove more precisely, the statement above `` when $y\to0$ in ${\rev}$, then $\tau(x,y)\to0$ in ${\fC}$''.
\end{exercise}

\begin{corollary}\label{XXX-cor}
For $f(q)\in C^\infty({\euc}^m)$, its Grassmann continuation
$\tilde{f}(x)\in {\mathfrak{F}}_{S\!D}^1({\fR}^{m|0})$.
\end{corollary}
\par
\par{\it Proof:} Putting $N=1$ in Taylor's expansion formula~\eqref{tay-f}, we get
$$
\tilde{f}(x+y)=\tilde{f}(x)+\sum_{{j}=1}^my_j\partial_{x_j}\tilde{f}(x)+\sum_{{j}=1}^my_j\rho_j(x,y)
$$
Remarking
\begin{equation}
\begin{gathered}
\partial_{x_j} \tilde{f}(x)=\widetilde{\partial_{q_j}f}(x)\in{\fC},\\
\rho_j(x,y)=
\sum_{k=1}^m y_k g_{j,k}(x,y),\;\;
g_{j,k}(x,y)=\int_0^1(1-t)\partial^2_{x_jx_{k}}\tilde{f}(x+ty)dt,
\end{gathered}
\label{f-smooth}
\end{equation}
we need to prove
\begin{claim}\label{1025}
When $y\to0$ in ${\fR}_{\mathrm{ev}}^m$, then $\rho_j(x,y)\to0$ for each $x\in{\fR}_{\mathrm{ev}}^m$ and $j=1,{\cdots},m$.
That is, for any $\epsilon>0$, $j$ and $x\in{\fR}_{\mathrm{ev}}^m$, there exists $\delta>0$ such that if $\dist_{m|0}(y)<\delta$, then $\dist_{1|0}\rho_j(x,y)<\epsilon$.
\end{claim}
\par{\it Proof}.
Take any ${\mathbf{I}}\in{\mathcal{I}}_{\mathrm{ev}}$ and decompose ${\mathbf{I}}={\mathbf{J}}+{\mathbf{K}}$.
Remarking Corollary~\ref{11.21.cor5.3}, we have %{cor3-1.3a}
$$
\partial^2_{x_jx_{k}}\tilde{f}(x+ty)=\sum_{|\alpha|=0}^{\infty}\frac{\partial^2_{q_jq_{k}}\partial_q^{\alpha}f(x_{\mathrm{B}}+ty_{\mathrm{B}})}{\alpha!}
(x_{\mathrm{S}}+ty_{\mathrm{S}})^{\alpha}.
$$
If ${\mathbf{I}}=\tilde0=(0,0,0,{\cdots})$, then
$$
|\proj_{\tilde0}\rho_j(x,y)|\le \sum_{k=1}^m |y_{k,\mathrm{B}}||\pi_{\mathrm{B}}g_{j,k}(x,y)|\to0\for y_{\mathrm{B}}\to0.
$$
For fixed ${\mathbf{I}}\neq\tilde0$, the family of index sets $\{{\mathbf{K}}\;|\;{\mathbf{K}}\subset {\mathbf{I}}\}$ has finite elements,
$\{x_{\mathrm{B}}+ty_{\mathrm{B}}\;|\; t\in[0,1],\;|y_{\mathrm{B}}|\le1\}$ is compact, if taking $\alpha$ such that $2|\alpha|> |{\mathbf{K}}|$, then ${\proj}_{\mathbf{K}}(x_{\mathrm{S}}+ty_{\mathrm{S}})^{\alpha}=0$. Therefore, there exists a constant $C_{\mathbf{K}}=C_{\mathbf{K}}(x_{\mathrm{S}},y_{\mathrm{S}})$ such that
$$
\begin{aligned}
&\bigg|{\proj}_{\mathbf{K}}\big(\sum_{|\alpha|=0}^{\infty}\frac{\partial^2_{q_jq_{k}}\partial_q^{\alpha}f(x_{\mathrm{B}}+ty_{\mathrm{B}})}{\alpha!}
(x_{\mathrm{S}}+ty_{\mathrm{S}})^{\alpha}\big)\bigg|\\
&\qquad\qquad
\le
\sum_{\alpha}\frac{\big|\partial^2_{q_jq_{k}}\partial_q^{\alpha}f(x_{\mathrm{B}}+ty_{\mathrm{B}})\big|}{\alpha!}\big|{\proj}_{\mathbf{K}}(x_{\mathrm{S}}+ty_{\mathrm{S}})^{\alpha}\big|\le C_{\mathbf{K}}.
\end{aligned}
$$
In fact, if $2|\alpha|> |{\mathbf{K}}|$ then ${\proj}_{\mathbf{K}}(x_{\mathrm{S}}+ty_{\mathrm{S}})^{\alpha}=0$, then
$$
\begin{aligned}
\big|{\proj}_{\mathbf{K}}(g_{j,k}(x,y))\big|&\le\int_0^1 dt\, (1-t)\big|{\proj}_{\mathbf{K}}(\partial^2_{x_jx_{k}}\tilde{f}(x+ty))\big|\\
&\le\sum_{2|\alpha|\le|{\mathbf{I}}|}\frac{\max_{t}\big|\partial^2_{q_jq_{k}}\partial_q^{\alpha}f(x_{\mathrm{B}}+ty_{\mathrm{B}})\big|}{\alpha!}
\int_0^1dt\,\big|{\proj}_{\mathbf{K}}(x_{\mathrm{S}}+ty_{\mathrm{S}})^{\alpha}\big|,
\end{aligned}
$$
and
$$
C_{\mathbf{K}}=C_{\mathbf{K}}(x_{\mathrm{S}},y_{\mathrm{S}})=\int_0^1dt\,\big|{\proj}_{\mathbf{K}}(x_{\mathrm{S}}+ty_{\mathrm{S}})^{\alpha}\big|\to0\when y_{\mathrm{S}}\to 0\quad\mbox{in}\quad{\mathfrak{C}},
$$
therefore
$$
\big|{\proj}_{\mathbf{I}}(\rho_j(x,y))\big|\le\sum_{k=1}^m\sum_{{\mathbf{I}}={\mathbf{J}}+{\mathbf{K}}}\big|{\proj}_{\mathbf{J}}(y_k)\big|\big|{\proj}_{\mathbf{K}}(g_{j,k}(x,y))\big|
\le \sum_{k=1}^m\sum_{{\mathbf{I}}={\mathbf{J}}+{\mathbf{K}}}\big|{\proj}_{\mathbf{J}}(y_k)\big|C_{\mathbf{K}}.
$$
This finite sum tends to $0$ when $y\to0$, this implies $\tilde{f}(x)\in {\mathfrak{F}}_{S\!D}^1({\fR}^{m|0})$. 
$\qquad/\!\!/$

\begin{remark}\label{dewitt-body}
By the way, concerning the Grassmann continuation $\tilde{f}$ of $f$, B.S. de Witt~\cite{DeW84-2} claimed in p.7 as follows:
{\begin{quotation}
``The presence of a soul in the independent variable evidently has little practical effect on the variety of functions with which one may work in applications of the theory. In this respect $\rev$ is a harmless generalization of its own subspace ${\euc}$, the real line."
\end{quotation}}
\end{remark}

Though he didn't give more explanation of this intuitional claim in ~\cite{DeW84-2}, but we interpret his saying as
\begin{proposition}\label{dew-cl}
Let $F\in C_G^{\infty}({\fR}^{m|0}:{\fC})$. Putting $f(q)=F(q)$ for $q\in{\euc}^m$, we have $\tilde{f}=F$.
\end{proposition}
We rephrase this as
\begin{claim}\label{1102}
Let a $C_G^{\infty}$ differentiable function $H=H(x)=\sum_{{\mathbf{J}}\in{\mathcal{I}}}H_{\mathbf{J}}(x)\sigma^{\mathbf{J}}$ be given as a map from ${\fR}^{m|0}$ to ${\fC}$ such that it is $0$ on ${\euc}^m$. 
That is, for any $\alpha$ and $q\in\euc^m$, if $\partial_q^{\alpha}H_{\mathbf{J}}(q)=0$, then $H$ equals to $0$ on ${\fR}^{m|0}$, i.e. for  any ${\mathbf{I}}\in{\mathcal{I}}$, ${\proj}_{\mathbf{I}}(H(x))=0$.
\end{claim}
\par
{\it Proof:} 
We apply Taylor's expansion formula~\eqref{tay-f} once more: For any $N$ and ${\mathbf{J}}$, remarking 
$\partial_q^{\alpha}H_{\mathbf{J}}(q)=0$, we have
$$
H_{\mathbf{J}}(x_{\mathrm{B}}+x_{\mathrm{S}})=\tau_N(H_{\mathbf{J}};x_{\mathrm{B}},x_{\mathrm{S}})
=\sum_{|\alpha|=N+1} x_{\mathrm{S}}^\alpha\int_0^1 dt\frac{1}{N!}(1-t)^N\partial_x^\alpha H_{\mathbf{J}}(x_{\mathrm{B}}+tx_{\mathrm{S}}).
$$
We need to show that for any ${\mathbf{I}}$, ${\proj}_{\mathbf{I}}(\tau_N(H_{\mathbf{J}};x_{\mathrm{B}},x_{\mathrm{S}}))=0$.
Since all terms consisting of $x_{\mathrm{S}}^{\alpha}$ have at least $2|\alpha|$ as the degree of Grassmann generators, if $2|\alpha|>|{\mathbf{I}}|\ge0$ then ${\proj}_{\mathbf{I}}(x_{\mathrm{S}}^{\alpha})=0$.
Taking $N$ sufficiently large such that ${\proj}_{\mathbf{I}}(x_{\mathrm{S}}^{\alpha})=0$ for all $\alpha$ with $|\alpha|=N+1$,
then for any  ${\mathbf{J}}$, ${\proj}_{\mathbf{I}}(\tau_N(H_{\mathbf{J}};x_{\mathrm{B}},x_{\mathrm{S}}))=0$. Therefore,
$$
{\proj}_{\mathbf{I}}(\sum_{{\mathbf{J}}\in{\mathcal{I}}}\sigma^{\mathbf{J}}\tau_N(H_{\mathbf{J}};x_{\mathrm{B}},x_{\mathrm{S}}))=0.
\qquad /\!\!/
$$
The proof of Proposition~\ref{dew-cl} is given by applying above Claim to $H(x)=F(x)-\tilde{f}(x)$. $\qquad\qed$

Following is claimed as (1.1.17) in \cite{DeW84-2} without proof and cited as Theorem 1 in \cite{MK86}.
\begin{claim}
Let $f$ be an analytic function from an open set $V\subset{\mathbb{C}}$ to $\mathbb{C}$.
Then, we have an unique Grassmann continuation $\tilde{f}:{\fC}^{1|0}\to{\fC}$ which is super analytic.
\begin{equation}
\tilde{f}(z)=\sum_{n=0}^{\infty}\frac{1}{n!}f^{(n)}(z_{\mathrm{B}})z_{\mathrm{S}}^n\quad
\mbox{for $z=z_{\mathrm{B}}+z_{\mathrm{S}}$ with $z_{\mathrm{B}}\in V$}.
\label{AGC}
\end{equation}
That is,
$$
{\tilde f}(z+w)={\tilde f}(z)+wF(z)
+w\rho_j(z,w)
\quad
\mbox{when $w\to0$ in ${\fC}^{1|0}$, then $\rho_j(z,w)\to 0$ in ${\fC}$}.
$$
\end{claim}
\begin{remark}
Above Claim itself is proved, since $F$-differentiability of $\tilde{f}\in {\mathfrak{F}}_{S\!D}^1({\fC}^{1|0}:{\fC})$ from $\fC$ to $\mathbb{C}$ is shown, by applying Corollary~\ref{XXX-cor}. Moreover, from Theorem~\ref{yag-thm3} below,
$\tilde{f}\in {\mathfrak{F}}_{S\!D}^{\infty}({\fC}^{1|0}:{\fC})$, i.e. $\tilde{f}$ is super analytic.
\par
But I want to point out the argument in the proof of Proposition~\ref{prop3-1.2} in S. Matsumoto and K. Kakazu~\cite{MK86} which seems not transparent. They claim the convergence of the right-hand side of \eqref{AGC} and using this, they proceed as follows:
\allowdisplaybreaks
\begin{align*}
{\tilde f}(z+w)&=\sum_{n=0}^\infty \frac{1}{n!}
f^{(n)}(z_{\mathrm{B}}+w_{\mathrm{B}})(z_{\mathrm{S}}+w_{\mathrm{S}})^n\\
&=\sum_{n=0}^\infty \frac{1}{n!}\bigg(\sum_{\ell=0}^\infty \frac{1}{\ell!}
f^{(\ell+n)}(z_{\mathrm{B}})w_{\mathrm{B}}^\ell\bigg)
\bigg(\sum_{k=0}^n
\frac{n!}{k!(n-k)!}z_{\mathrm{S}}^{n-k}w_{\mathrm{S}}^k\bigg)
\quad\text{(analyticity of $f$ on $\mathbb{C}$)}\\
&=\sum_{n=0}^\infty\bigg[\sum_{\ell=0}^\infty \frac{1}{\ell!}
f^{(\ell+j+k)}(z_{\mathrm{B}})w_{\mathrm{B}}^\ell 
\bigg(\sum_{k+j=n}
\frac{1}{k!j!}z_{\mathrm{S}}^{j}w_{\mathrm{S}}^k\bigg)\bigg]\quad\text{(renumbering)}\\
&=\sum_{n=0}^\infty\frac{1}{n!}\bigg[\sum_{{j}=0}^\infty \frac{1}{j!}
f^{(n+j)}(z_{\mathrm{B}})z_{\mathrm{S}}^{j}
\bigg(\sum_{\ell+k=n}
\frac{n!}{\ell!k!}w_{\mathrm{B}}^\ell w_{\mathrm{S}}^k\bigg)\bigg]\quad\text{(rearranging)}\\
&=\sum_{n=0}^\infty\frac{1}{n!}\bigg(\sum_{{j}=0}^\infty \frac{1}{j!}
f^{(n+j)}(z_{\mathrm{B}})z_{\mathrm{S}}^{j}\bigg)
(w_{\mathrm{B}}+w_{\mathrm{S}})^n
=\sum_{n=0}^\infty\frac{1}{n!}{\tilde f}^{(n)}(z)w^n.
\end{align*}
%}
From this expression, they conclude that $\tilde{f}$ is super analytic. Surely, from this expression, putting
$$
F(z)={\tilde f}^{(1)}(z),\;\;
\rho(z,w)=\sum_{n=2}^\infty\frac{1}{n!}{\tilde f}^{(n)}(z)w^{n-1},
$$
they have
$$
{\tilde f}(z+w)-{\tilde f}(z)
=F(z)w+w\rho(z,w).
$$
But we need to show that $F(z)$ is continuous w.r.t. $z$ and $\rho(z,w)$ is horizontal w.r.t $w$ to claim $\tilde{f}$ is super analytic. This horizontality is not so clear from their last argument. To clarify this,
I propose to use the analogous proof in Claim~\ref{1102}. % of the proof in Corollary~\ref{XXX-cor}.
\end{remark}
\begin{remark}
If $f$ is real analytic on ${\euc}^m$, there exists a function $\delta(q)>0$ such that
for $|q'|\le\delta(q)$, $f(q+q')$ has Taylor expansion at $q$. From above proof, $\tilde{f}(x+y)$ is Pringsheim regular w.r.t. $|y_{\mathrm{B}}|\le\delta(x_{\mathrm{B}})$. Here, those who is not familiar with Pringsheim regular, please check it in inter-net.
\end{remark}

\subsection{Super smooth functions on superdomain}
For future use, we prepare some algebraic lemmas.
\begin{lemma}\label{yg-vv}
Suppose that there exist elements $\{A_i\}_{i=1}^{\infty}\subset {\rod}$ satisfying
\begin{equation}
\sigma_j A_i+ \sigma_i A_j=0 \forany i,j\in{\mathbb{N}}
\label{VV2.19-1}
\end{equation}
Then there exists a unique element $F\in {\fR}$ such that $A_i=\sigma_i F$ for $i=1,{\cdots},{\infty}$.
\end{lemma}
\par{\it Proof}. We follow the argument in Lemma 4.4 of \cite{yag88}. Since $A_i$ is represented by $A_i=\sum_{{\mathbf{J}}\in{\mathcal{I}}}a_{\mathbf{J}}^i\sigma^{\mathbf{J}}$ with $a_{\mathbf{J}}^i\in{\mathbb{C}}$
and $\sigma_iA_i=0$, we have $\sum_{\{{\mathbf{J}}\,|\,j_i=0\}}a_{\mathbf{J}}^i\sigma^{\mathbf{J}}=0$.
Therefore, each $A_i$ can be written uniquely as
$A_i=(\sum_{\{{\mathbf{J}}\,|\,j_i=0\}}b_{\mathbf{J}}^i\sigma^{\mathbf{J}})\sigma_i$ for some $b_{\mathbf{J}}^i\in\mathbb{C}$.
From the condition \eqref{VV2.19-1}, we have $b_{\mathbf{J}}^i=b_{\mathbf{J}}^j$ for ${\mathbf{J}}$ with $j_i=j_j=0$.
Letting $b_{\mathbf{J}}=b_{\mathbf{J}}^i$ for ${\{{\mathbf{J}}\,|\,j_i=0\}}$, we put 
$$
F=\sum_{{\mathbf{J}}\in{\mathcal{I}}}b_{\mathbf{J}}\sigma^{\mathbf{J}}=\sum_{i=1}^{\infty}(\sum_{\{{\mathbf{J}}\,|\,j_i=0\}\in{\mathcal{I}}}b_{\mathbf{J}}^i\sigma^{\mathbf{J}})
$$
which is well-defined and further more $A_i=\sigma_ i F$ holds for each $i$. 
Since we may change the order of summation freely in $\fR$, we have
$$
F=\sum_{\{{\mathbf{J}}\,|\,j_i=0\}}b_{\mathbf{J}}\sigma^{\mathbf{J}}+\sum_{\{{\mathbf{J}}\,|\,j_i\neq{0}\}}b_{\mathbf{J}}\sigma^{\mathbf{J}}
=\sum_{\{{\mathbf{J}}\,|\,j_j=0\}}b_{\mathbf{J}}\sigma^{\mathbf{J}}+\sum_{\{{\mathbf{J}}\,|\,j_j\neq{0}\}}b_{\mathbf{J}}\sigma^{\mathbf{J}}. \qquad\qed
$$

Repeating above argument, we have
\begin{corollary}[Lemma 4.4 of \cite{yag88}]
Let $\{A_{\mathbf{J}}\in{\mathfrak{R}}\;|\; |{\mathbf{J}}|={\mathrm{od}}\}$ satisfy 
$$
\sigma^{\mathbf{K}}A_{\mathbf{J}}+\sigma^{\mathbf{J}}A_{\mathbf{K}}=0\for {\mathbf{J}},{\mathbf{K}}\in{\mathcal{I}}_{\mathrm{od}}.
$$
Then there exists a unique element $F\in {\fR}$ such that $A_{\mathbf{J}}=\sigma^{\mathbf{J}} F$ for ${\mathbf{J}}\in{\mathcal{I}}_{\mathrm{od}}$.
\end{corollary}

\begin{definition}
We denote the set of maps $f:{\rod}\to{\fR}$ which are 
continuous and ${\fR}_{\mathrm{ev}}$-linear
(i.e. $f(\lambda X)=\lambda f(X)$ for $\lambda\in {\rev}$, $X\in {\rod}$)
by $f\in {\mathbf{L}}_{\rev}({\rod}: {\fR})$.
\end{definition}
\begin{corollary}[The \underline{self-duality} of ${\fR}$]\label{sd-FA}
For $f\in {\mathbf{L}}_{\rev}({\rod}: {\fR})$, 
there exists an element $u_f\in{\fR}$ satisfying
$$
f(X)=X\!{\cdot} u_f\for X\in {\rod}.
$$
\end{corollary}
\par{\it Proof}. 
Since $f:{\rod}\to {\fR}$ is ${\rev}$-linear, we have $f(XYZ)=XYf(Z)=-XZf(Y)$ for any $X, Y, Z\in {\rod}$.
By putting $X=\sigma_k, Y=\sigma_j, Z=\sigma_i$ and $f_i=f(\sigma_i)\in{\fR}$ for $i=1,{\cdots},{\infty}$,
we have $\sigma_k(\sigma_jf_i+\sigma_if_j)=0$ for any $k$. 
Therefore, $\sigma_jf_i+\sigma_if_j=0$, and by Lemma above,
there exists $u_f\in {\mathfrak{R}}$ such that $f_i=\sigma_i u_f$ for $i=1,{\cdots},{\infty}$.
\par
For ${\mathbf{I}}=(i_1,{\cdots})\in \{0,1\}^{\mathbb{N}}$ and $|{\mathbf{I}}|={\mathrm{odd}}$, then $i_k=1$ for some $k$. 
Rewrite ${\mathbf{I}}=(-1)^{i_1+{\cdots}+i_{k-1}}\sigma^{\check{{\mathbf{I}}}_k}\sigma_k$ 
with ${\check{{\mathbf{I}}}_k}=({i_1,{\cdots},i_{k-1},0},i_{k+1},{\cdots})$,
by  ${\rev}$-linearity of $f$, then we have $f(\sigma^{\mathbf{I}})=(-1)^{i_1+{\cdots}+i_{k-1}}\sigma^{\check{{\mathbf{I}}}_k}f(\sigma_k)$. 
Then, this map is well-defined because of $\sigma_jf_i+\sigma_if_j=0$, that is, it doesn't depend on other decomposition of ${\mathbf{I}}$. 
\begin{quotation}
{\small
$\because)$ Put $\tilde{\mathbf{I}}=({{i_1,{\cdots},i_{j-1},0},i_{j+1},{\cdots},i_{k-1},0},i_{k+1},{\cdots})$, for ${\mathbf{I}}=({{i_1,{\cdots},i_{j-1},1},i_{j+1},{\cdots},i_{k-1},1},i_{k+1},{\cdots})$.
Then, for ${\ell}=i_1+{\cdots}+i_{j-1}+i_{j+1}+{\cdots}+i_{k-1}\in{\mathbb{N}}$, 
remarking $|\tilde{\mathbf{I}}|=$odd, we have
$$
{\sigma^{\mathbf{I}}=\begin{cases}
(-1)^{\ell}\sigma_k\sigma_j\sigma^{\tilde{\mathbf{I}}}\\
-(-1)^{\ell}\sigma_j\sigma_k\sigma^{\tilde{\mathbf{I}}},
\end{cases}}
\Longrightarrow
{\begin{cases}
f((-1)^{\ell}\sigma_k\sigma_j\sigma^{\tilde{\mathbf{I}}})=(-1)^{\ell}\sigma_j\sigma^{\tilde{\mathbf{I}}}f(\sigma_k),\\
f(-(-1)^{\ell}\sigma_j\sigma_k\sigma^{\tilde{\mathbf{I}}})=-(-1)^{\ell}\sigma_k\sigma^{\tilde{\mathbf{I}}}f(\sigma_j).
\end{cases}}
$$
By $\sigma_jf_k+\sigma_kf_j=0$, we have 
$$
f((-1)^{\ell}\sigma_k\sigma_j\sigma^{\tilde{\mathbf{I}}})-f(-(-1)^{\ell}\sigma_j\sigma_k\sigma^{\tilde{\mathbf{I}}})
=-(-1)^{\ell}\sigma^{\tilde{\mathbf{I}}}[\sigma_jf_k+\sigma_kf_j]=0.
$$}
\end{quotation}

We extend $\tilde{f}$ as
$\tilde{f}(X)=\sum_{{\mathbf{I}}\in{\mathcal{I}}} X_{\mathbf{I}}{f}(\sigma^{\mathbf{I}})$ for $X=\sum_{{\mathbf{I}}\in{\mathcal{I}}} X_{\mathbf{I}}\sigma^{\mathbf{I}}\in {\rod}$.
Then, since $X_{\mathbf{I}}\in\mathbb{C}$,
$\tilde{f}(X)=\sum_{{\mathbf{I}}\in{\mathcal{I}}} X_{\mathbf{I}} \sigma^{\mathbf{I}} u_f=X\!{\cdot} u_f$.
In fact, if ${\mathbf{I}}$ with $|{\mathbf{I}}|$=odd with $i_k\neq{0}$, then,
by $f_k=f(\sigma_k)=\sigma u_f$ and $\rev$-linearity,
$$
\tilde{f}(\sigma^{\mathbf{I}})={f}(\sigma^{\mathbf{I}})=(-1)^{i_1+{\cdots}+i_{k-1}}\sigma^{{\check{{\mathbf{I}}}_k}}f(\sigma_k)
=(-1)^{i_1+{\cdots}+i_{k-1}}\sigma^{{\check{\mathbf{I}}_k}}\sigma_k u_f=\sigma^{\mathbf{I}} u_f.
$$
Clearly $\tilde{f}(X)=f(X)$.  $\qquad\qed$

\begin{remark}
K. Masuda gives the following example which exhibits that \underline{${\mathfrak{B}}_L$ is not necessarily}\\
\underline{self-dual}.
\par{A counter-example\footnote{Though I don't recognize at first reading, but analogously examples are considered in \cite{BG84} or \cite{JaPi81}}}: {\it
Let $L=2$. Define a map $f$ as
$$
f(X_1\sigma_1+X_2\sigma_2)=X_1\sigma_2 \forany X_1, X_2\in\mathbb{R}.
$$
Then, remarking that $(b_0+b_1\sigma_1\sigma_2)(X_1\sigma_1+X_2\sigma_2)=b_0(X_1\sigma_1+X_2\sigma_2)$,
we have readily $f\in {\mathbf{L}}_{{\mathfrak{B}}_{2,{\mathrm{ev}}}}({\mathfrak{B}}_{2,{\mathrm{od}}}:{\mathfrak{B}}_{2})$.
If we assume that there exists a $u_f\in {\mathfrak{B}}_{2}$ such that $f(X)=X\!{\cdot} u_f$, then
$\sigma_1  f(\sigma_1)=\sigma_1\cdot\sigma_1\!{\cdot} u_f=0$ but
$\sigma_1f(\sigma_1)=\sigma_1\cdot\sigma_2\neq{0}$, contradiction!
Hence, there exists no $u_f\in {\mathfrak{B}}_{2}$ such that $f(X)=X\!{\cdot} u_f$.
}
\end{remark}

Repeating the argument in proving Corollary \ref{cor3-1.3}, 
we get 
\begin{equation}
\left.
 f\in{\CSS}({\mathfrak{U}}:{\fC})\Longrightarrow {\frac{d}{dt}}f(X+tY)\right|_{t=0}=
\sum_{{j}=1}^{m}
y_j{\frac{\partial}{\partial x_j}}f(X)
+\sum_{s=1}^{m}
\omega_s{\frac {\partial}{\partial \theta_s}}f(X)
\label{II-3-11-1}
\end{equation}
where $X=(x,\theta),Y=(y,\omega)\in \supermn$ 
such that $X+tY\in {\mathfrak{U}}$ for any $t\in[0,1]$. That is,
\begin{corollary}
${\CSS}({\mathfrak{U}}:{\fC})\Longrightarrow {\mathfrak{G}} _{S\!D}^1({\mathfrak{U}}:{\fC})$.
\end{corollary}

To relate the definitions ${\CSS}$ and ${\mathfrak{G}} _{S\!D}^{\infty}$ or ${\mathfrak{F}} _{S\!D}^{\infty}$, we need the following notion.
\begin{definition}[p.246 of \cite{yag88}]
Let ${\mathfrak{U}}$ be an open set in ${\fR}^{m|n}$ and  $f:{\mathfrak{U}}\to\euc $(or $\to\mathbb{C}$). 
$f$ is said to be \underline{admissible} on ${\mathfrak{U}}$ if there exists some $L\ge0$ and a $\euc$(or $\mathbb{C}$)-valued function $\phi$ defined on ${\mathfrak{U}}_{L}=p_{L}({\mathfrak{U}})$ such that $f(X)=\phi{\circ}p_{L}(X)=\phi(p_{L}(X))$.
For $r$ with $(0\le r\le{\infty})$,
$f$ is said to be admissible $C^r$ (or simply $f\in C_Y^r({\mathfrak{U}}:{\fC})$) if $\phi\in C^r({\mathfrak{U}}_{L}:\euc)$ or $C^r({\mathfrak{U}}_{L}:\mathbb{C})$.
\par
Let $f(X)=\sum_{{\mathbf{I}}\in{\mathcal{I}}}\sigma^{\mathbf{I}}{\cdot}f_{\mathbf{I}}(X)$ with $f_{\mathbf{I}}$ is $\euc$(or $\mathbb{C}$)-valued on ${\mathfrak{U}}$.
For each ${\mathbf{I}}\in{\mathcal{I}}$, if $f_{\mathbf{I}}$ is admissible $C^r$ (or simply $f\in C_Y^r$) on ${\mathfrak{U}}$, $f\in C_Y^r({\mathfrak{U}}:{\fC})$ is called \underline{admissible} on ${\mathfrak{U}}$.
More precisely, there exists some $L_{\mathbf{I}}\ge0$ and a $\euc$(or $\mathbb{C}$)-valued function $\phi_{\mathbf{I}}$ defined on ${\mathfrak{U}}_{L}=p_{L}({\mathfrak{U}})$ such that $f_{\mathbf{I}}(X)=\phi_{\mathbf{I}}{\circ}p_{L}(X)=\phi_{\mathbf{I}}(p_{L}(X))$.
Moreover, we define its partial derivatives by  %for $|A|=|L|$, 
$$
\frac{\partial f}{\partial X_{A,{\mathbf{K}}}}=\sum_{\mathbf{J}}\sigma^{\mathbf{J}}{\cdot}\frac{\partial f_{\mathbf{J}}}{\partial X_{A,{\mathbf{K}}}}=
\begin{cases}
|{\mathbf{K}}|={\mathrm{ev}} & \mbox{if $1\le A\le m$},\\
|{\mathbf{K}}|={\mathrm{od}} & \mbox{if $m+1\le A\le m+n$}
\end{cases}.
$$
\end{definition}
\begin{definition}[p.246 of \cite{yag88}]
A ${\fR}$ (or ${\fC}$)-valued function $f$ on ${\mathfrak{U}}$ is said to be  \underline{projectable} if for each $L\ge0$, 
there exists a ${\fR}$(or $\fC$)-valued function $f_{L}$ defined on ${\mathfrak{U}}_{L}\subset {\fR}^{m|n}_{L}$ such that $p_{L}{\circ}f=f_{L}{\circ}p_{L}$ on ${\mathfrak{U}}$.
\end{definition}
\begin{claim}
A projectable function on ${\mathfrak{U}}$ is also admissible on ${\mathfrak{U}}$.
\end{claim}
\par{\it Proof}.
We use the map ${\proj}_{\mathbf{I}}:{\fR}\ni X=\sum_{{\mathbf{I}}\in{\mathcal{I}}}X_I\sigma^{\mathbf{I}}\to X_{\mathbf{I}}\in{\euc}(\mbox{or ${\mathbb{C}}$})$ introduced in \S 2.
Then, for each ${\mathbf{I}}\in{\mathcal{I}}$, taking $L$ such that  ${\mathbf{I}}\in{\mathcal{I}}_{L}$, we have 
$$
{\begin{CD}
{\mathfrak{U}} @>f>> {\fR}\\
@V{p_{L}}V{}V  %
@V{}V{p_{L}}V\\
%@VV{\cong}V\\
{\mathfrak{U}}_{L} @>>f_{L}>{\fR}_{L}\\
\end{CD}}\quad
\Longrightarrow\quad
{\begin{CD}
{\mathfrak{U}} @>{\proj}_{\mathbf{I}}{\circ}f>>{\mathbb{C}}\\
@V{p_{L}}V{}V  %
@A{}A{\mbox{Id}}A\\
%@VV{\cong}V\\
{\mathfrak{U}}_{L} @>>{\proj}_{\mathbf{I}}{\circ}f_{L}> {\mathbb{C}}
\end{CD}}. \qquad\qed
$$

\begin{theorem}[Theorem 1 of \cite{yag88}]\label{yagi-thm1}
Let ${\mathfrak{U}}$ be a convex open set in ${\fR}^{m|n}$.
If $f:{\mathfrak{U}}\to{\euc}$ is in ${\mathfrak{G}} _{S\!D}^1$, then $f$ is projectable and $C_Y^1$ on ${\mathfrak{U}}$.
\end{theorem}
\par{\it Proof}.
Since
$\frac{d}{dt} f(X+tH)=\sum_{A=1}^{m+n} H_A F_A(X+tH)$, we have
$$
f(X+H)-f(X)=\int_0^1\frac{d}{dt} f(X+tH)dt=\sum_{A=1}^{m+n} H_A \int_0^1F_A(X+tH) dt.
$$
This means that if $p_{L}(H_A)=0$, then $p_{L}(f(X+H)-f(X))=0$. Therefore if we define $f_{L}:{\mathfrak{U}}_{L}\to {\fR}_{L}$ by $f_{L}(p_{L}(Z))=p_{L}(f(Z))$, then it implies that $f$ is projectable and so admissible.
For $E_{A,{\mathbf{K}}}=\sigma^{\mathbf{K}}{\mathbf{e}}_A\in {\euc}^{m|n}$ with
${\mathbf{e}}_A=(\underbrace{\overbrace{0,{\cdots},0,1}^{A},0,{\cdots},0}_{m+n})$, we have %{\mathfrak{e}}
$$
\frac{\partial}{\partial X_{A,{\mathbf{K}}}}f(X)=\frac{d}{dt} f(X+tE_{A,{\mathbf{K}}})\big|_{t=0}=\sigma^{\mathbf{K}} F_A(X), \quad |A|=|{\mathbf{K}}|.
$$
$f_{L}$ is $C^1$ on ${\mathfrak{U}}_{L}$, thus the function $f$ is admissible $C^1$ on ${\mathfrak{U}}$. $\qquad\qed$

\subsection{Cauchy-Riemann relation}

To understand the meaning of supersmoothness, we consider the dependence with
respect to the ``coordinate" more precisely.

\begin{proposition}[Theorem 2 of \cite{yag88}]\label{propII-3.5-1}
Let $ f(X)=\sum_{\mathbf{I}}f_{\mathbf{I}}(X)\sigma^{\mathbf{I}} \in G^{\infty}_{S\!D} ({\mathfrak{U}}:{\fC})$  %{\CSS}
where ${\mathfrak{U}}$ is a superdomain in $\supermn$.
Let $X=(X_{A})$ be represented by 
$X_{{A}}=\sum_{{\mathbf{I}}} X_{{A},{\mathbf{I}}}\sigma^{\mathbf{I}}$ where ${A}=1,{\cdots}, m+n$,
$ X_{{A},{\mathbf{I}}}\in {\mathbb C}$ for $|{\mathbf{I}}|\ne 0$ and $ X_{{A},\tilde 0}\in {\euc}$.
Then, $f(X)$, considered as a function 
of countably many variables $\{X_{{A},{\mathbf{I}}}\}$ with values in $\fC$,
satisfies the following (Cauchy-Riemann type) equations.
\begin{equation}
\left\{
{\begin{aligned}
&{\frac{\partial}{\partial  X_{{A},{\mathbf{I}}}}}f(X) =
\sigma^{\mathbf{I}} {\frac{\partial}{\partial X_{{A},\tilde 0}}}f(X)
 \for 1\le{A}\le m,\; |{\mathbf{I}}|={\mathrm{ev}},\\
&\sigma^{\mathbf{K}}{\frac{\partial}{\partial X_{{A},{\mathbf{J}}}}}f(X)
+\sigma^{\mathbf{J}}{\frac{\partial}{\partial X_{{A},{\mathbf{K}}}}}f(X)=0\for m+1\le{A}\le m+n,\;|{\mathbf{J}}|={\mathrm{od}}=|{\mathbf{K}}|.
\end{aligned}}
\right.
\label{II-3-12-1-1}
\end{equation}
Here, we define
\begin{equation}
{\frac{\partial}{\partial  X_{{A},{\mathbf{I}}}}}f(X) =
\left.{\frac{d}{dt}}f(X+tE_{{A},{\mathbf{I}}})\right|_{t=0}
\with
E_{{A},{\mathbf{I}}}=\sigma^{\mathbf{I}}{\mathfrak{e}}_A=
(\overbrace{0,{\cdots},0,\sigma^{\mathbf{I}}}^{A},0,{\cdots},0)\in{\supermn}.
\label{II-3-13-1}
\end{equation}
\par
Conversely, let a function $f(X)=\sum_{\mathbf{I}}f_{\mathbf{I}}(X)\sigma^{\mathbf{I}}$ be given such that
$f_{\mathbf{I}}(X+tY)\in C^\infty([0,1]:{\mathbb C})$ for each fixed $X,Y\in U$ and
$f(X)$ satisfies above \eqref{II-3-12-1-1} with \eqref{II-3-13-1}. 
Then,  $ f \in G^{\infty}_{S\!D}({\mathfrak{U}}:{\fC})$.
\end{proposition}

{\it Proof}. 
Replacing $Y$ with $E_{{A},{\mathbf{J}}}$
with $1\le{A}\le m$ and $|{\mathbf{J}}|=$even in \eqref{II-3-11-1},
we get readily the first equation of \eqref{II-3-12-1-1}. 
Here, we have used \eqref{II-3-5}.
Considering $E_{{A},{\mathbf{J}}}$ or $E_{{A},{\mathbf{K}}}$
for $m+1\le {A} \le m+n$ and $|{\mathbf{J}}|=\text{odd}=|{\mathbf{K}}|$ in \eqref{II-3-11-1}
and multiplying $\sigma^{\mathbf{K}}$ or $\sigma^{\mathbf{J}}$ from left, respectively,
we have the second equality in \eqref{II-3-12-1-1} readily. 	
\par
To prove the converse statement, we have to construct functions 
$F_A$($1\le A\le m+n$) which satisfies
\begin{equation}
{\dt}f(X+tH)\big|_{t=0}=\sum_{A=1}^{m+n} H_A F_A(X)
\label{fsd1}
\end{equation}
for $X\in U$ and $H=(H_A)\in\supermn$. 

For $1\le{A}\le m$, we put
$\displaystyle{
F_{A}(X)=\frac{\partial}{\partial X_{{A},\tilde 0}}f(X)
\for \;X\in {\mathfrak{U}}}$.

On the other hand,
from the second equation of \eqref{II-3-12-1-1} and Lemma \ref{yg-vv},
we have an element $F_{A}(X)$($m+1\le {A}\le m+n$) such that
$\displaystyle{
\sigma^{j}F_{A}(X)=\frac{\partial}{\partial X_{{A},{\mathbf{J}}}}f(X)}$.

Using these $\{F_A(X)\}$ defined above, we claim that \eqref{fsd1} holds following Yagi's argument.

Since $f$ is admissible, for any $L\ge0$, $p_{L}{\circ}f$ is so also, therefore there exist some $N\ge0$ and a ${\fR}_{L}$-valued $C^{\infty}$ function $f_N$ such that $p_{L}{\circ}f(X)=f_N{\circ}p_N(X)$ on $X\in {\mathfrak{U}}$.
By natural imbedding from ${\fR}_{L}$ to ${\fR}_N$, we may assume $N\ge L$.
Then, we can show that
$$
\frac{\partial}{\partial X_{A,{\mathbf{K}}}}f_N(p_N(X))=\begin{cases}
p_{L}\big(\displaystyle{\frac{\partial}{\partial X_{A,{\mathbf{K}}}}f(X)}\big) &\mbox{ if ${\mathbf{K}}\in{\mathcal{I}}_N$},\\
\quad 0 &\mbox{ if otherwise.}
\end{cases}
$$ 
Therefore, for any $L\ge 0$, 
{\allowdisplaybreaks
\begin{align*}
p_{L}\big(\dt f(X+tH)\bigg|_{t=0}\big) &=\dt p_{L}(f(X+tH))\bigg|_{t=0}
\big(\because)\, p_{L}(\dt g(t)\big|_{t=0})=\dt(p_{L}(g(t))\big|_{t=0}\big),\\
&=\dt f_N(p_N(X+tH))\bigg|_{t=0}
\big(\because)\, p_{L}(f(X))=f_N(p_N(X)) \big)\\
&=\sum_{A}\sum_{\mathbf{K}}(p_N(H))_{A,{\mathbf{K}}}{\cdot}\frac{\partial}{\partial X_{A,{\mathbf{K}}}}f_N(p_N(X)) \big(\because)\, \mbox{finite dimensional case}\big)\\
&=\sum_{A}\sum_{\mathbf{K}}(p_N(H))_{A,{\mathbf{K}}}{\cdot}p_{L}\big(\frac{\partial}{\partial X_{A,{\mathbf{K}}}}f(X)\big) \big(\because)\, p_{L}(g(X))=g_N(p_N(X)) \big)\\
&=\sum_{A}\sum_{\mathbf{K}}(p_N(H))_{A,{\mathbf{K}}}{\cdot}p_{L}(\sigma^{\mathbf{K}} F_A(X)) \big(\because)\, \mbox{by \eqref{II-3-12-1-1}}\big) \\
&=\sum_{A}\sum_{\mathbf{K}}(p_N(H))_{A,{\mathbf{K}}}{\cdot}p_{L}(\sigma^{\mathbf{K}}){\cdot}p_{L}(F_A(X))\\
&=\sum_{A}\big(\sum_{\mathbf{K}}(p_N(H))_{A,{\mathbf{K}}}{\cdot}p_{L}(\sigma^{\mathbf{K}})\big)p_{L}(F_A(X))\\
&=\sum_{A}p_{L}((p_N(H))_{A,{\mathbf{K}}}{\cdot}p_{L}(\sigma^{\mathbf{K}}))p_{L}(F_A(X))\\
&=\sum_{A}(p_{L}(H))_{A}{\cdot}p_{L}(F_A(X))
=p_{L}(\sum_{A}H_AF_A(X)).
\end{align*}}
Thus, we have \eqref{fsd1}. The continuity of $F_A(X)$ is clear. $\qquad\qed$

\begin{remark}
For function with finite number of independent variables, it is well-known how to define its partial derivatives.
But when that number is infinite, it is not so clear whether the change of order of differentiation affects the result, etc.
Therefore, we reduce the calculation to the cases with finite number $L$ of generators and making that $L$ to infinity.
\end{remark}

\begin{theorem}[Theorem 3 of \cite{yag88}]\label{yag-thm3}
Let $f$ be a $\fC$-valued $C^{\infty}$ function on an open set ${\mathfrak{U}}\subset \supermn$.
If $f$ is ${\mathfrak{G}} _{S\!D}^1({\mathfrak{U}}:\fC)$, then $f$ is ${\mathfrak{G}} _{S\!D}^{\infty}$ on ${\mathfrak{U}}$.
\end{theorem}
\par{\it Proof}.
Since $f\in {\mathfrak{G}} _{S\!D}^1$, it satisfies Cauchy-Riemann equation.
As $f$ is $C^{\infty}$ on ${\mathfrak{U}}$, 
$g(X)=\frac{\partial}{\partial X_{A,\tilde{0}}}f(X)$ also satisfies the C-R equation, for $1\le A\le m$.
In fact, for $1\le B\le m$, $|{\mathbf{J}}|=$even,
$$
\begin{aligned}
\frac{\partial}{\partial X_{B,{\mathbf{J}}}}g(X)&=\frac{\partial}{\partial X_{B,{\mathbf{J}}}}\frac{\partial}{\partial X_{A,\tilde{0}}}f(X)
=\frac{\partial}{\partial X_{A,\tilde{0}}}\frac{\partial}{\partial X_{B,{j}}}f(X)\\
&=\frac{\partial}{\partial X_{A,\tilde{0}}}\sigma^{\mathbf{J}}\frac{\partial}{\partial X_{B,\tilde{0}}}f(X)
=\sigma^{\mathbf{J}}\frac{\partial}{\partial X_{B,\tilde{0}}}\frac{\partial}{\partial X_{A,\tilde{0}}}f(X)
=\sigma^{\mathbf{J}}\frac{\partial}{\partial X_{B,\tilde{0}}}g(X).
\end{aligned}
$$
And for $m+1\le A\le m+n$, $|{\mathbf{J}}|=|{\mathbf{K}}|={\mathrm{odd}}$,
$$
\begin{aligned}
\sigma^{\mathbf{K}}{\frac{\partial}{\partial X_{{B},{\mathbf{J}}}}}g(X)
+\sigma^{\mathbf{J}}{\frac{\partial}{\partial X_{{B},{\mathbf{K}}}}}g(X)
&=\sigma^{\mathbf{K}}{\frac{\partial}{\partial X_{{B},{\mathbf{J}}}}}\frac{\partial}{\partial X_{A,\tilde{0}}}f(X)
+\sigma^{\mathbf{J}}{\frac{\partial}{\partial X_{{B},{\mathbf{K}}}}}\frac{\partial}{\partial X_{A,\tilde{0}}}f(X)\\
&=\frac{\partial}{\partial X_{A,\tilde{0}}}\bigg(\sigma^{\mathbf{K}}{\frac{\partial}{\partial X_{{B},{\mathbf{J}}}}}f(X)
+\sigma^{\mathbf{J}}{\frac{\partial}{\partial X_{{B},{\mathbf{K}}}}}f(X)\bigg)=0.
\end{aligned}
$$
Hence $\frac{\partial}{\partial X_A}f$(for $1\le A\le m$) is ${\mathfrak{G}} _{S\!D}^1$ on ${\mathfrak{U}}$. 

Analogously, for $m+1\le A\le m+n$, $\frac{\partial}{\partial X_{A,J}}f=\sigma^{\mathbf{J}}{\cdot}\frac{\partial}{\partial X_A}f$ is also ${\mathfrak{G}} _{S\!D}^1$ on ${\mathfrak{U}}$. 
In fact, we have, for $|{\mathbf{K}}|={\mathrm{even}}$,

\begin{lemma}[Lemma 5.1  of \cite{yag88}]\label{yagi5-1}
Let $f\in {\mathfrak{G}} _{S\!D}^{\infty}({\fR}^{0|n})$. Then
$$
f(\theta)=f(\theta_1,{\cdots},\theta_n)=\sum_{|a|\le n}\theta^a f_a\with f_a\in\fC.
$$
\end{lemma}
\par{\it Proof}.
For  $n=1$ and $|{\mathbf{J}}|={\mathrm{odd}}$, we have,
$$
\dt f(\theta+t\sigma^{\mathbf{J}})\big|_{t=0}=\frac{\partial}{\partial \theta_{\mathbf{J}}}f(\theta)=\sigma^{\mathbf{J}}{\cdot}\frac{d}{d\theta} f(\theta)\with
\theta=\sum_{{\mathbf{I}}\in{\mathcal{I}}_{\mathrm{od}}}\theta_{\mathbf{I}}\sigma^{\mathbf{I}},\; \theta_I\in{\mathbb{C}}.
$$
Hence
$$
\frac{\partial}{\partial \theta_{\mathbf{K}}}\frac{\partial}{\partial \theta_{\mathbf{J}}}f(\theta)
=\frac{\partial}{\partial s}\frac{\partial}{\partial t}f(\theta+t\sigma^{\mathbf{J}}+s\sigma^{\mathbf{K}})\bigg|_{t=s=0}
=\sigma^{\mathbf{K}}{\cdot}\sigma^{\mathbf{J}}{\cdot}
\frac{d}{d\theta} \frac{d}{d\theta} f(\theta).
$$
Since $|{\mathbf{J}}|, |{\mathbf{K}}|$ are odd, we have $\sigma^{\mathbf{J}}\sigma^{\mathbf{K}}=-\sigma^{\mathbf{K}}\sigma^{\mathbf{J}}$ and therefore
$$
\frac{\partial}{\partial \theta_{\mathbf{K}}}\frac{\partial}{\partial \theta_{\mathbf{J}}}f(\theta)
=\sigma^{\mathbf{K}}{\cdot}\sigma^{\mathbf{J}}{\cdot}
\frac{d}{d\theta} \frac{d}{d\theta} f(\theta)
=-\sigma^{\mathbf{J}}{\cdot}\sigma^{\mathbf{K}}{\cdot}\frac{d}{d\theta} \frac{d}{d\theta} f(\theta)
=-\frac{\partial}{\partial \theta_{\mathbf{J}}}\frac{\partial}{\partial \theta_{\mathbf{K}}}f(\theta).
$$
Since $f$ is $C^{\infty}$ as a function of infinite variables $\{\theta_{\mathbf{J}}\}$ 
and its higher derivatives are symmetric, we have therefore % by Proposition \ref{h3.5.3}
$$
\frac{\partial}{\partial \theta_{\mathbf{K}}}\frac{\partial}{\partial \theta_{\mathbf{J}}}f(\theta)=0.
$$
By representing $f(\theta)=\sum_{\mathbf{K}}\sigma^{\mathbf{K}} f_{\mathbf{K}}(\theta)$, the each component $f_{\mathbf{K}}(\theta)$ is a polynomial of degree $1$ with variables $\{\theta_{\mathbf{J}}\;|\; {\mathbf{J}}\in{\mathcal{I}}_{\mathrm{od}}\}$.
Then $\displaystyle{\sigma^{\mathbf{J}}{\cdot}\frac{d}{d\theta} f(\theta)=\frac{\partial}{\partial \theta_{\mathbf{J}}}f(\theta)}$ is constant for any $|{\mathbf{J}}|=$odd.
Thus $\displaystyle{\frac{d}{d\theta} f(\theta)}$ is constant denoted by $a\in{\fC}$.
Then, $\displaystyle{\frac{d}{d\theta}( f(\theta)-{\theta}a)=0}$. Therefore there exists $b\in\fC$ such that $f(\theta)={\theta}a+b$.
\par
We proceed by induction w.r.t. $n$.
Let $f$ be a ${\mathfrak{G}} _{S\!D}^{\infty}$ function on an open set $U\subset {\fR}^{0|n}$. Fixing $\theta_1,{\cdots}, \theta_{n-1}$,
$f(\theta_1,{\cdots}, \theta_{n-1},\theta_n)$ is a ${\mathfrak{G}} _{S\!D}^{\infty}$ function with one variable $\theta_n$.
Thus, we have
$$
f(\theta_1,{\cdots}, \theta_{n-1},\theta_n)=\theta_n g(\theta_1,{\cdots}, \theta_{n-1})+h(\theta_1,{\cdots}, \theta_{n-1})
\with
\frac{\partial}{\partial \theta_n}f(\theta)=g(\theta_1,{\cdots}, \theta_{n-1}).
$$
Therefore $g$ is ${\mathfrak{G}} _{S\!D}^{\infty}$ w.r.t. $(\theta_1,{\cdots}, \theta_{n-1})$, $h$ is also ${\mathfrak{G}} _{S\!D}^{\infty}$ w.r.t. $(\theta_1,{\cdots}, \theta_{n-1})$. $\qquad\qed$

\begin{remark}
Though this Lemma with a sketch of the proof is announced in \cite{DeW84-2} and is cited in \cite{MK86} without proof, but I feel some ambiguity of his proof. This point is ameliorated by \cite{yag88} as above.
\end{remark}

\begin{lemma}[Lemma 5.2 of \cite{yag88}]\label{yag-lem5-2}
Let $f\in {\mathfrak{G}} _{S\!D}^{\infty}({\fR}^{m|0})$ on a convex open set $\mathcal{U}\subset{\fR}^{m|0}$ which vanishes identically on ${\mathcal{U}}_{\mathrm{B}}=\pi_{\mathrm{B}}(\mathcal{U})$. Then, $f$ vanishes identically on $\mathcal{U}$.
\end{lemma}
\par{\it Proof}.
It is essential to prove the case $m=1$.
Take an arbitrary point $t\in {\mathcal{U}}_{\mathrm{B}}$ and we consider the behavior of $f$ on $\pi_{\mathrm{B}}^{-1}(t)$.
Let $X\in \pi_{\mathrm{B}}^{-1}(t)$ and $X_{L}=p_{L}(X)$.
Then $\{X_{\mathbf{K}}\;|\; {\mathbf{K}}\in{\mathcal{I}}_{L},\; |{\mathbf{K}}|={\mathrm{ev}}\ge 2\}$ is a coordinate for $(\pi_{\mathrm{B}}^{-1}(t))_{L}$ as the ordinary space ${\mathbb{C}}$.
Let $f_{L}$ be the $L$-th projection of $f$. Then,
$$
\frac{\partial}{\partial X_{\mathbf{K}}}f_N(X_{L})=\sigma^{\mathbf{K}}{\cdot}\frac{\partial}{\partial X_{\tilde{0}}}f_{L}(X_{L})\for {\mathbf{K}}\in{\mathcal{I}}_{L} \et |{\mathbf{K}}|={\mathrm{ev}}.
$$
If ${\mathbf{K}}_1, {\cdots}, {\mathbf{K}}_h\in {\mathcal{I}}_{L}$, $|{\mathbf{K}}_j|=$even$>0$ and $2h>L$, then $\sigma^{{\mathbf{K}}_1}{\cdots}\sigma^{{\mathbf{K}}_h}=0$
and $\displaystyle{\frac{\partial}{\partial x_{{\mathbf{K}}_1}}{\cdots}\frac{\partial}{\partial x_{{\mathbf{K}}_h}}f_{L}(X_{L})=0}$.
This implies that $f_{L}$ is a polynomial on $(\pi_{\mathrm{B}}^{-1}(t))_{L}$.
Moreover, for any $h\ge 0$,
$$
\frac{\partial}{\partial x_{{\mathbf{K}}_1}}{\cdots}\frac{\partial}{\partial x_{{\mathbf{K}}_h}}f_{L}(t)
=\sigma^{{\mathbf{K}}_1}{\cdots}\sigma^{{\mathbf{K}}_h}\bigg(\frac{\partial}{\partial X_{\tilde{0}}}\bigg)^h f_{L}(t).
$$
Since $f$ vanishes on ${\mathcal{U}}_{\mathrm{B}}$, we have
$$
\bigg(\frac{\partial}{\partial X_{\tilde{0}}}\bigg)^h f_{L}(t)=0 \quad \mbox{on}\quad {\mathcal{U}}_{\mathrm{B}}
$$
and hence
$$
\frac{\partial}{\partial x_{{\mathbf{K}}_1}}{\cdots}\frac{\partial}{\partial x_{{\mathbf{K}}_h}}f_{L}(t)=0 \forany h\ge 0 \et
{\mathbf{K}}_1, {\cdots}, {\mathbf{K}}_h\in {\mathcal{I}}_{L} \with |{\mathbf{K}}_j|={\mathrm{even}}>0.
$$
Thus the polynomial $f_{L}\big|_{\pi_{\mathrm{B}}^{-1}(t)}$ must vanish identically and hence $f_{L}\equiv 0$ on ${\mathcal{U}}_{L}$. This holds for any $L\ge0$. Thus $f\equiv 0$ on $\mathcal{U}$. $\qquad\qed$

\subsection{Proof of Main Theorem \ref{mthm}}  %{\F}={\mathfrak{F}}, {\G}={\mathfrak{G}}
\begin{theorem}\label{mthm}
Let ${\mathfrak{U}}$ be a superdomain in ${\fR}^{m|n}$ and 
let a function $f:{\mathfrak{U}}\to {\fC}$ be given.
Following conditions are equivalent:\\
(a) $f$ is  super Fr\'echet (F-, in short) differentiable on ${\mathfrak{U}}$, i.e. $f\in {\mathfrak{F}}_{S\!D}^{\infty}({\mathfrak{U}}:{\fC})$, \\
(b) $f$ is  super G\^ateaux (G-, in short) differentiable on ${\mathfrak{U}}$, i.e. $f\in {\mathfrak{G}}_{S\!D}^{\infty}({\mathfrak{U}}:{\fC})$, \\
(c) $f$ is $\infty$-times G-differentiable 
and $f\in {\mathfrak{G}}_{S\!D}^{1}({\mathfrak{U}}:{\mathfrak{C}})$,\\
(d) $f$ is $\infty$-times G-differentiable and its G-differential $df$ is ${\rev}$-linear,\\ 
(e) $f$ is $\infty$-times G-differentiable and its G-differential $df$ satisfies 
Cauchy-Riemann type  equations,\\
(f) $f$ is supersmooth, i.e. it has the following representation, called superfield expansion, 
such that
$$
f(x,\theta)=\sum_{|a|\le n}\theta^a \tilde{f}_a(x)\with
f_a(q)\in C^{\infty}(\pi_{\mathrm{B}}({\mathfrak{U}}))\et
 \tilde{f}_a(x)=\sum_{|\alpha|=0}^{\infty}\frac{1}{\alpha!}\frac{\partial ^{\alpha}f_a(q)}{\partial q^{\alpha}}\bigg|_{q=x_{\mathrm{B}}}
 x_{\mathrm{S}}^{\alpha}.
$$
\end{theorem}
\begin{remark}
In the above, (f) stands for the ``algebraic'' nature and (a) claims the ``analytic'' nature of ``superfields''.
Yagi~\cite{yag88} proves essentially the equivalence $(b)\Longleftrightarrow(e)\Longleftrightarrow(f)$.
\end{remark}

It is clear from outset that $(a)\Rightarrow (b)\Rightarrow  (c)\Rightarrow  (d)$.
From Proposition \ref{propII-3.5-1}, $(d)\Rightarrow (a)$. Lastly, the equivalence of $(d)$ and $(e)$ is given by

\begin{theorem}[Thorem 4 of \cite{yag88}]
Let $f$ be a ${\mathfrak{G}} _{S\!D}^{\infty}$ function on a convex open set ${\mathcal{U}}\subset\supermn$. Then, there exist ${\fR}$-valued $C^{\infty}$ functions $u_a$ on ${\mathcal{U}}_{\mathrm{B}}$ such that
$$
f(x,\theta)=\sum_{|a|\le n}\theta^a \tilde{u}_a(x).
$$
Moreover, this expression is unique.
\end{theorem}
\par
{\it Proof}.
$\Longrightarrow)$
For fixed $x$, by Lemma \ref{yagi5-1}, $f(x,\theta)$ has  the representation 
$f(x,\theta)=\sum_{|a|\le n}\theta^a\varphi_a(x)$
with $\varphi_a(x)\in {\fC}$.
Since $f\in {\mathfrak{G}} _{S\!D}^{\infty}$, it is clear that for each $a$, $\varphi_a(x)\in {\fC}$ is on ${\fR}^{m|0}$ and moreover
$\varphi_a(x_{\mathrm{B}})$ is in $C^{\infty}(\euc^m)$. Denoting the Grassmann continuation of it by $\tilde{\varphi}_a(x)$, we should have $\tilde{\varphi}_a(x)=f_a(x)$ by Lemma \ref{yag-lem5-2}.

$\Longleftarrow)$ Since the supersmoothness leads the C-R relation, we get the superdifferentiability. $\qquad\qed$

\section{Inverse and implicit function theorems}
\subsection{Composition of supersmooth functions}
Following is the slight modification of the arguments in Inoue and Maeda~\cite{IM91}.
%\subsection{Composition of supersmooth functions}
\begin{definition}\label{3.14b}
Let ${\mathfrak{U}}\subset\supermn$ and ${\mathfrak{V}}\subset {\fR}^{p|q}$
be superdomains 
and let $\varphi$ be a continuous mapping from ${\mathfrak{U}}$ to ${\mathfrak{V}}$, denoted by
$\varphi(X)=(\varphi_1(X),{\cdots}, \varphi_{p}(X),
\varphi_{p+1}(X),{\cdots}, \varphi_{p+q}(X))\in{\fR}^{p|q}$.
$\varphi$ is called a supersmooth mapping from ${\mathfrak{U}}$ to ${\mathfrak{V}}$
if each $\varphi_A(X)\in {\CSS}({\mathfrak{U}}:{\fR})$ for $A=1,{\cdots}, p+q$ and
$\varphi ({\mathfrak{U}})\subset {\mathfrak{V}}$.
\end{definition}

\begin{proposition}[Composition of supersmooth mappings]\label{prop-geofcomposition}%\label{3.15b}
Let ${\mathfrak{U}}\subset{\fR}^{m|n}$ and ${\mathfrak{V}}\subset {\fR}^{p|q}$
be superdomains and 
let $F: {\mathfrak{U}}\to  {\fR}^{p|q}$ and $G:{\mathfrak{V}}\to{\fR}^{r|s}$
be supersmooth mappings such that $F(\mathfrak{U})\subset {\mathfrak{V}}$.
Then, the composition $G\circ F:{\mathfrak{U}}\to{\fR}^{r|s}$ 
gives a supersmooth mapping and
\begin{equation}
d_X G(F(X))
=[d_X F(X)][d_Y G(Y)]\big|_{Y=F(X)}.
\label{3-23}
\end{equation}
Or more precisely, 
\begin{equation}
d_X G(F(X))=(\partial_{X_A}(G\circ F)_B(X))
=(\sum_{C=1}^{r+s}\partial_{X_A}G_C(X))(\partial_{Y_C}F_B(Y))\big|_{Y=F(X)}).
\label{3-23-1}
\end{equation}
\end{proposition}

\par{\it Proof}. 
Put $F(Y)=(G_B(Y))_{B=1}^{r+s}$, $F(X)=(F_C)_{C=1}^{p+q}$, $X=(X_A)_{A=1}^{m+n}$, and $Y=(Y_C)_{C=1}^{r+s}$.
By smooth G-differentiability of the composition of mappings between Fr\'echet spaces, we have the smoothness of
$\Phi(X+tH)$ w.r.t $t$.
Moreover, we have \eqref{3-23-1}.

By the characterization of supersmoothness, we need to say ${\rev}$-linearity of $d_F\Phi$, 
i.e. $d_F\Phi(X)(\lambda H)=\lambda d_F\Phi(X)(H)$ for $\lambda\in{\rev}$ which is obvious from
$$
\begin{aligned}
d(G\circ F)(X)(\lambda H)&=\dt G(F(X+t\lambda H))\bigg|_{t=0}\\
&=(\sum_{C=1}^{r+s}\lambda H_A\partial_{X_A}G_C(X))(\partial_{Y_C}F_B(Y))\big|_{Y=F(X)})\\
&=\lambda \dt G(F(X+t H))\bigg|_{t=0}=\lambda d(G\circ F)(X)(H). \qquad\qed
\end{aligned}
$$

\begin{definition}\label{3.16b}
Let ${\mathfrak{U}}\subset\supermn$ and
${\mathfrak{V}}\subset{\fR}^{p|q}$ be superdomains and
let $\varphi:{\mathfrak{U}}\to {\mathfrak{V}}$ be a supersmooth mapping represented by
$\varphi(X)=(\varphi_1(X),{\cdots},\varphi_{p+q}(X))$ with
$\varphi_A(X)\in {\CSS}({\mathfrak{U}}:{\fR})$.
\newline
(1) $\varphi$ is called a supersmooth diffeomorphism if 
\par(i) $\varphi$ is a homeomorphism between
${\mathfrak{U}}$ and ${\mathfrak{V}}$ and
\par(ii) $\varphi$ and $\varphi^{-1}$ are supersmooth mappings. 
\newline
(2) For any $f\in {\CSS}({\mathfrak{V}}:{\fR})$, 
$(\varphi^* f)(X)=(f\circ \varphi)(X)=f(\varphi(X))$,
called the {\it pull back} of $f$,
is well-defined and belongs to ${\CSS}({\mathfrak{U}}:{\fR})$.
\end{definition}

\begin{remark}
 It is easy to see that if $\varphi$ is a supersmooth diffeomorphism, then
$\varphi_{\mathrm{B}}=\pi_{\mathrm{B}} \circ \varphi$ is 
an (ordinary) $C^\infty$ diffeomorphism from ${\mathfrak{U}}_{\mathrm{B}}$ to ${\mathfrak{V}}_{\mathrm{B}}$.
\end{remark}
\begin{remark}
If we introduce the topologies in
${\CSS}({\mathfrak{V}}:{\fC})$ and ${\CSS}({\mathfrak{U}}:{\fC})$ properly,
$\varphi^*$ gives a continuous linear mapping from 
${\CSS}({\mathfrak{V}}:{\fC})$ to ${\CSS}({\mathfrak{U}}:{\fC})$.
Moreover, if $\varphi:{\mathfrak{U}}\to {\mathfrak{V}}$ is a supersmooth diffeomorphism,
then $\varphi^*$ defines an automorphism from
${\CSS}({\mathfrak{V}}:{\fR})$ to ${\CSS}({\mathfrak{U}}:{\fR})$.
\end{remark}

\subsection{Inverse and implicit function theorems}
We recall
\begin{proposition}[Inverse function theorem on ${\euc}^m$]
Let $U$ be an open set in ${\euc}^m$.
Let $f: U\ni x\to y=f(x)\in{\euc}^m$ be a $C^k$ $(k\ge1)$ mapping such that
$f'(x_0)\neq{0}$ for some $x_0\in U$.
\par
Then, there exist a neighbourhood $W$ of $y_0=f(x_0)$ and a neighbourhood $U_0\subset U$ of $x_0$, such that
$f$ maps $U_0$ injectively onto $W$. Therefore, $f|_{U_0}$ has its inverse $\phi=(f|_{U_0})^{-1}:W\to U_0\in C^k(W)$.
Moreover, for any $y=f(x)\in W$ with $x\in U_0$, we have
$\phi'(y)=f'(x)^{-1}$.
\end{proposition}

Applying this, we have
\begin{theorem}[Inverse function theorem on $ {\supermn}$]\label{3.17b} 
Let $F=(f,g): {\supermn}\ni X \rightarrow \,Y=F(X)\in {\supermn}$ 
be a supersmooth mapping on some superdomain containing $\tilde{X}$.
That is,
$$
f(X)=(f_i(X))_{i=1}^m\in{\fR}_{\mathrm{ev}}^m,\;\;
g(X)=(g_k(X))_{k=1}^n\in{\fR}_{\mathrm{od}}^n,
$$
More precisely, we put
$$
f(X)=(f_i(X))_{i=1}^m\in {\fR}_{\mathrm{ev}}^m,\;\;
g(X)=(g_k(X))_{k=1}^n\in {\fR}_{\mathrm{od}}^n,
$$
such that
\begin{equation}
\begin{gathered}
f_i(x,\theta)=\sum_{|a|={\mathrm{ev}}\le n}\theta^a f_{ia}(x),\;\;
g_k(x,\theta)=\sum_{|b|={\mathrm{od}}\le n} \theta^b g_{kb}(x) \in \ccsl_{S\!S}({\supermn})\\
\with
\begin{cases}
f_{ia}(x_{\mathrm{B}}), \;\; g_{kb}(x_{\mathrm{B}})\in {\mathbb{C}}&\mbox{if $|a|, |b|\neq{0}$},\\
f_{ia}(x_{\mathrm{B}})\in{\euc}&\mbox{if $|a|=0$}.
\end{cases}
\end{gathered}
\label{ccsl-assume}
\end{equation}
\par
We assume the super matrix $[d_X F(X)]$
is invertible at ${\tilde X}$, i.e. $\pi_{\mathrm{B}}(\sdet [d_X \Phi(X)]|_{X={\tilde X}})\neq{0}$. 
Then, there exist a superdomain ${\mathfrak{U}}$, a neighbourhood of ${\tilde X}$ and another  superdomain ${\mathfrak{V}}$, a neighbourhood of 
${\tilde Y} = F({\tilde X}) $ such that $F: {\mathfrak{U}}\to {\mathfrak{V}}$ has a unique supersmooth inverse $\Phi=F^{-1}=(\phi,\psi):{\mathfrak{V}}\to {\mathfrak{U}}$
satisfying 
$$
\phi(Y)=(\phi_{i'}(Y))_{i'=1}^m\in{\fR}_{\mathrm{ev}}^m,\;\;
\psi(Y)=(\psi_{k'}(Y))_{k'=1}^n\in{\fR}_{\mathrm{od}}^n
$$
and
\begin{equation} 
\Phi(F(X)) = X \for X\in {\mathfrak{U}}\et
F(\Phi(Y))=Y\for Y\in {\mathfrak{V}}.
\label{02-13-1}
\end{equation}
Moreover, we have
\begin{equation}
\left.{d_Y}\Phi(Y)= ({d_X}F(X))^{-1} \,\right|_{X=\Phi(Y)} 
\quad \text{in}\quad {\mathfrak{V}}.
\label{3-31}
\end{equation}
\end{theorem}

%%%

\begin{remark}
A question is posed on the meaning of ``the supermatrix $f|_{U_0}$ has inverse''.
If $G(X): {\mathfrak{U}} \subset {\supermn}\rightarrow \, {\supermn}$ is a super mapping represented by
$$
G(X)=(g_1(x,\theta),{\cdots},g_m(x,\theta),g_{m+1}(x,\theta),{\cdots},g_{m+n}(x,\theta))\in {\supermn},
$$
from the definition of $dg_j$, we get
$$
d_XG(X)=\begin{pmatrix}
A&C\\
D&B
\end{pmatrix}.
$$
Here,
$$
\begin{gathered}
A=\begin{pmatrix}
\frac{\partial g_1}{\partial x_1}&\frac{\partial g_2}{\partial x_1}&{\cdots}&\frac{\partial g_m}{\partial x_1}\\
\frac{\partial g_1}{\partial x_2}&\frac{\partial g_2}{\partial x_2}&{\cdots}&\frac{\partial g_m}{\partial x_2}\\
{\vdots}&{\vdots}&{\cdots}&{\vdots}\\
\frac{\partial g_1}{\partial x_m}&\frac{\partial g_2}{\partial x_m}&{\cdots}&\frac{\partial g_m}{\partial x_m}\\
\end{pmatrix},\quad
C=\begin{pmatrix}
\frac{\partial g_{m+1}}{\partial x_1}&\frac{\partial g_{m+2}}{\partial x_1}&{\cdots}&\frac{\partial g_{m+n}}{\partial x_1}\\
\frac{\partial g_{m+1}}{\partial x_2}&\frac{\partial g_{m+2}}{\partial x_2}&{\cdots}&\frac{\partial g_{m+n}}{\partial x_2}\\
{\vdots}&{\vdots}&{\cdots}&{\vdots}\\
\frac{\partial g_{m+1}}{\partial x_m}&\frac{\partial g_{m+2}}{\partial x_m}&{\cdots}&\frac{\partial g_{m+n}}{\partial x_m}\\
\end{pmatrix},\\
D=\begin{pmatrix}
\frac{\partial g_1}{\partial \theta_1}&\frac{\partial g_2}{\partial \theta_1}&{\cdots}&\frac{\partial g_m}{\partial \theta_1}\\
\frac{\partial g_1}{\partial \theta_2}&\frac{\partial g_2}{\partial \theta_2}&{\cdots}&\frac{\partial g_m}{\partial \theta_2}\\
{\vdots}&{\vdots}&{\cdots}&{\vdots}\\
\frac{\partial g_1}{\partial \theta_n}&\frac{\partial g_2}{\partial \theta_n}&{\cdots}&\frac{\partial g_m}{\partial \theta_n}\\
\end{pmatrix},\quad
B=\begin{pmatrix}
\frac{\partial g_{m+1}}{\partial \theta_1}&\frac{\partial g_{m+2}}{\partial \theta_1}&{\cdots}&\frac{\partial g_{m+n}}{\partial \theta_1}\\
\frac{\partial g_{m+1}}{\partial \theta_2}&\frac{\partial g_{m+2}}{\partial \theta_2}&{\cdots}&\frac{\partial g_{m+n}}{\partial \theta_2}\\
{\vdots}&{\vdots}&{\cdots}&{\vdots}\\
\frac{\partial g_{m+1}}{\partial \theta_n}&\frac{\partial g_{m+2}}{\partial \theta_n}&{\cdots}&\frac{\partial g_{m+n}}{\partial \theta_n}\\
\end{pmatrix}
\end{gathered}
$$
Therefore, $d_XG(X)$ gives an even super matrix.
\end{remark}

\par{\it Proof of Theorem ~\ref{3.17b} }.
(I) To make clear the point, we consider the case $m=1$, $n=2$, that is, ${\mathfrak{U}}, {\mathfrak{V}}\subset{\fR}^{1|2}$.
Let 
$$
F(X)=F(x,\theta)=({f}(x,\theta), {g}_1(x,\theta),{g}_2(x,\theta)):{\mathfrak{U}}\to {\mathfrak{V}}
$$ 
with
\begin{equation}
\left\{
\begin{aligned}
&{f}(x,\theta)=f_{(0)}(x)+f_{(12)}(x)\theta_1\theta_2, \\
&{g_1}(x,\theta)=g_{1(1)}(x)\theta_1+ g_{1(2)}(x)\theta_2,\\
&{g_2}(x,\theta)=g_{2(1)}(x)\theta_1+ g_{2(2)}(x)\theta_2,
\end{aligned}
\right.\et
f_{(0)}(x_{\mathrm{B}})\in\euc,\; f_{(12)}(x_{\mathrm{B}}),\, g_{k(l)}(x_{\mathrm{B}})\in{\mathbb{C}}.
\label{02-20-1}
\end{equation}
In this case, we have
$$
d_XF(X)=\begin{pmatrix}
f'_{(0)}(x)+f'_{(12)}(x)\theta_1\theta_2&g_{1(1)}'(x)\theta_1+ g_{1(2)}'(x)\theta_2&g_{2(1)}'(x)\theta_1+ g_{2(2)}'(x)\theta_2\\
f_{(12)}(x)\theta_2&g_{1(1)}(x)&g_{2(1)}(x)\\
-f_{(12)}(x)\theta_1&g_{1(2)}(x)&g_{2(2)}(x)
\end{pmatrix}
=\begin{pmatrix}
A&C\\
D&B
\end{pmatrix}
$$
with
$$
\sdet(d_XF(X))=\det[A-CB^{-1}D](\det B)^{-1}
\et
\det B=\beta(x)=g_{1(1)}(x)g_{2(2)}(x)-g_{1(2)}(x)g_{2(1)}(x).
$$
Therefore, 
\begin{equation}
\pi_{\mathrm{B}}(\sdet(d_XF(X)))=f'_{(0)}(x_{\mathrm{B}})\beta(x_{\mathrm{B}})^{-1}.
\label{12-sdet}
\end{equation}
We need to find $\Phi=(\phi,\psi_1,\psi_2)$ such that
\begin{equation}
{\left\{
\begin{aligned}
&\phi(f(x_{\mathrm{B}},\theta),g_{1}(x_{\mathrm{B}},\theta),g_{2}(x_{\mathrm{B}},\theta))=x_{\mathrm{B}},\\
&\psi_{1}(f(x_{\mathrm{B}},\theta),g_{1}(x_{\mathrm{B}},\theta),g_{2}(x_{\mathrm{B}},\theta))=\theta_1,\\
&\psi_{2}(f(x_{\mathrm{B}},\theta),g_{1}(x_{\mathrm{B}},\theta),g_{2}(x_{\mathrm{B}},\theta))=\theta_2
\end{aligned}
\right.}
\with
{\left\{
\begin{aligned}
&\phi(y,\omega)=\phi_{(0)}(y)+\phi_{(12)}(y)\omega_1\omega_2,\\
&\psi_1(y,\omega)=\psi_{1(1)}(y)\omega_1+\psi_{1(2)}(y)\omega_2,\\
&\psi_2(y,\omega)=\psi_{2(1)}(y)\omega_1+\psi_{2(2)}(y)\omega_2.
\end{aligned}
\right.}
\label{12-s-diff11}
\end{equation}
To state more precisely, we have
$$
\left\{
\begin{aligned}
&y_{\mathrm{B}}=f_{(0)}(x_{\mathrm{B}}),\quad
y_{\mathrm{S}}=f_{(12)}(x_{\mathrm{B}})\theta_1\theta_2,\\
&\phi_{(0)}(y_{\mathrm{B}}+y_{\mathrm{S}})=\phi_{(0)}(y_{\mathrm{B}})+\phi'_{(0)}(y_{\mathrm{B}})y_{\mathrm{S}},\\
&\phi_{(12)}(y_{\mathrm{B}}+y_{\mathrm{S}})=\phi_{(12)}(y_{\mathrm{B}})+\phi'_{(12)}(y_{\mathrm{B}})y_{\mathrm{S}},
\end{aligned}
\right.
$$
and from the first equation of \eqref{12-s-diff11}, 
\begin{equation}
\left\{
\begin{aligned}
&\phi_{(0)}(y_{\mathrm{B}})=\phi_{(0)}(f_{(0)}(x_{\mathrm{B}}))=x_{\mathrm{B}},\\
&\phi'_{(0)}(y_{\mathrm{B}})y_{\mathrm{S}}+\phi_{(12)}(y_{\mathrm{B}})\beta(x_{\mathrm{B}})\theta_1\theta_2=[\phi'_{(0)}(y_{\mathrm{B}})f_{(12)}(x_{\mathrm{B}})+\phi_{(12)}(y_{\mathrm{B}})\beta(x_{\mathrm{B}})]\theta_1\theta_2=0.
\end{aligned}
\right.
\label{03-14}
\end{equation}

Since $f'_{(0)}(\tilde{x}_{\mathrm{B}})\neq{0}$, there exists a neighborhood $U_0\subset\euc$ of $\tilde{x}_{\mathrm{B}}$ and $V_0\subset\euc$ of $f_{(0)}(\tilde{x}_{\mathrm{B}})$ where
we find a function $\phi_{(0)}(y_{\mathrm{B}})$ satisfying the first equation \eqref{03-14}.
Moreover, since $\beta(x_{\mathrm{B}})\neq{0}$, taking the smaller neighborhood if necessary,  we define
$$
\phi_{(12)}(y_{\mathrm{B}})=-\phi'_{(0)}(y_{\mathrm{B}})f_{(12)}(x_{\mathrm{B}})\beta(x_{\mathrm{B}})^{-1}\big|_{x_{\mathrm{B}}=\phi_{(0)}(y_{\mathrm{B}})}.
$$

On the other hand, putting
$$
\begin{aligned}
&\omega_1=g_{1(1)}(x_{\mathrm{B}})\theta_1+g_{1(2)}(x_{\mathrm{B}})\theta_2,\\
&\omega_2=g_{2(1)}(x_{\mathrm{B}})\theta_1+g_{2(2)}(x_{\mathrm{B}})\theta_2,
&\omega_1\omega_2=\beta(x_{\mathrm{B}})\theta_1\theta_2
\end{aligned}
$$
and remarking  $y_{\mathrm{S}}\omega_j=0$ for $j=1,2$, from the last two equations of \eqref{12-s-diff11}, we should have
$$
\begin{aligned}
&\psi_{1(1)}(y_{\mathrm{B}})
(g_{1(1)}(x_{\mathrm{B}})\theta_1+g_{1(2)}(x_{\mathrm{B}})\theta_2)
+\psi_{1(2)}(y_{\mathrm{B}})
(g_{2(1)}(x_{\mathrm{B}})\theta_1+g_{2(2)}(x_{\mathrm{B}})\theta_2)=\theta_1,\\
&\psi_{2(1)}(y_{\mathrm{B}})
(g_{1(1)}(x_{\mathrm{B}})\theta_1+g_{1(2)}(x_{\mathrm{B}})\theta_2)
+\psi_{2(2)}(y_{\mathrm{B}})
(g_{2(1)}(x_{\mathrm{B}})\theta_1+g_{2(2)}(x_{\mathrm{B}})\theta_2)=\theta_2,
\end{aligned}
$$
that is,
$$
\begin{pmatrix}
\psi_{1(1)}(y_{\mathrm{B}})&\psi_{1(2)}(y_{\mathrm{B}})\\
\psi_{2(1)}(y_{\mathrm{B}})&\psi_{2(2)}(y_{\mathrm{B}})
\end{pmatrix}
\begin{pmatrix}
g_{1(1)}(x_{\mathrm{B}})&g_{1(2)}(x_{\mathrm{B}})\\
g_{2(1)}(x_{\mathrm{B}})&g_{2(2)}(x_{\mathrm{B}})
\end{pmatrix}
=\begin{pmatrix}
1&0\\
0&1
\end{pmatrix}.
$$
Therefore, we have $\psi_*(y_{\mathrm{B}})$, which satisfy the desired property.\\
(II) Do analogously as above for general $m,n$ by putting
$$
f_i(x,\theta)=\sum_{|a|={\mathrm{ev}}\le n}f_{i,a}(x)\theta^a,\;\;
g_k(x,\theta)=\sum_{|b|={\mathrm{od}}\le n}g_{k,b}(x)\theta^b,
$$
and
$$
\phi_{i'}(y,\omega)=\sum_{|a'|={\mathrm{ev}}\le n}\phi_{i',a'}(y)\omega^{a'},\;\;
\psi_{k'}(y,\omega)=\sum_{|b'|={\mathrm{od}}\le n}\psi_{k',b'}(y)\omega^{b'},
$$
but with more patience.  $\qquad\qed$

\begin{remark}
Above theorem holds for functions $f_i\in {\mathcal{C}}_{S\!S}({\fR}^{m|n}:{\rev})$ and
$g_k\in {\mathcal{C}}_{S\!S}({\fR}^{m|n}:{\rod})$.
\end{remark}

Moreover, we have
\begin{proposition}[Implicit function theorem]\label{3.18b}
Let $\Phi (X,Y): {\mathfrak{U}} \times {\mathfrak{V}} \to {\fC}^{p|q}$ 
be a supersmooth mapping and $( {\tilde X }, {\tilde Y}) \in {\mathfrak{U}} \times {\mathfrak{V}}$,
where ${\mathfrak{U}}$ and ${\mathfrak{V}}$ are superdomains of ${\supermn}$ and ${\fR}^{p|q}$,
respectively.
Suppose $\Phi ( {\tilde X }, {\tilde Y} )=0$ and
$ {\partial}_Y \Phi=[\partial_{y_j} \Phi, \partial_{\omega_r} \Phi ]$
is a continuous and invertible supermatrix 
at $({\tilde X}_{\mathrm{B}},{\tilde Y}_{\mathrm{B}})
\in \pi_{\mathrm{B}}({\mathfrak{U}}) \times \pi_{\mathrm{B}}({\mathfrak{V}})$.
Then, there exist a superdomain ${\mathfrak{V}} \subset {\mathfrak{U}} $ 
satisfying ${\tilde X}_{\mathrm{B}}\in \pi_{\mathrm{B}}({\mathfrak{V}})$ 
and a unique supersmooth mapping $Y=f(X)$ on ${\mathfrak{V}}$
such that $ {\tilde Y} = f( {\tilde X} )$ and $ \Phi (X,f(X))=0 $ in ${\mathfrak{V}}$.
Moreover, we have
\begin{equation}
{\partial_X} f(X) 
= -\left.[\partial_X \Phi(X,Y)]{[ \partial_Y \Phi(X,Y)]^{-1}}\right|_{Y=f(X)}.
\label{3-46}
\end{equation}
\end{proposition}
\par
{\it Proof}.
\eqref{3-46} %(2.46) 
is easily obtained by
$$
0=\partial_X \Phi(X,f(X))=\left.
(\partial_X \Phi(X,Y)+\partial_Xf(X)\partial_Y\Phi(X,Y))
\right|_{Y=f(X)}.
$$
The existence proof is omitted here because the arguments in proving 
Proposition \ref{3.17b}  %2.16 
work well in this situation.	$ \qquad\qed$

\begin{problem}
Use Ekeland's idea ~\cite{eke11} to give another proof of above Theorems, if possible.
\end{problem}

\subsection{Global inverse function theorem}
We have the following theorem of Hadamard type:
\begin{proposition}[Global inverse function theorem on ${\euc}^m$]\label{3.17b-1} 
Let $f: {\euc}^m\ni x \rightarrow \,y=f(x)\in {\euc}^m$ 
be a smooth mapping on ${\euc}^m$.
We assume the Jacobian matrix $[d_x f(x)]$
is invertible on ${\euc}^m$, 
and $\Vert (\det [d_x f(x)])\Vert \ge \delta>0$ for any $x$.
Then, $f$ gives a smooth diffeomorphism from ${\euc}^m$ onto ${\euc}^m$.
\end{proposition}

\begin{proposition}[Global inverse function theorem on $\supermn$]\label{3.17b-2} 
Let $F=(f,g): {\supermn}\ni X \rightarrow \,Y=F(X)\in {\supermn}$ 
be a supersmooth mapping on ${\supermn}$.
We assume the super matrix 
$$
d_X F(X)=\begin{pmatrix}
\frac{\partial f_i}{\partial x_j}&\frac{\partial g_k}{\partial x_j}\\
\frac{\partial f_i}{\partial \theta_l} &\frac{\partial g_k}{\partial \theta_l} 
\end{pmatrix}
$$
is invertible at any ${X}\in \supermn$, 
and there exists $\delta>0$ such that  for any $x$
$$
\Vert \pi_{\mathrm{B}}(\sdet \frac{\partial f_i}{\partial x_j})\Vert \ge \delta>0,
\et
\{x\;|\; \pi_{\mathrm{B}}\det(\frac{\partial g_k}{\partial \theta_l} )=0\}=\emptyset.
$$
Then, $F$ gives a supersmooth diffeomorphism from ${\supermn}$ onto ${\supermn}$.
\end{proposition}
\par
{\it Proof}. From the proof  of above Theorem \ref{3.17b}, it is obvious.  $\qquad\qed$

\chapter{Elementary integral calculus on superspace}
As is well-known, to study a scalar PDE by applying functional analysis, 
we use essentially the following tools: Taylor expansion, 
integration by parts, the formula for the change of variables under integral sign and Fourier transformation.
Therefore, beside the elementary differential calculus, it is necessary to develop the elementary integral calculus on superspace ${\fR}^{m|n}$.
But as is explained soon later, we have the relations
\begin{equation}
\begin{cases}
dx_j\wedge dx_k=-dx_k\wedge dx_j&\mbox{for even variables $\{x_j\}_{j=1}^m$},\\
d\theta_j\wedge d\theta_k=d\theta_k\wedge d\theta_j &\mbox{for odd variables $\{\theta_k\}_{k=1}^n$, which differs from ordinary one}.
\end{cases}
\label{di}
\end{equation}
Therefore, the integration containing odd variables doesn't follow our conventional intuition.
%%%
\section{Integration w.r.t. odd variables -- Berezin integral}
It seems natural to put formally
$$
d\theta_j=\sum_{{\mathbf{I}}\in{\mathbf{I}},|{\mathbf{I}}|={\mathrm{od}}}d\theta_{j,{\mathbf{I}}}\,\sigma^{\mathbf{I}}
\for \theta_j=\sum_{{\mathbf{I}}\in{\mathbf{I}},|{\mathbf{I}}|={\mathrm{od}}}\theta_{j,{\mathbf{I}}}\,\sigma^{\mathbf{I}}.
$$
\begin{remark}
Since above sum $\sum_{\mathbf{I}}$ stands for the position in the sequence space $\omega$ of K\"othe and the element of it is given by $d\theta_{j,{\mathbf{I}}}$ for $|{\mathbf{I}}|$ is finite, we may give the meaning to $d\theta_j$.
\end{remark}

Then, rather formally, since for ${\mathbf{J}}$ and ${\mathbf{K}}$ with $|{\mathbf{J}}|={\mathrm{od}}=|{\mathbf{K}}|$,
$d\theta_{j,{\mathbf{J}}}$, $d\theta_{k,{\mathbf{K}}}$ and $\sigma^{\mathbf{J}}$, $\sigma^{\mathbf{K}}$ are anticommutative, minus signs cancel out each other and the second equality in \eqref{di} holds. Analogously the first one in \eqref{di} holds. 
This make us imagine that even if there exists the notion of integration,
it differs much from the standard one on $\euc^m$.

Here, we borrow explanation of Vladimirov and Volovich.
Since the supersmooth functions on ${\fR}^{0|n}$ are characterized as the polynomials with value in ${\fC}$, 
we need to define the integrability for those under the conditions that

\par
(i) integrability of all polynomials,  
\par 
(ii) linearity of an integral, and  
\par
(iii) invariance of the integral w.r.t. shifts.\\
Put ${\mathcal{P}}_n={\mathcal{P}}_n({\fC})=\{u(\theta)=\sum_{a\in\{0,1\}^n}\theta^au_a\;|\; u_a\in{\fC}\}$.

We say a mapping $J_n:{\mathcal{P}}_n\to{\fC}$ is an integral if it satisfies
%\par
%(1) ${\fR}$-linearity (from the right): $I(u\alpha+v\beta)=I(u)\alpha +I(v)\beta $ for $\alpha, \beta\in{\fR}$, $u,v\in{\mathcal{P}}_n$.
\par
(1) ${\fC}$-linearity (from the right): $J_n(u\alpha +v\beta )=J_n(u)\alpha+J_n(v)\beta$ for $\alpha, \beta\in{\fC}$, $u,v\in{\mathcal{P}}_n$.
\par
(2) translational invariance: $J_n(u({\cdot}+\omega))=J_n(u)$  for all $\omega\in {\fR}^{0|n}$ and $u\in{\mathcal{P}}_n$.

\begin{theorem}
For the existence of the integral $J_n$ satisfying above conditions (1) and (2), it is necessary and sufficient that
\begin{equation}
J_n(\phi_a)=0 \for \phi_a(\theta)=\theta^a,\;|a|\le n-1.
\label{VVII3.4}
\end{equation}
Moreover, we have
$$
J_n(u)=\frac{\partial}{\partial \theta_n}{\cdots}\frac{\partial}{\partial \theta_1}u(\theta)\bigg|_{\theta=0}J_n(\phi_{\tilde{1}})
\where \phi_{\tilde{1}}(\theta)=\theta^{\tilde{1}}=\theta_1{\cdots}\theta_n.
$$
\end{theorem}
\par
{\it Proof}.
If there exists $J_n$ satisfying (1) and (2), then we have
$$
J_n(v)=\sum_{|a|\le n}I(\phi_a) v_a\for v(\theta)=\sum_{|a|\le n}\theta^a v_a=\sum_{|a|\le n}\phi_a(\theta) v_a.
$$
As
$$
\begin{aligned}
&(\theta+\omega)^a=\theta^a+\sum_{|a-b|\ge 1,b\le a}(-1)^*\theta^b\omega^{a-b},\\
&J_n(v({\cdot}+\omega))=\sum_{|a|\le n}J_n(\phi_a({\cdot}+\omega))v_a=\sum_{|a|\le n}J_n(\phi_a)v_a
+\sum_{|a|\le n}\sum_{|a-b|\ge 1,b\le a}(-1)^*J_n(\phi_b)v_b\omega^{a-b},
\end{aligned}
$$
by virtue of (2), we have
$$
\sum_{|a|\le n}\sum_{|a-b|\ge 1,b\le a}(-1)^*J_n(\phi_b)v_b\omega^{a-b}=0.
$$
Here, $v_a\in{\fC}$ and $\omega\in\rod^n$ are arbitrary, we have \eqref{VVII3.4}.
Converse is obvious.  $\qquad\qed$

\begin{definition}\label{int5.5}
We put $J_n(\phi_{\tilde{1}})=1$, i.e.,
\begin{equation}
\int_{\superon}  d \theta_n \cdots  d \theta_1 
\,\theta_1 \cdots \theta_n =1.
\label{int5-8}
\end{equation}
Therefore,  we put, for any $ v=\sum_{|a|{\le} n}\theta^a v_a \in {\mathcal{P}}_n (\fC) $
\begin{equation}
\begin{aligned}
J_n(v)=\int_{\superon} d \theta \, v( \theta ) 
&= \int_{{\superon}}  
d \theta_n \cdots d \theta_1 \, v(\theta_1,\cdots ,\theta_n ) 
= (\partial_{\theta_n} \cdots 
\partial_{\theta_1} v)(0)\\
&=v_{\tilde{1}}=\int_{\mathrm{Berezin}} d^n\theta v(\theta).
\end{aligned}
\label{int5-7}
\end{equation}
This is called the (Berezin) integral of $v$ on  $ \superon$.
\end{definition}

Then, we have
\begin{proposition}\label{int5.6}
Given
$ v, \, w \in {\mathcal{P}}_n ({\fC}) $ , we have the following:
\par
(1) (${\fC}$-linearity) 
For any homogeneous $ \lambda ,\, \mu \in {\fC}$,
\begin{equation}
\int_{\superon}  d \theta( \lambda v + \mu w)( \theta ) =
(-1)^{np(\lambda)}\lambda \int_{\superon} 
 d\theta \,v( \theta ) +
(-1)^{np(\mu)}\mu \int_{\superon} d \theta \,w( \theta ) .
\label{int5-9}
\end{equation}
\par
(2) (Translational invariance) For any $\rho\in\superon$, we have
\begin{equation}
\int_{\superon}  d \theta \,v( \theta + \rho )=
\int_{\superon}  d \theta \,v( \theta ).
\label{int5-10}
\end{equation}
\par
(3) (Integration by parts)
For $ v \in {\mathcal{P}}_n ( {\fC})$ such that $p(v) = 1$ or 0, we have
\begin{equation}
\int_{\superon} d\theta 
\,v( \theta ) \partial_{\theta_s} w( \theta ) =
-(-1)^{p(v)}
\int_{\superon} d\theta \,(\partial_{\theta_s} v( \theta )) w( \theta ).
\label{int5-11}
\end{equation}
\par
(4) (Linear change of variables)
Let $A=(A_{jk})$ with $A_{jk}\in \rev$ be an invertible matrix.
Then, 
\begin{equation}
\int_{\superon} d \theta \,v( \theta )
= (\det A )^{-1} \int_{\superon} d\omega \,v(A\cdot\omega ).
\label{int5-12}
\end{equation}
\par
(5) (Iteration of integrals)
\begin{equation}
\int_{\superon}  d \theta \,v( \theta )
= \int_{{\mathfrak R}{}^{0|n-k}} \!\!\!\!d \theta_n  \cdots  d \theta_{k+1}
 \left( \int_{{\mathfrak R}{}^{0|k}} \!\!\!\!d \theta_{k}  \cdots d \theta_1
\,v(\theta_1, \cdots , \theta_k , \theta_{k+1} , \cdots , \theta_n ) \right).
\label{int5-13}
\end{equation} 
\par
(6) (Odd change of variables)
Let $\theta=\theta(\omega)$ be an odd change of variables such that
$\theta(0)=0$ and
$\det\left.
\displaystyle{{\frac{\partial\theta(\omega)}{\partial \omega}}}
\right|_{\omega=0}\ne 0$.
Then, for any  $ v \in {\mathcal{P}}_n ({\fC})$,
\begin{equation}
\int_{\superon}  d \theta \, v(\theta)=
\int_{\superon}  d \omega \, v(\theta(\omega))
\,{\det}^{-1}{\frac{\partial \theta(\omega)}{\partial \omega}}.
\label{int5-14}
\end{equation} 
\par
(7) For $ v \in {\mathcal{P}}_n ({\fC})$ and $ \omega \in \superon $,
\begin{equation}
\int_{\superon}  d \theta 
\,( \omega_1 - \theta_1 ){\cdots}( \omega_n -  \theta_n ) v( \theta ) 
= v( \omega ) .
\label{int5-15}
\end{equation} 
\end{proposition}
\par
{\it Proof}. We follow the arguments
in pp.755-757 of V.S. Vladimirov and I. V. Volovivh~\cite{VV84} with slight modifications if necessary.\\
By definition, we have (1).\\
(2) Remarking the top term of $v(\theta+\rho)$ containing the term $\theta_1{\cdots}\theta_n$ is same as $v(\theta)$, combining \eqref{VVII3.4} we get the result.\\
(3) Using $\partial_{\theta_j}(vw)=\partial_{\theta_j}v{\cdot}w+(-1)^{p(v)}v{\cdot}\partial_{\theta_j}w$ with (2), we get (3).\\
(4) As $(A\theta)_j=\sum_{k=1}^n a_{jk}\theta_k$ with $a_{jk}\in\rev$,
$$
\begin{aligned}
J_n(v(A\theta))&=\partial_{\theta_n}{\cdots}\partial_{\theta_1}v(A\theta)\big|_{\theta=0}
=\sum_{\sigma\in\wp_n}\partial_{\theta_{\sigma(n)}}{\cdots}\partial_{\theta_{\sigma(1)}}v(0) a_{\sigma(1) 1}{\cdots}a_{\sigma(n) n}\\
&=\partial_{\theta_n}{\cdots}\partial_{\theta_1}v(0)\sum_{\sigma\in\wp_n}\sgn(\sigma)a_{\sigma(1) 1}{\cdots}a_{\sigma(n) n}
=J_n(v)\det A.
\end{aligned}
$$
(5) Obvious.\\
(6) We prove by induction w.r.t. $n$.
When $n=1$, for $\theta_1=a\omega_1$ where $a\in\rev$ with $\pi_{\mathrm{B}}a\neq{0}$, it is clear that \eqref{int5-14} holds. In fact,
$$
J_1(v)=a^{-1}J(v(a\omega_1))=J_1(\big(\frac{\partial \theta_1}{\partial \omega_1}\big)^{-1}v(a\omega_1)).
$$
Assuming \eqref{int5-14} holds for $J_{n-1}$, we prove it for $J_n$.
Let $\theta_j=\theta_j(\omega_1,{\cdots},\omega_n)$ for $j=1,{\cdots},n$.
Without loss of generality, we may assume that $\displaystyle{\frac{\partial \theta_1(\omega)}{\partial \omega_1}\bigg|_{\omega=0}}$ is invertible. By the property (5), we have
$$
J_n(v)=J_1(J_{n-1}(v)).
$$
Putting $\tilde{\omega}=(\omega_2,{\cdots},\omega_n)$, 
we solve the equation $\theta_1=\theta_1(\omega_1,\tilde{\omega})$ w.r.t. $\omega_1$, having
\begin{equation}
\omega_1=\bar{\omega}_1(\theta_1,\tilde{\omega})\with
\theta_1=\theta_1(\bar{\omega}_1(\theta_1,\tilde{\omega}),\tilde{\omega}),\;
\omega_1=\bar{\omega}_1(\theta_1(\omega_1,\tilde{\omega}),\tilde{\omega}).
\label{VVII3.28}
\end{equation}
We put this relation into $\theta_j=\theta_j(\omega)$ to have
$$
\theta_j=\theta_j(\bar{\omega}_1(\theta_1,\tilde{\omega}),\tilde{\omega})=\theta'_j(\theta_1,\tilde{\omega})\for j=2,{\cdots},n.
$$
By (5) and the induction hypothesis, we have
$$
J_{n-1}(v)=J_{n-1}(\det{}^{-1}\frac{\partial\theta'}{\partial\tilde{\omega}}v(\theta_1,\theta'(\theta_1,\tilde{\omega}))).
$$ 
Changing the order of integration w.r.t. $\theta_1$ and $\tilde{\omega}$, we get
$$
J_n(v)=J_{n-1}(J_1(\det{}^{-1}\frac{\partial\theta'}{\partial\tilde{\omega}}v(\theta_1,\theta'(\theta_1,\tilde{\omega})))).
$$
Using $\theta_1=\theta_1(\omega_1,\tilde{\omega})$, we have
$$
\begin{aligned}
J_n(v)
&=J_{n-1}(J_1(
\det{}^{-1}\frac{\partial\theta'}{\partial\tilde{\omega}}\bigg(\frac{\partial \theta_1}{\partial \omega_1}\bigg)^{-1}
v(\theta_1(\omega_1,\tilde{\omega}),\theta'(\theta_1(\omega_1,\tilde{\omega}),\tilde{\omega}))))\\
&=J_n(
\bigg(\frac{\partial \theta_1}{\partial \omega_1}\bigg)^{-1}\det{}^{-1}\frac{\partial\theta'}{\partial\tilde{\omega}}\bigg|_{\theta_1=\theta_1(\omega)}
v(\theta(\omega))).
\end{aligned}
$$
On the other hand, if we have
$$
\bigg(\det \frac{\partial\theta'}{\partial\tilde{\omega}}\bigg)\frac{\partial \theta_1}{\partial \omega_1}=\det\frac{\partial\theta}{\partial\omega},
$$
then we prove \eqref{int5-14}.
In fact, from \eqref{VVII3.28}, we get
$$
\frac{\partial\theta_1}{\partial\omega_k}=\frac{\partial\bar\omega_1}{\partial\omega_k}\frac{\partial\theta_1}{\partial\omega_1}
+\frac{\partial\theta_1}{\partial\omega_k} \for k=2,{\cdots},n,
$$
and
$$
\frac{\partial\theta_i'}{\partial\omega_k}=\frac{\partial\theta_i}{\partial\omega_k}+\frac{\partial\theta_i}{\partial\omega_1}
\frac{\partial\bar\omega_1}{\partial\omega_k}
=\frac{\partial\theta_i}{\partial\omega_k}-\frac{\partial\theta_i}{\partial\omega_1}
\frac{\partial\theta_1}{\partial\omega_k}\bigg(\frac{\partial \theta_1}{\partial \omega_1}\bigg)^{-1}.
$$
Subtracting from $k$-row $(k\ge 2)$ in $(\frac{\partial\theta_i}{\partial\omega_k})$ the first row multiplied by
$\partial\theta_1/\partial \omega_k(\partial\theta_1/\partial \omega_1)^{-1}$, we obtain, using above relation,
$$
\det\frac{\partial\theta}{\partial\omega}
=\begin{vmatrix}
\frac{\partial\theta_1}{\partial\omega_1}&\frac{\partial\theta_2}{\partial\omega_1}&{\cdots}&\frac{\partial\theta_n}{\partial\omega_1}\\
0&\frac{\partial\theta_2'}{\partial\omega_2}&{\cdots}&\frac{\partial\theta'_n}{\partial\omega_2}\\
{\cdots}&{\cdots}&{\cdots}&{\cdots}\\
0&\frac{\partial\theta_2'}{\partial\omega_n}&{\cdots}&\frac{\partial\theta'_n}{\partial\omega_n}
\end{vmatrix}
=\frac{\partial\theta_1}{\partial\omega_1}{\cdot}\det\frac{\partial\theta'}{\partial\omega}\bigg|_{\theta_1=\theta_1(\omega)}. 
$$
(7) is clear. $\qquad\qed$

\begin{remark}
Above Berezin integration is defined without measure but using inner-multiplication in exterior algebra\footnote{See, Mini-column 2 in Chapter 3} such that
$$
\frac{\partial}{\partial z_j}\rfloor dz_k=\delta_{jk}\sim \int d\theta_j\,\theta_k=\delta_{jk}.
$$
\end{remark}

\begin{remark}
(i) We get the integration by parts formula, without the fundamental theorem of elementary analysis.\\
(ii) Moreover, since in conventional integration we get $\displaystyle{\int dy f(y)}=a\int dx f(ax)$, therefore the formula in \eqref{int5-12} is very different from usual one. Analogous difference appears in \eqref{int5-14}.\\
(iii) \eqref{int5-15} allows us to put 
$$ 
\delta ( \theta - \omega ) = ( \theta_1 - \omega_1 ) \cdots ( \theta_n - \omega_n ),
$$
though $\delta (- \theta )=(-1)^n \delta (\theta)$. \\

(iv) For the future use, we give a Lie group theoretic proof of \eqref{int5-14} due to Berezin~\cite{Ber87}.
Let a transformation $T$ may be included in a one-parameter family $T_t$ of transformations $\theta=\theta(\omega)$, that is, $T_1\omega=\theta(\omega)$ and $T_0\omega=\omega$ with $T_{t+s}=T_tT_s$.
Set
\begin{equation}
\theta_k(t)=(T_t\omega)_k \et
g(t)=\int_{\rod^n} d\omega\,\det{}^{-1}\frac{\partial \theta(t)}{\partial\omega}v(\theta(t)).
\label{ab1}
\end{equation}
{\begin{claim}
 $g(t)$ is an analytic function w.r.t. $t$ and $g(t)=g(0)$.  
 \end{claim}}
{\begin{quotation}
{\small
In fact, using the multiplicativity of determinant, 
we have
$$
g(t+s)=\int_{\rod^n} d\omega\,\det{}^{-1}\bigg(\frac{\partial \theta(t+s)}{\partial \theta(t)}\bigg){\cdot}
\det{}^{-1}\bigg(\frac{\partial \theta(t)}{\partial\omega}\bigg)v(\theta(t+s)).
$$
Putting
$R(s)=\bigg(\frac{\partial \theta(t+s)}{\partial \theta(t)}\bigg)\in\rev$, $\Delta(s)=\det{}^{-1}R(s)\in\rev$, we get
% and $\zeta_j(s)=\theta_j(t+s)$
$$
g'(t)=\ds g(t+s)\big|_{s=0}
=\int d\omega\, \det{}^{-1}\bigg(\frac{\partial \theta(t)}{\partial\omega}\bigg)\ds\big[\Delta(s)v(\theta(t+s))\big]_{s=0}.
$$

To continue calculation, we remark the following:
(i)Since $T_t$ is a one-parameter transformation group, $\theta_j(t)=\theta_j(t;\omega)$ satisfy an autonomaous system of differential equations
$$
\theta_k'(t)=-\Phi_k(\theta_1(t),{\cdots},\theta_n(t)).
$$
(ii) As
$$
\Delta(s)=\det R^{-1}(s)=\exp{(-\tr \log R(s))},
$$
we have
$$
\begin{aligned}
\frac{d\Delta(s)}{ds}\bigg|_{s=0}&=-\tr(R'(s)R^{-1}(s))\exp{(-\tr \log R(s))}\big|_{s=0}\\
&=-\tr R'(0)
=-\sum_k \frac{d}{ds}\frac{\partial\theta_k(t+s)}{\partial\theta_k(t)}\bigg|_{s=0}=\sum_k\frac{\partial\Phi_k(\theta(t))}{\partial\theta_k(t)}.
\end{aligned}
$$

Therefore, we have, using $\Phi_k(\theta(t))\in\rod$,
$$
\ds\big[\Delta(s)v(\theta(t+s))\big]_{s=0}=\sum_k\frac{\partial(\Phi_k(\theta(t))v(\theta(t)))}{\partial\theta_k(t)},
$$
and
\begin{equation}
g'(t)=\int_{\rod^n} d\omega\,\det{}^{-1}\bigg(\frac{\partial \theta(t)}{\partial\omega}\bigg)\sum_k\frac{\partial (\Phi_k(\theta(t)) v(\theta(t)))}{\partial\theta_k(t)}.
\label{ab2}
\end{equation}
Applying the reasoning of the proof of (2) (translational invariance), we have
$$
g'(0)=\int_{\rod^n} d\omega\,\sum_k\bigg(\frac{\partial}{\partial\omega_k}\psi_k\bigg)d\omega=0\with
\psi_k=\theta'_k(0)v(\omega).
$$
Since $g'(t)$ has the same form as $g(t)$, that is, \eqref{ab2} is obtained by replacing $v(\theta(t))$ with $\sum_k\frac{\partial (\Phi_k(\theta(t)) v(\theta(t)))}{\partial\theta_k(t)}$ in \eqref{ab1}, we get $g''(0)=0$. Repeating this procedure, we get $g^{(n)}(0)=0$ for $n\ge 1$.}
\end{quotation}}
\par
It follows from the Lie group theory that an arbitrary transformation $\theta=\theta(\omega)$ can be represented in the form of a product of a finite number of transformations $T=T_1{\cdots} T_r$, each of which is included in a one-parameter group.
\end{remark}
%\end{comment}

\section{Berezin integral w.r.t. even and odd variables}
\subsection{A naive definition and its problem}
Because of Remark~\ref{dewitt-body} in \S4 of Chapter 4, we are inclined to ``define''
$$
\int_a^b dq f(q)=\int_{\tilde{a}}^{\tilde{b}}dx \tilde{f}(x)\where 
\tilde{f}(x)=\sum_{n=0}^{\infty}\frac{1}{n!} \partial_q^n f(q) x_{\mathrm{S}}^n \with x=q+x_{\mathrm{S}}.
$$
Therefore,
\begin{definition}\label{def3-1.2}
For a set $U\subset\euc^m$, we define $\pi_{\mathrm{B}}^{-1}(U)=\{X\in \supermo\;|\; \pi_{\mathrm{B}}(X)\in U\}$.
A set ${\mathfrak{U}}_{\mathrm{ev}}\subset\supermo$ 
is called an ``even superdomain'' 
if $U=\pi_{\mathrm B}({\mathfrak{U}}_{\mathrm{ev}})
\subset {\mathbb R}^m$ is open, connected and
$\pi_{\mathrm B}^{-1}(U)={\mathfrak{U}}_{\mathrm{ev}}$.
$U$ is denoted also by ${\mathfrak{U}}_{\mathrm{ev},\mathrm B}$.
When $ {\mathfrak{U}} \subset \supermn$ is represented by
${\mathfrak{U}}={\mathfrak{U}}_{\mathrm{ev}} \times {\fR}_{\mathrm{od}}^n$ with an even superdomain 
${\mathfrak{U}}_{\mathrm{ev}}\subset\supermo$,
${\mathfrak{U}}$ is called a ``superdomain'' in ${\fR}^{m|n}$.
\end{definition}

\begin{definition}[A naive definition of Berezin integral]\label{NDI}　%\label{Berezin4.8}
For a super domain ${\mathfrak{U}}={\mathfrak{U}}_{\mathrm{ev}} \times {{\fR}^{0|n}}$
and a supersmooth function $u(x,\theta)=\sum_{|a|\le n}\theta^a u_a(x):{\mathfrak{U}}\to{\fR}$, we ``define'' its integral as
\begin{equation}
\begin{gathered}
\Biint_{\!\!\!\mathfrak{U}} dx d\theta\, u(x,\theta)=\int_{{\mathfrak{U}}_{\mathrm{ev}}}dx\bigg(\int_{{\fR}^{0|n}}d\theta\, u(x,\theta)\bigg)
=\int_{\pi_{\mathrm{B}}({\mathfrak{U}}_{\mathrm{ev}})}dq\,u_{\tilde{1}}(q),\\
\where
\int_{{\fR}^{0|n}}d\theta\, u(x,\theta)=\frac{\partial}{\partial\theta_n}{\cdots}
\frac{\partial}{\partial\theta_1}u(x,\theta)\bigg|_{\theta_1={\cdots}=\theta_n=0}=u_{\tilde{1}}(x)\et \tilde{1}=(\overbrace{1,{\cdots},1}^{n}).
\end{gathered}
\label{BF}
\end{equation}
In the above, $u_{\tilde{1}}(x)$ is the Grassmann continuation of $u_{\tilde{1}}(q)$.
\end{definition}
Desiring that the standard \underline{formula of the change of variables under integral sign}(={\bf {CVF}}) holds by replacing standard Jacobian with super Jacobian(= super determinant of Jacobian matrix) on ${\fR}^{m|n}$, we have 
\begin{theorem}\label{Ber-naive}
Let ${\mathfrak{U}}={\mathfrak{U}}_{\mathrm{ev}} \times {{\fR}^{0|n}}\subset{\fR}^{m|n}_X$ and 
${\mathfrak{V}}={\mathfrak{V}}_{\mathrm{ev}} \times {{\fR}^{0|n}}\subset{\fR}^{m|n}_Y$ be given. Let
\begin{equation}
\varphi:{\mathfrak{V}}\ni Y=(y,\omega)\to X=(x,\theta)=(\varphi_{\bar{0}}(y,\omega),\varphi_{\bar{1}}(y,\omega))\in{\mathfrak{U}}
\label{int5-17}
\end{equation}
be a {supersmooth} diffeomorphism from ${\mathfrak{V}}$ onto ${\mathfrak{U}}$,
that is, 
\begin{equation}
{\sdet} J(\varphi)(y,\omega)\neq{0}\et\varphi({\mathfrak{V}})={\mathfrak{U}}
\where
J(\varphi)(y,\omega)=
\begin{pmatrix}
\frac{\partial \varphi_{\bar{0}}(y,\omega)}{\partial y}&\frac{\partial\varphi_{\bar{1}}(y,\omega)}{\partial y}\\
\frac{\partial \varphi_{\bar{0}}(y,\omega)}{\partial \omega}&\frac{\partial\varphi_{\bar{1}}(y,\omega)}{\partial \omega}
\end{pmatrix}.
\label{int5-17-1}
\end{equation}
Then, for any function $u\in{\CSS}({\mathfrak{U}}:{\fC})$ with ``\underline{compact support}'', 
that is, $u(x,\theta)=\sum_{|a|\le n}\theta^a u_a(x)$ where
$u_a(x_{\mathrm{B}})\in C^{\infty}_0({\mathfrak{U}}_{\mathrm{ev, B}}:{\fR})$ for all $a\in\{0,1\}^n$ except $a=\tilde{1}$,
we have CVF %the change of variables formula
\begin{equation}
\Biint_{\!\!\!\mathfrak{U}}dxd\theta \, u(x,\theta)
=\Biint_{\!\!\!\varphi^{-1}(\mathfrak{U})}dyd\omega \,{\sdet} J(\varphi)(y,\omega) 
u(\varphi(y,\omega)).
\label{B2.2.12}
\end{equation}
\end{theorem}
\begin{remark}
Seemingly, this theorem implies that
Berezin ``measure'' $D_0(x,\theta)$ is transformed by $\varphi$ as
\begin{equation}
(\varphi^*D_0(x,\theta))(y,\omega)=D_0(y,\omega){\cdot}{\sdet}J(\varphi)(y,\omega),\\
\label{Ber-mea}
\end{equation}
where
$$
D_0(x,\theta)=dx_1\wedge{\cdots}\wedge dx_m\otimes
\frac{\partial}{\partial\theta_n}{\cdots}\frac{\partial}{\partial\theta_1}
=dx_1{\cdots}dx_m {\cdot}{\partial}_{\theta_n}{\cdots}{\partial}_{\theta_1}
=dx\partial_{\theta},\; D_0(y,\omega)=dy\partial_{\omega}.
$$
But this assertion is shown to be false in general by the following examples.
Moreover, we remark also that the condition of ``the compact supportness of integrands'' above seems not only
cumbersome from conventional point of view but also fatal in holomorphic category.
\end{remark}
\begin{remark}
Though we give some examples which show the immatureness of the above naive definition, but we give a precise proof of this theorem in \S1 of Chapter 9 for future use.
\end{remark}

\begin{example}\label{Example1} 
Let
${\mathfrak{U}}=\pi_{\mathrm B}^{-1}({\Omega})\times\rod^2\subset {\fR}^{1|2}$ 
with ${\Omega}=(0,1)$, $\pi_{\mathrm{B}}:{\fR}^{1|0}\to  \euc$ and let $u$ be supersmooth on ${\fR}^{1|2}$ with value in ${\fR}$ such that
$u(x,\theta)=u_{\tilde{0}}(x)+\theta_1\theta_2u_{\tilde{1}}(x)$.
Then, we have
$$
\Biint_{\!\!\!\mathfrak{U}} dx d\theta\,u(x,\theta)=
\int_{\Omega} dx\int d\theta\, u(x,\theta)=\int_{\pi_{\mathrm B}^{-1}({\Omega})}dx\,u_{\tilde{1}}(x)
=\int_{\Omega}dq\,u_{\tilde{1}}(q).
$$
But, if we use the coordinate change
\begin{equation}
\varphi:(y,\omega)\to (x,\theta)\with
x=y+\omega_1\omega_2\phi(y),\;\theta_k=\omega_k: {\mathfrak{U}}\to {\mathfrak{U}}
\label{ex-cv}
\end{equation}
whose Berezinian is
$$
{\Ber}(\varphi)(y,\omega)={\sdet}J(\varphi)(y,\omega)=1+\omega_1\omega_2\phi'(y)
\where
J(\varphi)(y,\omega)=\begin{pmatrix}
1+\omega_1\omega_2\phi'(y)&0&0\\
\omega_2\phi(y)&1&0\\
-\omega_1\phi(y)&0&1
\end{pmatrix},
$$
and if we assume that the formula \eqref{B2.2.12} holds,
then since
$$
\begin{gathered}
u(\varphi(y,\omega))=u_{\tilde{0}}(y+\omega_1\omega_2\phi(y))+\omega_1\omega_2u_{\tilde{1}}(y+\omega_1\omega_2\phi(y))
=u_{\tilde{0}}(y)+\omega_1\omega_2(\phi(y)u_{\tilde{0}}'(y)+u_{\tilde{1}}(y)),\\
\et 
(1+\omega_1\omega_2\phi'(y))u(\varphi(y,\omega))=u_{\tilde{0}}(y)+\omega_1\omega_2(\phi(y)u_{\tilde{0}}'(y)+\phi'(y)u_{\tilde{0}}(y)+u_{\tilde{1}}(y)),
\end{gathered}
$$
we have
$$
\Biint_{\!\!\!\varphi^{-1}(\mathfrak{U})} dy d\omega\,(1+\omega_1\omega_2\phi'(y))u(\varphi(y,\omega))
=\int_{\pi_{\mathrm B}^{-1}({\Omega})}dy\,(\phi(y)u_{\tilde{0}}(y))'+ \int_{\pi_{\mathrm B}^{-1}({\Omega})}dx\,u_{\tilde{1}}(x).
$$
Therefore, if $\int_{\pi_{\mathrm B}^{-1}({\Omega})}dy\,(\phi(y)u_{\tilde{0}}(y))'\neq 0$, then
$\iint_{\mathfrak{U}} D_0(x,\theta)\,u(x,\theta)\neq \iint_{\varphi^{-1}(\mathfrak{U})} D_0(y,\omega)\,u(\varphi(y,\omega))$. 
This implies that if we apply \eqref{BF} as definition, 
the change of variables formula doesn't hold when, for example, the integrand hasn't compact support.
\end{example}
\begin{example}\label{Example2}
[Inconsistency related to $Q$-integration where matrix $Q$ is mentioned in Chapter 3]
Let a set of matrix $Q$ be given by
$$
{\mathcal Q}=\bigg\{Q=\begin{pmatrix}
x_1&\theta_1\\
\theta_2&ix_2\end{pmatrix}\;\big|\; x_1, x_2\in\rev,\;\theta_1, \theta_2\in\rod
\bigg\}\cong {\fR}^{2|2}
$$
and let regard $Q$ as a variable with its ``volume element'' $dQ=\frac{dx_1dx_2}{2\pi}d\theta_2 d\theta_1$. Then, we have
\begin{equation}
\int_{\mathfrak Q}dQ\, e^{-\str Q^2}=\int_{\fR^{2|2}}
\frac{dx_1dx_2}{2\pi}d\theta_2 d\theta_1\, e^{-(x_1^2+x_2^2+2\theta_1\theta_2)}=1.
\label{C1}
\end{equation}

We apply change of variables to a super matrix $Q$ as
\begin{equation}
\left\{
{\begin{aligned}
&y_1=x_1+\frac{\theta_1\theta_2}{x_1-ix_2},\;
y_2=x_2-\frac{i\theta_1\theta_2}{x_1-ix_2},\\
&\omega_1=\frac{\theta_1}{x_1-ix_2},\;
\omega_2=-\frac{\theta_2}{x_1-ix_2},
\end{aligned}}\right.
\label{C2}
\end{equation}
or
\begin{equation}
\left\{
{\begin{aligned}
&x_1=y_1+{\omega_1\omega_2}(y_1-iy_2),\;
x_2=y_2-{i\omega_1\omega_2}(y_1-iy_2),\\
&\theta_1=\omega_1(y_1-iy_2),\;
\theta_2=-\omega_2(y_1-iy_2),
\end{aligned}}
\right.
\label{C3}
\end{equation}
to make it diagonal. Then,
\begin{equation}
GQG^{-1}=\begin{pmatrix}
y_1&0\\
0&iy_2
\end{pmatrix},\quad
GQ^2G^{-1}=\begin{pmatrix}
y_1^2&0\\
0&-y_2^2
\end{pmatrix}
\end{equation}
where
$$
G=\begin{pmatrix}
1+2^{-1}\omega_1\omega_2&\omega_1\\
\omega_2&1-2^{-1}\omega_1\omega_2
\end{pmatrix},\quad
G^{-1}=\begin{pmatrix}
1+2^{-1}\omega_1\omega_2&-\omega_1\\
-\omega_2&1-2^{-1}\omega_1\omega_2
\end{pmatrix}.
$$
Clearly,
$$
x_1-ix_2=y_1-iy_2,\et\str Q^2=x_1^2+x_2^2+2\theta_1\theta_2=y_1^2+y_2^2.
$$
and their Jacobian (called Berezian) is
$$
dQ=\frac{dx_1dx_2}{2\pi}d\theta_2 d\theta_1
=-\frac{dy_1dy_2}{2\pi}d\omega_2 d\omega_1\,(y_1-iy_2)^{-2}.
$$
This implies
$$
-\int \frac{dy_1dy_2}{2\pi}d\omega_2 d\omega_1
\,(y_1-iy_2)^{-2}e^{-(y_1^2+y_2^2)}=0
$$
which contradicts to \eqref{C1}.
\end{example}

\subsection{Integration of Gaussian type and Pfaffian}
In spite of above immatureness of the naive definition, we may give examples which
mention the relation of Gaussian type integral, determinant and Pfaffian.
\begin{definition}
For $n \times n$-anti symmetric matrix ${\tilde B}=({\tilde B}_{jk})$ with even elements,
we define the Pfaffian $\Pf({\tilde B})$ of $B$ as
\begin{equation}
\Pf({\tilde B})=
{1\over{(n/2)!}}\sum_{\rho \in \wp_n} 
\sgn (\rho )
{\tilde B}_{\rho (1)\,\rho (2)} \cdots 
{\tilde B}_{\rho (n-1)\,\rho (n)}.
\end{equation}
Here, $\wp_n$ is the permutation group of degree $n$, 
$\sgn (\rho )$ is the signature of $\rho \in \wp_n$.
\end{definition}

\begin{remark} Let $n$ be even, and let $A=(A_{ij})$ be anti symmeric matrix.
Then, we have
$$
\int d\theta \, \exp(-\frac{1}{2} A_{ij}\theta_i \theta_j)=\Pf(A).
$$
Moreover, $\Pf(A)^2=\det(A)$ holds.
\end{remark}

\begin{definition} \label{SP.5.0}
A even super matrix $M=
\begin{pmatrix}
A&C\\ 
D&B 
\end{pmatrix}$ is called positive-definite if the following conditions are satisfied:
\begin{description} 
\item[(gs.1)] $A$ has the body part $A_{\mathrm B}$ which is regular positive definite symmetric matrix.
\item[(gs.2)] $B$ is a regular anti-symmetric matrix.
\item[(gs.3)] $C$ and $D$ satisfies $\trp C+D=0$.
\end{description}
\end{definition} 

For above super matrix $M$, we define the corresponding bilinear form as
\begin{equation*}
\begin{aligned}
&\langle X,MX \rangle=\trp XMX \\
&\qquad=\sum_{j,k=1}^mx_jA_{jk}x_k
+\sum_{j=1}^m\sum_{s=1}^{n}x_jC_{j\,m+s}\theta_s
+\sum_{k=1}^m\sum_{t=1}^{n}\theta_tD_{m+t\,k}x_k
+\sum_{s,t=1}^{n}\theta_sB_{m+s\,m+t}\theta_t.
\end{aligned}
\label{GSBL-1}
\end{equation*}

\begin{lemma}\label{GS.5.2}
Let $M$ be a even, positive definite matrix. Then,
\begin{equation}
\begin{aligned}
G(\lambda,M)&=\int_{\supermn} dX 
e^{-\lambda^{-1}{2^{-1}}\langle X,MX \rangle}\for \lambda>0\\
&=
{\begin{cases} 
0 & \text{if $n$ is odd},\\
(2\pi \lambda)^{m/2}
(2\lambda)^{-n/2}
(\det~A)^{-1/2} \Pf(B-DA^{-1}C) & \text{if $n$ is even}.
\end{cases}}
\end{aligned}
\label{GS-5.2}
\end{equation}
\end{lemma}

\begin{comparison}
For a positive definite symmetric real matrix $H$, we have %正定値実対称行列とするとき
\begin{equation}
\int_{\euc^m}e^{-{\lambda}x{\cdot}Hx/2}dx
=\bigg(\frac{2\pi}{\lambda}\bigg)^{m/2}(\det H)^{-1/2}.
\label{EG1}
\end{equation}
\end{comparison}

\begin{exercise} Prove the following by Berezin integral:\\
(i) $\det(A)=\Pf(A)^2$ for antisymmeric matrix $A$,\\
(ii) For any $2n\times 2n$ antisymmeric matrix $A$ and any $2n\times 2n$ matrix $B$, $\det(B^tAB)=\det(B)\Pf(A)$,\\
(iii) For any $n\times n$ matrix $B$,
$$
\Pf\begin{pmatrix}
0&B\\
-B^t&0
\end{pmatrix}=(-1)^{n(n-1)/2}\det(B).
$$
\end{exercise}

\section{Contour integral w.r.t. an even variable}
To overcome the inconsistency in above examples, we need to reconsider the meaning of ``body part'',
that is, not to insist on Remark \ref{dewitt-body} of de Witt (\S4 of Chapter 4). 

We recall the idea of the contour integral noted in A. Rogers~\cite{rog86-2}.
\begin{quotation}
Contour integrals are a means of ``pulling back'' an integral in a space that is algebraically (as well as
possibly geometrically) more complicated than $\euc^m$. A familiar example, of course, is complex contour integration; 
if $\gamma:[0,1]\to{\mathbb{C}}$ is piecewise $C^1$ and $f:{\mathbb{C}}\to{\mathbb{C}}$, one has
the one-dimensional contour integral
$$
\int_{\gamma} f(z)dz=\int_0^1f(\gamma(t)){\cdot}\gamma'(t)dt=\int_0^1dt\,\gamma'(t){\cdot}f(\gamma(t)).
$$
This involves the algebraic structure of ${\mathbb{C}}$ because the right-hand side of above includes multiplication $\cdot$ of complex numbers.
\end{quotation}
We follow this idea to define the integral of a {supersmooth} function $u(x)$ 
on an even superdomain ${\mathfrak{U}}_{\mathrm{ev}} \subset \fR^{m|0}=\rev^m$
 (see also, Rogers~\cite{rog85-1,rog85-2,rog86-2} and Vladimirov and Volovich~\cite{VV84}). 

\begin{definition}\label{int5.1} 
Let $u(x)$ be a {supersmooth} function defined on an even superdomain 
${\mathfrak{U}}_{\mathrm{ev}}\subset{\fR}^{1|0}$ such that 
$[a,b]\subset \pi_{\mathrm{B}}({\mathfrak{U}}_{\mathrm{ev}})$.
Let $ \lambda=\lambda_{\mathrm B} + \lambda_{\mathrm S}$, 
$\mu =\mu_{\mathrm B} + \mu_{\mathrm S} \in {\mathfrak{U}}_{\mathrm{ev}}$ 
with $\lambda_{\mathrm B}=a$, $\mu_{\mathrm B}=b$,
and let a continuous and piecewise $C^1 $-curve 
$ \gamma : [a,b] \,\rightarrow \, {\mathfrak{U}}_{\mathrm{ev}} $
be given such that 
$\gamma(a) = \lambda $, $\gamma(b)  = \mu $.
We define
\begin{equation}
\int_{\gamma}  dx \, u(x)
=
\int_{a}^{b} dt\,{\dot \gamma}(t){\cdot}u(\gamma(t))\in{\fC}
\label{int5-1}
\end{equation}
and call it the {integral of $u$} along the curve $\gamma$.
\end{definition} 

Using the integration by parts for functions on $\euc$, we get the following fundamental result. 
\begin{proposition} [p.7 of de Witt~\cite{DeW84-2}]\label{Prop5.1.1}
Let $u(t) \in C^{\infty} ([{a},{b}]:{\fC})$ and $U(t) \in C^{\infty} ([{a},{b}]:{\fC}) $ be given such that $ U' (t)=u(t) $ on
$ [ {a} , {b} ] $. We denote the Grassmann continuations of them  as  $\tilde{u}(x)$ and  $\tilde{U}(x)$. %$u(t)$ and $U(t)$
Then, for any continuous and piecewise
$C^1 $-curve ${\gamma}:[ {a} , {b} ] 
\, \rightarrow \, {\mathfrak{U}}_{\mathrm{ev}}\subset{\fR}^{1|0}$
such that $[ {a} , {b} ] \subset \pi_{\mathrm{B}}({\mathfrak{U}}_{\mathrm{ev}})$ and 
${\gamma}( {a} ) = \lambda$, ${\gamma}( {b} )=\mu $ with $\lambda_{\mathrm B}=a$, $\mu_{\mathrm B}=b$,
we have
\begin{equation}
\int_{\gamma}  dx \, \tilde{u}(x)= \tilde{U}( \lambda)-  \tilde{U}(\mu).
\label{int5-2}
\end{equation}
\end{proposition}
\par
{\it Proof}.
By definition, we get
{\allowdisplaybreaks
\begin{align*}
\int_{{a}}^{{b}}dt\,{\dot{\gamma}}(t)u({\gamma}(t))
&=
\int_{{a}}^{{b}}dt\,({\dot{\gamma}_{\mathrm B}}(t)+{\dot{\gamma}_{\mathrm S}}(t))
\sum_{\ell\ge 0} {\frac 1{\ell!}}
u^{(\ell)}({\gamma}_{\mathrm B}(t)){\gamma}_{\mathrm S}(t)^\ell\\
{}
&=\int_{{a}}^{{b}}dt\,
{\dot{\gamma}_{\mathrm B}}(t)u({\gamma}_{\mathrm B}(t))
+\int_{{a}}^{{b}}dt\,
{\dot{\gamma}_{\mathrm B}}(t)\sum_{k\ge 1} {\frac 1{k!}}
u^{(k)}({\gamma}_{\mathrm B}(t)){\gamma}_{\mathrm S}(t)^k\\
&\qquad\qquad\qquad\qquad\qquad\qquad\qquad\qquad
+\int_{{a}}^{{b}}dt\,
\sum_{\ell\ge 0} {\frac 1{\ell!}}
u^{(\ell)}({\gamma}_{\mathrm B}(t)){\dot{\gamma}_{\mathrm S}}(t){\gamma}_{\mathrm S}(t)^\ell\\
&=
U({b})-U({a})+
\sum_{\ell\ge 0} {\frac 1{(\ell+1)!}}
\left\{
U^{(\ell+1)}({b})\mu_{\mathrm S}^{\ell+1}
-U^{(\ell+1)}({a})\lambda_{\mathrm S}^{\ell+1}
\right\}\\
&=\tilde{U}(\mu)-\tilde{U}(\lambda).	
\end{align*}
}
Here, we used the integration by parts formula for functions on $\euc$ with value in Fr\'echet space:
$$
\begin{aligned}
\int_{{a}}^{{b}}dt\,
&u^{(\ell)}({\gamma}_{\mathrm B}(t)){\dot{\gamma}_{\mathrm S}}(t){\gamma}_{\mathrm S}(t)^\ell\\
&=
\int_{{a}}^{{b}}dt\,
u^{(\ell)}({\gamma}_{\mathrm B}(t))
\dt \frac{{\gamma}_{\mathrm S}(t)^{\ell+1}}{\ell+1}\\
&=-\int_{{a}}^{{b}}dt\,
{\dot{\gamma}_{\mathrm B}}(t)u^{(\ell+1)}({\gamma}_{\mathrm B}(t))
\frac{{\gamma}_{\mathrm S}(t)^{\ell+1}}{\ell+1}
+U^{(\ell+1)}({{b}})\frac{{\mu_{\mathrm S}}^{\ell+1}}{\ell+1}
-U^{(\ell+1)}({{a}})\frac{{\lambda_{\mathrm S}}^{\ell+1}}{\ell+1}.   \qquad\qed
\end{aligned}
$$

\begin{problem}
How do we extend Proposition \ref{Prop5.1.1} to the case when
$u(t) \in C([{a},{b}]:{\fC}) ? $ % or ${}\in{\mathcal{D}}'([{a},{b}]:{\fC}) $?
\end{problem}

\begin{lemma}[Lemma 3.9 in \cite{rog85-1} on ${\mathfrak{B}}_L$]\label{rog3.9}
(a) (reparametrization of paths) Let $\gamma:[a,b]\to\rev$ be a path in $\rev$ and let $c,\,d\in\euc$.
Also let $\phi:[c,d]\to[a,b]$ be $C^1$ with $\phi(c)=a$, $\phi(d)=b$ and $\phi'(s)>0$ for all $s\in[c,d]$.
Then
$$
\int_{\gamma} dx\, u(x)=\int_{\gamma\circ\phi}dx\, u(x).
$$
(b)(sum of paths) Let $\gamma_1:[a,b]\to\rev$ and $\gamma_2:[c,d]\to\rev$ be two paths with $\gamma_1(b)=\gamma_2(c)$ Also define $\gamma_1+\gamma_2$ to be the path $\gamma_1+\gamma_2:[a,b+d-c]\to \rev$ defined by
$$
\gamma_1+\gamma_2(t)=
\begin{cases}
\gamma_1(t), &{a\le t\le b},\\
\gamma_2(t-b+c), & {b\le t\le b+d-c}.
\end{cases}
$$
Then if ${\mathfrak{U}}_{\mathrm{ev}}$ is open in $\rev$, $u:{\mathfrak{U}}_{\mathrm{ev}}\to{\fR}$ is in $\CSS$ and $\gamma_1([a,b])\subset {\mathfrak{U}}_{\mathrm{ev}}$, $\gamma_2([c,d])\subset {\mathfrak{U}}_{\mathrm{ev}}$,
$$
\int_{\gamma_1+\gamma_2}dx\,u(x) =\int_{\gamma_1}dx\,u(x)+\int_{\gamma_2}dx\,u(x).
$$
(c)(inverse of a path)  Let $\gamma:[a,b]\to\rev$ be a path in $\rev$.
Define the curve
$-\gamma:[a,b]\to\rev$ by
$$
-\gamma(t)=\gamma(a+b-t)
$$
Then if ${\mathfrak{U}}_{\mathrm{ev}}$ is open in $\rev$
 with $\gamma([a,b])\subset {\mathfrak{U}}_{\mathrm{ev}}$ and $u:{\mathfrak{U}}_{\mathrm{ev}}\to\fR$ is supersmooth,
 $$
 \int_{-\gamma} dx\, u(x)= -\int_{\gamma} dx\,u(x).
 $$
 \end{lemma}
 \par {\it Proof}.
Applying CVF on $\euc$ for $t=\phi(s)$ and $dt=\phi'(s)ds$, we have
 $$
 \begin{aligned}
\int_{\gamma\circ\phi}dx\, u(x)=\int_c^d ds\, (\gamma(\phi(s)))'u(\gamma(\phi(s)))
&=\int_c^d ds\, \phi'(s)[\gamma'(\phi(s))u(\gamma(\phi(s)))]\\
&=\int_a^b dt\,\gamma'(t)u(\gamma(t))=\int_{\gamma} dx\, u(x).
\end{aligned}
$$
Others are proved analogously. $\qquad\square$ 

\begin{corollary}[Corollary 3.7 in \cite{rog85-1} on ${\mathfrak{B}}_L$]\label{int5.3}
Let $u(x)$ be a {supersmooth} function defined on a even superdomain 
${\mathfrak{U}}_{\mathrm{ev}} \subset {\fR}^{1|0}$ into $\fC$.\\
(a) Let $ {\gamma}_1 , {\gamma}_2 $ be continuous and piecewise $C^1 $-curves from 
$[a,b] \, \rightarrow \, {\mathfrak{U}}_{\mathrm{ev}}$
such that $\lambda ={\gamma}_1(a)={\gamma}_2 (a) $ 
and $ \mu = {\gamma}_1 (b) ={\gamma}_2 (b) $.
If ${\gamma}_1 $ is homotopic to ${\gamma}_2 $, then 
\begin{equation}
\int_{{\gamma}_1}dx \, u(x)
= \int_{{\gamma}_2}dx \, u(x).
\label{int5-3}
\end{equation}
(b) If $u:\rev\to{\fR}$ is $\CSS$ on all $\rev$, one can write ``unambiguously''
$$
\int_{\lambda}^{\mu}dx\, u(x)=\int_{\gamma} dx\,u(x).
$$
Here, $\gamma:[a,b]\to \rev$ is any path in $\rev$ with $\gamma(a)={\lambda}$, $\gamma(b)={\mu}$.
\end{corollary}

\begin{proposition}
For a given change of variable $x=\varphi(y)$, 
we define the pull-back  of 1-form ${\mathfrak{v}}_x=dx\,\rho(x)$ 
by
$(\varphi^*{\mathfrak{v}})_y=dy\,\frac{\partial \varphi(y)}{\partial y}\,\rho(\varphi(y))$. 
Then, for paths $\gamma:[a,b]\to \fR^{1|0}_x$, $\varphi^{-1}\circ\gamma:[a,b]\to \fR^{1|0}_y$ and
$u$, we have
$$
\int_{\gamma}\,{\mathfrak{v}}u=\int_{\gamma}dx\,{\mathfrak{v}}_x\,\rho(x)u(x)
=\int_{\varphi^{-1}\circ\gamma}dy\,(\varphi^*{\mathfrak{v}})_y\,\rho(\varphi(y))u(\varphi(y))
=\int_{\varphi^{-1}\circ\gamma}\varphi^*{\mathfrak{v}}\,\varphi^*u.
$$
\end{proposition}
\par
{\it Proof}. 
By definition, we have not only
$$
\int_{\gamma} {\mathfrak{v}}_x u(x)=\int_a^b dt\, \gamma'(t)\rho(\gamma(t))u(\gamma(t)),
$$
but also
$$
\begin{aligned}
\int_{\varphi^{-1}\circ\gamma}(\varphi^*{\mathfrak{v}})_y\varphi^*u(y)
&=\int_{\varphi^{-1}\circ\gamma} dy\, \frac{\partial \varphi(y)}{\partial y} \rho(\varphi(y))u(\varphi(y))\\
&=\int_a^b dt\, (\varphi^{-1}(\gamma(t)))'\frac{\partial \varphi(y)}{\partial y}\rho(\varphi(y)) u(\varphi(y))\bigg|_{y=\varphi^{-1}\circ\gamma(t)}\\
&=\int_a^b dt\, \dot\gamma(t)\rho(\gamma(t))u(\gamma(t)).
\end{aligned}
$$
Here, we used $y=\varphi^{-1}(\varphi(y))$, $x=\gamma(t)$, $y=\varphi^{-1}(\gamma(t))$ with
$$
1=\dot{\varphi}(y)\dot{\varphi}^{-1}(\varphi(y)),
\quad
{\dot{\varphi}(y)}=\frac{1}{\dot{\varphi}^{-1}(\varphi(y))}=\frac{\partial \varphi(y)}{\partial y}. \qquad\qed
$$
\begin{example}[Translational invariance]\label{TI}
Let $I=(a,b)\subset \euc$. We identify $q\in I$ as $\gamma(q)=x\in\rev$.
We put
${\mathfrak{M}}=\gamma(I)=\{x\in\rev\;|\; \pi_{\mathrm{B}}(x)=q\in I\}\subset\rev$. %={\mathfrak{M}}(\iota, I)
Taking a non-zero nilpotent element $\nu\in\rev$, we put $\tau_{\nu}:\rev\ni y\to x=\varphi(y)=\tau_{\nu}(y)=y-\nu\in\rev$,
% and $\tau_{\nu}^{-1}(y)=y+\nu=\tau_{-\nu}(y)$,
$$
{\mathfrak{M}}_1=\tau_{\nu}^{-1}({\mathfrak{M}})=\{x+\nu\rev\;|\; \pi_{\mathrm{B}}(x)=q\in I\},\quad \gamma_1(q)=\tau_{\nu}^{-1}(\gamma(q)).
$$
Then, we have
$$
\int_{{\mathfrak{M}}}dx\,u(x)=\int_a^b dq\,\gamma'(q)u(\gamma(q))=\int_a^b dq \,\gamma_1'(q)u(\gamma(q))
=\int_{{\mathfrak{M}}_1}dy\,u(y-\nu).
$$
\end{example}
\begin{remark} %
Above identification $\gamma(q)=x\in\rev$ is obtained as the Grassmann continuation $\tilde{\iota}$ of 
a function $\iota(q)=q\in C^{\infty}(I:\euc)$. In fact,
$$
\tilde{\iota}(x)=\sum_{\alpha}\frac{\partial^{\alpha}\iota(q)}{\partial q^{\alpha}}(x_{\mathrm{B}})x_{\mathrm{S}}^{\alpha}
=x_{\mathrm{B}}+x_{\mathrm{S}}=x.
$$
\end{remark}

\section{Modification of Rogers, Vladimirov and Volovich's approach }
Now, we modify arguments of Vladimirov and Volovich~\cite{VV84}  suitably to get a new definition.
\begin{definition}[Parameter set, paths and integral]\label{4.9}
For any domain $\Omega$ in ${\euc}^m$, we denote $\widetilde{\Omega}=\Omega\times\rod^n$ as a parameter set.
\par
(1) %{}_{\scriptstyle{}\atop\scriptstyle{}}
A smooth map $\gamma$ from $\widetilde{\Omega}$ to ${\fR}^{m|n}$ belongs to
$C^\infty(\widetilde{\Omega}:{\fR}^{m|n})$ when
$$
\gamma(q,{\vartheta})=(\gamma_{\bar{0}}(q,{\vartheta}), \gamma_{\bar{1}}(q,{\vartheta}))=(\gamma_{\bar0,j}(q,{\vartheta}), \gamma_{\bar1,k}(q,{\vartheta}))_{\scriptstyle{j=1,{\cdots},m}\atop\scriptstyle{k=1,{\cdots},n}},
$$
and
$$
\gamma_{\bar0,j}(q,{\vartheta})=\sum_{|a|\le n}{\vartheta}^a\gamma_{\bar0,j,a}(q)\in{\rev},\quad
\gamma_{\bar1,k}(q,{\vartheta})=\sum_{|a|\le n}{\vartheta}^a\gamma_{\bar1,k,a}(q)\in{\rod}.
$$
Here,
$$
\begin{gathered}
\gamma_{\bar0,j,a}(q)=\sum_{|{\mathbf{I}}|=|a|({\mathrm{mod}}2)} \gamma_{\bar0,j,a,{\mathbf{I}}}(q)\sigma^{\mathbf{I}},\;\;
\gamma_{\bar1,k,a}(q)=\sum_{|{\mathbf{J}}|=|a|+1({\mathrm{mod}}2)} \gamma_{\bar1,k,a,{\mathbf{J}}}(q)\sigma^{\mathbf{J}} %,\\
\end{gathered}
$$
and
$$
\begin{gathered}
\gamma_{\bar0,a,{\mathbf{I}}}(q),\; \gamma_{\bar1,a,{\mathbf{J}}}(q)\in C^\infty(\Omega:{\mathbb{C}}^n)
 \et
\gamma_{\bar0,\bar{0},\tilde{0}}(q)\in C^\infty(\Omega:{\mathbb{R}}^m)\\
\with
\bar0={even}\;\;\mbox{or}\;\; \bar{0}=(0,{\cdots},0)\in\{0,1\}^n,\; \tilde{0}=(0,{\cdots})\in{\mathcal{I}}.
\end{gathered}
$$
Moreover, if
$$
\sdet J(\gamma)(q,{\vartheta})\neq 0  \where
J(\gamma)(q,{\vartheta})=\frac{\partial\gamma(q,{\vartheta})}{\partial(q,{\vartheta})}=
\begin{pmatrix}
\frac{\partial\gamma_{\bar{0}}(q,{\vartheta})}{\partial q}&\frac{\partial\gamma_{\bar{1}}(q,{\vartheta})}{\partial q}\\
\frac{\partial\gamma_{\bar{0}}(q,{\vartheta})}{\partial{\vartheta}}&\frac{\partial\gamma_{\bar{1}}(q,{\vartheta})}{\partial{\vartheta}}
\end{pmatrix}, 
$$
then, this $\gamma$ is called a path from $\widetilde{\Omega}$ to ${\fR}^{m|n}$, whose image is said to be \underline{FSM(=foliated singular mani}\\
\underline{-fold)} denoted by
$$
{\mathfrak{M}}={\mathfrak{M}}(\gamma,\Omega)=\gamma(\widetilde{\Omega})=\{(x,\theta)\in \supermn\;\big|\;
x=\gamma_{\bar{0}}(q,{\vartheta}), \theta=\gamma_{\bar{1}}(q,{\vartheta}), q\in\Omega, {\vartheta}\in\rod^n\}.
$$
\par
(2) Let ${\mathfrak{M}}$ be given as above. For a supersmooth function $u(x,\theta)=\sum_{|a|\le n}\theta^a u_a(x)$ defined on ${\mathfrak{M}}$, we define the integration of $u(x,\theta)$ on ${\mathfrak{M}}$ as follows:
\begin{equation}
\VViint_{\!\!\!\mathfrak{M}} dxd\theta\,u(x,\theta)
=\int_{\rod^n} d{\vartheta}\,\bigg[\int_{\Omega} dq\,\sdet J(\gamma)(q,{\vartheta})u(\gamma(q,{\vartheta}))\bigg].
\label{20100315}
\end{equation}
Here, we assume that for each ${\vartheta}\in\rod^n$, integrands in the bracket $[{\cdots}]$ above are integrable on $\Omega$.
\end{definition}

\begin{remark}
The reason for nomination ``singular'' is explained in A. Kharenikov~\cite{Khr99}.
It stems from the difference from the naive Definition~\ref{NDI} of Berezin
i.e., their definition domain are not ${\mathfrak{U}}_{\mathrm{ev}}=\{x\in{\fR}^{m|0}\;|\; \pi_{\mathrm{B}}(x)\in\Omega\}$ though defined via $\Omega$ in ${\euc}^m$.
\end{remark}

We need to check the well-definedness of \eqref{20100315} in Definition~\ref{4.9}.
First of all, we remark that by the algebraic nature of integration w.r.t. odd variables, we may interchange the order of integration as
$$
\begin{aligned}
\int_{\rod^n} d{\vartheta}\,\bigg[\int_{\Omega} dq\,\sdet J(\gamma)(q,{\vartheta}){\cdot}u(\gamma(q,{\vartheta}))\bigg]
&=\frac{\partial}{\partial{\vartheta}_n}{\cdots}\frac{\partial}{\partial{\vartheta}_1}\int_{\Omega} dq\,\sdet J(\gamma)(q,{\vartheta}){\cdot}u(\gamma(q,{\vartheta}))\bigg|_{{\vartheta}=0}\\
&=\int_{\Omega} dq\,\frac{\partial}{\partial{\vartheta}_n}{\cdots}\frac{\partial}{\partial{\vartheta}_1}\big(\sdet J(\gamma)(q,{\vartheta}){\cdot}u(\gamma(q,{\vartheta}))\big)\bigg|_{{\vartheta}=0}\\
&=\int_{\Omega} dq\,\bigg[\int_{\rod^n} d{\vartheta}\,\sdet J(\gamma)(q,{\vartheta}){\cdot}u(\gamma(q,{\vartheta}))\bigg]. 
\end{aligned}
$$
In case when $\gamma_{\bar{0}}(q,{\vartheta})$ doesn't depend on ${\vartheta}$,
putting $\bar{\vartheta}=\gamma_{\bar{1}}(q,{\vartheta})$ and $\bar q=\gamma_{\bar{0}}(q)$, we get
$$
\begin{aligned}
\int_{\Omega} dq\,&\frac{\partial}{\partial{\vartheta}_n}{\cdots}\frac{\partial}{\partial{\vartheta}_1}\big(\sdet J(\gamma)(q,{\vartheta}){\cdot}u(\gamma(q,{\vartheta}))\big)\bigg|_{{\vartheta}=0}\\
&=\int_{\Omega} dq\,\det\bigg(\frac{\partial \gamma_{\bar{0}}(q)}{\partial q}\bigg)
\bigg[\int_{\rod^n} d{\vartheta}\,\det{}^{-1}\bigg(\frac{\partial \gamma_{\bar{1}}(q,{\vartheta})}{\partial{\vartheta}}\bigg){\cdot}u(\gamma_{\bar{0}}(q), \gamma_{\bar{1}}(q,{\vartheta}))\bigg]\\
&=\int_{\Omega} dq\,\det\bigg(\frac{\partial \gamma_{\bar{0}}(q)}{\partial q}\bigg)
\bigg[\int_{\rod^n} d\bar{{\vartheta}}\,u(\gamma_{\bar{0}}(q),\bar{{\vartheta}})\bigg]\\
&=\int d\bar{q}\int d\bar{{\vartheta}}\, u(\bar{q},\bar{{\vartheta}})
=\int_{\gamma_{\bar{0}}(\Omega)}dx\bigg[\int_{\rod^n} d\theta\, u(x,\theta)\bigg]. 
\end{aligned}
$$
That is, putting $\tilde\Omega=\Omega\times\rod^n$ and ${\mathfrak{M}}=\gamma(\tilde\Omega)$, we have $\gamma(q,{\vartheta})=(\gamma_{\bar{0}}(q),\gamma_{\bar{1}}(q,{\vartheta}))$ and
\begin{equation}
\begin{aligned}
{\VViint}_{\mathfrak{M}} dxd\theta\,u(x,\theta)
&=\int_{\rod^n} d\theta\,\bigg[\int_{\gamma_{\bar{0}}(\Omega)}dx\, u(x,\theta)\bigg]=\int d\theta\bigg(\int dx \,u(x,\theta)\bigg)\\
&=\int_{\gamma_{\bar{0}}(\Omega)} dx\,\bigg[\int_{\rod^n} d\theta\,u(x,\theta)\bigg]
=\int dx\bigg(\int d\theta \,u(x,\theta)\bigg).
\end{aligned}
\label{VV4.1}
\end{equation}

Moreover, we need the following definition:
\begin{definition}
Let two FSM ${\mathfrak{M}}=\gamma(\widetilde{\Omega})$ and ${\mathfrak{M}}'=\gamma'(\widetilde{\Omega}')$ be given. They are called superdiffeomorphic each other
if there exist diffeomorphisms
$\phi:\widetilde{\Omega}'\to \widetilde{\Omega}$ and $\varphi:{\mathfrak{M}}'\to{\mathfrak{M}}$,
such that the following diagram holds with $\gamma'=\varphi^{-1}\circ \gamma\circ\phi$:
$$
{\begin{CD}
{\widetilde{\Omega}}@>\gamma>>{\mathfrak{M}}=\gamma({\widetilde{\Omega}})\\ 
@A{\phi}AA  
@AA{\varphi}A\\
{\widetilde{\Omega}}'@>\gamma_1>>{\mathfrak{M}}'=\gamma'({\widetilde{\Omega}}').
\end{CD}}
$$
\end{definition}

Using this notion, we have the desired result:
\begin{proposition}[Reparametrization invariance]
Let $\Omega$ and $\Omega'$ be domains in $\euc^m$ and we put $\tilde\Omega$ and $\tilde\Omega'$ as above.
We assume $\tilde\Omega$ and $\tilde\Omega'$ are superdiffeomorphic each other, that is, there exist a diffeomorphism
$\phi_{\bar{0}}:\Omega'\to \Omega$ such that $\frac{\partial\phi_{\bar{0}}(q')}{\partial q'}$ which is continuous in $\Omega'$ and $\det(\frac{\partial\phi_{\bar{0}}(q')}{\partial q'})>0$ and a map
$\phi_{\bar{1}}:\Omega'\times \rod^n\ni(q',\eta')\to\phi_{\bar{1}}(q',\eta')\in\rod^n$ which is supersmooth w.r.t. $\eta'$ with $\det(\frac{\partial\phi_{\bar{1}}(q',\eta')}{\partial\eta'})\neq{0}$.
Put
$$
{\mathfrak{M}}'=\{X'=(x',\theta')\;|\; X'=\gamma\circ\phi(q',\eta'),\;\; (q',\eta')\in \tilde\Omega'\}\where
\phi(q',\eta')=(\phi_{\bar{0}}(q'),\phi_{\bar{1}}(q',\eta')).
$$
For a given path $\gamma:\tilde\Omega\to{\fR}^{m|n}$, we define a path $\gamma\circ\phi:\tilde\Omega'\to{\fR}^{m|n}$.
Then, we have
$$
\VViint_{\gamma(\tilde\Omega)} dxd\theta\,u(x,\theta)
={\VViint}_{\gamma\circ\phi(\tilde\Omega')}dx'd\theta'\,u(x',\theta').
$$
\end{proposition}
\par{\it Proof}. By definition, we have
$$
\VViint_{\gamma(\tilde\Omega)} dxd\theta\,u(x,\theta)
=\int_{\rod^n}d\eta\,\bigg(\int_{\Omega}dq\,\sdet J(\gamma)(q,\eta)
u(\gamma(q,\eta))\bigg)
$$
and
$$
{\VViint}_{\gamma\circ\phi(\tilde\Omega')}dx'd\theta'\,u(x',\theta')
=\int_{\rod^n}d\eta'\,\bigg(\int_{\Omega'}dq'\,
\sdet J(\gamma\circ\phi)(q',\eta')
u(\gamma\circ\phi(q',\eta'))\bigg).
$$
Using 
$$
\begin{gathered}
\gamma\circ\phi(q',\eta')=(\gamma_{\bar{0}}(\phi_{\bar{0}}(q'),\phi_{\bar{1}}(q',\eta')), \gamma_{\bar{1}}(\phi_{\bar{0}}(q'),\phi_{\bar{1}}(q',\eta'))),
\\
J(\gamma\circ\phi)(q',\eta')=J(\gamma)(\phi(q',\eta') J(\phi)(q',\eta'),\\
\sdet J(\phi)(q',\eta')=\det{}^{-1}\bigg(\frac{\partial\phi_{\bar{1}}(q',\eta')}{\partial\eta'}\bigg)\det\bigg(\frac{\partial\phi_{\bar{0}}(q')}{\partial q'}\bigg),
\end{gathered}
$$
we have
$$
\sdet J(\gamma\circ\phi)(q',\eta')
=\det\bigg(\frac{\partial\phi_{\bar{0}}(q')}{\partial q'}\bigg)\det{}^{-1}\bigg(\frac{\partial\phi_{\bar{1}}(q',\eta')}{\partial\eta'}\bigg)
\sdet J(\gamma)(q,\eta)\bigg|{}_{\scriptstyle{q=\phi_{\bar{0}}(q')}\atop\scriptstyle{\eta=\phi_{\bar{1}}(q',\eta')}}.
$$

Remarking the order of integration, we have
$$
\begin{aligned}
\int_{\rod^n} & d\eta'\,\bigg(\int_{\Omega'}dq'\,\sdet J(\gamma\circ\phi)(q',\eta')u(\gamma\circ\phi(q',\eta'))\bigg)\\
=&\int_{\Omega'}dq'\,\det\bigg(\frac{\partial\phi_{\bar{0}}(q')}{\partial q'}\bigg)
\bigg[\int_{\rod^n}d\eta'\,
\det{}^{-1}\bigg(\frac{\partial\phi_{\bar{1}}(q',\eta')}{\partial\eta'}\bigg)\big[\sdet J(\gamma)(q,\eta) u(\gamma(q,\eta))\big]
\bigg|{}_{\scriptstyle{q=\phi_{\bar{0}}(q')}\atop\scriptstyle{\eta=\phi_{\bar{1}}(q',\eta')}}\bigg]\\
=&
\int_{\Omega}dq\,
\bigg[\int_{\rod^n}d\eta\,\sdet J(\gamma)(q,\eta)u(\gamma(q,\eta))\bigg]
=\iint_{\tilde\Omega}dqd\eta\,\sdet J(\gamma)(q,\eta)u(\gamma(q,\eta))\bigg].  \qquad\qed
\end{aligned}
$$

Finally, we prove our goal:
\begin{theorem}[CVF=change of variable formula]\label{CVF-VV}
Let a supersmooth diffeomorphism $\varphi$ be given from 
a foliated singular manifold ${\mathfrak{N}}(\delta,\Omega)\subset \supermn$ onto a neighbourhood ${\mathfrak{O}}$
of another foliated singular manifold ${\mathfrak{M}}(\gamma,\Omega)\subset \supermn$:
\begin{equation}
\varphi:(y,\omega)\to (x,\theta)\with
x=\varphi_{\bar{0}}(y,\omega), \quad \theta=\varphi_{\bar{1}}(y,\omega)
\label{VV4.6}
\end{equation}
That is, 
${\mathfrak{M}}=\varphi({\mathfrak{N}})$ and $\sdet J(\varphi)\neq{0}$.
Moreover, we assume that $\delta=\varphi^{-1}\circ\gamma$ is $\sdet J(\gamma)\neq{0}$. 
\par
Then, for any integrable  function $u\in{\CSS}({\mathfrak{O}}:{\fR})$, CVF holds.
\begin{equation}
\VViint_{\!\!\!\mathfrak{M}} dxd\theta\,u(x,\theta)=
\VViint_{\!\!\!\varphi^{-1}({\mathfrak{M}})} dyd\omega\,{\sdet}J(\varphi)(y,\omega){\cdot}u(\varphi(y,\omega)).
\label{VV4.10-1}
\end{equation}
\end{theorem}

\begin{remark} Analogous result is proved on  superspace ${\mathfrak{B}}_{L}^{m|n}$ based on Banach-Grassmann  algebra assuming the set
$\{x\in{\mathfrak{B}}_{L\bar0}^m\;\big|\; x=\gamma(q,{\vartheta}),\,q\in\Omega\}$ is independent from each
${\vartheta}\in{\mathfrak{B}}_{L\bar1}^n$ in \cite{VV84}.
\end{remark}
\begin{remark} Formulas \eqref{B2.2.12} and \eqref{VV4.10-1} have the same form but their definitions 
\eqref{4.9} and \eqref{20100315} are very different each other!  
This difference is related to the primitive question  ``How to consider the body of supermanifolds?'' (see, R. Catenacci, C. Reina and P. Teofilatto~\cite{CRT85}). Though we don't develop supermanifolds theory with charts based on $\supermn$ in this note, but in Chapter 8, we consider the simplest case $\supermn$.
\end{remark}

\subsubsection{Proof of Theorem \ref{CVF-VV} -- change of variable formula under integral sign}
We want to prove the following diagram:
$$
\begin{CD}
(\tilde\Omega,dqd\eta) @>\gamma>>({\mathfrak{M}},dxd\theta)@>{u(x,\theta)}>>
{\mathrm{R}}\!{\mathrm{V}}\!{\mathrm{V}}\!\!\!-\!\!{\int\!\!\!\int}_{\!\!\mathfrak{M}} dxd\theta\,u(x,\theta)\in {{\fR}}\\
@|@A{\varphi}A{}A @|{}\\
(\tilde\Omega,dqd\eta) @>>\delta>({\mathfrak{N}},dyd\omega)@>>\varphi^*u(y,\omega)>
{\mathrm{R}}\!{\mathrm{V}}\!{\mathrm{V}}\!\!\!-\!\!{\int\!\!\!\int}_{\!\!\varphi^{-1}({\mathfrak{M}})} dyd\omega\,{\sdet}J(\varphi)(y,\omega)u(\varphi(y,\omega))\in {{\fR}}.
\end{CD}
$$

By definition, we have paths
$$
\begin{gathered}
\Omega\times\rod^n\ni(q,\eta)\to\gamma(q,\eta)=(x,\theta),\\
\Omega\times\rod^n\ni(q,\eta)\to\gamma_1(q,\eta)=(y,\omega),
\end{gathered}
$$
which are related each other
$$
(x,\theta)=\gamma(q,\eta)=\varphi(y,\omega)=\varphi(\gamma_1(q,\eta)),\quad \gamma_1=\varphi^{-1}\circ\gamma.
$$
We define pull-back of a ``superform''  as
$$
{\mathfrak{v}}=dx d\theta\, u(x,\theta)\to \varphi^*{\mathfrak{v}}=dy d\omega \,\sdet J(\varphi)(y,\omega)\,
u(\varphi(y,\omega)).
$$

Then, we have
\begin{claim}
\begin{equation}
{\VViint}_{\varphi^{-1}\circ\gamma(\tilde{\Omega})}\varphi^*{\mathfrak{v}}={\VViint}_{\gamma(\tilde{\Omega})}{\mathfrak{v}}.
\label{10-2-25-4.9}
\end{equation}
\end{claim}
\par
{\it {Proof}}. Since $J(\varphi^{-1}\circ\gamma)=J(\gamma){\cdot}J(\varphi^{-1})$ which yields
$$
\sdet J(\varphi^{-1}\circ\gamma)(q,\eta)(\sdet J(\varphi)(y,\omega)\bigg|_{(y,\omega)=\varphi^{-1}\circ\gamma(q,\eta)}
=\sdet J(\gamma)(q,\eta),
$$
and by the definitions of path(contour) and integral, we have
$$
\begin{aligned}
&{\VViint}_{\!\!\!\varphi^{-1}\circ\gamma(\tilde{\Omega})}{\varphi^*}{\mathfrak{v}}\\
&\qquad\qquad=\int_{\rod^n}d{\vartheta}
\bigg[\int_{\Omega}dq\, \sdet J({\varphi^{-1}}\circ\gamma)(q,{\vartheta})(\sdet J(\varphi)(y,\omega){\cdot}u(\varphi(y,\omega))%
\bigg|_{(y,\omega)=\varphi^{-1}\circ\gamma(q,{\vartheta})}\bigg]\\
&\qquad\qquad=\int_{\rod^n}d{\vartheta}
\bigg[\int_{\Omega}dq\, \sdet J(\gamma)(q,{\vartheta}){\cdot}u(\gamma(q,{\vartheta}))\bigg]
={\VViint}_{\!\!\!\gamma(\tilde{\Omega})}{\mathfrak{v}}.
\end{aligned}
$$
we have the claim. $\qquad /\!\!/$

Now, we interpret  \eqref{10-2-25-4.9} as change of variables:
Since we may denote integrals as
$$
{\VViint}_{\gamma(\tilde{\Omega})}{\mathfrak{v}}={\VViint}_{\mathfrak{M}} dxd\theta\, u(x,\theta),
$$
and
$$
{\VViint}_{\varphi^{-1}\circ\gamma(\tilde{\Omega})}\varphi^*{\mathfrak{v}}
={\VViint}_{\varphi^{-1}\mathfrak{M}} dyd\omega\,\sdet J(\varphi)(y,\omega) u(\varphi(y,\omega)),
$$
we have
$$
{\VViint}_{\mathfrak{M}} dxd\theta\, u(x,\theta)
={\VViint}_{\varphi^{-1}\mathfrak{M}} dyd\omega\,\sdet J(\varphi)(y,\omega) u(\varphi(y,\omega)). \qquad\qed
$$

\begin{remark}
It is fair to say that new definitions without inconsistency for CVF are introduced M.J. Rothstein~\cite{roth87} or M.R. Zirnbauer~\cite{zir96}, but they are not so easy to understand at least for me.
\end{remark}

\paragraph{\bf Resolution of  inconsistency:} 
Here, we resolve the inconsistency derived from the naive definition of Berezin integral by applying 
modified Rogers, Vladimirov and Volovich's definition above.
\par
\underline{Resolution of  inconsistency in Example~\ref{Example1}}:
From Theorem \ref{CVF-VV}, we interpret as follows:
\par
For $\Omega=(0,1)$, we are given $\widetilde\Omega=\Omega\times\rod^2$, Defining a map
$\gamma:\widetilde{\Omega}\to{\mathfrak{M}}$ as
$$
\gamma:\widetilde\Omega\ni(q,{\vartheta})\to(x,\theta)=(\gamma_{\bar{0}}(q,{\vartheta}), \gamma_{\bar{1}}(q,{\vartheta}))=\gamma(q,{\vartheta}),
$$
then we may regard ${\mathfrak{M}}=\{(x,\theta)\in{\fR}^{1|2}\;\big|\; \pi_{\mathrm{B}}(x)\in\Omega,\;\theta\in\rod^2\}$ as
a foliated singular manifold $\gamma(\widetilde{\Omega})$ in ${\fR}^{1|2}$.
We are given another foliated singular manifold ${\mathfrak{N}}=\delta(\widetilde{\Omega})$ in ${\fR}^{1|2}$ such that
they are super-differentiably isomorphic
$$
\varphi:\delta(\widetilde\Omega)\ni(y,\omega)\to\varphi(y,\omega)=(x,\theta)\in\gamma(\widetilde\Omega),
$$
with
$$
\begin{cases}
x=\varphi_{\bar{0}}(y,\omega)=y+\omega_1\omega_2\phi(y),\\
\theta_1=\varphi_{\bar1,1}(y,\omega)=\omega_1,\, \theta_2=\varphi_{\bar1,2}(y,\omega)=\omega_2,
\end{cases}
$$ 
and moreover
$$
\delta=\varphi^{-1}\circ\gamma:(q,{\vartheta})\to (q-{\vartheta}_1{\vartheta}_2\phi(q),{\vartheta})
=(\delta_{\bar{0}}(q,{\vartheta}),\delta_{\bar{1}}(q,{\vartheta}))=(y,\omega).
$$
Then, ${\mathfrak{N}}=\varphi^{-1}({\mathfrak{M}})$ and
$$
\begin{gathered}
J(\varphi)(y,\omega)=\begin{pmatrix}
1+\omega_1\omega_2\phi'(y)&0&0\\
\omega_2\phi(y)&1&0\\
-\omega_1\phi(y)&0&1
\end{pmatrix},\;\;
J(\gamma)(q,{\vartheta})=
\begin{pmatrix}
1&0&0\\
0&1&0\\
0&0&1
\end{pmatrix},\\
J(\delta)(q,{\vartheta})=\begin{pmatrix}
1-{\vartheta}_1{\vartheta}_2\phi'(q)&0&0\\
-{\vartheta}_2\phi(q)&1&0\\
{\vartheta}_1\phi(q)&0&1
\end{pmatrix}.
\end{gathered}
$$
\par
In this case, for $u(x,\theta)=u_{\bar{0}}(x)+\theta_1\theta_2u_{\bar{1}}(x)$, we have
$$
\begin{aligned}
\VViint_{\!\!\!\mathfrak{M}} dxd\theta\, u(x,\theta)
&=\int_{\rod^2} d{\vartheta}\,\bigg[\int_\Omega dq \, {\sdet}J(\gamma)(q,{\vartheta})u(\gamma(q,{\vartheta}))\bigg]\\
&=\int_0^1dq\,\int_{\rod^2} d{\vartheta}\, u(q,{\vartheta})
=\int_0^1 dq\,
\frac{\partial}{\partial{\vartheta}_2}\frac{\partial}{\partial{\vartheta}_1}u(q,{\vartheta})
\bigg|_{{\vartheta}=0}\\
&=\int_0^1 dq\,u_{\bar{1}}(q),
\end{aligned}
$$
and
\begin{equation}
\begin{aligned}
\VViint_{\!\!\!{\mathfrak{N}}}dyd\omega&\,(\varphi^*u)(y,\omega)\\
&=\iint_{\widetilde{\Omega}} dq d{\vartheta}\, \sdet J(\delta)(q,{\vartheta})
\big[\sdet J(\varphi)(y,\omega)u(\varphi(y,\omega))\big]_{(y,\omega)=\delta(q,{\vartheta})}\\
&=\int_0^1dq\,\bigg[\int_{\rod^2} d{\vartheta}\,\sdet J(\delta)(q,{\vartheta})
\big[\sdet J(\varphi)(y,\omega)u(\varphi(y,\omega))\big]_{(y,\omega)=\delta(q,{\vartheta})}\bigg]\\
&=\int_0^1dq\,\int_{\rod^2} d{\vartheta}\, u(q,{\vartheta}). 
\end{aligned}
\label{VV-C}
\end{equation}
Therefore, without any condition on support of integrand $u$, we get the following:
$$
\begin{aligned}
\VViint_{\!\!\!\mathfrak{M}} dxd\theta\,  &u(x,\theta)\\
&=\iint_{\widetilde{\Omega}}dq d{\vartheta}\,\sdet J(\gamma)(q,{\vartheta}){\cdot}u(\gamma(q,{\vartheta}))\\
&=\iint_{\widetilde{\Omega}}dq d{\vartheta}\,
\sdet J(\delta)(q,{\vartheta})\big[\sdet J(\varphi)(y,\omega){\cdot}u(\varphi(y,\omega))\big]\bigg|_{(y,\omega)=\delta(q,{\vartheta})}\\
&=\VViint_{\!\!\!\varphi^{-1}({\mathfrak{N}})}dyd\omega\,\sdet J(\varphi)(y,\omega){\cdot}u(\varphi(y,\omega))\\
&=\VViint_{\!\!\!{\mathfrak{N}}}dyd\omega\,(\varphi^*u)(y,\omega).\qquad\qed
\end{aligned}
$$

\begin{remark}
In order to recognize this phenomena and for future use, we calculate more precisely (please sensitive to the underlined parts):
\begin{equation}
\begin{aligned}
\sdet J(\varphi)(y,\omega){\cdot}&u(\varphi(y,\omega))\\
&=(1+\omega_1\omega_2\phi'(y))[u_{\bar{0}}(y+\omega_1\omega_2\phi(y))+\omega_1\omega_2u_{\bar{1}}(y+\omega_1\omega_2\phi(y))]\\
&=(1+\omega_1\omega_2\phi'(y))[u_{\bar{0}}(y)+\omega_1\omega_2(\phi(y)u_{\bar{0}}'(y)+u_{\bar{1}}(y))]\\
&=u_{\bar{0}}(y)+\omega_1\omega_2[\underline{(\phi(y)u_{\bar{0}}(y))'}+u_{\bar{1}}(y)],
\end{aligned}
\label{VV-in}
\end{equation}
and putting $(y,\omega)=\delta(q,{\vartheta})$, then we have
\begin{equation}
\begin{aligned}
\sdet J(\delta)(q,{\vartheta})
&\big[\sdet J(\varphi)(y,\omega){\cdot}u(\varphi(y,\omega))\big]_{(y,\omega)=\delta(q,{\vartheta})}\\
&=(1-{\vartheta}_1{\vartheta}_2\phi'(q))\big(u_{\bar{0}}(y)+\omega_1\omega_2[(\phi(y)u_{\bar{0}}(y))'+u_{\bar{1}}(y)]
\big)\big|_{\scriptstyle{y=q-{\vartheta}_1{\vartheta}_2\phi(q),}\atop\scriptstyle{\omega_1={\vartheta}_1,\,\omega_2={\vartheta}_2}}\\
&={\underline{(1-{\vartheta}_1{\vartheta}_2\phi'(q))}}\big[{\underline{u_{\bar{0}}(q)-{\vartheta}_1{\vartheta}_2\phi(q)u_{\bar{0}}'(q)}}+{\vartheta}_1{\vartheta}_2
[(\phi(q)u_{\bar{0}}(q))'+u_{\bar{1}}(q)]\big]\\
&=u_{\bar{0}}(q)+{\underline{{\vartheta}_1{\vartheta}_2}}[(\phi(q)u_{\bar{0}}(q))'+u_{\bar{1}}(q){\underline{-(\phi(q)u_{\bar{0}}(q))'}}]
=u_{\bar{0}}(q)+{\vartheta}_1{\vartheta}_2u_{\bar{1}}(q).
\end{aligned}
\end{equation}
Or, since $u(\varphi(y,\omega))_{(y,\omega)=\delta(q,{\vartheta})}=u(q,{\vartheta})$ and
$$
\sdet J(\delta)(q,{\vartheta}){\cdot}\sdet J(\varphi)(\delta(q,{\vartheta}))=(1-{\vartheta}_1{\vartheta}_2\phi'(q))(1+{\vartheta}_1{\vartheta}_2\phi'(q))=1,
$$
we get the result.
\par
From these, one reason of inconsistent term in \eqref{VV-in} comes from $\omega_1\omega_2(\phi(y)u_{\bar{0}}(y))'$.
\end{remark}

\par
\underline{Resolution of  inconsistency in Example~\ref{Example2}}:
[An incosistency derived from the diagonalization of matrix $Q$ mentioned in Chapter 3]
From \eqref{LAC2}, we define a path $\gamma$ from $(q,\eta)$ to $(x,\theta)$
and $\tilde\gamma$ from
$(q,\eta)$ to $(y,\omega)$
$$
(x,\theta)=(\gamma_{\bar{0}}(q,\eta),\gamma_{\bar{1}}(q,\eta))=\gamma(q,\eta)
\et
(y,\omega)=\tilde\gamma(q,\eta)=\varphi^{-1}\circ\gamma(q,\eta).
$$
Then, 
$$
\left\{
{\begin{aligned}
&y_1=q_1+\frac{\eta_1\eta_2}{q_1-iq_2},\;
y_2=q_2-\frac{i\eta_1\eta_2}{q_1-iq_2},\\
&\omega_1=-i\frac{\eta_1}{q_1-iq_2},\;
\omega_2=i\frac{\eta_2}{q_1-iq_2},
\end{aligned}}\right.
$$
we have 
$$
J(\tilde\gamma)(q,\eta)=\begin{pmatrix}
1-\eta_1\eta_2(q_1-iq_2)^{-2}&i\eta_1\eta_2(q_1-iq_2)^{-2}&i\eta_1(q_1-iq_2)^{-2}&-i\eta_2(q_1-iq_2)^{-2}\\
i\eta_1\eta_2(q_1-iq_2)^{-2}&1+\eta_1\eta_2(q_1-iq_2)^{-2}&\eta_1(q_1-iq_2)^{-2}&-\eta_2(q_1-iq_2)^{-2}\\
\eta_2(q_1-iq_2)^{-1}&-i\eta_2(q_1-iq_2)^{-1}&-i(q_1-iq_2)^{-1}&0\\
-\eta_1(q_1-iq_2)^{-1}&i\eta_1(q_1-iq_2)^{-1}&0&i(q_1-iq_2)^{-1}
\end{pmatrix}
$$
 and $\sdet J(\tilde\gamma)(q,\eta)=-(q_1-iq_2)^{-2}$, therefore
$$
\begin{aligned}
&\iint dQ\, e^{-(x_1^2+x_2^2-2\theta_1\theta_2)}=\iint dqd\eta\,\sdet J(\gamma)(q,\eta)e^{-(q_1^2+q_2^2-2\eta_1\eta_2)},\\
&\iint d\tilde{Q}\,e^{-(y_1^2+y_2^2)}=\iint dqd\eta\,\sdet J(\tilde\gamma)(q,\eta)\bigg(\sdet(\varphi)(y,\omega)\,e^{-(y_1^2+y_2^2)}\bigg|_{(y,\omega)=\tilde{\gamma}(q,\eta)}\bigg).
\end{aligned}
$$
By this calculation, we have no inconsistency. 

\section[Supersymmetric transformation in superspace ]{Supersymmetric transformation in superspace -- as an example of change of variables}
Stimulated probably from the success of QED, Berezin and Marinov claim in their paper~\cite{BM77}  
``\underline{Treat bosons and fermions on equal footing}'', as mentioned before.
As the object of this lecture notes, we give a basic idea of this ``equal footing''.
To do this, we need to develop an integration theory which admits the change of variables under integral sign.
Especially we use transformations of mixing even and odd variables.

In the following, we assume $f_a(x_{\mathrm{B}})\in\mathbb{C}$, that is,$f_a(x)\in\cev$, therefore, we denote also $\theta^a f_a(x)$ by $f_a(x)\theta^a$.

Let $x=(x_1,{\cdots},x_m)\in  \euc^m\subset {\fR}^{m|0}$ and $({\theta}_1,{\theta}_2)\in \cod^2$.
Taking ${\varepsilon}\in\cod$, $\gamma>0$ and ${\lambda},{\mu}\in \euc^m\subset {\fR}^{m|0}$, we define transformation $\tau({\lambda},{{\mu}})$ as
$$
\left\{
\begin{aligned}
&x\longrightarrow y=x+2{{\mu}}{\varepsilon}{\theta}_1+2{\lambda}{\varepsilon}{\theta}_2,\;\mbox{that is,}\;
y_j=x_j+2{{\mu}}{\varepsilon}{\theta}_1+2{\lambda}_j{\varepsilon}{\theta}_2\\
&{\theta}_1\longrightarrow \omega_1={\theta}_1+\gamma {\lambda}{\cdot}x{\varepsilon},\with
\;\mbox{${\lambda}{\cdot}x=\sum_{j=1}^m\lambda_jx_j$}\;\\
&{\theta}_2\longrightarrow \omega_2={\theta}_2-\gamma{{\mu}}{\cdot}x{\varepsilon}
\end{aligned}
\right.
$$
By the definition of Grassmann continuation, for any $f\in C^{\infty}(\euc^m:{\mathbb{C}})$, we get
$$
f(x+2{{\mu}}{\varepsilon}{\theta}_1+2{\lambda}\omega{\theta}_2)=f(x)+2{\mu}{\cdot}\nabla f{\varepsilon}{\theta}_1+2{\lambda}{\cdot}\nabla f{\varepsilon}{\theta}_2,\with
{\mu}{\cdot}\nabla=\sum_{j=1}^m\mu_j\frac{\partial}{\partial x_j}.
$$
Therefore, (for $a=(a_1,a_2)\in\{0,1\}^2$, putting $\bar{0}=(0,0), \bar{1}=(1,0), \bar{2}=(0,1), \bar{3}=(1,1)$),
for 
$$
u(y,\omega)=\sum_{|a|\le 2}u_a(y)\omega^a=u_{\bar{0}}(y)+u_{\bar{1}}(y)\omega_1+u_{\bar{2}}(y)\omega_2+u_{\bar{3}}(y)\omega_1\omega_2,
$$
remarking
$$
({\theta}_1+\gamma {\lambda}{\cdot}x{\varepsilon})({\theta}_2-\gamma{{\mu}}{\cdot}x{\varepsilon})
={\theta}_1{\theta}_2+\gamma {\lambda}{\cdot}x{\varepsilon}{\theta}_2-{\theta}_1\gamma{{\mu}}{\cdot}x{\varepsilon}
$$
and calculating slightly, we have
\begin{equation}
\begin{aligned}
(\tau^*({\lambda},{{\mu}})u)(x,\theta)&=u(\tau({\lambda},{{\mu}})(x,\theta))\\
&=u(x,\theta)+[(\gamma {\lambda}{\cdot}x\,u_{\bar{1}}-\gamma{\mu}{\cdot}x\,u_{\bar{2}}(x))\\
&\qquad
+(-2\nabla u_{\bar{0}}(x){\cdot}{\mu}+\gamma{\mu}{\cdot}xu_{\bar{3}}(x)){\theta}_1
+(-2\nabla u_{\bar{0}}(x){\cdot}{\lambda}+\gamma{\lambda}{\cdot}xu_{\bar{3}}(x)){\theta}_2\\
&\qquad\qquad
+2(\nabla u_{\bar{1}}(x){\cdot}{\lambda}-\nabla u_{\bar{2}}(x){\cdot}{\mu}){\theta}_2{\theta}_1]\varepsilon.
\end{aligned}
\label{KLP4.2}
\end{equation}

\begin{definition}
A function $u\in\ccsl_{\mathrm{SS}}({\fR}^{m|2}:{\fC})$ is called supersymmetric, if for any ${\lambda},{\mu}\in \euc^m\subset {\fR}^{m|0}$, it satisfies
$$
(\tau^*({\lambda},{\mu})u)(x,\theta)=u(x,\theta)(=(\tau^*({0},{0})u)(x,\theta)).
$$
\end{definition}
\begin{proposition}[Proposition 4.1 of KLP~\cite{KLP84}]
Following conditions are equivalent  for $u\in\ccsl_{\mathrm{SS}}({\fR}^{m|2}:{\fC})$:
\newline
(i) $u\in\ccsl_{\mathrm{SS}}({\fR}^{m|2}:{\fC})$ is supersymmetric.
\newline
(ii) $u_{\bar{1}}(x)=u_{\bar{2}}(x)=0$ and moreover
\begin{equation}
\frac{2}{\gamma}\nabla u_{\bar{0}}(x)=x u_{\bar{3}}(x),\;\mbox{i.e.}\; \frac{2}{\gamma}\frac{\partial}{\partial x_j}u_{\bar{0}}(x)=x_ju_{\bar{3}}(x).
\label{KLP4.3}
\end{equation}
\newline
(iii) There exists a function $\phi({\cdot})\in C^{\infty}([0,{\infty}):{\mathbb{C}})$ satisfying
$$
u(x,\theta)=\phi\bigg(x^2-\frac{4}{\gamma}{\theta}_1{\theta}_2\bigg)=\phi(x^2)-\frac{4}{\gamma}\phi'(x^2){\theta}_1{\theta}_2.
$$
\end{proposition}

{\it Proof. }
[$(i)\Longrightarrow (ii)$] If $u$ is supersymmetric, then the coefficient of $\varepsilon$ of the right-hand side of \eqref{KLP4.2} should be $0$ for any ${\lambda}, {\mu}\in \euc^m$. This implies \eqref{KLP4.3}.
\newline
[$(ii)\Longrightarrow (iii)$] Restricting \eqref{KLP4.3} to $\euc^m$,  we get  that $u_{\bar{0}}(q)$ depends only on $|q|^2=q{\cdot}q$, that is, there exists a function $\phi({\cdot})\in C^{\infty}([0,{\infty}):{\mathbb{C}})$ such that $u_{\bar{0}}(q)=\phi(|q|^2)$. Since the derivative of the Grassmann continuated function equals to the Grassmann continuated of the derivative, therefore (iii) follows.
\newline
[$(iii)\Longrightarrow (i)$] Obvious. $\qquad\square$

\begin{proposition}
Let $u\in\ccsl_{\mathrm{SS}}({\fR}^{m|2}:{\fC})$ with $u_a({\cdot})$ is integrable for each $a$.
Then, for any $\tau=\tau({\lambda},{\mu})$, we have
$$
\int_{{\fR}^{m|2}} dx d{\theta}_2 d{\theta}_1 \, (\tau^*u)(x,{\theta}_1,{\theta}_2)=\int_{{\fR}^{m|2}} dx d{\theta}_2 d{\theta}_1  \, u(x,{\theta}_1,{\theta}_2).
$$
\end{proposition}

{\it Proof. }
Integrating w.r.t. $\theta$, we have
$$
\begin{aligned}
&\int_{{\fR}^{m|2}} dx d{\theta}_2 d{\theta}_1  \, u(x,{\theta}_1,{\theta}_2)=-\int_{{\fR}^{m|0}}dx\,u_{\bar{3}}(x),\\
&\int_{{\fR}^{m|2}} dx d{\theta}_2 d{\theta}_1 \, (\tau^*u)(x,{\theta}_1,{\theta}_2)=-\int_{{\fR}^{m|0}}dx\,u_{\bar{3}}(x)
-2\int_{{\fR}^{m|0}}dx\,(\nabla u_{\bar{1}}(x){\cdot}{\lambda}-\nabla u_{\bar{2}}(x){\cdot}{\mu})\varepsilon.
\end{aligned}
$$
On the other hand, from integrability,
$$
\int_{{\fR}^{m|0}}dx\,\nabla u_{\bar{j}}(x)=0,\quad {\bar{j}}=\bar{1}, \bar{2}
$$
we get the result. $\qquad\square$

\begin{lemma}
Let $u\in\ccsl_{\mathrm{SS}}({\fR}^{2|2}:{\fC})$ be supersymmetric and integrable. Then,
$$
\int_{{\fR}^{2|2}} dx d{\theta}_2 d{\theta}_1 \, u(x,{\theta}_1,{\theta}_2)=\frac{4\pi}{\gamma}u_{\bar{0}}(0).
$$
\end{lemma}

{\it Proof. } From previous Proposition, there exists a function $\phi({\cdot})\in C^{\infty}([0,{\infty}):{\mathbb{C}})$ such that  
$u(x,{\theta}_1,{\theta}_2)=\phi(x^2+\frac{4}{\gamma}\bartheta\theta)$,
and especially $u_{\bar{0}}(x)=\phi(x^2)$. Since this is integrable, it implies
$\lim_{t\to{\infty}}\phi(t)=0$  and
$$
\begin{aligned}
\int_{{\fR}^{2|2}} dx d{\theta}_2 d{\theta}_1 \, u(x,{\theta}_1,{\theta}_2)&
=-\frac{4}{\gamma}\int_{{\fR}^{2|0}}dx\,\phi'(x^2)\\
&=-\frac{8}{\gamma}\int_0^{\infty}rdr\,\phi'(r^2)=-\frac{4}{\gamma}\int_0^{\infty}ds\,\phi'(s)
=\frac{4}{\gamma}\phi(0)=\frac{4}{\gamma}u_{\bar{0}}(0).\qquad\qed
\end{aligned}
$$

\chapter[Efetov's method in RMT and beyond] {Efetov's method in Random Matrix Theory and beyond} 

In 1950s, physicists get so many experimental data which are obtained after making neutron collided with Uranium 238, etc. and checking these data, they are embarrassed so much to grasp the meaning of them. At that time, E. Wigner poses a working hypothesis or Ansatz that high resonance of these experiments behaves like eigenvalues of large size matrix. 
In other word,  we quote from M.L. Mehta~\cite{Meh91} 
\begin{quotation}
Consider a large matrix whose elements are random variables with given probability laws. Then, what can one say about the probabilities of a few of its eigenvalues or of a its eigenvectors?
\end{quotation}
We give another quotation from Y.V. Fyodorov~\cite{fyo94}:
\begin{quotation}
Wigner suggested that fluctuations in positions of compound nuclei resonances can be described in terms of statistical properties of eigenvalues of very large real symmetric matrices with independent, identically distributed entries. The rational behind such a proposal was the idea that in the situation when it is hardly possible to understand in detail individual spectra associated with any given nucleus composed of many strongly interacting quantum particles, it may be reasonable to look at the corresponding systems as "black boxes" and adopt a kind of statistical description, not unlike thermodynamics approach to classical matter. 
\end{quotation}

We should remark that there exist mathematical works concerning the distribution of the zeros of the Riemann zeta function from this point of view.

In any way, in Random Matrix Theory(=RMT), Wigner's semi-circle law gives 
a well-known corner-stone.

We report mathematical refinements of
this law as an application of superanalysis.
That is, using Efetov's idea, we rewrite the average of the empirical measure
of the eigenvalue distribution of the Hermitian matrices in a compact form.
Careful calculations give not only the precise convergence rate of that law,
but also the precise rate of the edge mobility.

\begin{remark}
The usage of the method of steepest descent by physicists are not so mathematically rigorous in the sense of de Bruijn's criteria, because we have no general method of choosing the steepest descent path for the integral considered.
\end{remark}

\section{Results -- outline} 
Let ${\mathfrak{U}}_N$ be a set of Hermitian $N\times N$ matrices,
which is identified with ${\mathbb{R}}^{N^2}$ as a topological space.
In this set, we introduce a probability measure $d\mu_N({H})$ by 
\begin{equation}
\begin{gathered}
d\mu_N({H})=
\prod_{k=1}^N d(\Re H_{kk})\prod_{j<k}^N d(\Re H_{jk}) d(\Im H_{jk})
P_{N,J}(H),\\
P_{N,J}(H)=Z^{-1}_{N,J}\exp\big[-{\frac{N}{2J^2}}\tr {H}^*{H}\big]
\end{gathered}
\label{wig1}
\end{equation}
where ${H}=(H_{jk})$, 
$H^*=(H^*_{jk})=(\overline{H}_{kj})={}^t\overline{H}$,
$\prod_{k=1}^N d(\Re H_{kk})\prod_{j<k}^N d(\Re H_{jk}) d(\Im  H_{jk})$ 
being the Lebesgue measure on ${\mathbb{R}}^{N^2}$,
and $Z^{-1}_{N,J}$ is the normalizing constant given by
$Z_{N,J}=2^{N/2}(J^2\pi/N)^{3N/2}$.

Let $E_\alpha=E_\alpha({H})$ ($\alpha=1,\cdots,N$) 
be real eigenvalues of ${H}\in{{\mathfrak{U}}_N}$.

We put
\begin{equation}
\rho_N(\lambda)=\rho_N(\lambda;{H})=
N^{-1}\sum_{\alpha=1}^N \delta(\lambda-E_\alpha(H)),
\label{wig2}
\end{equation}
where $\delta$ is the Dirac's delta. Denoting for a function $f$ on ${\mathfrak{U}}_N$,
$$
\big\langle f\big\rangle_N=\big\langle f(\cdot)\big\rangle_N
=\int_{{\mathfrak{U}}_N}d\mu_N({H})\,f({H}),
$$
we get
\begin{theorem}[Wigner's semi-circle law]
\begin{equation}
\lim_{N\to\infty}\big\langle \rho_N(\lambda)\big\rangle_N=w_{sc}(\lambda)=
\begin{cases}
(2{\pi}J^2)^{-1}\sqrt{4J^2-\lambda^2}&\for |\lambda|<2J,\\
0 & \for |\lambda|>2J.
\end{cases}
\label{wig3}
\end{equation}
\end{theorem}
{\it Remark. }
By definition,
the limit \eqref{wig3} is interpreted as
$$
\lim_{N\to\infty}
\langle\phi(\cdot),\int_{{\mathfrak{U}}_N}d\mu_N({H})\,
N^{-1}\sum_{\alpha=1}^N\delta(\cdot-E_\alpha(H))\rangle
=\langle\phi, w_{sc}\rangle
=\int_{\euc}d\lambda\,\phi(\lambda) w_{sc}(\lambda)
$$
for any $\phi\in C_0^\infty({\euc})={\mathcal{D}}({\euc})$.
$\langle\cdot,\cdot\rangle$ stands for the duality between ${\mathcal{D}}(\euc)$
and ${\mathcal{D}}'(\euc)$.
We need more interpretation to give the meaning to 
$\int_{{\mathfrak{U}}_N}d\mu_N({H})\,
N^{-1}\sum_{\alpha=1}^N\delta(\cdot-E_\alpha(H))$, which will be given in \S2???.
%\begin{comment}??%%%%%
Or, $H\to \rho_N(\lambda;H)d\lambda$ is considered a measure (on $\euc$)-valued
random variable on $\frak{U}_N$ and 
$\big\langle \rho_N(\lambda)\big\rangle_N d\lambda$ is considered a family of
probability measures on $\euc$ which is tight.
%??\end{comment}%%%%%

Though there exist several methods to prove this fact,
we explain a new derivation of this fact
using odd variables obtained by K. B. Efetov \cite{efe83}.
We follow mainly Fyodorov \cite{fyo94}
and E. Br\'{e}zin \cite{bre83}
(see also, P.A. Mello \cite{mel94}, M.R. Zirnbauer \cite{zir96}).

Moreover, we get, as a byproduct of this new treatise,
\begin{theorem}[A refined version of Wigner's semi-circle law] 
For each $\lambda$ with $|\lambda|<2J$, when $N\to\infty$, we have 
\begin{equation}
\begin{aligned}
\big\langle \rho_N(\lambda)\big\rangle_N=&
\frac{\sqrt{4J^2-\lambda^2}}{2\pi J^2}\\
&-\frac{(-1)^NJ}{\pi(4J^2-\lambda^2)}
\cos(N[\frac{\lambda\sqrt{4J^2-\lambda^2}}{2J^2}
+2\arcsin(\frac{\lambda}{2J})])N^{-1}
+O(N^{-2}).
\end{aligned}
\label{wig4}
\end{equation}
\par
When $\lambda$ satisfies $|\lambda|>2J$,
there exist constants $C_{\pm}(\lambda)>0$
and $k_{\pm}(\lambda)>0$ such that
\begin{equation}
\bigg|\big\langle \rho_N(\lambda)\big\rangle_N\bigg|
\le C_{\pm}(\lambda)\exp{[-k_{\pm}(\lambda)N]}
\label{wig5}
\end{equation}
with $k_{\pm}(\lambda)\to0$ and $C_{\pm}(\lambda)\to\infty$ 
for $\lambda\searrow 2J$ or $\lambda\nearrow -2J$, respectively.
\end{theorem}
\begin{theorem}[The spectrum edge problem] 
Let $z\in[-1,1]$. We have
\begin{equation}
\begin{aligned}
&\big\langle \rho_N(2J-zN^{-2/3})\big\rangle_N=
N^{-1/3}f(z/J)+O(N^{-2/3})\qquad \mbox{as $N\to\infty$},\\
&\big\langle \rho_N(-2J+zN^{-2/3})\big\rangle_N=
-N^{-1/3}f(z/J)+O(N^{-2/3})\qquad\mbox{as $N\to\infty$},
\end{aligned}
\label{wig6}
\end{equation}
where 
$$
f(w)=\frac{1}{4\pi^2 J}(\Ai'(w)^2-\Ai^{\prime\prime}(w)\Ai(w)),
\quad
\Ai(w)=\int_{\euc}dx\,\exp{[-\frac{i}{3}x^3+iwx]}.
$$
\end{theorem}

(A) One of the key expression 
obtained by \underline{introducing new auxiliary variables}, is
\begin{equation}
\big\langle \rho_N(\lambda)\big\rangle_N=\pi^{-1}\Im
\int_{\mathfrak{Q}} dQ\,\big(\{(\lambda-i0) {\mathbb{I}}_2-Q\}^{-1}\big)_{bb}
\exp{[-N{\mathcal{L}(Q)}]}
\label{wig7}%??{wig4}
\end{equation}
where ${\mathbb{I}}_n$ stands for $n\times n$-identity matrix and
\begin{equation}
\begin{gathered}
{\mathcal{L}(Q)}=\str[(2J^2)^{-1}Q^2+\log((\lambda-i0){\mathbb{I}}_2-Q)],\\
{\mathfrak{Q}}=\big\{Q=\begin{pmatrix}
x_1&\rho_1\\
\rho_2&ix_2
\end{pmatrix}\,\big|\, x_1, x_2\in{\rev},\; \rho_1,\rho_2\in{\rod}\big\}
\cong{\mathfrak{R}}^{2|2},
\;\;
dQ=\frac{dx_1 dx_2}{2\pi} d\rho_1 d\rho_2,\\
\big(((\lambda-i0){\mathbb{I}}_2-Q)^{-1}\big)_{bb}
=\frac{(\lambda-i0-x_1)(\lambda-i0-ix_2)+\rho_1\rho_2}
{(\lambda-i0-x_1)^2(\lambda-i0-ix_2)}.
\end{gathered}
\label{wig8}
\end{equation}
Here in \eqref{wig4}, \underline{the parameter $N$ appears only in one place}.
This formula is formidably charming but \underline{not yet directly justified},
like Feynman's expression
of certain quantum objects applying his notorious measure. % notorious

(B) In physics literatures, for example in \cite{fyo94}, \cite{zir96}, 
they claim without proof that they may apply 
the method of steepest descent to \eqref{wig7} when $N\to\infty$. 
More precisely, as
$$
\delta{\mathcal{L}}(Q)\tilde{Q}
=\frac{d}{d\epsilon}{\mathcal{L}}(Q+\epsilon\tilde{Q})\bigg|_{\epsilon=0},
$$
they seek solutions of
$$
\delta{\mathcal{L}}(Q)=\str \big(\frac{Q}{J^2}-\frac{1}{\lambda-Q}\big)=0.
$$
As a candidate of effective saddle points, they \underline{take}\footnote{for what reason?}
$$
Q_c=(\frac12\lambda+\frac12\sqrt{\lambda^2-4J^2}){\mathbb{I}}_2,
$$
and they have
$$
\lim_{N\to\infty}\big\langle \rho_N(\lambda)\big\rangle_N
=\pi^{-1}\Im (\lambda-Q_c)^{-1}_{bb}=w_{sc}(\lambda).  \qquad\qed
$$

\begin{problem}
Not only the expression \eqref{wig7}
nor the applicability of the saddle point method to it
are not so clear. 
Though these formulas obtained are not yet well-defined mathematically but so charming. Unfortunately for the time being, we mathematicians may not use them directly. For example,  the expression \eqref{wig7} is mathematically verified only for $\lambda-i\epsilon\;(\epsilon>0)$, but we need probably some new integration theory which admits taking limit $\epsilon\to 0$ under integral sign. Can we justify the physicist procedures by using this new integration theory?
More precisely, under what condition, do we have the following equality?
\begin{equation}
\lim_{N\to\infty}\lim_{\epsilon\to0}
\big\langle\frac1N\tr \frac{1}{(\lambda-i\epsilon){\mathbb{I}}_N-H}\big\rangle_N
=\lim_{\epsilon\to0}\lim_{N\to\infty}
\big\langle\frac1N\tr \frac{1}{(\lambda-i\epsilon){\mathbb{I}}_N-H}\big\rangle_N.
\end{equation}
If this assertion is true, may we justify the physicists argument
of ``saddle point method"?
\end{problem}

\begin{remark}
(i) For mathematical rigour,
we \underline{dare to loose such a beautiful expression}
like \eqref{wig7},
but we have the two formulae \eqref{wig2.17} and \eqref{wig2.18} below
which lead to our conclusion.\\
(ii) It is not so simple even in the integral on $\euc^m$ to apply the saddle point method, that is, to take an appropriately deformed ``path'' in ${\mathbb{C}}^m$,
as is explained in de Bruijn~\cite{deB61}.\\
(iii) The set $({\mathfrak{U}}_N,\;d\mu_N(\cdot))$ is called 
GUE=the Gaussian Unitary Ensemble.
Other ensembles may be treated analogously as indicated in \cite{zir96} but
are not treated here.
\end{remark}

\section[The derivation of (6.1.7) with $\lambda-{i}\epsilon$] {The derivation of \eqref{wig7} with $\lambda-i\epsilon$($\epsilon>0$ fixed) and its consequences}

It is well-known that
$$
\delta(q)=\frac{1}{\pi}\lim_{\epsilon\to0}\Im \frac{1}{q-i\epsilon}
=\frac{1}{2\pi{i}}\lim_{\epsilon\to0}
\bigg[\frac{1}{q-i\epsilon}-\frac{1}{q+i\epsilon}\bigg]
=\frac{1}{\pi}\lim_{\epsilon\to0}\frac{\epsilon}{q^2+\epsilon^2}
\quad\mbox{in ${\mathcal{D}}'(\euc)$},
$$
that is,
for any $\phi\in C_0^\infty(\euc)$,
$$
\pi^{-1}\Im\int_{\euc}dq \, \frac {\phi(q)}{q-i\epsilon}
=\pi^{-1}\int_{\euc}dq \, \frac {\epsilon \phi(q)}{q^2+\epsilon^2}
=\pi^{-1}\int_{\euc}dq \, \frac {\phi(\epsilon q)}{1+q^2}\to \phi(0)
=\langle\phi,\delta\rangle
\quad\text{as $\epsilon\to 0$}.
$$

Therefore, for any fixed $\phi\in C_0^\infty(\euc)$, we have
$$
\begin{aligned}
\int_{{\mathfrak{U}}_N} & d\mu_N(H)
\langle\phi(\cdot),\frac1N\sum_{\alpha=1}^N\delta(\cdot-E_\alpha(H))\rangle
{\overset{\mathrm{def}}{=}}
\int_{{\mathfrak{U}}_N}d\mu_N(H)
\lim_{\epsilon\to0}\int_\euc d\lambda\,\phi(\lambda)
\frac1{\pi N}\sum_{\alpha=1}^N
\frac{\epsilon}{(\lambda-E_\alpha(H))^2+\epsilon^2}\\
=&\lim_{\epsilon\to0}\int_{{\mathfrak{U}}_N}d\mu_N(H)
\int_\euc d\lambda\,\phi(\lambda)\frac1{\pi N}\sum_{\alpha=1}^N
\frac{\epsilon}{(\lambda-E_\alpha(H))^2+\epsilon^2}
\quad\text{by Lebesgue's dom.conv.theorem}\\
=&\lim_{\epsilon\to0}\int_\euc d\lambda\,\phi(\lambda)
\int_{{\mathfrak{U}}_N}d\mu_N(H)\frac1{\pi N}\sum_{\alpha=1}^N
\frac{\epsilon}{(\lambda-E_\alpha(H))^2+\epsilon^2}
\quad\text{by Fubini's theorem.}\\
\end{aligned}
$$
The second equality is guaranteed by the fact that 
for any $\phi\in C_0^\infty(\euc)$, we have, 
for any $\epsilon>0$ and $H\in{\mathfrak{U}}_N$,
$$
\big|\int_\euc d\lambda\,\phi(\lambda)\frac1{\pi N}\sum_{\alpha=1}^N
\frac{\epsilon}{(\lambda-E_\alpha(H))^2+\epsilon^2}\big|
\le\max|\phi(\lambda)|.
$$
Here, we used the fact 
$\int_\euc d\lambda\,{\epsilon}{(\lambda^2+\epsilon^2)^{-1}}=\pi$.
The third equality holds because we have
$$
\big|\phi(\lambda)\frac1{\pi N}\sum_{\alpha=1}^N
\frac{\epsilon}{(\lambda-E_\alpha(H))^2+\epsilon^2}\big|
\le \epsilon^{-1}|\phi(\lambda)|\quad
\bigg(\because) \frac{N}{\pi N}\frac{\epsilon}{a^2+\epsilon^2}\le\frac{1}{\epsilon}\bigg),
$$
and the right hand side is integrable w.r.t. 
the product measure $d\lambda\,d\mu_N(H)$ for any fixed $\epsilon>0$.

In order to check whether we may take the limit 
before integration w.r.t. $d\lambda$ in the last line above,
we calculate the following quantity as explicitly as possible:
\begin{equation}
g(\lambda,\epsilon,N)=\int_{{\mathfrak{U}}_N}d\mu(H)
\frac{1}{\pi N}\Im\sum_{\alpha=1}^N
\frac{1}{\lambda-i\epsilon-E_\alpha(H)}.
\label{wig2b}
\end{equation}
We claim in this section that\\
(i) $g(\lambda,\epsilon,N)$ exists as a function of $\lambda$ 
for any $\epsilon>0$ and $N\in{\mathbb{N}}$ 
and\\
 (ii) $\lim_{\epsilon\to0}g(\cdot,\epsilon,N)$ exists in 
${\mathcal{D}}'(\euc)$ for any $N\in{\mathbb{N}}$ and 
it is denoted by $\big\langle\rho_N(\lambda)\big\rangle_N$.

Now, we put
$$
\begin{gathered}
z_j=x_j+iy_j,\;\overline{z}_j={x}_j-i{y}_j,\; 
x_j,\,y_j\in\rev;\;
\theta_k,\overline{\theta}_k\in{\rod}={\mathfrak{C}}_{\mathrm{od}},\\
X={}^t(z,\theta),\;z={}^t(z_1,\cdots,z_N),\;
\theta={}^t(\theta_1,\cdots,\theta_N),\\
X^*=(z^*,\theta^*),\;
z^*=(\overline{z}_1,\cdots,\overline{z}_N),\;
\theta^*=(\overline{\theta}_1,\cdots,\overline{\theta}_N).
\end{gathered}
$$
Here, $\theta_k$ and $\overline{\theta}_k$ 
are considered as two different odd variables.

The following is the key formula which is well known:
\begin{lemma} \label{lem:wig2.1}
Put $\mu=\lambda-i\epsilon$ $(\epsilon>0)$.
\begin{equation}
\begin{aligned}
\tr \frac{1}{{\mu} {\mathbb{I}}_N-H}
&=\sum_{\alpha=1}^N \frac{1}{{\mu}-E_\alpha(H)}\\
&=i\int_{{\fC}^{N|2N}}
\prod_{j=1}^N\frac{d\overline{z}_j\,dz_j}{2{\pi}i}
\prod_{k=1}^N d\overline{\theta}_k d\theta_k\,
(z^*\!\!\cdot\! z) 
\exp{[{-}iX^*( {\mathbb{I}}_2\otimes({\mu} {\mathbb{I}}_N-H))X]}.
\end{aligned}
\label{wig2-1}
\end{equation}
\end{lemma}

To prove this lemma, we need the following lemma.
\begin{lemma}\label{lem:wig2.2}
Let $\Gamma$=the diagonal matrix with diagonal given by
$(\gamma_1,\cdots,\gamma_N)$ where $\gamma_j\in\euc$.
Putting $(z^*\!\!\cdot\! z)=\sum_{j=1}^N\overline{z}_jz_j=|z|^2$, we have
\begin{equation}
i\int_{{\fC}^{N|2N}}\prod_{j=1}^N\frac{d\overline{z}_j\,dz_j}{2{\pi}i}
\prod_{k=1}^N d\overline{\theta}_k d\theta_k\,
(z^*\!\!\cdot\! z)\exp{[{-}iX^*( {\mathbb{I}}_2\otimes(\Gamma{-}i\epsilon {\mathbb{I}}_N))X]}
=\sum_{j=1}^N\frac{1}{\gamma_j{-}i\epsilon}.
\label{wig2-2}\end{equation}
\end{lemma}

{\it Proof. }
Identifying $z_j=x_j+iy_j$, $\overline{z}_j=x_j-iy_j$,
$d\overline{z}_j\wedge dz_j=2idx_j\wedge dy_j$, using the polar coordinates
$(r_j,\omega_j)\in \euc_+\times S^{1}$ and denoting $|z_j|^2=x_j^2+y_j^2=r_j^2$, $0\le \omega_j\le 2\pi$, we get
$$
\begin{aligned}
i\int_{{\fC}^{1|0}}\frac{d\overline{z}_j\,dz_j}{2{\pi}i}\,|z_j|^2e^{-i(\gamma_j-i\epsilon)|z_j|^2}
&=i\int_0^{\infty}\!\!\!\int_0^{2\pi} r_jdr_j d\omega_j\, r_j^2e^{-i(\gamma_j-i\epsilon)r_j^2}\\
&=\frac{i}{(\epsilon+i\gamma_j)^2}=\frac{1}{\epsilon+i\gamma_j}{\cdot}\frac{1}{\gamma_j-i\epsilon}.
\end{aligned}
$$
Analogously,
$$
\int_{{\fC}^{N-1|0}}\prod_{k\neq j}^N[\frac{d\overline{z}_k\,dz_k}{2{\pi}i}e^{-i(\gamma_k-i\epsilon)|z_k|^2}]
=\prod_{k\neq j}\frac{1}{\epsilon+i\gamma_k}=(\epsilon+i\gamma_j)\prod_{k=1}^N\frac{1}{\epsilon+i\gamma_k}.
$$

Remarking
$$
i\int_{{\fC}^{N|0}}\prod_{j=1}^N\frac{d\overline{z}_j\,dz_j}{2{\pi}i}\,
(z^*\!\!\cdot\! z)\exp{[{-}iz^*(\Gamma{-}i\epsilon  {\mathbb{I}}_N)z]}
={\bigg(}\sum_{j=1}^N\frac{1}{\gamma_j{-}i\epsilon}{\bigg)}
\prod_{j=1}^N\frac{1}{\epsilon{+}i\gamma_j},
$$
and
$$
\int_{\fC^{0|2N}}\prod_{k=1}^N d\overline{\theta}_k d\theta_k\,
\exp{[{-}i{\theta}^*(\Gamma{-}i\epsilon  {\mathbb{I}}_N)\theta]}
=\prod_{k=1}^N(\epsilon{+}i\gamma_k),
$$
we get the result \eqref{wig2-2} readily. $\qquad\qed$

{\it Proof of Lemma~\ref{lem:wig2.1}.}
By diagonalization of $\lambda  {\mathbb{I}}_N-H$, we reduce Lemma~\ref{lem:wig2.1} to Lemma ~\ref{lem:wig2.2}. 
In fact, taking $G$ such that $GG^*=G^*G={\mathbb{I}}_N$で$GHG^*=\Gamma$, and defining a change of variables
$$
\begin{aligned}
\tilde{z}=G^*z,\;\; \bar{\tilde{z}}=z^*G,\;\;\tilde{\theta}=G^*\theta,\;\; \bar{\tilde{\theta}}=\theta^*G,\;\;&
dz=d\tilde{z},\;\; 
d\bar{z}=d\bar{\tilde{z}},\;\; d\theta=d{\tilde{\theta}},\;\;d\bar{\theta}=d\bar{\tilde{\theta}},\\
(z^*,\theta^*)\begin{pmatrix}
G(\mu{\mathbb{I}}_N-H)G^*&0\\
0&G(\mu{\mathbb{I}}_N-H)G^*
\end{pmatrix}
\begin{pmatrix}
z\\
\theta
\end{pmatrix}
&=z^*G(\mu{\mathbb{I}}_N-H)G^*z+\theta^*G(\mu{\mathbb{I}}_N-H)G^*\theta\\
&=\tilde{z}^*(\mu{\mathbb{I}}_N-\Gamma)\tilde{z}+\tilde{\theta}^*(\mu{\mathbb{I}}_N-\Gamma)\tilde{\theta},
\end{aligned}
$$
we reduce it to the diagonal case 
(Even if we use the naive definition of integration, every linear transformation is permitted under integral sign, see \S1 of the last chapter). $\qquad\qed$

\begin{lemma} For $\mu=\lambda-i\epsilon$ $(\epsilon>0)$,
\begin{equation}
\begin{aligned}
\big\langle\tr \frac{1}{{\mu} {\mathbb{I}}_N-H}\big\rangle_N
=i\int\prod_{j=1}^N\frac{d\overline{z}_j\,dz_j}{2{\pi}i} &
\prod_{k=1}^N d\overline{\theta}_k d\theta_k\,
(z^*\!\!\cdot\! z) \exp{[{-}iX^*( {\mathbb{I}}_2\otimes{\mu} {\mathbb{I}}_N)X]}\\
&\qquad\times
\exp{[-\frac{J^2}{2N}
\sum_{j,k=1}^N(\overline{z}_jz_k+\overline{\theta}_j\theta_k)
(\overline{z}_kz_j+\overline{\theta}_k\theta_j)]}.
\end{aligned}
\label{wig2-3}
\end{equation}
\end{lemma}

{\it Proof. }
By definition, we have
\begin{equation}
\begin{aligned}
\langle\tr \frac{1}{{\mu} {\mathbb{I}}_N-H}\rangle_N
=\int_{{\mathfrak{U}}_N} d\mu_N(H) &
\big[i\int\prod_{j=1}^N\frac{d\overline{z}_j\,dz_j}{2{\pi}i}
\prod_{k=1}^N d\overline{\theta}_k d\theta_k\,\\
&\qquad\times
(z^*\!\!\cdot\! z) 
\exp{[{-}iX^*( {\mathbb{I}}_2\otimes({\mu} {\mathbb{I}}_N-H)X]}
\big].
\end{aligned}
\label{wig2-4}
\end{equation}

As
$
X^*( {\mathbb{I}}_2\otimes H)X=
H_{jk}(\overline{z}_jz_k+\overline{\theta}_j\theta_k),
$
we have
\begin{equation}
\big\langle \exp{[\pm i\sum_{j,k=1}^N H_{jk}
(\overline{z}_jz_k+\overline{\theta}_j\theta_k)
]}\big\rangle_N
=\exp{\big[-\frac{J^2}{2N}\sum_{j,k=1}^N
(\overline{z}_jz_k+\overline{\theta}_j\theta_k)
(\overline{z}_kz_j+\overline{\theta}_k\theta_j)\big]}.
\label{wig2-5}\end{equation}

After changing the order of integration
and substituting \eqref{wig2-5} into \eqref{wig2-4},
we get \eqref{wig2-3}. $\qquad\qed$

There are at least two approach from \eqref{wig2-3} to Wigner's law:
The method (I) permits us to make $\epsilon\to0$ rather easily
and leads us to a not so simple looking formula but which is calculable,
the other one (II) yields the beautiful formula \eqref{wig4} formally,
but in order to make $\epsilon\to0$ in that formula rigorously,
we reform it until it is represented by Hermite polynomials, 
and at that time the beauty of the formula \eqref{wig4} is lost.

(I) The following calculation is proposed by Br\'{e}zin \cite{bre83,BK90}:

Using
$$
%(\str A_X^2=)
\sum_{j,k=1}^N (\overline{z}_jz_k+\overline{\theta}_j\theta_k)
(\overline{z}_kz_j+\overline{\theta}_k\theta_j)
=(z^*\!\!\cdot\! z)^2+2(\theta^*\!\!\cdot\! z)(z^*\!\!\cdot\!\theta)
-(\theta^*\!\!\cdot\!\theta)^2,
$$
we have
\begin{equation}
\exp{[\frac{J^2}{2N}(\theta^*\!\!\cdot\!\theta)^2]}
=\bigg(\frac{N}{2\pi J^2}\bigg)^{1/2}
\int_{-\infty}^\infty d\tau\,\exp{[-\tau(\theta^*\!\!\cdot\!\theta)
-\frac{N}{2J^2}\tau^2]}.
\label{gauss-type}
\end{equation}
\begin{quotation}
{\small
$\because)$ This equality is non-trivial if we adopt the naive definition of integration. Formally, putting $\displaystyle{\gamma=\frac{N}{J^2}}$, by
$$
-\tau(\theta^*\!\!\cdot\!\theta)-\frac{\gamma}{2}\tau^2
=-\frac{\gamma}{2}[(\tau+\gamma^{-1}(\theta^*\!\!\cdot\!\theta))^2-\gamma^{-2}(\theta^*\!\!\cdot\!\theta)^2]
$$
we get
$$
\begin{aligned}
\int_{-\infty}^\infty d\tau\,\exp{[-\tau(\theta^*\!\!\cdot\!\theta)
-\frac{\gamma}{2}\tau^2]}
&=\exp{[\frac{(\theta^*\!\!\cdot\!\theta)^2}{2\gamma}]}
\int_{-\infty}^\infty d\tau\,\exp{[-\frac{\gamma}{2}(\tau+\frac{\theta^*\!\!\cdot\!\theta}{\gamma})^2]}\\
&=\int_{-\infty}^\infty ds\,\exp{[-\frac{\gamma}{2}s^2]}
\end{aligned}
$$
and therefore it seems the above equality \eqref{gauss-type} follows. But
for $s=\tau+\gamma^{-1}(\theta^*\!\cdot\theta)$, do we consider as $``d\tau=ds"$?
(Though RVV-integral is applicable, we dare to prove this fact as follows only with naive definition.)

We may regard the integrand as ${\fC}$-valued real $\tau$-variable function, that is, enclosing $(\theta^*\!\!\cdot\theta)$,
we may integrate. In fact, since $(\theta^*\!\!\cdot\!\theta)^{N+1}=0$,
$$
\sqrt{\frac{\gamma}{2\pi}}\int_{-\infty}^\infty d\tau\,e^{-\gamma \tau^2/2}\sum_{j=0}^N\frac{(-\tau(\theta^*\!\!\cdot\!\theta))^{j}}{j!}
=\sqrt{\frac{\gamma}{2\pi}}\sum_{j=0}^N\int_{-\infty}^\infty d\tau\,e^{-\gamma \tau^2/2}(-\tau)^{j}
\frac{(\theta^*\!\!\cdot\!\theta)^{j}}{j!}.
$$
Using
$$
\int_{-{\infty}}^{\infty}d\tau\,\tau^{2\ell+1}e^{-\gamma \tau^2/2}=0,\quad
\int_{-{\infty}}^{\infty}d\tau\,\tau^{2\ell}e^{-\gamma \tau^2/2}
=\sqrt{\frac{2\pi}{\gamma}}\frac{2(2\ell-1)!}{2^{\ell}\gamma^{\ell}(\ell-1)!},
$$
the right-hand side of above equation,
$$
\sum_{\ell=0}^{[N/2]}\sqrt{\frac{\gamma}{2\pi}}\int_{-\infty}^\infty d\tau\,e^{-\gamma \tau^2/2}(-\tau)^{2\ell}
\frac{(\theta^*\!\!\cdot\!\theta)^{2\ell}}{(2\ell)!}=\exp{[\frac{1}{2\gamma}(\theta^*\!\!\cdot\!\theta)^2]}.\qquad/\!\!/
$$
\par
For comparison, we remember a little for ordinary integral case: By $x^2+2ix\xi=(x+i\xi)^2+\xi^2$, we have
$$
\begin{aligned}
\int_{\euc}dx\,e^{-ix\xi}e^{-x^2/2}&=e^{-\xi^2/2}\int_{\euc}dx\,e^{-(x+i\xi)^2/2}=e^{-\xi^2/2}\int_{\euc}dy\,e^{-y^2/2}\\
&=\int_{\euc}dx\,(\sum_{n=0}^{\infty}\frac{(-ix\xi)^n}{n!})e^{-x^2/2}
=\sum_{m=0}^{\infty}\frac{(-1)^m\xi^{2m}}{(2m)!}\int_{\euc}dx\,x^{2m}e^{-x^2/2}.
\end{aligned}
$$
Even in this case, we need to explain why the second equality and the last one holds  in the above, but we take these admitted.
}
\end{quotation}

Substituting this relation into \eqref{wig2-4}, we get
$$
\begin{aligned}
\big\langle\tr \frac{1}{{\mu} {\mathbb{I}}_N-H}\big\rangle_N
=&i\int\prod_{j=1}^N\frac{d\overline{z}_j\,dz_j}{2{\pi}i}
\prod_{k=1}^N d\overline{\theta}_k d\theta_k\,
(z^*\!\!\cdot\! z)\\
&\qquad\times
\exp{[{-}iX^*( {\mathbb{I}}_2\otimes{\mu} {\mathbb{I}}_N)X
-\frac{J^2}{2N}
((z^*\!\!\cdot\! z)^2
+2(\theta^*\!\!\cdot\! z)(z^*\!\!\cdot\!\theta))]}\\
&\qquad\qquad\qquad\times
\bigg(\frac{N}{2\pi J^2}\bigg)^{1/2}
\int_{-\infty}^\infty d\tau\,\exp{[-\tau(\theta^*\!\!\cdot\!\theta)
-\frac{N}{2J^2}\tau^2]}.
\end{aligned}
$$

\begin{proposition} We have the following formula:
\begin{equation}
\begin{aligned}
\big\langle&\tr \frac{1}{(\lambda{-}i0) {\mathbb{I}}_N-H}\big\rangle_N\\
&=i\frac{1}{(N-1)!}\bigg(\frac{N}{2\pi J^2}\bigg)^{1/2}
\bigg(\frac{N}{J^2}\bigg)^{N+1}
\int_0^\infty d{s}\,{s}^N
\exp{[-\frac{N}{2J^2}(2i\lambda{s}+{s}^2)]}\\%{-}i\epsilon
&\qquad\qquad\qquad\times
\int_{-\infty}^\infty d\tau\,(\tau+i\lambda)^{N-1}
(\tau+i\lambda+{s})
\exp{[-\frac{N}{2J^2}\tau^2]}\\
&=i\frac{1}{(N-1)!}\bigg(\frac{N}{2\pi J^2}\bigg)^{1/2}
\bigg(\frac{N}{J^2}\bigg)^{N+1}
\iint_{\euc_+\times\euc} d{s} d\tau \,
(1+(\tau+i\lambda)^{-1}{s})\\
&\qquad\qquad\qquad\times 
\exp[-N(\frac{1}{2J^2}(\tau^2+2i\lambda{s}+{s}^2)
-\log s(\tau+i\lambda))].
\end{aligned}
\label{wig2-11}
\end{equation}
\end{proposition}
\par
{\it Proof. } %We put $|z|^2=z^*\!\!\cdot\! z$.
%For notational simplicity, we put $\mu=\lambda-i\epsilon$.
Since
$$
-i{\mu}(\theta^*\!\!\cdot\!\theta)
-\frac{J^2}{N}(\theta^*\!\!\cdot\! z)(z^*\!\!\cdot\!\theta)
-\tau(\theta^*\!\!\cdot\!\theta)
=-\sum_{a,b}\overline\theta_a((\tau+i{\mu})\delta_{ab}
+\frac{J^2}{N}z_a\overline{z}_b)\theta_b,
$$
using Lemma 2.3 and Lemma \ref{l2-8} below, we get
\begin{equation}
\int\prod_{k=1}^N d\overline{\theta}_k d\theta_k\,
\exp{[-i{\mu}(\theta^*\!\!\cdot\!\theta)
-\frac{J^2}{N}(\theta^*\!\!\cdot\! z)(z^*\!\!\cdot\!\theta)
-\tau(\theta^*\!\!\cdot\!\theta)]}
=(\tau+i{\mu})^{N-1}
(\tau+i{\mu}+\frac{J^2}{N}(z^*\!\!\cdot\! z)).
\label{wig2-12}
\end{equation}
%Now, we continue our proof of \eqref{wig2-11}.
Using the expression \eqref{wig2-12}, we have
$$
\begin{aligned}
\big\langle\tr \frac{1}{{\mu} {\mathbb{I}}_N-H}\big\rangle_N
=&i\int\prod_{j=1}^N\frac{d\overline{z}_j\,dz_j}{2{\pi}i}
(z^*\!\!\cdot\! z)
\exp{[-i{\mu}(z^*\!\!\cdot\! z)-\frac{J^2}{2N}
(z^*\!\!\cdot\! z)^2]}\\
&\qquad\times
\bigg(\frac{N}{2\pi J^2}\bigg)^{1/2}
\int_{-\infty}^\infty d\tau\,
(\tau+i{\mu})^{N-1}
(\tau+i{\mu}+\frac{J^2}{N}(z^*\!\!\cdot\! z))
\exp{[-\frac{N}{2J^2}\tau^2]}.
\end{aligned}
$$

Identifying ${\mathbb{C}}^N=\euc^{2N}$ by $z_j=x_j+iy_j$, $\overline{z}_j=x_j-iy_j$,
$d\overline{z}_j\wedge dz_j=2idx_j\wedge dy_j$ 
and using the polar coordinate
$(r,\omega)\in \euc_+\times S^{2N-1}$ with
${\displaystyle{\prod_{j=1}^N\frac{d\overline{z}_j\,dz_j}{2{\pi}i}
=\prod_{j=1}^N\frac{dx_jdy_j}\pi
=\pi^{-N}r^{2N-1}dr\,d\omega}}$,  
$\int_{S^{2N-1}}d\omega={\mathrm {vol}}(S^{2N-1})=2\pi^N/(N-1)!$,
we get,
$$
\begin{aligned}
\big\langle\tr \frac{1}{{\mu} {\mathbb{I}}_N-H}\big\rangle_N
&=i\frac{1}{(N-1)!}
\int_0^\infty dr^2\, r^{2N}
\exp{[-i{\mu}r^2-\frac{J^2}{2N}r^4]}\\
&\qquad\times
\bigg(\frac{N}{2\pi J^2}\bigg)^{1/2}
\int_{-\infty}^\infty d\tau\,
(\tau+i{\mu})^{N-1}
(\tau+i{\mu}+\frac{J^2}{N}r^2)
\exp{[-\frac{N}{2J^2}\tau^2]}.
\end{aligned}
$$
Changing the independent variables as $r^2=({N}/{J^2}){\tilde r}$
and making $\epsilon\to0$, i.e. $\mu\to\lambda-i0$,
we get the result. 
Here, this procedure of making $\epsilon\to0$ under integral sign is admitted
because of Lebesgue's dominated convergence theorem.   $\qquad\qed$

\begin{remark} The formula \eqref{wig2-11} equals 
to Br\'{e}zin's equality (2.16) of \cite{bre83}
where he takes $\lambda+i\epsilon$ instead of our choice $\lambda-i\epsilon$.
On the other hand, probably, he miscopies his equality in (44) of \cite{BK90},
more precisely, the term $(1+xy)$ in (44) should be $(1+(x-iz)^{-1}y)$.
\end{remark}

Now, we prepare a technical lemma:
\begin{lemma}
For any $\ell>0$ and $n=0,1,2,3,\cdots$,
\begin{align}
    & \int_{-\infty}^{\infty}dt \,e^{-\ell t^{2}}t^{2n+1}=0,\quad
\int_{0}^{\infty}dt \,e^{-\ell t^{2}}t^{2n+1} =\frac{n!}{2\ell^{n+1}},
\label{A13}\\
     & \int_{-\infty}^{\infty}dt \,e^{-\ell t^{2}}t^{2n} =\ell^{-n-\frac{1}{2}}
               \Gamma\left(\frac{2n+1}{2}\right)=  \ell^{-n-\frac{1}{2}}
                 \frac{(2n)!}{n!2^{2n}}\pi^{\frac{1}{2}}.
            \label{A14}%\tag 3.18
\end{align}
Let $\delta_0>0,\;d_0>0$. For $\tau_{N}=d_0N^{-\gamma}$ 
such that $0\le\gamma\le1/2$, we have 
\begin{equation}
    \int_{\tau_{N}}^{\infty}dt \,e^{-\delta_0 Nt^{2}}<
          (2\delta_0d_0N^{1-\gamma})^{-1}e^{-\delta_0d_0^2 N^{1-2\gamma}}. 
\label{A11}
\end{equation}
\end{lemma}
\par{\it Proof. }
The first two are well known.
As $t>\tau_N$, we have
$$
\delta N(t^2-\tau_N^2)=\delta N(t-\tau_N)(t+\tau_N)>2\delta N\tau_N(t-\tau_N).
$$
Therefore, we get
$$
\int_{\tau_N}^\infty dt\,e^{-\delta_0 Nt^2}
<e^{-\delta_0 N\tau_N^2}\int_{\tau_N}^\infty dt\,
e^{-2\delta_0d_0N^{1-\gamma}(t-\tau_N)}
=(2\delta_0d_0N^{1-\gamma})^{-1}e^{-\delta_0d_0^2 N^{1-2\gamma}}.   \qquad\qed
$$
Then, we have
\begin{corollary} For $\lambda=0$, we get readily
\begin{equation}
\big\langle\rho_N(0)\big\rangle_N
=\frac{1}{\pi{J}}[1-(-1)^N\frac{1}{4}N^{-1}+\frac{1}{32}N^{-2}
+(-1)^N\frac{5}{128}N^{-3}
+O(N^{-4})].
\end{equation}
\end{corollary}
\par{\it Proof. } 
Using \eqref{A13}, \eqref{A14} to have
$$
\begin{aligned}
\big\langle\rho_N(0)\big\rangle_N&=\frac{1}{\pi N}
\Im\big\langle\tr \frac{1}{{-i0} {\mathbb{I}}_N-H}\big\rangle_N\\
&=\frac{1}{\pi N!}\bigg(\frac{N}{2\pi J^2}\bigg)^{1/2}
\bigg(\frac{N}{J^2}\bigg)^{N+1}
\int_0^\infty d{s}\,{s}^N
\exp{[-\frac{N}{2J^2}{s}^2]}
\int_{-\infty}^\infty d\tau\,\tau^{N-1}
(\tau+{s})
\exp{[-\frac{N}{2J^2}\tau^2]}\\
&=\frac{1}{\pi N!}\bigg(\frac{N}{2\pi J^2}\bigg)^{1/2}
\bigg(\frac{N}{J^2}\bigg)^{N+1}\times
{
\begin{cases}
{\displaystyle{
\frac{1}{2}\bigg(\frac{N}{2J^2}\bigg)^{-N-1}
\Gamma\bigg(\frac{N+1}{2}\bigg)^2}},
&\mbox{$N$=even},\\
{\displaystyle{\frac{1}{2}\bigg(\frac{N}{2J^2}\bigg)^{-N-1}
\Gamma\bigg(\frac{N}{2}\bigg)
\Gamma\bigg(\frac{N+2}{2}\bigg)}},
&\mbox{$N$=odd},
\end{cases}
}
\end{aligned}
$$
we may calculate explicitly by the Stirling formula. $\qquad\qed$

In proving \eqref{wig2-12} above, we have also used
\begin{lemma}\label{l2-8}
Let $M=(M_{ab})$ with $M_{ab}=\alpha\delta_{ab}+\beta z_a\overline{z}_b$.
Then, we have
$$
\det M=\alpha^{N-1}(\alpha+\beta|z|^2),\quad |z|^2=z^*\!\!\cdot\! z.
$$
\end{lemma}
\par
{\it Proof. }
Let $u$ satisfy $Mu=\gamma u$. Then, we have
\begin{equation}
\overline{z}\cdot Mu= \gamma\overline{z}\cdot u,\quad
\overline{u}\cdot Mu= \gamma\overline{u}\cdot u.
\label{aa}
\end{equation}
From the first equation above, we get
$$
(\gamma-\alpha-\beta|z|^2)\sum_{j=1}^N u_j\overline{z}_j=0.
$$
If $\sum_{j=1}^N u_j\overline{z}_j\neq{0}$, $\gamma=\alpha+\beta|z|^2$.
On the other hand, if $\sum_{j=1}^N u_j\overline{z}_j=0$,
the second one in \eqref{aa} implies that $\gamma=\alpha$.
Taking into account the multiplicity, we have the desired result. $\qquad\qed$

(II) In order to get the expression \eqref{wig4},
we proceed as follows:
Putting
$$
A_X=\begin{pmatrix}
\sum_{j=1}^N \overline{z}_jz_j & \sum_{j=1}^N \overline{\theta}_jz_j\\
\sum_{j=1}^N \theta_j\overline{z}_j & \sum_{j=1}^N \overline{\theta}_j\theta_j
\end{pmatrix},
$$
we have
$$
\str A_X^2=\sum_{j,k=1}^N (\overline{z}_jz_k+\overline{\theta}_j\theta_k)
(\overline{z}_kz_j+\overline{\theta}_k\theta_j).
$$

On the other hand, the following is known as the
Hubbard-Stratonovich formula:
\begin{lemma} Let $A$ be any even $2\times2$ supermatrix.
For $Q\in{\frak Q}$ given in \eqref{wig5}, we have
\begin{equation}
\exp{[-\frac{J^2}{2N}\str A^2]}
=\int_{\frak Q}dQ\,\exp{[-\frac{N}{2J^2}\str Q^2\pm i\str (QA)]}.
\label{wig2-6}\end{equation}
\end{lemma}

{\it Proof. }
Let $A=\begin{pmatrix} a&\theta_1\\\theta_2&b\end{pmatrix}$ 
with $a,\,b\in\rev$ and $\theta_1,\,\theta_2\in\rod$.
For any $\gamma>0$, we claim
$$
\int_{\frak Q}dQ\,\exp{[-\frac12\str(\gamma Q\pm i\gamma^{-1}A)^2]}=1.
$$
As we have readily
$$
\str(\gamma Q\pm i\gamma^{-1}A)^2=
\gamma^2(x_1^2+x_2^2+2\rho_1\rho_2)\pm 
2i(x_1a+\rho_1\theta_2-\rho_2\theta_1-ix_2b)
-\gamma^{-2}(a^2-b^2+2\theta_1\theta_2),
$$
we get
$$
\begin{gathered}
\int d\rho_1d\rho_2\,
\exp{[-\gamma^2\rho_1\rho_2
\mp i(\rho_1\theta_2-\rho_2\theta_1)+\gamma^{-2}\theta_1\theta_2]}
=(\gamma^2-\theta_1\theta_2)(1+\gamma^{-2}\theta_1\theta_2)=\gamma^2,\\
\hspace{-5mm}\int \frac{dx_1 dx_2}{2\pi}\,
\exp{[-\frac{\gamma^2}{2}(x_1^2+x_2^2)
\mp i(x_1a-ix_2b)+\frac{\gamma^{-2}}{2}(a^2-b^2)]}\qquad\\
\qquad\qquad=\int \frac{dx_1 dx_2}{2\pi}\,
\exp{[-\frac{1}{2}(\gamma x_1\pm i\gamma^{-1}a)^2-
\frac{1}{2}(\gamma x_2\pm\gamma^{-1}b)^2]}=
\gamma^{-2}. \qquad\qed
\end{gathered}
$$

Substituting  \eqref{wig2-6} with $A=A_X$ into \eqref{wig2-3}, 
noting 
$\str(QA_X)=X^*(Q\otimes {\mathbb{I}}_N)X$,
taking the part of integral and
changing the order of integration,
we have 
$$
\begin{aligned}
&i\int\prod_{j=1}^N\frac{d\overline{z}_j\,dz_j}{2{\pi}i}
\prod_{k=1}^N d\overline{\theta}_k d\theta_k\,
(z^*\!\!\cdot\! z) \exp{[{-}iX^*
\big(({\mu}{\mathbb{I}}_2- Q)\otimes {\mathbb{I}}_N\big)X]}\\
&\qquad\qquad=\sum_{j=1}^N(\{{\mu} {\mathbb{I}}_2- Q)\otimes {\mathbb{I}}_N\}^{-1})_{bb,jj}
\sdet^{-1}(i({\mu} {\mathbb{I}}_2- Q)\otimes {\mathbb{I}}_N).
\end{aligned}
$$
Therefore, we have
\begin{lemma} For $\mu=\lambda-i\epsilon$ $(\epsilon>0)$, 
\begin{equation}
\big\langle \tr\frac{1}{{\mu}{\mathbb{I}}_N-H}\big\rangle_N
=\int_{\frak Q}dQ\,
\sum_{j=1}^N(\{({\mu}{\mathbb{I}}_2- Q)\otimes {\mathbb{I}}_N\}^{-1})_{bb,jj}
\sdet^{-1}(i({\mu} {\mathbb{I}}_2- Q)\otimes {\mathbb{I}}_N).
\end{equation}
Here, $(C)_{bb,jj}$ is the $j$-th diagonal element of
the boson-boson block of the (even) supermatrix $C$.
\end{lemma}

Remarking
$$
\begin{gathered}
(\{{\mu} {\mathbb{I}}_2- Q)\otimes {\mathbb{I}}_N\}^{-1})_{bb,jj}=
(\{{\mu} {\mathbb{I}}_2- Q\}^{-1})_{bb}\forany j=1,2,\cdots,N,\\
\sdet^{-1}(i({\mu}{\mathbb{I}}_2- Q)\otimes {\mathbb{I}}_N)=
\sdet^{-N}({\mu}{\mathbb{I}}_2- Q),\\
\str(AB)=\str(BA),\quad \str(A+B)=\str A+\str B,\quad
\log(\sdet^\ell A)=\ell\str(\log A) \for \ell\in{\Bbb Z},
\end{gathered}
$$
we have
\begin{equation}
\big\langle\tr\frac{1}{{\mu} {\mathbb{I}}_N-H}\big\rangle_N=
\int_{\frak Q} dQ\, N\big(\{{\mu} {\mathbb{I}}_2-Q\}^{-1}\big)_{bb}
\exp{[-N{\mathcal{L}}(\mu;Q)]}
\label{wig2-9}\end{equation}
with
$$
\begin{gathered}
{\mathcal{L}}(\mu;Q)=\str[(2J^2)^{-1}Q^2+\log({\mu} {\mathbb{I}}_2-Q)],\\
\big(\{{\mu} {\mathbb{I}}_2-Q)\}^{-1}\big)_{bb,11}=
\frac{{\mu}-ix_2}
{({\mu}-x_1)({\mu}-ix_2)-\rho_1\rho_2}
=\frac{({\mu}-x_1)({\mu}-ix_2)+\rho_1\rho_2}
{({\mu}-x_1)^2({\mu}-ix_2)},\\
{\frak Q}=\big\{Q=\begin{pmatrix}
x_1&\rho_1\\
\rho_2&ix_2
\end{pmatrix}\,\big|\, x_1, x_2\in{\rev},\; 
\rho_1,\rho_2\in{\rod}\big\}.
\end{gathered}
$$
\begin{remark}
If we could make directly $\epsilon\to0$ in \eqref{wig2-9}, 
we had the formula \eqref{wig4}.
We claim at least symbolically we do that.
\end{remark}

\begin{lemma}  For $\mu=\lambda-i\epsilon$ $(\epsilon>0)$, 
\begin{equation}
\frac{1}{N}\langle\tr\frac{1}{{\mu} {\mathbb{I}}_N-H}\rangle_N
=\int_{\euc^2}\frac{dx_1 dx_2}{2\pi}
\frac{N({\mu}-x_1-ix_2)}
{J^2({\mu}-x_1)({\mu}-ix_2)}
\exp{[-N\Phi(x_1,x_2;{\mu})]},
\label{wig2-10}
\end{equation}
where
$$
\Phi(x_1,x_2;{\mu})=\frac{x_1^2+x_2^2}{2J^2}
+\log \frac{{\mu}-x_1}{{\mu}-ix_2}.
$$
\end{lemma}
\par{\it Proof. }
As the integrand in \eqref{wig2-9} is represented by
$$
\frac{({\mu}-x_1)({\mu}-ix_2)+\rho_1\rho_2}
{({\mu}-x_1)^2({\mu}-ix_2)}
\exp{[-N\{\frac{1}{2J^2}(x_1^2+x_2^2+2\rho_1\rho_2)
+\log\frac{{\mu}-x_1}{{\mu}-ix_2}
-\frac{\rho_1\rho_2}{({\mu}-x_1)({\mu}-ix_2)}\}]},
$$
we have
\begin{equation}
\begin{aligned}
&\int d\rho_1d\rho_2\,
\frac{({\mu}-x_1)({\mu}-ix_2)+\rho_1\rho_2}
{({\mu}-x_1)^2({\mu}-ix_2)}
\exp{[-N\{\frac{1}{J^2}
-\frac{1}{({\mu}-x_1)({\mu}-ix_2)}\}\rho_1\rho_2]}\\
&\qquad\qquad\qquad\qquad
=\frac{-1}{({\mu}-x_1)^2({\mu}-ix_2)}
+\frac{N}{{\mu}-x_1}\{\frac{1}{J^2}
-\frac{1}{({\mu}-x_1)({\mu}-ix_2)}\}.
%\exp{[-N\Phi(x_1,x_2;{\mu})]}.
\end{aligned}\label{wig2*}
\end{equation}
Remarking $({\mu}-x_1)^{-2}
=\partial_{x_1}({\mu}-x_1)^{-1}$, 
%integrating with respect to $\frac{dx_1 dx_2}{2\pi}$
by integration by parts, 
we have
$$
\begin{aligned}
&\int \frac{dx_1 dx_2}{2\pi}
\frac{-1}{({\mu}-x_1)^2({\mu}-ix_2)}
\exp{[-N\Phi(x_1,x_2;{\mu})]}\\
&\qquad\qquad=\int \frac{dx_1 dx_2}{2\pi}
\frac{-N}{({\mu}-x_1)({\mu}-ix_2)}
{\big\{}\frac{x_1}{J^2}-\frac{1}{{\mu}-x_1}{\big\}}
\exp{[-N\Phi(x_1,x_2;{\mu})]},
\end{aligned}
$$
which yields \eqref{wig2-10}.   $\qquad\qed$

\begin{remark} As the right-hand side of \eqref{wig2-10} is rewritten
\begin{equation}
\frac{1}{N}\langle\tr\frac{1}{{\mu} {\mathbb{I}}_N-H}\rangle_N
=\frac{N}{2\pi J^2}\int_{\euc^2}{dx_1 dx_2}
\frac{({\mu}-x_1-ix_2)({\mu}-ix_2)^{N-1}}
{({\mu}-x_1)^{N+1}}
\exp{[-N\frac{x_1^2+x_2^2}{2J^2}]},
\label{wig2-11a}\end{equation}
there is no singularity in the integrand when $\Im\mu\neq{0}$. $\qquad\qed$
\end{remark}

Using the fact that for any real smooth integrable function $f$, 
$$
\lim_{\epsilon\to0}\Im\int_{\euc}{dx}\,(\lambda-i\epsilon-x)^{-1}f(x)
=\lim_{\epsilon\to0}\int_{\euc}{dx}\frac{\epsilon}{(\lambda-x)^2+\epsilon^2}f(x)
=\pi f(\lambda),
$$
and integrating by parts based on
$\partial_{x_1}^{\ell}(\mu-x_1)^{-1}={\ell}!(\mu-x_1)^{-{\ell}-1}$, 
we have,
$$
\lim_{\epsilon\to0}\Im\int_{\euc}{dx_1}\,(\lambda-i\epsilon-x_1)^{-{\ell}-1}
\exp[{-N\frac{x_1^2}{2J^2}}]
=\pi\frac{(-1)^\ell}{\ell!}\partial_{\lambda}^\ell
\exp[{-N\frac{\lambda^2}{2J^2}}].
$$
Using the Hermite polynomial ${H}_\ell(x)$ defined by
$$
{H}_\ell(x)=(-1)^\ell e^{x^2/2}\partial_x^\ell e^{-x^2/2}
=\sum_{k=0}^{[\ell/2]}\frac{(-1)^k\ell! x^{\ell-2k}}{2^k k!(\ell-2k)!}
=\frac{1}{\sqrt{2\pi}}\int_{-\infty}^\infty dt\, e^{-t^2/2}(x{\mp}it)^\ell,
$$
with
$$
{H}_\ell(\gamma x)
=\frac{\gamma^{\ell+1}}{\sqrt{2\pi}}
\int_{-\infty}^\infty dt \, e^{-\gamma^2t^2/2}(x{\mp}it)^\ell,
$$
we have
\begin{lemma}
\begin{equation}
\pi\frac{(-1)^\ell}{\ell!}\partial_{\lambda}^\ell
\exp[{-N\frac{\lambda^2}{2J^2}}]
=\frac{\pi}{\ell!}\bigg(\frac{N}{J^2}\bigg)^{(2\ell+1)/2}
\exp[{-N\frac{\lambda^2}{2J^2}}]
\int_{\euc}dt\,(\lambda{\mp}it)^\ell
\exp[{-N\frac{t^2}{2J^2}}].   
\label{wig2.17}
\end{equation}
\end{lemma}
\par{\it Proof. } Using Bell's polynomial, we have
$$
\begin{aligned}
\pi\frac{(-1)^\ell}{\ell!}\partial_{\lambda}^\ell
\exp[{-N\frac{\lambda^2}{2J^2}}]
&=\pi\sum_{k=0}^{[\ell/2]}\frac{(-1)^k2^{-k}}{k!(\ell-2k)!}
\bigg(\frac{N}{J^2}\bigg)^{\ell-k}\lambda^{\ell-2k}
\exp[{-N\frac{\lambda^2}{2J^2}}]\\
&=\frac{\pi}{\ell!}\bigg(\frac{N}{J^2}\bigg)^{\ell/2}
{H}_\ell\bigg(\bigg(\frac{N}{J^2}\bigg)^{1/2}\lambda\bigg)
\exp[{-N\frac{\lambda^2}{2J^2}}].  \qquad\qed
\end{aligned}
$$

On the other hand,
$$
\int_{\euc}{dx_2}\,
({\mu}-ix_2)^\ell\exp{[-N\frac{x_2^2}{2J^2}]}
=\sqrt{2\pi}\bigg(\frac{N}{J^2}\bigg)^{-(\ell+1)/2}
{H}_\ell\bigg(\bigg(\frac{N}{J^2}\bigg)^{1/2}\mu\bigg),
$$
and
$$
\begin{aligned}\int_{\euc}{dx_2}\,
(-ix_2)({\mu}-ix_2)^\ell\exp{[-N\frac{x_2^2}{2J^2}]}
&=-\ell\frac{J^2}{N}\int_{\euc}{dx_2}\,
({\mu}-ix_2)^{\ell-1}\exp{[-N\frac{x_2^2}{2J^2}]}\\
&=-\ell \sqrt{2\pi}\bigg(\frac{N}{J^2}\bigg)^{-(\ell+2)/2}
{H}_{\ell-1}\bigg(\bigg(\frac{N}{J^2}\bigg)^{1/2}\mu\bigg).
\end{aligned}
$$

Therefore, we have
$$
\big\langle\rho_N(\lambda)\big\rangle_N=
\frac{N}{2\pi^2 J^2}\lim_{\epsilon\to0}(\Im K_1+\Im K_2),
$$
where
$$
\begin{aligned}
K_1&=\int_{\euc^2}{dx_1}{dx_2}\,(\mu-x_1)^{-N}(\mu-ix_2)^{N-1}
\exp[{-N\frac{x_1^2+x_2^2}{2J^2}}],\\
K_2&=\int_{\euc^2}{dx_1}{dx_2}\,(\mu-x_1)^{-N-1}(-ix_2)(\mu-ix_2)^{N-1}
\exp[{-N\frac{x_1^2+x_2^2}{2J^2}}].
\end{aligned}
$$
Moreover, we get
$$
\begin{aligned}
\lim_{\epsilon\to0}\Im K_1
&=\frac{\pi}{(N-1)!}\bigg(\frac{N}{J^2}\bigg)^{(N-1)/2}
{H}_{N-1}\bigg(\bigg(\frac{N}{J^2}\bigg)^{1/2}\lambda\bigg)
\exp[{-N\frac{\lambda^2}{2J^2}}]\\
&\qquad\qquad\qquad\qquad\qquad
\times
\sqrt{2\pi}\bigg(\frac{N}{J^2}\bigg)^{-N/2}
{H}_{N-1}\bigg(\bigg(\frac{N}{J^2}\bigg)^{1/2}\lambda\bigg),\\
\lim_{\epsilon\to0}\Im K_2
&=\frac{\pi}{N!}\bigg(\frac{N}{J^2}\bigg)^{N/2}
{H}_{N}\bigg(\bigg(\frac{N}{J^2}\bigg)^{1/2}\lambda\bigg)
\exp[{-N\frac{\lambda^2}{2J^2}}]\\
&\qquad\qquad\qquad\qquad\qquad
\times
(-1)(N-1)\sqrt{2\pi}\bigg(\frac{N}{J^2}\bigg)^{-(N+1)/2}
{H}_{N-2}\bigg(\bigg(\frac{N}{J^2}\bigg)^{1/2}\lambda\bigg).
\end{aligned}
$$

Combining these, we have proved
\begin{proposition} For any $\lambda\in\euc$, we have
\begin{equation}
\begin{aligned}
\big\langle\rho_N(\lambda)\big\rangle_N
&=
\frac{1}{\sqrt{2\pi} J(N-1)!}\exp{[-\frac{N}{2J^2}\lambda^2]}\\
&\qquad\qquad\times
\bigg[
\sqrt{N}
{H}_{N-1}^2\bigg(\bigg(\frac{N}{J^2}\bigg)^{1/2}\lambda\bigg)-\frac{N-1}{\sqrt{N}}{H}_{N}\bigg(\bigg(\frac{N}{J^2}\bigg)^{1/2}\lambda\bigg)\\
&\qquad\qquad\qquad\qquad\times
{H}_{N-2}\bigg(\bigg(\frac{N}{J^2}\bigg)^{1/2}\lambda\bigg)
\bigg]\\
&=\bigg(\frac{N}{2\pi J^2}\bigg)^{1/2}
\frac{1}{2\pi(N-1)!}
\bigg(\frac{N}{J^2}\bigg)^N\\
&\qquad\qquad\times
\iint_{\euc^2} dt\,ds\,\exp{[-{N}\phi_{\pm}(t,s,\lambda)]}a_{\pm}(t,s,\lambda;N),
\end{aligned}\label{wig2.18}
\end{equation}
where
\begin{equation}
\begin{aligned}
\phi_{\pm}(t,s,\lambda)&=\frac{1}{2J^2}(t^2+s^2+\lambda^2)
-\log(\lambda{\mp}it)(\lambda{\mp}is),\\
a_{\pm}(t,s,\lambda:N)&=\frac{1}{(\lambda{\mp}it)(\lambda{\mp}is)}-\frac12(1-N^{-1})
\bigg[\frac{1}{(\lambda{\mp}it)^2}+\frac{1}{(\lambda{\mp}is)^2}\bigg].
\end{aligned}
\end{equation}
\end{proposition}

\section{The proof of semi-circle law and beyond that}

To prove semi-circle law, we need to apply the method of saddle point which is only proved in the ordinary space $\euc^m$. As we mentioned before, it seems difficult even ordinary case to assure whether we may apply that method suitably. And this point is stressed by the old book of de Bruijn \cite{deB61}, pp.77-78 :
\begin{quotation}
The saddle point method, due to B. Riemann and P. Debye, 
is one of the most important and most powerful methods
in asymptotics.  $\cdots $ (omission) $\cdots $

Any special application of the saddle point method consists of two stages.
\par
(i) The stage of exploring, conjecturing and scheming,
which is usually the most difficult one. 
It results in choosing a new integration path, 
made ready for application of (ii).
\par
(ii) The stage of carrying out the method. 
Once the path has been suitably chosen, 
this second stage is, as a rule, rather a matter of routine,
although it may be complicated. 
It essentially depends on the Laplace method
of Chapter 4.

$\cdots$ (omission) $\cdots $
Most authors dealing with special applications do not go into the trouble 
of explaining what arguments led to their choice of path. 
The main reason is that it is always very difficult to say why a certain
possibility is tried and others are discarded, especially since
this depends on personal imagination and experience.
\end{quotation}

In the following, we only explain the rough idea to prove certain claims whose precise proofs are shown in A. Inoue and Y. Nomura~\cite{IN00}. 

Now, we study the asymptotic behavior of the following integral w.r.t. $N$:
\begin{equation}
\begin{aligned}
&\big\langle\tr \frac{1}{(\lambda{-}i0) {\mathbb{I}}_N-H}\big\rangle_N
=i\frac{1}{(N-1)!}\bigg(\frac{N}{2\pi J^2}\bigg)^{1/2}
\bigg(\frac{N}{J^2}\bigg)^{N+1}(I_1+ I_2),\\
&\qquad\qquad\quad
I_1=\int_0^\infty d{s}\,{s}^N
\exp{[-\frac{N}{2J^2}(2i\lambda{s}+{s}^2)]}
\int_{-\infty}^\infty d\tau\,(\tau+i\lambda)^{N}
\exp{[-\frac{N}{2J^2}\tau^2]},\\%{-}i\epsilon
&\qquad\qquad\quad
I_2=\int_0^\infty d{s}\,{s}^{N+1}
\exp{[-\frac{N}{2J^2}(2i\lambda{s}+{s}^2)]}
\int_{-\infty}^\infty d\tau\,(\tau+i\lambda)^{N-1}
\exp{[-\frac{N}{2J^2}\tau^2]}.
\end{aligned}
\label{3-1}
\end{equation}
\begin{theorem} 
Let $|\lambda|<2J$.
Putting $\theta=-\arg\tau_+$, 
$\tau_+=2^{-1}(-i\lambda+\sqrt{4J^2-\lambda^2})$, we have
\begin{equation}
I_1+ I_2=2\pi e^{-N}J^{2(N+1)}\bigg[
e^{-i\theta}N^{-1}+
\frac{1}{12}
\bigg(e^{-i\theta}
-(-1)^N\frac{3e^{-iN(\sin2\theta+2\theta)}}{\cos^2\theta}\bigg)
N^{-2}+O(N^{-3})\bigg].
\end{equation}
\end{theorem}
For the proof, see Appendix A.4 of \cite{IN00}.

Remarking the Stirling formula 
$$
(N-1)!=e^{-N}N^{N-1/2}\sqrt{2\pi}(1+\frac{1}{12}N^{-1}+
\frac{1}{288}N^{-2}-\frac{139}{51840}N^{-3}-\frac{571}{2488320}N^{-4}+O(N^{-5})),
$$
that is,
\begin{equation}
\frac{1}{(N-1)!}=\frac{e^{N}N^{-N+1/2}}{\sqrt{2\pi}}(1-\frac{1}{12}N^{-1}-
\frac{1}{96}N^{-2}+O(N^{-3})),
\end{equation}
we get
$$
\begin{aligned}
\big\langle\tr &\frac{1}{(\lambda{-}i0) {\mathbb{I}}_N-H}\big\rangle_N
=i\frac{1}{(N-1)!}\bigg(\frac{N}{2\pi J^2}\bigg)^{1/2}
\bigg(\frac{N}{J^2}\bigg)^{N+1}(I_1+I_2),\\
&
=i\frac{N^2}{J}\bigg(e^{-i\theta}N^{-1}+\frac{1}{12}[e^{-i\theta}-(-1)^N
\frac{3e^{-iN(\sin2\theta+2\theta)}}{\cos^2\theta}]N^{-2}
+O(N^{-3})\bigg)\\
&\qquad\qquad\qquad
\times
(1-\frac{1}{12}N^{-1}+O(N^{-2}))\\
&
=i\frac{N}{J}\bigg(e^{-i\theta}-\frac{(-1)^N}4
\frac{e^{-iN(\sin2\theta+2\theta)}}{\cos^2\theta}N^{-1}
+O(N^{-2})\bigg).
\end{aligned}
$$
Therefore, we proved the first part of Theorem 1.2.   $\qquad\qed$

The relation \eqref{wig5} for $|\lambda|\ge 2J$ is proved analogously:
That is, we have
\begin{theorem}
Let $\lambda>2J$. There exists constsnt $k(\lambda)>0$ and $C(\lambda)>0$ 
such that
$$
I_1+ I_2= J^{2N}e^{-N}K(N)+\text{pure imaginary part}\with
\big| K(N)\big|\le C(\lambda)N^{-\frac{1}{2}}e^{-k(\lambda)N}.
$$
\end{theorem} 
See, Appendix A.4 of \cite{IN00}, for the proof.

Substituting this estimate into the definition of 
$\big\langle\rho_N(\lambda)\big\rangle_N$, we get
$$
\begin{aligned}
\big\langle\rho_N(\lambda)\big\rangle_N
&=\Im\frac{1}{\pi N}\big\langle\tr\frac{1}{(\lambda{-}i0) {\mathbb{I}}_N-H}\big\rangle_N\\
&=\Im i\frac{1}{\pi N!}\bigg(\frac{N}{2\pi J^2}\bigg)^{1/2}
\bigg(\frac{N}{J^2}\bigg)^{N+1}(I_1+ I_2),\\
&=\Im i\frac{1}{\pi N!}\bigg(\frac{N}{2\pi J^2}\bigg)^{1/2}
\bigg(\frac{N}{J^2}\bigg)^{N+1}J^{2N}e^{-N}
\left(K(N)+\text{pure imaginary part}\right)\\
&=\frac{1}{\pi N!}\bigg(\frac{N}{2\pi J^2}\bigg)^{1/2}
\bigg(\frac{N}{J^2}\bigg)^{N+1}J^{2N}e^{-N}K(N).
\end{aligned}
$$
Applying the Stirling formula to the last line of the above, 
we get the estimate \eqref{wig4}. $\qquad\qed$

\section{Edge mobility}

To study the asymptotic behavior of
$\big\langle\rho_N(2J-zN^{-2/3})\big\rangle_N$, or
$\big\langle\rho_N(-2J+zN^{-2/3})\big\rangle_N$   for $|z|\le1$
%when $\lamda\to \lambda^{\pm}_{c}$ 
as $N\to\infty$, we use in this section the formula \eqref{wig2.18}:
\begin{equation}
\begin{aligned}
\big\langle\rho_N(2J-zN^{-2/3})\big\rangle_N
&=\frac{N^{N+1/2}}{(2\pi)^{3/2}(N-1)! J^{2N+1}}
\iint_{\euc^2} dt\,ds\,a_+(t,s,2J-zN^{-2/3};N)\\
&\qquad\qquad\qquad\qquad\qquad\qquad\qquad\times\exp{[-{N}\phi_+(t,s,2J-zN^{-2/3})]},
\end{aligned}
\end{equation}
\begin{equation}
\begin{aligned}
\big\langle\rho_N(-2J+zN^{-2/3})\big\rangle_N
&=\frac{N^{N+1/2}}{(2\pi)^{3/2}(N-1)! J^{2N+1}}
\iint_{\euc^2} dt\,ds\,a_-(t,s,-2J+zN^{-2/3};N)\\
&\qquad\qquad\qquad\qquad\qquad\qquad\;\;\times\exp{[-{N}\phi_-(t,s,-2J+zN^{-2/3})]},
\end{aligned}
\end{equation}
where
\begin{equation}
\begin{aligned}
\phi_{\pm}(t,s,\lambda)&=\frac{1}{2J^2}(t^2+s^2+\lambda^2)
-\log(\lambda{\mp}it)(\lambda{\mp}is),\\
a_{\pm}(t,s,\lambda;N)
&=\frac{2(\lambda{\mp}it)(\lambda{\mp}is)-(1-N^{-1})
[(\lambda{\mp}it)^2+(\lambda{\mp}is)^2]}
{2(\lambda{\mp}it)^2(\lambda{\mp}is)^2}.
\end{aligned}
\end{equation}

\begin{proposition} For $|z|\le1$, we have
\begin{equation}
\begin{aligned}
\big\langle\rho_N(2J-zN^{-2/3})\big\rangle_N
=\frac{N^{-1/3}}{8\pi^2 J^5}
\iint_{\euc^2} dx\,dy\,{(x-y)^2}&\exp{[-\frac{i}{3J^3}(x^3+y^3-3zJ(x+y))]}\\
&\qquad\qquad\qquad
+O(N^{-2/3}).
\end{aligned}
\end{equation}
The right-hand integral above, should be interpreted as the oscillatory one.
\end{proposition}
\par{\it Proof. }
In this proof, we abbreviate the subscript $+$ of $\phi_+$ and $a_+$.

Put $u=N^{-1/3}$.
For $\lambda=2J-zu^2$,
using the change of variables 
$s=-iJ+yu$, $t=-iJ+xu$, we have
$$
\varphi(x,y,z;u)=\phi(-iJ+xu,-iJ+yu,2J-zu^2)
=h(u)-\log g(u) ,
$$
where
$$
\begin{aligned}
h(u)&=\frac{1}{2J^2}((-iJ+xu)^2+(-iJ+yu)^2+(2J-zu^2)^2)\\
&=1-\frac{i(x+y)}{J}u+\frac{x^2+y^2-4zJ}{2J^2} u^2+\frac{z^2}{2J^2} u^4,\\
g(u)&=(2J-zu^2-i(-iJ+xu))(2J-zu^2-i(-iJ+yu))\\
&=J^2-iJ(x+y)u-(xy+2zJ)u^2+iz(x+y)u^3+z^2u^4.
\end{aligned}
$$
Analogously, we put
$$
\begin{aligned}
\alpha(x,y,z;u)&=a(-iJ+xu,-iJ+yu,2J-zu^2;u^{-3})\\
&=\frac{(x-y)^2u^2+u^3[2J^2-2iJ(x+y)u
-(x^2+y^2+4zJ)u^2+2i(x+y)zu^3+2z^2u^4]}{2g^2(u)}.
\end{aligned}
$$

Using Taylor's expansion of $\varphi(x,y,z;u)$ w.r.t. $u$ at $u=0$,
we get
$$
\varphi(x,y,z;u)=1-\log J^2+\frac{i}{3J^3}[x^3+y^3-3zJ(x+y)]u^3+R(u),
$$
with
$$
R(u)=\frac{u^4}{3!}\int_0^1 d\tau\,(1-\tau)^3
\varphi^{(4)}(x,y,z;\tau u),
$$
$$
\varphi^{(4)}(x,y,z;u)=\frac{4!z^2}{g(u)}
-\frac{3(g^{\prime\prime}(u)^2+g'(u)g^{(3)}(u))}{g(u)^2}
+\frac{12g'(u)^2g^{\prime\prime}(u)}{g(u)^3}-\frac{6g'(u)^4}{g(u)^4},
$$
$$
|\partial_x^k\partial_y^\ell R(u)|\le C_{k,\ell}u^4
\quad\mbox{for $u\ge0$, $x,\,y\in\euc$, $|z|\le 1$, $k+\ell\le2$}.
$$
Moreover,
$$
e^{-NR(u)}=1+S(u),\quad
S(u)=-u^{-2}\int_0^1d\tau\, R'(\tau u),
$$
$$
R'(u)=\frac{2u^3}{3}\int_0^1 d\tau\,(1-\tau)^3
\varphi^{(4)}(x,y,z;\tau u)+\frac{u^5}{6}\int_0^1 d\tau\,(1-\tau)^3
\varphi^{(5)}(x,y,z;\tau u).
$$
Therefore, we have
$$
\exp{[-N\varphi(x,y,z;N^{-1/3})]}=e^{-N}J^{2N}
\exp{[-\frac{i}{3J^3}(x^3+y^3-3zJ(x+y))]}e^{-NR(N^{-1/3})}.
$$

On the other hand, as we have
$$
g(u)^{-2}=J^{-4}-2u\int_0^1 d\tau\,g'(\tau u)g(\tau u)^{-3},
$$
we get
$$ 
\alpha(x,y,z;u)=\frac{(x-y)^2}{2J^4}u^2+A(u),
\quad
A(u)=-u^3(x-y)^2\int_0^1 d\tau\,g'(\tau u)g(\tau u)^{-3}.
$$
with
$$
|\partial_x^k\partial_y^\ell A(u)|\le C_{k,\ell}u^3
\quad\mbox{for $u\ge0$, $x,\,y\in\euc$, $|z|\le 1$, $k+\ell\le2$}.
$$

Combining these, we get
\begin{equation}
\begin{aligned}
&\iint_{\euc^2} dt\,ds\,\exp{[-{N}\phi(t,s,2J-zN^{-2/3})]}
\,a(t,s,2J-zN^{-2/3};N)\\
&\qquad
=N^{-2/3}\iint_{\euc^2} dx\,dy\,
\exp{[-{N}\varphi(x,y,z;N^{-1/3})]}\alpha(x,y,z;N^{-1/3})\\
&\qquad
=e^{-N}J^{2N}N^{-2/3}
\iint_{\euc^2} dx\,dy\,
\exp{[-\frac{i}{3J^3}(x^3+y^3-3zJ(x+y))]}\\
&\qquad\qquad\qquad\times
(1+S(N^{-1/3}))
\bigg(\frac{(x-y)^2}{2J^4}N^{-2/3}+A(N^{-1/3})\bigg)\\
&\qquad
=e^{-N}J^{2N}\bigg[N^{-4/3}
\iint_{\euc^2} dx\,dy\,\frac{(x-y)^2}{2J^4}
\exp{[-\frac{i}{3J^3}(x^3+y^3-3zJ(x+y))]}\\
&\qquad\qquad\qquad\qquad\qquad\qquad\qquad\qquad\qquad
+O(N^{-5/3})\bigg].
\end{aligned}\label{aaa}
\end{equation}
Here, we applied the lemma below to
$$ 
f(x,y)=A(u),\; A(u)S(u),\; S(u)\frac{(x-y)^2}{2J^2}u^2,
$$
for getting the last term $O(N^{-5/3})$.   

Moreover, we may rewrite the above \eqref{aaa} using the Stirling formula to get
$$
\begin{aligned}
\big\langle\rho_N(2J-zN^{-2/3})\big\rangle_N
&=\frac{N^{-1/3}}{8\pi^2J^5}\iint_{\euc^2} dx\,dy\,
\times\exp{[-\frac{i}{3J^3}(x^3+y^3-3zJ(x+y))]}(x-y)^2\\
&\qquad\qquad\qquad\qquad\qquad\qquad\qquad\qquad\qquad\qquad
+O(N^{-2/3}).\qquad\qed
\end{aligned}
$$

\begin{lemma}
If $f$ satisfies
$$
|\partial_x^k\partial_y^\ell f(x,y)|\le C_{k,\ell},
$$
we have
$$
\bigg|\iint_{\euc^2}dxdy\, f(x,y)\exp{[-\frac{i}{3J^3}(x^3+y^3-3zJ(x+y))]}\bigg|
\le C<\infty.
$$
\end{lemma}
\par{\it Proof. }
We use Lax's technique (combining integration by parts with $\partial_qe^{i\phi(q)}=i\partial_q\phi(q)e^{i\phi(q)}$ where $|\partial_q\phi(q)|\neq{0}$) to estimate the oscillatory integrals noting
$$
(1-\partial_x^2-\partial_y^2)\exp{[-\frac{i}{3J^3}(x^3+y^3-3zJ(x+y))]}
=\Phi(x,y)\exp{[-\frac{i}{3J^3}(x^3+y^3-3zJ(x+y))]}
$$
with
$$
\Phi(x,y)=1+18z^2J^2+\frac{2i(x+y)}{J^3}+\frac{6z(x^2+y^2)}{J^2}
+\frac{x^4+y^4}{J^6}.
$$
Therefore, we have
$$
\begin{aligned}
&\iint_{\euc^2}dxdy\, f(x,y)\exp{[-\frac{i}{3J^3}(x^3+y^3-3zJ(x+y))]}\\
&\qquad
=\iint_{\euc^2}dxdy\, (1-\partial_x^2-\partial_y^2)\frac{f(x,y)}{\Phi(x,y)}
\exp{[-\frac{i}{3J^3}(x^3+y^3-3zJ(x+y))]}.
\end{aligned}
$$
By the assumption,
$(1-\partial_x^2-\partial_y^2)({f(x,y)}/{\Phi(x,y)})$ 
is integrable w.r.t. $dxdy$, we get the desired result. $\qquad\qed$

Using the Airy function defined by
$$
\Ai(z)=\int_{\euc} dx\,\exp{[-\frac{i}{3}x^3+izx]}
=\int_{\euc} dx\,\exp{[\frac{i}{3}x^3-izx]}=\overline{\Ai(z)}
\quad\mbox{for $z\in\euc$},
$$
we have
$$
\begin{aligned}
\int_{\euc}dx\,\exp{[-\frac{ix^3}{3J^3}+\frac{izx}{J^2}]}
&=J\Ai(\frac{z}{J}),\\
\int_{\euc}dx\,x\,\exp{[-\frac{ix^3}{3J^3}+\frac{izx}{J^2}]}
&=-iJ^2\Ai'(\frac{z}{J}),\\
\int_{\euc}dx\,x^2\,\exp{[-\frac{ix^3}{3J^3}+\frac{izx}{J^2}]}
&=-J^3\Ai^{\prime\prime}(\frac{z}{J}).
\end{aligned}
$$
And we get
$$
\big\langle\rho_N(2J-zN^{-2/3})\big\rangle_N
=N^{-1/3}f(zJ^{-1})+O(N^{-2/3}),
$$
where
$$
f(z)=\frac{1}{4\pi^2J}(\Ai'(z)\Ai'(z)-\Ai^{\prime\prime}(z)\Ai(z)).  \qquad\qed
$$

\begin{corollary} For $|z|\le 1$, we have
$$
\big\langle\rho_N(-2J+zN^{-2/3})\big\rangle_N
=-N^{-1/3}f(zJ^{-1})+O(N^{-2/3}).
$$
\end{corollary}
\begin{remark}
Though Br\'{e}zin and Kazakov applied the Br\'ezin formula (2.7) %\eqref{}
to obtain the analogous statement, 
but we can't follow their proof (48) of \cite{BK90}.
\end{remark}

\begin{center}

============   Mini Column 3:  {\bf On GOE, GUE and GSE} ============

\end{center}

In the following, to expect giving a trigger to audience having curiosity to RMT, we mention here Gaussian ensembles classified by F. Dyson. The Dyson index $\beta$ is defined by numbers of real components in matrix elements belonging to each ensemble.
Though I give a look to not only a book by Mehta~\cite{Meh91} but also survey paper by J. Verbaarschot~\cite{ver09}, it seems not sufficient to comprehend this interesting branch only ``a look''. 

\begin{definition}
A set of $N\times N$ Hermitian matrix $H=(H_{jk})$ whose matrix element is distributed with the probability
$$
P(H)dH=Z_{\beta N}^{-1}e^{-(N\beta/4)\tr H^2}dH %,\quad C_{1N}=\int_{E_{1G}}dH\,e^{-(N/4)\tr H^2}
$$
is called Wigner-Dyson Ensemble. Especially 
\begin{enumerate}
\item When $H=(H_{jk})$ is real symmetric matrix, then it's index is $\beta=1$ with volume element $dH=\prod_{k\le j}dH_{kj}$. This is called Gaussian Orthogonal Ensemble(=GOE).
\item When $H=(H_{jk})$ is complex Hermite matrix with $H_{kj}=H_{kj}^{(0)}+iH_{kj}^{(1)},\;H_{kj}^{(0)},H_{kj}^{(1)}\in\euc$, then it's index is $\beta=2$ with volume element $dH=\prod_{k\le j}dH_{kj}^{(0)}\prod_{k< j}dH_{kj}^{(1)}$.
This is called Gaussian Unitary Ensemble(=GUE).
\item  Let (real) quaternion with base $(1,{\pmb{i}},{\pmb{j}},{\pmb{k}})$ be decomposed as $H_{jk}=H_{jk}^{(0)}+H_{jk}^{(1)}{\pmb{i}}+H_{jk}^{(2)}{\pmb{j}}+H_{jk}^{(3)}{\pmb{k}}$, $H_{jk}^{(*)}\in\euc$.  When $H=(H_{jk})$ is Hermite whose matrix element is quaternion, then $\beta=4$ with volume element $dH=\prod_{k\le j}dH_{kj}^{(0)}\prod_{{\ell}=1}^3\prod_{k< j}dH_{kj}^{({\ell})}$. This is called Gaussian Symplectic Ensemble(=GSE).
\end{enumerate}
\end{definition}

\begin{remark}
We prepare letters ${\pmb{i}},\,{\pmb{j}},\,{\pmb{k}}$ with relations
${\pmb{i}}^2={\pmb{j}}^2={\pmb{k}}^2={\pmb{i}}{\pmb{j}}{\pmb{k}}=-1$.
For example, using Pauli matrices
$\{{\boldsymbol{\sigma}}_j\}_{j=1}^3$, we may put ${\pmb{i}}=i{\boldsymbol{\sigma}}_3$, ${\pmb{j}}=i{\boldsymbol{\sigma}}_2$, ${\pmb{k}}=i{\boldsymbol{\sigma}}_1$, then
we have a matrix representation
$$
q=q^{(0)}+q^{(1)}{\pmb{i}}+q^{(2)}{\pmb{j}}+q^{(3)}{\pmb{k}}\sim
\begin{pmatrix}
q^{(0)}+q^{(1)}i&q^{(2)}+q^{(3)}i\\
-q^{(2)}+q^{(3)}i&q^{(0)}-q^{(1)}i
\end{pmatrix}.
$$
\end{remark}

\begin{remark} Following facts are known:
\begin{enumerate}
\item GOE is invariant unde the similarity transformation ${\mathcal{T}}_1(O):H\mapsto {}^tOHO$ by $O$ belonging to $O(N)$=real orthogonal group,
\item GUE is invariant under the similarity transformation ${\mathcal{T}}_2(U):H\mapsto U^{-1}HU$ by $U$ belonging to $U(N)$=unitary group,
\item GSE is invariant under the similarity transformation ${\mathcal{T}}_4(S):H\mapsto {}^tSHS$ by $S$ belonging to 
$Sp(N)$=symplectic group.
\end{enumerate}
\end{remark}

Let $\lambda_1,{\cdots},\lambda_N$ be eigenvalues of $N\times N$ Hermite matrix. We check the relation between their differential $d\lambda_j$ and Lebesgue measure $dH$.
In case $H\in\mbox{GOE}$, since the number of independent components of $H_{jk}$ equals to $N(N+1)/2$,
 we have $\ell=N(N+1)/2-N=N(N-1)/2$ independent variables $\mu_m$ except $\{\lambda_j\}$.
Because
$$
\tr H^2=\sum_{j=1}^N\lambda_j^2
$$
and putting Jacobian of change of variables as
$$
J(\lambda,\mu)=\bigg|\det\bigg(\frac{\partial(H_{11},H_{12},{\cdots}, H_{NN})}{\partial(\lambda_1,{\cdots}\lambda_N,\mu_1,{\cdots},\mu_{\ell})}\bigg)|,
$$
we have
$$
dH=J(\lambda,\mu)\prod_{j=1}^Nd\lambda_j\prod_{k=1}^{\ell}d\mu_k.
$$
We need to calculate
$$
\iint d\lambda d\mu J(\lambda,\mu)=\int d\lambda \bigg[\int d\mu  J(\lambda,\mu)\bigg]?
$$
In other word, find the reason for the appearance of difference product of eigenvalues (van der Monde determinant)?

{\small [Report problem 6-1]: Give a precise description for above representation using eigenvalues (see, pp. 55-69
in Mehta~\cite{Meh91} or Theorem 5.22 of Deift~\cite{Dei00} for $\beta=2$\footnote{Frankly speaking, atlom has not enough patience to understand these facts, therefore he leaves these explanation for younger people expecting young gives stimulation to old}.}

\begin{theorem}[Theorem 3.3.1 of ~\cite{Meh91}]
Let $x_1,{\cdots},x_N$ be eigenvalues for a hermite matrix belonging to GOE $(\beta=1)$, GUE $(\beta=2)$ or GSE $(\beta=4)$.
Then, the joint probablity density of $x_1,{\cdots},x_N$ is given
$$
P_{N\beta}(x_1,{\cdots},x_N)=Z_{N\beta}^{-1}e^{-(\beta/2)\sum_{j=1}^Nx_j^2}\prod_{j<k}|x_j-x_k|^{\beta}.
$$
Here,
$$
Z_{N\beta}=(2\pi)^{N/2}\beta^{-N/2-\beta N(N-1)/4}[\Gamma(1+\beta/2)]^{-N}\prod_{j=1}^N\Gamma(1+\beta j/2).
$$
\end{theorem}

\begin{theorem}[Theorem 17.1 of Mehta~\cite{Meh91}] For any given positive integer $N$, we put $dx=dx_1{\cdots} dx_N$,
$$
\Delta(x)=\Delta(x_1,{\cdots},x_N)
=\begin{cases}
\prod_{1\le j<\ell\le N}(x_j-x_{\ell})&\mbox{if $N>1$},\\
1&\mbox{if $N=1$}
\end{cases}
$$ 
and
$$
\Phi(x)=\Phi(x_1,{\cdots},x_N)=|\Delta(x)|^{2\gamma}\prod_{j=1}^Nx_j^{\alpha-1}(1-x_j)^{\beta-1}.
$$
Then, for
$$
\Re \alpha>0,\; \Re \beta>0,\; \Re\gamma>-\min\bigg(\frac{1}{N}, \frac{\Re \alpha}{N-1}, \frac{\Re \beta}{N-1}\bigg),
$$
we have
$$
I(\alpha, \beta, \gamma, N)=\int_0^1{\cdots}\int_0^1dx \Phi(x)
=\prod_{j=0}^{N-1}\frac{\Gamma(1+\gamma+j\gamma)\Gamma(\alpha+j\gamma)\Gamma(\beta+j\gamma)}{\Gamma(1+\gamma)\Gamma(\alpha+\beta+(N+j-1)\gamma)}.
$$
\end{theorem}

\begin{center}
 
 =========== End of Mini Column 3 ========
 
\end{center}

\section{Relation between RMT and Painlev\'e transcendents}
%\subsection{Tracy and Widom's discovery}
It is shown rather recently that there is a mysterious connection between RMT, combinatorics and Painlev\'e functions. Borrowing the description of Tracy and Widom~\cite{TW99}, ~\cite{TW02ICM}, we explain our problem.

Let ${\mathfrak{U}}_{N}$ be a set of unitary ${N}\times {N}$ matrices with Haar measure.
Denoting (real) eigenvalues of $U\in{\mathfrak{U}}_{N}$ as $\{\lambda_1\le\lambda_2\le{\cdots}\le \lambda_N\}$
with its maximum $\lambda_N=\lambda_N(U)$, we consider  $P_2(\lambda_N(U)<t)$.    Then, we have                                                                                                              
\begin{theorem}[\cite{TW02ICM}]
$$
\lim_{N\to\infty}P_2\bigg(\frac{\lambda_N(U)-2N}{N^{-1/6}}<s\bigg)=F_2(s)=\exp\bigg(-\int_s^{\infty}ds\,(x-s)q^2(x)\bigg).
$$
Here, $q(s)$ is  Painleve II function satisfying
$$
q''=sq+2q^3\quad q(s)\sim \Ai(s)=\int_{\euc} dx\,\exp{[-\frac{i}{3}x^3+isx]}\quad\mbox{when $s\to\infty$}.
$$
\end{theorem}

Following theorem, which has curious resemblance to above, is proved by  J. Baik, P. Deift and K. Johansson~\cite{BDJ99}:
Putting uniform probability measure $P$ on symmetric group ${\mathcal{S}}_{N}$, we denote$\ell_N=\ell_N(\sigma)$ the length of the longest increasing subsequence for each $\sigma\in{\mathcal{S}}_{N}$. Then, we have 
\begin{theorem}[\cite{BDJ99}]
$$
\lim_{N\to\infty}P\bigg(\frac{\ell_N-2N}{N^{1/6}}\le s\bigg)=F_2(s).
$$
\end{theorem}

\paragraph{Table of Painlev\'e equations}:
$$
\begin{aligned}
(\mathrm{I})&\quad w''=6w^2+s,\\
(\mathrm{II})&\quad w''=2w^3+sw+\alpha,\\
(\mathrm{III})&\quad w''=\frac{{w'}^2}{w}-\frac{w'}{s}+\frac{\alpha w^2+\beta}{s}+\gamma w^3+\frac{\delta}{w},\\
(\mathrm{IV})&\quad w''=\frac{{w'}^2}{2w}+\frac{3w^3}{2}+4sw^2+2(s^2-\alpha)w+\frac{\beta}{w},\\
(\mathrm{V})&\quad w''={{w'}^2}\bigg(\frac{1}{2w}+\frac{1}{w-1}\bigg)-\frac{w'}{s}+\frac{(w-1)^2}{s^2}\bigg(\alpha w+\frac{\beta}{w}\bigg)+\frac{\gamma w}{s}+\frac{\delta w(w+1)}{w-1},\\
(\mathrm{VI})&\quad w''=\frac{{w'}^2}{2}\bigg(\frac{1}{w}+\frac{1}{w-1}+\frac{1}{w-s}\bigg)
-w'\bigg(\frac{1}{s}+\frac{1}{s-1}+\frac{1}{w-s}\bigg)\\
&\qquad\qquad\qquad\qquad\qquad\qquad
+\frac{w(w-1)(w-s)}{s^2(s-1)^2}\bigg(\alpha+\frac{\beta s}{w^2}+\frac{\gamma(s-1)}{(w-1)^2}+
\frac{\delta s(s-1)}{(w-s)^2}\bigg).
\end{aligned}
$$

\begin{problem} (i) Can we find another explanation of above theorems by  finding ``slowness variables'' as analogous as Efetov's reproof of Wigner's semi-circle law?\\
(ii) Moreover, the results, for example, C. Itzykson and J.B. Zuber~\cite{IZ80},  D. Bessis, C. Itzykson and J.B. Zuber~\cite{BIZ80} or A. Matytsin~\cite{mat94}, should be viewed from our point of view, but I have not enough intelligence to appreciate these works. By the way, the article by A. Zvonkin~\cite{zvo97} seems a nice guide in this direction.
\end{problem}

\begin{remark}
(i) To finish this lecture notes, I am almost drowned by checking by internet searching papers on RMT. 
For young researchers, T.Tao~\cite{Tao11} may be recommended. But this is also thick to begin with, therefore take a look to X. Zeng and Z. Hou~\cite{ZH12}.\\
(ii) There are also papers concerned about. Whether Matytsin's procedure~\cite{mat94} which tries to generalize Itzykson-Zuber formula has some relations to Functional Derivative Equation or not.
At least, my life in heaven or hell will be full of mathematical problems considered.
\end{remark}

\chapter{Fundamental solution of Free Weyl equation \`{a} la Feynman}
Because of the integration theory is not yet completed when lecture is delivered,  this chapter is chosen because here we only use the ``naive'' definition of integral on ${\fR}^{3|2}$.

Let $V$ be a representation space and let a function $\psi(t,q):\euc\times \euc^3\to V$ be given satisfying
\begin{equation}
\left\{
\begin{aligned}
&i\hbar \pdt \psi(t,q)={\hat{\mathbb{H}}}\psi(t,q), \quad
\hat{\mathbb{H}}=\hat{\mathbb{H}}(-i\hbar\partial_q)=c\boldsymbol\sigma_j\frac{\hbar}{i}\frac{\partial}{\partial q_j},\\
&\psi(0,q)=\underline{\psi}(q).
\end{aligned}
\right.
\label{fW-1.1}
\end{equation}
Here, %
$c$ and $\hbar$ are positive constants,
the summation with respect to  $j=1,2,3$ is abbreviated.
Put $\bf{I}$ as an identity map from $V$ to $V$, and define maps 
$\{\boldsymbol{\sigma}_i:V\to V\}$ satisfying
\begin{gather}
\boldsymbol\sigma_j\boldsymbol\sigma_k+\boldsymbol\sigma_k\boldsymbol\sigma_j
=2\delta_{jk}{\bf{I}} \quad\mbox{for $j,k=1,2,3,$} 
\quad (\mbox{Clifford relation})
\label{fW-1.3}\\
\boldsymbol\sigma_1\boldsymbol\sigma_2=i\boldsymbol\sigma_3,\quad 
\boldsymbol\sigma_2\boldsymbol\sigma_3=i\boldsymbol\sigma_1,\quad
\boldsymbol\sigma_3\boldsymbol\sigma_1=i\boldsymbol\sigma_2. \label{fW-1.4} 
\end{gather}
Especially when we put $V={\mathbb{C}}^2$ and $\psi(t,q)={}^t(\psi_1(t,q),\psi_2(t,q))$, we have so-called Pauli matrices:
\begin{equation}
\boldsymbol\sigma_1=\begin{pmatrix} 0&1\\ 1&0
\end{pmatrix},\quad
\boldsymbol\sigma_2=\begin{pmatrix} 0&-i\\ i&0
\end{pmatrix},\quad
\boldsymbol\sigma_3=\begin{pmatrix} 1&0\\ 0&-1
\end{pmatrix},\quad
{\bf{I}}={\mathbb{I}}_2=\begin{pmatrix} 1&0\\ 0&1
\end{pmatrix}.
\label{fW-1.2}
\end{equation}

This equation~\eqref{fW-1.1} is called free Weyl equation and we want to construct a fundamental solution of this modifying Feynman's procedure. (Modification is really necessary because Feynman doesn't proceed as he wants at that time).

Before this, we recall a very primitive and well-known method mentioned in Chapter 1.
Applying formally the Fourier transformation (which contains a parameter $\hbar$) 
with respect to  $q\in\euc^3$
to \eqref{fW-1.1}, we get 
$$
i\hbar\pdt{\hat\psi}(t,p)={\mathbb{H}}{\hat\psi}(t,p)
\where {\mathbb{H}}={\mathbb{H}}(p)=c\boldsymbol\sigma_jp_j=c
\begin{pmatrix}
p_3&p_1-ip_2\\
p_1+ip_2&-p_3
\end{pmatrix}.
$$
As ${\mathbb{H}}^2=c^2|p|^2{\mathbb{I}}_2$ by \eqref{fW-1.3}, we easily have
\begin{equation}
e^{-i\hbar^{-1}t {\mathbb{H}}}\hat{\underline{\psi}}(p)
=\big[\cos(c\hbar^{-1}t|p|){\mathbb{I}}_2-ic^{-1}|p|^{-1}\sin(c\hbar^{-1}t|p|) {\mathbb{H}}\big]
\hat{\underline{\psi}}(p).
\label{fW-matrix-sol}
\end{equation}
In other word, denoting $\gamma_t=c\hbar^{-1}t|p|$,
\begin{equation}
\begin{aligned}
&\begin{pmatrix}
\hat{\psi}_1(t,p)\\
\hat{\psi}_2(t,p)
\end{pmatrix}
=e^{-i\hbar^{-1}t {\mathbb{H}}}\hat{\underline{\psi}}(p)\\
&\quad=\frac{1}{|p|}
\begin{pmatrix}
|p|\cos\gamma_t-ip_3\sin\gamma_t &-i(p_1-ip_2)\sin\gamma_t \\
-i(p_1+ip_2)\sin\gamma_t &|p|\cos\gamma_t+ip_3\sin\gamma_t
\end{pmatrix}
\begin{pmatrix}
\hat{\underline{\psi}_1}(p)\\
\hat{\underline{\psi}_2}(p)
\end{pmatrix}.
\end{aligned}
\label{fW-matrix-sol2}
\end{equation}

Therefore, we have
\begin{proposition}
For any $t\in\euc$,
\begin{equation}
e^{-i\hbar^{-1}t\hat{\mathbb{H}}}{\underline{\psi}}(q)=(2\pi\hbar)^{-3/2}\int_{\euc^3}dp\,
e^{i\hbar^{-1}qp}e^{-i\hbar^{-1}t\hat{\mathbb{H}}}\hat{\underline{\psi}}(p)
=\int_{\euc^3}dq'\,{\mathbb{E}}(t,q,q'){\underline{\psi}}(q')
\label{fW-1.5}
\end{equation}
with
\begin{equation}
{\mathbb{E}}(t,q,q')=
(2\pi\hbar)^{-3}\int_{\euc^3}dp\,e^{i\hbar^{-1}(q-q')p}
\big[\cos(c\hbar^{-1}t|p|){\mathbb{I}}_2-ic^{-1}|p|^{-1}\sin(c\hbar^{-1}t|p|){\mathbb{H}}(p)\big].
\label{fW-1.6}
\end{equation}
\end{proposition}

\begin{remark}
This calculation doesn't work when \eqref{fW-1.1} is changed to
\begin{equation}
\hat{\mathbb{H}}(t)=\sum_{j=1}^3c\pmb\sigma_j\bigg(\frac{\hbar}{i}\frac{\partial}{\partial q_j}-\frac{e}{c}A_j(t,q)\bigg)
+eA_0(t,q).
\label{eW-1.1}
\end{equation}
We don't insist on this equation having physical meaning because we add minimally electro-magnetic potential to \eqref{fW-1.1} very formally. But \underline{it seems mathematically interesting to seek a solution of this}\\
\underline{toy model represented as Feynman suggested like below}.
\end{remark}

Here, we give another formula for a solution of \eqref{fW-1.1}.
\begin{theorem}[Path-integral representation of a solution for the free Weyl equation]\label{fW-pi-rep}
\begin{equation}
\psi(t,q)=\left.\flat\Big((2\pi\hbar)^{-3/2}\hbar\iint_{{\mathfrak{R}}^{3|2}} d{\unbxi}d{\unbpi}\,
{\mathcal{D}}^{1/2}(t,{\mixxdata})
e^{i\hbar^{-1}{{\mathcal{S}}}(t,{\mixxdata})}
{\mathcal{F}}(\sharp{\underline{\psi}})({\unbxi},{\unbpi})\Big)\right|_{\barx_{\mathrm B}=q}.
\end{equation}
Here, ${\mathcal{S}}(t,{\mixxdata})$ and ${\mathcal{D}}(t,{\mixxdata})$ are solutions 
of the Hamilton-Jacobi and continuity equations, 
\eqref{fW-1.16} and \eqref{fW-1.18} respectively, and $\sharp$, $\flat$ are given in \eqref{sharp-flat} below.
\end{theorem}

\section{A strategy of constructing parametrices} % for Dirac \eqref{eD1.1-1} or Weyl \eqref{eW-1.1}}
Taking the free Weyl equation as a simplest model of constructing parametrices for Dirac \eqref{eD1.1-1} in Chapter 1 or Weyl \eqref{eW-1.1} above, we explain our strategy which is a superly extended version of what we explained before.
\noindent
\newline
(1) Is it possible to extract ``symbol'' corresponding to a given system of PDO's?
\newline
(2) To define ``symbol'', we need to represent the matrix structure as the differential operators acting on superspace.
For example, to regard $2\times 2$ matrix structure as differential operators, we decompose
$$
\begin{pmatrix}
a&c\\
d&b
\end{pmatrix}
=\frac{a+b}{2}\begin{pmatrix}
1&0\\
0&1
\end{pmatrix}
+\frac{a-b}{2}
\begin{pmatrix}
1&0\\
0&-1
\end{pmatrix}
+\frac{c+d}{2}
\begin{pmatrix}
0&1\\
1&0
\end{pmatrix}
+i\frac{c-d}{2}
\begin{pmatrix}
0&-i\\
i&0
\end{pmatrix},
$$
and attach differential operators for each $\boldsymbol{\sigma}_j$ and we get the symbol of that system of PDO's.
\newline
(3) For the Hamiltonian ${\mathcal{H}}(x,\xi,\theta,\pi)$, we construct a solution ${\mathcal{S}}(t,x,\xi,\theta,\pi)$ of the Hamilton-Jacobi equation with given initial data.
\newline
(4) Calculating the Hessian of ${\mathcal{S}}(t,x,\xi,\theta,\pi)$ w.r.t. $\xi, \pi$, we define
its determinant below, called super Van Vleck determinant:
$$
{\mathcal{D}}(t,x,\xi,\theta,\pi)=\sdet 
\begin{pmatrix}
\frac{\partial^2{\mathcal{S}}}{\partial x_j\partial \xi_k}&\frac{\partial^2{\mathcal{S}}}{\partial x_j\partial {\pi_b}}\\
\frac{\partial^2{\mathcal{S}}}{\partial \theta_a\partial \xi_k}&\frac{\partial^2{\mathcal{S}}}{\partial \theta_a\partial {\pi_b}}
\end{pmatrix}
\with j.k=1,2,3, \; a,b=1,2.
$$
\newline
(5) Finally, defining an operator
$$
%u(t,x,\theta)=
({\mathcal{U}}(t){\unbu})(x,\theta)=
(2\pi\hbar)^{-3/2}\hbar\iint_{{\mathfrak{R}}^{3|2}} d{\unbxi}d{\unbpi}\,
{\mathcal{D}}^{1/2}(t,{\mixxdata})
e^{i\hbar^{-1}{{\mathcal{S}}}(t,{\mixxdata})}
{\mathcal{F}}{\unbu}({\unbxi},{\unbpi}),
$$
we check its properties and show that it's the desired parametrix (or fundamental solution for this free Weyl case).

\section{Sketchy proofs of the procedure mentioned  above} 
\noindent
\par
(1)  A ``spinor"
$\psi(t,q)={}^t(\psi_1(t,q),\psi_2(t,q))
:{\euc}\times{\euc}^3\to {\mathbb{C}}^2$ is identified with an even supersmooth function $u(t,x,\theta)=u_0(t,x)+u_1(t,x)\theta_1\theta_2:
{\euc}\times{\mathfrak{R}}^{3|2}\to \cev$ as follows:
\begin{equation}
\begin{gathered}
{\mathbb{C}}^2\ni\begin{pmatrix}
\psi_1\\
\psi_2\end{pmatrix}
\;\begin{matrix}{\overset{\sharp}{\rightarrow}}\\[-9pt]
{\underset{\flat}{\leftarrow}}\end{matrix}\;
u(\theta)=u_0+u_1\theta_1\theta_2\in{\cev}\\
\with u_0=u(0)=\psi_1,\;u_1=\partial_{\theta_2}\partial_{\theta_1}u(\theta)\big|_{\theta=0}=\psi_2.
\end{gathered}
\label{sharp-flat}
\end{equation}
Here, functions $u_0(t,x)$, $u_1(t,x)$
are the Grassmann continuation of $\psi_1(t,q)$,
$\psi_2(t,q)$, respectively.
\par
(2) Pauli matrices $\{\boldsymbol\sigma_j\}$ have differential operator representations
\begin{equation}
\begin{aligned}
&\sigma_1\Big(\theta,\fraccli
\frac{\partial}{\partial\theta}\Big)=
i\clifpin^{-1}\Big(\theta_1\theta_2
+\clifpin^2\frac{\partial^2}{\partial\theta_1\partial\theta_2}\Big),\\
&\sigma_2\Big(\theta,\fraccli
\frac{\partial}{\partial\theta}\Big)=
-\clifpin^{-1}\Big(\theta_1\theta_2
-\clifpin^2\frac{\partial^2}{\partial\theta_1\partial\theta_2}\Big),\\
&\sigma_3\Big(\theta,\fraccli
\frac{\partial}{\partial\theta}\Big)=
1-\theta_1\frac{\partial}{\partial\theta_1}
-\theta_2\frac{\partial}{\partial\theta_2},
\end{aligned}\label{fW-1.7}
\end{equation}
satisfying \eqref{fW-1.3} and \eqref{fW-1.4}.
Here, we take an arbitrarily chosen parameter $\clifpin\in {\mathbb{C}}^{\times}={\mathbb{C}}-\{0\}$.
\begin{remark} Only for $|\clifpin|=1$,
$\{\flat\sigma_j(\theta,-i\clifpin\partial_\theta)\sharp\}$ are unitary matrices.
From here, we take \fbox{$\clifpin=i$} to have the matrix representation \eqref{fW-1.2} for \eqref{fW-1.7}.
\end{remark}

(3)  Since, using \eqref{fW-1.7}, the differential operator ${\mathbb{H}}$ given by \eqref{fW-1.1} is identified with
\begin{equation}
\begin{aligned}
\hat{\mathcal{H}}_0\Big(\frachi \pdx,\theta,{\pdtheta}\Big)
=-ic\hbar
\Big(\theta_1\theta_2-\frac{\partial^2}{\partial\theta_1\partial\theta_2}\Big)
\frac{\partial}{\partial x_1}
+& c\hbar
\Big(\theta_1\theta_2+\frac{\partial^2}{\partial\theta_1\partial\theta_2}\Big)
\frac{\partial}{\partial x_2}\\
&\quad
-ic\hbar\Big(1-\theta_1\frac{\partial}{\partial\theta_1}
-\theta_2\frac{\partial}{\partial\theta_2}\Big)
\frac{\partial}{\partial x_3}
\end{aligned}
\label{fW-1.8}
\end{equation}
where the ``ordinary symbol'' of the operator \eqref{fW-1.8} is considered as
 \begin{equation}
{\mathcal{H}}_0(\xi,\theta,\pi)=c(\xi_1+i\xi_2)\theta_1\theta_2
+c\spin^{-2}(\xi_1-i\xi_2)\pi_1\pi_2
+c\xi_3[1-i\spin^{-1}(\theta_1\pi_1+\theta_2\pi_2)].
\label{fW-1.8bis}
\end{equation}

Therefore, the super-version of Weyl equation is given
\begin{equation}
\left\{
\begin{aligned}
&i\hbar \pdt u(t,x,\theta)
=\hat{\mathcal{H}}_0\Big(\frachi \pdx,\theta,{\pdtheta}\Big)u(t,x,\theta),\\
&u(0,x,\theta)=u(x,\theta).
\end{aligned}
\right.
\label{fW-1.9}
\end{equation}
On the other hand, ``complete Weyl symbol'' of the right-hand side of the differential operator \eqref{fW-1.8}
is given by
\begin{equation}
{\mathcal{H}}(\xi,\theta,\pi)
=c(\xi_1+i\xi_2)\theta_1\theta_2
+c\spin^{-2}(\xi_1-i\xi_2)\pi_1\pi_2
-ic\spin^{-1}\xi_3(\theta_1\pi_1+\theta_2\pi_2).
\label{fW-1.10} 
\end{equation}
Here, $\spin\in {\euc}^{\times}$ or $\in i{\euc}^{\times}$ 
(${\euc}^{\times}={\euc}\setminus\{0\}$) is a parameter introduced for Fourier transformations of even and odd variables:
$$
\begin{gathered}
(F_{\mathrm{e}} v)(\xi) =(2\pi\hbar)^{-m/2}\int_{\supermo} dx\,
e^{-i\hbar^{-1}\langle x|\xi\rangle}v(x),\\
({\bar{F}}_e w)(x) =(2\pi\hbar)^{-m/2}\int_{\supermo} d\xi\,
e^{i\hbar^{-1}\langle x|\xi\rangle}w(\xi),\\
(F_o v)(\pi) =\kbar^{n/2} \iota_n
\int_{\superon} d\theta\,
e^{-i\kbar^{-1}\langle\theta|\pi\rangle}v(\theta),\\
({\bar{F}}_o w)(\theta) =
\kbar^{n/2}\iota_n
\int_{\superon} d\pi\,
e^{i\kbar^{-1}\langle\theta|\pi\rangle}w(\pi),
\end{gathered}
$$ 
which are explained more precisely later in \S2 of Chapter 9. Here,
$$
\langle\eta|y\rangle = \sum_{j=1}^m \eta_j y_j,
\quad
\langle\rho|\omega\rangle = \sum_{k=1}^n \rho_k \omega_k ,
\quad
\iota_n=e^{-\pi{i}n(n-2)/4}.
$$ 
Moreover,
$$ 
\begin{gathered}
({\mathcal F}u)(\xi,\pi)
= c_{m,n}
\int_{\supermn}\!\!dX\,e^{-i\hbar^{-1}\langle X|\Xi\rangle} u(X) 
=\sum_a[(F_{\mathrm{e}}u_a)(\xi)][(F_o\theta^a)(\pi)],\\
({\bar {\mathcal F}}v)(x,\theta)
= c_{m,n}
\int_{\supermn}\!\! d\Xi\,e^{i\hbar^{-1}\langle X|\Xi\rangle }v(\Xi)
=\sum_a[({\bar{F}}_ev_a)(x)][({\bar{F}}_o\pi^a)(\theta)]
\end{gathered}
$$
with
$$
\langle X|\Xi\rangle=\langle x|\xi\rangle
+\hbar\kbar^{-1}\langle\theta|\pi\rangle\in\rev,\quad 
c_{m,n}=(2\pi\hbar )^{-m/2}\kbar^{n/2}\iota_n.
$$
More explicitly, we have
\begin{example}[$n=1$]
$$
\left\{
\begin{aligned}
&e^{i\pi/4}\spin^{1/2}\int_{{\fR}^{0|1}}d\theta\,e^{-i{\spin}^{-1}\theta\pi}(u_0+u_1\theta)
=e^{i\pi/4}\spin^{1/2}(u_1-i{\spin}^{-1}u_0\pi),\\
&e^{i\pi/4}\spin^{1/2}\int_{{\fR}^{0|1}}d\pi\,e^{i{\spin}^{-1}\theta\pi}(u_1-i{\spin}^{-1}u_0\pi)
=u_0+u_1\theta.
\end{aligned}
\right.
$$
\end{example}
\begin{example}[$n=2$]
For $u(\theta)=u_0+\theta_1\theta_2u_1$ and $v(\pi)=\pi_1 v_1+\pi_2 v_2$ with
$u_0, u_1, v_1, v_2\in\fC$, we have
$$
\begin{aligned}
({F}_{\mathrm{o}}u)(\pi)&={\spin}\int_{{\fR}^{0|2}}d\theta\,e^{-i{\spin}^{-1}\langle \theta|\pi\rangle}u(\theta)
={\spin}(u_1+{\spin}^{-2}\pi_1\pi_2 u_0),\\
({\bar{F}}_{\mathrm{o}}v)(\theta)&={\spin}\int_{{\fR}^{0|2}}d\pi\,e^{i{\spin}^{-1}\langle \theta|\pi\rangle}v(\pi)
={\spin}(-i{\spin}^{-1}\theta_2v_1+i{\spin}^{-1}\theta_1v_2),\\
{\bar{F}}_{\mathrm{o}}({F}_{\mathrm{o}}u)(\theta)&=
{\spin}\int_{{\fR}^{0|2}}d\pi\,e^{i{\spin}^{-1}\langle \theta|\pi\rangle}[{\spin}(u_1+{\spin}^{-2}\pi_1\pi_2 u_0)]
=u_0+\theta_1\theta_2u_1=u(\theta),\\
{F}_{\mathrm{o}}({\bar{F}}_{\mathrm{o}}v)(\pi)&=
{\spin}\int_{{\fR}^{0|2}}d\theta\,e^{-i{\spin}^{-1}\langle \theta|\pi\rangle}[{\spin}(-i{\spin}^{-1}\theta_2v_1+i{\spin}^{-1}\theta_1v_2)]
=\pi_1 v_1+\pi_2 v_2=v(\pi).
\end{aligned}
$$
\end{example}

(A) From here, we mention \fbox{Jacobi's method} to construct a solution of Hamilton-Jacobi equation:

(4) Consider classical mechanics or Hamilton flow corresponding to ${\mathcal{H}}(\xi,\theta,\pi)$
\begin{equation}
\left\{
\begin{aligned}
&\dt x_j=\frac{\partial{\mathcal{H}}(\xi,\theta,\pi)}{\partial \xi_j},\quad
\dt \xi_k=-\frac{\partial{\mathcal{H}}(\xi,\theta,\pi)}{\partial x_k}=0
\for j,k=1,2,3,\\
&\dt \theta_a=-\frac{\partial{\mathcal{H}}(\xi,\theta,\pi)}{\partial \pi_a},\quad
\dt \pi_b=-\frac{\partial{\mathcal{H}}(\xi,\theta,\pi)}{\partial \theta_b}
\for a,b=1,2,3.
\end{aligned}
\right.
\label{fW-1.11}
\end{equation}

\begin{proposition}[existence]\label{prop:existence}
Under above setting, for any initial data $(x(0),\xi(0),\theta(0),\pi(0))=({\unbx,\unbxi,\unbtheta,\unbpi})
\in{\fR}^{6|4}$, \eqref{fW-1.11} has a unique solution $(x(t),\xi(t),\theta(t),\pi(t))$.
\end{proposition}

\begin{remark}
(i) The solution of above Proposition is denoted by $x(t)$ or 
$x(t,{\unbx,\unbxi,\unbtheta,\unbpi})$, etc.
\newline
(ii) Instead of ${\fR}^{3|2}\times{\fR}^{3|2}$, {we} regard ${\fR}^{6|4}$ as  the
cotangent space ${\mathcal T}^*{\fR}^{3|2}$ of ${\fR}^{3|2}$.
\end{remark}

\paragraph{\bf Inverse mapping:} 
\begin{proposition}[inverse]\label{prop:inverse}
For any $(t,{\unbxi},{\unbpi})$, the map defined by
$$
({\unbx},{\unbtheta})\mapsto
(\barx=x(t,{\unbx,\unbxi,\unbtheta,\unbpi}),\bartheta=\theta(t,{\unbx,\unbxi,\unbtheta,\unbpi}))
$$ 
is supersmooth from ${\fR}^{3|2}$ to ${\fR}^{3|2}$. The inverse map of this, defined by
$$
({\barx},{\bartheta})\to
(\unbx=y(t,{\mixdata}), \unbtheta=\omega(t,{\mixdata})), 
$$
satisfies
\begin{equation}
\left\{
\begin{aligned}
&{\barx}=x(t,y(t,{\mixdata}),\unbxi,\omega(t,{\mixdata}),\unbpi),\quad
{\bartheta}=\theta(t,y(t,{\mixdata}),\unbxi,\omega(t,{\mixdata}),\unbpi),\\
&{\unbx}=y(t,x(t,{\unbx,\unbxi,\unbtheta,\unbpi}),\unbxi,\theta(t,{\unbx,\unbxi,\unbtheta,\unbpi}),\unbpi),\quad
{\unbtheta}=\omega(t,x(t,{\unbx,\unbxi,\unbtheta,\unbpi}),\unbxi,\theta(t,{\unbx,\unbxi,\unbtheta,\unbpi}),\unbpi)
\end{aligned}
\right.
\label {fW-1.12}
\end{equation}
\end{proposition}

\paragraph{\bf Action integral:}
For notational simplicity, we introduce the following short-hand symbols in this chapter:
\begin{equation}
\begin{gathered}
\gamma_t=\gamma(t,\unbxi)=c\spin^{-1}t|\unbxi|,\;\;\zeta=\unbxi_1+i\unbxi_2,\;\;\bar{\zeta}=\unbxi_1-i\unbxi_2,\;\;
|\unbxi|^2=\sum_{j=1}^3{\unbxi}_j^2=|{\zeta}|^2+\unbxi_3^2, \\
\delta(t)=\delta(t,\unbxi)=|\unbxi|\cos\gamma_t+i\unbxi_3\sin\gamma_t,\;\;
\bar{\delta}(t)=\bar{\delta}(t,\unbxi)=|\unbxi|\cos\gamma_t-i\unbxi_3\sin\gamma_t.
\end{gathered}
\label{fW-shn} %short-hand notation
\end{equation}
Putting
\begin{equation}
{\mathcal{S}}_0(t,{\unbx,\unbxi,\unbtheta,\unbpi})=\int_0^t ds\,\{\langle \dot x(s)|\xi(s) \rangle
+\langle \dot\theta(s)|\pi(s)\rangle -{\mathcal{H}}(\xi(s),\theta(s),\pi(s))\},
\label {fW-1.13}
\end{equation}
and
\begin{equation}
{\mathcal{S}}(t,{\mixdata})=\langle {\unbx}|{\unbxi}\rangle+
\hbar\spin^{-1}\langle {\unbtheta}|{\unbpi}\rangle
+{\mathcal{S}}_0(t,{\unbx,\unbxi,\unbtheta,\unbpi})\Bigg|
_{\stackrel{\scriptstyle{{\unbx}=y(t,{\mixdata})}}{{\unbtheta}=\omega(t,{\mixdata})}},\\
\label {fW-1.14}
\end{equation}
{we} have
\begin{proposition}[phase function]\label{prop:HJ}
Above defined ${\mathcal{S}}(t,{\mixdata})$ is represented
\begin{equation}
{\mathcal{S}}(t,{\mixdata})=\langle\barx|\unbxi\rangle
+{\bar{\delta}}(t)^{-1}\big[\hbar\spin^{-1}|\unbxi|\langle\bartheta|\unbpi\rangle-\spin\zeta \sin\gamma_t
\bartheta_1\bartheta_2
-\spin^{-1}(2\hbar\spin^{-1}-1)\bar{\zeta}\sin\gamma_t
\unbpi_1\unbpi_2\big].
\label{fW-7.2.9bis}
\end{equation}
Moreover, if  $\hbar=\spin$, then it satisfies Hamilton-Jacobi equation:
\begin{equation}
\left\{
\begin{aligned}
&\pdt {\mathcal{S}}(t,{\mixdata})+{\mathcal{H}}
\Big(\frac{\partial {\mathcal{S}}}{\partial\bar{x}},
\bar{\theta},\frac{\partial {\mathcal{S}}}{\partial \bar{\theta}}\Big)=0,\\
&{\mathcal{S}}(0,{\mixdata})=\langle \bar{x}|{\unbxi}\rangle+
\langle\bar{\theta}|{\unbpi}\rangle. 
\end{aligned}
\right.
\label {fW-1.16}
\end{equation}
\end{proposition}

Then, {we}  put
\begin{equation}
{\mathcal D}(t,{\mixdata})=\sdet 
\begin{pmatrix}
\frac{\partial^2{\mathcal{S}}}{\partial \bar{x}\,\partial {\unbxi}}&
\frac{\partial^2{\mathcal{S}}}{\partial \bar{x}\,\partial {\unbpi}}\\
\frac{\partial^2{\mathcal{S}}}{\partial \bar{\theta}\,\partial {\unbxi}}&
\frac{\partial^2{\mathcal{S}}}{\partial \bar{\theta}\,\partial {\unbpi}}
\end{pmatrix}.
\label {fW-1.17}
\end{equation}
Here, ``$\sdet$"  stands for super-determinant.

\begin{proposition}[amplitude function]\label{prop:continuity} By calculation, we have
\begin{equation}
{\mathcal D}(t,{\mixdata})=(\hbar^{-1}\spin)^2|\unbxi|^{-2}\bar{\delta}(t)^2.
\label {fW-1.17b}
\end{equation}
Moreover, if $\hbar=\spin$, then it satisfies the continuity equation (or the 0th part of transport equation):
\begin{equation}
\left\{
\begin{aligned}
&\pdt {\mathcal D}
+\frac{\partial}{\partial\bar{x}}
\Big({\mathcal D}
\frac{\partial{\mathcal{H}}}{\partial {\xi}}\Big)
+\frac{\partial}{\partial \bar{\theta}}
\Big({\mathcal D}
\frac{\partial{\mathcal{H}}}{\partial {\pi}}\Big)=0,\\
&{\mathcal D}(0,{\mixdata})=1.
\end{aligned}
\right.
\label {fW-1.18}
\end{equation}
In the above, the independent variables of ${\mathcal D}$ are $(t,{\mixdata})$,
those of ${\partial{\mathcal{H}}}/{\partial \xi}$ or
${\partial{\mathcal{H}}}/{\partial \pi}$ are
$({\mathcal{S}}_{\barx}, \bar\theta,{\mathcal{S}}_{\bar{\theta}})$.
\end{proposition}

(B) There exists another method (in the next section) to solve \eqref{fW-1.16} with the initial data
$$
{\mathcal{S}}(0,{\mixdata})=\langle \bar{x}|{\unbxi}\rangle+\hbar\spin^{-1}
\langle\bar{\theta}|{\unbpi}\rangle,
$$
for any $\hbar$ and $\spin$. Curiously, by that method, we have a solution slightly different from \eqref{fW-7.2.9bis} but coincides when $\hbar=\spin$ (I haven't found the reason why so!).

(5) Quantization:
Using these classical quantities ${\mathcal{S}}$ and ${\mathcal{A}}$, we have a new representation of solution desired as follows:
In this paragraph, we rewrite  variables from 
$$
{\fR}^{6|4}\ni(\barx,\unbxi,\bartheta,\unbpi)\to
(\barx,\bartheta,\unbxi,\unbpi)\in {\fR}^{3|2}\times{\fR}^{3|2}={\mathcal{T}}^*{\fR}^{3|2}.
$$
We define an operator
\begin{equation}
\begin{aligned}
({\mathcal{U}}(t)u)(\bar{x},\bar\theta)&=(2\pi\hbar)^{-3/2}\spin\iint d{\unbxi}d{\unbpi}\,
{\mathcal{D}}^{1/2}(t,{\mixxdata})
e^{i\hbar^{-1}{{\mathcal{S}}}(t,{\mixxdata})}
{\mathcal{F}}u({\unbxi},{\unbpi}),
\end{aligned}
\label {fw-1.19}
\end{equation}
where ${\mathcal{F}}$ stands for the Fourier transformation defined 
for functions on the superspace.
The function
\underline{$u(t,\bar{x},\bar\theta)=({\mathcal{U}}(t){\underline{u}})(\bar{x},\bar\theta)$ 
will be shown as a desired solution for \eqref{fW-1.9} if $\hbar=\spin$}.

It is shown that
\begin{equation}
\hat{\mathcal{H}}_0\Big({\frachi}{\pdx},\theta,{\pdtheta}\Big)={\hat{\mathcal{H}}}^{W}
={\hat{\mathcal{H}}}^{W}\Big({\frachi}{\pdx},\theta,{\pdtheta}\Big)
\end{equation}
where the left-hand side is given \eqref{fW-1.8} with $\clifpin=i$ and ${\hat{\mathcal{H}}}^{W}$ is a (Weyl type) pseudo-differential operator with
symbol ${\mathcal{H}}(\xi,\theta,\pi)$ defined by 
\begin{equation} 
({\hat{\mathcal{H}}}^{W}u)(x,\theta)=(2\pi\hbar)^{-3}\spin^2\iint d\xi d\pi dyd\omega \, 
e^{i\hbar^{-1}
\langle x-y|\xi\rangle +i\spin^{-1}\langle \theta-\omega|\pi\rangle}
{\mathcal{H}}\bigg(\xi,\frac{\theta+\omega}2,\pi\bigg)u(y,\omega).
\end{equation}

\begin{proposition}
(1) For $t\in\euc$,
${\mathcal{U}}(t)$ is a well defined  unitary operator in 
$\clsl_{\mathrm{SS,ev}}^2({\fR}^{3|2})$ if $\hbar=\spin$.
\newline
(2) 
(i) $\euc\ni t\mapsto {\mathcal{U}}(t)\in {\Bbb
B}(\,\clsl_{\mathrm{SS,ev}}^2({\fR}^{3|2}),\,\clsl_{\mathrm{SS,ev}}^2({\fR}^{3|2}))$ 
is continuous.
\newline
(ii) ${\mathcal{U}}(t){\mathcal{U}}(s)={\mathcal{U}}(t+s)$ for any $t,s\in\euc$.
\newline
(iii) 
For ${\underline{u}}\in \ccsl_{\mathrm{SS,ev},0}({\mathfrak{R}}^{3|2})$, we put 
$u(t,\barx,\bartheta)=({\mathcal{U}}(t){\underline{u}})(\barx,\bartheta)$.
Then, it satisfies
\begin{equation}
\left\{
\begin{aligned}
&i\hbar\pdt u(t,\barx,\bartheta)={\hat{\mathcal{H}}}^{W} u(t,\barx,\bartheta),\\
& u(0,\barx,\bartheta)={\underline{u}}(\barx,\bartheta).
\end{aligned}
\right.
\end{equation}
\end{proposition}

Finally, we interprete the above theorem with $\hbar=\spin$ using
the identification maps
\begin{equation}
\sharp:L^2(\euc^3:{\mathbb{C}}^2)\to \clsl_{\mathrm{SS,ev}}^2({\mathfrak{R}}^{3|2}) \et 
\flat:\clsl_{\mathrm{SS,ev}}^2({\mathfrak{R}}^{3|2})\to L^2(\euc^3:{\mathbb{C}}^2).
\end{equation}
That is, remarking 
$\flat\,{\hat{\mathcal{H}}^W}\,\sharp \psi=\hat{\mathbb{H}}\psi$
and putting ${\mathbb{U}}(t)\psi=\flat\,{\mathcal{U}}(t)\,\sharp\psi$,
we have
\begin{proposition}
(1) For $t\in\euc$,
${\mathbb{U}}(t)$ is a well defined unitary operator in $L^2(\euc^3:{\mathbb{C}}^2)$.
\newline
(2) 
(i) $\euc\ni t\mapsto {\mathbb{U}}(t)\in {\Bbb
B}(L^2(\euc^3:{\mathbb{C}}^2),L^2(\euc^3:{\mathbb{C}}^2))$ is continuous. 
\newline
(ii) ${\mathbb{U}}(t){\mathbb{U}}(s)={\mathbb{U}}(t+s)$ for any $t,s\in\euc$.
\newline
(iii) Put $\clifpin=i$.
For ${\underline{\psi}}\in C_0^\infty(\euc^3:{\mathbb{C}}^2)$,
we put 
$\psi(t,q)=\flat\big({\mathcal{U}}(t)\sharp{\underline{\psi}}\big)\big|_{\barx_{\mathrm B}=q}$. 
Then, it satisfies 
\begin{equation}
\left\{
\begin{aligned}
&i\hbar\pdt \psi(t,q)=\hat{\mathbb{H}}\psi(t,q),\\
&\psi(0,q)={\underline{\psi}}(q).
\end{aligned}
\right.
\end{equation}
\end{proposition}

\begin{corollary}
${\mathbb{H}}$ is an essentially self-adjoint operator in $L^2(\euc^3:{\mathbb{C}}^2)$.
\end{corollary}

\begin{remark}
Since the free Weyl equation is simple, it is not necessary to use Fr\'echet-Grassmann algebra with countably infinite
Grassmann generators, because I can construct ``classical quantities'' explicitly. This point is exemplified in the last chapter
for the construction of Hamilton flow for the Weyl equation with electro-maganetic potentials.
Odd variables are symbolically important for presenting the position in matrix structure and for operations being consistent. In this case, odd and even variables are rather separated without interaction, therefore simple!
\end{remark}
\begin{claim} \label{fW-claim7-2-1}
\eqref{fw-1.19} reduces to \eqref{fW-matrix-sol} after integration w.r.t. $d\unbpi$.
\end{claim}

Proof of this claim will be given later.

\section{Another construction of the solution for H-J equation}
In the above explanation, most essential part is to define ``phase'' function satisfying Hamilton-Jacobi equation.
We introduce another new method here to seek  the reason why and when we need to put $\hbar=\spin$ (\underline{$\hbar$ has a physical meaning but $\spin$ is artificially introduced}).

\begin{claim}
Let ${\mathcal{H}}(\xi,\theta,\pi)$ be given in \eqref{fW-1.10} as a
Hamiltonian function corresponding to free Weyl equation.
A solution of the Hamilton-Jacobi equation with ${{a}}\neq0$
\begin{equation}
{\mathcal{S}}_t
+{\mathcal{H}}({\mathcal{S}}_{x},\theta,{\mathcal{S}}_{\theta})=0\with
{\mathcal{S}}(0,x,\xi,\theta,\pi)=\langle x|\xi\rangle+{{a}}\langle \theta|\pi\rangle
\label{HJ-Weylfree-bis} %{{a}}={\hbar}{\spin}^{-1}
\end{equation}
is constructed without solving Hamilton equation.
\end{claim}
\begin{remark}
Though to quantize, we need to take ${{a}}={\hbar}{\spin}^{-1}$ and finally ${\hbar}{\spin}^{-1}=1$, to clarify the dependence of $\hbar$ in ``super version of classical mechanics'', we prefer to introduce parameter ${{a}}$.
\end{remark}

Assuming that the solution ${\mathcal{S}}(t,x,\xi,\theta,\pi)$ of \eqref{HJ-Weylfree-bis} is even supersmooth, we expand it
as
\begin{equation}
\begin{aligned}
{\mathcal{S}}(t,x,\xi,\theta,\pi)=&{\mathcal{S}}_{\bar0\bar0}+
{\mathcal{S}}_{\theta_2\theta_1}\theta_1\theta_2
+{\mathcal{S}}_{\pi_1\theta_1}\theta_1\pi_1
+{\mathcal{S}}_{\pi_2\theta_2}\theta_2\pi_2\\
&+{\mathcal{S}}_{\pi_2\theta_1}\theta_1\pi_2
+{\mathcal{S}}_{\pi_1\theta_2}\theta_2\pi_1
+{\mathcal{S}}_{\pi_2\pi_1}\pi_1\pi_2
+{\mathcal{S}}_{\pi_2\pi_1\theta_2\theta_1}\theta_1\theta_2\pi_1\pi_2.
\end{aligned}\label{fW-7.6}
\end{equation}
Here, $(f_{\theta_1})_{\theta_2}=f_{\theta_2\theta_1}$,
${\mathcal{S}}_{\theta_2\theta_1}={\mathcal{S}}_{\theta_2\theta_1}(t,x,\xi)={\mathcal{S}}_{\theta_2\theta_1}(t,x,\xi,0,0)$, etc.
$\bar 0=(0,0),\;\bar 1=(1,1)\in\{0,1\}^2$,
${\mathcal{S}}_{\bar1\bar1}={\mathcal{S}}_{\pi_2\pi_1\theta_2\theta_1}$ 

\begin{lemma} 
Let ${\mathcal{S}}$ be a solution of \eqref{HJ-Weylfree-bis}. Put
$$
\tilde{\mathcal   H}_*
={\mathcal   H}_*({\mathcal   S}_x(t, x,\xi,\theta,\pi),\theta,{\mathcal   S}_\theta(t, x,\xi,\theta,\pi)) \et
\tilde{\mathcal   H}_*^0=\tilde{\mathcal   H}_*\big|_{\theta=\pi=0},
$$
 then terms ${\mathcal   S}_{**}$ in \eqref{fW-7.6} satisfy
\begin{gather}
{\mathcal   S}_{\bar0\bar0,t}+\tilde{\mathcal   H}^0=0\with 
{\mathcal   S}_{\bar0\bar0}(0,x,\xi)=\langle x|\xi\rangle,\label{fW3-71}\\
{\mathcal   S}_{\theta_2\theta_1,t}+\tilde{\mathcal   H}_{\pi_2\pi_1}^0{\mathcal   S}_{\theta_2\theta_1}^2
+(\tilde{\mathcal   H}_{\pi_1\theta_1}^0+\tilde{\mathcal   H}_{\pi_2\theta_2}^0){\mathcal   S}_{\theta_2\theta_1}
+\tilde{\mathcal   H}_{\theta_2\theta_1}^0=0\with
{\mathcal   S}_{\theta_2\theta_1}(0,x,\xi)=0,\label{fW3-72}\\
{\mathcal   S}_{\pi_1\theta_1,t}+
(\tilde{\mathcal   H}_{\pi_1\theta_1}^0
+{\mathcal   S}_{\theta_2\theta_1}\tilde{\mathcal   H}_{\pi_2\pi_1}^0){\mathcal   S}_{\pi_1\theta_1}=0
\with {\mathcal   S}_{\pi_1\theta_1}(0,x,\xi)={{a}},\label{fW3-73}\\
{\mathcal   S}_{\pi_2\theta_2,t}+
(\tilde{\mathcal   H}_{\pi_2\theta_2}^0
+{\mathcal   S}_{\theta_2\theta_1}\tilde{\mathcal   H}_{\pi_2\pi_1}^0){\mathcal   S}_{\pi_2\theta_2}=0
\with {\mathcal   S}_{\pi_2\theta_2}(0,x,\xi)={{a}},\label{fW3-74}\\
{\mathcal   S}_{\pi_2\theta_1,t}+
(\tilde{\mathcal   H}_{\pi_1\theta_1}^0
+{\mathcal   S}_{\theta_2\theta_1}\tilde{\mathcal   H}_{\pi_2\pi_1}^0){\mathcal   S}_{\pi_2\theta_1}=0
\with {\mathcal   S}_{\pi_2\theta_1}(0,x,\xi)=0,\label{fW3-75}\\
{\mathcal   S}_{\pi_1\theta_2,t}+
(\tilde{\mathcal   H}_{\pi_2\theta_2}^0
+{\mathcal   S}_{\theta_2\theta_1}\tilde{\mathcal   H}_{\pi_2\pi_1}^0){\mathcal   S}_{\pi_1\theta_2}=0
\with {\mathcal   S}_{\pi_1\theta_2}(0,x,\xi)=0,\label{fW3-76}
\end{gather}
\end{lemma}

\par{\it Proof:} Restricting \eqref{HJ-Weylfree-bis} to $\theta=\pi=0$, we get \eqref{fW3-71}.
Differentiating \eqref{HJ-Weylfree-bis} w.r.t  $\theta_1$ and then $\theta_2$ and restricting to $\theta=\pi=0$,
we get Riccati type ODE \eqref{fW3-72} with parameter  $(x, \xi)$. In fact, from the differential formula for composite functions and remarking $\partial_{\theta_j}\partial_{\theta_j}{\mathcal   S}=0$, ${\mathcal{H}}_{\xi_k\xi_j}=0$,
we get
$$
\begin{gathered}
\partial_{\theta_1}\tilde{\mathcal   H} 
=\frac{\partial{\mathcal   S}_{x_j}}{\partial\theta_1}\tilde{\mathcal   H}_{\xi_j}
+\tilde{\mathcal   H}_{\theta_1}
+\frac{\partial{\mathcal   S}_{\theta_2}}{\partial\theta_1}\tilde{\mathcal   H}_{\pi_2},\\
\partial_{\theta_2}\partial_{\theta_1}\tilde{\mathcal   H}
=\frac{\partial^2{\mathcal   S}_{x_j}}{\partial\theta_2\partial\theta_1}\tilde{\mathcal   H}_{\xi_j}
-\frac{\partial{\mathcal   S}_{x_j}}{\partial\theta_1}
\partial_{\theta_2}\tilde{\mathcal   H}_{\xi_j}
+\partial_{\theta_2}\tilde{\mathcal   H}_{\theta_1}
+\frac{\partial{\mathcal   S}_{\theta_2}}{\partial\theta_1}
\partial_{\theta_2}\tilde{\mathcal   H}_{\pi_2}
\end{gathered}
$$
and 
$$
\begin{gathered}
\partial_{\theta_2}\tilde{\mathcal   H}_{\xi_j}
=\tilde{\mathcal   H}_{\theta_2\xi_j}+\frac{\partial{\mathcal   S}_{\theta_1}}{\partial\theta_2}
\tilde{\mathcal   H}_{\pi_1\xi_j},\\
\partial_{\theta_2}\tilde{\mathcal   H}_{\theta_1}
=\frac{\partial{\mathcal   S}_{x_j}}{\partial\theta_2}\tilde{\mathcal   H}_{\xi_j\theta_1}
+\tilde{\mathcal   H}_{\theta_2\theta_1}
+\frac{\partial{\mathcal   S}_{\theta_1}}{\partial\theta_2}\tilde{\mathcal   H}_{\pi_1\theta_1},\\
\partial_{\theta_2}\tilde{\mathcal   H}_{\pi_2}
=\frac{\partial{\mathcal   S}_{x_j}}{\partial\theta_2}\tilde{\mathcal   H}_{\xi_j\pi_2}
+\tilde{\mathcal   H}_{\theta_2\pi_2}
+\frac{\partial{\mathcal   S}_{\theta_1}}{\partial\theta_2}\tilde{\mathcal   H}_{\pi_1\pi_2}.
\end{gathered}
$$
Remarking also $\tilde{\mathcal   H}_{\xi_j}^0=0$, $\tilde{\mathcal   H}_{\xi_j\theta_k}^0=0$, etc,
 and restricting $\partial_{\theta_2}\partial_{\theta_1}\tilde{\mathcal   H}$  
to $\theta=\pi=0$, we have \eqref{fW3-72}.

Analogously, from
$$
\partial_{\pi_1}\partial_{\theta_1}\tilde{\mathcal   H}
=\frac{\partial^2{\mathcal   S}_{x_j}}{\partial\pi_1\partial\theta_1}\tilde{\mathcal   H}_{\xi_j}
-\frac{\partial{\mathcal   S}_{x_j}}{\partial\theta_1}
\partial_{\pi_1}\tilde{\mathcal   H}_{\xi_j}
+\partial_{\pi_1}\tilde{\mathcal   H}_{\theta_1}
+\frac{\partial{\mathcal   S}_{\theta_2}}{\partial\theta_1}
\partial_{\pi_1}\tilde{\mathcal   H}_{\pi_2}+\frac{\partial^2{\mathcal   S}_{\theta_2}}{\partial\pi_1\partial\theta_1}\tilde{\mathcal   H}_{\pi_2},
$$
and ${\mathcal{H}}_{\pi_2\theta_1}=0={\mathcal{H}}_{\pi_1\theta_2}$ with
$$
\begin{gathered}
\partial_{\pi_1}\tilde{\mathcal   H}_{\xi_j}
=\frac{\partial{\mathcal   S}_{\theta_k}}{\partial\pi_1}\tilde{\mathcal   H}_{\pi_k\xi_j},\quad
\partial_{\pi_1}\tilde{\mathcal   H}_{\theta_1}
=\frac{\partial{\mathcal   S}_{x_j}}{\partial\pi_1}\tilde{\mathcal   H}_{\xi_j\theta_1}
+\frac{\partial{\mathcal   S}_{\theta_k}}{\partial\pi_1}\tilde{\mathcal   H}_{\pi_k\theta_1},\\
\partial_{\pi_1}\tilde{\mathcal   H}_{\pi_2}
=\frac{\partial{\mathcal   S}_{x_j}}{\partial\pi_1}\tilde{\mathcal   H}_{\xi_j\pi_2}
+\frac{\partial{\mathcal   S}_{\theta_1}}{\partial\pi_1}\tilde{\mathcal   H}_{\pi_1\pi_2},
\end{gathered}
$$
we get \eqref{fW3-73}. Other equations \eqref{fW3-74}-\eqref{fW3-76} are obtained analogously.  $\qquad\qed$

\begin{lemma} Regarding $x$ and $\xi$ as parameters in \eqref{fW-shn} and
solving ODEs in the above lemma, we get
\begin{align}
&{\mathcal{S}}_{\bar0\bar0}(t,x,\xi)=\langle x|\xi\rangle,\label{fW3-71-sol2}\\
&{\mathcal{S}}_{\theta_2\theta_1}=\frac{-{{\spin}}{{\zeta}}\sin\,\gamma_t}{\bar{\delta}(t)},\label{fW3-72-sol2}\\
&{\mathcal{S}}_{\pi_1\theta_1}={\mathcal{S}}_{\pi_2\theta_2}
={{a}}\frac{|{\xi}|}{\bar{\delta}(t)},\label{fW3-73-sol2}\\
&{\mathcal{S}}_{\pi_1\theta_2}={\mathcal{S}}_{\pi_2\theta_1}=0.
\end{align}
\end{lemma}

\par{\it Proof.}
From \eqref{fW-7.6}, using ${\mathcal{S}}_{\theta_j}\big|_{\theta=\pi=0}=0$, we get $\tilde{\mathcal   H}^0=0$ 
and so \eqref{fW3-71-sol2} is obvious. Putting this into \eqref{fW3-72}, we get
$$
\begin{aligned}
&\tilde{\mathcal   H}_{\pi_2\pi_1}^0={\spin}^{-2}({\mathcal   S}_{x_1}-i{\mathcal   S}_{x_2})\big|_{\theta=\pi=0}
={\spin}^{-2}(\xi_1-i\xi_2)={\spin}^{-2}\bar{\zeta},\\
&\tilde{\mathcal   H}_{\pi_1\theta_1}^0=-i{\spin}^{-1}{\mathcal   S}_{x_3}\big|_{\theta=\pi=0}=\tilde{\mathcal   H}_{\pi_2\theta_2}^0=-i{\spin}^{-1}\xi_3,\\
&\tilde{\mathcal   H}_{\theta_2\theta_1}^0=({\mathcal   S}_{x_1}+i{\mathcal   S}_{x_2})\big|_{\theta=\pi=0}=\xi_1+i\xi_2
={\zeta}
\end{aligned}
$$
and 
\eqref{fW3-72} becomes
\begin{equation}
{\mathcal   S}_{\theta_2\theta_1,t}+{\spin}^{-2}\bar{\zeta}{\mathcal   S}_{\theta_2\theta_1}^2
-2i{\spin}^{-1}\xi_3{\mathcal   S}_{\theta_2\theta_1}
+{\zeta}=0\with
{\mathcal   S}_{\theta_2\theta_1}(0,x,\xi)=0.
\label{N7.34}
\end{equation}
\begin{quotation}
{\bf Solving ODE of Riccati type}:
For a given ODE
$$
y'=q_0(t)+q_1(t)y+q_2(t)y^2,
$$
assuming $q_2\neq{0}$ and putting $v=q_2(t)y$, we define
$$
P=q_1+\frac{q_2'}{q_2},\quad Q=q_2 q_0
$$
then $v'=(q_2(t)y)'$ is calculated as
$$
v'=v^2+P(t)v+Q(t).
$$
Moreover, differentiating $v=-\frac{u'}{u}$ w.r.t. $t$, we get
$$
u''-P(t)u'+Q(t)u=0.
$$
Solving this and using $u$, we put $\displaystyle{y=-\frac{u'}{q_2 u}}$, then this is a solution of
ODE of Riccati type.
\begin{problem}
If $q_2(t)$ has $0$-point, then is there any explicit formula for solution for Riccati ODE? 
\end{problem}
\end{quotation}
Using this, we calculate \eqref{N7.34}. Putting $q_0=-{\zeta}$, $q_1=2i{\spin}^{-1}$, $q_2=-{\spin}^{-2}\bar{\zeta}$, we get
$$
\ddot{u}-2i{\spin}^{-1}\dot{u}+{\spin}^{-2}|{\zeta}|^2u=0
$$
and defining 
$\lambda_{\pm}=i{\spin}^{-1}(\xi_3\pm |\xi|)$, we have
$$
u(t)=\alpha e^{\lambda_+t}+\beta e^{\lambda_-t}.
$$
Therefore, using
$$
{\mathcal   S}_{\theta_2\theta_1}(0)=\frac{\dot{u}}{{\spin}^{-2}\bar{\zeta}{u}}=0,
$$
and
$$
\dot{u}(0)=0=i{\spin}^{-1}[\alpha(\xi_3+|\xi|)+\beta(\xi_3-|\xi|)]
$$
we finally get the desired result \eqref{fW3-72-sol2},
$$
{\mathcal   S}_{\theta_2\theta_1}(t)=\frac{\alpha(\xi_3+|\xi|)e^{i{\spin}^{-1}|\xi|t}+\beta(\xi_3-|\xi|)e^{-i{\spin}^{-1}|\xi|t}}
{\alpha e^{i{\spin}^{-1}|\xi|t}+\beta e^{-i{\spin}^{-1}|\xi|t}}
=\frac{-{\spin}{\zeta}\sin{\gamma_t}}{|\xi|\cos{\gamma_t}-i\xi_3\sin{\gamma_t}}.
$$
Using this result and putting
\begin{equation}
w_0(t, x,\xi)
=\tilde{\mathcal   H}_{\pi_1\theta_1}^0+{\mathcal   S}_{\bar1\bar0}\tilde{\mathcal   H}_{\pi_2\pi_1}^0
=\tilde{\mathcal   H}_{\pi_2\theta_2}^0+{\mathcal   S}_{\bar1\bar0}\tilde{\mathcal   H}_{\pi_2\pi_1}^0
=\frac{\bar{\delta}(t)'}{\bar{\delta}(t)}=(\log \bar{\delta}(t))',
\label{w0}
\end{equation}
we get
\begin{gather}
{\mathcal   S}_{\pi_1\theta_1}(t,x,\xi)={{a}}|\xi|e^{-\int_s^t dr\,w_0(r, x,\xi)}={{a}}\frac{|\xi|}{\bar{\delta}(t)}
={\mathcal   S}_{\pi_2\theta_2}(t, x,\xi),\label{fW3-78}\\
{\mathcal   S}_{\pi_2\theta_1}(t, x,\xi)={\mathcal   S}_{\pi_1\theta_2}(t, x,\xi)=0.\qquad\qed
\label{fW3-79}
\end{gather}

\begin{lemma} Terms ${\mathcal   S}_{\pi_2\pi_1}$ and ${\mathcal   S}_{\bar1\bar1}$, satisfy the following:
\begin{gather}
{\mathcal   S}_{\pi_2\pi_1,t}
+{\mathcal   S}_{\pi_1\theta_1}{\mathcal   S}_{\pi_2\theta_2}\tilde{\mathcal   H}_{\pi_2\pi_1}^0=0
\with {\mathcal   S}_{\pi_2\pi_1}(0,x,\xi)=0,\label{fW3-77}\\
{\mathcal   S}_{\bar1\bar1,t}+2w_0{\mathcal   S}_{\bar1\bar1}+w_1=0
\with  {\mathcal   S}_{\bar1\bar1}(0,x,\xi)=0.\label{fW3-78-2} %-i{{b}}{\hbar}
\end{gather}
Here, we put
$$
\begin{aligned}
w_1=&w_1(t,x,\xi)\\
=&({\mathcal   S}_{\theta_2\theta_1}{\mathcal   S}_{\pi_2\pi_1,x_j}
-{\mathcal   S}_{\pi_1\theta_1}{\mathcal   S}_{\pi_2\theta_2,x_j})\tilde{\mathcal{H}}^0_{\xi_j\pi_1\theta_1}
+({\mathcal   S}_{\theta_2\theta_1}{\mathcal   S}_{\pi_2\pi_1,x_j}
-{\mathcal   S}_{\pi_1\theta_1,x_j}{\mathcal   S}_{\pi_2\theta_2})\tilde{\mathcal{H}}^0_{\xi_j\pi_2\theta_2}\\
&+[{\mathcal   S}_{\theta_2\theta_1}^2{\mathcal   S}_{\pi_2\pi_1,x_j}
+{\mathcal   S}_{\pi_1\theta_1}{\mathcal   S}_{\pi_2\theta_2}{\mathcal   S}_{\theta_2\theta_1,x_j}
-{\mathcal   S}_{\theta_2\theta_1}({\mathcal   S}_{\pi_1\theta_1}{\mathcal   S}_{\pi_2\theta_2})_{x_j}]\tilde{\mathcal{H}}^0_{\xi_j\pi_2\pi_1}+{\mathcal   S}_{\pi_2\pi_1,x_j}
\tilde{\mathcal{H}}^0_{\xi_j\theta_2\theta_1}. 
\end{aligned}
$$
\end{lemma}

\par{\it Proof}. Differentiating \eqref{HJ-Weylfree-bis} w.r.t. $\pi_1$ and then $\pi_2$, restricting to $\theta=\pi=0$, we have \eqref{fW3-71} and  \eqref{fW3-77} is obtained, using ${\mathcal   S}_{\pi_1\theta_2}=0$ and restricting to $\theta=\pi=0$ and
$$
\partial_{\pi_2}\partial_{\pi_1}\tilde{\mathcal{H}}
=\partial_{\pi_2}
\bigg(\frac{\partial{\mathcal   S}_{x_j}}{\partial\pi_1}\tilde{\mathcal   H}_{\xi_j}
+\frac{\partial{\mathcal   S}_{\theta_1}}{\partial\pi_1}\tilde{\mathcal   H}_{\pi_1}\bigg)
=\frac{\partial{\mathcal   S}_{\theta_1}}{\partial\pi_1}\frac{\partial{\mathcal   S}_{\theta_2}}{\partial\pi_2}\tilde{\mathcal   H}_{\pi_2\pi_1}
$$
\eqref{fW3-78-2} is get, since
$$
\begin{aligned}
\partial_{\pi_2}\partial_{\pi_1}\partial_{\theta_2}\partial_{\theta_1}\tilde{\mathcal{H}}
&=\frac{\partial^4{\mathcal   S}_{x_j}}{\partial\pi_2\partial\pi_1\partial\theta_2\partial\theta_1}\tilde{\mathcal   H}_{\xi_j}
+\frac{\partial^2{\mathcal   S}_{x_j}}{\partial\theta_2\partial\theta_1}
\partial_{\pi_2}\partial_{\pi_1}\tilde{\mathcal   H}_{\xi_j}
-\frac{\partial^2{\mathcal   S}_{x_j}}{\partial\pi_1\partial\theta_1}\partial_{\pi_2}\bigg(\tilde{\mathcal   H}_{\theta_2\xi_j}+\frac{\partial{\mathcal   S}_{\theta_1}}{\partial\theta_2}\tilde{\mathcal   H}_{\pi_1\xi_j}\bigg)\\
&\qquad
-\frac{\partial^2{\mathcal   S}_{x_j}}{\partial\pi_2\partial\theta_2}\partial_{\pi_1}\tilde{\mathcal   H}_{\xi_j\theta_1}
+\partial_{\pi_2}\partial_{\pi_1}\tilde{\mathcal   H}_{\theta_2\theta_1}
+\frac{\partial^3{\mathcal   S}_{\theta_1}}{\partial\pi_2\partial\pi_1\partial\theta_2}\tilde{\mathcal   H}_{\pi_1\theta_1}
+\frac{\partial{\mathcal   S}_{\theta_1}}{\partial\theta_2}\partial_{\pi_2}\partial_{\pi_1}\tilde{\mathcal   H}_{\pi_1\theta_1}\\
&\qquad\qquad
+\frac{\partial^3{\mathcal   S}_{\theta_2}}{\partial\pi_2\partial\pi_1\partial\theta_1}
\bigg(\frac{\partial{\mathcal   S}_{x_j}}{\partial\theta_2}\tilde{\mathcal   H}_{\xi_j\pi_2}
+\tilde{\mathcal   H}_{\theta_2\pi_2}
+\frac{\partial{\mathcal   S}_{\theta_1}}{\partial\theta_2}\tilde{\mathcal   H}_{\pi_1\pi_2}\bigg)\\
&\qquad\qquad\qquad
+\frac{\partial{\mathcal   S}_{\theta_2}}{\partial\theta_1}
\partial_{\pi_2}\partial_{\pi_1}\bigg(\frac{\partial{\mathcal   S}_{x_j}}{\partial\theta_2}\tilde{\mathcal   H}_{\xi_j\pi_2}
+\tilde{\mathcal   H}_{\theta_2\pi_2}
+\frac{\partial{\mathcal   S}_{\theta_1}}{\partial\theta_2}\tilde{\mathcal   H}_{\pi_1\pi_2}\bigg)
+{\cdots},
\end{aligned}
$$
then remarking
$$
\begin{aligned}
&\partial_{\pi_2}\partial_{\pi_1}\tilde{\mathcal   H}_{\xi_j}
={\mathcal   S}_{\pi_1\theta_1}{\mathcal   S}_{\pi_2\theta_2}\tilde{\mathcal   H}_{\pi_2\pi_1\xi_j},\\
&\partial_{\pi_2}(\tilde{\mathcal   H}_{\theta_2\xi_j}+\frac{\partial{\mathcal   S}_{\theta_1}}{\partial\theta_2}\tilde{\mathcal   H}_{\pi_1\xi_j})=
\frac{\partial{\mathcal   S}_{\theta_2}}{\partial\pi_2}\tilde{\mathcal   H}_{\pi_2\theta_2\xi_j}
+\frac{\partial{\mathcal   S}_{\theta_1}}{\partial\theta_2}\frac{\partial{\mathcal   S}_{\theta_2}}{\partial\pi_2}
\tilde{\mathcal   H}_{\pi_2\pi_1\xi_j}+{\cdots},\\
&\partial_{\pi_1}\tilde{\mathcal   H}_{\xi_j\theta_1}=\frac{\partial{\mathcal   S}_{\theta_1}}{\partial\pi_1}
\tilde{\mathcal   H}_{\pi_1\xi_j\theta_1},\\
&\partial_{\pi_2}\partial_{\pi_1}\tilde{\mathcal   H}_{\theta_2\theta_1}=
\frac{\partial^2{\mathcal   S}_{x_j}}{\partial\pi_2\partial\pi_1}\tilde{\mathcal   H}_{\xi_j\theta_2\theta_1},\\
&\partial_{\pi_2}\partial_{\pi_1}\tilde{\mathcal   H}_{\pi_1\theta_1}=
\frac{\partial^2{\mathcal   S}_{x_j}}{\partial\pi_2\partial\pi_1}\tilde{\mathcal   H}_{\xi_j\pi_1\theta_1},\\
&\partial_{\pi_2}\partial_{\pi_1}\bigg(\frac{\partial{\mathcal   S}_{x_j}}{\partial\theta_2}\tilde{\mathcal   H}_{\xi_j\pi_2}
+\tilde{\mathcal   H}_{\theta_2\pi_2}
+\frac{\partial{\mathcal   S}_{\theta_1}}{\partial\theta_2}\tilde{\mathcal   H}_{\pi_1\pi_2}\bigg)\\
&\quad
=-\frac{\partial^2{\mathcal   S}_{x_j}}{\partial\pi_2\partial\theta_2}\frac{\partial{\mathcal   S}_{\theta_1}}{\partial\pi_1}
\tilde{\mathcal   H}_{\pi_1\xi_j\pi_2}
+\frac{\partial^2{\mathcal   S}_{x_j}}{\partial\pi_2\partial\pi_1}\tilde{\mathcal   H}_{\xi_j\theta_2\pi_2}
+\frac{\partial^3{\mathcal   S}_{\theta_1}}{\partial\pi_2\partial\pi_1\partial\theta_2}\tilde{\mathcal   H}_{\pi_1\pi_2}
+\frac{\partial{\mathcal   S}_{\theta_1}}{\partial\theta_2}\frac{\partial^2{\mathcal   S}_{x_j}}{\partial\pi_2\partial\pi_1}
\tilde{\mathcal   H}_{\xi_j\pi_1\pi_2}+{\cdots},
\end{aligned}
$$
and restricting to $\theta=\pi=0$, we have
$$
\begin{aligned}
\tilde{\mathcal   H}_{\bar1\bar1}^0
&={\mathcal   S}_{\theta_2\theta_1x_j}{\mathcal   S}_{\pi_1\theta_1}{\mathcal   S}_{\pi_2\theta_2}\tilde{\mathcal   H}_{\pi_1\pi_2\xi_j}
-{\mathcal   S}_{\pi_1\theta_1x_j}({\mathcal   S}_{\pi_2\theta_2}\tilde{\mathcal   H}_{\pi_2\theta_2\xi_j}
+{\mathcal   S}_{\theta_2\theta_1}{\mathcal   S}_{\pi_2\theta_2}\tilde{\mathcal   H}_{\pi_2\pi_1\xi_j})\\
&\qquad
-{\mathcal   S}_{\pi_2\theta_2 x_j}{\mathcal   S}_{\pi_1\theta_1}\tilde{\mathcal   H}_{\pi_1\xi_j\theta_1}
+{\mathcal   S}_{\pi_2\pi_1 x_j}\tilde{\mathcal   H}_{\xi_j\theta_2\theta_1}
+{\mathcal   S}_{\bar1\bar1}\tilde{\mathcal   H}_{\pi_1\theta_1}
+{\mathcal   S}_{\theta_2\theta_1}{\mathcal   S}_{\pi_2\pi_1x_j}\tilde{\mathcal   H}_{\xi_j\pi_1\theta_1 }\\
&\qquad
-{\mathcal   S}_{\bar1\bar1}
(\tilde{\mathcal   H}_{\theta_2\pi_2}+{\mathcal   S}_{\theta_2\theta_1}\tilde{\mathcal   H}_{\pi_1\pi_2})\\
&\qquad
-{\mathcal   S}_{\theta_2\theta_1}(-{\mathcal   S}_{\pi_2\theta_2 x_j}{\mathcal   S}_{\pi_1\theta_1}\tilde{\mathcal   H}_{\pi_1\xi_j\pi_2}
+{\mathcal   S}_{\pi_2\pi_1 x_j}\tilde{\mathcal   H}_{\xi_j\theta_2\pi_2}
+{\mathcal   S}_{\bar1\bar1}\tilde{\mathcal   H}_{\pi_1\pi_2}
+{\mathcal   S}_{\theta_2\theta_1}{\mathcal   S}_{\pi_2\pi_1x_j}\tilde{\mathcal   H}_{\xi_j\pi_1\pi_2})\\
&={\mathcal   S}_{\bar1\bar1}[\tilde{\mathcal   H}_{\pi_1\theta_1}
-(\tilde{\mathcal   H}_{\theta_2\pi_2}+{\mathcal   S}_{\theta_2\theta_1}\tilde{\mathcal   H}_{\pi_1\pi_2})-{\mathcal   S}_{\theta_2\theta_1}\tilde{\mathcal   H}_{\pi_1\pi_2}]\\
&\qquad
+{\mathcal   S}_{\theta_2\theta_1}[-{\mathcal   S}_{\pi_1\theta_1x_j}\tilde{\mathcal   H}_{\pi_2\pi_1\xi_j}
+{\mathcal   S}_{\pi_2\theta_2 x_j}{\mathcal   S}_{\pi_1\theta_1}\tilde{\mathcal   H}_{\pi_1\xi_j\pi_2}
-{\mathcal   S}_{\pi_2\theta_2 x_j}\tilde{\mathcal   H}_{\xi_j\theta_2\pi_2}
+{\mathcal   S}_{\pi_2\pi_1x_j}\tilde{\mathcal   H}_{\xi_j\pi_1\theta_1 }\\
&\qquad\qquad\qquad\qquad\qquad\qquad\qquad\qquad\qquad\qquad\qquad\qquad\qquad\qquad\qquad%\quad
-{\mathcal   S}_{\theta_2\theta_1}{\mathcal   S}_{\pi_2\pi_1x_j}\tilde{\mathcal   H}_{\xi_j\pi_1\pi_2}]\\
&\qquad
+{\mathcal   S}_{\theta_2\theta_1x_j}\tilde{\mathcal   H}_{\pi_1\pi_2\xi_j}
-{\mathcal   S}_{\pi_1\theta_1x_j}{\mathcal   S}_{\pi_2\theta_2}\tilde{\mathcal   H}_{\pi_2\theta_2\xi_j}
-{\mathcal   S}_{\pi_2\theta_2 x_j}{\mathcal   S}_{\pi_1\theta_1}\tilde{\mathcal   H}_{\pi_1\xi_j\theta_1}
+{\mathcal   S}_{\pi_2\pi_1 x_j}\tilde{\mathcal   H}_{\xi_j\theta_2\theta_1}
.\qquad\qed
\end{aligned}
$$

Remarking ${\mathcal{S}}_{**x_j}=0$ for $*=\theta_a, \pi_b$, we have $w_2=0$. From \eqref{fW3-77} and \eqref{fW3-78-2},
\begin{equation}
{\mathcal   S}_{\pi_2\pi_1}(t, x,\xi)=-{{a}}^2{\spin}^{-1}\bar{\zeta}\frac{\sin\gamma_t}{\bar{\delta}(t)},\quad
{\mathcal   S}_{\bar1\bar1}(t, x,\xi)=0. 
\end{equation}

Finally we have
\begin{equation}
{\mathcal{S}}(t,x,\theta,\xi,\pi)=\langle x|\xi\rangle
+\bar{\delta}(t)^{-1}
\big[{{a}}|\xi|\langle\theta|\pi\rangle
-{\spin}\zeta\sin\gamma_t
\theta_1\theta_2
-{{a}}^2{\spin}^{-1}\bar{\zeta}\sin\gamma_t
\pi_1\pi_2\big]. 
\label{fW-3-20}
\end{equation}

\begin{claim}
$\mathcal{S}(t,x,\theta,\xi,\pi)$ defined above satisfies \eqref{HJ-Weylfree-bis}.
\end{claim}

\par{\it Proof}.
Indeed, as
$$
\begin{gathered}
{\mathcal{S}}_{x_j}=\xi_j\for j=1,2,3, \et \zeta={\mathcal{S}}_{x_1}+i{\mathcal{S}}_{x_2},\;
\bar{\zeta}={\mathcal{S}}_{x_1}-i{\mathcal{S}}_{x_2},\\
{\mathcal{S}}_{{\theta}_1}=\bar{\delta}(t)^{-1}
\big[{{a}}|\xi|{\pi}_1
-\spin\zeta\sin\gamma_t{\theta}_2\big],\quad
{\mathcal{S}}_{{\theta}_2}=\bar{\delta}(t)^{-1}
\big[{{a}}|\xi|{\pi}_2
+\spin\zeta\sin\gamma_t{\theta}_1\big],
\end{gathered}
$$
we get
$$
\begin{gathered}
{\mathcal{S}}_{{\theta}_1}{\mathcal{S}}_{{\theta}_2}
=\bar{\delta}(t)^{-2}
\big[({{a}}|\xi|)^2{\pi}_1{\pi}_2
-\spin{a}|\xi|\zeta\sin\gamma_t
\langle{\theta}|{\pi}\rangle+(\spin\zeta\sin\gamma_t)^2
{\theta}_1{\theta}_2\big],\\
{\theta}_1{\mathcal{S}}_{{\theta}_1}+{\theta}_2{\mathcal{S}}_{{\theta}_2}
=\bar{\delta}(t)^{-1}
\big[{{a}}|\xi|\langle{\theta}|{\pi}\rangle
-2\spin\zeta\sin\gamma_t{\theta}_1{\theta}_2\big].
\end{gathered}
$$
Substituting these into 
${\mathcal{H}}({\mathcal{S}}_{x},{\theta},{\mathcal{S}}_{{\theta}})$, we have
\begin{equation}
\begin{aligned}
&{\mathcal{H}}({\mathcal{S}}_{x},{\theta},{\mathcal{S}}_{{\theta}})
=c\zeta{\theta}_1{\theta}_2
+c\spin^{-2}{\bar{\zeta}}{\mathcal{S}}_{{\theta}_1}{\mathcal{S}}_{{\theta}_2}
-ic\spin^{-1}\xi_3
({\theta}_1{\mathcal{S}}_{{\theta}_1}+{\theta}_2{\mathcal{S}}_{{\theta}_2})\\
&\qquad\qquad
=c|\xi|^2{\bar{\delta}}^{-2}
\big[\zeta{\theta}_1{\theta}_2
-{{a}}\spin^{-1}(|\xi|\sin\gamma_t+i\xi_3\cos\gamma_t)
\langle{\theta}|{\pi}\rangle
+{{a}}^2\spin^{-2}{\bar{\zeta}}{\pi}_1{\pi}_2\big].
\end{aligned}\label{fW-2.31}
\end{equation}
On the other hand, since we get easily
$$
\partial_t({\bar{\delta}}(t)^{-1})=c\spin^{-1}|{\xi}|\bar{\delta}(t)^{-2}
(|{\xi}|\sin\gamma_t+i{\xi}_3\cos\gamma_t) \et
\partial_t(\bar{\delta}(t)^{-1}\sin\gamma_t)=c\spin^{-1}|{\xi}|^2\bar{\delta}(t)^{-2},
$$
we have
\begin{equation}
%\begin{aligned}
{\mathcal{S}}_t=c{{a}}\spin^{-1}|{\xi}|^2\bar{\delta}(t)^{-2}
(|{\xi}|\sin\gamma_t+i{\xi}_3\cos\gamma_t)\langle\theta|\pi\rangle
%\\&\qquad\qquad\qquad\qquad
+c\spin^{-1}|{\xi}|^2{\bar{\delta}}(t)^{-2}\big[-\spin\zeta\theta_1\theta_2
-{{a}}^2\spin^{-1}{\bar{\zeta}}\pi_1\pi_2\big].
%\end{aligned}
\label{fW-2.33}
\end{equation}
Therefore, we have ${\mathcal{S}}_t+\tilde{\mathcal{H}}=0$. $\qquad\qed$

\begin{remark}
(1) From above calculation, we needn't assume that ${{a}}=1$ to make $\mathcal{S}$ satisfy \eqref{HJ-Weylfree-bis}. 
Since it seems preferable to calculate ``classical quantities'' without mentioning ${\hbar}$, this construction is better than Jacobi's method mentioned before.
But to get the desired object after quantization, we need some relation between ${\hbar}$ and ${\spin}$.\\
(2) Theoretically, we may calculate even when $A_j(q)$ are added, but we mayn't have such an explicit formula
of ${\mathcal   S}_{\bar1\bar1}(t, x,\xi)$ as \eqref{fW-3-20}. 
Especially when $A_j(t,q)$ depends also on $t$, we have only the existence of the solution of ODE of Riccati type.
In this case, we need to construct a solution following Jacobi's method explained in the previous section.
Therefore, it is mysterious why by such method, we need the condition $\hbar=\kbar$ even in ``classical mechanics''?!
\end{remark}

From this, we get
\begin{proposition}\label{prop:continuity*}
Van Vleck determinant is calculated as
$$
{\mathcal{D}}(t,{x,\xi,\theta,\pi})=\sdet %(-1)^{3+2}\,
\begin{pmatrix}
\frac{\partial^2{\mathcal{S}}}{\partial{x}\,\partial{\xi}}&
\frac{\partial^2{\mathcal{S}}}{\partial{x}\,\partial{\pi}}\\
\frac{\partial^2{\mathcal{S}}}{\partial{\theta}\,\partial{\xi}}&
\frac{\partial^2{\mathcal{S}}}{\partial{\theta}\,\partial{\pi}}
\end{pmatrix}={{a}}^{-2}|\unbxi|^{-2}\bar{\delta}(t)^2.
$$
Moreover, for any ${{a}}\neq0$, it satisfies
\begin{equation}
\left\{
\begin{aligned}
&\pdt {\mathcal{D}}
+\frac{\partial}{\partial{x}}
\Big({\mathcal{D}}
\frac{\partial{\mathcal{H}}}{\partial {\xi}}\Big)
+\frac{\partial}{\partial{\theta}}
\Big({\mathcal{D}}
\frac{\partial{\mathcal{H}}}{\partial {\pi}}\Big)=0,\\
&{\mathcal{D}}(0,{x,\xi,\theta,\pi})={{a}}^{-2}.
\end{aligned}
\right.
\label {fW-1.18-2}
\end{equation}
Here, independent variables in ${\mathcal{D}}$ is $(t,{x,\xi,\theta,\pi})$, those for ${\partial{\mathcal{H}}}/{\partial \xi}$ and
${\partial{\mathcal{H}}}/{\partial \pi}$ are $({\mathcal{S}}_{x}, {\theta},{\mathcal{S}}_{\bar{\theta}})$.
\end{proposition}
\begin{corollary}
Putting ${\mathcal{A}}$ as the square root of ${\mathcal{D}}$, we have
$$
\partial_t{\mathcal{A}}+{\mathcal{A}}_x\tilde{\mathcal{H}}_{\xi}+{\mathcal{A}}_{\theta}\tilde{\mathcal{H}}_{\pi}
+\frac{1}{2}{\mathcal{A}}[\partial_x(\tilde{\mathcal{H}}_{\xi})+\partial_{\theta}(\tilde{\mathcal{H}}_{\pi})]=0.
$$
In our case, we have
\begin{equation}
\partial_t{\mathcal{A}}+\frac{1}{2}{\mathcal{A}}\partial_{\theta}(\tilde{\mathcal{H}}_{\pi})=0.
\label{transport0}
\end{equation}
\end{corollary}
\par{\it Proof}.
Since ${\mathcal{A}}={\bar{\delta}(t)}/|\xi|$, 
${\mathcal{A}}_{x_j}=0$, ${\mathcal{A}}_{\theta_k}=0$ and
$$
\begin{aligned}
\partial_{x_k}(\tilde{\mathcal{H}}_{\xi_k})&={\mathcal{S}}_{x_{k}x_j}\tilde{\mathcal{H}}_{\xi_j\xi_k}+{\mathcal{S}}_{x_{k}\theta_b}\tilde{\mathcal{H}}_{\pi_b\xi_k}=0,\\
\partial_{\theta_a}(\tilde{\mathcal{H}}_{\pi_a})&={\mathcal{S}}_{\theta_a x_j}\tilde{\mathcal{H}}_{\xi_j\pi_a}+\tilde{\mathcal{H}}_{\theta_a\pi_a}+{\mathcal{S}}_{\theta_a\theta_b}\tilde{\mathcal{H}}_{\pi_b\pi_a},
\end{aligned}
$$
we  get the desired result. $\qquad\qed$

\section{Quantization} % and fundamental solution}
To compare with the Jacobi's method in A. Inoue~\cite{ino98-1}. we reproduce some calculations therein.
\subsection{Feynman's Quantization}
To quantize, it seems better to change the order of independent variables from $({x,\xi,\theta,\pi})$ to $({x,\theta,\xi,\pi})$.
We ``define'' an operator 
\begin{equation}
\begin{aligned}
({\mathcal U}(t)u)({x},{\theta})&=(2\pi\hbar)^{-3/2}\spin\iint d{\xi}d{\pi}\,
{\mathcal{A}}(t,\xi)
e^{i\hbar^{-1}{{\mathcal{S}}}(t,{x,\theta,\xi,\pi})}
{\mathcal{F}}u({\xi},{\pi})\\
&\qquad\qquad\qquad\qquad
\with {\mathcal{A}}(t,\xi)=\frac{ \bar{\delta}(t)}{{{a}}|\xi|}.
\end{aligned}
\label{fW-1.19}
\end{equation}

Following Feynman's idea, it seems natural to have
\begin{claim}[From ${\mathcal{H}}(\xi,\theta,\pi)$ to $\hat{\mathcal{H}}^{W}(-i\hbar\partial_x,\theta,\partial_{\theta})$]
If $a=\hbar\spin^{-1}=1$, we have
\begin{equation}
i\hbar\frac{\partial}{\partial t}({\mathcal U}(t)u)({x},{\theta})\bigg|_{t=0}=\hat{\mathcal{H}}_0(-i\hbar\partial_x,\theta,\partial_{\theta})u(x,\theta)=\hat{\mathcal{H}}^{W}(-i\hbar\partial_x,\theta,\partial_{\theta})u(x,\theta)
\end{equation}
where $\hat{\mathcal{H}}_0$ is given in \eqref{fW-1.8} with $\clifpin=i$. Moreover, for ${\mathcal{H}}(\xi,\theta,\pi)$ in \eqref{fW-1.10}, we put
$$
\hat{\mathcal{H}}^{W}(-i\hbar\partial_x,\theta,\partial_{\theta})u(x,\theta)
=(2\pi\hbar)^{-3}\spin^2\iint d\xi d\pi dx' d\theta'\,
e^{i\hbar^{-1}\langle X-X'|\Xi\rangle}{\mathcal{H}}\bigg(\xi,\frac{\theta+\theta'}{2},\pi\bigg)(u_0(x)+u_1\theta_1'\theta_2'),
$$
for $X=(x,\theta)$, $X'=(x',\theta')$, $\Xi=(\xi,\pi)$, $\langle X|\Xi\rangle=\langle x|\xi\rangle+\hbar\spin^{-1}\langle\theta|\pi\rangle$.
\end{claim}

\par{\it Proof}.
Remarking
$$
\partial_{x}\big({\mathcal{A}}(t,\xi)
{\mathcal{H}}_\xi({\mathcal{S}}_{x},{\theta},{\mathcal{S}}_{\theta})\big)=0,
$$
and applying Hamilton-Jacobi~\eqref{HJ-Weylfree-bis} and continuity~\eqref{transport0} equations,
we have
\begin{equation}
\pdt({\mathcal{A}}e^{i\hbar^{-1}{\mathcal{S}}})
=({\mathcal{A}}_t+i\hbar^{-1}{\mathcal{A}}{\mathcal{S}}_t)e^{i\hbar^{-1}{\mathcal{S}}}
=-{\mathcal{A}}[\cdots]\,e^{i\hbar^{-1}{\mathcal{S}}},
\label{fW7-4-3}
\end{equation}
where
\begin{equation}
\begin{aligned}
{}[\cdots]&=\big[\frac12 \partial_{{\theta}_a}\tilde{\mathcal{H}}_{\pi_a}
+i\hbar^{-1}\tilde{\mathcal{H}}\big]\\
&=c\spin^{-1}\xi_3\frac{|\xi|\sin\gamma_t+i\xi_3\cos\gamma_t}{\bar{\delta}(t)}\\
&\qquad
+ic\hbar^{-1}\frac{|\xi|^2}{{\bar{\delta}}(t)^{2}}[\zeta\theta_1\theta_2-{{a}}\spin^{-1}(|\xi|\sin\gamma_t+i\xi_3\cos\gamma_t)\langle\theta|\pi\rangle+{{a}}^2\spin^{-2}\bar{\zeta}\pi_1\pi_2].
\end{aligned}
\label{fW-3.4bis}
\end{equation}
Therefore,
$$
\pdt({\mathcal{A}} e^{i\hbar^{-1}{\mathcal{S}}})\bigg|_{t=0}
=-\frac{ic\spin^{-1}}{{{a}}}(1-i{{a}}\hbar^{-1}\langle\theta|\pi\rangle)\xi_3-\frac{ic\hbar^{-1}\zeta}{{{a}}}\theta_1\theta_2
-ic{{a}}^2\hbar^{-1}\spin^{-2}\bar{\zeta}\pi_1\pi_2.
$$
Since ${\mathcal{S}}(0,x,\theta,\xi,\pi)=\langle x|\xi\rangle+a\langle\theta|\pi\rangle$, we have
$$
\begin{gathered}
\int d\pi\,e^{i\hbar^{-1}{\mathcal{S}}(0,x,\theta,\xi,\pi)}(1-i{{a}}\hbar^{-1}\langle\theta|\pi\rangle)(\spin^2 \hat{u}_1+\hat{u}_0\pi_1\pi_2)
=e^{i\hbar^{-1}\langle x|\xi\rangle}[\hat{u}_0-{{a}}^2\hbar^{-2}\spin^2\hat{u}_1\theta_1\theta_2],\\
\int d\pi\, e^{i\hbar^{-1}{\mathcal{S}}(0,x,\theta,\xi,\pi)}\theta_1\theta_2(\spin^2 \hat{u}_1+\hat{u}_0\pi_1\pi_2)
=e^{i\hbar^{-1}\langle x|\xi\rangle}\hat{u}_0\theta_1\theta_2,\\
\int d\pi\, e^{i\hbar^{-1}{\mathcal{S}}(0,x,\theta,\xi,\pi)}\pi_1\pi_2(\spin^2 \hat{u}_1+\hat{u}_0\pi_1\pi_2)
=e^{i\hbar^{-1}\langle x|\xi\rangle}\spin^2 \hat{u}_1.
\end{gathered}
$$

These imply that
$$
\int d\pi\,\pdt({\mathcal{A}} e^{i\hbar^{-1}{\mathcal{S}}})\bigg|_{t=0}(\spin^2\hat{u}_1+\hat{u}_0\pi_1\pi_2)
=
-\frac{ic\spin^{-1}}{{{a}}}\xi_3(\hat{u}_0-{{a}}^2\hbar^{-2}\spin^2\hat{u}_1\theta_1\theta_2)-\frac{ic\hbar^{-1}\zeta}{{{a}}}\hat{u}_0\theta_1\theta_2-i{{a}}c\hbar^{-1}\bar{\zeta}\hat{u}_1,
$$
which proves, if ${{a}}=\hbar\spin^{-1}=1$,
$$
\begin{aligned}
i\hbar\pdt({\mathcal{U}}(t)u)(x,\theta)\bigg|_{t=0}
&=(2\pi\hbar)^{-3/2}\int d\xi\, e^{i\hbar\langle x|\xi\rangle}c[\xi_3(\hat{u}_0-\hat{u}_1\theta_1\theta_2)+\zeta\hat{u}_0\theta_1\theta_2+\bar{\zeta}\hat{u}_1]\\
&=(\hat{\mathcal{H}}_0(-i\hbar\partial_x,\theta,\partial_{\theta})u)(x,\theta). \qquad\qed
\end{aligned}
$$

More generally, we have
\begin{claim} For any $t$, we have
$$
i\hbar\pdt({\mathcal{U}}(t)u)(x,\theta)=\hat{\mathcal{H}}_0(-i\hbar\partial_{x},{\theta},\partial_{{\theta}})({\mathcal{U}}(t)u)(x,\theta)=\hat{\mathcal{H}}^{W}(-i\hbar\partial_{x},{\theta},\partial_{{\theta}})({\mathcal{U}}(t)u)(x,\theta).
$$
\end{claim}
\par{\it Proof 1}.[Reduction to matrix form]
Putting
$$
v_0(t,x)=({\mathcal U}(t)u)(x,\theta)\big|_{\theta=0},\quad
v_1(t,x)=\partial_{\theta_2}\partial_{\theta_1}({\mathcal U}(t)u)(x,\theta)\big|_{\theta=0},
$$
we get
\begin{equation}
{\mathcal U}(t)u(x,\theta)=v_0(t,x)+v_1(t,x)\theta_1\theta_2
\label{fW-3.1}\end{equation}
where
\begin{equation}
\begin{aligned}
v_0(t,x)
&=(2\pi\hbar)^{-3/2}\int d\xi\,e^{i{\hbar}^{-1}\langle x|\xi\rangle}V_0(t,\xi),\\
v_1(t,\xi)
&=(2\pi\hbar)^{-3/2}\int d\xi\,e^{i{\hbar}^{-1}\langle x|\xi\rangle}V_1(t,\xi),
\end{aligned}\label{fW-3.2bis}
\end{equation}
with
\begin{equation}
\begin{pmatrix}
V_0(t,\xi)\\
V_1(t,\xi)
\end{pmatrix}
=\frac{1}{{{a}}|\xi|}
\begin{pmatrix}
\bar{\delta}(t)& -i\hbar^{-1}\spin{{a}}\bar\zeta \sin\gamma_t\\
-i\hbar^{-1}\spin\zeta \sin\gamma_t&{{a}}^2\hbar^{-2}\spin^2{\delta}(t)
\end{pmatrix}
\begin{pmatrix}
\hat{u}_0(\xi)\\
\hat{u}_1(\xi)
\end{pmatrix}.
\label{fW-svmf}
\end{equation}
Therefore, when ${{a}}=1=\hbar\spin^{-1}$, above reduces to the matrix representation \eqref{fW-matrix-sol2}.

In fact, rewriting
$$
\begin{aligned}
{\mathcal U}(t)u(x,\theta)
&=
(2\pi\hbar)^{-3/2}\int_{{\fR}^{3|0}}d\xi\, \frac{\bar{\delta}(t)}{{{a}}|\xi|}\exp{\{i\hbar^{-1}(\langle x|\xi\rangle-\spin\bar{\delta}(t)^{-1}\zeta\sin\gamma_t\theta_1\theta_2)\}}\\
&\qquad
\times \int_{{\fR}^{0|2}}d\pi\,\exp{\{i\hbar^{-1}\bar{\delta}(t)^{-1}({{a}}|\xi|\langle\theta|\pi\rangle-{{a}}^2\spin^{-1}\bar{\zeta}\sin\gamma_t\pi_1\pi_2)\}}
(\spin^2{\hat{\unbu}}_1+{\hat{\unbu}}_0\pi_1\pi_2),
\end{aligned}
$$
and since
$$
\begin{aligned}
&\int_{{\fR}^{0|2}}d\pi\,\exp{\{i\hbar^{-1}\bar{\delta}(t)^{-1}({{a}}|\xi|\langle\theta|\pi\rangle-{{a}}^2\spin^{-1}\bar{\zeta}\sin\gamma_t\pi_1\pi_2)\}}(\spin^2{\hat{\unbu}}_1+{\hat{\unbu}}_0\pi_1\pi_2)\\
&\qquad
={\hat{\unbu}}_0+\spin^2{\hat{\unbu}}_1\partial_{\pi_2}\partial_{\pi_1}\exp{\{i\hbar^{-1}\bar{\delta}(t)^{-1}({{a}}|\xi|\langle\theta|\pi\rangle-{{a}}^2\spin^{-1}\bar{\zeta}\sin\gamma_t\pi_1\pi_2)\}}\bigg|_{\pi=0}\\
&\qquad
={\hat{\unbu}}_0+{\hat{\unbu}}_1(-i\hbar^{-1}\spin{{a}}^2\bar{\delta}(t)^{-1}\bar{\zeta}\sin\gamma_t+{{a}}^2\hbar^{-2}\spin^2\bar{\delta}(t)^{-2}|\xi|^2\theta_1\theta_2),
\end{aligned}
$$
we have
$$
\begin{aligned}
({\mathcal U}(t)u)(x,\theta)&=
(2\pi\hbar)^{-3/2}\int_{{\fR}^{3|0}}d\xi\,\frac{1}{{{a}}|\xi|} \exp{\{i\hbar^{-1}(\langle x|\xi\rangle-\spin\bar{\delta}(t)^{-1}\zeta\sin\gamma_t\theta_1\theta_2)\}}\\
&\qquad\qquad\qquad\qquad
\times
(\bar{\delta}(t){\hat{\unbu}}_0+{\hat{\unbu}}_1(-i\hbar^{-1}\spin{{a}}^2\bar{\zeta}\sin\gamma_t+{{a}}^2\hbar^{-2}\spin^2\bar{\delta}(t)^{-1}|\xi|^2\theta_1\theta_2)).\qquad/\!\!/
\end{aligned}
$$
\begin{remark} 
Above formula \eqref{fW-svmf} with ${{a}}=1=\hbar\spin^{-1}$ gives the proof for Claim~\ref{fW-claim7-2-1}.
\end{remark}

\par{\it Proof 2}.[Another direct calculation]
Interchanging the order of differentiation and derivation under integral sign, we get
$$
\begin{aligned}
i\hbar\pdt({\mathcal{U}}(t)u)(x,\theta)&=i\hbar(2\pi\hbar)^{-3/2}\hbar\iint d\xi d\pi\,\pdt({\mathcal{A}}e^{i\hbar^{-1}{\mathcal{S}}})
({\mathcal{F}}u)(\xi,\pi)\\
&=-i\hbar(2\pi\hbar)^{-3/2}\int d\xi\bigg[\int d\pi {\mathcal{A}}[\cdots]e^{i\hbar^{-1}{\mathcal{S}}}(\hbar^2\hat{u}_1+\hat{u}_0\pi_1\pi_2)\bigg]
\end{aligned}
$$
where $[\cdots]$ is given in \eqref{fW-3.4bis}.
On the other hand, remarking Weyl quantization of $\langle\theta|\pi\rangle$ gives $1-\theta_a\partial_{\theta_a}$,
$$
\begin{aligned}
\hat{\mathcal{H}}^W(-i\hbar\partial_{x},{\theta},\partial_{{\theta}})&({\mathcal{U}}(t)u)(x,\theta)\\
&=
(2\pi\hbar)^{-3/2}\hbar\iint d\xi d\pi\,
{\mathcal{H}}(\xi,{\theta},\partial_{\theta})[{\mathcal{A}}e^{i\hbar^{-1}{\mathcal{S}}}](\hbar^2\hat{u}_1+\hat{u}_1\pi_1\pi_2)\\
&=(2\pi\hbar)^{-3/2}\int d\xi\bigg[\int d\pi\,{\mathcal{A}}\{\cdots\}e^{i\hbar^{-1}{\mathcal{S}}}(\hbar^2\hat{u}_1+\hat{u}_1\pi_1\pi_2)\bigg],
\end{aligned}
$$
where
$$
{\mathcal{H}}(\xi,{\theta},\partial_{\theta})e^{i\hbar^{-1}{\mathcal{S}}}
=\{\cdots\}e^{i\hbar^{-1}{\mathcal{S}}}
$$
with
$$
{\mathcal{H}}(\xi,{\theta},\partial_{{\theta}})
=c\zeta\theta_1\theta_2-c\bar{\zeta}\frac{\partial^2}{\partial\theta_1\partial\theta_2}
+c\xi_3\bigg(1-\theta_1\frac{\partial}{\partial\theta_1}-\theta_2\frac{\partial}{\partial\theta_2}\bigg)
$$
and
\begin{equation}
\begin{aligned}
\{\cdots\}
&=c\zeta\theta_1\theta_2+c\bar{\zeta}\bigg[
-\frac{ia^2\hbar^{-1}\spin\zeta\sin\gamma_t}{\bar{\delta}(t)}+
\frac{\hbar^{-2}}{\bar{\delta}(t)^2}\big(a^2|\xi|^2\pi_1\pi_2-a\spin\zeta\sin\gamma_t|\xi|\langle\theta|\pi\rangle\\
&\qquad\qquad\qquad\qquad
+\spin^2\zeta^2\sin^2\gamma\theta_1\theta_2\big)\bigg]
+c\xi_3\bigg[1-\frac{i\hbar^{-1}}{\bar{\delta}(t)}\big(|\xi|\langle\theta|\pi\rangle-2\spin\zeta\sin\gamma_t\theta_1\theta_2\big)\bigg]\\
&=c|\xi|\bar{\delta}(t)^{-1}(\xi_3\cos\gamma_t-i|\xi|\sin\gamma_t)\\
&\qquad\qquad
+c\bar{\delta}(t)^{-2}|\xi|^2
\big[\zeta\,{\theta}_1{\theta}_2
+\hbar^{-2}{\bar{\zeta}}\,{\pi}_1{\pi}_2\\
&\qquad\qquad\qquad
+i\hbar^{-1}(-\xi_3\cos\gamma_t+i|\xi|\sin\gamma_t)
\langle{\theta}|{\pi}\rangle\big]\when {{a}}=\hbar\spin^{-1}=1.
\end{aligned}\label{fW-3.5bb}
\end{equation}
Comparing \eqref{fW-3.4bis} and \eqref{fW-3.5bb}, we have
$-i\hbar[\cdots]=\{\cdots\}$ when ${{a}}=\hbar\spin^{-1}=1$, which implies
$$
i\hbar\pdt{\mathcal{U}}(t)u(x,\theta)=\hat{\mathcal{H}}^{W}
\Big(\frac{\hbar}{i}\frac{\partial}{\partial x},\theta,\frac{\partial}{\partial \theta}\Big)
({\mathcal{U}}(t)u)(x,\theta).   \qquad\qed
$$
\begin{remark}
In calculating the right-hand side above, here we used the closed explicit form.
But, in general, we must establish the composition formula of
pseudo-differential operators of Weyl type and 
Fourier integral operators of above type (see, Theorem 4.5 of Inoue and Maeda~\cite{IM02}). 
\end{remark}

\subsection{$L^2$ boundedness}
In order to prove the unitarity of the operator ${\mathcal{U}}(t)$, rewriting \eqref{fW-3.2bis} with $a=\hbar\spin^{-1}=1$,
using the Parceval equality only w.r.t. $x$ or $\xi$, we have
\begin{proposition}
$$
||{\mathcal{U}}(t)u||=||u|| \quad\mbox{in $\clsl_{S\!S}^2({\fR}^{3|2})$}.
$$
\end{proposition}
\par{\it Proof}. 
From above calculation, we have
$$
\begin{aligned}
|V_0(t,\xi)|^2&=\frac{|\bar{\delta}(t)|^2}{|\xi|^2}\overline{(\hat{u}_0(\xi)
-i\bar{\delta}(t)^{-1}\bar\zeta\sin\gamma_t\hat{u}_1(\xi)
)}(\hat{u}_0(\xi)
-i\bar{\delta}(t)^{-1}\bar\zeta\sin\gamma_t\hat{u}_1(\xi)
),\\
|V_1(t,\xi)|^2&=\frac{|\bar{\delta}(t)|^2}{|\xi|^2}\overline{(
-i\bar{\delta}(t)^{-1}\zeta\sin\gamma_t\hat{u}_0(\xi)
+\bar{\delta}(t)^{-1}{\delta}(t)\hat{u}_1(\xi))}(
-i\bar{\delta}(t)^{-1}\zeta\sin\gamma_t\hat{u}_0(\xi)
+\bar{\delta}(t)^{-1}{\delta}(t)\hat{u}_1(\xi)),
\end{aligned}
$$ 
therefore
$$
||{\mathcal{U}}(t)u||^2=\int d\xi 
(|V_0(t,\xi)|^2+|V_1(t,\xi)|^2)
=\int d\xi(|\hat{u}_0(\xi)|^2+|\hat{u}_1(\xi)|^2)=||u||^2.
 \qquad\qed
$$

\subsection{Continuity}
\begin{proposition}
$$
\lim_{t \to 0}\Vert {\mathcal U}(t)u-u\Vert=0
$$
\end{proposition}
\par{\it Proof}. Since
$$
({\mathcal{U}}(t)u)(x,\theta)-u(x,\theta)
=(2\pi\hbar)^{-3/2}\int d\xi \,e^{i\hbar^{-1}\langle x|\xi\rangle}
[V_0(t,\xi)-V_0(0,\xi)+(V_1(t,\xi)-V_1(0,\xi))\theta_1\theta_2],
$$
we have
$$
\Vert {\mathcal U}(t)u-u\Vert^2
=\Vert V_0(t,\xi)-V_0(0,\xi)\Vert^2 +\Vert V_1(t,\xi)-V_1(0,\xi)\Vert ^2\to 0\when t\to0.  \qquad\qed
$$
\begin{problem}
Prove above Proposition using
$$
\begin{aligned}
&({\mathcal{U}}(t)u)(x,\theta)-u(x,\theta)\\
&\qquad
=(2\pi\hbar)^{-3/2}\hbar\iint d\xi d\pi\,[\mathcal{A}(t,\xi)e^{i\hbar^{-1}{\mathcal{S}}(t,x,\xi,\theta,\pi)}-\mathcal{A}(0,\xi)e^{i\hbar^{-1}{\mathcal{S}}(0,x,\xi,\theta,\pi)}]({\mathcal{F}}{\unbu})(\xi,\pi)\\
&\qquad
=(2\pi\hbar)^{-3/2}\hbar\iint d\xi d\pi
\bigg[\int_0^t ds \frac{d}{ds}(\mathcal{A}(s,\xi)e^{i\hbar^{-1}{\mathcal{S}}(s,x,\xi,\theta,\pi)})\bigg]({\mathcal{F}}{\unbu})(\xi,\pi)\\
&\qquad
=(2\pi\hbar)^{-3/2}\int d\xi \int_0^t ds\bigg[\int d\pi\,\frac{d}{ds}(\mathcal{A}(s,\xi)e^{i\hbar^{-1}{\mathcal{S}}(s,x,\xi,\theta,\pi)})({\mathcal{F}}{\unbu})(\xi,\pi)\bigg].
\end{aligned}
$$
\end{problem}

\subsection{Evolutional property}
\begin{proposition}
$$
{\mathcal{U}}(t){\mathcal{U}}(s)u={\mathcal{U}}(t+s)u \forany u\in
\ccsl_{\mathrm{SS,0}}({\fR}^{3|2}).
$$
\end{proposition}
\par
{\it Proof. }
Denoting ${\mathcal{U}}(s)u=u(s,x,\theta)=v_0(s,x)+v_2(s,x)\theta_1\theta_2$, we have
$$
\begin{aligned}
{\mathcal{U}}(t) & u(s,x,\theta)\\
&=(2\pi\hbar)^{-3/2}
\int d\xi \,e^{i\hbar^{-1}\langle x|\xi\rangle}{\mathcal{A}}(t,\xi)\,{\mathcal{A}}(s,\xi)
\big[(\hat u_0-i\delta(s)^{-1}\bar\eta\sin\gamma_s \hat u_1)\\
&\qquad\qquad\qquad\qquad\qquad\qquad\qquad\qquad
-i\delta(t)^{-1}\bar\zeta\sin\gamma_t\delta(s)^{-1}
(-i\zeta\sin\gamma_s\hat u_0+\bar\delta(s)\hat u_1)\big]\\
&+(2\pi\hbar)^{-3/2}
\int d\xi \,e^{i\hbar^{-1}\langle x|\xi\rangle}{\mathcal{A}}(t,\xi)\,{\mathcal{A}}(s,\xi)
\big[-i\delta(t)^{-1}\zeta\sin\gamma_t
(\hat u_0-i\delta(s)^{-1}\bar\zeta\sin\gamma_s \hat u_1)\\
&\qquad\qquad\qquad\qquad\qquad\qquad\qquad\qquad
+\delta(t)^{-1}\bar\delta(t)\delta(s)^{-1}
(-i\zeta\sin\gamma_s\hat u_0+\bar\delta(s)\hat u_1)\big]
\theta_1\theta_2.
\end{aligned}
$$
By simple calculation, we get
$$
|\xi|^{-2}[(\delta(t)\delta(s)-|\zeta|^2\sin\gamma_t\sin\gamma_s)\hat u_0
=|\xi|^{-1}\delta(t+s)\hat u_0,
$$
$$
-i(\delta(t)\bar\zeta\sin\gamma_s+\bar\delta(s)\bar\zeta\sin\gamma_t)\hat u_1
=-i\bar\zeta|\xi|^{-1}\sin\gamma_{t+s}\hat u_1,
$$
and
$$
|\xi|^{-1}\delta(t+s)\hat u_0
-i|\xi|^{-1}\bar\zeta\sin\gamma_{t+s}\hat u_1
={\mathcal{A}}(t+s,\xi)[\hat u_0-i\delta(t+s)^{-1}\bar\zeta\sin\gamma_{t+s}\hat u_1].
$$
Analogously, the coefficient of $\theta_1\theta_2$ is calculated as
$$
|\xi|^{-1}[-i\zeta\sin\gamma_{t+s}\hat u_0+\bar\delta(t+s)\hat u_1].
$$
Therefore, we have the evolutional property.   $\qquad\qed$

\section{Fundamental solution}
Though we have proved ${\mathcal{U}}(t_2){\mathcal{U}}(t_1)={\mathcal{U}}(t_1+t_2)$ by using explicit representation in this case, we reconsider this evolutional property as products formula for FIOp modifying arguments in Chapter 10 of H. Kumano-go~\cite{Kum82}.

\subsection{Super version of sharp products of phase functions}

For the future study of resolution of Feynman's problem, we proceed without using explicit formula for the solution of H-J equation as far as possible:

Let $H(t,X,\Xi)$ be given as $C^{\infty}(\euc\times {\mathcal{T}}^*{\fR}^{3|2}:\fC)$. \\
\fbox{Assumption (A)} \; There exists $T>0$ such that for any $-T<s, t<T$, any $\underline{X}=(\unbx,\unbtheta)\in {\fR}^{3|2}$ and
$\underline{\Xi}=(\unbxi,\unbpi)\in{\fR}^{3|2}$, there exists a unique solution $(X(t,s),\Xi(t,s))$ satisfying
\begin{equation}
\left\{
\begin{aligned}
\dt X(t,s)&=(-1)^{p(\Xi)}\partial_{\Xi}{\mathcal{H}}(t,X(t,s),\Xi(t,s)),\\
\dt \Xi(t,s)&=-\partial_{X}{\mathcal{H}}(t,X(t,s),\Xi(t,s)),
\end{aligned}
\right.
\with
\begin{pmatrix}
X(s,s)\\
\Xi(s,s)
\end{pmatrix}
=
\begin{pmatrix}
\underline{X}\\
\underline{\Xi}
\end{pmatrix}.
\end{equation}
More precisely, we denote these as
$$
X(t,s)=X(t,s,\underline{X},\underline{\Xi}),\quad
\Xi(t,s)=\Xi(t,s,\underline{X},\underline{\Xi}).
$$
\fbox{Assumption (B)}\; For any $-T<s, t<T$, any ${X}=(x,\theta)\in {\fR}^{3|2}$ and
${\Xi}=(\xi,\pi)\in{\fR}^{3|2}$,
there exists a solution ${\mathcal{S}}(t,s,X,\Xi)$ satisfying
\begin{equation}
\left\{
\begin{aligned}
&\partial_t {\mathcal{S}}(t,s,X,\Xi)+{\mathcal{H}}(t,X,\partial_{X}{\mathcal{S}}(t,s,X,\Xi))=0,\\
&{\mathcal{S}}(s,s,X,\Xi)=\langle X|\Xi\rangle.
\end{aligned}
\right.
\label{HK4.20}
\end{equation}
Moreover, it satisfies
\begin{equation}
\partial_s{\mathcal{S}}(t,s,X,\Xi)-{\mathcal{H}}(s,(-1)^{p(\Xi)}\partial_{\Xi}{\mathcal{S}}(t,s,X,\Xi),\Xi)=0.
\label{HK4.22}
\end{equation}

Putting
\begin{equation}
\varphi(t_2,t_1,t_0,X^2,\Xi^1,X^1,\Xi^0)={\mathcal{S}}(t_2,t_1,X^2,\Xi^1)-\langle X^1|\Xi^1\rangle+{\mathcal{S}}(t_1,t_0,X^1,\Xi^0),
\label{HK5.2-0}
\end{equation}
we need to get critical points $(\tilde{X}^1,\tilde{\Xi}^1)$ of $\varphi(t_2,t_1,t_0,X^2,\Xi^1,X^1,\Xi^0)$ w.r.t. $\Xi^1,\; X^1$ which satisfy
$$
\left\{
\begin{aligned}
\partial_{X^1}\varphi(t_2,t_1,t_0,X^2,\tilde{\Xi}^1,\tilde{X}^1,\Xi^0)=0,\\
\partial_{\Xi^1}\varphi(t_2,t_1,t_0,X^2,\tilde{\Xi}^1,\tilde{X}^1,\Xi^0)=0.
\end{aligned}
\right.
$$
\fbox{Assumption (C)} \;
For given $(t_2,t_1,t_0,X^2,\Xi^0)$, there exists a solution $(\tilde{X}^1,\tilde{\Xi}^1)$, denoted by 
$$
(\tilde{X}^1,\tilde{\Xi}^1)=(\tilde{X}^1(t_2,t_1,t_0,X^2,\Xi^0),\tilde{\Xi}^1(t_2,t_1,t_0,X^2,\Xi^0)),
$$
which satisfies
\begin{equation}
\left\{
\begin{aligned}
&\partial_{X^1}{\mathcal{S}}(t_1,t_0,\tilde{X}^1,\Xi^0)-\tilde{\Xi}^1=0,\\
&\partial_{\Xi^1}{\mathcal{S}}(t_2,t_1,X^2,\tilde{\Xi}^1)-(-1)^{p({\Xi}^1)}\tilde{X}^1=0.
\end{aligned}
\right.
\label{ASMC}
\end{equation}

\begin{proposition}
Defining $\#$-(called  sharp) product as
$$
[{\mathcal{S}}(t_2,t_1)\#{\mathcal{S}}(t_1,t_0)](X^2,\Xi^0)
=\varphi(t_2,t_1,t_0,X^2,\Xi^1,X^1,\Xi^0)\bigg|_{\stackrel{\scriptstyle X^1=\tilde{X}^1(t_2,t_1,t_0,X^2,\Xi^0)}{\Xi^1=\tilde{\Xi}^1(t_2,t_1,t_0,X^2,\Xi^0)}},
$$
then
\begin{align}
&[\langle X|\Xi\rangle\#{\mathcal{S}}(t,s)](X^2,\Xi^0)=[{\mathcal{S}}(t,s)\#\langle X|\Xi\rangle](X^2,\Xi^0)={\mathcal{S}}(t,s,X^2,\Xi^0),\label{HKi}\\
&[{\mathcal{S}}(t_2,t_1)\#{\mathcal{S}}(t_1,t_0)](X^2,\Xi^0)={\mathcal{S}}(t_2,t_0,X^2,\Xi^0)\forany t_1,\; t_0\le t_1\le t_2.\label{HKii}
\end{align}
\end{proposition}
\par{\it Proof of {\eqref{HKi}}}: Put
$$
\varphi=\langle X^2|\Xi^1\rangle-\langle X^1|\Xi^1\rangle+{\mathcal{S}}(t,s,X^1,\Xi^0).
$$
Since
$$
\left\{
\begin{aligned}
\partial_{X^1}\varphi(X^2,\Xi^1,X^1,\Xi^0)&=-\Xi^1+\partial_{X^1}{\mathcal{S}}(t,s,X^1,\Xi^0)=0,\\
\partial_{\Xi^1}\varphi(X^2,\Xi^1,X^1,\Xi^0)&=(-1)^{p(\Xi^1)}(X^2-X^1)=0,
\end{aligned}
\right.
$$
we have
$$
(\tilde{X}^1,\tilde{\Xi}^1)=(X^2,\partial_{X^1}{\mathcal{S}}(t,s,X^2,\Xi^0)).
$$
Therefore
$$
\varphi(X^2,\Xi^1,X^1,\Xi^0)\bigg|_{\stackrel{\scriptstyle{X^1=X^2}}{{\Xi^1=\tilde{\Xi}^1}}}={\mathcal{S}}(t,s,X^2,\Xi^0).\qquad/\!\!/
$$

\par{\it Proof of {\eqref{HKii}}}: Substituting the relation derived from \eqref{ASMC}, i.e.
\begin{equation}
\tilde{X}^1=(-1)^{p(\Xi^1)}\partial_{\Xi^1}{\mathcal{S}}(t_2,t_1,X^2,\tilde{\Xi}^1), \quad
\tilde{\Xi}^1=\partial_{X^1}{\mathcal{S}}(t_1,t_0,\tilde{X}^1,\Xi^0),
\label{HK5.8}
\end{equation}
into the definition, we have
\begin{equation}
[{\mathcal{S}}(t_2,t_1)\#{\mathcal{S}}(t_1,t_0)](X^2,\Xi^0)
={\mathcal{S}}(t_2,t_1,X^2,\tilde{\Xi}^1)-\langle \tilde{X}^1|\tilde{\Xi}^1\rangle+{\mathcal{S}}(t_1,t_0,\tilde{X}^1,\Xi^0).
\label{HK5.9}
\end{equation}
Differentiating \eqref{HK5.9} w.r.t. $t_1$, we have
$$
\begin{aligned}
\partial_{t_1}[{\mathcal{S}}(t_2,t_1)\#{\mathcal{S}}(t_1,t_0)]=&(\partial_{t_1}{\mathcal{S}}(t_2,t_1,X^2,\tilde{\Xi}^1)
+\partial_{t_1}\tilde{\Xi}^1\cdot\partial_{\Xi}{\mathcal{S}}(t_2,t_1,X^2,\tilde{\Xi}^1))\\
&-(\partial_{t_1} \tilde{X}^1\cdot \tilde{\Xi}^1+ \tilde{X}^1\cdot \partial_{t_1}\tilde{\Xi}^1)\\
&\quad
+\partial_{t_1}{\mathcal{S}}(t_1,t_0, \tilde{X}^1,\Xi^0)
+\partial_{t_1} \tilde{X}^1\cdot \partial_{X}{\mathcal{S}}(t_1,t_0, \tilde{X}^1,\Xi^0).
\end{aligned}
$$
Remarking \eqref{HK5.8}, we get
$$
\partial_{t_1}\tilde{\Xi}^1\cdot\partial_{\Xi}{\mathcal{S}}(t_2,t_1,X^2,\tilde{\Xi}^1)
-(\partial_{t_1}\tilde{X}^1\cdot \tilde{\Xi}^1+\tilde{X}^1\cdot \partial_{t_1}\tilde{\Xi}^1)
+\partial_{t_1}\tilde{X}^1\cdot \partial_{X}{\mathcal{S}}(t_1,t_0,\tilde{X}^1,\Xi^0)=0.
$$
From  \eqref{HK4.20} and  \eqref{HK4.22}, that is,
$$
\begin{aligned}
\partial_{t_1}{\mathcal{S}}(t_2,t_1,X^2,\tilde{\Xi}^1)&={\mathcal{H}}(t_1,(-1)^{p(\Xi)}\partial_{\Xi^1}{\mathcal{S}}(t_2,t_1,X^2,\tilde{\Xi}^1),\tilde{\Xi}^1),\\
\partial_{t_1}{\mathcal{S}}(t_1,t_0,\tilde{X}^1,\Xi^0)&=-{\mathcal{H}}(t_1,\tilde{X}^1,\Xi^0),
\end{aligned}
$$ 
applying \eqref{HK5.8} once more, we have
$$
\begin{aligned}
\partial_{t_1}&[{\mathcal{S}}(t_2,t_1)\#{\mathcal{S}}(t_1,t_0)](X^0,\Xi^0)=\partial_{t_1}{\mathcal{S}}(t_2,t_1,X^2,\tilde{\Xi}^1)+\partial_{t_1}{\mathcal{S}}(t_1,t_0,\tilde{X}^1,\Xi^0)\\
&={\mathcal{H}}(t_1,(-1)^{p(\Xi^1)}\partial_{\Xi}{\mathcal{S}}(t_2,t_1,X^2,\tilde{\Xi}^1),\tilde{\Xi}^1)-{\mathcal{H}}(t_1,\tilde{X}^1,\partial_{X}{\mathcal{S}}(t_1,t_0,\tilde{X}^1,\Xi^0))=0.
\end{aligned}
$$
Since $t_1$ is arbitrary, taking $t_1=t_0$, 
then the relations
$$
{\mathcal{S}}(t_0,t_0,\tilde{X}^1,\Xi^0)=\langle \tilde{X}^1|\Xi^0\rangle,\; \tilde{\Xi}^1=\partial_{X^1}{\mathcal{S}}(t_1,t_0,\tilde{X}^1,\Xi^0)\big|_{t_1=t_0}=\Xi^0,
$$
we have %
$$
[{\mathcal{S}}(t_2,t_1)\#{\mathcal{S}}(t_1,t_0)](X^2,\Xi^0)\big|_{t_1=t_0}={\mathcal{S}}(t_2,t_0,X^2,\Xi^0).\qquad /\!\!/
$$

\subsection{Superversion of products of FIOp}

Denoting
$$
{\mathcal{K}}(t_1,t_0,X^1,\Xi^0)={\mathcal{A}}(t_1,t_0,X^1,\Xi^0)e^{i\hbar^{-1}{\mathcal{S}}(t_1,t_0,X^1, \Xi^0)},
$$
we put, for $\unbv(X^1)=({\mathcal{U}}(t_1,t_0){\unbu})(X^1)$, 
$$
\begin{aligned}
({\mathcal{U}}(t_1,t_0){\unbu})(X^1)
&=c_{3,2}\iint_{{\fR}^{3|2}} d{\Xi^0}\,{\mathcal{K}}(t_1,t_0,X^1,\Xi^0)
{\mathcal{F}}{\unbu}(\Xi^0),\\
({\mathcal{U}}(t_2,t_1){\unbv})(X^2)
&=c_{3,2}\iint_{{\fR}^{3|2}} d{\Xi^1}\,{\mathcal{K}}(t_2,t_1,X^2,\Xi^1)
{\mathcal{F}}{\unbv}(\Xi^1).
\end{aligned}
$$
Changing the order of integration rather freely, we have
$$
\begin{aligned}
{}[\,{\mathcal{U}}(t_2,t_1)({\mathcal{U}}(t_1,t_0){\unbu})](X^2)
&=c_{3,2}^3\iint_{{\fR}^{3|2}} d{\Xi^1}\,{\mathcal{K}}(t_2,t_1,X^2,\Xi^1)\\
&\qquad\quad
\times
\iint_{{\fR}^{3|2}} d{X^1}e^{-i\hbar^{-1}\langle X^1|\Xi^1\rangle}
\bigg[\iint_{{\fR}^{3|2}} d{\Xi^0}\,{\mathcal{K}}(t_1,t_0,X^1,\Xi^0)
{\mathcal{F}}{\unbu}(\Xi^0)\bigg]\\
&=c_{3,2}\iint_{{\fR}^{3|2}} d{\Xi^0}\,[{\mathcal{K}}(t_2,t_1){\star}{\mathcal{K}}(t_1,t_0)](X^2,\Xi^0)
{\mathcal{F}}{\unbu}(\Xi^0)
\end{aligned}
$$
where we put
\begin{equation}
\begin{aligned}
{}[{\mathcal{K}}(t_2,t_1)&{\star}{\mathcal{K}}(t_1,t_0)](X^2,\Xi^0)\\
&=
c_{3,2}^2\iint_{{\fR}^{6|4}}d{\Xi^1}d{X^1}\,
{\mathcal{K}}(t_2,t_1,X^2,\Xi^1)e^{-i\hbar^{-1}\langle X^1|\Xi^1\rangle}{\mathcal{K}}(t_1,t_0,X^1,\Xi^0)\\
&=c_{3,2}^2\iint_{{\fR}^{6|4}}d{\Xi^1}d{X^1}\,
{\mathcal{A}}(t_2,t_1,X^2,\Xi^1){\mathcal{A}}(t_1,t_0,X^1,\Xi^0)e^{i\hbar^{-1}\varphi(t_2,t_1,t_0,X^2,\Xi^1,X^1,\Xi^0)}
\end{aligned}
\label{CF-FIO}
\end{equation}
with $\varphi(t_2,t_1,t_0,X^2,\Xi^1,X^1,\Xi^0)$ defined in \eqref{HK5.2-0}.

Though we want to show, by direct calculation with certain remainder term if necessary, 
\begin{equation}
[{\mathcal{K}}(t_2,t_1){\star}{\mathcal{K}}(t_1,t_0)](X^2,\Xi^0)={\mathcal{K}}(t_2,t_0,X^2,\Xi^0),
\end{equation}
this procedure is simple philosophically (because we know how to proceed in ordinary Schr\"odinger case) but seems complicated technically even for  the free Weyl equation case (because odd variables stay as ``variable coefficients'').

To prove this as generally as possible, we need to calculate Taylor expansion of $\varphi$ defined in \eqref{HK5.2-0} at $(\tilde{X}^1,\tilde{\Xi}^1)$.
Putting as
$$
\begin{gathered}
(\eta_j,\varpi_a)=\Eta=\Xi^1-\tilde{\Xi}^1=(\xi^1_j-\tilde{\xi}^1_j,\tilde{\pi}^1-\varpi_a),\; 
(y_k,\vartheta_b)=Y=X^1-\tilde{X}^1=(x_k^1-\tilde{x}_k^1,\theta_b^1-\tilde{\theta}_b),\\
{\mathfrak{a}}=(\alpha,a),\;\; \alpha=(\alpha_1,\alpha_2,\alpha_3),\quad a=(a_1,a_2),
\;\; |{\mathfrak{a}}|=|\alpha|+|a|,\;\; {\mathfrak{b}}=(\beta,b),\;\;\mbox{etc},\\
\partial_{X}^{\mathfrak{a}}\partial_{\Xi}^{\mathfrak{b}}
=\partial_{x^1}^{\alpha}\partial_{{\xi}^1}^{\beta}\partial_{\theta^1}^{a}\partial_{\pi^1}^{b}\sim
\partial_{Y}^{\mathfrak{a}}\partial_{\Eta}^{\mathfrak{b}}
=\partial_{y}^{\alpha}\partial_{\eta}^{\beta}\partial_{\vartheta}^{a}\partial_{\varpi}^{b},\\
\Phi(s\Eta,sY)=\Phi(s\eta_j, sy_k, s\varpi_a,s\vartheta_b)=\varphi(t_2,t_1,t_0,X^2,\tilde{\Xi}^1+s\Eta,\tilde{X}^1+sY,\Xi^0),
\end{gathered}
$$
we have, abbreviating the dependence on $(X^2,\Xi^0)$,
$$
\begin{gathered}
\Phi(\Eta,Y)=\Phi(0,0)+\frac{d}{ds}\Phi(0,0)+\frac{1}{2!}\frac{d^2}{ds^2}\Phi(0,0)+R,\\
\with
R=R(\Eta,Y)=\int_0^1 ds\frac{(1-s)^2}{2!}\frac{d^3}{ds^3}\Phi(s\Eta,sY).
\end{gathered}
$$

\begin{claim}\label{TE} By the definition of $(\tilde{X}^1,\tilde{\Xi}^1)$, we have 
{\allowdisplaybreaks
\begin{align}
&\Phi(0,0)={\mathcal{S}}(t_2,t_0,X^2,\Xi^0),\label{TE-0}\\
&\frac{d}{ds}\Phi(0,0)=\Eta\partial_{\Eta}\Phi(0,0)
+Y\partial_{Y}\Phi(0,0),\label{TE-1}\\
&\frac{d^2}{ds^2}\Phi(0,0)=\sum_{|{\mathfrak{a}}|+|\mathfrak{b}|=2}\Eta^{{\mathfrak{a}}}Y^{\mathfrak{b}}\partial_{Y}^{\mathfrak{b}}\partial_{\Eta}^{{\mathfrak{a}}}\Phi(0,0),\label{TE-2}\\
&\frac{d^3}{ds^3}\Phi(s\Eta,sY)=\sum_{|{\mathfrak{a}}|+|\mathfrak{b}|=3}
\Eta^{{\mathfrak{a}}}Y^{\mathfrak{b}}\partial_{Y}^{\mathfrak{b}}\partial_{\Eta}^{{\mathfrak{a}}}\Phi(s\Eta,sY).
\label{TE-3}
\end{align}}
\end{claim}

 {\it Proof of Claim \ref{TE}.}
By definition of $\#$-product, we have \eqref{TE-0}.\\
By \eqref{ASMC}, \eqref{TE-1} is obtained from
$$
\frac{d}{ds}\Phi=\eta_j\Phi_{\eta_j}+y_k\Phi_{y_k}+\varpi_a\Phi_{\varpi_a}+\vartheta_b\Phi_{\vartheta_b}.
$$
\eqref{TE-2} is rewritten as
$$
\begin{aligned}
\frac{d^2}{ds^2}\Phi=&\eta_j[\eta_{j'}\Phi_{\eta_{j'}\eta_j}+y_{k'}\Phi_{y_{k'}\eta_j}+\varpi_{a'}\Phi_{\varpi_{a'}\eta_j}
+\vartheta_{b'}\Phi_{\vartheta_{b'}\eta_j}]\\
&+y_k[\eta_{j'}\Phi_{\eta_{j'}y_k}+y_{k'}\Phi_{y_{k'}y_k}+\varpi_{a'}\Phi_{\varpi_{a'}y_k}+\vartheta_{b'}\Phi_{\vartheta_{b'}y_k}]\\
&\quad+\varpi_a[\eta_{j'}\Phi_{\eta_{j'}\varpi_a}+y_{k'}\Phi_{y_{k'}\varpi_a}+\varpi_{a'}\Phi_{\varpi_{a'}\varpi_a}+\vartheta_{b'}\Phi_{\vartheta_{b'}\varpi_a}]\\
&\qquad+\vartheta_b[\eta_{j'}\Phi_{\eta_{j'}\vartheta_b}+y_{k'}\Phi_{y_{k'}\vartheta_b}+\varpi_{a'}\Phi_{\varpi_{a'}\vartheta_b}+\vartheta_{b'}\Phi_{\vartheta_{b'}\vartheta_b}]\\
&=(\eta_j,y_k,\varpi_a,\vartheta_b){\cdot} N %{{}^t\! M}
\begin{pmatrix}
\eta_{j'}\\
y_{k'}\\
\varpi_{a'}\\
\vartheta_{b'}
\end{pmatrix}
\end{aligned}
$$
where
\begin{equation}%{{}^t\! M}
N=\begin{pmatrix}
\Phi_{\eta_{j'}\eta_j}&\Phi_{y_{k'}\eta_j}&-\Phi_{\varpi_{a'}\eta_j}&-\Phi_{\vartheta_{b'}\eta_j}\\
\Phi_{\eta_{j'}y_k}&\Phi_{y_{k'}y_k}&-\Phi_{\varpi_{a'}y_k}&-\Phi_{\vartheta_{b'}y_k}\\
\Phi_{\eta_{j'}\varpi_a}&\Phi_{y_{k'}\varpi_a}&\Phi_{\varpi_{a'}\varpi_a}&\Phi_{\vartheta_{b'}\varpi_a}\\
\Phi_{\eta_{j'}\vartheta_b}&\Phi_{y_{k'}\vartheta_b}&\Phi_{\varpi_{a'}\vartheta_b}&\Phi_{\vartheta_{b'}\vartheta_b}
\end{pmatrix}
=\begin{pmatrix}
N_{\mathrm{BB}}&N_{\mathrm{BF}}\\
N_{\mathrm{FB}}&N_{\mathrm{FF}}
\end{pmatrix}.
\label{JM}
\end{equation}
More precisely,
$$
N_{\mathrm{BB}}=
\begin{pmatrix}
\Phi_{\eta_{j'}\eta_j}&\Phi_{y_{k'}\eta_j}\\
\Phi_{\eta_{j'}y_k}&\Phi_{y_{k'}y_k}
\end{pmatrix}
=\begin{pmatrix}
\Phi_{\eta_1\eta_1}&\Phi_{\eta_2\eta_1}&\Phi_{\eta_3\eta_1}&\Phi_{y_1\eta_1}&\Phi_{y_2\eta_1}&\Phi_{y_3\eta_1}\\
\Phi_{\eta_1\eta_2}&\Phi_{\eta_2\eta_2}&\Phi_{\eta_3\eta_2}&\Phi_{y_1\eta_2}&\Phi_{y_2\eta_2}&\Phi_{y_3\eta_2}\\
\Phi_{\eta_1\eta_3}&\Phi_{\eta_2\eta_3}&\Phi_{\eta_3\eta_3}&\Phi_{y_1\eta_3}&\Phi_{y_2\eta_3}&\Phi_{y_3\eta_3}\\
\Phi_{\eta_1y_1}&\Phi_{\eta_2y_1}&\Phi_{\eta_3y_1}&\Phi_{y_1y_1}&\Phi_{y_2y_1}&\Phi_{y_3y_1}\\
\Phi_{\eta_1y_2}&\Phi_{\eta_2y_2}&\Phi_{\eta_3y_2}&\Phi_{y_1y_2}&\Phi_{y_2y_2}&\Phi_{y_3y_2}\\
\Phi_{\eta_1y_3}&\Phi_{\eta_2y_3}&\Phi_{\eta_3y_3}&\Phi_{y_1y_3}&\Phi_{y_2y_3}&\Phi_{y_3y_3}
\end{pmatrix},
$$
$$
N_{{\mathrm{B}}{\mathrm{F}}}
=\begin{pmatrix}
-\Phi_{\varpi_{a'}\eta_j}&-\Phi_{\vartheta_{b'}\eta_j}\\
-\Phi_{\varpi_{a'}y_k}&-\Phi_{\vartheta_{b'}y_k}
\end{pmatrix}
=\begin{pmatrix}
-\Phi_{\varpi_{1}\eta_1}&-\Phi_{\varpi_{2}\eta_1}&-\Phi_{\vartheta_{1}\eta_1}&-\Phi_{\vartheta_{2}\eta_1}\\
-\Phi_{\varpi_{1}\eta_2}&-\Phi_{\varpi_{2}\eta_2}&-\Phi_{\vartheta_{1}\eta_2}&-\Phi_{\vartheta_{2}\eta_2}\\
-\Phi_{\varpi_{1}\eta_3}&-\Phi_{\varpi_{2}\eta_3}&-\Phi_{\vartheta_{1}\eta_3}&-\Phi_{\vartheta_{2}\eta_3}\\
-\Phi_{\varpi_{1}y_1}&-\Phi_{\varpi_{2}y_1}&-\Phi_{\vartheta_{1}y_1}&-\Phi_{\vartheta_{2}y_1}\\
-\Phi_{\varpi_{1}y_2}&-\Phi_{\varpi_{2}y_2}&-\Phi_{\vartheta_{1}y_2}&-\Phi_{\vartheta_{2}y_2}\\
-\Phi_{\varpi_{1}y_3}&-\Phi_{\varpi_{2}y_3}&-\Phi_{\vartheta_{1}y_3}&-\Phi_{\vartheta_{2}y_3}
\end{pmatrix},
$$
$$
N_{\mathrm{FB}}
=\begin{pmatrix}
\Phi_{\eta_{j'}\varpi_a}&\Phi_{y_{k'}\varpi_a}\\
\Phi_{\eta_{j'}\vartheta_b}&\Phi_{y_{k'}\vartheta_b}
\end{pmatrix}
=\begin{pmatrix}
\Phi_{\eta_1\varpi_1}&\Phi_{\eta_2\varpi_1}&\Phi_{\eta_3\varpi_1}&\Phi_{y_1\varpi_1}&\Phi_{y_2\varpi_1}&\Phi_{y_3\varpi_1}\\
\Phi_{\eta_1\varpi_2}&\Phi_{\eta_2\varpi_2}&\Phi_{\eta_3\varpi_2}&\Phi_{y_1\varpi_2}&\Phi_{y_2\varpi_2}&\Phi_{y_3\varpi_2}\\
\Phi_{\eta_1\vartheta_1}&\Phi_{\eta_2\vartheta_1}&\Phi_{\eta_3\vartheta_1}&\Phi_{y_1\vartheta_1}&\Phi_{y_2\vartheta_1}&\Phi_{y_3\vartheta_1}\\
\Phi_{\eta_1\vartheta_2}&\Phi_{\eta_2\vartheta_2}&\Phi_{\eta_3\vartheta_2}&\Phi_{y_1\vartheta_2}&\Phi_{y_2\vartheta_2}&\Phi_{y_3\vartheta_2}
\end{pmatrix},
$$
$$
N_{\mathrm{FF}}
=\begin{pmatrix}
\Phi_{\varpi_{a'}\varpi_a}&\Phi_{\vartheta_{b'}\varpi_a}\\
\Phi_{\varpi_{a'}\vartheta_b}&\Phi_{\vartheta_{b'}\vartheta_b}
\end{pmatrix}
=\begin{pmatrix}
\Phi_{\varpi_{1}\varpi_1}&\Phi_{\varpi_{2}\varpi_1}&\Phi_{\vartheta_{1}\varpi_1}&\Phi_{\vartheta_{2}\varpi_1}\\
\Phi_{\varpi_{1}\varpi_2}&\Phi_{\varpi_{2}\varpi_2}&\Phi_{\vartheta_{1}\varpi_2}&\Phi_{\vartheta_{2}\varpi_2}\\
\Phi_{\varpi_{1}\vartheta_1}&\Phi_{\varpi_{2}\vartheta_1}&\Phi_{\vartheta_{1}\vartheta_1}&\Phi_{\vartheta_{2}\vartheta_1}\\
\Phi_{\varpi_{1}\vartheta_2}&\Phi_{\varpi_{2}\vartheta_2}&\Phi_{\vartheta_{1}\vartheta_2}&\Phi_{\vartheta_{2}\vartheta_2}
\end{pmatrix}.
$$
Concerning \eqref{TE-3}, we don't mention precisely here. \qquad /\!\!/

In the above, to get an even super-matrix corresponding to the Jacobian for $\Phi(\Eta,Y)$, we have changed the order of vector notation as
$$
\begin{gathered}
{\fR}^{3|2}\times {\fR}^{3|2}\ni (\Xi^1,X^1)=(\xi^1_k,\pi^1_b,x^1_j,\theta^1_a)\to (\xi^1_k,x^1_j,\pi^1_b,\theta^1_a)=(\Xi^1_{\mathrm{B}},X^1_{\mathrm{B}},\Xi^1_{\mathrm{F}},X^1_{\mathrm{F}})\in{\fR}^{6|4},\\
{\fR}^{3|2}\times {\fR}^{3|2}\ni (\Eta,Y)=(\eta_j, \varpi_a,y_k,\vartheta_b)\to(\eta_j, y_k,\varpi_a,\vartheta_b)=(\Eta_{\mathrm{B}},Y_{\mathrm{B}},\Eta_{\mathrm{F}},Y_{\mathrm{F}})\in{\fR}^{6|4}.
\end{gathered}
$$

Then, we need to calculate
$$
\begin{aligned}
c_{3,2}^2&\iint_{{\fR}^{6|4}}d{\Xi^1}d{X^1}\,{\mathcal{A}}(t_2,t_1,\Xi^1){\mathcal{A}}(t_1,t_0,\Xi^0)e^{i\hbar^{-1}\varphi(t_2,t_2,t_0,X^2, \Xi^1,X^1,\Xi^0)}\\
&=c_{3,2}^2\iint_{{\fR}^{6|4}}d{\Eta}d{Y}\,{\mathcal{A}}(t_2,t_1,\tilde{\Xi}^1+\Eta){\mathcal{A}}(t_1,t_0,\Xi^0)e^{i\hbar^{-1}[\Phi(0,0)+(1/2)\ddot{\Phi}(0,0)+R]}.
\end{aligned}
$$

\paragraph{\bf For free Weyl case}
Now, we return to our special case such that
$$
\begin{aligned}
\varphi&={\mathcal{S}}(\tau_2,x^2,\theta^2,\xi^1,\pi^1)-\langle x^1|\xi^1\rangle-\langle \theta^1|\pi^1\rangle
+{\mathcal{S}}(\tau_1,x^1,\theta^1,\xi^0,\pi^0)\\
&=\langle x^2|\xi^1\rangle
+{\mathrm{C}}^1\langle\theta^2|\pi^1\rangle
-{\mathrm{D}}^1\theta^2_1\theta^2_2
-{\mathrm{E}}^1\pi^1_1\pi^1_2\\
&\qquad\qquad
-\langle x^1|\xi^1\rangle-\langle \theta^1|\pi^1\rangle%\\
+\langle x^1|\xi^0\rangle
+{\mathrm{C}}^0\langle\theta^1|\pi^0\rangle
-{\mathrm{D}}^0\theta^1_1\theta^1_2
-{\mathrm{E}}^0\pi^0_1\pi^0_2
\end{aligned}
$$
where $\tau_2=t_2-t_1,\; \tau_1=t_1-t_0$, $\zeta^*=\xi_1^*+i\xi_2^*$, $\bar{\zeta}^*=\xi_1^*-i\xi_2^*$ with $*=0,1$ and
$$
\begin{gathered}
{\mathrm{C}}^1=\frac{|\xi^1|}{\bar{\delta}(\tau_2,|\xi^1|)},\quad
{\mathrm{D}}^1=\frac{{\hbar}\zeta^1 \sin\gamma(\tau_2,|\xi^1|)}{\bar{\delta}(\tau_2,|\xi^1|)},\quad
{\mathrm{E}}^1=\frac{\bar{\zeta}^1\sin\gamma(\tau_2,|\xi^1|)}{{\hbar}\bar{\delta}(\tau_2,|\xi^1|)},\\
{\mathrm{C}}^0=\frac{|\xi^0|}{\bar{\delta}(\tau_1,|\xi^0|)},\quad
{\mathrm{D}}^0=\frac{{\hbar}\zeta^0\sin\gamma(\tau_1,|\xi^0|)}{\bar{\delta}(\tau_1,|\xi^0|)},\quad
{\mathrm{E}}^0=\frac{\bar{\zeta}^0\sin\gamma(\tau_1,|\xi^0|)}{\hbar\bar{\delta}(\tau_1,|\xi^0|)}.
\end{gathered}
$$

From \eqref{ASMC}, we may define 
$\tilde{\Xi}^1=(\tilde{\xi}_k^1,\tilde{\pi}_b^1)$, $\tilde{X}^1=(\tilde{x}_j^1,\tilde{\theta}_a^1)$ satisfying
$$
\left\{
\begin{aligned}
\varphi_{x_j^1}&=\partial_{x_j^1}{\mathcal{S}}(t_1,t_0,\tilde{X}^1,\Xi^0)-\tilde{\xi}_j^1=-\tilde{\xi}_j^1+\xi_j^0=0,\\
\varphi_{\xi_k^1}&=-\tilde{x}_k^1+\partial_{\xi_k^1}{\mathcal{S}}(t_1,t_0,\tilde{X}^1,\Xi^0)
={\mathrm{C}}^1_{\xi_k^1}\langle\theta^2|\pi^1\rangle-{\mathrm{D}}^1_{\xi_k^1}\theta^2_1\theta^2_2
-{\mathrm{E}}^1_{\xi_k^1}\pi^1_1\pi^1_2-\tilde{x}_k^1=0,\\
\varphi_{\theta_1^1}&=\partial_{\theta_1^1}{\mathcal{S}}(t_1,t_0,\tilde{X}^1,\Xi^0)-\tilde{\pi}_2^1
=-\tilde{\pi}_1^1+{\mathrm{C}}^0\pi_1^0-{\mathrm{D}}^0\tilde{\theta}_2^1=0,\\
\varphi_{\theta_2^1}&=\partial_{\theta_2^1}{\mathcal{S}}(t_1,t_0,\tilde{X}^1,\Xi^0)-\tilde{\pi}_1^1
=-\tilde{\pi}_2^1+{\mathrm{C}}^0\pi_2^0+{\mathrm{D}}^0\tilde{\theta}_1^1=0,\\
\varphi_{\pi_1^1}&=\partial_{\pi_1^1}{\mathcal{S}}(t_2,t_1,X^2,\tilde{\Xi}^1)+\tilde{\theta}_1^1
=-{\mathrm{C}}^1\theta_1^2-{\mathrm{E}}^1\tilde{\pi}_2^1+\tilde{\theta}_1^1=0,\\
\varphi_{\pi_2^1}&=\partial_{\pi_2^1}{\mathcal{S}}(t_2,t_1,X^2,\tilde{\Xi}^1)+\tilde{\theta}_2^1
=-{\mathrm{C}}^1\theta_2^2+{\mathrm{E}}^1\tilde{\pi}_1^1+\tilde{\theta}^1_2=0.
\end{aligned}
\right.
$$
From these, we have
$$
1-D^0{\mathrm{E}}^1_0=\frac{|\xi^0|\bar{\delta}(\tau_1+\tau_2,|\xi^0|)}{\bar{\delta}(\tau_1,|\xi^0|)\bar{\delta}(\tau_2,|\xi^0|)}
\with
{\mathrm{C}}^1_0=\frac{|{\xi}^0|}{\bar{\delta}(\tau_2,|{\xi}^0|)},\quad
{\mathrm{E}}^1_0=\frac{\bar{\zeta}^0
\sin \gamma(\tau_2,|\xi^0|)}{{\hbar}\bar{\delta}(\tau_2,|{\xi}^0|)}.
$$
Therefore,
$$
\left\{
\begin{aligned}
\begin{pmatrix}
\tilde{\theta}_1^1\\
\tilde{\pi}_2^1
\end{pmatrix}
&=(1-{\mathrm{D}}^0{\mathrm{E}}^1_0)^{-1}
\begin{pmatrix}
-{\mathrm{E}}^1_0&1\\
-1&{\mathrm{D}}^0
\end{pmatrix}
\begin{pmatrix}
-{\mathrm{C}}^0\pi_2^0\\
{\mathrm{C}}^1_0\theta_1^2
\end{pmatrix},\\
\begin{pmatrix}
\tilde{\theta}_2^1\\
\tilde{\pi}_1^1
\end{pmatrix}
&=(1-{\mathrm{D}}^0{\mathrm{E}}^1_0)^{-1}
\begin{pmatrix}
{\mathrm{E}}^1_0&1\\
-1&-{\mathrm{D}}^0
\end{pmatrix}
\begin{pmatrix}
-{\mathrm{C}}^0\pi_1^0\\
{\mathrm{C}}^1_0\theta_2^2
\end{pmatrix},
\end{aligned}
\right.
$$
\begin{equation}
\left\{
\begin{aligned}
\tilde{\theta}_1^1&=\tilde{\theta}_1^1(\tau_2,\tau_1,\theta^2,\xi^0,\pi^0)=\frac{1}{\bar{\delta}(\tau_1+\tau_2,|\xi^0|)}[\bar{\delta}(\tau_1,|\xi^0|)\theta_1^2+\hbar^{-1}\bar{\zeta}^0\sin\gamma(\tau_2,|\xi^0|)\pi_2^0],\\
\tilde{\pi}_2^1&=\tilde{\pi}_2^1(\tau_2,\tau_1,\theta^2,\xi^0,\pi^0)=\frac{1}{\bar{\delta}(\tau_1+\tau_2,|\xi^0|)}[\hbar{\zeta}^0\sin\gamma(\tau_1,|\xi^0|)\theta_1^2+\bar{\delta}(\tau_2,|\xi^0|)\pi_2^0],\\
\tilde{\theta}_2^1&=\tilde{\theta}_2^1(\tau_2,\tau_1,\theta^2,\xi^0,\pi^0)=\frac{1}{\bar{\delta}(\tau_1+\tau_2,|\xi^0|)}[\hbar^{-1}\bar{\zeta}^0\sin\gamma(\tau_2,|\xi^0|)\pi_1^0+\bar{\delta}(\tau_1,|\xi^0|)\theta_2^2],\\
\tilde{\pi}_1^1&=\tilde{\pi}_1^1(\tau_2,\tau_1,\theta^2,\xi^0,\pi^0)=\frac{1}{\bar{\delta}(\tau_1+\tau_2,|\xi^0|)}[\bar{\delta}(\tau_2,|\xi^0|)\pi_1^0-\hbar\zeta^0\sin\gamma(\tau_1,|\xi^0|) \theta_2^2].
\end{aligned}
\right.
\label{SP12}
\end{equation}

For notational simplicity, we put
$$
\begin{aligned}
\Phi(s\Eta, sY)&=\langle x^2|\tilde{\xi}^1+s\eta\rangle
+{\mathrm{C}}^1(s)\langle\theta^2|\tilde{\pi}^1+s\varpi\rangle
-{\mathrm{D}}^1(s)\theta^2_1\theta^2_2
-{\mathrm{E}}^1(s)(\tilde{\pi}^1_1+s\varpi_1)(\tilde{\pi}^1_2+s\varpi_2)\\
&\qquad\qquad
-\langle \tilde{x}^1+sy|\tilde{\xi}^1+s\eta\rangle-\langle \tilde{\theta}^1+s\vartheta|\tilde{\pi}^1+s\varpi\rangle\\
&\qquad
+\langle \tilde{x}^1+sy|\xi^0\rangle
+{\mathrm{C}}^0\langle \tilde{\theta}^1+s\vartheta|\pi^0\rangle
-{\mathrm{D}}^0(\tilde{\theta}^1_1+s\vartheta_1)(\tilde{\theta}^1_2+s\vartheta_2)
-{\mathrm{E}}^0\pi^0_1\pi^0_2
\end{aligned}
$$
where
$$
\begin{gathered}
{\mathrm{C}}^1(s)=\frac{|{\xi}^0+s\eta|}{\bar{\delta}(\tau_2,|{\xi}^0+s\eta|)},\quad
{\mathrm{D}}^1(s)=\frac{{\hbar}({\xi}_1^0+s\eta_1+i{\xi}_2^0+is\eta_2)\sin\gamma(\tau_2,|{\xi}^0+s\eta|)}{\bar{\delta}(\tau_2,|{\xi}^0+s\eta|)},\\
{\mathrm{E}}^1(s)=\frac{({\xi}_1^0+s\eta_1-i{\xi}_2^0-is\eta_2)\sin\gamma(\tau_2,|{\xi}^0+s\eta|)}{{\hbar}\bar{\delta}(\tau_2,|{\xi}^0+s\eta|)}.
\end{gathered}
$$
Therefore, we have
$$
C^1(0)=C_0^1,\;\; D^1(0)=D_0^1,\;\; E^1(0)=E_0^1.
$$

By definition above, we have
\begin{equation}
\begin{aligned}
\frac{d}{ds}\Phi(s\Eta, sY)&=\langle x^2|\eta\rangle
+\dot{\mathrm{C}}^1(s)\langle\theta^2|\tilde{\pi}^1+s\varpi\rangle
+{\mathrm{C}}^1(s)\langle\theta^2|\varpi\rangle
-\dot{\mathrm{D}}^1(s)\theta^2_1\theta^2_2\\
&-\dot{\mathrm{E}}^1(s)(\tilde{\pi}^1_1+s\varpi_1)(\tilde{\pi}^1_2+s\varpi_2)
-{\mathrm{E}}^1(s)[\varpi_1(\tilde{\pi}^1_2+s\varpi_2)+(\tilde{\pi}^1_1+s\varpi_1)\varpi_2]\\
&-\langle y|\tilde{\xi}^1+s\eta\rangle-\langle \tilde{x}^1+sy|\eta\rangle
-\langle \vartheta|\tilde{\pi}^1+s\varpi\rangle
-\langle \tilde{\theta}^1+s\vartheta|\varpi\rangle\\
&+\langle y|\xi^0\rangle
+{\mathrm{C}}^0\langle \vartheta|\pi^0\rangle
-{\mathrm{D}}^0[\vartheta_1(\tilde{\theta}^1_2+s\vartheta_2)
+(\tilde{\theta}^1_1+s\vartheta_1)\vartheta_2],\\
%%%
\frac{d^2}{ds^2}\Phi(s\Eta, sY)&=
\ddot{\mathrm{C}}^1(s)\langle\theta^2|\tilde{\pi}^1+s\varpi\rangle
+2\dot{\mathrm{C}}^1(s)\langle\theta^2|\varpi\rangle-\ddot{{\mathrm{D}}}^1(s)\theta^2_1\theta^2_2\\
&-\ddot{\mathrm{E}}^1(s)(\tilde{\pi}^1_1+s\varpi_1)(\tilde{\pi}^1_2+s\varpi_2)
-2\dot{\mathrm{E}}^1(s)[\varpi_1(\tilde{\pi}^1_2+s\varpi_2)+(\tilde{\pi}^1_1+s\varpi_1)\varpi_2]\\
&\qquad
-2{\mathrm{E}}^1(s)\varpi_1\varpi_2-2\langle y|\eta\rangle
-2\langle \vartheta|\varpi\rangle-2{\mathrm{D}}^0\vartheta_1\vartheta_2,\\
%%%
\frac{d^3}{ds^3}\Phi(s\Eta, sY)&=
\dddot{\mathrm{C}}^1(s)\langle\theta^2|\tilde{\pi}^1+s\varpi\rangle
+3\ddot{\mathrm{C}}^1(s)\langle\theta^2|\varpi\rangle
-\dddot{\mathrm{D}}^1(s)\theta^2_1\theta^2_2\\
&-\dddot{\mathrm{E}}^1(s)(\tilde{\pi}^1_1+s\varpi_1)(\tilde{\pi}^1_2+s\varpi_2)
-3\ddot{\mathrm{E}}^1(s)[\varpi_1(\tilde{\pi}^1_2+s\varpi_2)+(\tilde{\pi}^1_1+s\varpi_1)\varpi_2]\\
&\qquad
-6\dot{\mathrm{E}}^1(s)\varpi_1\varpi_2\\
&=r_0(s)+r^1_1(s)\varpi_1+r^1_2(s)\varpi_2+r_2(s)\varpi_1\varpi_2
\end{aligned}\label{XXX}
\end{equation}
where
$$
\begin{aligned}
r_0(s)&=\dddot{{\mathrm{C}}}^1(s)\langle\theta^2|\tilde{\pi}^1\rangle-\dddot{{\mathrm{D}}}^1(s)\theta^2_1\theta^2_2
-\dddot{{\mathrm{E}}}^1(s)\tilde{\pi}^1_1\tilde{\pi}^1_2,\\
r_1^1(s)&=s\dddot{{\mathrm{C}}}^1(s)\theta_1^2+3\ddot{{\mathrm{C}}}^1(s)\theta_1^2
+s\dddot{{\mathrm{E}}}^1(s)\tilde{\pi}^1_2+3\ddot{{\mathrm{E}}}^1(s)\tilde{\pi}^1_2,\\
r_2^1(s)&=s\dddot{{\mathrm{C}}}^1(s)\theta_2^2+3\ddot{{\mathrm{C}}}^1(s)\theta_2^2
-s\dddot{{\mathrm{E}}}^1(s)\tilde{\pi}^1_1-3\ddot{{\mathrm{E}}}^1(s)\tilde{\pi}^1_1,\\
r_2(s)&=-s^2\dddot{{\mathrm{E}}}^1(s)-6s\ddot{{\mathrm{E}}}^1(s)-6\dot{{\mathrm{E}}}^1(s).
\end{aligned}
$$

\begin{exercise}
Show, by plugging \eqref{SP12} into $\Phi(0,0)$,
$$
\begin{aligned}
\Phi(0,0)
&=\langle x^2|\tilde{\xi}^1\rangle
+{\mathrm{C}}^1(0)\langle\theta^2|\tilde{\pi}^1\rangle
-{\mathrm{D}}^1(0)\theta^2_1\theta^2_2
-{\mathrm{E}}^1(0)\tilde{\pi}^1_1\tilde{\pi}^1_2\\
&\qquad
-\langle \tilde{x}^1|\tilde{\xi}^1\rangle-\langle \tilde{\theta}^1\tilde{\pi}^1\rangle+\langle \tilde{x}^1|\xi^0\rangle
+{\mathrm{C}}^0\langle \tilde{\theta}^1|\pi^0\rangle
-{\mathrm{D}}^0\tilde{\theta}^1_1\tilde{\theta}^1_2
-{\mathrm{E}}^0\pi^0_1\pi^0_2\\
&={\mathcal{S}}(t_2-t_0,X^2,\Xi^0).
\end{aligned}
$$
Proceed analogously to prove,
$$
\frac{d}{ds}\Phi(0,0)=0.
$$
\end{exercise}

For terms in \eqref{XXX}, using integration by parts, we get
$$
\int_0^1ds\frac{(1-s)^2}{2!}\frac{d^3}{ds^3}\Phi(s\Eta, sY)
=R_{0}+R^1_1\varpi_1+R^1_2\varpi_2+R_{2}\varpi_1\varpi_2
$$
with
$$
\begin{aligned}
R_{0}&=R_0(X^2,\eta,\Xi^0)=\int_0^1ds\frac{(1-s)^2}{2!}
(\dddot{{\mathrm{C}}}^1(s)\langle\theta^2|\tilde{\pi}^1\rangle-\dddot{{\mathrm{D}}}^1(s)\theta^2_1\theta^2_2
-\dddot{{\mathrm{E}}}^1(s)\tilde{\pi}^1_1\tilde{\pi}^1_2\\
&=({\mathrm{C}}(1)-{\mathrm{C}}(0)-\dot{{\mathrm{C}}}(0)-\frac{1}{2}\ddot{{\mathrm{C}}}(0))\langle\theta^2|\tilde{\pi}^1\rangle
-({\mathrm{D}}(1)-{\mathrm{D}}(0)-\dot{{\mathrm{D}}}(0)-\frac{1}{2}\ddot{{\mathrm{D}}}(0))\theta^2_1\theta^2_2\\
&\qquad\qquad
-({\mathrm{E}}(1)-{\mathrm{E}}(0)-\dot{{\mathrm{E}}}(0)-\frac{1}{2}\ddot{{\mathrm{E}}}(0))\tilde{\pi}^1_1\tilde{\pi}^1_2,\\
\end{aligned}
$$
$$
\begin{aligned}
R^1_1&=R^1_1(X^2,\eta,\Xi^0)=\int_0^1ds\frac{(1-s)^2}{2!}(s\dddot{{\mathrm{C}}}^1(s)\theta_1^2+3\ddot{{\mathrm{C}}}^1(s)\theta_1^2
+s\dddot{{\mathrm{E}}}^1(s)\tilde{\pi}^1_2+3\ddot{{\mathrm{E}}}^1(s)\tilde{\pi}^1_2)\\
&=({\mathrm{C}}(1)-{\mathrm{C}}(0)-\dot{{\mathrm{C}}}(0))\theta_1^2+({\mathrm{E}}(1)-{\mathrm{E}}(0)-\dot{{\mathrm{E}}}(0))\tilde{\pi}_2^1,\\
R^1_2&=R^1_2(X^2,\eta,\Xi^0)=\int_0^1ds\frac{(1-s)^2}{2!}(s\dddot{{\mathrm{C}}}^1(s)\theta_2^2+3\ddot{{\mathrm{C}}}^1(s)\theta_2^2
-s\dddot{{\mathrm{E}}}^1(s)\tilde{\pi}^1_1-3\ddot{{\mathrm{E}}}^1(s)\tilde{\pi}^1_1)\\
&=({\mathrm{C}}(1)-{\mathrm{C}}(0)-\dot{{\mathrm{C}}}(0))\theta_2^2-({\mathrm{E}}(1)-{\mathrm{E}}(0)-\dot{{\mathrm{E}}}(0))\tilde{\pi}_1^1,\\
R_2&=R_2(X^2,\eta,\Xi^0)=-\int_0^1ds\frac{(1-s)^2}{2!}(s^2\dddot{{\mathrm{E}}}^1(s)+6s\ddot{{\mathrm{E}}}^1(s)+6\dot{{\mathrm{E}}}^1(s))\\
&=-{\mathrm{E}}^1(1)+{\mathrm{E}}^1(0).
\end{aligned}
$$
From \eqref{SP12},  we have
$$
\begin{aligned}
\langle\theta^2|\tilde{\pi}^1\rangle&=\frac{1}{\bar{\delta}(\tau_1+\tau_2,|\xi^0|)}(\bar{\delta}(\tau_2,|\xi^0|)\langle\theta^2|{\pi}^0\rangle
-2\hbar\zeta^0\sin\gamma(\tau_1,|\xi^0|)\theta_1^2\theta_2^2),\\
\tilde{\pi}^1_1\tilde{\pi}^1_2&=\frac{1}{\bar{\delta}^2(\tau_1+\tau_2,|\xi^0|)}(\bar{\delta}^2(\tau_2,|\xi^0|)\pi^0_1\pi^0_2
-\hbar\zeta^0\sin\gamma(\tau_2,|\xi^0|)\bar{\delta}(\tau_2,|\xi^0|)\langle\theta^2|{\pi}^0\rangle\\
&\qquad\qquad\qquad\qquad\qquad\qquad\qquad\qquad\qquad\qquad\qquad
-\hbar^2(\zeta^0)^2\sin^2\gamma(\tau_2,|\xi^0|)\theta_1^2\theta_2^2.
\end{aligned}
$$
Using above, we have
\begin{equation}
\begin{aligned}
\frac{d^2}{ds^2}\Phi(0,0)&=
\ddot{{\mathrm{C}}}^1(0)\langle\theta^2|\tilde{\pi}^1\rangle
+2\dot{{\mathrm{C}}}^1(0)\langle\theta^2|\varpi\rangle-\ddot{{\mathrm{D}}}^1(0)\theta^2_1\theta^2_2\\
&\qquad-\ddot{{\mathrm{E}}}^1(0)\tilde{\pi}^1_1\tilde{\pi}^1_2
-2\dot{{\mathrm{E}}}^1(0)[\varpi_1\tilde{\pi}^1_2+\tilde{\pi}^1_1\varpi_2]\\
&\qquad\qquad
-2{\mathrm{E}}^1(0)\varpi_1\varpi_2-2\langle y|\eta\rangle
-2\langle \vartheta|\varpi\rangle-2{\mathrm{D}}^0\vartheta_1\vartheta_2\\
&=\Phi_2(X^2,\eta,\Xi^0)-2\langle y|\eta\rangle+\Psi_2(\theta^2,\eta,\varpi,\vartheta,\pi^0,\xi^0)
\end{aligned}
\label{phi2*}
\end{equation}
with
$$
\begin{aligned}
\Phi_2(X^2,\eta,\Xi^0)&=\ddot{{\mathrm{C}}}^1(0)\langle\theta^2|\tilde{\pi}^1\rangle
-\ddot{{\mathrm{D}}}^1(0)\theta^2_1\theta^2_2-\ddot{{\mathrm{E}}}^1(0)\tilde{\pi}^1_1\tilde{\pi}^1_2,\\
\Psi_2(\varpi,\vartheta,\theta^2,\eta,\pi^0,\xi^0)&=2\dot{{\mathrm{C}}}^1(0)\langle\theta^2|\varpi\rangle-2\dot{{\mathrm{E}}}^1(0)[\varpi_1\tilde{\pi}^1_2+\tilde{\pi}^1_1\varpi_2]-2{\mathrm{E}}^1(0)\varpi_1\varpi_2-2\langle \vartheta|\varpi\rangle-2{\mathrm{D}}^0\vartheta_1\vartheta_2.
\end{aligned}
$$

From \eqref{phi2*}, we have
$$
\begin{aligned}
c_{3,2}^2\iint_{{\fR}^{6|4}}&d{\Eta}d{Y}\,{\mathcal{A}}(\tau_2,\tilde{\Xi}^1+\Eta){\mathcal{A}}(\tau_1,\Xi^0)e^{i\hbar^{-1}[\Phi(0,0)+2^{-1}\ddot{\Phi}(0,0)+R]}\\
&=e^{i\hbar^{-1}\Phi(0,0)}\hbar^2\iint d\varpi d\vartheta
\bigg[(2\pi\hbar)^{-3}\iint d\eta dy \frac{\bar{\delta}(\tau_1,|\xi^0|)}{|\xi^0|}\frac{\bar{\delta}(\tau_2,|\xi^0+\eta|)}{|\xi^0+\eta|}\\
&\qquad\qquad\times
e^{i\hbar^{-1}(-\langle y|\eta\rangle+2^{-1}\Phi_2(X^2,\eta,\Xi^0)+2^{-1}\Psi_2(\varpi,\vartheta,\eta,\pi^0,\xi^0)+R^1_2\varpi_2+R_{2}(\eta)\varpi_1\varpi_2+R_{0})}\bigg].
\end{aligned}
$$
Remarking
$$
(2\pi\hbar)^{-3}\iint dy d\eta\, e^{-i\hbar^{-1}\langle y|\eta\rangle}g(\eta)=g(0),
$$
holds for suitable  function $g$, and
\begin{equation}
\begin{gathered}
\dot{\mathrm{C}}^1(0)\big|_{\eta=0}=0,\quad \ddot{\mathrm{C}}^1(0)\big|_{\eta=0}=0,
\quad \dddot{\mathrm{C}}^1(0)\big|_{\eta=0}=0,\\
\dot{\mathrm{D}}^1(0)\big|_{\eta=0}=0,\quad \ddot{\mathrm{D}}^1(0)\big|_{\eta=0}=0,
\quad \dddot{\mathrm{D}}^1(0)\big|_{\eta=0}=0,\\
\dot{\mathrm{E}}^1(0)\big|_{\eta=0}=0,\quad \ddot{\mathrm{E}}^1(0)\big|_{\eta=0}=0,
\quad \dddot{\mathrm{E}}^1(0)\big|_{\eta=0}=0,
\end{gathered}
\label{A11-19bis}
\end{equation}
we have
$$
\frac{d^2}{ds^2}\Phi(0,0)(X^2,\eta,\Xi^0)\big|_{\eta=0}=-2{\mathrm{E}}^1(0)\varpi_1\varpi_2-2\langle \vartheta|\varpi\rangle-2{\mathrm{D}}^0\vartheta_1\vartheta_2\et
R\big|_{\eta=0}=0.
$$
Therefore
\begin{equation}
\begin{aligned}
c_{3,2}^2&\iint_{{\fR}^{6|4}}d{\Eta}d{Y}\,{\mathcal{A}}(\tau_2,\tilde{\Xi}^1+\Eta){\mathcal{A}}(\tau_1,\Xi^0)e^{i\hbar^{-1}[\Phi(0,0)+2^{-1}\ddot{\Phi}(0,0)+R]}\\
&=e^{i\hbar^{-1}\Phi(0,0)}\hbar^2\frac{\bar{\delta}(\tau_1,|\xi^0|)}{|\xi^0|}\frac{\bar{\delta}(\tau_2,|\xi^0|)}{|\xi^0|}\iint d\varpi d\vartheta
\exp{[-{\mathrm{E}}^1(0)\varpi_1\varpi_2-\langle \vartheta|\varpi\rangle-{\mathrm{D}}^0\vartheta_1\vartheta_2]}\\
&=e^{i\hbar^{-1}{\mathcal{S}}(\tau_1+\tau_2,X^2\Xi^0)}\hbar^2\frac{\bar{\delta}(\tau_1,|\xi^0|)}{|\xi^0|}\frac{\bar{\delta}(\tau_2,|\xi^0|)}{|\xi^0|}(1-D^0E^1(0))\\
&=e^{i\hbar^{-1}{\mathcal{S}}(\tau_1+\tau_2,X^2\Xi^0)}\hbar^2\frac{\bar{\delta}(\tau_1+\tau_2,|\xi^0|)}{|\xi^0|}.
\end{aligned}
\label{A11-19}
\end{equation}

In fact, to prove \eqref{A11-19bis},
$$
\frac{d}{ds}|{\xi}^0+s\eta|=\frac{\langle \xi^0+s\eta|\eta\rangle}{|{\xi}^0+s\eta|},\quad
\frac{d^2}{ds^2}|{\xi}^0+s\eta|=\frac{|\eta|^2|\xi^0+s\eta|^2-\langle \xi^0+s\eta|\eta\rangle^2}{|{\xi}^0+s\eta|^3},
$$
and putting $\gamma_{\tau}(s)=c\hbar^{-1}\tau|{\xi}^0+s\eta|$, we have
$$
\frac{d}{ds}\gamma_{\tau}(s)
=c\hbar^{-1}\tau\frac{d}{ds}|{\xi}^0+s\eta|,\;\;
\frac{d^2}{ds^2}\gamma_{\tau}(s)
=c\hbar^{-1}\tau\frac{d^2}{ds^2}|{\xi}^0+s\eta|,
$$
therefore,
$$
\begin{gathered}
\frac{d}{ds}|{\xi}^0+s\eta|\big|_{\eta=0}=0,\quad \frac{d^2}{ds^2}|{\xi}^0+s\eta|\big|_{\eta=0}=0,\\
\frac{d}{ds}\gamma_{\tau}(s)\big|_{\eta=0}=0,\quad \frac{d^2}{ds^2}\gamma_{\tau}(s)\big|_{\eta=0}=0.
\end{gathered}
$$

Using above
$$
\begin{aligned}
&\frac{d}{ds}\bar{\delta}(\tau,|{\xi}^0+s\eta|)
=\frac{d}{ds}|{\xi}^0+s\eta|\cos\gamma_{\tau}(s)-|{\xi}^0+s\eta|\dot{\gamma}_{\tau}(s)\sin\gamma_{\tau}(s)\\
&\qquad
-i\eta_3\sin\gamma_{\tau}(s)-i(\xi^0_3+s\eta_3)\dot{\gamma}_{\tau}(s)\cos\gamma_{\tau}(s),\\
&\frac{d}{ds}\bar{\delta}(\tau,|{\xi}^0+s\eta|)\big|_{\eta=0}=0,\\
&\frac{d^2}{ds^2}\bar{\delta}(\tau,|{\xi}^0+s\eta|)
=\frac{d^2}{ds^2}|{\xi}^0+s\eta|\cos\gamma_{\tau}(s)-2\frac{d}{ds}|{\xi}^0+s\eta|\dot{\gamma}_{\tau}(s)\sin\gamma_{\tau}(s)\\
&\qquad
-|{\xi}^0+s\eta|[\ddot{\gamma}_{\tau}(s)\sin\gamma_{\tau}(s)+\dot{\gamma}^2_{\tau}(s)\cos\gamma_{\tau}(s)]\\
&\qquad
-2i\eta_3\dot{\gamma}_{\tau}(s)\cos\gamma_{\tau}(s)-i(\xi^0_3+s\eta_3)[\ddot{\gamma}_{\tau}(s)\cos\gamma_{\tau}(s)
-\dot{\gamma}^2_{\tau}(s)\sin\gamma_{\tau}(s)],\\
&\frac{d^2}{ds^2}\bar{\delta}(\tau,|{\xi}^0+s\eta|)\bigg|_{s=0}=0.
\end{aligned}
$$ 
From these, we get
$$
\begin{aligned}
&\dot{\mathrm{C}}^1(s)=\frac{d}{ds}|\xi^0+s\eta|\bar{\delta}^{-1}(\tau,|\xi^0+s\eta|)-|\xi^0+s\eta|\bar{\delta}^{-2}(\tau,|\xi^0+s\eta|)
\frac{d}{ds}\bar{\delta}(\tau,|\xi^0+s\eta|),\\
&\ddot{\mathrm{C}}^1(s)=\frac{d^2}{ds^2}|{\xi}^0+s\eta|\bar{\delta}^{-1}(\tau,|\xi^0+s\eta|)-2\frac{d}{ds}|{\xi}^0+s\eta|\frac{d}{ds}\bar{\delta}^{-1}(\tau,|\xi^0+s\eta|)+|{\xi}^0+s\eta|\frac{d^2}{ds^2}\bar{\delta}^{-1}(\tau,|\xi^0+s\eta|),\\
&\dot{\mathrm{C}}^1(0)\bigg|_{\eta=0}=0,\quad
\ddot{\mathrm{C}}^1(0)\bigg|_{\eta=0}=0,\\
\end{aligned}
$$
Analogously, we prove other equalities in \eqref{A11-19bis}.\quad/\!\!/

\chapter[SUSYQM and Its Applications] {Supersymmetric Quantum Mechanics and Its Applications}
\section{What is SUSYQM}
\subsection{Another interpretation of the Atiyah-Singer index theorem}
Seemingly, being stimulated by a physicist E. Witten's paper~\cite{witt82-1}, a mathematician E. Gezler declared 
in the introduction of his paper~\cite{get83} that 
\begin{quote}
\underline{The Atiyah-Singer index theorem is nothing but the superversion of the Weyl's}\\
\underline{theorem on the asymptotic behavior w.r.t. time $t$ for $e^{t\Delta_g/2}$}.
\end{quote}
Here, $(M,g)$ is a compact $d$-dimensional Riemannnian manifold, $\Delta_g$ is the Laplace-Beltrami operator corresponding to the Riemannnian metric $g=g_{jk}(q)dq^jdq^k$.
Though, he declared this, but he didn't try to demonstrate this assertion directly in that paper.

\paragraph{Our goal in this chapter} \underline{We interprete his declaration and calculate the index for the simplest}\\ \underline{example following prescription of Witten and Getzler.}

Roughly speaking, his declaration is sketched as follows:
Let $K(t,q,q')$ be the kernel of the fundamental solution of IVP
$$
\pdt v(t,q)=\frac{1}{2}\Delta_g v(t,q)\with \lim_{t\to0}v(t,q)=\underline{v}(q).
$$
That is,
$$
v(t,q)=\int_{M}d_gq' K(t,q,q')\underline{v}(q')=(e^{t\Delta_g/2}\underline{v})(q)
$$
where
$$
g(q)=\det(g_{ij}(q)),\quad d_gq=\sqrt{g(q)}dq,\quad
\Delta_g=d^* d=\frac{1}{\sqrt{g(q)}}\partial_{q^i}(\sqrt{g(q)}g^{ij}(q)\partial_{q^j}).
$$
Then, the Weyl's theorem states that $e^{t\Delta_g/2}$ belongs to trace class and
$$
\tr(e^{t\Delta_g/2})=\int_{M}d_gq\, K(t,q,q)\to Ct^{-d/2}\int_{M}d_gq \,1 \when t\to0.
$$
Here, $d$ is the exterior differential, $d^*$ is the adjoint of $d$ w.r.t. $d_gq$ and $\Delta_g=d^*d$ which is desired to be derived from $\sum_{j,k=1}^d g^{jk}(q)p_jp_k\in C^{\infty}(T^*M:\euc)$ by ``quantization''.
\par
His claim goes as follows: Extend\footnote{for simplicity, we only consider the case $M=\euc^d$ in this section, and concerning Riemann metric $g=g_{ij}(q)dq^idq^j$ on $\euc^d$} $(M,g)$ superly denoted by $(\tilde{M}, \tilde{g})$ where $\tilde{M}$ is a supermanifold corresponding to $M$ and $\tilde{g}$ is a super Riemann metric of $g$.
In this case, $\Delta_{\tilde{g}}$ corresponds to  the form Laplacian
$dd^*+d^*d$ acting on differential forms on $(M,g)$, moreover, it has the supersymmetric structure. 

Therefore, calculating the trace of the kernel for  ``$e^{t(dd^*+d^*d)/2}$'' , we get the Witten index which gives us new proof of Atiyah-Singer index theorem.
\begin{remark}
(0) For a given  Riemann metric $g$ on $\euc^d$, we calculate its super-extension $\tilde{g}$ in \S6 of Chapter 9 which corresponds to the symbol of $dd^*+d^*d$. \\
(i) What occurs when we quantize Lagrangian $(1/2)\sum_{j,k=1}^dg_{jk}(q)\dot{q}^j\dot{q}^k$ on $(M,g)$?
In case if we quantize following Feynman's prescription with purely imaginary time, the quantized object deviate $(1/12)R$ from $(1/2)\Delta_g$ with $R$=the scalar curvature (see  B. DeWitt~\cite{deW84}, Inoue-Maeda~\cite{IM85}).\\
(ii) See also, the recent work of Y. Miyanishi ~\cite{miy13}, where he constructs a parametrix for the Schr\"odinger equation on $S^2$ with action integral deformed with $(1/12)R$ from $(1/2)\Delta_g$, but $R=2$ for $S^2$. 
More precisely, he goes as follows;
\par
Let $\unbq, \barq$ be 2 points on $S^2$, let $\gamma_0\in C_{t,\unbq,\barq}$ be the shortest path between them with length $d(\unbq,\barq)$. Taking a bump function $\chi$ with compact support contained in $d(\unbq,\barq)<\pi$,
he defines an integral operator
$$
U(t)\unbu(\barq)=\frac{1}{2\pi i\hbar}\int_{S^2}d_g\unbq\, \chi(d(\unbq,\barq))A(t,\unbq,\barq)e^{i\hbar (S(t,\unbq,\barq)+iRt/12)}\unbu(\unbq)
$$
where %$g(q)=\det g_{ij}(q)$, $d_gq=\sqrt{g(q)}dq$,
$$
S(t,\unbq,\barq)=\frac{d(\unbq,\barq)^2}{2t}
\et
A(t,\unbq,\barq)=\bigg[g^{-1/2}(\unbq)g^{-1/2}(\barq)\det\bigg(\frac{\partial^2S(t,\unbq,\barq)}{\partial\unbq \partial\barq}\bigg)\bigg]^{1/2}.
$$
Then, he asserts that taking the suitable products of these operators corresponding to time slicing method and restricting it to ``lower energy'' part of $(1/2)\Delta_g$, then it converges to the solution of
$$
i\hbar\frac{\partial u(t,\barq)}{\partial t}=\frac{\hbar^2}{2}\Delta_g u(t,\barq) \with \lim_{t\to0}u(t,\barq)=\unbu(\barq).
$$

His definition of the integral operator is different from ours because he needs to introduce additionally the cut off and to use not only the action integral but also van Vleck determinant corresponding to the shortest path between two points
\footnote{In our case considered, we only have the unique classical trajectory!}.
By the way, how to recognize the claim ``put equal weights for every possible paths'' in physics literature?
From my point of view, if we consider ``weights'' as amplitude, we need to use
$$
\bigg[\det\bigg(\frac{\partial^2S(\gamma)}{\partial\unbq \partial\barq}\bigg)\bigg]^{1/2}
%e^{i\hbar^{-1}S(\gamma)} 
\with \forany\gamma\in C_{t,\unbq,\barq}
$$
or need another phase factor for each path as proposed in L. Schulman~\cite{schul68}.

The usage of  the projection to low energy part corresponding to the spectral decomposition for $(1/2)\Delta_g$ make us suspicious ``Is his procedure truely quantization?'', because the quantization should be carried out only using classical quantities. 
To overcome this point, it seems reasonable to present such projector using classical objects like Fujiwara ~\cite{fuj74}.f
Moreover, he adds also the factor
$g^{-1/2}(\unbq)g^{-1/2}(\barq)$ to permit Copenghagen interpretation, that is, consider time evolution in the intrinsic Hilbert space (=half density bundle).

In spite of these, we have\footnote{no problem for integrability and differentiation under integral sign for this case}
$$
i\hbar{\frac{\partial}{\partial t}} \bigg[\int_{S^2}d_g\unbq\,
\bigg[\det\bigg(\frac{\partial^2S(t,\unbq,\barq)}{\partial\unbq \partial\barq}\bigg)\bigg]^{1/2}
e^{i\hbar^{-1}S(t,\unbq,\barq)}\unbu(\unbq)\bigg]\bigg|_{t=0}=\hbar^2\bigg[\frac{1}{2}\Delta_g+\frac{1}{12} R\bigg]\unbu(\barq).
$$
That is, this guarantees us Feynman's picture of quantization. 
Therefore, it seems more natural
to consider separately\footnote{From Feynman's introduction and Fujiwara's procedure, we, at least myself, insist too much to get quantized object from Feynman picture implies also to have directly fundamental solution of Schr\"odinger equation by his method}
two things, one is quantization process and another is construction of  the fundamental solution for evolution equation corresponding to that quantized object having as infinitesimal operator.
\end{remark}
\begin{remark}
As mentioned before, how do we interpret the saying ``put equal weights for every possible paths'': it is explained, for example, in
D. V. Perepelitsa~\cite{per06} as follows(with slight modification):
\begin{quotation}
Feynman~\cite{fey48} posits that the contribution to the propagator from a particular trajectory is 
$\exp [i\hbar^{-1}S(\gamma)]$ where $\gamma=\gamma(\cdot)\in C_{t,\unbq,\barq}$.
That is, every possible path contributes with equal amplitude to the propagator, but with a phase related to the classical action. Summing over all possible trajectories, we arrive at the propagator. The normalization constant $A(t)$ is independent of any individual path and therefore depends only on time.
$$
U(\barq,t; \unbq,0)=A(t)\sum_{\gamma\in C_{t,\unbq,\barq}}e^{i\hbar^{-1}S(\gamma)}.
$$
\end{quotation}
Since I explained in sectiion 1.3 of Chapter 1, as  there \underline{doesn't exist full Feynman measure}\footnote{Recall also, there \underline{doesn't exist ``functor'' called full quantization}, see Abraham-Marsden~\cite{AM80}}, we ``approximate''  $D_F\gamma$ on $C_{t,\unbq,\barq}$ by the measure on $M$ with some density function, that is,
$$
\begin{gathered}
{D}(\gamma)=\bigg[\det\bigg(\frac{\partial^2\tilde{S}(t,\unbq,\barq)}{\partial\unbq \partial\barq}\bigg)\bigg]^{1/2}
={D}(t,\unbq,\barq),\\
\where {S}(\gamma)={S}(t,\unbq,\barq)=\int_0^t dt\,L(\gamma(s), \dot{\gamma}(s)),
\end{gathered}
$$
but even taking the classical trajectory $\gamma_c\in C_{t,\unbq,\barq}$ in ${D}(\gamma_c)$, it generally depends not only on $t$ but also $(\unbq,\barq)$?
\end{remark}

\subsection{What is SUSYQM?}
In order to make clear what should be calculated, we cite the definition.
\begin{definition}[p.120, H.L. Cycon, R.G. Froese, W. Kirsh and B. Simon~\cite{CFKS}]
Let ${\mathfrak{H}}$ be a Hilbert space and let ${\mathbf{H}}$ and ${\mathbf{Q}}$ be selfadjoint operators, 
and
${\mathbf{P}}$ be a bounded self-adjoint operator in ${\mathfrak{H}}$ such that
$$
{\mathbf{H}}={\mathbf{Q}}^2\ge0,\quad {\mathbf{P}}^2={\mathbf{I}},\quad [{\mathbf{Q}},{\mathbf{P}}]_+={\mathbf{Q}}{\mathbf{P}}+{\mathbf{P}}{\mathbf{Q}}=0.
$$ Then, we say that the system $({\mathbf{H}},{\mathbf{P}},{\mathbf{Q}})$ has supersymmetry or it defines 
a SUSYQM(=supersymmetric Quantum Mechanics).
\end{definition}
Under this circumstance, we may decompose
$$
{\mathfrak{H}}={\mathfrak{H}}_{\mathrm b}\oplus{\mathfrak{H}}_{\mathrm f}
\where
{\mathfrak{H}}_{\mathrm f}=\{u\in{\mathfrak{H}}\,|\, {\mathbf{P}}u=-u\},\quad
{\mathfrak{H}}_{\mathrm b}=\{u\in{\mathfrak{H}}\,|\, {\mathbf{P}}u=u\}.
$$
Using this decomposition and identifying an element 
$u=u_{\mathrm b}+u_{\mathrm f}\in{\mathfrak{H}}$
as a vector $\begin{pmatrix} u_{\mathrm b}\\u_{\mathrm f} \end{pmatrix}$, 
we have a representation
$$
{\mathbf{P}}=\begin{pmatrix}
{\mathbf{I}}_{\mathrm b}&0\\
0&-{\mathbf{I}}_{\mathrm f}
\end{pmatrix}=\text{(or simply denoted by)}
\begin{pmatrix}
1&0\\
0&-1
\end{pmatrix}.
$$
Since ${\mathbf{P}}$ and ${\mathbf{Q}}$ anti-commute and ${\mathbf{Q}}$ is self-adjoint, ${\mathbf{Q}}$ has always the form
\begin{equation}
{\mathbf{Q}}=\begin{pmatrix}
0&{\mathbf{A}}^*\\
{\mathbf{A}}&0
\end{pmatrix} 
\et
{\mathbf{H}}=\begin{pmatrix}
{\mathbf{A}}^*{\mathbf{A}}&0\\
0&{\mathbf{A}}{\mathbf{A}}^*
\end{pmatrix},
\label{rep}\end{equation}
where ${\mathbf{A}}$, called the annihilation operator, 
is an operator which maps ${\mathfrak{H}}_{\mathrm b}$ into ${\mathfrak{H}}_{\mathrm f}$, 
and its adjoint ${\mathbf{A}}^*$, called the creation operator, 
maps ${\mathfrak{H}}_{\mathrm f}$ into ${\mathfrak{H}}_{\mathrm b}$.
Thus, ${\mathbf{P}}$ commutes with ${\mathbf{H}}$, 
and ${\mathfrak{H}}_{\mathrm b}$ and ${\mathfrak{H}}_{\mathrm f}$ are invariant under ${\mathbf{H}}$,
i.e. ${\mathbf{H}}{\mathfrak{H}}_{\mathrm b}\subset{\mathfrak{H}}_{\mathrm b}$ 
and ${\mathbf{H}}{\mathfrak{H}}_{\mathrm f}\subset{\mathfrak{H}}_{\mathrm f}$.
That is, there is a one-to-one correspondence between densely defined closd operators ${\mathbf{A}}$
and self-adjoint operators ${\mathbf{Q}}$ ({\bf supercharges}) of the above form.

\begin{definition} We define a supersymmetric index of ${\mathbf{H}}$ if it exists by
$$
\ind_s({\mathbf{H}})\equiv\dim(\Ker({\mathbf{H}}\big|{\mathfrak{H}}_{\mathrm b}))
-\dim(\Ker({\mathbf{H}}\big|{\mathfrak{H}}_{\mathrm f}))
\in \bar{\mathbb{Z}}={\mathbb{Z}}\cup\{\pm\infty\}.
$$
\end{definition}

On the other hand, we have
\begin{definition}
Let $X$, $Y$ be two Banach spaces and let ${\mathcal{C}}(X,Y)$ be a set of
densely defined closed operators from $X$ to $Y$.
$T\in{\mathcal{C}}(X,Y)$ is called Fredholm iff the range of $T$, $R(T)$, 
is closed in $Y$ and
both $ker T$ and $Y/R(T)$ are finite-dimensional.
$T\in{\mathcal{C}}(X,Y)$ is called semi-Fredholm iff $R(T)$ 
is closed in $Y$ and
at least one of $ker T$ and $Y/R(T)$ is finite-dimensional.
If the operator is semi-Fredholm,
then the Fredholm index $\ind_F(T)=\dim(\ker T)-dim(Y/R(T))$ exists in 
$\bar{\mathbb{Z}}$.
\end{definition}

\begin{corollary}
If the operator $A$ is semi-Fredholm, we have the relation
$$
\ind_s(H)=\text{ind}_F(A)\equiv\dim(\Ker A)-\dim(\Ker A^*).
$$
\end{corollary}

In order to check whether the supersymmetry is broken or unbroken,
E. Witten introduced the so-called Witten index.
\begin{definition} Let $({\mathbf{H}},{\mathbf{P}},{\mathbf{Q}})$ be SUSYQM with \eqref{rep}.
\newline
(I) Putting, for $t>0$ 
$$
\Delta_t({\mathbf{H}})=\tr (e^{-t{\mathbf{A}}^*{\mathbf{A}}}-e^{-t{\mathbf{A}}{\mathbf{A}}^*})=\str e^{-t{\mathbf{H}}},
$$
we define, if the limit exists, 
the (heat kernel regulated) Witten index ${\mathcal{W}}_H$ of $({\mathbf{H}},{\mathbf{P}},{\mathbf{Q}})$ by
$$
{\mathcal{W}}_H=\lim_{t\to\infty}\Delta_t({\mathbf{H}}).
$$
We define also the (heat kernel regulated) 
axial anomaly ${\mathcal{A}}_H$ of $({\mathbf{H}},{\mathbf{P}},{\mathbf{Q}})$ by
$$
{\mathcal{A}}_H=\lim_{t\to0}\Delta_t({\mathbf{H}}).
$$
\newline
(II) Putting, for $z\in{\mathbb{C}}\setminus[0,\infty)$,
$$
\Delta_z({\mathbf{H}})=-z\tr[({\mathbf{A}}^*{\mathbf{A}}-z)^{-1}-({\mathbf{A}}{\mathbf{A}}^*-z)^{-1}]=-z\str ({\mathbf{H}}-z)^{-1},
$$
we define 
the (resolvent regulated) Witten index ${\mathcal{W}}_R$ of $({\mathbf{H}},{\mathbf{P}},{\mathbf{Q}})$, if the limit exists, by
$$
{\mathcal{W}}_R=\lim_{{\scriptstyle{z\to0,}}\atop{\scriptstyle{|\Re z|\le C_0|\Im z|}}}
\Delta_z({\mathbf{H}})
\quad\mbox{for some $C_0>0$}.
$$
Similarly, 
we define the (resolvent regulated) axial anomaly ${\mathcal{A}}_R$ by
$$
{\mathcal{A}}_R=-\lim
_{{\scriptstyle{z\to\infty}}\atop{\scriptstyle{|\Re z|\le C_1|\Im z|}}}
\Delta_z(H)\quad\mbox{for some $C_1>0$}.
$$
\end{definition}

We have
\begin{theorem}
Let ${\mathbf{Q}}$ be a supercharge on ${\mathfrak{H}}$.
If $\exp(-t{\mathbf{Q}}^2)$ is trace class for some $t>0$, then ${\mathbf{Q}}$ is Fredholm and
$$
\ind_t({\mathbf{Q}})(\mbox{independent of $t$})=\ind_F({\mathbf{Q}})=\ind_s(H).
$$
If $({\mathbf{Q}}^2-z)^{-1}$ is trace class for some $z\in{\mathbb{C}}\setminus[0,\infty)$,
then ${\mathbf{Q}}$ is Fredholm and
$$
\ind_z({\mathbf{Q}})(\mbox{independent of $z$})=\ind_F({\mathbf{Q}})=\ind_s(H).
$$
\end{theorem}
\begin{quotation}
{\small
Concerning definitions used in above theorem:
\begin{definition}
Let $X$, $Y$ be two Banach spaces and let ${\mathcal C}(X,Y)$ be the set of all densely  defined closed operators from 
$X$ to $Y$. $T\in{\mathcal C}(X,Y)$ is called Fredholm if it has closed range $R(T)$ in $Y$ and $\ker T$ and $Y/R(T)$
are finite dimension.
$T\in{\mathcal C}(X,Y)$ is called semi-Fredholm if $R(T)$ is closed in $Y$ and if at least one of $\ker T$ or $Y/R(T)$
is finite dimension.
\par
If $T\in{\mathcal C}(X,Y)$ is semi-Fredholm, then  Fredholm index $\ind_F(T)=\dim(\ker T)-\dim(Y/R(T))$
exists in $\bar{\mathbb Z}$.
\end{definition}

\begin{corollary}
If an operator $A$ is semi-Fredholm, we have
$$
\ind_s({\mathbb{H}})={\ind}_F(A)\equiv\dim(\ker A)-\dim(\ker A^*).
$$
\end{corollary}
}
\end{quotation}

\subsection{Examples of SUSYQM}
\paragraph{\bf Example 1} [Witten~\cite{witt82-1}]
Let $(M,g)$, $g=\sum_{i,j=1}^d g_{ij}(q)dq^idq^j$ 
be a $d$-dimensional smooth Riemannian manifold.
We put $\Lambda(M)=\cup_{k=0}^d \Lambda^k(M)$ or 
$\Lambda_0(M)=\cup_{k=0}^d \Lambda_0^k(M)$,
where 
$$
\begin{gathered}
\Lambda^k(M)=\{\omega=\sum_{1\le i_1<\cdots<i_k\le m}
\omega_{i_1\,\cdots\,i_k}(q)dq^{i_1}\wedge\cdots\wedge dq^{i_k}\,|\,
\omega_{i_1\,\cdots\,i_k}(q)\in C^\infty(M:{\mathbb{C}})\},\\
\Lambda_0^k(M)=\{\omega\in \Lambda^k(M)\,|\,
\omega_{i_1\,\cdots\,i_k}(q)\in C_0^\infty(M:{\mathbb{C}})\},\quad
\tilde\Lambda^k(M)=\{\omega\in \Lambda^k(M)\,|\,
\Vert\omega\Vert<\infty\}.
\end{gathered}
$$
Let $d$ be an exterior differential acting on 
$\omega_{i_1\,\cdots\,i_k}(q)dq^{i_1}\wedge\cdots\wedge dq^{i_k}$ as
$$
d\omega
=\sum_{j=1}^d\frac{{\partial}\omega_{i_1\,\cdots\,i_k}(q)}{{\partial}q^j}
dq^j{\wedge} dq^{i_1}\wedge\cdots\wedge dq^{i_k}.
$$
${\mathbf{P}}$ is defined by ${\mathbf{P}}\omega=(-1)^k\omega$ for $\omega\in\Lambda^k(M)$.

Put ${\mathfrak{H}}=\overline{\Lambda(M)}$ where 
$\overline{\Lambda(M)}=\cup_{k=0}^d \overline{\Lambda^k(M)}$
with $\overline{\Lambda^k(M)}$ is the closure of 
$\tilde\Lambda^k(M)$ in $L^2$-norm $\Vert\cdot\Vert$.
Denoting the adjoint of $d$ in ${\mathfrak{H}}$ by $d^*$ and
putting
$$
{\mathbf{Q}}_1=d+d^*,\quad {\mathbf{Q}}_2=i(d-d^*),\quad {\mathbf{H}}={\mathbf{Q}}_1^2={\mathbf{Q}}_2^2=dd^*+d^*d,
$$
we have that $({\mathbf{H}},{\mathbf{Q}}_\alpha,{\mathbf{P}})$ has the supersymmetry on ${\mathfrak{H}}$ 
for each $\alpha=1,2.$

\paragraph{\bf Example 1${}'$} [Witten's deformed Laplacian~\cite{witt82-1}]
For any real-valued function $\phi$ on $M$, we put
$$
d_{\lambda}=e^{-{\lambda}{\phi}}de^{{\lambda}{\phi}},\quad
d_{\lambda}^*=e^{{\lambda}{\phi}}d^*e^{-{\lambda}{\phi}}
$$
where $\lambda$ is a real parameter. We have
$d_{\lambda}^2=0={d_{\lambda}^*}^2$.
$$
{\mathbf{Q}}_{1{\lambda}}=d_{\lambda}+d_{\lambda}^*,\quad
{\mathbf{Q}}_{2{\lambda}}=i(d_{\lambda}-d_{\lambda}^*),\quad
{\mathbf{H}}_{\lambda}=d_{\lambda}d_{\lambda}^*+d_{\lambda}^*d_{\lambda}.
$$
Defining ${\mathbf{P}}$ as before, we have the supersymmetric system 
$({\mathbf{H}}_{\lambda},{\mathbf{Q}}_\alpha,{\mathbf{P}})$ on ${\mathfrak{H}}$ for each $\alpha=1,2.$

Now, we calculate ${\mathbf{H}}_{\lambda}$ more explicitly:
$$
{\mathbf{H}}_{\lambda}=dd^*+d^*d+{\lambda}^2(d{\phi})^2+
\sum_{i,j=1}^d{\lambda}\frac{D^2{\phi}}{D{q^i}D{q^j}}[a^{i*},a^j]_-.
$$
Here, 
the annihilation and creation operators $a^j$ and $a^{j*}$, respectively,
are defined as follows:
For any $0\le \ell\le d$ and $q\in M$,
$$
\left\{
\begin{aligned}
&a_q^{i*} dq^{j_1}\wedge\cdots\wedge dq^{j_\ell}
= dq^i\wedge dq^{j_1}\wedge\cdots\wedge dq^{j_\ell},\\
&a_q^i dq^{j_1}\wedge\cdots\wedge dq^{j_\ell} 
=
\sum_{k=1}^\ell (-1)^k g^{ij_k}(q)dq^{j_1}\wedge\cdots\wedge dq^{j_{k-1}}\wedge 
dq^{j_{k+1}}\wedge\cdots\wedge dq^{j_\ell}.
\end{aligned}
\right.
$$
Then, these give mappings from $\Lambda(T^*M)\to\Lambda(T^*M)$, and we get
$$
[a_q^i,a_q^j]_+=0,\quad
[a_q^i,a_q^{j*}]_+=g^{ij}(q),\quad
[a_q^{i*},a_q^{j*}]_+=0.
$$
Moreover,
$$
(d{\phi})^2=g^{ij}\frac{{\partial}{\phi}}{{\partial}q^i}
\frac{{\partial}{\phi}}{{\partial}q^j},\quad
\frac{D^2{\phi}}{D{q^i}D{q^j}}=\nabla_i\nabla_j\phi,\quad
\frac{D\psi^i}{Dt}=\frac{d\psi^i}{dt}-\Gamma^i_{jk}\dot{q}^j\psi^k.
$$
[Notation]:
For $\omega=\sum_{1\le i_1<\cdots<i_k\le d}
\omega_{i_1\,\cdots\,i_k}(q)dq^{i_1}\wedge\cdots\wedge dq^{i_k}$, we put
$$
(\nabla_j\omega)_{i_1\,\cdots\,i_k}=
\frac{\partial\omega_{i_1\,\cdots\,i_k}}{\partial q^j}
-\sum_{r,l}\Gamma^l_{ji_r}\omega_{i_1\cdots i_{r-1}li_{r+1}\cdots i_k}.
$$
Then, we have
$$
\nabla_j=\partial_{q^j}-\sum_{k,l,m}\Gamma^{k}_{jl} g_{km} a^{l*} a^m,\quad
d=\sum_{l=1}^d a^{l*}\nabla_l=-\sum_{l=1}^d \nabla_l^*a^{l*},\quad
d^*=-\sum_{l=1}^d a^l\nabla_l,
$$
$$
\begin{aligned}
d^*\omega
=\sum_{j,r}(-1)^{r-1}g^{ji_r}
[\frac{{\partial}\omega_{i_1\,\cdots\,i_k}(q)}{{\partial}q^j}
&-\sum_{l,s}\Gamma^l_{j\,i_s}
\omega_{i_1\,\cdots\,i_{s-1}\,l\,i_{s+1}\,\cdots\,i_k}(q)]\\
&\qquad\times dq^{i_1}\wedge\cdots\wedge dq^{i_{r-1}}
\wedge dq^{i_{r+1}}\wedge\cdots\wedge dq^{i_k}.
\end{aligned}
$$

The most important thing is to consider the operator ${\mathbf{H}}_{\lambda}$
as the quantized one from the Lagrangian
$$
{\mathcal{L}}_{\lambda}=\frac12
\int dt\,\bigg[g_{ij}\big(\frac{d q^i}{dt}\frac{d q^j}{dt}
+i\bar{\psi}^i\frac{D{\psi}^j}{Dt}\big)
+\frac14 R_{ijkl}\bar{\psi}^i{\psi}^k\bar{\psi}^j{\psi}^l
-{\lambda}^2g^{ij}\frac{{\partial}{\phi}}{{\partial}q^i}
\frac{{\partial}{\phi}}{{\partial}q^j}
-{\lambda}\frac{D^2{\phi}}{D{q^i}D{q^j}}\bar{\psi}^i{\psi}^j\bigg].
$$
Here, we used the summation convention and 
${\psi}^i$ and $\bar{\psi}^i$ are anti-commuting fields tangent to $M$, 
which becomes
the creation and annihilation operators after quantization.
After representing the solution of
$$
\lambda^{-1}\frac{\partial}{\partial t}u(t)=\lambda^{-2}{\mathbf{H}}_{\lambda}u(t),
$$
and applying the SUSYQM structure, he concludes that the principal term above when $\lambda\to\infty$(${\lambda}^{-1}\sim{\hbar}$) is governed by the instantons or tunneling paths corresponding to ${\mathcal{L}}_{\lambda}$. That is, those paths are  
defined by Lagrangean below;
$$
\begin{aligned}
\bar{\mathcal{L}}_{\lambda}&=\frac12
\int dt\,\bigg(g_{ij}\frac{d q^i}{dt}\frac{d q^j}{dt}
+{\lambda}^2g^{ij}\frac{{\partial}{\phi}}{{\partial}q^i}
\frac{{\partial}{\phi}}{{\partial}q^j}\bigg)\\
&=\frac12
\int dt\,\bigg|\frac{d q^i}{dt}
{\pm}{\lambda}g^{ij}\frac{{\partial}{\phi}}{{\partial}q^j}\bigg|^2
{\mp}{\lambda}\int dt\,\frac{d{\phi}}{dt}.
\end{aligned}
$$
Using these paths with physicists'  steepest descent method, he may calculate Witten index.

\paragraph{\bf Example 2} (Deift~\cite{dei78} in p.123 Cycon et al.~\cite{CFKS}).
Let ${\mathfrak{H}}=L^2(\euc)\otimes {\mathbb{C}}^2=L^2(\euc:{\mathbb{C}}^2)=L^2(\euc)^2$, 
and $\phi$ be a polynomial in $q$.
Set $A=d/dq+\phi(q)$ and $A^*=-d/dq+\phi(q)$ with domains
$ D(A)=D(A^*)=\{u\in H^1(\euc)\,|\,\phi u\in L^2(\euc)\}.$
$$
{\mathbf{Q}}=\begin{pmatrix}
0&-\frac{d}{dq}+\phi\\
\frac{d}{dq}+\phi&0
\end{pmatrix},\quad
{\mathbf{P}}=\begin{pmatrix}
1&0\\
0&-1
\end{pmatrix}.
$$
Then,
$ D(AA^*)=D(A^*A)=\{u\in L^2(\euc)\,|\,u^{\prime\prime},\;\phi^2u,\;\phi' u\in L^2(\euc)\}$,
$$
A^*A=-\frac{d^2}{dq^2}+\phi^2(q)-\phi'(q),\quad
AA^*=-\frac{d^2}{dq^2}+\phi^2(q)+\phi'(q),
$$
and
$$
{\mathbf{H}}={\mathbf{Q}}^2=\begin{pmatrix}
-\frac{d^2}{dq^2}+\phi^2(q)-\phi'(q)&0\\
0&-\frac{d^2}{dq^2}+\phi^2(q)+\phi'(q)
\end{pmatrix}.
$$
This $({\mathbf{H}},{\mathbf{Q}},{\mathbf{P}})$ forms a SUSYQM in ${\mathfrak{H}}$.

(i) Especially $\phi(q)=q$, we have
$$
\dim(\Ker A)=1,\quad \dim(\Ker A^*)=0,\quad \text{ind}_F(A)=1=\ind_{s}({\mathbf{H}}).
$$

(ii) (Boll\'e et al. \cite{BoGGSS}).
For real-valued $\phi,\phi'\in L^\infty(\euc)$, we assume that 
$$
\begin{gathered}
\lim_{q\to\pm\infty}\phi(q)=\phi_{\pm}\in\euc,\quad \phi_-^2\le\phi_+^2,\\
\int_{\euc}dq\,(1+|q|^2)|\phi'(q)|<\infty \et
\int_{\euc}dq\,(1+|q|^2)|\phi(q)-\phi_{\pm}|<\infty.
\end{gathered}
$$
Then, they assert that, for $z\in{\mathbb{C}}\setminus[0,\infty)$,
$$
\Delta_z({\mathbf{H}})=\frac12[\phi_+(\phi_+^2-z)^{-1/2}-\phi_-(\phi_-^2-z)^{-1/2}],\quad
W_R=\frac12[\sign(\phi_+)-\sign(\phi_-)]\et
{\mathcal{A}}_R=0.
$$

\begin{remark}
Especially, physicists have the above result by calculating the quantity
$$
\Delta_t({\mathbf{H}})=\int dqd{\psi}d{\bar\psi}
\Big[\int_{\{t-periodic\}}[dq][d{\psi}][d{\bar\psi}]
e^{-\int_0^t ds{\mathcal{L}}(q(s),\dot q(s),\psi(s),\bar\psi(s))}
\Big],
$$
but it seems difficult to make rigorous their procedure mathematically.
\end{remark}

\section{Wick rotation and PIM}
%\section{Super extension of harmonic oscillator and its index}
Though arguments so far in this chapter mainly treat (abstract) heat equation, but we need special trick if we use Fourier
transformation within path-integral method. Because, if we naively proceed as before, we may have 
$$
v(t,{\barq})=\frac{1}{\sqrt{2\pi}}\int d{\unbp}\, D^{1/2}(t,{\barq},{\unbp})e^{S(t,{\barq},{\unbp})}\hat{\underline{v}}(\unbp)
$$
and by Fourier inversion formula, we need to have at $t=0$,
$$
D^{1/2}(t,{\barq},{\unbp})\bigg|_{t=0}=1,\quad
e^{S(t,{\barq},{\unbp})}\bigg|_{t=0}=e^{i\barq\unbp}.
$$
Since the second requirement above seems strange because $S(t,{\barq},{\unbp})\in\euc$, therefore we need to reconsider the procedure from scratch.
\subsection{Quantization and Path-Integral Method}
From a Lagrangian
\begin{equation}
L(q,\dot{q})=\frac{m}{2}\dot{q}^2+A(q)\dot{q}-V(q)\in C^{\infty}(T{\mathbb{R}}),
\label{orig-lag}
\end{equation}
by Legendre transformation, we get a Hamiltonian function
\begin{equation}
H(q,p)=\frac{1}{2m}(p-A(q))^2+V(q)\in C^{\infty}(T^*{\mathbb{R}})
\label{orig-hamil}
\end{equation}
whose hamilton flow is defined by
\begin{equation}
\left\{
\begin{aligned}
\dot{q}&=H_p,\\
\dot{p}&=-H_q,
\end{aligned}
\right.
\with
\begin{pmatrix}
q(0)\\
p(0)
\end{pmatrix}
=\begin{pmatrix}
\unbq\\
\unbp
\end{pmatrix}.
\label{h-flow}
\end{equation}
\paragraph{\bf{Quantum mechanics}}
Substituting $-i\hbar\partial_q$ into $p$ in \eqref{orig-hamil}, we ``quantize'' this as
\begin{equation}
\left\{
\begin{aligned}
&i\hbar\frac{\partial u(t,q)}{\partial t}=H(q,-i\hbar\partial_q)u(t,q),\\
&u(t,q)\big|_{t=0}=\underline{u}(q),
\end{aligned}
\right.
\label{sch-eq}
\end{equation}
with
\begin{equation}
H(q,-i\hbar\partial_q)=\frac{1}{2m}\bigg(-i\hbar\frac{\partial}{\partial q}-A(q)\bigg)^2+V(q).
\label{sch0}
\end{equation}
\subsection{LPIM and HPIM} Feynman feels not good at this ``quantization'' because Bohr's corresponding principle is not seen transparently. %ignored.
Therefore, he introduces
LPI=Lagrangian  Path-Integral as
\begin{equation}
\begin{gathered}
K^L(t,\barq,\unbq)=\int_{{\mathcal{C}}^L_{t,\barq,\unbq}}d_F q\, e^{i\hbar^{-1}\int_0^t ds \, L(q(s),\dot{q}(s))},\\
\with
{\mathcal{C}}^L_{t,\barq,\unbq}=\{q(\cdot)\in{\mathcal{P}}([0,t]:\euc)\;|\; q(0)=\unbq,\; q(t)=\barq\}
\end{gathered}
\label{LPIM}
\end{equation}
which gives the solution of \eqref{sch-eq} by
$$
u(t,q)=\int d\unbq\,K^L(t,q,\unbq)\underline{u}(\unbq).
$$
%We should remark that ${\mathcal{P}}([0,t]:\euc)$ stands for path space with suitable topology which has no Feynman measure in ordinary sense.

How about HPI=Hamiltonian  Path-Integral?
\begin{equation}
\begin{gathered}
K^H(t,\barq,\unbq)=\int_{{\mathcal{C}}^H_{t,\barq,\unbq}}d_Fq\,d_Fp\,
e^{i\hbar^{-1}\int_0^t ds [\dot{q}(s)p(s)-H(q(s),p(s))]},\\
\with
{\mathcal{C}}^H_{t,\barq,\unbq}=\{(q(\cdot),p(\cdot))\in{\mathcal{P}}([0,t]:T^*\euc)\;|\; 
q(0)=\unbq, q(t)=\barq, p(\cdot): \mbox{arbitrary}\}.
\end{gathered}
\label{HPIM}
\end{equation}

Though Hamilton mechanics enlarges the scope of Lagrange mechanics, is such ``extension''  truely necessary for PI?
From above representation,
check at least when $t\to0$ whether 
\begin{equation}
\lim_{t\to0}\int d\unbq\, K^H(t,\barq,\unbq)\underline{u}(\unbq)=\underline{u}(\barq).
\label{HPIM-delta}
\end{equation}
In other word,
$$
\int_{{\mathcal{C}}^H_{t,\barq,\unbq}}d_Fq\,d_Fp\,
e^{i\hbar^{-1}\int_0^t ds [\dot{q}(s)p(s)-H(q(s),p(s))]}=\delta(\barq-\unbq)?
$$
But this is doubtful because if paths $q(\cdot)$ or $p(\cdot)$ are smooth and $t\to0$, then seemingly
$$
\int_0^t ds [\dot{q}(s)p(s)-H(q(s),p(s))]\to 0 ?
$$

\begin{remark}
Therefore in stead of above \eqref{HPIM}, Kumano-go and Fujiwara~\cite{KF08} put
$$
\int_{[0,t)} ds [p(s)\,dq(s)-H(q(s),p(s))]=\mbox{to gain ``jump'' $p(0)(q(t)-q(0))$}
$$
using Lebesgue-Stieltjes integral for left-continuous $q(\cdot)$ on $(0,t]$ and right-continuous $p(\cdot)$ on $[0,t)$.
That is, by definition,
$$
\begin{aligned}
\int_{[0,t)} ds\,p(s)\,dq(s)
&=\lim_{N\to\infty}\sum_{j=1}^N p(\theta_j)[q(t_{j})-q(t_{j-1})]\where \theta_j\in[t_{j-1},t_j)\\
&=\lim_{N\to\infty}\sum_{j=1}^N[p(t_{j-1}+0)(q(t_{j})-q(t_{j-1}))+\int_{t_{j-1}}^{t_j}ds\, p(s)\dot{q}(s)].
\end{aligned}
$$
Though it seems necessary to include discontinuous paths in ${\mathcal{P}}$,
but I'm not sure such choice of paths are adequate or not. This will be touched on at the last section.
\end{remark}

By the way, since there exists no rigorous Feynman measure, \underline{one may define ``path-integral like}\\
\underline{object" rather freely}. I declare my opinion concerning Path-Integral Mehod (not Path-Integral).

\fbox{My dogmatic opinion 1}: Path-Integral Method, if it is the one related to quantum mechanics, should be the representation of solution which exhibits the Bohr correspondence transparently!

\fbox{My dogmatic opinion 2}: Broken line approximation is a sort of Trotter-Kato formula far from PIM idea! 
Schematically,
$$
 e^{i\hbar^{-1}\int_0^t ds L(\gamma(s),\dot{\gamma}(s))}d_F\gamma
\not\sim
e^{i\hbar^{-1}\int_0^t ds V(\gamma(s))} e^{i\hbar^{-1}\int_0^t ds \frac{1}{2}\dot{\gamma}^2(s)}d_F\gamma.
$$
But formulation applying this procedure seems meaningful to construct a parametrix or a fundamental solution for a already known PDE.
Feynman's procedure is not to derive a PDE rather to offer a black box to yield physical explanation for experimental facts.

\begin{claim}
Under these dogmas, we explain why we need HPIM if the order of spacial derivatives are one.
This point of view is already explained in \S 2 of Chapter 1 concerning IVP for \eqref{sqm}.
\end{claim}

\begin{problem}
1. Reinterprete the papers N. Kumano-go~\cite{n-kum95}, J. Le Rousseau~\cite{LR06} from above point of view.\\
2. How to find the representation of solution for Dirac equation under my dogma, we need HPIM with superanalysis.
$$
i\hbar\frac{\partial}{\partial t}\psi(t,q)=[c{\pmb{\alpha}}_j(\frac{\hbar}{i}\frac{\partial}{\partial q_j}-\frac{e}{c}A_j(t,q))+mc^2{\pmb{\beta}}]
\psi(t,q).
$$
\end{problem}

\subsection{Relation between quantum and heat}
\subsubsection{Purely imaginary $\hbar$}
It seems well-known that 
for a solution $u(\hbar: t ,q)$ of IVP of Schr\"odinger equation \eqref{sch-eq},
bringing $\hbar$ to $-i$, that is, putting $v(t,q)=u(-i: t ,q)$, we have
\begin{equation}
\frac{\partial}{\partial t}v(t,q)=[\frac{(\partial_{q}+A(q))^2}{2m}+V(q)]v(t,q).
\label{h-heat}
\end{equation}
We need to pay something making Schr\"odinger to heat. In fact,
comparing this \eqref{h-heat} with \eqref{heat} below, we need to take $+A$ and $+V$ instead of $-A$ and $-V$!
By the way, from heat to Schr\"odinger, we pay more, because in general analytic semi-group is not extendable to imaginary time!

\subsubsection{Purely imaginary mass}
As another way connecting Schr\"odinger to heat, E. Nelson\cite{nel64} proposed to complexify the mass $m$ to imaginary one $i\mu$ to have
\begin{equation}
\hbar\pdt v(t,q)=[-\frac{(-i\hbar\partial_{q}-A(q))^2}{2\mu}-iV(q)]v(t,q)=[\frac{(\hbar\partial_{q}-iA(q))^2}{2\mu}-iV(q)]v(t,q)
\label{m-heat}
\end{equation}
with $\hbar=1$ and $A=0$.

\subsubsection{Wick rotation}[Purely imaginary time]
Though $\hbar$ and $m$ may be considered as parameters, but time $t$ is a dependent variable. Therefore we need some care to make time imaginary.

For a solution $u(t,q)$ of \eqref{sch-eq}, taking $\theta\in[0,\pi/2]$ and $\tau\in{\mathbb{R}}$,
we put
\begin{equation}
{w}(\tau,q)=u(t,q)\bigg|_{t=e^{-i\theta}\tau}=u(e^{-i\theta}\tau,q).
\label{wick0}
\end{equation}
Then it satisfies at least formally
\begin{equation}
 \left\{
\begin{aligned}
&i\hbar\frac{\partial w(\tau,q)}{\partial \tau}=e^{-i\theta}H(q,-i\hbar\partial_q)w(\tau,q),\\
&w(\tau,q)\big|_{\tau=0}=\underline{u}(q).
\end{aligned}
\right.
\label{wick}
\end{equation}
This implies that for initial data ${w}(0,q)=\underline{u}(q)$,
\begin{equation}
 \left\{
\begin{aligned}
i\hbar\frac{\partial {w}(\tau,q)}{\partial \tau}&=H(q,-i\hbar\partial_q){w}(\tau,q)
\;\;\;\mbox{with}\;\; \theta=0,\\
\hbar\frac{\partial {w}(\tau,q)}{\partial \tau}&=-H(q,-i\hbar\partial_q){w}(\tau,q)\;\;\;\mbox{with}\;\;
\theta=\pi/2.
\end{aligned}
\right.
\label{wick1}
\end{equation}
For heat type equation, we ``assume in general'' that all coefficients are real-valued and it satisfies
\begin{equation}
 \frac{\partial}{\partial t}v(t,q)=[\frac{(\partial_{q}-A(q))^2}{2m}-V(q)]v(t,q),
\label{heat}
\end{equation}
but by Wick rotation from $u(t,q)$ of \eqref{sch-eq}, we get
\begin{equation}
\hbar\frac{\partial}{\partial t}v(t,q)
=[-\frac{(-i\hbar\partial_{q}-A(q))^2}{2m}-V(q)]v(t,q)
=[\frac{(\hbar\partial_{q}-iA(q))^2}{2m}-V(q)]v(t,q).
\label{wick-heat}
\end{equation}
I feel discrepancy! Because even if $\hbar=1$, we must change $-A\to -iA$!

Therefore, starting from \eqref{orig-hamil}, after quantization we get \eqref{sch0}, from it we have three slightly different heat type equations.

\begin{problem}
From the right-hand operator in \eqref{wick}, we define its symbol 
\begin{equation}
\tilde{H}(q,p)=e^{-i\theta}H(q,p)=e^{-i\hbar^{-1}qp}[e^{-i\theta}H(q,-i\hbar\partial_q)]e^{i\hbar^{-1}qp}\big|_{\hbar=0}.
\label{symbol}
\end{equation}
Can we quantize $\tilde{H}(q,p)$, so I called,  by PIM?
And ask whether a functor-like procedure PIM (Lagrangian or Hamiltonian) and a restriction of $\theta$ to $\theta=0$ or $\theta=\pi/2$ are commutative under what condition?  
\end{problem}

\paragraph{\bf A tentative conclusion} Applying Lagrangian PIM, answers are affirmative.

But for $\theta=\pi/2$ with Hamiltonian PIM, something curious occurs, in the sense that it seems impossible to use Fourier inversion formula and by a device, we may have heat type equation with not real coefficients when $A(q)$ exists.

\paragraph{\bf Question} Is that impossible in general to apply HPIM for heat type equation?
Especially, is it possible to include vector potential in Riemann metric by considering $A(q)$ stemming from the connection?

\subsection{H-J equation}
\subsubsection{Classical-flow}
Putting $\tilde{H}(q,p)=e^{-i\theta}{H}(q,p)$, we have
\begin{equation} \tag{CM}
\left\{
\begin{aligned}
\dot{q}_j&=\frac{\partial \tilde{H}(q,p)}{\partial p_j}=e^{-i\theta}\frac{1}{m}(p_j-A_j(q)),\\
\dot{p}_j&=-\frac{\partial \tilde{H}(q,p)}{\partial q_j}=e^{-i\theta}\big[\frac{1}{m}(p_j-A_j(q))\partial_{q_j}A_j(q)-\partial_{q_j}V(q)\big]
\end{aligned}
\right.
\with
\begin{pmatrix}
{q}_j(0)=\unbq_j\\
{p}_j(0)=\unbp_j
\end{pmatrix}.
\label{hamiltonflow}
\end{equation}
\subsubsection{Action integral(L)}
For the solution $(q(\sigma,\unbq,\unbp),p(\sigma,\unbq,\unbp))$ of \eqref{hamiltonflow}, we put
\begin{equation}
{\tilde{S}}_0(\tau,\unbq,\unbp)
=\int_0^\tau\,d\sigma[\dot{q}(\sigma,\unbq,\unbp)p(\sigma,\unbq,\unbp)
-\tilde{H}(q(\sigma,\unbq,\unbp),p(\sigma,\unbq,\unbp))].
\label{action0}
\end{equation}

Solving as $\unbp_j=\xi_j(\tau,\barq,\unbq)$ from $\barq_j={q}_j(\tau,\unbq,\unbp)$, by inverse function theorem, 
we define
\begin{equation}
{\tilde{S}}^L(\tau,\barq,\unbq)={\tilde{S}}^L_0(\tau,\unbq,\unbp)\bigg|_{\unbp=\xi(\tau,\barq,\unbq)}.
\label{actionL}
\end{equation}
Then, this satisifies
\begin{equation}
{\tilde{S}}^L_{\tau}+\tilde{H}({\barq},{\tilde{S}}^L_{\barq})=0\;\;\mbox{with}\;\; \lim_{{\tau}t\to0}2{\tau}{\tilde{S}}^L({\tau},{\barq},{\unbq})= |{\barq}-{\unbq}|^2.
\label{HJL}
\end{equation}
\subsubsection{Continuity equation (L)}
For $%\displaystyle{
{\tilde{D}}^L({\tau},{\barq},{\unbq})=\det\Big(-{\displaystyle{\frac{\partial^2 {\tilde{S}}^L}{\partial{\barq}\partial{\unbq}}}}\Big)%}
$, we have
\begin{equation}
{\tilde{D}}^L_{\tau}+\partial_{\barq}({\tilde{D}}^L\tilde{H}_p({\barq},{\tilde{S}}^L_{\barq}(\tau,{\barq},{\unbq})))=0\;\;\mbox{with}\;\; \lim_{{\tau}\to0}{\tau}{\tilde{D}}^L({\tau},{\barq},{\unbq})=1.%,\;\;
%{\tilde{D}}^L\widetilde{\tilde{H}_p}={\tilde{D}}^L({\tau},{\barq},{\unbq})\tilde{H}_p({\barq},{\tilde{S}}^L_{\barq}(\tau,{\barq},{\unbq})).
\label{CEL}
\end{equation}
\subsubsection{Action integral (H)}
Getting $\unbq_j=x_j(\tau,\barq,\unbp)$ as before, we put,
\begin{equation}
{\tilde{S}}^H(\tau,\barq,\unbp)=[\unbq\unbp+{\tilde{S}}_0(\tau,\unbq,\unbp)]\bigg|_{\unbq=x(\tau,\barq,\unbp)}
\label{actionH}
\end{equation}
which satisfies
\begin{equation}
{\tilde{S}}^H_t+\tilde{H}({\barq},{\tilde{S}}^H_{\barq})=0\;\;\mbox{with}\;\; {\tilde{S}}^H(0,{\barq},{\unbp})= \langle \barq |\unbp \rangle.
\label{HJH}
\end{equation}

\paragraph{Continuity equation (H)}
Putting
$$
{\tilde{D}}^H({\tau},{\barq},{\unbp})=\det\Big(\frac{\partial^2 {\tilde{S}}^H}{\partial{\barq}\partial{\unbp}}\Big),
$$
we have
$$
{\tilde{D}}^H_{\tau}+\partial_{\barq}({\tilde{D}}^H\tilde{H}_p({\barq},{\tilde{S}}_{\barq}))=0\;\;\mbox{with}\;\;  {\tilde{D}}^H({\tau},{\barq},{\unbp})\big|_{{\tau}=0}=1.
$$
Remark: No singularity at $\tau=0$!

\subsection{Example: Harmonic Oscillator type}
\subsubsection{Lagrangian and Hamiltonian functions}
From Wick rotated Schr\"odinger equation, we have the symbol \eqref{symbol} $\tilde{H}(q,p)$ from which
we have the Hamilton flow:
Putting
$$
\begin{gathered}
\beta^2=a^2+m^2\omega^2,\\
\cos  (e^{-i\theta}\omega\sigma)={\mathrm{C}}_{\sigma},\quad
\sin (e^{-i\theta}\omega\sigma)={\mathrm{S}}_{\sigma},
\end{gathered}
$$
we have Hamilton flow
$$
\left\{
\begin{aligned}
q(\sigma)
&=\frac{1}{m\omega}[({m\omega}{\mathrm{C}}_{\sigma}-a{\mathrm{S}}_{\sigma})\unbq+{\mathrm{S}}_{\sigma}\unbp],\\
p(\sigma)
&=\frac{1}{m\omega}[-\beta^2{\mathrm{S}}_{\sigma}\unbq
+({m\omega}{\mathrm{C}}_{\sigma}+a{\mathrm{S}}_{\sigma})\unbp].
\end{aligned}
\right.
\with
\barq=\frac{1}{m\omega}[({m\omega}{\mathrm{C}}_{\tau}-a{\mathrm{S}}_{\tau})\unbq+{\mathrm{S}}_{\tau}\unbp].
%p(\sigma)
$$
We put
$$
\begin{aligned}
\tilde{S}_0(\tau,\unbq,\unbp)&=\int_0^{\tau}d\sigma [\dot q(\sigma)p(\sigma)-\tilde{H}(q(\sigma),p(\sigma))]\\
&=\frac{{\mathrm{S}}_{\tau}}{2m^2\omega^2}\big[-\beta^2(m\omega {\mathrm{C}}_{\tau}-a{\mathrm{S}}_{\tau})\unbq^2
+(m\omega {\mathrm{C}}_{\tau}+a{\mathrm{S}}_{\tau})\unbp^2
-2\beta^2{\mathrm{S}}_{\tau}\unbq\unbp\big].
\end{aligned}
$$

\subsubsection{Wick rotated Harmonic Oscillator (L)}
From $\barq=q(\tau,\unbq,\unbp)$, getting 
$$
\unbp=\xi(\tau,\barq,\unbq)=\frac{m\omega\barq-(m\omega{\mathrm{C}}_{\mathrm{\tau}}-a{\mathrm{S}}_{\mathrm{\tau}})\unbq}{{\mathrm{S}}_{\tau}},
$$
we define
$$
\begin{aligned}
\tilde{S}^L(\tau,\barq,\unbq)&=\tilde{S}_0(\tau,\unbq,\xi(\tau,\barq,\unbq))\\
%&=\frac{1}{2{\mathrm{S}}_{\tau}}\big[m\omega{\mathrm{C}}_{\tau}(\barq^2+\unbq^2)-{2}{m\omega}\barq\unbq+a{\mathrm{S}}_{\tau}(\barq^2-\unbq^2)\big]\\
&={\left\{
\begin{aligned}
&\frac{m\omega(\cos{\omega\tau}\,(\barq^2+\unbq^2)-{2}\barq\unbq)}{2 \sin{\omega\tau}}+\frac{a}{2}(\barq^2-\unbq^2), \mbox{($\theta=0$)},\\
&\frac{im\omega[\cosh{\omega\tau}\,(\barq^2+\unbq^2)-{2}\barq\unbq]}{2\sinh{\omega\tau}}+\frac{a}{2}(\barq^2-\unbq^2),\mbox{($\theta=\pi/2$)},
\end{aligned}
\right.}
\end{aligned}
$$
which satisfies
$$
{\tilde{S}}^L_{\tau}+{\tilde{H}}(\barq, {\tilde{S}}^L_{\barq})=0\;\;\mbox{with}\;\;
\lim_{\tau\to0}\bigg(\tau {\tilde{S}}^L(\tau,\barq,\unbq)-\frac{m(\barq-\unbq)^2}{2}\bigg)=0.
$$
Defining
$$
{\tilde{D}}^L(\tau,\barq,\unbq)=-\frac{\partial^2 {\tilde{S}}^L(\tau,\barq,\unbq)}{\partial\barq\,\partial\unbq}
=\frac{m\omega}{{\mathrm{S}}_{\tau}}
={\left\{
\begin{aligned}
&\frac{m\omega}{\sin{\omega\tau}},\; \; \mbox{($\theta=0$)},\\
&\frac{im\omega}{\sinh{\omega\tau}},\; \mbox{($\theta=\pi/2$)},
\end{aligned}
\right.}
$$
we have
$$
{\tilde{D}}^L_{\tau}+\partial_{\barq}({\tilde{D}}^L\widetilde{{\tilde{H}}_p})=0\;\;\mbox{with}\;\;
\lim_{\tau\to0}\tau {\tilde{D}}^L(\tau,\barq,\unbq)=m,
\widetilde{{\tilde{H}}_p}={\tilde{H}}_p(\barq, {\tilde{S}}^L_{\barq}).
$$
Finally, we put
$$
[{\tilde{U}}_{\tau}^L\,{\underline{u}}](\barq)=\frac{1}{\sqrt{2\pi{i} \hbar}}
\int_\euc d\unbq\,{\tilde{A}}^L(\tau,\barq,\unbq)
e^{i\hbar^{-1}{\tilde{S}}^L(\tau,\barq,\unbq)}{\underline{u}}(\unbq)
=\int_\euc d\unbq\,{\tilde{K}}^L(\tau,\barq,\unbq){\underline{u}}(\unbq).
$$
$$
{\tilde{K}}^L(\tau,\barq,\unbq)=\frac{1}{\sqrt{2\pi{i} \hbar}}{\tilde{A}}^L(\tau,\barq,\unbq)
e^{i\hbar^{-1}{\tilde{S}}^L(\tau,\barq,\unbq)}.
$$
Then, not only
$$
i\hbar\frac{\partial}{\partial \tau}{\tilde{U}}_{\tau}^L\,{\underline{u}}\bigg|_{\tau=0}
=[\hat{\tilde{H}}(q,-i\hbar\partial_q)]{\underline{u}}.\quad\mbox{(Feynman quantization)},
$$
but also
$$
i\hbar\frac{\partial}{\partial \tau}{\tilde{U}}_{\tau}^L\,{\underline{u}}=[\hat{\tilde{H}}(q,-i\hbar\partial_q)]{\tilde{U}}_{\tau}^L\,{\underline{u}}.
$$

\subsubsection{Wick rotated Harmonic Oscillator(H)}
Analogously defining
$$
\unbq=x(\tau,\barq,\unbp)=\frac{m{\omega}\barq-{\mathrm{S}}_{\tau}\unbp}{m\omega{\mathrm{C}}_{\tau}-a{\mathrm{S}}_{\tau}},
$$
we get
$$
\begin{aligned}
\tilde{S}^H(\tau,\barq,\unbp)&=[\unbq\unbp+\tilde{S}_0(\tau,\unbq,\unbp)]\bigg|_{\unbq=x(\tau,\barq,\unbp)}\\
&={\left\{
\begin{aligned}
&-\frac{\sin{\omega\tau}(\beta^2\barq^2+{\unbp^2})}{2(m\omega\cos{\omega\tau}-a\sin{\omega\tau})}
+\frac{m\omega}{m\omega\cos{\omega\tau}-a\sin{\omega\tau}}\barq\unbp,
\qquad\;\mbox{($\theta=0$)},\\
&\frac{i\sinh{\omega\tau}(\beta^2\barq^2+{\unbp^2})}{2[m\omega\cosh{\omega\tau}+ia\sinh{\omega\tau}]}
+\frac{m\omega}{m\omega\cosh{\omega\tau}+ia\sinh{\omega\tau}}\barq\unbp,
\quad\mbox{($\theta=\pi/2$)},
\end{aligned}
\right.}
\end{aligned}
$$
which satisfies
$$
{\tilde{S}}^H_{\tau}+{\tilde{H}}(\barq, {\tilde{S}}^H_{\barq})=0\;\;\mbox{with}\;\;
\lim_{\tau\to0}{\tilde{S}}^H(\tau,\barq,\unbp)=\barq\unbp.
$$
Defining
$$
{\tilde{D}}^H(\tau,\barq,\unbp)=\frac{\partial^2 {\tilde{S}}^H(\tau,\barq,\unbp)}{\partial{\barq}\,\partial\unbp}
=\frac{m\omega}{m\omega{\mathrm{C}}_{\tau}-a{\mathrm{S}}_{\tau}}\\
={\left\{
\begin{aligned}
&\frac{m\omega}{m\omega\cos{\omega\tau}-a\sin{\omega\tau}},\quad &\mbox{($\theta=0$)},\\
&\frac{m\omega}{m\omega\cosh{\omega\tau}+ia\sinh{\omega\tau}},\quad&\mbox{($\theta=\pi/2$)},
\end{aligned}
\right.}
$$
we have
$$
{\tilde{D}}^H_{\tau}+\partial_{\barq}({\tilde{D}}^H\widetilde{{\tilde{H}}_p})=0
\;\;\mbox{with}\;\;\lim_{\tau\to0}{\tilde{D}}^H(\tau,\barq,\unbp)=1\;\;\mbox{where}\;\;
\widetilde{{\tilde{H}}_p}={\tilde{H}}_p(\barq, {\mathbb{S}}^H_{\barq}).
$$

Then, we put
$$
\begin{aligned}
{}
[{\tilde{U}}^H_{\tau}\,{\underline{u}}](\barq)&=\frac1{\sqrt{2\pi\hbar}}
\int_\euc d\unbp\,{\tilde{A}}^H({\tau},\barq,\unbp)e^{i\hbar^{-1}{\tilde{S}}^H({\tau},\barq,\unbp)}
{\hat{\underline{u}}}(\unbp)\\
&=\frac1{2\pi\hbar}
\iint_{\euc^2} d\unbp\, d\unbq\,{\tilde{A}}^H({\tau},\barq,\unbp)
\,e^{i\hbar^{-1}({\tilde{S}}^H({\tau},\barq,\unbp)-\unbq\unbp)}
{\underline{u}}(\unbq)=\int_\euc d\unbp\,{\tilde{K}}^H({\tau},\barq,\unbp)\,{\hat{\underline{u}}}(\unbp),
\end{aligned}
$$
where
$$
{\tilde{K}}^H({\tau},\barq,\unbp)=\frac1{\sqrt{2\pi\hbar}}
\sqrt{\frac{m\omega}{m\omega{\mathrm{C}}_{\tau}-a{\mathrm{S}}_{\tau}}}
\,e^{i\hbar^{-1}{\tilde{S}}^H({\tau},\barq,\unbp)}.
$$
Simply we have
$$
\frac1{2\pi\hbar}\int d\unbp\,{\tilde{K}}^H({\tau},\barq,\unbp)e^{-i\hbar^{-1}\unbq\unbp}
={\tilde{K}}^L({\tau},\barq,\unbq).
$$

\subsection{Heat type}
Try to get the PIM-like solution for \eqref{heat}, \eqref{wick-heat}, \eqref{h-heat} or \eqref{m-heat} from \eqref{orig-hamil} or else by HPIM! 
\begin{problem}
Since this symbol \eqref{orig-hamil} is obtained by putting $p$ in \eqref{heat} of $\partial_q$,
we want to ask what is the natural definition of symbol for \eqref{heat}?
\end{problem}

\subsubsection{LPIM works from \eqref{orig-hamil}}
Solving Hamilton equation corresponding to \eqref{orig-hamil} with $q(s)=q(s,\unbq,\unbp)$ and $p(s)=p(s,\unbq,\unbp)$, we put
\begin{equation}
\begin{aligned}
\phi^L(t,\barq,\unbq)&=\int_0^t \,ds[\dot{q}(s)p(s)-H(q(s),p(s))]\bigg|_{\unbp=\xi(t,\barq,\unbq)}\\
&=\frac{m\omega[\cos{\omega{t}}\,(\barq^2+\unbq^2)-{2}\barq\unbq]}{2 \sin{\omega{t}}}+\frac{a(\barq^2-\unbq^2)}{2}\\
& \et
\alpha^L(t,\barq,\unbq)=\sqrt{-\frac{\partial^2 \phi^L(t,\barq,\unbq)}{\partial \barq\partial \unbq}}=\sqrt{\frac{m\omega}{\sin{\omega t}}}.
\end{aligned}
\label{heatL}
\end{equation}
Putting operators as
\begin{equation}
[T_t^L\underline{v}](\barq)=\frac{1}{\sqrt{2\pi}}\int d\unbq\, \alpha^L(t,\barq,\unbq) e^{-\phi^L(t,\barq,\unbq)}\underline{v}(\unbq)
\label{heatLsol}
\end{equation}
then
$$
\lim_{t\to0}[T_t^L\underline{v}](\barq)\\
=\lim_{t\to0}\frac{1}{\sqrt{2{\pi}}}\int d\unbq\,\sqrt{\frac{m}{t}}
\exp{\{-\frac{m(\barq-\unbq)^2}{2t}\}} 
\underline{v}(\unbq)=\underline{v}(\barq),
$$
and
$$
\frac{\partial}{\partial t}[T_t^L\underline{v}](\barq)\bigg|_{t=0}
=[\frac{1}{2m}(\partial_{\barq}-a{\barq})^2-\frac{m\omega^2}{2}{\barq}^2]\underline{v}(\barq).
$$
Moreover, we have
$$
\frac{\partial}{\partial t}[T_t^L\underline{v}](\barq)=[\frac{1}{2m}(\partial_{\barq}-a{\barq})^2-\frac{m\omega^2}{2}{\barq}^2][T_t^L\underline{v}](\barq).
$$
\subsubsection{HPIM doesn't work directly from \eqref{orig-hamil}}
Analogously, calculating
\begin{equation}
\begin{gathered}
\phi^H(t,\barq,\unbp)=-\frac{\sin{\omega{t}}(\beta^2\barq^2+{\unbp^2})}{2(m\omega\cos{\omega{t}}-a\sin{\omega{t}})}
+\frac{m\omega\barq\unbp}{m\omega\cos{\omega{t}}-a\sin{\omega{t}}},\\
\;\;\mbox{with}\;\;\alpha^H(t,\barq,\unbp)=\sqrt{\frac{\partial^2 \phi^H(t,\barq,\unbp)}{\partial \barq\partial \unbp}}=\sqrt{\frac{1}{\cos{\omega t}}},
\end{gathered}
\label{heatH}
\end{equation}
we put
\begin{equation}
[T_t^H\underline{v}](\barq)=\frac{1}{\sqrt{2\pi}}\int d\unbp\, \alpha^H(t,\barq,\unbp) e^{-\phi^H(t,\barq,\unbp)}{\hat{\underline{v}}}(\unbp).
\label{heatHsol}
\end{equation}
Unfortunately,
$$
\lim_{t\to0}[T_t^H\underline{v}](\barq)=\lim_{t\to0}\frac{1}{2\pi}\int d\unbp\,
\exp{\{-t\frac{\beta^2\barq^2+\unbp^2}{2m}-\barq\unbp\}}\,{\hat{\underline{v}}}(\unbp)\;{\neq}\;\underline{v}(\barq),
$$
therefore, we need to modify above procedure from scratch.

\paragraph{\bf {Devices}} (0) We propose to consider as Hamilton function, instead of \eqref{orig-hamil},
$$
\tilde{H}(q,p)=i\bigg(\frac{(p-aq)^2}{2m}+\frac{m\omega^2}{2}q^2\bigg).
$$
Then Hamilton flow is given by
$$
\tilde{H}_p=i\frac{p-aq}{m},\quad
-\tilde{H}_q=-i[\frac{-a(p-aq)}{m}+\frac{m\omega^2}{2}q],
$$
with $\beta^2=a^2+m^2\omega^2$,
$$
\tilde{\mathbb{X}}=\begin{pmatrix}
\frac{-ia}{m}&\frac{i}{m}\\
\frac{-i\beta^2}{m}&\frac{ia}{m}
\end{pmatrix} \et \tilde{\mathbb{X}}^2=\omega^2.
$$

\paragraph{\bf {Devices}} (I) We propose to consider, instead of \eqref{heat},
\begin{equation}
 i\frac{\partial}{\partial t}v(t,q)=i[-\frac{(-i\partial_{q}+iaq)^2}{2m}-\frac{m\omega^2}{2}q^2]v(t,q),
\label{iheat}
\end{equation}
with symbol
$$
\tilde{H}(q,p)=-i\bigg(\frac{(p+iaq)^2}{2m}+\frac{m\omega^2}{2}q^2\bigg).
$$
Now, we put
$$
\gamma=a^2-m^2\omega^2,\quad{\coshsl}_{\mathrm{s}}=\cosh{\omega s}, 
\quad 
\,{\sinhsl}_{\mathrm{s}}=\sinh{\omega s},
$$
and we have
$$
{\left\{
\begin{aligned}
&q(s)=\frac{(m\omega\,{\coshsl}_{\mathrm{t}}+a\,{\sinhsl}_{\mathrm{t}})\unbq-i\,{\sinhsl}_{\mathrm{t}}\unbp}{m\omega},\\
&p(s)=\frac{-i\gamma\,{\sinhsl}_{\mathrm{t}}\unbq+(m\omega\,{\coshsl}_{\mathrm{t}}-a\,{\sinhsl}_{\mathrm{t}})\unbp}{m\omega}.
\end{aligned}
\right.}
$$
Therefore
\begin{equation}
\begin{aligned}
\tilde{\phi}_0(t,\unbq,\unbp)&=-\frac{i}{2m}\int_0^t ds\,[p(s)^2+\gamma q(s)^2]\\
&=-\frac{i\,{\sinhsl}_{\mathrm{t}}}{2m^2\omega^2}[\gamma(m\omega\,{\coshsl}_{\mathrm{t}} +a\,{\sinhsl}_{\mathrm{t}})\unbq^2
+(m\omega\,{\coshsl}_{\mathrm{t}} -a\,{\sinhsl}_{\mathrm{t}})\unbp^2-2i\gamma\,{\sinhsl}_{\mathrm{t}}\unbq\unbp].
\end{aligned}
\label{iH0action}
\end{equation}

From
\begin{equation}
\barq=\,{\coshsl}_{\mathrm{t}}\unbq-\frac{i\,{\sinhsl}_{\mathrm{t}}}{m\omega}(ia\unbq+\unbp),\quad
\unbq=x(t,\barq,\unbp)=\frac{m\omega\barq+i\,{\sinhsl}_{\mathrm{t}}\unbp}{m\omega\,{\coshsl}_{\mathrm{t}}+a\,{\sinhsl}_{\mathrm{t}}},
\label{iHpath}
\end{equation}
we have
\begin{equation}
\begin{aligned}
\tilde{\phi}^H&(t,\barq,\unbp)=[\unbq\unbp+\phi_0(t,\unbq,\unbp)]\bigg|_{\unbq=x(t,\barq,\unbp)}\\
&=-\frac{i\,{\sinhsl}_{\mathrm{t}}}{2m^2\omega^2}[\gamma(m\omega\,{\coshsl}_{\mathrm{t}} +a\,{\sinhsl}_{\mathrm{t}})\unbq^2
+(m\omega\,{\coshsl}_{\mathrm{t}} -a\,{\sinhsl}_{\mathrm{t}})\unbp^2]+\frac{m^2\omega^2\,{\coshsl}_{\mathrm{t}}^2-a^2\,{\sinhsl}_{\mathrm{t}}^2}{m^2\omega^2}\unbq\unbp\bigg|_{\unbq=x(t,\barq,\unbp)}\\
&
=\frac{m\omega}{m\omega\,{\coshsl}_{\mathrm{t}} +a\,{\sinhsl}_{\mathrm{t}}}\barq\unbp+
\frac{i\,{\sinhsl}_{\mathrm{t}}(m^2\omega^2\unbp^2-\gamma\barq^2)}{2m^2\omega^2(m\omega \,{\coshsl}_{\mathrm{t}}+a\,{\sinhsl}_{\mathrm{t}})}.
\end{aligned}
\label{iHHaction}
\end{equation}

Now define
\begin{equation}
\begin{aligned}
v^H(t,\barq)&=\frac{1}{\sqrt{2\pi}}\int d\unbp\, \tilde{\alpha}^H(t,\barq,\unbp)e^{i\tilde{\phi}^H(t,\barq,\unbp)}\hat{\underline{v}}(\unbp)
\with
\tilde{\alpha}^H(t,\barq,\unbp)=\sqrt{\frac{m\omega}{m\omega\,{\coshsl}_{\mathrm{t}}}}\\
%&\quad
&={\sqrt{\frac{m\omega}{2\pi \coshsl_{\mathrm{t}}}}}
\int d\unbq\,
\exp{\{-\frac{m\omega}{{\sinhsl}_{\mathrm{t}}}({\coshsl}_{\mathrm{t}}(\barq^2+\unbq^2)-2\barq\unbq)+\frac{a}{2}(\barq^2-\unbq^2)\}}{\underline{v}}(\unbq).
\end{aligned}
\label{iHsol}
\end{equation}
%\end{document}

Since
$$
\lim_{t\to0}\tilde{\phi}^H(t,\barq,\unbp)=\barq\unbp, \quad \lim_{t\to0}\frac{m\omega}{m\omega\,{\coshsl}_{\mathrm{t}}}=1,
$$
we get
$$
\lim_{t\to 0}v^H(t,\barq)=\frac{1}{\sqrt{2\pi}}\int d\unbp\, e^{i\barq\unbp}\hat{\underline{v}}(\unbp)={\underline{v}}(\barq)
$$
and
$$
\frac{\partial}{\partial t}v^H(t,q)=[\frac{(\partial_{q}-aq)^2}{2m}+\frac{m\omega^2}{2}q^2]v^H(t,q).
$$

Therefore, in this case, we get the desired quantization if $\omega$ is replaced by $\pm i\omega$. 
But concerning kernel convergence, I have't checked it!

\paragraph{\bf {Devices}}(II) We multiply $i$ to both sides of \eqref{heat} and replacing $\partial_q$ with $p$, we get\\
 \fbox{something-like symbol} as
$$
\tilde{H}(q,p)=i(\frac{(p-aq)^2}{2m}-\frac{m\omega^2}{2}q^2),
$$
from which we may define Hamilton flow defined by
$$
\left\{
\begin{aligned}
\dot{q}&=\tilde{H}_p=i\frac{p-aq}{m},\\
\dot{p}&=-\tilde{H}_q=-i[\frac{-a(p-aq)}{m}-m\omega^2q],
\end{aligned}
\right.\quad
\;\;\mbox{or}\;\;
\begin{pmatrix}
\dot{q}\\
\dot{p}
\end{pmatrix}
=\tilde{\mathbb{X}}
\begin{pmatrix}
q\\
p
\end{pmatrix},
\quad
\tilde{\mathbb{X}}=\frac{i}{m}\begin{pmatrix}
-a&1\\
-\gamma&a
\end{pmatrix}
$$
where
$$
\gamma=a^2-m^2\omega^2,\quad \tilde{\mathbb{X}}^2=-\omega^2.
$$
Therefore, we get
$$
\begin{aligned}
q(s)&=C_s\unbq+\frac{iS_s}{m\omega}(-a\unbq+\unbp)=\frac{m\omega C_s-iaS_s}{m\omega}\unbq+\frac{iS_s}{m\omega}\unbp,\\
p(s)&=C_s\unbp+\frac{iS_s}{m\omega}(-\gamma\unbq+a\unbp)=\frac{-i\gamma S_s}{m\omega}\unbq
+\frac{m\omega C_s+iaS_s}{m\omega}\unbp,
\end{aligned}
$$
and
$$
\begin{aligned}
\tilde{\phi}_0(t,\unbq,\unbp)&=\int_0^t ds[\dot{q}p-\tilde{H}(q,p)]=\frac{i}{2m}\int_0^tds\,(p(s)^2-\gamma q(s)^2)\\
&=\frac{iS_t}{2m^2\omega^2}[-\gamma(m\omega C_t-iaS_t)\unbq^2+(m\omega C_t+iaS_t)\unbp^2-2i\gamma S_t\unbq\unbp]
\end{aligned}
$$
because
$$
\int _0^t ds\,\cos(2\omega s)=\frac{S_t C_t}{\omega},\quad
\int _0^t ds\,\sin(2\omega s)=\frac{S_t ^2}{\omega}.
$$
From
$$
\barq=\frac{m\omega C_t-iaS_t}{m\omega}\unbq+\frac{iS_t}{m\omega}\unbp,\quad
\unbq=\frac{m\omega\barq-iS_t\unbp}{m\omega C_t-iaS_t}=x(t,\barq,\unbp)
$$
we put
$$
\begin{aligned}
\tilde{\phi}^{H}(t,\barq,\unbp)&=[\unbq\unbp+\tilde{\phi}_0(t,\unbq,\unbp)]\big|_{\unbq=x(t,\barq,\unbp)}\\
&=\{\frac{iS_t}{2m^2\omega^2}[-\gamma(m\omega C_t-iaS_t)\unbq^2+(m\omega C_t+iaS_t)\unbp^2]+\frac{m^2\omega^2C_t^2+a^2S_t^2}{m^2\omega^2}\unbq\unbp\}\big|_{\unbq=x(t,\barq,\unbp)}\\
&=\frac{-iS_t(\gamma\barq^2+\unbp^2)}{2(m\omega C_t-iaS_t)}+\frac{m\omega}{m\omega C_t-iaS_t}\barq\unbp.\\
\tilde{\alpha}^H(t,\barq,\unbp)&=\sqrt{\frac{m\omega}{m\omega C_t-iaS_t}}.
\end{aligned}
$$
Then, we define
$$
\begin{aligned}
\tilde{v}^H(t,\barq)=(T_t^H\underline{v})(\barq)&=\frac{1}{\sqrt{2\pi}}\int d\unbp\, \tilde{\alpha}^H\, e^{i\tilde{\phi}^{H}(t,\barq,\unbp)}({\mathcal{F}}\underline{v})(\unbp)\\
&=\frac{1}{2\pi}\iint d\unbp d\unbq \tilde{\alpha}^H\, e^{-i\tilde{\phi}^{H}(t,\barq,\unbp)-i\unbq\unbp}\underline{v}(\unbq).
\end{aligned}
$$
Then, we get
$$
\begin{aligned}
\lim_{t\to0}\tilde{v}^H(t,q)&=\tilde{v}^H(0,q)=\underline{v}(q),\\
\pdt\tilde{v}^H(t,q)\bigg|_{t=0}&=[\frac{(\partial_q-aq)^2}{2m}-\frac{m\omega^2}{2}q^2]\underline{v}(q),\\
[T_s^H(T_t^H\underline{v})](\barq)&=(T_{t+s}^H\underline{v})(\barq),\\
\pdt\tilde{v}^H(t,q)&=[\frac{(\partial_q-aq)^2}{2m}-\frac{m\omega^2}{2}q^2]\tilde{v}^H(t,q).
\end{aligned}
$$

\subsection{Quantum field theory}
Though N. Kumano-go and D. Fujiwara~\cite{KF08} is so interesting to give meaning not only to the formula of integration by parts but also the stationary method, etc, in $\infty$-dimensional space. But I can't appreciate the reason why they take as candidate paths left continuous or right-continuous one. For example, to apply this idea to field theory after replacing time-sclicing as standard subdivision, how to settle left or right continuity at each boundary of subdivision.

For reader's sake, I cite a typical example of the so-called functional method of field theory.

It is known that for Lagrangian
$$
I(u)=\iint_{\euc\times \euc^3}dt dx\,\bigg[\frac{1}{2}|u_t(t,x)|^2-\frac{1}{2}|\nabla u(t,x)|^2-\frac{m^2}{2}|u(t,x)|^2-\frac{1}{4}|u(t,x)|^4\bigg]
$$
whose Euler equation is
$$
(\square+m^2) u(t,x)-|u(t,x)|^2u(t,x)=0,
$$
we consider
$$
Z(\phi)=\int d_Fu\,e^{i\langle u|\phi\rangle+L(u)}\with \langle u|\phi\rangle=\iint_{\euc\times \euc^3}dt dx\,u(t,x)\phi(t,x)
$$
which satisfies FDE(=Functional Derivative Equation)
\begin{equation}
\bigg[(\square +m^2)\frac{\delta}{i{\delta \phi(t,x)}}+\frac{\delta^3}{(i\delta \phi(t,x))^3}\bigg]Z(\phi)=0.
\label{qft-model}
\end{equation}
How to give the meaning to $Z(\phi)$ above by PIM?
Time slicing method for 1-dimension will be generalized to standard subdivision like Dodziuk procedure ~\cite{dod76}, but in this case, how we define ``right-continuous or left continuous paths''?
\begin{remark}
To treat FDE like \eqref{qft-model}, the first difficulty stems from giving the meaning for the higher order derivative at the point $(t,x)$.
$$
\frac{\delta^2}{{\delta \phi(t,x)}^2}Z(\phi).
$$
See, an example treated in Inoue~\cite{ino87-2} and also \cite{ino86, ino90}.
But to develop functional analysis method (which gives the large success to PDE) to FDE, we need integration by parts but without standard measure.
\end{remark}

\subsubsection{[H]: Path-integral method of heat type equation under H-formulation is impossible?}
\begin{problem}
Whether we may represent the solution of a given PDE by using classical mechanical objects corresponding to it, is our problem. 
\par
On the other hand, we may compare this with the problem posed by H. Widom~\cite{wido80}:
Let $A^W$ be the self-adjoint operator obtained from a real valued symbol $A(q,p)$ by Weyl quantization whose spectral resolution is given by $\displaystyle{A^W=\int dE_{\lambda}\,\lambda}$.
Taking a function $f$ in the suitable class, we may define, by functional calculus method, $f(A^W)=\int dE_{\lambda}f(\lambda)$.
In this case, whether $f(A^W)$ is a pseudo-differential operator and how its symbol is represented by, are discussed there. Moreover, let  two self-adjoint operators $A^W$ and $B^W$ be given with a function with two variables. Whether
$f(A^W,B^W)$ gives a pseudo-differential operator, this is considered in R.S. Strichartz~\cite{stri72}.
These consideration is applied to to the system version of Egorov's theorem in \S 4 of Chapter 9.
\par
We need to remark the physicist's usage of analytic continuation w.r.t. time $t$, because it is not so obvious whether the operator like $e^{tA^W}$ is analytically continued to $e^{itA^W}$.
\newline
(1) Let a non-negative function $H(q,p)$ with order $2$ w.r.t. $p$ having functions $S^L(t,\barq,\unbq)$ and $D^L(t,\barq,\unbq)$ on configuration space
as solutions of Hamilton-Jacobi and continuity equation, respectively.
Taking the normalization constant $C_L$, and defining 
$$
(T^L_t{\underline{v}}) (\barq)=v^L(t,\barq)=C_L\int_{\euc}d\unbq\, \sqrt{D^L(t,\barq,\unbq)}e^{S^L(t,\barq,\unbq)}{{\underline{v}}}(\unbq),
$$
have we a parametrix of corresponding heat type equation 
$$
\pdt v^L(t,\barq)=H^{W}(\barq,\partial_{\barq})v^L(t,\barq)\with v^L(0,\unbq)={\underline{v}}(\unbq)?
$$
(2) Under the same setting as (1), we define functions $S^H(t,\barq,\unbp)$ and $D^H(t,\barq,\unbp)$ on phase space as solutions of Hamilton-Jacobi and continuity equation, respectively.
Defining 
$$
(T^H_t{\underline{v}}) (\barq)=v^H(t,\barq)=C_H\int_{\euc}d\unbp\, \sqrt{D^H(t,\barq,\unbp)}e^{S^H(t,\barq,\unbp)}\hat{{\underline{v}}}(\unbp)
$$
whether we have a parametrix of corresponding heat type equation with suitable devices?
\end{problem}

\begin{remark}
For (1), I show not only simple examples in the previous paragraph, but also construct a quantized operator of Riemann metric in Inoue and Maeda~\cite{IM85}.
Concerning (2), I give two simple examples showing it seems hard to get the desired results without some devices.
\end{remark}

$\bullet$ As is mentioned at the very beginning of this section, taking the Fourier transformation of
$$
v_t=\frac{1}{2}\partial_q^2v\with v(0)={\underline{v}},
$$
we get readily
$$
v(t,\barq)=\frac{1}{\sqrt{2}}\int d\unbp\,e^{i\barq\unbp-\frac{1}{2}\unbp^2t}\hat{{\underline{v}}}(\unbp).
$$
But, except normalization constant, it seems impossible to have a solution $S(t,\barq,\unbp)$ of Hamilton-Jacobi equation satisfying
$$
e^{i\barq\unbp-\frac{1}{2}\unbp^2t}=A(t,\barq,\unbp)e^{S(t,\barq,\unbp)}\with
A\sim\bigg(\frac{\partial^2 S}{\partial\barq\partial\unbp}\bigg)^{1/2}.
$$

$\bullet$ Therefore, multiplying imaginary unit ``$i$'' to both sides of heat equation,
$$
iv_t=i\frac{1}{2}\partial_q^2v\with v(0)={\underline{v}}
$$
and substituting $p$ into $-i\partial_q$, we assign the Hamilton function $\tilde{H}(q,p)=-\frac{i}{2}p^2$.
Then, we have
$$
\begin{gathered}
\tilde{S}_0(t,\unbq,\unbp)=-\frac{i}{2}\unbp^2t, \quad\barq=\unbq-i\unbp t,\\
\tilde{S}(t,\barq,\unbp)=\unbq\unbp+\tilde{S}_0(t,\unbq,\unbp)\bigg|_{\unbq=\barq+i\unbp t}=\barq\unbp+\frac{i}{2}\unbp^2t.
\end{gathered}
$$
Putting 
$$
v(t,\barq)=\frac{1}{\sqrt{2\pi}}\int_{\euc} d\unbp\,e^{i\tilde{S}(t,\barq,\unbp)}\hat{{\underline{v}}}(\unbp)=\frac{1}{\sqrt{2\pi}}\int_{\euc} d\unbp\,e^{i\barq\unbp-\frac{1}{2}\unbp^2t}\hat{{\underline{v}}}(\unbp),
$$
we have the desired expression. See also, Qi's equation in \S3, Chapter 9.

$\bullet$ Unfortunately, this procedure doesn't work for harmonic oscillator. In fact, since we have
$$
\tilde{H}(q,p)=-\frac{i}{2}(p^2+\omega^2 q^2),
$$
solutions of Hamilton equation are given by
$$
\left\{
\begin{aligned}
&\dot{q}=\tilde{H}_p(q,p)=-ip,\\
&\dot{p}=-\tilde{H}_q(q,p)=i\omega^2q
\end{aligned}
\right.
\quad
\begin{pmatrix}
q(s)\\ p(s)
\end{pmatrix}
=e^{sX}
\begin{pmatrix}
\unbq\\
\unbp
\end{pmatrix}
\with
X=\begin{pmatrix}
0&-i\\
i\omega^2&0
\end{pmatrix},
$$
there appeared the terms with $\cosh \omega s$, $\sinh \omega s$. To get the desired
one with $\cos\omega s$, $\sin\omega s$, we need to complexify $\omega$, but no philosophical evidence to do so.
Only when $\omega\to 0$, we have the desired result obtained before.

\section{Spin addition} 
Preparing a representation space $V$ with the scalar product $(\cdot,\cdot)$ and two bounded operators ${\mathbf{b}}$ and ${\mathbf{b}}^*$ such that
$$
{\mathbf{b}}^2=({\mathbf{b}}^*)^2=0,\quad
{\mathbf{b}}{\mathbf{b}}^*+{\mathbf{b}}^*{\mathbf{b}}=0.
$$
In stead of ODE ${\mathbb{H}}(q,\partial_q)$ \eqref{PIM-heat0} in $L^2(\euc)$, we put
$$
{\mathbb{Q}}_-=\frac{1}{\sqrt{2}}(\partial_q-\omega q){\mathbf{b}},\quad
{\mathbb{Q}}_+=\frac{1}{\sqrt{2}}(\partial_q+\omega q){\mathbf{b}}^*
$$
and we define
$$
\begin{gathered}
{\mathbb{H}}={\mathbb{H}}(q,\partial_q)={\mathbb{Q}}^2=H(q,\partial_q){\mathbb{I}}+\frac{\omega}{2}[{\mathbf{b}},{\mathbf{b}}^*],\\
{\mathbb{Q}}={\mathbb{Q}}_-+{\mathbb{Q}}_+,\quad {\mathbb{P}}=[{\mathbf{b}},{\mathbf{b}}^*],
\end{gathered}
$$
on $\mathfrak{H}=L^2(\euc)\otimes V$.
Since
$$
{\mathbb{P}}^2={\mathbb{I}},\quad [{\mathbb{Q}},{\mathbb{P}}]_+={\mathbb{Q}}{\mathbb{P}}+{\mathbb{P}}{\mathbb{Q}}=0,
$$
$({\mathbb{H}},{\mathbb{P}},{\mathbb{Q}})$ gives SUSYQM on $\mathfrak{H}=L^2(\euc)\otimes V$.
Especially, taking
$$
\begin{gathered}
V={\mathbb{C}}^2\with (u,v)=u_1\bar{v}_1+u_2\bar{v}_2 \for u={}^t(u_1,u_2), v={}^t(v_1,v_2)\in{\mathbb{C}}^2,\\
{\mathbf{b}}^* %(\sim\theta)
=\begin{pmatrix}
0&0\\
1&0 
\end{pmatrix},
\quad
{\mathbf{b}} %(\sim\frac{\partial}{\partial\theta})
=\begin{pmatrix}
0&1\\
0&0
\end{pmatrix},
\quad
[{\mathbf{b}},{\mathbf{b}}^*]={\mathbf{b}}{\mathbf{b}}^*-{\mathbf{b}}^*{\mathbf{b}}
=\begin{pmatrix}
1&0\\
0&-1
\end{pmatrix},
\end{gathered}
$$
we get an ordinary matrix representation of ${\mathbb{H}}$ in $\mathfrak{H}=L^2(\euc)\otimes V$.

We represent these by using an odd variable $\theta$.

Putting $\Lambda=\{u_0+u_1\theta\;|\; u_0, u_1\in{\mathbb{C}}\}$,
we decompose
$$
\Lambda_b=\{u_0\;|\; u_0\in{\mathbb{C}}\},\quad \Lambda_f=\{u_1\theta \;|\; u_1\in{\mathbb{C}}\},\quad
\Lambda=\Lambda_b\oplus \Lambda_f.
$$
Then
$$
\theta(u_0+u_1\theta)=u_0\theta\sim\begin{pmatrix}
0&0\\
1&0 
\end{pmatrix}
\begin{pmatrix}
u_0\\
u_1
\end{pmatrix},\quad
\frac{\partial}{\partial\theta}(u_0+u_1\theta)=u_1\sim
\begin{pmatrix}
0&1\\
0&0
\end{pmatrix}
\begin{pmatrix}
u_0\\
u_1
\end{pmatrix},
$$
and
$$
[{\mathbf{b}},{\mathbf{b}}^*]\sim \frac{\partial}{\partial\theta}\theta-\theta\frac{\partial}{\partial\theta}
\sim
\begin{pmatrix}
0&1\\
0&0
\end{pmatrix}
\begin{pmatrix}
0&0\\
1&0 
\end{pmatrix}
-\begin{pmatrix}
0&0\\
1&0 
\end{pmatrix}
\begin{pmatrix}
0&1\\
0&0
\end{pmatrix}
=\begin{pmatrix}
1&0\\
0&-1
\end{pmatrix}.
$$
\begin{remark}
Since
$$
({\bf b}u,v)=(u,{\bf b}^*v)\for u=\begin{pmatrix}
u_0\\u_1
\end{pmatrix},
v=\begin{pmatrix}
v_0\\v_1
\end{pmatrix}
\in{\mathbb{C}}^2,
$$
we may have scalar product $({\cdot},{\cdot})$ in $\mathfrak{H}=L^2(\euc)\otimes V$ such that
\begin{equation}
(u_0+u_1\theta,\theta(v_0+v_1\theta))=(\frac{\partial}{\partial\theta}(u_0+u_1\theta),v_0+v_1\theta)
\label{ibp}
\end{equation}
which permits integration by parts w.r.t. $\theta$.
Please refer \eqref{FS-2.18}, \S 2 in Chapter 9.
\end{remark}

\subsection{Classical Mechanics corresponding to ${\mathbb{H}}(q,\partial_{q})$}
Prepare ${{\kbar}}\in{\mathbb{C}}\setminus\{0\}$ and define Fourier transformation for functions  of $\theta\in{\fR}^{0|1}$ as
$$
\begin{cases}
\int_{{\fR}^{0|1}}d\theta\,e^{-{{\kbar}}^{-1}\theta\pi}(u_0+u_1\theta)
=u_1-{{\kbar}}^{-1}u_0\pi,\\
\int_{{\fR}^{0|1}}d\pi\,e^{{{\kbar}}^{-1}\theta\pi}(u_1-{{\kbar}}^{-1}u_0\pi)
=-{{\kbar}}^{-1}(u_0+u_1\theta),\\
-{{\kbar}}\iint_{{\fR}^{0|2}}d\pi\,d\theta'\,e^{{{\kbar}}^{-1}(\theta-\theta')\pi}(u_0+u_1\theta')=u_0+u_1\theta
\end{cases}
$$
Then the Weyl symbol of ${\mathbb{H}}(q,\partial_q)$ is given by
\begin{equation}
{\mathcal{H}}(x,\xi,\theta,\pi)=-\frac{1}{2}\xi^2-\frac{{\omega}^2}{2}x^2-{{\kbar}}^{-1}{\omega}\theta\pi.
\end{equation}
In fact, since
$$
-{{\kbar}}\iint d\pi d\theta' e^{{{\kbar}}^{-1}(\theta-\theta')\pi}\frac{\theta+\theta'}{2}\pi (u_0+u_1\theta')
=\frac{1}{2}(u_0-u_1\theta),
$$
we get
$$
\begin{aligned}
\hat{\mathcal{H}}(x,\partial_x,\theta,\partial_{\theta})&u(x,\theta)
=\frac{{-\kbar}}{2\pi\hbar}\iint d\xi dx' d\pi d\theta' e^{i\hbar^{-1}(x-x')\xi+{{\kbar}}^{-1}(\theta-\theta')\pi}\\
&\qquad\qquad\qquad\qquad\qquad\qquad\times
{\mathcal{H}}(\frac{x+x'}{2},\xi,\frac{\theta+\theta'}{2},\pi)(u_0(x')+u_1(x')\theta')\\
&=\frac{1}{2\pi\hbar}\iint_{{\fR}^{2|0}}e^{i\hbar^{-1}(x-x')\xi}H(\frac{x+x'}{2},\xi)(u_0(x')+u_1(x')\theta)
+{\omega}(u_0(x)-u_1(x)\theta)\\
&\qquad\qquad\qquad\qquad\qquad\qquad\sim\bigg({H}(q,\partial_q){\mathbb{I}}+\frac{{\omega}}{2}[{\bf b},{\bf b}^*] \bigg)u.
\end{aligned}
$$

\subsection{LH-formulation}
In this case, since the Hamilton equation w.r.t. odd variables is given by
$$
\dot\theta=-{{\kbar}}^{-1}\omega\theta,\quad \dot{\pi}={{\kbar}}^{-1}\omega\pi,
$$
we have
$$
\theta(s)=e^{-{{\kbar}}^{-1}\omega s}\unbtheta,\quad \pi(s)=e^{{{\kbar}}^{-1}\omega s}\unbpi.
$$
Without using Fourier transform w.r.t. even variables, we may put
\begin{equation}
{\mathcal{S}}(t,\barx,\bartheta,\unbx,\unbpi)=
[\unbtheta\unbpi+S_0(t,\unbx,\unbxi)]\bigg|_{\unbxi=\cdots,\unbtheta=\cdots}
=e^{{{\kbar}}^{-1}\omega t}\bartheta\unbpi+S(t,\barx,\unbx),
\label{susyqm3.12bis}
\end{equation}
where
\begin{equation}
{S}(t,\barx,\unbx)=\frac{{\omega}\,\ccsl}{2\cssl}\barx^2
-\frac{{\omega}}{\cssl}\barx\unbx
+\frac{{\omega}\,\ccsl}{2\cssl}\unbx^2.
\label{ss4.4bis} %3.22
\end{equation}
Then \eqref{susyqm3.12bis} satisfies
\begin{equation}
{\mathcal{S}}_t+{\mathcal{H}}(\barx,{\mathcal{S}}_{\barx},
\bartheta,{\mathcal{S}}_{\bartheta})=0
\with \lim_{t\to0}({\mathcal{S}}-\frac{(\barx-\unbx)^2}{2t}-\bartheta\unbpi)=0.
\label{susyqm3.13bis}
\end{equation}

Moreover, putting
$$
{\mathcal{D}}(t,\barx,\bartheta,\unbx,\unbpi)=\sdet
\begin{pmatrix}
-\frac{\partial^2 {\mathcal{S}}}{\partial\barx\partial\unbx}&0\\
0&-\frac{\partial^2 {\mathcal{S}}}{\partial\bartheta\partial\unbpi}
\end{pmatrix}=\frac{\omega}{\sinh(\omega t)}e^{\omega t},
\label{susyqm3.14bis}
$$
we have
\begin{equation}
{\mathcal{D}}_t+\partial_{\barx}({\mathcal{D}}{\mathcal{H}}_\xi)
+\partial_{\bartheta}({\mathcal{D}}{\mathcal{H}}_\pi)=0
\with \lim_{t\to0}t{\mathcal{D}}(t,\barx,\bartheta,\unbx,\unbpi)=1
\label{susyqm3.15bis}
\end{equation}
where
${\mathcal{H}}_\xi={\mathcal{H}}_\xi(\barx,{\mathcal{S}}_{\barx},
\bartheta,{\mathcal{S}}_{\bartheta})$ and 
${\mathcal{H}}_\pi={\mathcal{H}}_\pi(\barx,{\mathcal{S}}_{\barx},
\bartheta,{\mathcal{S}}_{\bartheta})$.

We put
\begin{equation}
\begin{aligned}
{\mathcal V}_t^{\mathrm{LH}}\,{\underline{u}}(t,\barx,\bartheta)&=
\frac{1}{\sqrt{2\pi}}\int_{{\fR}^{1|1}}d\unbx d\unbpi\,
{\mathcal{D}}^{1/2}e^{-{\mathcal{S}}}
({\underline{u}}_1-{{\kbar}}^{-1}{\underline{u}}_0\unbpi)\\
&=e^{\omega t/2}V_t^L{\underline{u}}_0(\barx)
+e^{-\omega t/2}V_t^L{\underline{u}}_1(\barx)\bartheta.
\end{aligned}
\label{susyqm3.16bis}
\end{equation}

%\paragraph{\bf HH-formulation:} 
\subsection{HH-formulation}
We define
\begin{equation}
{\mathcal{S}}(t,\barx,\bartheta,\unbxi,\unbpi)=
[\unbtheta\unbpi+S_0(t,\unbx,\unbxi)]\bigg|_{\unbx=\cdots,\unbtheta=\cdots}
=e^{-\omega t}\bartheta\unbpi+S(t,\barx,\unbxi),
\label{susyqm3.12}
\end{equation}
which satisfies
\begin{equation}
{\mathcal{S}}_t+{\mathcal{H}}(\barx,{\mathcal{S}}_{\barx},
\bartheta,{\mathcal{S}}_{\bartheta})=0
\with \lim_{t\to0}{\mathcal{S}}=\barx\unbxi+\bartheta\unbpi.
\label{susyqm3.13}
\end{equation}
Moreover, putting
$$
{\mathcal{D}}(t,\barx,\bartheta,\unbxi,\unbpi)=\sdet
\begin{pmatrix}
\frac{\partial^2 {\mathcal{S}}}{\partial\barx\partial\unbxi}&0\\
0&-\frac{\partial^2 {\mathcal{S}}}{\partial\bartheta\partial\unbpi}
\end{pmatrix}=\frac{1}{\cosh(\omega t)}e^{\omega t},
\label{susyqm3.14}
$$
we have
\begin{equation}
{\mathcal{D}}_t+\partial_{\barx}({\mathcal{D}}{\mathcal{H}}_\xi)
+\partial_{\bartheta}({\mathcal{D}}{\mathcal{H}}_\pi)=0
\with \lim_{t\to0}{\mathcal{D}}(t,\barx,\bartheta,\unbxi,\unbpi)=1
\label{susyqm3.15}
\end{equation}
where
${\mathcal{H}}_\xi={\mathcal{H}}_\xi(\barx,{\mathcal{S}}_{\barx},
\bartheta,{\mathcal{S}}_{\bartheta})$ and 
${\mathcal{H}}_\pi={\mathcal{H}}_\pi(\barx,{\mathcal{S}}_{\barx},
\bartheta,{\mathcal{S}}_{\bartheta})$.

\begin{equation}
{\mathcal{S}}(t,\barx,\bartheta,\unbxi,\unbpi)
=[-\unbtheta\,\unbpi+S_0(t,\unbx,\unbxi)]\bigg|_{\unbx=\cdots,\unbtheta=\cdots}
=-e^{-\omega t}\bartheta\unbpi+S(t,\barx,\unbxi)
\label{susyqm3.21}  %{susyqm3.21bis}
\end{equation}
we put
$$
{\mathcal{V}}_t\,{\underline{v}}(t,\barx,\bartheta)
=e^{\omega t/2}V_t^L{\underline{v}}_0(\barx)
+e^{-\omega t/2}V_t^L{\underline{v}}_1(\barx)\bartheta.
\label{susyqm3.19}$$
Putting $v(t,x,\theta)={\mathcal{V}}_t\,{\underline{v}}(t,x,\theta)$,
we have
\begin{equation}
\pdt v(t,x,\theta)=\bigg(\frac12\frac{\partial^2}{\partial x^2}
-\frac12\omega^2 x^2
+\frac{\omega}{2}\big[\frac{\partial}{\partial\theta},\theta\big]_-\bigg)v(t,x,\theta)
\with
v(0,x,\theta)={\underline{v}}(x,\theta).
\label{susyqm3.20}
\end{equation}
From the kernel of ${\mathcal{V}}_t$, we may calculate the Witten index which
is shown soon later.

\subsection{A generalization}
For
$$
H(q,p)=\frac12\sum_{j=1}^dp_j^2-\sum_{j=1}^d\frac{\omega_j^2q_j^2}{2}
\in C^\infty(T^*\euc^d:\euc),
$$
we may extend it as
$$
{\mathcal{H}}(x,\theta,\xi,\pi)=\frac12\sum_{j=1}^d\xi_j^2
-\sum_{j=1}^d\frac{\omega_j^2x_j^2}{2}+\sum_{j=1}^d\omega_j\theta_j\pi_j.
$$
\begin{equation}
{\mathcal{S}}(t,\barx,\bartheta,\unbx,\unbpi)
=[-\langle\unbtheta|\unbpi\rangle
+S_0(t,\unbx,\unbxi)]\bigg|_{\unbtheta=\cdots,\unbxi=\cdots}
=-\sum_{j=1}^d e^{-\omega_j t}\bartheta_j\unbpi_j+S(t,\barx,\unbx)
\label{susyqm3.21bis}
\end{equation}
with
$$
S(t,\barx,\unbx)=\sum_{j=1}^d\bigg[
\frac{\omega_j\cosh(\omega_jt)}{2\sinh(\omega_j t)}(\barx_j^2+\unbx_j^2)
-\frac{\omega_j}{\sinh(\omega_j t)}\barx_j\unbx_j\bigg].
\label{susyqm3.22bis}$$
We define
$$
{\mathcal{D}}(t,\barx,\bartheta,\unbxi,\unbpi)=\sdet\begin{pmatrix}
-\frac{\partial^2 {\mathcal{S}}}{\partial\barx\partial\unbx}&0\\
0&\frac{\partial^2 {\mathcal{S}}}{\partial\bartheta\partial\unbpi}
\end{pmatrix}
=\prod_{j=1}^d\frac{\omega_j}{\sinh(\omega_jt)}e^{\omega_j t}
\label{susyqm3.23bis}$$
and we have
$$
{\mathcal{D}}^{1/2}e^{-{\mathcal{S}}}=e^{\omega t/2}
\sqrt{\prod_{j=1}^d\frac{\omega_j}{\sinh(\omega_jt)}}
e^{\sum_{j=1}^d e^{-\omega_j t}\bartheta_j\unbpi_j-S(t,\barx,\unbx)},
\quad
\omega=\sum_{j=1}^d \omega_j.
\label{susyqm3.24bis}$$
\subsubsection{[H]:LH-formulation for the case $d=2$ with 
$u(t,x,\theta)=u_0(t,x)+u_1(t,x)\theta_1
+u_2(t,x)\theta_2+u_3(t,x)\theta_1\theta_2$}
\par
Define the Fourier transformations by
$$
\begin{cases}
\int d\theta\,e^{\langle\theta|\pi\rangle}
(u_0+u_1\theta_1+u_2\theta_2+u_3\theta_1\theta_2)
=u_3-u_2\pi_1+u_1\pi_2-u_0\pi_1\pi_2,\\
\int d\pi\,e^{-\langle\theta|\pi\rangle}
(u_3-u_2\pi_1+u_1\pi_2-u_0\pi_1\pi_2)
=-(u_0+u_1\theta_1+u_2\theta_2+u_3\theta_1\theta_2).
\end{cases}
$$
Then, we have
$$
\begin{aligned}
&{\mathcal{V}}_t{\underline{u}}(\barx,\bartheta)=
\frac{-1}{2\pi}\int d\unbx d\unbpi\,{\mathcal{D}}^{1/2}(t,\barx,\unbx,\bartheta,\unbpi)
e^{-{\mathcal{S}}(t,\barx,\unbx,\bartheta,\unbpi)}
({\underline{u}}_3(\unbx)-{\underline{u}}_2(\unbx)\unbpi_1
+{\underline{u}}_1(\unbx)\unbpi_2-{\underline{u}}_0(\unbx)\unbpi_1\unbpi_2)\\
&\qquad\qquad
=\int d\unbx\,\sqrt{\frac{\omega_1}{2\pi\sinh(\omega_1t)}
\frac{\omega_2}{2\pi\sinh(\omega_2t)}}e^{-S(t,\barx,\unbx)}
e^{\omega t/2}\\
&\qquad\qquad\qquad\qquad\times
({\underline{u}}_0(\unbx)
-e^{-\omega_1t}{\underline{u}}_1(\unbx)\bartheta_1
-e^{-\omega_2t}{\underline{u}}_2(\unbx)\bartheta_2
+e^{-\omega t}{\underline{u}}_3(\unbx)\bartheta_1\bartheta_2).
\end{aligned}
$$
Since
$$
\begin{aligned}
&\int d\unbtheta\,
(e^{\omega t/2}\unbtheta_1\unbtheta_2
+e^{(-\omega_1+\omega_2)t/2}\bartheta_1\unbtheta_2
+e^{(\omega_1-\omega_2)t/2}\unbtheta_1\bartheta_2
+e^{-\omega t/2}\bartheta_1\bartheta_2)
(u_0+u_1\unbtheta_1+u_2\unbtheta_2+u_3\unbtheta_1\unbtheta_2)\\
&\qquad=
e^{\omega t/2}u_0
-e^{(-\omega_1+\omega_2)t/2}u_1\bartheta_1
-e^{(\omega_1-\omega_2)t/2}u_2\bartheta_2
+e^{-\omega t/2}u_3\bartheta_1\bartheta_2
\end{aligned}
$$
and
$$
Pu=\bigg(1-2\theta_1\frac{\partial}{\partial\theta_1}\bigg)
\bigg(1-2\theta_2\frac{\partial}{\partial\theta_2}\bigg)
(u_0+u_1\theta_1+u_2\theta_2+u_3\theta_1\theta_2)
=u_0-u_1\theta_1-u_2\theta_2+u_3\theta_1\theta_2,
$$
we have the corresponding SUSYQM with
$$
\begin{gathered}
W(x)=\frac12(\omega_1x_1^2+\omega_2x_2^2),\quad 
b_j=\theta_j,\;b_k^*=\frac{\partial}{\partial \theta_k}\\
\et
-\sum_{j,k=1}^2\frac{\partial^2 W}{\partial x_j\partial x_k}[b_j,b_k^*]_-
=\omega_1\bigg(1-2\theta_1\frac{\partial}{\partial\theta_1}\bigg)
+\omega_2\bigg(1-2\theta_2\frac{\partial}{\partial\theta_2}\bigg)
=\sum_{j=1}^2\omega_j[\frac{\partial}{\partial\theta_j},\theta_j]_-.
\end{gathered}
$$
Moreover, we have
$$
\begin{aligned}
\str {\mathcal{V}}_t&= \int d\unbx d\unbtheta\,
(e^{\omega t/2}-e^{(-\omega_1+\omega_2)t/2}
-e^{(\omega_1-\omega_2)t/2}+e^{-\omega t/2})
\unbtheta_1\unbtheta_2
e^{-S(t,\unbx,\unbx)}\\
&=(e^{\omega_1 t/2}-e^{-\omega_1t/2})(e^{\omega_2 t/2}-e^{-\omega_2t/2})
\frac{1}{\sqrt{2(\cosh(\omega_1 t)-1)}\sqrt{2(\cosh(\omega_2 t)-1)}}= 1.
\end{aligned}
$$

\subsection{[H]:LH-formulation for general $d$ with 
$u(t,x,\theta)=\sum_{|a|\le d}u_a(t,x)\theta^a$}
\par
Since we have
$$
\begin{gathered}
\prod_{j=1}^d(\theta_j\pi_j)^{b_j}=
(-1)^{|b|(|b|-1)/2}\theta^b\pi^b\with
b=(b_1,\cdots,b_d)\in\{0,1\}^d,\quad |b|=\sum_{j=1}^d b_j,\\
\theta^a\theta^b=(-1)^{\tau(a,b)}\theta^{a+b}
\with
\tau(b,a) \equiv |a||b|+\tau(a,b)  \mod2,
\end{gathered}
$$
we get, with $b_j=1-a_j$, $a+b=\tilde{1}=(1,\cdots,1)$, $d=|a|+|b|$,
$$
\begin{aligned}
&\int d\theta\,e^{\langle\theta|\pi\rangle}\theta^a
=\int d\theta\,\prod_{j=1}^d(1+\theta_j\pi_j)\theta^a
=\int d\theta\,\theta^a\prod_{j=1}^d(\theta_j\pi_j)^{b_j}
=(-1)^{|b|(|b|-1)/2+\tau(a,b)}\pi^b,\\
&\int d\pi\,e^{-\langle\theta|\pi\rangle}
\pi^b=(-1)^{|a|(|a|-1)/2+\tau(b,a)}\theta^a,
\quad
\int d\pi\,e^{-\langle\theta|\pi\rangle}\bigg(
\int d\theta'\,e^{\langle\theta'|\pi\rangle}\theta^{\prime a}
\bigg)
=(-1)^{d(d-1)/2}\theta^a.
\end{aligned}
$$
Therefore
$$
\begin{aligned}
&\int d\theta\,e^{\langle\theta|\pi\rangle}
\sum_{|a|\le d}u_a\theta^a
=\int d\theta\,
\sum_{|a|\le d}u_a\theta^a\prod_{j=1}^d(\theta_j\pi_j)^{b_j}
=\sum_{|a|\le d}(-1)^{|b|(|b|-1)/2+\tau(a,b)}u_a\pi^b,\\
&\int d\pi\,e^{-\langle\theta|\pi\rangle}
\sum_{|a|\le d}(-1)^{|b|(|b|-1)/2+\tau(a,b)}u_a\pi^b
=(-1)^{d(d-1)/2}\sum_{|a|\le d}u_a\theta^a.
\end{aligned}
$$
On the other hand, as
$$
\begin{aligned}
\int d\unbpi\,e^{\sum_{j=1}^d e^{-\omega_jt}\bartheta_j\unbpi_j}
(-1)^{|b|(|b|-1)/2+\tau(a,b)}\unbpi^b
&=\int d\unbpi\,
(-1)^{|b|(|b|-1)/2+\tau(a,b)}\unbpi^b\prod_{j=1}^d
(e^{-\omega_jt}\bartheta_j\unbpi_j)^{a_j}\\
&=\prod_{j=1}^de^{-a_j\omega_jt}(-1)^{d(d-1)/2}(-1)^{|a|+\tau(a,b)}\bartheta^a,
\end{aligned}
$$
we have
$$
\begin{aligned}
&{\mathcal{V}}_t{\underline{u}}(\barx,\bartheta)=
\frac{(-1)^{d(d-1)/2}}{(2\pi)^{d/2}}
\int d\unbx d\unbpi\,{\mathcal{D}}^{1/2}(t,\barx,\unbx,\bartheta,\unbpi)
e^{-{\mathcal{S}}(t,\barx,\unbx,\bartheta,\unbpi)}
\sum_{|a|\le d}(-1)^{|b|(|b|-1)/2+\tau(a,b)}{\underline{u}}_a(\unbx)\unbpi^b\\
&\qquad\qquad
=\int d\unbx\,\prod_{j=1}^d
\sqrt{\frac{\omega_j}{2\pi\sinh(\omega_jt)}}e^{-S(t,\barx,\unbx)}
e^{\omega t/2}
\sum_{|a|\le d}\prod_{j=1}^d e^{-a_j\omega_jt}(-1)^{|a|+\tau(a,b)}
{\underline{u}}_a(\unbx)\bartheta^a.
\end{aligned}
$$
Since
$$
\begin{aligned}
&\int d\unbtheta\,
\sum_{|a'|+|b'|=d}e^{\sum_{j=1}^d(a'_j-b'_j)\omega_jt/2}
\unbtheta^{a'}\bartheta^{b'}
\sum_{|a|\le d}u_a\unbtheta^a\\
&\qquad=
\sum_{|a|\le d}\prod_{j=1}^d e^{(b_j-a_j)\omega_jt/2}(-1)^{|a|+\tau(a,b)}
{\underline{u}}_a(\unbx)\bartheta^a.
\end{aligned}
$$
and
$$
Pu=\prod_{j=1}^d\bigg(1-2\theta_j\frac{\partial}{\partial\theta_j}\bigg)
\sum_{|a|\le d}u_a\theta^a
=\sum_{|a|\le d}(-1)^{|b|}u_a\theta^a,
$$
we have the corresponding SUSYQM as before.

Moreover, we have
$$
\begin{aligned}
\str {\mathcal{V}}_t&= \int d\unbx d\unbtheta\,
\prod_{j=1}^d(e^{\omega_j t/2}-e^{-\omega_jt/2})\unbtheta_1\cdots\unbtheta_d
e^{-S(t,\unbx,\unbx)}\\
&=\prod_{j=1}^d(e^{\omega_j t/2}-e^{-\omega_jt/2})
\frac{1}{\sqrt{\prod_{j=1}^d2(\cosh(\omega_j t)-1)}}= 1.
\end{aligned}
$$

Putting $v(t,x,\theta)={\mathcal{V}}_t\,{\underline{v}}(t,x,\theta)$,
we have
\begin{equation}
\pdt v(t,x,\theta)=\bigg(\frac12\sum_{j=1}^d\frac{\partial^2}{\partial x_j^2}
-\frac12\sum_{j=1}^d\omega_j^2 x_j^2
+\sum_{j=1}^d\frac{\omega_j}{2}\big[\frac{\partial}{\partial\theta_j},\theta_j\big]_-\bigg)v(t,x,\theta)
\label{susyqm3.20??}
\end{equation}
with
$v(0,x,\theta)={\underline{v}}(x,\theta)$.
Here, we put $\omega=\sum_{j=1}^d\omega_j$.

\begin{problem}
Extend the procedure in this chapter to the operator posed by M.S. Abdalla and U.A.T. Ramjit~\cite{AR89}:
$$
H(q,p,t)=\frac{1}{2m(t)}p^2+\frac{m(t)}{2}\omega_0^2q^2, \quad m(t)=m_0e^{2\Gamma(t)},
$$
Show the difference of solution of above operator  from harmonic oscillator represented by $\Gamma(t)$ and how it changes adding spin?
\end{problem}

\section{A simple example of supersymmetric extension}

We consider the simplest $1$-dimensional example.
Let 
\begin{equation}
H(q,p)=\frac 1{2}(p-A(q))^2+V(q)\in C^\infty(T^*\euc:\euc)
\label{susyqm4.1}
\end{equation}
be given with $A(q),V(q)\in C^\infty(\euc:\euc)$. 
Using Legendre transformation as
\begin{equation}
\dot q=\frac{\partial H}{\partial p}=p-A(q), \quad
L(q,\dot q)=\dot q p-H(q,p),
\label{susyqm4.2}
\end{equation}
we get a Lagrangian
$$
L(q,\dot q)=\frac 12 \dot q^2+A(q)\dot q-V(q)\in C^\infty(T\euc:\euc).
\label{susyqm4.3}
$$

Instead of a path $q:\euc\ni t\to q(t)\in\euc$, we consider
a generalized path
\begin{equation}
\Phi:\euc\ni t\to \Phi(t)=x(t)
+i(\rho_1\psi_2(t)-\rho_2\psi_1(t))+i\rho_1\rho_2F(t)\in\rev
\label{susyqm4.4}
\end{equation}
with $\rho_\alpha\in\rod$ $(\alpha=1,2)$ being odd parameters and
\begin{equation}
\left\{
\begin{aligned}
&x:\euc\ni t\to x(t)\in\rev,\\
&\psi_\alpha:\euc\ni t\to \psi_\alpha(t)\in\rod
\with \alpha=1,2,\\
&F:\euc\ni t\to F(t)\in\rev.
\end{aligned}
\right.
\label{susyqm4.5}
\end{equation}

Introducing operators
$$
{\mathcal{D}}_\alpha={\frac{\partial}{\partial\rho_\alpha}}
-i\rho_\alpha\pdt  \with \alpha=1,2,
$$
and $\epsilon_{\alpha\beta}=-\epsilon_{\beta\alpha}$, $\epsilon_{12}=1$,
we extend $L(q,\dot q)$ as
\begin{equation}
\tilde{\mathcal{L}}_0
=-\frac 14 
({\mathcal{D}}_\alpha\Phi)\epsilon_{\alpha\beta}({\mathcal{D}}_\beta\Phi)
+\frac i2  A \rho_\alpha\epsilon_{\alpha\beta}{\mathcal{D}}_\beta\Phi
-iW(\Phi).
\label{susyqm4.6}
\end{equation}
In the above,
$A(q)$ is extended from $q\in\euc$
to $\Phi=x+i(\rho_1\psi_2-\rho_2\psi_1)+i\rho_1\rho_2 F\in \rev$ by he Grassmann extension
as
\begin{equation}
A(\Phi)
=A(x)+iA'(x)(\rho_1\psi_2-\rho_2\psi_1+\rho_1\rho_2 F)
+A''(x)\rho_1\rho_2\psi_1\psi_2.
\label{susyqm4.7}
\end{equation}
and $W(\Phi)$ is analogously extended from $W(q)$ to $W(\Phi)$
whose relation to $V(q)$ will be given later.
\begin{remark} The following relation will be worth noticing:
\begin{equation}
\left({\frac{\partial}{\partial\rho_\alpha}}
-i\rho_\alpha\pdt\right)^2=-i\pdt \quad\text{for each $\alpha=1,2$.}
\end{equation}
\end{remark}

Now, we have
\begin{equation}
\begin{aligned}
{\mathcal{L}}'_0
&\defeq\int d\rho_2 d\rho_1\, \tilde{\mathcal{L}}_0\\
&=\frac 12 \dot x^2+A(x)\dot x+\frac12 F^2
+\frac i2\left(\psi_2\dot\psi_2-\dot\psi_1\psi_1\right)
+W'(x)F-iW''(x)\psi_1\psi_2.
\end{aligned}
\label{susyqm4.8}
\end{equation}
Assuming that the ``auxilliary field $F$" should satisfy
\begin{equation}
0=\frac{\delta {\mathcal{L}}'_0}{\delta F}=F+W',
\label{susyqm4.9}
\end{equation}
we arrived at
\begin{equation}
{\mathcal{L}}_0=\frac 12\dot x^2+A(x)\dot x
+\frac i2\left(\psi_2\dot\psi_2-\dot\psi_1\psi_1\right)
-\frac 12 W'(x)^2-iW''(x)\psi_1\psi_2.
\label{susyqm4.10}
\end{equation}
This is the desired Lagrangian 
with variables $x,\dot x,\psi_\alpha,\dot\psi_\alpha$, but variables
$\psi_\alpha,\dot\psi_\alpha$ are not independent each other.
In fact, they satisfy
$$
\{\psi_\alpha,\psi_\beta\}=\psi_\alpha\psi_\beta+\psi_\beta\psi_\alpha=0, \quad
\{\psi_\alpha,\dot\psi_\beta\}=0 \et
\{\dot\psi_\alpha,\dot\psi_\beta\}=0.
$$

To find out ``independent Grassmann variables" in \eqref{susyqm4.10}, 
we introduce new variables by the following two methods:

(I) Defining new variables as
\begin{equation}
\left\{
\begin{aligned}
&\xi=\frac{\delta{\mathcal{L}}_0}{\delta \dot x}=\dot x+ax,\\
&\phi_\alpha=\frac{\delta{\mathcal{L}}_0}{\delta \dot \psi_\alpha}=-\frac i2\psi_\alpha
\for \alpha=1,2,
\end{aligned}
\right.
\label{susyqm4.11}
\end{equation}
we put
$$
\begin{aligned}
{\mathcal{H}}(x,\xi,\psi_1,\psi_2)&=\dot x\xi+\dot\psi_\alpha\phi_\alpha-{\mathcal{L}}_0\\
&=\frac 12(\xi-ax)^2+\frac 12 W'(x)^2
+\frac i2 W''(x)\psi_\alpha\epsilon_{\alpha\beta}\psi_\beta.
\end{aligned}
$$
Rewriting the variables $\psi_1,\psi_2$ as $\theta,\pi$, respectively, we get
\begin{equation}
{\mathcal{H}}(x,\xi,\theta,\pi)
=\frac 12(\xi-ax)^2+\frac 12 W'(x)^2
+i W''(x)\theta\pi.
\label{susyqm4.12}
\end{equation}

(II) In the above, we use the ``real" odd variables $\psi_\alpha$.
We ``complexify" these variables by putting
\begin{equation}
\psi=\frac 1{\sqrt 2}(\psi_1+i\psi_2), \quad
\bar\psi=\frac 1{\sqrt 2}(\psi_1-i\psi_2),
\quad i.e.\quad
\psi_1=\frac 1{\sqrt 2}(\psi+\bar\psi), \quad
\psi_2=\frac 1{\sqrt 2 i}(\psi-\bar\psi),
\label{susyqm4.13}
\end{equation}
and then we rewrite ${\mathcal{L}}_0$ as
\begin{equation}
\bar{\mathcal{L}}_0=\frac 12\dot x^2+A(x)\dot x
+\frac i2 (\psi\dot{\bar\psi}+\bar\psi\dot\psi)
-\frac 12 W'(x)^2-W''(x)\bar\psi\psi.
\label{susyqm4.14}
\end{equation}
Introducing new variables as
\begin{equation}
\xi=\frac{\delta \bar{\mathcal{L}}_0}{\delta \dot x}=\dot x+A(x),\quad
\phi=\frac{\delta \bar{\mathcal{L}}_0}{\delta \dot\psi}=-\frac i2\bar\psi,\quad
\bar\phi=\frac{\delta \bar{\mathcal{L}}_0}{\delta \dot{\bar\psi}}=-\frac i2\psi,
\label{susyqm4.15}
\end{equation}
we put
$$
\begin{aligned}
{\mathcal{H}}(x,\xi,\psi,\bar\psi)
&\defeq\dot x\xi+\dot\psi\phi+\dot{\bar\psi}\bar\phi
-\bar{\mathcal{L}}_0\\
&=\frac 12(\xi-A(x))^2+\frac 12 W'(x)^2+W''(x)\bar\psi\psi.
\end{aligned}
$$
Rewriting $\psi$ and $\bar\psi$ by $\theta$ and $\pi$, respectively, 
we get finally a function
\begin{equation}
{\mathcal{H}}(x,\xi,\theta,\pi)
=\frac 12(\xi-A(x))^2+\frac 12 W'(x)^2-W''(x)\theta\pi
\in {\mathcal{C}}_{S\!S}({\mathfrak{R}}^{2|2}:\rev).
\label{susyqm4.16}
\end{equation}
Here, $(x,\theta)\in {\mathfrak{R}}^{1|1}$, $(\xi,\pi)\in {\mathfrak{R}}^{1|1}$.

\begin{remark}
(0) The difference between \eqref{susyqm4.12} and \eqref{susyqm4.16} is the existence of $i$ in front of
the term $ W''(x)\theta\pi$.
This difference is rather significant when we consider Witten index 
for supersymmetric quantum mechanics
using the kernel representation of the corresponding evolution operator.
\newline
(1) As there is no preference at this stage to take $\pi$ and $\theta$
instead of $\theta$ and $\pi$, there is no significance of the sign $\pm$
in front of the terms $iW''(x)\theta\pi$ in \eqref{susyqm3.12} or $W''(x)\theta\pi$ in \eqref{susyqm4.16}
in these cases.
\newline
(2) We may regard ${\mathcal{H}}(x,\xi,\theta,\pi)$ as a Hamiltonian in
${\mathcal{C}}_{S\!S}(T^*{\mathfrak{R}}^{1|1}:\rev)$.
\newline
(3) These Hamiltonians \eqref{susyqm4.12} and \eqref{susyqm4.16} are called supersymmetric extensions of \eqref{susyqm4.1}
because they give supersymmetric quantum mechanics after quantization (see \S4).
The procedure above is author's unmatured understanding of amalgam of physics papers
such as
F. Cooper and B. Freedman~\cite{CF}, A.C. Davis, A.J. Macfarlane, P.C. Popat and T.W. van Holten~\cite{D-M-P-V} etc.
But supersymmetry in superspace ${\fR}^{m|n}$ will be studied separately.
\newline
(4) On the other hand, using the identification (1.13), we have
$$
{\mathcal{H}}^\hbar_{\pm}
(x,\partial_x,\theta,\partial_\theta)=\#
\begin{pmatrix} H^\hbar_{\pm} -\frac \hbar{2}{b} & 0\\
0& H^\hbar_{\pm} +\frac \hbar{2}{b}
\end{pmatrix}
\flat=\#{\mathbf H}^{\hbar,b}_{\pm}\flat .
$$
Moreover, in this case, the ``complete Weyl symbol of 
the above ${\mathcal{H}}^\hbar(x,\partial_x,\theta,\partial_\theta)$" 
is calculated by
\begin{equation}
{\mathcal{H}}_\pm(x,\xi,\theta,\pi)
=(e^{-i\hbar^{-1}(x\xi+\theta\pi)}{\mathcal{H}}^\hbar_{\pm}  %{\mathbf H}^\hbar
(x,\partial_x,\theta,\partial_\theta)
e^{i\hbar^{-1}(x\xi+\theta\pi)}\Big|_{\hbar=0}
=\frac 12 (\xi-ax)^2\pm\frac 12 b^2x^2+{i}b\theta\pi.
\label{susyqm4.17}
\end{equation}
${\mathcal{H}}_+$ equals to \eqref{susyqm4.12} when $A(q)=aq$ and $W(x)=\frac 12 bx^2$,
and ${\mathcal{H}}_-$ is obtained from \eqref{susyqm4.16} with 
$A(q)=aq$ and $W(x)=-\frac i2 bx^2$.
These give the relation between $W(q)$ and $V(q)$.
(See SUSYQM defined in \S 1.)
\newline
(5) Witten~\cite{witt82-1} considered as a quantum mechanical operator
\begin{equation}
{\mathbf H}(q,\partial_q)=\left(-\frac 12 \partial_q^2+v(q)\right)
\begin{pmatrix}
1&0\\
0&1
\end{pmatrix}
-\frac 12 w(q)
\begin{pmatrix}
1&0\\
0&-1
\end{pmatrix}.
\label{susyqm4.18}
\end{equation}
This operator is supersymmetric when there exists a function $\psi(q)$
such that
$$
v(q)=\frac 12 \psi'(q)^2,\quad w(q)=\psi''(q).
$$
\end{remark}

\begin{problem}
Though a trial to prove Atiyah-Singer index theorem applying super analysis by S. Rempel and T. Schmitt~\cite{RS83}, was informed by A. Rogers long-time ago, but I feel shame that I haven't comprehend well yet. 
Moreover, because I have stumbled before appreciate the naturalness of definition of  weights of Douglis-Nirenberg for system of PDE, therefore I'm far from Rempel and Schmitt's reformulation.
\par
More recently, I find the paper of F.F. Voronov ~\cite{vor93} though I haven't appreciated it yet. Because Voronov~\cite{Vor91, vor93} uses Banach-Grassmann algebra, therefore besides algebraic calculation, estimates by inequalities seems far from my understanding.
\end{problem}

\begin{problem}
The proofs of ``analytic torsion=Reidemeister torsion'' by  J. Cheeger~\cite{che79},  D. Burghelea, L. Friedlander and T. Kappeler~\cite{BFK98} are given. Reprove this in our context.
\end{problem}
%\end{comment}

\chapter{Miscellaneous}
\section[Proof of Berezin's Theorem~5.2.1]{Proof of Berezin's Theorem~\ref{Ber-naive}}%[
To be self-contained, we give a precise proof following F.A. Berezin~\cite{Ber87} and A. Rogers~\cite{Rog07} because their proofs are not so easy to understand at least for a tiny little old mathematician.

First of all, we prepare
\begin{lemma}\label{primitiveFubini}
Let $u(x,\theta)=\sum_{|a|\le n}\theta^a u_a(x)$ be supersmooth on ${\mathfrak{U}}={\mathfrak{U}}_{\mathrm{ev}}\times\rod^n$.
If $\int_{{\mathfrak{U}}_{\mathrm{ev}}} dx \,u_a(x)$ exists for each $a$,  then
we have
$$
{\Biint}_{\mathfrak{U}} dxd\theta\,u(x,\theta)=\int_{{\mathfrak{U}}_{\mathrm{ev}}} dx\bigg[\int_{\rod^n} d\theta \,u(x,\theta)\bigg]
=\int_{\rod^n} d\theta\bigg[\int_{{\mathfrak{U}}_{\mathrm{ev}}} dx \,u(x,\theta)\bigg].
$$
\end{lemma}
\par{\it Proof}. By the primitive definition of integral, we have
$$
{\Biint}_{\mathfrak{U}} dxd\theta\,u(x,\theta)=\int_{{\mathfrak{U}}_{\mathrm{ev}}} dx\bigg[\int_{\rod^n}  d\theta \,u(x,\theta)\bigg]
=\int_{{\mathfrak{U}}_{\mathrm{ev}}} dx \,u_{\tilde1}(x),
$$
and
$$
\int_{\rod^n}  d\theta\sum_{|a|\le n}\bigg[\int_{{\mathfrak{U}}_{\mathrm{ev}}} dx\, \theta^a\,u_a(x)\bigg]
=\int_{\rod^n}  d\theta\sum_{|a|\le n}\theta^a\bigg[\int_{{\mathfrak{U}}_{\mathrm{ev}}} dx\, u_a(x)\bigg]=\int_{{\mathfrak{U}}_{\mathrm{ev}}} dx\, u_{\tilde1}(x). \qquad\qed
$$

(I) Now, we consider a simple case:
Let a linear coordinate change be given by
$$
(x,\theta)=(y,\omega)M,\quad M=\begin{pmatrix}
A&C\\
D&B
\end{pmatrix},
$$
that is,  
$$
x_i=\sum_{k=1}^m y_kA_{ki}+\sum_{\ell=1}^n \omega_\ell D_{\ell i}
=x_i(y,\omega),\quad 
\theta_j=\sum_{k=1}^m y_k C_{kj}+\sum_{\ell=1}^n \omega_\ell B_{\ell j}
=\theta_j(y,\omega)
$$
with $A_{ki},B_{\ell j}\in\cev$ and $C_{\ell i}, D_{kj}\in\cod$,
and we have 
\begin{equation}
\sdet\bigg(\frac{\partial(x,\theta)}{\partial(y,\omega)}\bigg)=\det A\,
{\det}^{-1}(B-DA^{-1}C)={\det}(A-CB^{-1}D){\det}^{-1}B={\sdet}M.
\label{sdet}
\end{equation}

Interchanging the order of integration, putting $\omega^{(1)}=\omega B$ and $y^{(1)}=yA$, we get
$$
\begin{aligned}
{\Biint} dyd\omega\,&u(y A+ \omega D,y C+\omega B)
=\int  dy \big[\int d\omega\,u(y A+ \omega D,y C+\omega B)\big]\\%\;(y\to Ay,\, \omega\to B\omega)\\
&=\int dy\big[\int d\omega^{(1)}\,\det B{\cdot} u(y A+ \omega^{(1)} B^{-1} D,y C+\omega^{(1)})\big]\\
&=\int d\omega^{(1)}\,\det B\big[\int dy\,u(y A+ \omega^{(1)} B^{-1} D,y C+\omega^{(1)})\big]\\
&=\int d\omega^{(1)}\,\det B\big[\int dy^{(1)}\,\det A^{-1}{\cdot}u(y^{(1)}+ \omega^{(1)} B^{-1} D,y^{(1)}A^{-1}C+\omega^{(1)})\big],
\end{aligned}
$$
that is, since
$$
\frac{\partial(y,\omega)}{\partial(y^{(1)}, \omega^{(1)})}=\begin{pmatrix}
A^{-1}&0\\
0&B^{-1}
\end{pmatrix},\quad
\sdet\bigg(\frac{\partial(y,\omega)}{\partial(y^{(1)}, \omega^{(1)})}\bigg)=\det A^{-1}{\cdot}\det B,
$$
we have
\begin{equation}
\begin{aligned}
{\Biint} dyd\omega\, &u(y A+ \omega D,y C+\omega B)\\
&=
{\Biint} dy^{(1)} d\omega^{(1)}\,
\sdet\bigg(\frac{\partial(y,\omega)}{\partial(y^{(1)}, \omega^{(1)})}\bigg)
u(y^{(1)}+ \omega^{(1)} B^{-1} D,y^{(1)}A^{-1}C+\omega^{(1)}).
\end{aligned}
\label{LCV1}
\end{equation}

Analogously, using Lemma \ref{primitiveFubini} and by introducing change of variables as
$$
y^{(2)}=y^{(1)},\omega^{(2)}=\omega^{(1)}+y^{(1)}A^{-1}C\Longrightarrow
\sdet\bigg(\frac{\partial(y^{(1)},\omega^{(1)})}{\partial(y^{(2)}, \omega^{(2)})}\bigg)
={\sdet
\begin{pmatrix}
1&-A^{-1}C\\
0&1
\end{pmatrix}}
=1,
$$
we get
\begin{equation}
\begin{aligned}
{\Biint} dy^{(1)}& d\omega^{(1)}\,  %\sdet\bigg(\frac{\partial(y,\omega)}{\partial(y^{(1)}, \omega^{(1)})}\bigg) 
u(y^{(1)}+\omega^{(1)}B^{-1}D,y^{(1)}A^{-1}C+\omega^{(1)})\\
&=
{\Biint}dy^{(2)}d\omega^{(2)}\,\sdet\bigg(\frac{\partial(y^{(1)},\omega^{(1)})}{\partial(y^{(2)}, \omega^{(2)})}\bigg)
u(y^{(2)}+(\omega-y^{(2)}A^{-1}C)B^{-1}D,\omega^{(2)}).
\end{aligned}
\label{LCV2}
\end{equation}

Then by
$$
\begin{aligned}
&y^{(3)}=y^{(2)}(1-A^{-1}CB^{-1}D),\, \omega^{(3)}=\omega^{(2)}\\
&\Longrightarrow
\sdet\bigg(\frac{\partial(y^{(2)},\omega^{(2)})}{\partial(y^{(3)}, \omega^{(3)})}\bigg)
={\sdet
\begin{pmatrix}
(1-A^{-1}CB^{-1}D)^{-1}&0\\
0&1
\end{pmatrix}}=\det{}^{-1} (1-A^{-1}CB^{-1}D),
\end{aligned}
$$
we have
\begin{equation}
\begin{aligned}
{\Biint} dy^{(2)}d\omega^{(2)}\,& %\sdet\bigg(\frac{\partial(y^{(1)},\omega^{(1)})}{\partial(y^{(2)}, \omega^{(2)})}\bigg)
u(y^{(2)}+(\omega-y^{(2)}A^{-1}C)B^{-1}D,\omega^{(2)})\\
&= %\det{}^{-1} (1-A^{-1}CB^{-1}D)
{\Biint} dy^{(3)}d\omega^{(3)}\,\sdet\bigg(\frac{\partial(y^{(2)},\omega^{(2)})}{\partial(y^{(3)}, \omega^{(3)})}\bigg)
u(y^{(3)}+\omega^{(3)}B^{-1}D,\omega^{(3)}).
\end{aligned}
\label{LCV3}
\end{equation}

Finally by
$$
x=y^{(3)}+\omega^{(3)}B^{-1}D,\,\theta=\omega^{(3)}\Longrightarrow
\sdet\bigg(\frac{\partial(y^{(3)},\omega^{(3)})}{\partial(x, \theta)}\bigg)
={\sdet
\begin{pmatrix}
1&0\\
-B^{-1}D&1
\end{pmatrix}}=1,
$$
using $\det B\,{\det}^{-1}(A-CB^{-1}D){\cdot}
(\det A\,{\det}^{-1}(B-DA^{-1}C))=1$ from \eqref{sdet},  we have,
\begin{equation}
{\Biint} dyd\omega\,u(y A+ \omega D,y C+\omega B)
=\sdet M^{-1}{\cdot} %\det B\,{\det}^{-1}(A-CB^{-1}D)
{\Biint} dxd\theta\,
\sdet\bigg(\frac{\partial(y^{(3)},\omega^{(3)})}{\partial(x, \theta)}\bigg)
u(x,\theta).
\label{LCV4}
\end{equation}

\begin{remark}
For the linear change of  variables, it is not necessary to assume the compactness of support for integrand using primitive definition of integration.
\end{remark}
(II) (ii-a) If $H_1$ and $H_2$ are superdiffeomorphisms of open subsets of $\supermn$ with the image of $H_1$ 
equals to the domain of $H_2$, then
$$
\Ber(H_1){\cdot}\Ber(H_2)=\Ber(H_2\circ H_1) \where \Ber(H)(y,\omega)=\sdet J(H)(y,\omega).
$$
Here, for $H(y,\omega)=(x_k(y,\omega),\theta_l(y,\omega)):\supermn\to\supermn$, we put 
$$
J(H)(y,\omega)=\begin{pmatrix}
\frac{\partial x_k(y,\omega)}{\partial y_i}&\frac{\partial \theta_l(y,\omega)}{\partial y_i}\\
\frac{\partial x_k(y,\omega)}{\partial \omega_j}&\frac{\partial \theta_l(y,\omega)}{\partial \omega_j}
\end{pmatrix}=\frac{\partial(x,\theta)}{\partial(y,\omega)}.
$$
(ii-b) Any  superdiffeomorphism of an open subset of $\supermn$ may be decomposed as $H=H_2\circ H_1$ where
\begin{equation}
\left\{
\begin{aligned}
&H_1(y,\omega)=(h_1(y,\omega),\omega)=(\tilde{y},\tilde{\omega})\with h_1:\supermn\to{\fR}^{m|0},\\
&H_2(\tilde{y},\tilde{\omega})=(\tilde{y},h_2(\tilde{y},\tilde{\omega}))\with h_2:\supermn\to{\fR}^{0|n}.
\end{aligned}
\right.
\label{20100413-1}
\end{equation}
\begin{remark}
(i) If $H(y,\omega)=(h_1(y,{\omega}),h_2(y,\omega))$ is given by $h_1(y,\omega)=yA+{\omega} D$ 
and $h_2(y,\omega)=yC+{\omega} B$ as above,
putting $H_1(y,\omega)=(yA+{\omega} D,\omega)=(\tilde{y},{\omega})$ and $H_2(\tilde{y},{\omega})=(\tilde{y},\tilde{y}A^{-1}C+{\omega}(B-DA^{-1}C))$, we have $H=H_2\circ H_1$.
In this case, we rewrite the procedures \eqref{LCV1}--\eqref{LCV4} as 
$$
\begin{aligned}
{\Biint} & dyd\omega\,u(yA+{\omega}D, yC+{\omega}B)\\
&={\Biint} d\tilde{y}d\tilde{\omega}\,\sdet\bigg(\frac{\partial(y,\omega)}{\partial(\tilde{y},{\omega})}\bigg)
u(\tilde{y},(\tilde{y}-{\omega}D)A^{-1}C+{\omega}B)\with \tilde{y}=yA+{\omega}D\\
&=\det A^{-1}{\cdot}{\Biint}  d{x}d{\theta}\,\sdet\bigg(\frac{\partial(\tilde{y},{\omega})}{\partial(x,\theta)}\bigg)u(x,\theta)
\with x=\tilde{y},\;\theta=\tilde{y}A^{-1}C+{\omega}(B-DA^{-1}C)\\
&=\det A^{-1}{\cdot}{\det}(B-DA^{-1}C){\cdot}{\Biint} dxd\theta\,u(x,\theta).
\end{aligned}
$$
(ii) Analogously, putting $H_1(y,\omega)=(y,yC+{\omega}B)=(y,{\theta})$ and 
$H_2(y,{\theta})=(y(A-CB^{-1}D)+{\theta}B^{-1}D,{\theta})$, 
we have $H=H_2\circ H_1$, and
$$
\begin{aligned}
{\Biint} & dyd\omega\,u(yA+{\omega}D,yC+{\omega}B)\\
&={\Biint} dyd{\theta}\,\sdet\bigg(\frac{\partial(y,\omega)}{\partial(y,{\theta})}\bigg)
u(y(A-CB^{-1}D)+{\theta}B^{-1}D, {\theta})\with  {\theta}=yC+{\omega}B\\
&=\det B{\cdot}{\Biint}  d{x}d{\theta}\,\sdet\bigg(\frac{\partial(y,{\theta})}{\partial(x,\theta)}\bigg)u(x,\theta)
\with x=y(A-CB^{-1}D)+CB^{-1}{\theta}\\
&=\det B{\cdot}{\det}^{-1}(A-CB^{-1}D){\cdot}{\Biint} dxd\theta\,u(x,\theta).
\end{aligned}
$$
\end{remark}
(iii) For any given superdiffeomorphism $H(y,\omega)=(h_1(y,\omega),h_2(y,\omega))$, defining
$H_1(y,\omega)=(h_1(y,\omega),\omega)=(\tilde{y},{\omega})$ and 
$H_2(\tilde{y},{\omega})=(\tilde{y},\tilde{h}_2(\tilde{y},{\omega}))$
such that $h_2(y,\omega)=\tilde{h}_2(h_1(y,\omega),\omega)$, we have $H=H_2\circ H_1$.
Using the inverse function $y=g(\tilde{y},\omega)$ of $\tilde{y}=h_1(y,\omega)$, we put
$\tilde{h}_2(\tilde{y},\omega)=h_2(g(\tilde{y},\omega),\omega)$. We denote
$h_1(y,\omega)=(h_{1a}(y,\omega))=(h_{11},{\cdots}, h_{1n})$, etc.
Then, for $k, \ell=1,{\cdots},n$,
$$
\frac{\partial \tilde{h}_{2{\ell}}}{\partial \omega_k}=\frac{\partial h_{2{\ell}}}{\partial \omega_k}
+\sum_{i=1}^m\frac{\partial g_i}{\partial \omega_k}\frac{\partial h_{2{\ell}}}{\partial y_i}
$$
with
$$
0=\frac{\partial \tilde{y}_j}{\partial \omega_k}=\frac{\partial h_{1j}(g(y,\omega),\omega)}{\partial \omega_k}
=\sum_{j=1}^m\frac{\partial g_i}{\partial \omega_k}\frac{\partial h_{1j}}{\partial y_i}+\frac{\partial h_{1j}}{\partial \omega_k},
$$
we get
$$
\frac{\partial \tilde{h}_{2{\ell}}}{\partial \omega_k}=\frac{\partial h_{2{\ell}}}{\partial \omega_k}
-\sum_{i,j=1}^m\frac{\partial h_{1j}}{\partial \omega_k}\bigg(\frac{\partial h_{1j}}{\partial y_i}\bigg)^{-1}\frac{\partial h_{2{\ell}}}{\partial y_i}.
$$

Therefore, %taking the transposition of matrices, we have
$$
\Ber H=\sdet
\begin{pmatrix}
\frac{\partial h_1}{\partial y}&\frac{\partial h_2}{\partial y}\\
\frac{\partial h_1}{\partial \omega}&\frac{\partial h_2}{\partial \omega}
\end{pmatrix}
=\det \frac{\partial h_1}{\partial y}{\cdot}
\det{}^{-1}\bigg(
\frac{\partial h_2}{\partial \omega}
-\frac{\partial h_1}{\partial \omega}\bigg(\frac{\partial h_1}{\partial y}\bigg)^{-1}\frac{\partial h_2}{\partial y}\bigg)
=\det \frac{\partial h_1}{\partial y}{\cdot}\det{}^{-1}\frac{\partial \tilde{h}_2}{\partial \omega}.
$$
(III) For each type of superdiffeomorphisms $H_1$ and $H_2$, we prove the formula.\\
(III-1) Let $H(y,\omega)=(h(y,\omega),\omega)$ where $h=(h_j)_{j=1}^m:\supermn\to{\fR}^{m|0}$.
Then it is clear that
$$
\Ber(H)(y,\omega)=\det\bigg(\frac{\partial h_j(y,\omega)}{\partial y_i}\bigg)
=\sum_{\sigma\in{\wp}_m}\sgn(\sigma)\prod_{i=1}^m\frac{\partial h_{\sigma(i)}(y,\omega)}{\partial y_i}.
$$
For any $u(x,\theta)=\sum_{|a|\le n}\theta^a u_a(x)$, % with $f_a(q)$ having compact support,
we put
$$
{\Biint}_{\!\!\!{\mathfrak{U}}}dx d\theta\,u(x,\theta)=\int_{{\mathfrak{U}}_{\mathrm{ev, B}}}dx\bigg(\int_{{\fR}^{0|n}}d\theta\,u(x,\theta)\bigg)
=\int_{{\mathfrak{U}}_{\mathrm{ev, B}}}dx u_{\tilde 1}(x).
$$
On the other hand, we have
\begin{equation}
\begin{aligned}
{\Biint}_{\!\!\!\mathfrak{V}}& dyd\omega\,\Ber(H)(y,\omega)(u\circ H)(y,\omega)\\
&=\int_{\pi_{\mathrm{B}}(\mathfrak{V})}dy \,\frac{\partial}{\partial\omega_n}{\cdots}\frac{\partial}{\partial\omega_1}
\bigg(\det\bigg(\frac{\partial h_j(y,\omega)}{\partial y_i}\bigg)u(h(y,\omega),\omega)\bigg)\bigg|_{\omega=0}\\
&=\int_{\pi_{\mathrm{B}}(\mathfrak{V})}dy
\bigg(\det\bigg(\frac{\partial h_j(y,0)}{\partial y_i}\bigg) u_{\tilde 1}(h(y,0))\bigg)\\
&\qquad\qquad+\int_{\pi_{\mathrm{B}}(\mathfrak{V})}dy\,\frac{\partial}{\partial\omega_n}{\cdots}\frac{\partial}{\partial\omega_1}
\bigg(\sum_{|a|<n}\omega^a u_a(h(y,\omega))\det\bigg(\frac{\partial h_j(y,\omega)}{\partial y_i}\bigg)\bigg)\bigg|_{\omega=0}.
\end{aligned}
\label{PCV}
\end{equation}
Applying the standard integration on $\euc^m$ to the first term of the rightest hand side above, we have
$$
\int_{\pi_{\mathrm{B}}(\mathfrak{V})}dy\bigg( \det\bigg(\frac{\partial h_j(y,0)}{\partial y_i}\bigg) u_{\tilde 1}(h(y,0))\bigg)
=\int_{{\mathfrak{U}}_{\mathrm{ev, B}}}dx\,u_{\tilde 1}(x)\where {\mathfrak{U}}=H(\mathfrak{V}).
$$

\begin{claim}
The second term of the right hand side of \eqref{PCV} 
equals to the total derivatives of even variables.
More precisely, we have, for $u(x,\theta)=\sum_{|a|\le n}\theta^au_a(x)$,
$$
%\begin{aligned}
\frac{\partial}{\partial\omega_n}{\cdots}\frac{\partial}{\partial\omega_1}
\bigg(\sum_{|a|< n}\omega^a u_a(h(y,\omega))\Ber(H)(y,\omega)\bigg)\bigg|_{\omega=0}
%\\&=\sum_{\scriptstyle{|a+b|=n, |a|<n}\atop\scriptstyle{|b|={\mathrm{even}}}}
%(-1)^{\tau(a,b)}\partial_{\omega}^b[ f_a(h(y,\omega))\Ber(H)(y,\omega)]\big|_{\omega=0}
=\sum_{j=1}^m\frac{\partial}{\partial y_j}(*)
%\end{aligned}
$$
\end{claim}
%\newpage

As $h_j(y,\omega)\in\rev$, we have
$$
\begin{aligned}
&h_j(y,\omega)=h_{j\tilde{0}}(y)+\sum_{|c|={\mathrm{ev}}\ge2}\omega^c h_{j,c}(y),\\
&u_a(h(y,\omega))=u_a(h_{\tilde{0}}(y))+\sum_{|c|={\mathrm{ev}}\ge2}\omega^c h_{j,c}(y)u_{a,x_j}(h_{\tilde{0}}(y))
+\sum_{|\alpha|\ge2}\frac{\partial_x^{\alpha}u_a(h_{\tilde{0}}(y))}{\alpha!}(\sum_{|c|={\mathrm{ev}}\ge2}\omega^c h_{j,c}(y))^{\alpha},\\
&
{\begin{aligned}
\Ber(H)(y,\omega)&=\det\bigg(\frac{\partial h_j(y,\omega)}{\partial y_i}\bigg)
=\sum_{\sigma\in{\wp}_m}\sgn(\sigma)\prod_{i=1}^m\frac{\partial h_{\sigma(i)}(y,\omega)}{\partial y_i}\\
&=\det\bigg(\frac{\partial h_{j,\tilde{0}}(y)}{\partial y_i}\bigg)
+\sum_{\sigma\in{\wp}_m}\sgn(\sigma)
\sum_{j=1}^m\sum_{|c|={\mathrm{ev}}\ge2}\omega^c \frac{\partial h_{\sigma(j),c}(y)}{\partial y_j}
\prod_{i=1,i\neq j}^m\frac{\partial h_{\sigma(i),\tilde{0}}(y)}{\partial y_i}\\
&\qquad
+\sum_{\sigma\in{\wp}_m}\sgn(\sigma)
\sum_{j,k=1}^m\sum_{\scriptstyle{|c_j|={\mathrm{ev}}}\atop\scriptstyle{|c_1+c_2|=|c|\ge 4}}
\omega^c \frac{\partial h_{\sigma(j),c_1}(y)}{\partial y_j}
\frac{\partial h_{\sigma(k),c_2}(y)}{\partial y_k}
\prod_{i=1,i\neq j,k}^m\frac{\partial h_{\sigma(i),\tilde{0}}(y)}{\partial y_i}
+\mbox{etc}.
\end{aligned}}
\end{aligned}
$$

Putting $\tilde{1}-a=b$ or $=c_1+c_2$, $=c_1+c_2+c_3$, etc, we have
\begin{equation}
\mbox{the coefficient of $\omega^b$ of}\, f_a(h(y,\omega))\Ber(H)(y,\omega)={\mathrm{I}}+{\mathrm{II}}+{\mathrm{III}}
\label{100112}
\end{equation}
where
$$
\begin{aligned}
{\mathrm{I}}&=\sum_{j=1}^m h_{j,b}(y)u_{a,x_j}(h_{\tilde{0}}(y))\sum_{\sigma\in{\wp}_m}\sgn(\sigma)\prod_{i=1}^m\frac{\partial h_{\sigma(i),\tilde{0}}(y)}{\partial y_i},\\
{\mathrm{II}}&=u_a(h_{\tilde{0}}(y))
\sum_{\sigma\in{\wp}_m}\sgn(\sigma)
\sum_{j=1}^m \frac{\partial h_{\sigma(j),b}(y)}{\partial y_j}
\prod_{i=1,i\neq j}^m\frac{\partial h_{\sigma(i),\tilde{0}}(y)}{\partial y_i},\\
{\mathrm{III}}&=u_a(h_{\tilde{0}}(y))
\sum_{\sigma\in{\wp}_m}\sgn(\sigma)
\sum_{j,k=1}^m \sum_{b=c_1+c_2}\frac{\partial h_{\sigma(j),c_1}(y)}{\partial y_j}\frac{\partial h_{\sigma(k),c_2}(y)}{\partial y_k}
\prod_{i=1,i\neq j,k}^m\frac{\partial h_{\sigma(i),\tilde{0}}(y)}{\partial y_i}+\mbox{etc}.
\end{aligned}
$$
The term ${\mathrm{II}}$ is calculated as
$$
{\mathrm{II}}=
\sum_{j=1}^m \frac{\partial}{\partial y_j}\bigg[f_a(h_{\tilde{0}}(y))
\sum_{\sigma\in{\wp}_m}\sgn(\sigma)
h_{\sigma(j),b}(y)\prod_{i=1,i\neq j}^m\frac{\partial h_{\sigma(i),\tilde{0}}(y)}{\partial y_i}\bigg]-A-B
$$
where
$$
\begin{aligned}
A&=
\sum_{j=1}^m \bigg(\sum_{k=1}^m\frac{\partial h_{k,\tilde{0}}(y)}{\partial y_j}u_{a,x_k}(h_{\tilde{0}}(y))\bigg)
\sum_{\sigma\in{\wp}_m}\sgn(\sigma)h_{\sigma(j),b}(y)\prod_{i=1,i\neq j}^m\frac{\partial h_{\sigma(i),\tilde{0}}(y)}{\partial y_i},\\
B&=
\sum_{j=1}^m u_a(h_{\tilde{0}}(y))
\sum_{\sigma\in{\wp}_m}\sgn(\sigma)
h_{\sigma(j),b}(y)\frac{\partial}{\partial y_j}\bigg(\prod_{i=1, i\neq j}^m\frac{\partial h_{\sigma(i),\tilde{0}}(y)}{\partial y_i}\bigg).
\end{aligned}
$$
We want to prove
\begin{claim}
(i) $A={\mathrm{I}}$, (ii) $B=0$ and (iii) ${\mathrm{III}}=0$.
\end{claim}
(i) To prove $A={\mathrm{I}}$, for each $k=1,{\cdots},m$, we take all sums w.r.t. $\sigma\in{\wp}_m$ and $j$ such that $\sigma(j)=k$. Then, relabeling in $A$, we have
$$
\begin{aligned}
\sum_{\sigma\in{\wp}_m}\sum_{j=1}^m&
\frac{\partial h_{\sigma(j),\tilde{0}}(y)}{\partial y_j}
u_{a,x_k}(h_{\tilde{0}}(y))\sgn(\sigma)h_{k,b}(y)\prod_{i=1,i\neq j}^m\frac{\partial h_{\sigma(i),\tilde{0}}(y)}{\partial y_i}\\
&\qquad\qquad\qquad\qquad
=u_{a,x_k}(h_{\tilde{0}}(y))h_{k,b}(y)
\sum_{\sigma\in{\wp}_m}\sgn(\sigma)\prod_{i=1}^m\frac{\partial h_{\sigma(i),\tilde{0}}(y)}{\partial y_i}.
\end{aligned}
$$
(ii) Take two permutations $\sigma$ and $\tilde{\sigma}$ in $\wp_m$ such that
$$
\sigma(i)=\tilde{\sigma}(j),\;\; \sigma(j)=\tilde{\sigma}(i),\;\;,\sigma(k)=\tilde{\sigma}(k) \for k\neq i, j,\et \sgn(\sigma)\sgn(\tilde{\sigma})=-1.
$$
Then,
$$
\sgn(\sigma) h_{\sigma(j),b}(y)\frac{\partial}{\partial y_j}\bigg(\prod_{i=1, i\neq j}^m\frac{\partial h_{\sigma(i),\tilde{0}}(y)}{\partial y_i}\bigg)
+\sgn(\tilde{\sigma}) h_{\tilde{\sigma},b}(y)\frac{\partial}{\partial y_j}\bigg(\prod_{i=1, i\neq j}^m\frac{\partial h_{\tilde{\sigma}(i),\tilde{0}}(y)}{\partial y_i}\bigg)
=0.
$$
(iii) Interchanging the role of $j$, $k$ and $c_1$, $c_2$ in ${\mathrm{III}}$, we have ${\mathrm{III}}=0$.
Others are treated analogously.

Therefore,
$$
{\mathrm{I}}+{\mathrm{II}}+{\mathrm{III}}=A+B=\sum_{j=1}^m \frac{\partial}{\partial y_j} (*)
$$
and we have proved the claim above. 

Now, if we assume the compactness of the support of $u_a(x)$ for $|a|\neq \tilde{1}$,  then
$$
\int_{\pi_{\mathrm{B}}(\mathfrak{U})}dy \frac{\partial}{\partial y_i}\big(u_{a}(h(y,\omega))\partial_{\omega}^{\tilde{1}-a}\Ber(H)(y,\omega)\big)\big|_{\omega=0}=0.
$$

(III-2) For $H(y,\omega)=(y,\phi(y,\omega))$ with $\phi(y,\omega)=(\phi_1(y,\omega),{\cdots},\phi_n(y,\omega))\in{\fR}^{0|n}$, 
we want to claim
\begin{equation}
{\Biint}_{\mathfrak{V}}dxd\theta \,u(x,\theta)={\Biint}_{\mathfrak{U}}dyd\omega \bigg(\det\bigg(\frac{\partial\phi_i}{\partial\omega_j}\bigg)\bigg)^{-1}u(y,\phi(y,\omega)).
\label{R11.37}
\end{equation}
By the analogous proof of \eqref{int5-14} in Proposition~\ref{int5.6}, i.e. odd change of variables formula,
we have the above readily. $\qquad\qed$

\begin{remark}
How we decompose a given superdiffeomorphism?
Though Berezin decomposes it as ~\eqref{20100413-1} but  M. J. Rothstein~\cite{roth87} introduces another decomposition
and arguments which are outside of my comprehention\footnote{
Though following the principle of ``Quick response to feel strange'', I asked him by mail, but no response. I don't know exactly, but he maybe changes from mathematician to different occupation}.
Moreover, there is another trial by Zirnbauer~\cite{zir96} which is not appreciated for me.
I should be ashamed?!
Therefore, I take the understandable arguments of Rogers, Vladimirov and Volvich with slight modification in Chapter 5.
In any way, the following Rothstein's decomposition is a key which is not proved here:
{\small
\begin{quotation}
\begin{proposition}[Proposition 3.1 of Rothstein~\cite{roth87}]
Let superdiffeomorphism $\varphi$ from ${\mathfrak{V}}$ to ${\mathfrak{U}}=\varphi({\mathfrak{V}})$ be given as
\begin{equation}
x=x(y,\omega)=\varphi_{\bar{0}}(y,\omega), \quad \theta=\theta(y,\omega)=\varphi_{\bar{1}}(y,\omega).
\label{R4-17}
\end{equation}
We assume that the following:
$$
{\pi_{\mathrm{B}}}({\sdet}J(\varphi)(y,\omega))\neq{0}
\with
J(\varphi)(y,\omega)=\begin{pmatrix}
\frac{\partial \varphi_{\bar{0}}(y,\omega)}{\partial y}&\frac{\partial \varphi_{\bar{1}}(y,\omega)}{\partial y}\\
\frac{\partial \varphi_{\bar{0}}(y,\omega)}{\partial\omega}&\frac{\partial \varphi_{\bar{1}}(y,\omega)}{\partial\omega}
\end{pmatrix}
=\begin{pmatrix}
\frac{\partial x_j}{\partial y_i}&\frac{\partial \theta_l}{\partial y_i}\\
\frac{\partial x_j}{\partial \omega_k}&\frac{\partial \theta_l}{\partial \omega_k}
\end{pmatrix}.
$$
\par
Then, $\varphi$ is decomposed uniquely by
$\varphi^{(1)}:(y,\omega)\to(\tilde{y},\tilde{\omega})$ and $\varphi^{(2)}:(\tilde{y},\tilde{\omega})\to (x,\theta)$
satisfying
\par
(i) $\varphi^{(1)}$ endows identical ${\mathbb{Z}}_2$-gradings, that is,
$\varphi^{(1)}:(y,\omega)\to (\tilde{y},\tilde{\omega})=(\tilde\varphi^{(1)}_{\bar{0}}(y),\tilde\varphi^{(1)}_{\bar{1}}(y,\omega))$, and
\par
(ii) $\varphi^{(2)}$ is derived from the following even and degree increasing derivation ${\mathcal{Y}}_{(\tilde{y},\tilde{\omega})}$ by
$$
(x,\theta)=e^{{\mathcal{Y}}_{(\tilde{y},\tilde{\omega})}}(\tilde{y},\tilde{\omega})=\varphi^{(2)}(\tilde{y},\tilde{\omega}) \et
(x,\theta)=\varphi^{(2)}(\varphi^{(1)}(y,\omega))=\varphi^{(2)}\circ\varphi^{(1)}(y,\omega).
$$
Here, $\tilde\lambda_j\in\CSS(\varphi({\mathfrak{V}}):\rev^m),\; \tilde\gamma_k\in\CSS(\varphi({\mathfrak{V}}):\rod^n)$ and
\begin{equation}
{\mathcal{Y}}_{(\tilde{y},\tilde{\omega})}=\sum_{j=1}^m\tilde\lambda_j(\tilde{y},\tilde{\omega})\frac{\partial}{\partial \tilde{y}_j}+\sum_{k=1}^n\tilde\gamma_k(\tilde{y},\tilde{\omega})\frac{\partial}{\partial\tilde{\omega}_k}.
\label{20100413}
\end{equation}
\end{proposition}
\end{quotation}}
\end{remark}

%\newpage %\vspace{2cm}
\section{Function spaces and Fourier transformations}
\subsection{Function spaces}
Let $U\subset\euc^m$ be a domain. We introduce following function spaces:
\begin{definition}[Function spaces on $U$ with values in ${\mathfrak{C}}$] \label{fs-simple}
$$
\begin{aligned}
C^\infty (U:{\mathfrak{C}}) &=
\{u(q)=\sum_{{\mathbf{I}}\in{\mathcal{I}}}u_{\mathbf{I}}(q)\sigma^{\mathbf{I}} \,\big|\,u_{\mathbf{I}}(q)\in 
C^\infty(U:{\mathbb{C}}) \forany {\mathbf{I}}\in{{\mathcal{I}}}\},\\
C^\infty_0(U:{\mathfrak{C}}) &=
\{u(q)\in C^\infty(U:{\mathfrak{C}}) \,\big|\, u_{\mathbf{I}}(q)\in 
C^\infty_0(U:{\mathbb{C}}) \forany {\mathbf{I}}\in{{\mathcal{I}}} \},\\
{\mathcal{B}}(U:{\mathfrak{C}}) &=
\{u(q)\in C^\infty(U:{\mathfrak{C}}) \,\big|\, u_{\mathbf{I}}(q)\in 
{\mathcal{B}}(U:{\mathbb{C}}) \forany {\mathbf{I}}\in{{\mathcal{I}}} \},\\
\dot{\mathcal{B}}(\eucm:{\mathfrak{C}}) &=
\{u(q)\in C^\infty(U:{\mathfrak{C}}) \,\big|\,u_{\mathbf{I}}(q)\in 
\dot{\mathcal{B}}(\eucm:{\mathbb{C}}) \forany {\mathbf{I}}\in{{\mathcal{I}}} \},\\
{\mathcal{S}}(\eucm:{\mathfrak{C}})&=
\{u(q)\in C^\infty(\eucm:{\mathfrak{C}}) \,\big|\, u_{\mathbf{I}}(q)\in 
{\mathcal{S}}(\eucm:{\mathbb{C}})\},\\
{\mathcal{O}}_M(\eucm:{\mathfrak{C}})&=
\{u(q)\in C^\infty(\eucm:{\mathfrak{C}}) \,\big|\, u_{\mathbf{I}}(q)\in 
{\mathcal{O}}_M(\eucm:{\mathbb{C}})\}.
\end{aligned}
$$
\end{definition}

\begin{remark} In the above, we use the notation given in L.Schwartz~\cite{Sch-l70}
That is,
$$
\begin{aligned}
\dot{\mathcal{B}}(\eucm:{\mathbb{C}}) &=
\{u(q)\in {\mathcal{B}}(\eucm:{\mathbb{C}}) \,\big| 
\lim_{|q|\to\infty}|\partial_q^\alpha u(q)|=0 \forany \alpha\},\\
{\mathcal{O}}_M(\eucm:{\mathbb{C}}) &=
\{u(q)\in C^\infty(\eucm:{\mathbb{C}}) \,\big|\, 
|\partial_q^\alpha u(q)|\le C(1+|q|^2)^{k/2}\,\\
&\qquad\qquad\qquad\qquad\qquad\qquad
\text{for some constants $C>0$
and $k>0$} \}.
\end{aligned}
$$
\end{remark}

For a superdomain ${\mathfrak{U}}={\mathfrak{U}}_{\mathrm{ev}}\times{\fR}_{\mathrm{od}}^n\subset{\supermn}$
with $U=\pi_{\mathrm B}({\mathfrak{U}})=\pi_{\mathrm B}({\mathfrak{U}}_{\mathrm{ev}})\subset\eucm$, 
we put:

\begin{definition}[Function spaces on ${\mathfrak{U}}$ with values in ${\mathfrak{C}}$] \label{} %\proclaim{Definition 2.2 ()}
Putting $X=(x,\theta)$, $X_{\mathrm B}=q\in U$,
$\partial_\theta^a u(x,0)=u_a(x)$, 
$$
\begin{aligned}
{\mathcal{C}}_{S\!S}({\mathfrak{U}}:{\mathfrak{C}}) &=
\{u(X)=\sum_{|a|\le n}u_a(x)\theta^a \,\big|\,
u_a(q)\in  C^\infty(U:{\mathfrak{C}}) \forany
a\},\\
{\mathcal{C}}_{S\!S,0}({\mathfrak{U}}:{\mathfrak{C}}) &=
\{u(X)\in{\CSS}({\mathfrak{U}}:{\mathfrak{C}}) \,\big|\, u_a(q)\in 
C^\infty_0(U:{\mathfrak{C}}) \forany a\},\\
{\BSS}({\mathfrak{U}}:{\mathfrak{C}}) &=
\{u(X)\in{\CSS}({\mathfrak{U}}:{\mathfrak{C}}) \,\big|\,u_a(q)\in 
{\mathcal{B}}(U_{\mathrm B}:{\mathfrak{C}}) \forany a\},\\
\dot{\mathcal{B}}_{S\!S}(\supermn:{\mathfrak{C}}) &=
\{u(X)\in{\CSS}(\supermn:{\mathfrak{C}}) \,\big|\,u_a(q)\in 
\dot{\mathcal{B}}(\eucm:{\mathfrak{C}}) \forany a\},\\
{\SSS}(\supermn:{\mathfrak{C}}) &=
\{u(X)\in{\CSS}({\supermn}:{\mathfrak{C}}) \,\big|\, u_a(q)
\in  {\mathcal{S}}(\eucm:{\mathfrak{C}}) \forany a\},\\
{\mathcal{O}}_{M\!,S\!S}(\supermn:{\mathfrak{C}})&=
\{u(X)\in \CSS(\supermn:{\mathfrak{C}}) \,\big|\, u_a(q)
\in  {\mathcal{O}}_M(\eucm:{\mathfrak{C}}) \forany a\}.
\end{aligned}
$$
\end{definition}

\begin{remark}
If we consider function spaces whose member
are homogeneous, we denote them by adding subindeces $\mathrm{ev}$ or $\mathrm{od}$, i.e.
${\mathcal{C}}_{S\!S,\mathrm{ev}}({\mathfrak{U}}:{\mathfrak{C}})={\CSS}({\mathfrak{U}}:{\mathfrak{C}}_{\mathrm{ev}})$ or
${\mathcal{C}}_{S\!S,\mathrm{od}}({\mathfrak{U}}:{\mathfrak{C}})={\CSS}({\mathfrak{U}}:{\mathfrak{C}}_{\mathrm{od}})$ etc.
\end{remark}

\begin{definition}[Function spaces on ${\mathfrak{U}}$ with value `real']\label{} 
$$
\begin{aligned}
{\ccsl}_{S\!S}({\mathfrak{U}}_{\mathrm{ev}}) &=
\{u(x)\in{\CSS}({\mathfrak{U}}_{\mathrm{ev}}:{\mathfrak{C}}) \,\big|\,
u(x_{\mathrm B})\in C^\infty({\mathfrak{U}}_{\mathrm B}:{\mathbb{C}})\},\\
{\ccsl}_{S\!S}({\mathfrak{U}}) &=
\{u(X)\in{\CSS}({\mathfrak{U}}:{\mathfrak{C}}) \,\big|\, 
\partial_\theta^a u(x,0)\in {\ccsl }_{S\!S}({\mathfrak{U}}_{\mathrm{ev}}) \forany a\},\\
{\ccsl}_{S\!S,0}({\mathfrak{U}}) &=
\{u(X)\in{\CSS}({\mathfrak{U}}:{\mathfrak{C}}) \,\big|\, \partial_\theta^a u(X_{\mathrm B})\in 
C^\infty_0({\mathfrak{U}}_{\mathrm B}:{\mathbb{C}})\forany a\},\\
{\cbsl}_{S\!S}({\mathfrak{U}}) &=
\{u(X)\in{\BSS}({\mathfrak{U}}:{\mathfrak{C}}) \,\big|\, \partial_\theta^a u(X_{\mathrm B})\in 
{\mathcal{B}}({\mathfrak{U}}_{\mathrm B}:{\mathbb{C}})\forany a\},\\
\dot{\cbsl}_{S\!S}(\supermn) &=
\{u(X)\in{\CSS}(\supermn:{\mathfrak{C}}) \,\big|\, \partial_\theta^a u(X_{\mathrm B})\in 
\dot{\mathcal{B}}(\eucm:{\mathbb{C}})\forany a\},\\
{\cssl}_{S\!S}(\supermn) &=
\{u(X)\in{\SSS}({\supermn}:{\mathfrak{C}}) \,\big|\,\partial_\theta^a u(X_{\mathrm B})
\in  {\mathcal{S}}(\eucm:{\mathbb{C}})\forany a\},\\
{\comsl}_{\!,S\!S}(\supermn) &=
\{u(X)\in{\CSS}(\supermn:{\mathfrak{C}}) \,\big|\, \partial_\theta^a u(X_{\mathrm B})\in 
{\mathcal{O}}_M(\eucm:{\mathbb{C}})\forany a\}.
\end{aligned}
$$
\end{definition}

\begin{definition}[Topology of function spaces]\label{}
\hspace{1mm}
\begin{itemize} 
\item
We introduce seminorms in ${\CSS}({\mathfrak{U}}:{\mathfrak{C}}) $ for any integer $k$, 
${\mathbf{I}}\in{\mathcal{I}}$ and a compact set $K\subset {\mathfrak{U}}_{\mathrm B}$,
by defining
$$
p_{k,K,{\mathbf{I}}}(u)=\sup_{X_{\mathrm B}\in K,|{\mathfrak{a}}|\le k}
\left|\proj_{\mathbf{I}}(D_X^{\mathfrak{a}}u(X_{\mathrm B}))\right|.
\label{FS-2.1}
$$
${\CSS}({\mathfrak{U}}:{\mathfrak{C}}) $ with this topology will be denoted by 
${\ESS}({\mathfrak{U}}:{\mathfrak{C}})$.
\item
We say that $u_j\to 0$ in ${\mathcal{C}}_{S\!S,0}({\mathfrak{U}}:{\mathfrak{C}})$ when $j\to{\infty}$
iff for any ${\mathbf{I}}\in {\mathcal{I}}$, there exists a compact set 
$K_{\mathbf{I}}\subset U_{\mathrm B}$ such that \\
(i) the support of $\proj_{\mathbf{I}}(\partial_\theta^a
u_j(\cdot,0))$ is contained in $ K_{\mathbf{I}}$ for any $a$
and $j$, and \\
(ii) $\sup_{X_{\mathrm B}\in K_{\mathbf{I}}} 
\left|\proj_{\mathbf{I}}( D_X^{\mathfrak{a}}u_j(X_{\mathrm B}))\right|\to 0$
as $j\to\infty$ for any ${\mathfrak{a}}=(\alpha,a)$.\\
We denote ${{\mathcal{D}}}_{S\!S}({\mathfrak{U}}:{\mathfrak{C}})$
the set ${\mathcal{C}}_{S\!S,0}({\mathfrak{U}}:{\mathfrak{C}})$ with this topology 
and call it as the
space of test functions on $U$. 
\item
We say that $u_j\to 0$ in ${\BSS}({\mathfrak{U}}:{\mathfrak{C}})$ 
iff for any ${\mathbf{I}}\in {\mathcal{I}}$ and ${\mathfrak{a}}$, $\proj_{\mathbf{I}}(D_X^{\mathfrak{a}}u_j(X_{\mathrm 
B}))$ converges uniformly to $0$ on any compact set $K_{\mathbf{I}}$
and they are bounded on $\eucm$.
\item
For $u\in{\SSS}(\supermn:{\mathfrak{C}}) $ and any integer $k$ and 
${\mathbf{I}}\in{\mathcal{I}}$,   we put 
\begin{equation}
p_{k,{\mathbf{I}}}(u)=\sup_{X_{\mathrm B}\in{\eucm},|{\mathfrak{a}}|+j\le k}
(1+|X_{\mathrm B}|^2)^{j/2}\left|\proj_{\mathbf{I}}(D_X^{\mathfrak{a}}u(X_{\mathrm B}))\right|.
\label{FS-2.2}
\end{equation}
\item
The topology of spaces
$\cesl_{S\!S}({\mathfrak{U}})$, $\cdsl_{S\!S}({\mathfrak{U}})$, $\cbsl_{S\!S}({\mathfrak{U}})$ and 
$\cssl_{S\!S}(\supermn)$ is defined accordingly as above.
\end{itemize} %roster 
\end{definition}

\subsection{Scalar products and norms}
\begin{definition}[Hermitian conjugation]\label{HC} 
\hspace{1mm}
\begin{itemize} 
\item
For $x=(x_1, \cdots,x_m)\in {\mathfrak{R}}_{\mathrm{ev}}^m$ and
$\theta=(\theta_1,\cdots,\theta_n)\in {\mathfrak{R}}_{\mathrm{od}}^n$,
we put
$$
\bar{x}=({\bar{x}}_1,\cdots,{\bar{x}}_m) \with
\bar{x}_j=\sum_{|{\mathbf{I}}|={\mathrm{ev}}} {\overline {x_{j,{\mathbf{I}}}}}\overline{\sigma^{\mathbf{I}}}
$$
where
$\overline{\sigma^{\mathbf{I}}}$ is defined in (2.1.15) and
${\overline {x_{j,{\mathbf{I}}}}}=$the complex conjugate of $x_{j,{\mathbf{I}}}$ 
in ${\mathbb{C}}$.
\begin{equation}
{\overline{\theta^a}}={\bar\theta}_n^{a_n}\cdots{\bar\theta}_1^{a_1}
=(-1)^{{|a|(|a|-1)}/{2}}{\bar\theta}^a 
\et {\bar\theta}_k=\sum_{|{\mathbf{I}}|={\mathrm{od}}} {\overline{\theta_{k,{\mathbf{I}}}}}\overline{\sigma^{\mathbf{I}}}
\label{FS-2.2bis}
\end{equation}
where
${\overline {\theta_{k,{\mathbf{I}}}}}=$ the complex conjugate of $\theta_{k,{\mathbf{I}}}$ 
in ${\mathbb{C}}$.
\item
Let $w(q)=\sum_{{\mathbf{I}}\in{\mathcal{I}}}{w_{\mathbf{I}}(q)}\sigma^{\mathbf{I}}\in C^\infty(\eucm:{\mathfrak{C}})$. 
We put
\begin{equation}
{\overline {w(q)}}=\sum_{{\mathbf{I}}\in{\mathcal{I}}}{\overline {w_{\mathbf{I}}(q)}}\overline{\sigma^{\mathbf{I}}}
\with {\overline {w_{\mathbf{I}}(q)}}=\text{the complex conjugate of $w_{\mathbf{I}}(q)$ 
in ${\mathbb{C}}$}. 
\label{FS-2.3}
\end{equation}
\item
For ${\tilde w}(x)\in {\mathcal{C}}_{S\!S}(U_{\mathrm{ev}}:{\mathfrak{C}})$ with 
$w(q)\in C^\infty(\eucm:{\mathfrak{C}})$, we put
\begin{equation}
{\overline{\tilde w (x)}}\equiv{\tilde{\overline w}}(\bar{x}).
\label{FS-2.4}
\end{equation}
\item
For $v(\theta)=\sum_{|a|\le n}\theta^a v_a\in{\mathcal{P}}_n({\mathfrak{C}})$, we
put
\begin{equation}
{\overline{v(\theta)}}=\sum_{|a|\le n}\bar\theta^a v_a^*
=\bar v(\bar\theta)\with
v_a^*=(-1)^{{|a|(|a|-1)}/2}
((\overline{v_a})_{\mathrm{ev}}+(-1)^{|a|}(\overline{v_a})_{\mathrm{od}}).
\label{FS-2.5}
\end{equation}
\item
For $u(X)\in{\CSS}({\mathfrak{U}}:{\mathfrak{C}})$, we put
\begin{equation}
{\overline{u(X)}}=\sum_{|a|\le n} {\bar\theta}^a u_a^*(\bar{x})
=\bar{u}(\bar{X})\where \bar{X}=(\bar{x},\bar{\theta})
\label{FS-2.6}
\end{equation}
with
$$
\begin{gathered}
u_a^*(\bar{x})=(-1)^{|a|(|a|-1)/2}
((\overline{u_a(x)})_{\mathrm{ev}}+(-1)^{|a|}(\overline{u_a(x)})_{\mathrm{od}}),\\
\overline{u_a(x)}=\sum_{\alpha}{\frac 1{\alpha!}}\partial_q^{\alpha}{\overline{u_a(x_{\mathrm{B}})}}{\overline{x_{\mathrm{S}}^{\alpha}}}
=\sum_{\alpha}{\frac 1{\alpha!}}\partial_q^{\alpha}{\overline{u_a(x_{\mathrm{B}})}}{\barx_{\mathrm{S}}^{\alpha}},\\
\where \barx=x_{\mathrm{B}}+\barx_{\mathrm{S}}.
\end{gathered}
$$
\end{itemize}
\end{definition}

\begin{remark}
(1) The Grassmann continuation ${\tilde w}(x)$ is expanded as
${\tilde w}(x)=\sum_{\mathbf{H}}{\tilde w}_{\mathbf{H}}(x)\sigma^{\mathbf{H}} $ with
$$
{\tilde w}_{\mathbf{H}}(x)=\!\!\sum\limits_{
\scriptstyle {\mathbf{H}}={\mathbf{J}}+{\mathbf{I}}_1^{(1)}+\cdots+{\mathbf{I}}_m^{(\alpha_m)} 
\atop
\scriptstyle
{\mathbf{I}}_j={\mathbf{I}}_j^{(1)}+\cdots+{\mathbf{I}}_j^{(\alpha_j)},\, \alpha=(\alpha_1,\cdots,\alpha_m)}
\!\!(-1)^{*}{\frac 1{\alpha!}}\partial_q^\alpha w_J(x_{\mathrm B})
x_{1,{\mathbf{I}}_1^{(1)}}\cdots x_{1,{\mathbf{I}}_1^{(\alpha_1)}}x_{2,{\mathbf{I}}_2^{(1)}}\cdots
x_{m,{\mathbf{I}}_m^{(\alpha_m)}},
$$ 
we get 
$$
{\overline{\tilde w(x)}}=\sum_{\mathbf{H}}{\overline {\tilde w_{\mathbf{H}}(x)}}\sigma^{\mathbf{H}}
=\sum {\frac 1{\alpha!}}\partial_q^\alpha {\overline{ w (x_{\mathrm B})}}
{\overline{x_{\mathrm S}^\alpha}}={\widetilde{\overline {w(\cdot)}}}(\bar{x}).
$$
This guarantees the naturalness of \eqref{FS-2.3}.
\newline
(2) The definiton of \eqref{FS-2.4} and \eqref{FS-2.5} follows from the relation
$\overline{\theta^av_a}={\overline{v_a}}{\overline{\theta^a}}$
for any $v_a\in{\mathfrak{C}}$.
\end{remark}

\begin{definition}[$L^2$ spaces]\label{L2spaces} 
\hspace{1mm}
\begin{itemize} %roster
\item
For $u,\, w\in {\mathcal{C}}_{S\!S,0}({\mathfrak{U}}_{\mathrm{ev}}:{\mathfrak{C}})$, 
we define the scalar product and the $L^2$-norm by 
\begin{equation} 
(u,w)
=\int_{\pi_{\mathrm B}({\mathfrak{U}}_{\mathrm{ev}})}dx_{\mathrm B}\,
{\overline{u(x_{\mathrm B})}} w(x_{\mathrm B})\in{\mathfrak{C}}  \et \|u\|^2 = (u,u)\in{\fR}.
\label{FS-2.7}
\end{equation}
\item
For $v,\,w\in {\mathcal{P}}_n({\mathfrak{C}})$, 
we put
\begin{equation} 
(v,w) = \sum_{|a|\leq n} \overline{v_a}w_a
\et  {\| v\|}^2 = (v,v). \label{FS-2.8}
\end{equation} 
\item
For $u,\,w\in{\mathcal{C}}_{S\!S,0}({\mathfrak{U}}:{\mathfrak{C}})$,
we define also
\begin{equation} 
(u,w)= \sum_{|a| \le n}
\int_{{\mathfrak{U}}_{\mathrm{ev}}}dx\,
\overline{u_a(x)}w_a(x)
=\sum_{|a| \le n}
\int_{\pi_{\mathrm B}({\mathfrak{U}}_{\mathrm{ev}})}dx_{\mathrm B}\,
\overline{u_a(x_{\mathrm B})}w_a(x_{\mathrm B})\with {\| u \|}^2=(u,u).
\label{FS-2.9}
\end{equation} 
\end{itemize}
\end{definition}

\begin{remark}
(1) It is clear that
if $u,\, v$ are in $\ccsl_{S\!S,0}({\mathfrak{U}})$ or
$\cssl_{S\!S}(\supermn)$, $(u,v)\in{\mathbb{C}}$ and $(u,u)=\|u\|^2\ge 0$.
\newline
(2) In the following, we explain the derivation of the above scalar products: \\
(i) If $f(q),\,g(q)\in C^\infty(U_{\mathrm B}:{\mathbb{C}})$, using $\delta$-function symbolically,
we may consider the standard scalar product as 
$$
(f,g)=\int_{U_{\mathrm B}}dq\,{\overline{f(q)}}g(q)
=\iint_{U_{{\mathrm B},q'}\times U_{{\mathrm B},q}}dq'dq\,
\delta(q-q'){\overline{f(q)}} g(q')\with U_{\mathrm B}=\pi_{\mathrm{B}}(\mathfrak{U}).
$$ 
On the other hand, for $u(x)\in {\mathcal{C}}_{S\!S,0}({\mathfrak{U}}_{\mathrm{ev},x}:{\mathfrak{C}})$ and 
$w(y)\in {\mathcal{C}}_{S\!S,0}({\mathfrak{U}}_{\mathrm{ev},y}:{\mathfrak{C}})$,
we may regard ${\overline{u(x)}}w(y)$ as a function 
${\overline u}(\bar{X}) w(y)
\in {\mathcal{C}}_{S\!S,0}({\mathfrak{U}}_{\mathrm{ev},\bar{X}}\times {\mathfrak{U}}_{\mathrm{ev}, y}:{\mathfrak{C}})$. 
Therefore, remarking
$({\overline u}w)(x_{\mathrm B})
={\overline {u(x_{\mathrm B})}}w(x_{\mathrm B})
=\sum_{{\mathbf{I}}={\mathbf{J}}+{\mathbf{K}}\in{\mathcal{I}}}
(-1)^{\tau({\mathbf{I}};{\mathbf{J}},{\mathbf{K}})}{\overline{u_{\mathbf{J}}(q)}}w_{\mathbf{K}}(q)\sigma^{\mathbf{I}}$ with 
$\sigma^{\mathbf{J}}\sigma^{\mathbf{K}}=(-1)^{\tau({\mathbf{I}};{\mathbf{J}},{\mathbf{K}})}\sigma^{\mathbf{I}}$ for ${\mathbf{I}}={\mathbf{J}}+{\mathbf{K}}$,
we may define 
\begin{equation}
(u,w)=\iint_{{\mathfrak{U}}_{\mathrm{ev},y}\times {\mathfrak{U}}_{\mathrm{ev},\bar{X}}} dyd{\bar
x}\, \delta({\bar{x}}-y) {\overline{u(x)}} w(y)
 =\int_{{\euc}^m} dx_{\mathrm B}\,
{\overline {u(x_{\mathrm B})}}w(x_{\mathrm B})\in{\mathfrak{C}}.
\label{FS-2.10}
\end{equation} 
(ii) We introduce here a constant $\tau(a,b)$
for any multi-indeces $a$, $b$ by 
\begin{equation}
\theta^a\theta^b=(-1)^{\tau(a,b)}\theta^{a+b}
\label{FS-2.11}
\end{equation}
from which we get easily
\begin{equation}
\tau(b,a)\equiv |a||b|+\tau(a,b) \mod 2.
\label{FS-2.12}
\end{equation}
Moreover, for any $b$ and $\theta,\, {\bar\theta},\,\pi \in {\mathfrak{R}}_{\mathrm{od}}$, we get, by
induction with respect to $|b|$, 
\begin{equation}
\prod_{j=1}^n(\pi_j\theta_j)^{b_j}=
(-1)^{{|b|(|b|-1)}/2} \pi^b \theta^b\et
\prod_{j=1}^n({\bar\theta}_j\theta_j)^{b_j}=
(-1)^{{|b|(|b|-1)}/2} {\bar\theta}^b \theta^b.
\label{FS-2.13}
\end{equation}
(iii) By putting $\langle\bar\theta|{\theta}\rangle=
\sum_{j=1}^n \bar\theta_j\theta_j$
and
\begin{equation}
\overline{d\theta}=d\bar\theta_1\cdots d\bar\theta_n=(-1)^{n(n-1)/2}d\bar\theta_n\cdots d\bar\theta_1, 
\label{FS-2.13bis}
\end{equation}
we get \eqref{FS-2.8} from %(2.8) from 
\begin{equation}
(u,w)=\iint_{{\fR}^{0|n}_\theta\times{\fR}^{0|n}_{\bar\theta}}d\theta
\overline{d\theta}\, e^{\langle\bar\theta|{\theta}\rangle}
\sum_{a}{\overline{\theta^au_a}} {\sum_b \theta^bw_b}
=\sum_a ((\overline {u_a})_{\mathrm{ev}}+(\overline {u_a})_{\mathrm{od}})w_a=\sum_a\overline {u_a}w_a.
\label{FS-2.14}
\end{equation}
Here, we used equalities below and \eqref{FS-2.13}:
\begin{equation}
\left\{
\begin{aligned}
&\overline{\theta^au_a}\theta^bw_b
={\overline{\theta^a}}\theta^b
(({\overline{u_a}})_{\mathrm{ev}}+(-1)^{|a|+|b|}({\overline{u_a}})_{\mathrm{od}})w_b,\\
&\iint_{{\fR}^{0|n}_\theta\times{\fR}^{0|n}_{\bar\theta}}d\theta
\overline{d\theta}\, e^{\langle\bar\theta|{\theta}\rangle} 
{\overline{\theta^a}}\theta^b=\delta_{ab}=\prod_{k=1}^n\delta_{a_kb_k}.
\end{aligned}
\right.
\label{FS-2.15}
\end{equation} 
\begin{quotation}
{\small
In fact, taking up the top term w.r.t. $\bar{\theta}$, 
$$
\begin{aligned}
e^{\langle\bar\theta|{\theta}\rangle} {\overline{\theta^a}}\theta^b
&\sim(-1)^{|a|(|a|-1)/2}\bartheta^a\prod_{j=1}^n(\bartheta_j\theta_j)^{{\bar{a}}_j}\theta^b
=(-1)^{|a|(|a|-1)/2}\bartheta^a(-1)^{|\bar{a}|(|\bar{a}|-1)/2}\bartheta^{\bar{a}}\theta^{\bar{a}}\theta^b\\
&=(-1)^{|a|(|a|-1)/2+|\bar{a}|(|\bar{a}|-1)/2+\tau(a,\bar{a})}\bartheta^{\tilde{1}}\theta^{\bar{a}}\theta^b,
\end{aligned}
$$
and using \eqref{FS-2.13bis}, we have
$$
\begin{aligned}
\iint_{{\fR}^{0|n}_\theta\times{\fR}^{0|n}_{\bar\theta}}d\theta
\overline{d\theta}\, e^{\langle\bar\theta|{\theta}\rangle} 
{\overline{\theta^a}}\theta^b
&=(-1)^{n(n-1)/2+|a|(|a|-1)/2+|\bar{a}|(|\bar{a}|-1)/2+\tau(a,\bar{a})}\int_{{\fR}^{0|n}}d\theta\, \theta^{\bar{a}}\theta^b\\
&=(-1)^{n(n-1)/2+|a|(|a|-1)/2+|\bar{a}|(|\bar{a}|-1)/2+\tau(a,\bar{a})+\tau({\bar{a},a)}}\delta_{ab}=\delta_{ab}
\end{aligned}
$$
since
$n(n-1)/2+|a|(|a|-1)/2+|\bar{a}|(|\bar{a}|-1)/2+\tau(a,\bar{a})+\tau(\bar{a},a)\equiv 0$$\mod2$.$\qquad/\!\!/$
}
\end{quotation}

(iv) Finally, as we have
\begin{equation}
\begin{aligned} 
&\iint_{{\mathfrak{U}}_X\times {\mathfrak{U}}_{\bar{X}}}dX\overline{dX}\,  
\delta(\bar{x}-x) 
e^{\langle\bar\theta|{\theta}\rangle} 
{\overline{u(X)}}w(X)\\
&\quad=
\iint_{\pi_{\mathrm B}({\mathfrak{U}}_{\mathrm{ev}})_x\times \pi_{\mathrm B}({\mathfrak{U}}_{\mathrm{ev}})_{\bar{x}}}
dxd{\bar{x}}\,\delta(\bar{x}-x)
\left\{\iint_{{\fR}^{0|n}_\theta\times{\fR}^{0|n}_{\bar\theta}} d\theta
\overline{d\theta} \,
e^{\langle\bar\theta|{\theta}\rangle}
{\overline{u(x,\theta)}}w(x,\theta)\right\}\\ 
&\quad=\sum_{|a|\le n} 
\int_{\pi_{\mathrm B}({\mathfrak{U}}_{\mathrm{ev}})} dx_{\mathrm B}
\overline{u_a(x_{\mathrm B})} w_a(x_{\mathrm B})=\sum_{|a|\le n}(u_a,w_a) , 
\end{aligned} 
\label{FS-2.16}
\end{equation}
our definition \eqref{FS-2.9} seems canonical.
\end{remark}
\begin{lemma}\label{} %\proclaim{Lemma 2.7} 
Denoting by $A^*$ the dual of the operator $A$ in the above
scalar product, we get easily
\begin{gather}
\Big({\frac \hbar i}\partial_{x_j}\Big)^*={\frac \hbar i}\partial_{x_j},
\quad
(x_j)^*=x_j, 
\label{FS-2.17}\\
(\partial_{\theta_k})^*=\theta_k,\quad
(\theta_k)^*=\partial_{\theta_k},\quad
(D_\theta^a)^*=\theta^a,\quad
(\partial_\theta^a)^*={}^{t\!}(\theta^a).
\label{FS-2.18}
\end{gather}
\end{lemma}
\par
{\it Proof. }
For $a,b\in\{0,1\}^n$, we put $\check a_k=(a_1,\cdots,a_k-1,\cdots,a_n),\,
\hat b_k=(b_1,\cdots,b_k+1,\cdots,b_n)$ where $\partial_{\theta_k}\theta^a=0$ with $a$ for $a_k=0$.
Let $v,\,w\in {\mathcal{P}}_n({\mathfrak{C}})$. 
Remarking \eqref{FS-2.15}, we have %(2.15) and (1.21), we have
$$
\begin{aligned}
{\overline{\partial_{\theta_k}(\theta^a u_a)}} (\theta^bv_b)
&=\overline{u_a} (-1)^{\ell_k(a)} {\overline{\theta^{\check a_k}}} (\theta^bv_b)\\
&={\overline{\theta^{\check a_k}}}{\theta^b}(-1)^{\ell_k(a)}(({\overline{u_a}})_{\mathrm{ev}}+(-1)^{|\check a_k|+|b|}({\overline{u_a}})_{\mathrm{od}})v_b,\\
\end{aligned}
$$ 
and
$$
{\overline{\theta^au_a}}{\theta_k \theta^b v_b}
={\overline{u_a}}{\overline{\theta^a}}(-1)^{\ell_k(b)}\theta^{\hat b_k}v_b
=(-1)^{\ell_k(b)}{\overline{\theta^a}}\theta^{\hat b_k}(({\overline{u_a}})_{\mathrm{ev}}+(-1)^{|a|+|\hat b_k|}({\overline{u_a}})_{\mathrm{od}})v_b,
$$
which yield
$$
\begin{aligned}
(\partial_{\theta_k}u,v)
&=\iint_{{\fR}^{0|n}_\theta\times{\fR}^{0|n}_{\bar\theta}}d\theta
\overline{d\theta}\, e^{\langle\bar\theta|{\theta}\rangle}
\sum_{a}{\overline{\partial_{\theta_k}(\theta^au_a) }} {\sum_b v_b\theta^b}\\
&=\sum_{a,b}(-1)^{l_k(a)}
\delta_{\check a_k b}
(({\overline{u_a}})_{\mathrm{ev}}+(-1)^{|a|-1+|b|}({\overline{u_a}})_{\mathrm{od}})v_b\\
&=\sum_{a,b}(-1)^{l_k(b)} \delta_{a\hat b_k}
(({\overline{u_a}})_{\mathrm{ev}}
+(-1)^{|a|+|b|+1}({\overline{u_a}})_{\mathrm{od}})v_b\\
&=\iint_{{\fR}^{0|n}_\theta\times{\fR}^{0|n}_{\bar\theta}}d\theta \overline{d\theta}\,
e^{\langle\bar\theta|{\theta}\rangle} \sum_{a}{\overline{u_a\theta^a }} {\sum_b
{\theta_k} v_b\theta^b} =(u, \theta_k v).
\end{aligned}
$$
Repeating this arguments, we have other equalities readily.  $\qquad\qed$

\subsection{Distributions}

\begin{definition}[Distributions on $U\subset{\euc}^m$ with values in ${\mathfrak{C}}$]\label{}
\hspace{5mm}
\begin{itemize}
\item
Let $\Phi$ be a linear functional defined on ${\mathcal{D}}_{S\!S}(U:{\mathfrak{C}}) $ such
that $\Phi(u_j)\to 0$ in ${\mathfrak{C}}$ iff
$u_j\to 0$ in ${\mathcal{C}}_{S\!S,0}(U:{\mathfrak{C}}) $.
Then, we call this functional as a distribution on $U$, and the set composed
of these is denoted by ${\mathcal{D}}'_{S\!S}(U:{\mathfrak{C}}) $.
\item
${\mathcal{E}}'_{S\!S}(U:{\mathfrak{C}}) $ stands for the set consisting of 
continuous linear functionals
on ${\mathcal{E}}_{S\!S}(U:{\mathfrak{C}}) $.
\item
${\mathcal{B}}'_{S\!S}(U:{\mathfrak{C}}) $ stands for the set consisting of 
continuous linear functionals
on ${\mathcal{B}}_{S\!S}(U:{\mathfrak{C}}) $.
\item
$\Phi\in{\mathcal{S}}'_{S\!S}(\supermn:{\mathfrak{C}}) $ iff $\Phi$ is a continuous
linear functional on ${\mathcal{S}}_{S\!S}(\supermn:{\mathfrak{C}}) $.
\item
$\cdsl'_{S\!S}(U)$, $\cesl'_{S\!S}(U)$, 
$\cbsl'_{S\!S}(U)$ and $\cssl'_{S\!S}(\supermn)$ 
are defined analogously.
\end{itemize}
$$
\begin{aligned}
\DSS'(U:{\mathfrak{C}}) &=
\{ \Phi(q)=\sum_{{\mathbf{I}}\in{\mathcal{I}}}\Phi_{\mathbf{I}}(q)\sigma^{\mathbf{I}} \,\big|\,\Phi_{\mathbf{I}}(q)
\in{\mathcal{D}}'(U:{{\mathbb{C}}}) \forany {\mathbf{I}}\in{\mathcal{I}}\},\\
\ESS'(U:{\mathfrak{C}}) &=
\{ \Phi(q)\in \DSS'(U:{\mathfrak{C}}) \,\big|\,
\Phi_{\mathbf{I}}(q)=\proj_{\mathbf{I}}(\Phi(q))\in{\mathcal{E}}'(U:{{\mathbb{C}}})
\forany {\mathbf{I}}\in{\mathcal{I}}\},\\
\SSS'(\eucm:{\mathfrak{C}}) &=
\{ \Phi(q)\in \DSS'(\supermn:{\mathfrak{C}}) \,\big|\,
\Phi_{\mathbf{I}}(q)\in{\mathcal{S}}'(\eucm:{{\mathbb{C}}})
\forany {\mathbf{I}}\in{\mathcal{I}}\}.
\end{aligned}
$$
\end{definition}

Here, $\Phi\in\DSS'(U:{\mathfrak{C}}) $ acts on $u\in\DSS(U:{\mathfrak{C}}) $
by
$$
\langle\Phi,u\rangle=\sum_{{\mathbf{I}}\in{\mathcal{I}}}
\big(\sum_{{\mathbf{I}}={\mathbf{J}}+{\mathbf{K}}}\langle\Phi_{\mathbf{J}},u_{\mathbf{K}}\rangle\big)\sigma^{\mathbf{I}}.
$$
Other dualities are defined analogously.

\begin{definition}[Distributions on ${\mathfrak{U}}$ with values in ${\mathfrak{C}}$]\label{}
%\proclaim{Definition 2.9 ()}
$$
\begin{aligned}
\DSS'({\mathfrak{U}}:{\mathfrak{C}}) &=
\{
\Phi(X)=\sum_{a}\Phi_a(x)\theta^a\,|\,\Phi_a(X_{\mathrm B})
=\partial_\theta^a\Phi(x_{\mathrm B},0) 
\in{\DSS}'({\mathfrak{U}}_{\mathrm B}:{\mathfrak{C}}) \forany a\},\\ 
\ESS'({\mathfrak{U}}:{\mathfrak{C}}) &= \{ \Phi(X)\in \DSS'({\mathfrak{U}}:{\mathfrak{C}})\,|\,
\Phi_a(X_{\mathrm B})\in{\ESS}'({\mathfrak{U}}_{\mathrm B}:{\mathfrak{C}})
\forany a\},\\
\SSS'(\supermn:{\mathfrak{C}}) &=
\{ \Phi(X)\in \DSS'(\supermn:{\mathfrak{C}})\,|\,
\Phi_a(X_{\mathrm B})\in{\SSS}'(\eucm:{\mathfrak{C}})
\forany a\}.
\end{aligned}
$$
\end{definition}

Action of $\Phi\in\DSS'({\mathfrak{U}}:{\mathfrak{C}})$ on $u\in \DSS({\mathfrak{U}}:{\mathfrak{C}})$ are defined by
$$
\langle \Phi,u\rangle=\sum_{|a|\le n}
\langle \partial_\theta^a\Phi(x,\theta),\partial_\theta^a u(x,\theta)\rangle.
$$

\begin{proposition} \label{} %\proclaim{Proposition 2.10}
Let $\Phi$ be a continuous linear functional on ${\mathcal{D}}_{S\!S}({\mathfrak{U}}:{\mathfrak{C}})$.
Then, $\Phi(X)$ is represented by
$$
\Phi(X)=\sum_{|a|\le n}\Phi_a(x)\theta^a
\where \proj_{\mathbf{I}}(\Phi_a(x_{\mathrm B}))
\in{\mathcal{D}}'({\mathfrak{U}}_{\mathrm{B}}:{{\mathbb{C}}})
\qquad\text{for each ${\mathbf{I}}\in{\mathcal{I}}$}.
$$ 
Analogous results hold for any element of ${\mathcal{E}}'_{S\!S}({\mathfrak{U}}:{\mathfrak{C}})$,
${\mathcal{B}}'_{S\!S}({\mathfrak{U}}:{\mathfrak{C}})$ or ${\mathcal{S}}'_{S\!S}(\supermn:{\mathfrak{C}})$.
%(or ${\mathcal{E}}'_{S\!S}(U_{\mathrm B}:{\mathfrak{C}})$,
%${\mathcal{B}}'_{S\!S}(U_{\mathrm B}:{\mathfrak{C}})$, ${\mathcal{S}}'_{S\!S}(\eucm:{\mathfrak{C}})$).
\end{proposition}
{\it Proof } is omitted here.

\begin{definition}[Sobolev spaces]\label{}
%\proclaim{Definition 2.11 ()}
Let $k$ be a non-negative integer.
\begin{itemize} %roster
\item
We define, for  $u,\, v\in {\mathcal{C}}_{S\!S,0}({\mathfrak{U}}:{\mathfrak{C}})$,
\begin{equation} 
{(\!(}u,v{)\!)}_k=\sum_{|{\mathfrak{a}} |\le k } ( D_X^{\mathfrak{a}}u, D_X^{\mathfrak{a}}v)
\et {\Vert} u{\Vert}_k^2 ={(\!(} u,u {)\!)} _k.
\label{FS-2.19}
\end{equation}
\item
For any  $u ,\, v \in $
${\mathcal{S}}_{S\!S}({\supermn}:{\mathfrak{C}})$, we define 
\begin{equation} 
\begin{gathered}
{(\!(\!(} u,v {)\!)\!)}_k=\sum_{|{\mathfrak{a}} |+l\le k } 
( (1+|X_{\mathrm B} |^2 )^{l/2 }D_X^{\mathfrak{a}}u, 
(1+|X_{\mathrm B} |^2 )^{l/2 }D_X^{\mathfrak{a}}v)\\
\qquad\qquad\qquad\qquad\qquad
\et
{|||}  u {|||} _k^2={(\!(\!(} u,u{)\!)\!)}_k.
\end{gathered}
\label{FS-2.20}
\end{equation} 
\end{itemize}
\end{definition}

Now, we put
$$
\begin{aligned}
{\tilde {\mathcal{L}}}_{S\!S}^2 ({\mathfrak{U}}:{\mathfrak{C}})&=
\{ u \in {\mathcal{C}}_{S\!S,0}({\mathfrak{U}}:{\mathfrak{C}}) \,\big|\, \|u\| < \infty \}, \\
{\tilde {\mathcal{H}}}_{S\!S}^k ({\mathfrak{U}}:{\mathfrak{C}})&=
\{ u \in {\mathcal{C}}_{S\!S}({\mathfrak{U}}:{\mathfrak{C}}) \,\big|\, \|u\|_k < \infty \}
\end{aligned}
$$
and taking the completion of these spaces with respect to
corresponding norms, we get the desired spaces ${\mathcal{L}}_{S\!S}^2 ({\mathfrak{U}}:{\mathfrak{C}})$
and ${\mathcal{H}}_{S\!S}^k({\mathfrak{U}}:{\mathfrak{C}})$. The closure of ${\mathcal{C}}_{S\!S,0}({\mathfrak{U}}:{\mathfrak{C}})$ in
${\mathcal{H}}_{S\!S}^k({\mathfrak{U}}:{\mathfrak{C}})$ is denoted by ${\mathcal{H}}_{S\!S,0}^k({\mathfrak{U}}:{\mathfrak{C}})$.
The spaces ${\clsl}_{S\!S}^2({\mathfrak{U}})$ and $\chsl_{S\!S}^k({\mathfrak{U}})$ are defined
analogously.

{\it Remark}.
We should consider the integral in the last form as the one in the Lebesgue
sense.

\begin{definition} \label{} %\proclaim{Definition 2.12}
Let $1\le r\le\infty$.
$$
{\cdsl}_{L^r,S\!S}(\supermn)=
\{u(X)\in\cssl_{S\!S}(\supermn) \,\big|\, \partial_\theta^a u(X_{\mathrm B})
\in  {\mathcal{D}}_{L^r}(\eucm:{{\mathbb{C}}}) \quad\text{for any $a$}\}
$$
where
$$
{\mathcal{D}}_{L^r}(\eucm:{{\mathbb{C}}})=
\{u(q)\in C^\infty(\eucm:{{\mathbb{C}}}) \,\big|\,\partial_q^\alpha u(q)\in 
L^r(\eucm:{{\mathbb{C}}}) \quad\text{for any $\alpha$} \}.
$$
The topology on ${\cdsl}_{L^r,S\!S}(\supermn)$ is defined by seminorms
$$
p_{k,{\mathbf{I}}:r}(u)=\sum_{|{\mathfrak{a}}|\le k} \Vert \proj_{\mathbf{I}}
(D_X^{\mathfrak{a}}u(X_{\mathrm B}) )\Vert_{L^r}.
$$
\end{definition}

\begin{definition} \label{} %\proclaim{Definition 2.13}
$$
\begin{aligned}
{\mathcal{O}}'_{C}(\eucm:{{\mathbb{C}}}) &=
\{
\Phi(q)\in{\mathcal{D}}'(\eucm:{{\mathbb{C}}})\,\big|\,
(1+|q|^2)^{k/2}D_q^{\alpha}\Phi(q)\in
{\mathcal{B}}'(\eucm:{{\mathbb{C}}}) \\
&\qquad\qquad\qquad\qquad\qquad\qquad\qquad\qquad
\text{for any $\alpha$ and some $k$}
\},\\
{\mathcal{O}}'_{C}(\eucm:{\mathfrak{C}}) &=
\{
\Phi\in{\mathcal{D}}'_{S\!S}(\eucm:{\mathfrak{C}})\,\big|\,
\proj_{\mathbf{I}}(\phi(q))\in {\mathcal{O}}'_{C}(\eucm:{{\mathbb{C}}})
\},\\
\COCSS'(\supermn:{\mathfrak{C}}) &=
\{
\Phi\in{\mathcal{D}}'_{S\!S}(\supermn:{\mathfrak{C}})\,\big|\,
D_X^{\mathfrak{a}}\Phi(X_{\mathrm B})\in
{\mathcal{O}}'_{C}(\eucm:{\mathfrak{C}})
\}.
\end{aligned}
$$
\end{definition}

\begin{remark}
(1) The following assertions follow directly from definitions above and the
standard distribution theory of ~\cite{Sch-l70}:\\
(i) $\Phi\in\COCSS'(\supermn:{\mathfrak{C}})$ iff for any
$\varphi\in\DSS(\supermn:{\mathfrak{C}})$, $\Phi*\varphi\in\SSS(\supermn:{\mathfrak{C}})$. \\
(ii) $({\cdsl}_{L^1,S\!S}(\supermn))'=\cbsl'(\supermn)$.
\newline
(2) Though we don't mention other properties on function or
distribution spaces on ${\mathfrak{U}}\subset \supermn$, but they will be almost comparable
to those in standard case treated in ~\cite{Sch-l70}.
\newline
(3) Sobolev inequalities should be studied separately.
\end{remark}

\subsection{Fourier transformations, definitions and their basic properties}

In this section, we borrow ideas from \cite{{Ber87}, {BM77}, {DeW84-2}, {MZ86}} with necessary modifications.

\subsubsection{Fourier transformations (even case)}
We introduce the Fourier and inverse Fourier transformations
of functions with  even variables.
For $u(x),\, v(\xi)\in {\mathcal{S}}_{S\!S}(\supermo:{\mathfrak{C}}) $, we define
\begin{align}
(F_{\mathrm{e}}u)( \xi ) &=(2\pi\hbar)^{-m/2} \int_{\supermo}dx\, 
e^{ -i\hbar^{-1}\langle x|\xi \rangle} u(x),\label{F-4.1}\\
({\bar{F} }_{\mathrm{e}} v)( \xi ) &= (2\pi \hbar)^{-m/2}
\int_{\supermo} d\xi\,
e^{i\hbar^{-1}\langle y|\xi \rangle} v( \xi ).\label{F-4.2}
\end{align}
\begin{remark}
If $F$ stands for the standard Fourier
transformation on ${\mathcal{S}}({\euc}^m:{\mathbb{C}})$,
then it acts on
${\mathcal{S}}({\euc}^m:{\mathfrak{C}})$ by $(Fu)(p)=\sum_{{\mathbf{I}}\in{\mathcal{I}}}
(Fu_{\mathbf{I}})(p)\sigma^{\mathbf{I}}$
for
$u(q)=\sum_{{\mathbf{I}}\in{\mathcal{I}}}u_{\mathbf{I}}(q)\sigma^{\mathbf{I}}$ with 
$u_{\mathbf{I}}(q)\in {\mathcal{S}}(\eucm:{\mathbb{C}})$.
As $u(x)\in {\mathcal{S}}_{S\!S}(\supermo:{\mathfrak{C}}) $ is 
the Grassmann continuation of $u(q)\in {\mathcal{S}}({\euc}^m:{\mathbb{C}})$,
we need to say
$$
(\widetilde{Fu})(\xi)=(F_{\mathrm{e}}\tilde u)( \xi ) .
$$
\end{remark}

\begin{proposition} \label{F.4.1}%\proclaim{Proposition 4.1}
Let $ u,\,v \in {\mathcal{S}}_{S\!S}(\supermo:{\mathfrak{C}})$.
\begin{gather}
(F_{\mathrm{e}}( \partial_x^{\alpha} u ))( \xi ) 
= (i\hbar^{-1})^{|\alpha|} {\xi}^{\alpha} ( F_{\mathrm{e}} u)( \xi ), \quad
( F_{\mathrm{e}}( x^{\alpha} u))( \xi )
= (i\hbar)^{|\alpha|} 
\partial_{\xi}^{\alpha} ( F_{\mathrm{e}} u)( \xi ).
\label{F-4.3}\\
(F_{\mathrm{e}} (e^{i\hbar^{-1} \langle  x|\xi^{\prime}  \rangle} u))(\xi ) 
= (F_{\mathrm{e}} u)( \xi -\xi^{\prime} ), \quad
(F_{\mathrm{e}} (u( x -x^{\prime})))( \xi ) 
= e^{-i\hbar^{-1} \langle x^{\prime}|\xi \rangle}(F_{\mathrm{e}} u)( \xi ). 
\label{F-4.4}\\
(F_{\mathrm{e}} (u(t x )))( \xi ) = |t|^{-m}  (F_{\mathrm{e}} u)(t^{-1} \xi ) 
\for t \in {\euc}^{\times}={\euc}\setminus\{0\} . 
\label{F-4.5}
\end{gather}
\begin{gather}
{\bar{F}}_{\mathrm{e}}F_{\mathrm{e}} u = u 
\et 
F_{\mathrm{e}} {\bar{F}}_{\mathrm{e}} v = v.  
\label{F-4.6}\\
(u,v)=(F_{\mathrm{e}}u,F_{\mathrm{e}}v) \et \| F_{\mathrm{e}}u \| =\| u {\|}.
\label{F-4.7}
\end{gather}
\begin{equation} 
(F_{\mathrm{e}}\delta)(\xi )= (2\pi\hbar)^{-m/2}
\int_{\supermo}dx\,\delta (x)e^{-i\hbar^{-1}\langle x|\xi \rangle}
=(2\pi\hbar )^{-m/2}.
\label{F-4.8}
\end{equation} 
Moreover,
$F_{\mathrm{e}} : {\mathcal{S}}_{S\!S}(\supermo:{\mathfrak{C}})\to {\mathcal{S}}_{S\!S}(\supermo:{\mathfrak{C}})$
 mapping satisfying
\begin{equation}
{|||} \proj_{\mathbf{I}}(( F_{\mathrm{e}} u )(\xi_{\mathrm B})){|||} _k^2 \le\,
C_{m,\hbar}{|||}  \proj_{\mathbf{I}}(u(x_{\mathrm B}))
{|||} _{k}^2 
\qquad\text{for any ${\mathbf{I}}\in{\mathcal{I}}$}. 
\label{F-4.9}
\end{equation} 
\end{proposition} 

{\it Proof}.
As $\partial_x^\alpha{\tilde u}(x)$
$={\widetilde{(\partial_q^\alpha u)}}(x)$, if $\xi=p\in\eucm$, we
get the first part of \eqref{F-4.3} by
$$
\begin{aligned}
(F_{\mathrm{e}}( \partial_x^{\alpha} {\tilde u} ))( \xi ) |_{\xi=p}
&=(2\pi\hbar )^{-m/2}\int_{\eucm}dq\,e^{-i\hbar^{-1}\langle q|p \rangle}
{\partial_q^\alpha u}(q)\\
&=(i\hbar^{-1})^{|\alpha|} {p}^{\alpha} ( Fu)(p)
=
(i\hbar^{-1})^{|\alpha|} {\xi}^{\alpha} ( F_{\mathrm{e}} {\tilde
u})(\xi)|_{\xi=p}. 
\end{aligned}
$$
The second equality in \eqref{F-4.3} is proved analogously and which
shows that 
$(F_{\mathrm{e}}{\tilde u})(\xi)\in{\mathcal{S}}_{S\!S}(\supermn:{\mathfrak{C}})$. 
Other equalities in \eqref{F-4.3}-\eqref{F-4.7} are
proved as same as the standard case. \eqref{F-4.8} follows by defining
$\langle F_{\mathrm{e}}\delta,\tilde u\rangle
=\langle\delta, F_{\mathrm{e}}\tilde u\rangle
=\int_{\eucm} dp\,\delta(p)(Fu)(p)
=(Fu)(0)
=\int_{\eucm} dq\,u(q)
=\int_{\supermo}dx\,\tilde u(x)
=(F_{\mathrm{e}}\tilde u )(0)$.
\eqref{F-4.9} is a direct consequence 
of the standard theory of Fourier transformation.
$\qquad\qed$
\begin{remark}
The Plancherel formula \eqref{F-4.7} stands for
$F_{\mathrm{e}}^*=\bar{F}_{\mathrm{e}}=F_{\mathrm{e}}^{-1}$.
\end{remark}

\subsubsection{Fourier transformations (odd case)}
For $v(\theta),\,w(\pi)\in {\mathcal{P}}_n({\mathfrak{C}})$,
we define Fourier transformations with $\spin\in {\mathbb{C}}^\times={\mathbb{C}}\setminus\{0\}$ as 
\begin{align}
(F_{\mathrm{o}} v)(\pi) 
&=\spin^{n/2} {\iota}_n
\int_{\superon} d\theta\,
e^{-i\spin^{-1}\langle\theta|\pi\rangle}v(\theta),
\label{F-4.10}\\
({\bar{F}}_{\mathrm{o}} w)(\theta) 
&=\spin^{n/2}{\iota}_n
\int_{\superon} d\pi\,
e^{i\spin^{-1}\langle\theta|\pi\rangle}w(\pi)
\label{F-4.11}
\end{align}
where we put
$$
{\iota}_n=e^{-i{\pi}n(n-2)/4}\et
{\iota}_n^2=i^n(-1)^{{n(n-1)}/{2}}=\begin{cases} 
i & \when n\equiv 1 \mod 4,\\
1 & \when n\equiv 2 \mod 4,\\
i & \when n\equiv 3 \mod 4,\\
1 & \when n\equiv 0 \mod 4.
\end{cases}
$$
\begin{remark}
(1) Clearly, in \eqref{F-4.11}, if we change the role of the variables $\pi$
and $\theta$, we get $F_{\mathrm{o}}={\bar{F}}_{\mathrm{o}}$.\\
%\eqref{F-4.10}. 
(2) Moreover, we may put differently as
\begin{equation}
\begin{aligned}
({\tilde{F}}_{\mathrm{o}} v)(\pi)
&={\spin}^{n/2}\jmath_n 
\int_{\superon} d\theta\,
e^{-\spin^{-1}\langle\theta|\pi\rangle}v(\theta),\\
(\bar{{\tilde{F}}}_{\mathrm{o}} w)(\theta) 
&=\spin^{n/2}\jmath_n
\int_{\superon} d\pi\,
e^{\spin^{-1}\langle\theta|\pi\rangle}w(\pi).
\end{aligned}
\label{F-2.32-bis}
\end{equation}
with
\begin{equation}
\jmath_n=e^{i\pi n(n-1)/2},\quad %/4
\jmath_n^2=(-1)^{n(n-1)/2}
=\begin{cases} 
1 & \when n\equiv 1 \mod 4,\\
-1 & \when n\equiv 2 \mod 4,\\
-1 & \when n\equiv 3 \mod 4,\\
1 & \when n\equiv 0 \mod 4.
\end{cases}
\label{F-4.12}
\end{equation}
\end{remark}

\begin{proposition} \label{F.4.2}
Putting $ \tilde 1= \overbrace{(1,\cdots ,1)}^{n} $ and
$\bar{a}=\tilde 1-a$,
we have, for $v_a\in{\mathfrak{C}}$,
\begin{equation}
\int_{{\fR}^{0|n}} d\theta\, 
e^{-i\spin^{-1}\langle\theta |\pi \rangle}
\theta^a v_a 
=(-i{\spin}^{-1})^{|\bar{a}|}
(-1)^{{|\bar{a}|(|\bar{a}|-1)}/2+\tau(a,\bar{a})}\pi^{\bar{a}} v_a.
\label{F-4.13}
\end{equation}
Moreover, for $ v, \, w \in {\mathcal{P}}_n({\mathfrak{C}})$, we have the following:
\begin{equation} 
\begin{gathered}
(F_{\mathrm{o}}(\partial_\theta^a v))(\pi )=(i\spin^{-1})^{|a|}
(-1)^{n|a|} \pi^a (F_{\mathrm{o}} v)(\pi ),\\
(F_{\mathrm{o}}(\theta^a v))(\pi )=(i\spin^{-1})^{|a|}(-1)^{n|a|} \partial_\pi^a
(F_{\mathrm{o}} v)(\pi ). 
\end{gathered}
\label{F-4.14}
\end{equation}
\begin{equation}
\begin{gathered}
(F_{\mathrm{o}} (e^{i\spin^{-1}\langle \theta|\pi^{\prime} \rangle} v))(\pi ) 
=(F_{\mathrm{o}} v)( \pi -\pi^{\prime} ), \\
(F_{\mathrm{o}} (v( \theta -\theta^{\prime})))( \pi) 
= e^{-i\spin^{-1}\langle \theta^{\prime}|\pi \rangle}(F_{\mathrm{o}} v)( \pi ). 
\end{gathered}
\label{F-4.15}
\end{equation}
\begin{equation}
(F_{\mathrm{o}} (v(s\theta)))(\pi) = s^n  (F_{\mathrm{o}} v)(s^{-1} \pi )
\for s \in {\mathbb{C}}^{\times}.
\label{F-4.16}
\end{equation}
\begin{equation}
F_{\mathrm{o}}{\bar{F}}_{\mathrm{o}} w=w 
\et 
{\bar{F}}_{\mathrm{o}} F_{\mathrm{o}} v=v .
\label{F-4.17}
\end{equation}
\begin{equation}
(v,w)=(F_{\mathrm{o}} v, F_{\mathrm{o}} w) \et \| F_{\mathrm{o}} v \| =\| v {\|}\quad\mbox{if ${\spin}=1$}.
\label{F-4.18}
\end{equation}
\begin{equation}
(F_{\mathrm{o}}\delta)(\pi )=\spin^{n/2} {\iota}_n
\for \delta(\theta)=\theta^{\tilde 1}.
\label{F-4.19}
\end{equation} 
\end{proposition} %proclaim 

{\it Proof}.  
We get, by the definition of integration w.r.t. $\theta$ and \eqref{FS-2.13}, %(2.13),
$$
\begin{aligned}
\int_{{\fR}^{0|n}} d\theta \,
e^{-i\spin^{-1}\langle\theta|\pi \rangle}
\theta^a v_a
&=\int_{{\fR}^{0|n}} d\theta \,
\theta^av_a
\prod(-i\spin^{-1}\theta_j\pi_j)^{{\bar{a}}_j}\\
&=(-i{\spin}^{-1})^{|\bar{a}|}(-1)^{{|\bar{a}|(|\bar{a}|-1)}/2+\tau(a,\bar{a})}\pi^{\bar{a}}v_a.
\end{aligned}
$$
Moreover, we get
$$
\begin{aligned}
\int_{{\fR}^{0|n}} d\pi \,
e^{i\spin^{-1}\langle \theta |\pi \rangle}
&\left\{
\int_{{\fR}^{0|n}} d\theta' \,
e^{-i\spin^{-1}\langle \theta' |\pi \rangle}
\theta^{\prime a} v_a\right\}\\
&=\int_{{\fR}^{0|n}} d\pi \,
e^{i\spin^{-1}\langle \theta |\pi \rangle}
\big[(-i\spin^{-1})^{|\bar{a}|}(-1)^{{|\bar{a}|(|\bar{a}|-1)}/2+\tau(a,\bar{a})}\pi^{\bar{a}}v_a\big]\\
&=(-i\spin^{-1})^n(-1)^{{|a|(|a|-1)}/{2}+{|\bar{a}|(|\bar{a}|-1)}/{2}+\tau(a,\bar{a})+\tau(\bar{a},a)}\theta^a v_a,
\end{aligned}
$$
where we used \eqref{FS-2.13} with $b=\bar{a}$.
By the definition of ${\iota}_n$,   
${\iota}_n^2 (-1)^{n(n-1)/2}=1$ and \eqref{FS-2.12},
we have the Fourier inversion formula \eqref{F-4.17}.
Or, we may prove directly this by changing the order of integration:
$$
\begin{aligned}
{\bar{F}}_{\mathrm{o}} F_{\mathrm{o}} w (\theta) 
&=\spin^n{\iota}_n^2\iint d\pi d\theta'
e^{-i\spin^{-1}\langle \theta'-\theta|\pi \rangle}w(\theta')
=\spin^n{\iota}_n^2 (-1)^{n}\int d\theta'\left\{\int d\pi\,
e^{-i\spin^{-1}\langle\theta'-\theta|\pi\rangle}\right\}w(\theta')\\
&=\spin^n{\iota}_n^2 (-1)^{n}\int d\theta' (i\spin^{-1})^{n}(-1)^{n(n-1)/2}(\theta'-\theta)^{\tilde{1}}w(\theta')
=\int d\theta' \delta(\theta'-\theta)w(\theta').
\end{aligned}
$$
Remarking 
$
\partial_{\theta_j}(e^{-i\spin^{-1}\langle\pi|\theta\rangle}v)
=
e^{-i\spin^{-1}\langle\pi|\theta\rangle}
(-i\spin^{-1}\pi_jv+\partial_{\theta_j}v),
$
we get, after integration with respect to $\theta$,
$$
(F_{\mathrm{o}}\partial_{\theta_j}v)(\pi)=i\spin^{-1}(-1)^n\pi_j F_{\mathrm{o}}v(\pi)
$$
which proves the first equality of \eqref{F-4.14} when $|a|=1$.
Assuming the first equality of \eqref{F-4.14} holds for any $a$ 
satisfying $|a|=l$, we apply
the above for $w(\theta)=(\partial_{\theta}^av)(\theta)$. 
Then, we get
$$
\begin{aligned}
(-1)^{\ell_j(a)}F_{\mathrm{o}}\partial_{\theta}^{\hat a_j}w
=F_{\mathrm{o}}\partial_{\theta_j}w
&=i\spin^{-1}(-1)^n\pi_j F_{\mathrm{o}} w\\
&=(i\spin^{-1})^{|a|+1}(-1)^{n(|a|+1)}\pi_j\pi^aF_{\mathrm{o}}v
=(i\spin^{-1})^{|\hat a_j|}(-1)^{\ell_j(a)+n|{\hat a_j}|}\pi^{\hat a_j}F_{\mathrm{o}} v.
\end{aligned}
$$ 
As before, we put 
${\hat{a}}_j=(a_1,\cdots,a_{j-1},a_j+1,a_{j+1},\cdots, a_n)$
and $\ell_j(a)=\sum_{k=1}^{j-1}a_k$.\\
To prove the Plancherel formula \eqref{F-4.18}, remarking that
$$
\begin{gathered}
\overline{(F_{\mathrm{o}} u)(\pi)}
=\spin^{n/2}{\overline{{\iota}_n}}
\sum_{|a|\le n}(i\spin^{-1})^{|\bar{a}|}(-1)^{{|\bar{a}|(|\bar{a}|-1)}/{2}+\tau(a,{\bar{a}})}\overline{\pi^{\bar{a}}u_a},\\
(F_{\mathrm{o}} u)(\pi)
=\spin^{n/2}{{\iota}_n}
\sum_{|b|\le n}(-i\spin^{-1})^{|\bar{b}|}(-1)^{{|\bar{b}|(|\bar{b}|-1)}/{2}+\tau(b,{\bar{b}})}\pi^{\bar{b}}u'_b,
\end{gathered}
$$
we have
$$
\begin{aligned}
(F_{\mathrm{o}} u,F_{\mathrm{o}} u')
&=\spin^{n}\sum_{|a|\le n,|b|\le n}
\iint  d\pi \overline{d\pi}\,e^{\langle\bar{\pi}|\pi\rangle}
(i\spin^{-1})^{|\bar{a}|}(-1)^{{|\bar{a}|(|\bar{a}|-1)}/{2}+\tau(a,{\bar{a}})}\overline{\pi^{\bar{a}}u_a}\\
&\qquad\qquad\qquad\qquad\qquad
\times
(-i\spin^{-1})^{|\bar{b}|}(-1)^{{|\bar{b}|(|\bar{b}|-1)}/{2}+\tau(b,{\bar{b}})}\pi^{\bar{b}}u'_b\\
&=\sum_{|a|\le n}{\spin}^{n-2|\bar{a}|}\overline{u_a}v_a.
\end{aligned}
%=\sum_{|a|\le n}(u_a,v_a)=(u,v).
$$
This implies \eqref{F-4.18} for ${\spin}=1$. Especially in case $n=2$, $|a|=1$, above holds for any $\spin$.
Other equalities are proved by the analogous
fashion so omitted. $\qquad\qed$

\begin{example}[$n=2$]
For $u(\theta)=u_0+\theta_1\theta_2u_1$ and $v(\pi)=\pi_1 v_1+\pi_2 v_2$ with
$u_0, u_1, v_1, v_2\in\fC$, we have
$$
\begin{aligned}
({F}_{\mathrm{o}}u)(\pi)&={\spin}\int_{{\fR}^{0|2}}d\theta\,e^{-i{\spin}^{-1}\langle \theta|\pi\rangle}u(\theta)
={\spin}(u_1+{\spin}^{-2}\pi_1\pi_2 u_0),\\
({\bar{F}}_{\mathrm{o}}v)(\theta)&={\spin}\int_{{\fR}^{0|2}}d\pi\,e^{i{\spin}^{-1}\langle \theta|\pi\rangle}v(\pi)
={\spin}(-i{\spin}^{-1}\theta_2v_1+i{\spin}^{-1}\theta_1v_2),\\
{\bar{F}}_{\mathrm{o}}({F}_{\mathrm{o}}u)(\theta)&=
{\spin}\int_{{\fR}^{0|2}}d\pi\,e^{i{\spin}^{-1}\langle \theta|\pi\rangle}[{\spin}(u_1+{\spin}^{-2}\pi_1\pi_2 u_0)]
=u_0+\theta_1\theta_2u_1=u(\theta),\\
{F}_{\mathrm{o}}({\bar{F}}_{\mathrm{o}}v)(\pi)&=
{\spin}\int_{{\fR}^{0|2}}d\theta\,e^{-i{\spin}^{-1}\langle \theta|\pi\rangle}[{\spin}(-i{\spin}^{-1}\theta_2v_1+i{\spin}^{-1}\theta_1v_2)]
=\pi_1 v_1+\pi_2 v_2=v(\pi).
\end{aligned}
$$
Therefore, we get
$$
\begin{aligned}
({F}_{\mathrm{o}}u, {F}_{\mathrm{o}}u')
&=\int d\pi\overline{d\pi}\,e^{\langle\bar{\pi}|\pi\rangle}\overline{{F}_{\mathrm{o}}u(\pi)}{F}_{\mathrm{o}}u'(\pi)\\
&=\frac{\partial}{\partial\pi_2}\frac{\partial}{\partial\pi_1}\frac{\partial}{\partial\bar{\pi}_1}\frac{\partial}{\partial\bar{\pi}_2}
[e^{\langle\bar{\pi}|\pi\rangle}({\spin}{\overline{u_1}}-{\spin}^{-1}{\overline{u_0}}{\bar\pi}_1{\bar\pi}_2)
({\spin}u_1+{\spin}^{-1}\pi_1\pi_2 u_0)\bigg|_{\pi=\bar{\pi}=0}\\
&={\spin}^2{\overline{u_1}}u_1+{\spin}^{-2}{\overline{u_0}}{u_0},%={\overline{u_1}}u_1+{\overline{u_0}}{u_0},
\end{aligned}
$$
which implies that the Plancherel formula $({F}_{\mathrm{o}}u, {F}_{\mathrm{o}}u')=(u,u')$ for above $u, u'$ holds only when ${\spin}=1$.\\
Analogously but for any $\spin\neq0$,
$$
\begin{aligned}
({F}_{\mathrm{o}}v, {F}_{\mathrm{o}}v')
&=\int d\pi\overline{d\pi}\,e^{\langle\bar{\pi}|\pi\rangle}\overline{{F}_{\mathrm{o}}v(\pi)}{F}_{\mathrm{o}}v'(\pi)\\
&=\frac{\partial}{\partial\pi_2}\frac{\partial}{\partial\pi_1}\frac{\partial}{\partial\bar{\pi}_1}\frac{\partial}{\partial\bar{\pi}_2}
[e^{\langle\bar{\pi}|\pi\rangle}(\overline{v_1}\bar{\pi}_1-\overline{v_2}\bar{\pi}_2)(\pi_1 v'_1-\pi_2 v'_2)]\bigg|_{\pi=\bar{\pi}=0}\\
&=\overline{v_1}v'_1+\overline{v_2}v'_2=(v,v').\qquad/\!\!/
\end{aligned}
$$
\end{example}

\begin{example}[$n=3$]
Let $u(\theta)=u_0+\theta_1\theta_2 u_1+\theta_2\theta_3 u_2+\theta_1\theta_3 u_3$,
$v(\theta)=\theta_1v_1+\theta_2v_2+\theta_3v_3+\theta_1\theta_2\theta_3v_4$ with
$u_0, u_1, u_2, u_3, v_1, v_2, v_3, v_4\in\fC$, 
we have the following inversion formula:
$$
\begin{aligned}
({F}_{\mathrm{o}}u)(\pi)&={\spin}^{3/2}\iota_3\int_{{\fR}^{0|3}}d\theta\,e^{-i{\spin}^{-1}\langle \theta|\pi\rangle}u(\theta)\\
&={\spin}^{3/2}\iota_3
[i{\spin}^{-3}\pi_3\pi_2\pi_1u_0-i{\spin}^{-1}\pi_3u_1-i{\spin}^{-1}\pi_1u_2+i{\spin}^{-1}\pi_2u_3],\\
{\bar{F}}_{\mathrm{o}}({F}_{\mathrm{o}}u)(\theta)
&={\spin}^{3/2}\iota_3\int_{{\fR}^{0|3}}d\pi\,e^{i{\spin}^{-1}\langle \theta|\pi\rangle}({F}_{\mathrm{o}}u)(\pi)\\
&={\spin}^{3}\iota_3^2\int_{{\fR}^{0|3}}d\pi\,e^{i{\spin}^{-1}\langle \theta|\pi\rangle}
[i{\spin}^{-3}\pi_3\pi_2\pi_1u_0-i{\spin}^{-1}\pi_3u_1-i{\spin}^{-1}\pi_1u_2+i{\spin}^{-1}\pi_2u_3]\\
&={\spin}^{3}\iota_3^2[-i{\spin}^{-3}u_0+i{\spin}^{-3}\theta_2\theta_1u_1+i{\spin}^{-3}\theta_3\theta_2u_2+i{\spin}^{-3}
\theta_3\theta_1u_3]\\
&=u_0+\theta_1\theta_2 u_1+\theta_2\theta_3 u_2+\theta_1\theta_3 u_3.
\end{aligned}
$$

Since
$$
\begin{gathered}
\overline{({F}_{\mathrm{o}}u)(\bar{\pi})}={\overline{({F}_{\mathrm{o}}u)}}(\bar{\pi})={\spin}^{3/2}{\overline{\iota_3}}
[-i{\spin}^{-3}\overline{u_0}\bar{\pi}_1\bar{\pi}_2\bar{\pi}_3
+i{\spin}^{-1}\overline{u_1}\bar{\pi}_3+i{\spin}^{-1}\overline{u_2}\bar{\pi}_1-i{\spin}^{-1}\overline{u_3}\bar{\pi}_2],\\
({F}_{\mathrm{o}}u')(\pi)={\spin}^{3/2}\iota_3
[i{\spin}^{-3}\pi_3\pi_2\pi_1u'_0-i{\spin}^{-1}\pi_3u'_1-i{\spin}^{-1}\pi_1u'_2+i{\spin}^{-1}\pi_2u'_3],
\end{gathered}
$$
and
$$
\begin{aligned}
{\overline{({F}_{\mathrm{o}}u)}}(\bar{\pi})({F}_{\mathrm{o}}u')(\pi)
&={\spin}^{3}
[{\spin}^{-6}\overline{u_0}\bar{\pi}_1\bar{\pi}_2\bar{\pi}_3\pi_3\pi_2\pi_1u'_0
+{\spin}^{-2}\overline{u_1}\bar{\pi}_3\pi_3u'_1\\
&\qquad\qquad\qquad\qquad\qquad\qquad
+{\spin}^{-2}\overline{u_2}\bar{\pi}_1\pi_1u'_2
+{\spin}^{-2}\overline{u_3}\bar{\pi}_2\pi_2u'_3]
+{\cdots},
\end{aligned}
$$
where the term $(\cdots)$ vanishes after integration w.r.t. $d\pi{\overline{d\pi}}$,
we get
$$
\begin{aligned}
({F}_{\mathrm{o}}u, {F}_{\mathrm{o}}u')
&=\iint_{{\fR}^{0|3}\times{\fR}^{0|3}}d\pi{\overline{d\pi}}e^{\langle\bar{\pi}|\pi\rangle}
{\overline{({F}_{\mathrm{o}}u)}}(\bar\pi)({F}_{\mathrm{o}}u')(\pi)\\
&={\spin}^{-3}\overline{u_0}u_0'+{\spin}(\overline{u_1}u_1'+\overline{u_2}u_2'+\overline{u_3}u_3')
=\sum_{|a|=0,2}{\spin}^{3-2|\bar{a}|}\overline{u_a}u_a'.
\end{aligned}
$$
Therefore, the Plancherel formula holds for ${\spin}=1$.
\end{example}

\subsubsection{Fourier transformations (mixed case)}
Putting 
$$
c_{m,n}=(2\pi\hbar )^{-m/2}\spin^{n/2}{\iota}_n \et \langle X|\Xi \rangle=\hbar^{-1}\langle x|\xi\rangle+\spin^{-1}\langle\theta|\pi \rangle\in\rev,
$$
for any $u(X)=\sum_a\theta^a u_a(x)$,
$v(\Xi)=\sum_b\pi^b v_b(\xi)\in {\SSS}({\supermn}:{\mathfrak{C}})$,
we define %Fourier transformations as
\begin{equation}
\begin{aligned}
({\mathcal{F}}u)(\xi,\pi)
&= c_{m,n}
\int_{\supermn}dX\,e^{-i\langle X|\Xi \rangle} u(X)\\
&=(2\pi\hbar )^{-m/2}\spin^{n/2}{\iota}_n
\iint_{{\fR}^{m|0}_x\times{\fR}^{0|n}_\theta}dx d\theta\,e^{-i\hbar^{-1}\langle x|\xi \rangle-i{\spin}^{-1}\langle \theta|\pi \rangle} \sum_{a}\theta^au_a(x)\\
&=\sum_{|a|\le n}(2\pi{\hbar})^{-m/2}\int_{{\fR}^{m|0}_x}dx\,e^{-i\hbar^{-1}\langle x|\xi \rangle}
\bigg[
\spin^{n/2}{\iota}_n\int_{{\fR}^{0|n}_\theta}d\theta\,e^{-i{\spin}^{-1}\langle \theta|\pi \rangle} \theta^a\bigg]u_a(x)\\
&=\sum_{|a|\le n}
[(F_{\mathrm{o}}\theta^a)(\pi)](F_{\mathrm{e}} u_a)(\xi).
%&=\spin^{n/2}{\iota}_n\sum_a
%(-1)^{{\frac {|\bar{a}|(|\bar{a}|-1)}2}+\tau(a,\bar{a})} \\
%&\qquad\qquad\qquad\qquad
%\times
%(((F_{\mathrm{e}} u_a)(\xi))_{\mathrm{ev}}+(-1)^n((F_{\mathrm{e}} u_a)(\xi))_{\mathrm{od}})\pi^{\bar{a}}.
\end{aligned}
\label{F-4.20} 
\end{equation}
\begin{equation}
\begin{aligned}
({\bar {\mathcal{F}}}v)(x,\theta)
&= c_{m,n}
\int_{\supermn} d\Xi\,e^{i\langle X|\Xi \rangle}v(\Xi)\\
&= (2\pi\hbar )^{-m/2}\spin^{n/2}{\iota}_n
\iint_{{\fR}^{m|0}_{\xi}\times{\fR}^{0|n}_{\pi}}d{\xi} d\pi\,e^{i\hbar^{-1}\langle x|\xi \rangle+i{\spin}^{-1}\langle \theta|\pi \rangle} \sum_{b}\pi^b v_b(x)\\
&=\sum_{|b|\le n}(2\pi{\hbar})^{-m/2}\int_{{\fR}^{m|0}_{\xi}}d{\xi}\,e^{i\hbar^{-1}\langle x|\xi \rangle}
\bigg[
\spin^{n/2}{\iota}_n\int_{{\fR}^{0|n}_\pi}d\pi\,e^{i{\spin}^{-1}\langle \theta|\pi \rangle} \pi^b\bigg]v_b(\xi)\\
&=\sum_{|b|\le n}
[({\bar{F}}_{\mathrm{o}}\pi^b)(\theta)]({\bar{F}}_{\mathrm{e}} v_b)(x).
\end{aligned}
\label{F-4.21}
\end{equation}

\begin{proposition} %\proclaim{Proposition 4.3}
For any $u,\, v \in {{\mathcal{S}}}_{S\!S}({\supermn}:{\mathfrak{C}})$,
\begin{equation}
\begin{gathered}
({\mathcal{F}}(D_X^{\mathfrak{a}}u))(\xi)
= (i\hbar^{-1})^{|\alpha|} (i\spin^{-1})^{|a|}(-1)^{n|a|}
{\Xi}^{\mathfrak{a}} 
( {\mathcal{F}} u)( \Xi),\\
( {\mathcal{F}}(X^{\mathfrak{a}}u))(\Xi)
= (i\hbar)^{|\alpha|}(i \spin)^{|a|}(-1)^{n|a|}
D_{\Xi}^{\mathfrak{a}} 
({\mathcal{F}} u)( \Xi).
\end{gathered}
\label{F-4.22}
\end{equation}
\begin{equation}
\begin{gathered}
({\mathcal{F}}(e^{i\langle X|\Xi' \rangle}u))(\Xi) 
= ({\mathcal{F}} u)( \Xi -\Xi^{\prime}),\\
({\mathcal{F}} (u(X -X')))(\Xi)
= e^{-i\langle \Xi |X' \rangle}
({\mathcal{F}} u)( \Xi).
\end{gathered}
\label{F-4.23}
\end{equation}
\begin{equation} 
({\mathcal{F}} u)( t\xi, s\pi )
= |t|^{-m} s^{n} 
({\mathcal{F}} u)(t^{-1} \xi, s^{-1} \pi )
\for t\in {\mathbb{R}}^{\times},s \in {\mathbb{C}}^{\times} . 
\label{F-4.24}
\end{equation}
\begin{equation}
{\mathcal{F}}{\bar {\mathcal{F}}}u =u \et 
{\bar {\mathcal{F}}}{\mathcal{F}}v =v.
\label{F-4.25}
\end{equation}
\begin{equation}
({\mathcal{F}}u , {\mathcal{F}}v)=(u,v) \et \| {\mathcal{F}}u \| =\| u {\|}\for {\spin}=1.
%\qquad\text{if $\spin=\pm1$ and $n$ is even}.
\label{F-4.26}
\end{equation} 
If we define $\delta(X)=\delta(x)\delta(\theta)$, then
\begin{equation} 
({\mathcal{F}} \delta)(\Xi)= (F_{\mathrm{e}}\delta)(\xi) (F_{\mathrm{o}}\delta)(\pi )
=c_{m,n}.
\label{F-4.27}
\end{equation}
${\mathcal{F}}:{\mathcal{S}}_{S\!S}({\supermn}:{\mathfrak{C}}) \to {\mathcal{S}}_{S\!S}({\supermn}:{\mathfrak{C}})$ 
gives a continuous linear mapping satisfying
\begin{equation}
{|||}  \proj_{\mathbf{I}}((\partial_\pi^{\bar{a}}{\mathcal{F}} u)(\Xi_{\mathrm B})){|||}  _k 
\le  C_{m,n}^{\prime}
{|||}  \proj_{\mathbf{I}}((\partial_\theta^a u)(X_{\mathrm B})){|||}  _{k}
\qquad\text{for each \;${\mathbf{I}}\in{\mathcal{I}}$}. 
\label{F-4.28}
\end{equation}
\end{proposition} 
\par
{\it Proof}. Combining above results, we have readily these statements. $\qquad\qed$

\begin{remark}
Since by the formal definition of $\delta$-function, we have
$$
\begin{aligned}
\int_{\euc^m}dp\,\overline{{F}\bar{u}}(p){F}v(p)
&=(2\pi\hbar)^{-m}\int_{\euc^m}dp\,\overline{[\int_{\euc^m}dq\,e^{-i\hbar^{-1}qp}\overline{u(q)}]}[\int_{\euc^m}dq'\,e^{-i\hbar^{-1}q'p}v(q')]\\
&=\iint dq dq'[(2\pi\hbar)^{-m}\int dp\,e^{-i{\hbar}^{-1}\langle q-q'|p\rangle}]u(q)v(q')
=\int_{\euc^m}dq\,u(q)v(q),
\end{aligned}
$$
and
for $A=(A_{jk})$, putting
$$
\langle p,Ap \rangle=\sum_{j,k=1}^m p_jA_{jk}p_k,\quad
\langle D_q, AD_q\rangle=\sum_{j,k=1}^m (-i\hbar\partial_{q_j})A_{jk}(-i\hbar\partial_{q_k})\with
D_{q_j}=-i\hbar\partial_{q_j},
$$
we have
$$
(2\pi\hbar)^{-m/2}\int_{\euc^m} dp\,\langle p,Ap \rangle^{\ell}\overline{\hat{\bar{u}}}(p)
=\langle D_q, AD_q\rangle^{\ell}u(q).
$$

Therefore, we want to ask whether following claim holds or not:%\eqref{F-4.18}, 
\begin{claim}\label{claim2-1} %\proclaim{Lemma 4.4}
Let $u,v\in{{\mathcal{S}}}_{S\!S}(\supermn:{\mathfrak{C}})$. Then, we have
$$
\int_{\supermn} dX\, u(X)v(X)=
\int_{\supermn} d\Xi \,\overline{({\mathcal{F}}\bar{u})}(\Xi)
({\mathcal{F}}v)(\Xi).
\label{F-4.29}
$$
\end{claim}

But Claim \ref{claim2-1} doesn't hold in general:
For example, take $u(\theta)=u_0+\theta_1\theta_2 u_1$, $u'(\theta)=u'_0+\theta_1\theta_2 u'_1$, then
$$
\int_{\fR^{0|2}}d\theta (u_0+\theta_1\theta_2 u_1)(u'_0+\theta_1\theta_2 u'_1)=
-\int_{\fR^{0|2}}d\pi \overline{\hat{\bar{u}}}(\pi)\hat{u'}(\pi).
$$
\begin{quotation}
{\small
In fact
$$
\int_{\fR^{0|2}}d\theta (u_0+\theta_1\theta_2 u_1)(u'_0+\theta_1\theta_2 u'_1)=u_0u'_1+u_1u'_0,
$$
$$
\begin{gathered}
\overline{u(\cdot)}(\bar\theta)=\overline{u_0+\theta_1\theta_2 u_1}=\overline{u_0}+\overline{u_1}\overline{\theta_1\theta_2}
=\overline{u_0}-\bar{\theta}_1\bar{\theta}_2\overline{u_1}=\bar{u}(\bar\theta),\\
{\spin}\int_{\fR^{0|2}}d{\bar{\theta}}\,e^{-i{\spin}^{-1}\langle\bar\theta|\bar\pi\rangle}\overline{u(\cdot)}(\bar\theta)
={\spin}^{-1}\bar\pi_1\bar\pi_2\overline{u_0}-{\spin}\overline{u_1},\\
\overline{\hat{\overline{u}}(\bar\pi)}
=-{\spin}^{-1}\pi_1\pi_2{u_0}-{\spin}{u_1},\\
\int_{\fR^{0|2}}d\pi\,(-{\spin}^{-1}\pi_1\pi_2{u_0}-{\spin}{u_1})({\spin}^{-1}\pi_1\pi_2u'_0+{\spin}u'_1)
=-{u_0}u'_1-u_1u'_0.\qquad/\!\!/
\end{gathered}
$$
}
\end{quotation}

But, for $v(\theta)=\theta_1 v_1+\theta_2 v_2$, $v'(\theta)=\theta_1 v'_1+\theta_2 v'_2$,
Claim \ref{claim2-1} does hold:
$$
\int_{\fR^{0|2}}d\theta(\theta_1 v_1+\theta_2 v_2)(\theta_1 v'_1+\theta_2 v'_2)
=\int_{\fR^{0|2}}d\pi\,\overline{\hat{\overline{v}}(\cdot)}(\pi)\hat{v'}(\pi).
$$
\begin{quotation}
{\small
In fact
$$
\int_{\fR^{0|2}}d\theta(\theta_1 v_1+\theta_2 v_2)(\theta_1 v'_1+\theta_2 v'_2)
=((v_1)_{\mathrm{ev}}-(v_1)_{\mathrm{od}})v'_2-((v_2)_{\mathrm{ev}}-(v_2)_{\mathrm{od}})v'_1,
$$
$$
\begin{gathered}
\overline{\theta_1 v_1+\theta_2 v_2}=\overline{v_1}\bar\theta_1+\overline{v_2}\bar\theta_2,\\
{\spin}\int_{\fR^{0|2}}d{\bar{\theta}}\,e^{-i{\spin}^{-1}\langle\bar\theta|\bar\pi\rangle}
(\overline{v_1}\bar\theta_1+\overline{v_2}\bar\theta_2)
=-i\bar\pi_2((\overline{v_1})_{\mathrm{ev}}-(\overline{v_1})_{\mathrm{od}})+i\bar\pi_1
((\overline{v_2})_{\mathrm{ev}}-(\overline{v_2})_{\mathrm{od}}),\\
\overline{\hat{\overline{v}}(\cdot)}(\pi)
=i(({v_1})_{\mathrm{ev}}-({v_1})_{\mathrm{od}})\pi_2-i(({v_2})_{\mathrm{ev}}-({v_2})_{\mathrm{od}})\pi_1,\\
{\begin{aligned}
\int_{\fR^{0|2}}d\pi\,&[i(({v_1})_{\mathrm{ev}}-({v_1})_{\mathrm{od}})\pi_2-i(({v_2})_{\mathrm{ev}}-(\overline{v_2})_{{od}})\pi_1](-i{\pi}_2v'_1+i{\pi}_1v'_2)\\
&\qquad\qquad
=(({v_1})_{\mathrm{ev}}-({v_1})_{\mathrm{od}})v'_2-(({v_2})_{\mathrm{ev}}-({v_2})_{\mathrm{od}})v'_1.\qquad/\!\!/
\end{aligned}}
\end{gathered}
$$
}
\end{quotation}
\end{remark}

%\newpage %\vspace{2cm}
\section{Qi's example of weakly hyperbolic equation}
In 1958, M-y. Qi considered the following IVP:
\begin{equation}
v_{tt}-L(t,\partial_q)v=0\with
 L(t)=L(t,\partial_q)=t^2\partial_q^2+(4k+1)\partial_q \et v(0,q)=\varphi(q),\; v_t(0,q)=0.
\label{qi1}
\end{equation}
In Dreher and Witt~\cite{DW05}, following claim is cited from Qi~\cite{qi58}:
\begin{claim}
For suitably chosen $c_{jk}\neq{0}$, 
$$
v(t,q)=\sum_{j=0}^k c_{jk}t^{2j}\varphi^{(j)}\bigg(q+\frac{t^2}{2}\bigg)
$$
gives the solution of \eqref{qi1}.
\end{claim}

Though Qi uses the knowledge of Euler-Poisson equation and the Riemann-Louville fraction integral, we generalize the method of characteristics to a system of PDOp using superanalysis to have readily
\begin{theorem}
$$
v(t,q)
=(2\pi)^{-1/2}\int d{p}\,e^{iqp}e^{it^2p/2}\sum_{\ell=0}^k\frac{2^{2\ell}k!}{(2\ell)!(k-\ell)!}t^{2\ell}(i{p})^{\ell}\hat{\varphi}(p)
=\sum_{\ell=0}^k\frac{2^{2\ell}k!}{(2\ell)!(k-\ell)!}t^{2\ell}\varphi^{(\ell)}\bigg(q+\frac{t^2}{2}\bigg).
$$
\end{theorem}

\subsection{A systemization and superspace setting}
Putting $w=v_t-tv_q$, we have
$$
\partial_tw=v_{tt}-v_q-tv_{tq}=t^2v_{qq}+(4k+1)v_q-v_q-tv_{tq}
=-t w_q+4kv_q,
$$
then
\begin{equation}
i\partial_t U={\mathbb{H}}(t,\partial_q)U, \quad 
U=U(t,q)=\begin{pmatrix}
v(t,q)\\
w(t,q)
\end{pmatrix} \et
U(0,q)={\underline{U}}(q)
=\begin{pmatrix}
{\underline{v}}(q)\\
{\underline{w}}(q)
\end{pmatrix},
\label{ord-Qi}
\end{equation}
with
$$
{\mathbb{H}}(t,\partial_q)=i\begin{pmatrix}
t\partial_q&1\\
4k\partial_q&-t\partial_q
\end{pmatrix}.
$$

Preparing odd variables $\theta_1, \theta_2$, we define an operator
\begin{equation}
{\mathcal{H}}(t,\partial_x,\theta,\partial_\theta)
=it\partial_x\bigg(1-\theta_1\frac{\partial}{\partial\theta_1}-\theta_2\frac{\partial}{\partial\theta_2}\bigg)
+4ik\partial_x\theta_1\theta_2
-i\frac{\partial^2}{\partial\theta_1\theta_2}
\end{equation}
which acts on $u(t,x,\theta)=v(t,x)+w(t,x)\theta_1\theta_2$.
Then, we reformulate \eqref{ord-Qi} as follows:
\begin{equation}
i\pdt u(t,x,\theta)={\mathcal{H}}(t,\partial_x,\theta,\partial_\theta)u(t,x,\theta)
\with
u(0,x,\theta)={{\underline{v}}}(x)+{{\underline{w}}}(x)\theta_1\theta_2.
\label{superQi}
\end{equation}

Introducing Fourier transformation w.r.t. odd variables, we have a supersmooth function
\begin{equation}
{\mathcal{H}}(t,\xi,\theta,\pi)
=it\xi\langle\theta|\pi\rangle-4k{\xi}\theta_1\theta_2+i\pi_1\pi_2,
\label{SH-Qi}
\end{equation}
which is the Hamilton function corresponding to ${\mathcal{H}}(t,\partial_x,\theta,\partial_\theta)$.

\subsection{A solution of the Hamilton-Jacobi equation by direct method}
We solve the following Hamilton-Jacobi equation directly:
\begin{equation}
\left\{
\begin{aligned}
&{\mathcal{S}}_t+it{\mathcal{S}}_x\langle\theta|{\mathcal{S}}_{\theta}\rangle-4k{\mathcal{S}}_x\theta_1\theta_2
+i{\mathcal{S}}_{\theta_1}{\mathcal{S}}_{\theta_2}=0,\\
&{\mathcal{S}}(0,x,\xi,\theta,\pi)=\langle x|\xi\rangle+\langle \theta|\pi\rangle.
\end{aligned}
\right.
\label{HJ}
\end{equation}
Decomposing ${\mathcal{S}}(t,x,\xi,\theta,\pi)$ as
$$
\begin{aligned}
{\mathcal{S}}(t,x,\xi,\theta,\pi)={\mathcal{S}}(t,x,\xi,0,0)
+&X(t,x,\xi)\theta_1\theta_2+Y(t,x,\xi)\theta_1\pi_1+\tilde{Y}(t,x,\xi)\theta_2\pi_2\\
&+V(t,x,\xi)\theta_1\pi_2+\tilde{V}(t,x,\xi)\theta_2\pi_1
+Z(t,x,\xi)\pi_1\pi_2+W(t,x,\xi)\theta_1\theta_2\pi_1\pi_2,
\end{aligned}
$$
we calculate $X,Y, \tilde{Y}, V, \tilde{V}, Z, W$ which are shown also independent of $x$.

(0) Taking $\theta=0$, $\pi=0$ in \eqref{HJ}, we have 
$$
{\mathcal{S}}(t,x,\xi,0,0)_t=0\with {\mathcal{S}}(0,x,\xi,0,0)=\langle x|\xi\rangle,
$$
which gives
$$
{\mathcal{S}}(t,x,\xi,0,0)=\langle x|\xi\rangle.
$$

(1) Differentiating \eqref{HJ} w.r.t $\theta_1$ and $\theta_2$ and restricting to $\theta=\pi=0$, 
we have
\begin{equation}
X_t=4k\xi-2it\xi X-iX^2\with X(0)=0\where X=\partial_{\theta_2}\partial_{\theta_1}{\mathcal{S}}(t,x,\xi,\theta,\pi)\big|_{\theta=\pi=0}.
\label{815-1}
\end{equation}
Moreover, differentiating \eqref{815-1} once more w.r.t $x$, we put $\tilde{X}=X_x$ which gives
$$
\tilde{X}_t+2it{\xi}\tilde{X}+2iX\tilde{X}=0\with \tilde{X}(0)=0.
$$
Therefore $\tilde{X}(t)=0$ which implies $X(t,x,{\xi})$ is independent of $x$.

To solve \eqref{815-1}, we may associate the 2nd order ODE:
\begin{equation}
\ddot\phi+2it\xi\dot\phi-4ik\xi\phi=0\with \dot\phi(0)=0,
\label{2*15}
\end{equation}
from which we have a solution $X(t,{\xi})=-i\frac{\dot{\phi}(t)}{\phi(t)}$ of \eqref{815-1}.

For the sake of notational simplicity, we rewrite \eqref{2*15} as
\begin{equation}
\ddot\phi+\alpha t\dot \phi+\beta\phi=0\with \dot\phi(0)=0.\quad \alpha=2i\xi,\; \beta=-i4k\xi.
\end{equation}
This equation is solvable in polynomial w.r.t. $t$:
Putting $\phi(t)=\sum_{j=0}^{\infty}c_jt^j$, we have
$$
\dot\phi=\sum_{j=1}^{\infty}jc_jt^{j-1},\quad
\ddot\phi=\sum_{j=2}^{\infty}j(j-1)c_jt^{j-2}.
$$
Then, the coefficients of $t^j$ are given by
{\allowdisplaybreaks
\begin{align*}
t^0: &2c_2+\beta c_0=0 \Longrightarrow c_2=-\frac{\beta}{2}c_0,\\
t^1: &3{\cdot}2c_3+\alpha c_1+\beta c_1=0 \Longrightarrow c_3=-\frac{1}{3{\cdot}2}(\alpha+\beta)c_1,\\
t^2: &4{\cdot}3c_4+2\alpha c_2+\beta c_2=0 \Longrightarrow c_4=-\frac{1}{4{\cdot}3}(2\alpha+\beta)c_2=(-1)^2\frac{1}{4!}(2\alpha+\beta){\beta}c_0,\\
t^3: &5{\cdot}4c_5+3\alpha c_3+\beta c_3=0,\\
&{\cdots}\\
t^{2\ell-2}: &{2\ell}(2\ell-1)c_{2\ell}+(2\ell-2)\alpha c_{2\ell-2}+\beta c_{2\ell-2}=0,\\
t^{2\ell-1}: &(2\ell+1)2\ell c_{2\ell+1}+(2\ell-1)\alpha c_{2\ell-1}+\beta c_{2\ell-1}=0, \quad\mbox{etc.}
\end{align*}}
Since $\dot\phi(0)=0$ implies $c_1=0$, we have $c_{2\ell+1}=0$ for any $\ell$.
Moreover, putting $c_0=1$, i .e. $\phi(0)=1$, we have
$$
c_{2\ell}=\frac{(-1)^{\ell}(2\alpha)^{\ell}}{(2\ell)!}\bigg(\frac{\beta}{2\alpha}\bigg)_{\ell}
\where (x)_{\ell}=x(x+1){\cdots}(x+\ell-1).
$$
In our case, we have
$\beta(2\alpha)^{-1}=-k$ and $(-k)_{\ell}=(-1)^{\ell}\frac{k!}{(k-\ell)!}$, therefore
$$
\phi(t)=\sum_{\ell=0}^{k}c_{2\ell}t^{2\ell}=\sum_{\ell=0}^{k}\frac{(-1)^{\ell}(4i{\xi})^{\ell}\big(-k\big)_{\ell}}{(2\ell)!}t^{2\ell}
=\sum_{\ell=0}^{k}\frac{4^{\ell}k!}{(2\ell)!(k-\ell)!}(i\xi)^{\ell}t^{2\ell}.
$$

(2) $Y=Y(t,{\xi})=\partial_{\pi_1}\partial_{\theta_1}{\mathcal{S}}(t,x,\xi,0,0)$ satisfies
\begin{equation}
Y_t +it{\xi}Y+iXY=0\with Y(0)=1.
\end{equation}
From this, we have
$$
\dt(\log (Y{\phi}))=-it{\xi},\quad Y(t)=\frac{e^{-it^2{\xi}/2}}{\phi(t)}.
$$
Same relation holds for $\tilde{Y}=\partial_{\pi_2}\partial_{\theta_2}{\mathcal{S}}(t,x,\xi,0,0)$. 

(3) $V=\partial_{\pi_2}\partial_{\theta_1}{\mathcal{S}}(t,x,\xi,0,0)$ satisfies
$$
V_t+it{\xi}V+XV=0  \with V(0)=0.
$$
From this, $V(t)=0$. Analogously, $\tilde{V}=\partial_{\pi_1}\partial_{\theta_2}{\mathcal{S}}(t,x,\xi,0,0)=0$.

(4) For $Z=Z(t,{\xi})=\partial_{\pi_2}\partial_{\pi_1}{\mathcal{S}}(t,x,\xi,0,0)$, we have 
\begin{equation}
Z_t +iY^2=0\with Z(0)=0.
\end{equation}
Therefore, 
$$
Z(t)=-i\int_0^t ds\,Y^2(s)=-i\int_0^t ds\,\frac{e^{-is^2{\xi}}}{\phi^2(s)}.
$$

(5) Putting 
$$
W=W(t,{\xi})=\partial_{\pi_2}\partial_{\pi_1}\partial_{\theta_2}\partial_{\theta_1}{\mathcal{S}}(t,x,\xi,0,0),
$$
and remarking 
$Z_x=\partial_x\partial_{\pi_2}\partial_{\pi_1}{\mathcal{S}}(t,x,\xi,0,0)=0$,
we have
\begin{equation}
W_t +2it{\xi}W+2iXW =0\with W(0)=0.
\end{equation}
Therefore, $W=0$.

Now, we define
$$
{\mathcal{D}}(t,{\xi},\theta,\pi)=\sdet
\begin{pmatrix}
{\mathcal{S}}_{x{\xi}}&{\mathcal{S}}_{x\pi_1}&{\mathcal{S}}_{x\pi_2}\\
{\mathcal{S}}_{\theta_1{\xi}}&{\mathcal{S}}_{\theta_1\pi_1}&{\mathcal{S}}_{\theta_1\pi_2}\\
{\mathcal{S}}_{\theta_2{\xi}}&{\mathcal{S}}_{\theta_2\pi_1}&{\mathcal{S}}_{\theta_2\pi_2}
\end{pmatrix}
=Y^{-2}(t,{\xi})={\mathcal{A}}^2(t,{\xi},\theta,\pi).
$$

\subsection{Quantization}
Using above defined ${\mathcal{S}}$ and ${\mathcal{A}}$, we construct a function
$$
u(t,x,\theta)=(2\pi)^{-1/2}\int d{\xi} d\pi\, {\mathcal{A}}(t,x,{\xi},\theta,\pi)e^{i{\mathcal{S}}(t,x,{\xi},\theta,\pi)}\hat{\unbu}(\xi,\pi).
$$
It is shown that this gives a solution of \eqref{superQi}.

Since
$$
\int d\pi \, e^{iY\langle\theta|\pi\rangle +iZ\pi_1\pi_2}(\hat{{\underline{v}}}(\xi)\pi_1\pi_2+\hat{{\underline{w}}}(\xi))
=\hat{{\underline{v}}}(\xi)+(Y^2\theta_1\theta_2+iZ)\hat{{\underline{w}}}(\xi),
$$
we have
$$
\begin{aligned}
(2\pi)^{-1/2}\int d{\xi} d\pi\, &{\mathcal{A}}e^{i{\mathcal{S}}}(\hat{{\underline{v}}}(\xi)\pi_1\pi_2+\hat{{\underline{w}}}(\xi))\\
&=(2\pi)^{-1/2}\int d{\xi}\,{\mathcal{A}}e^{i\langle x|{\xi}\rangle+iX\theta_1\theta_2}
[\hat{{\underline{v}}}(\xi)+(Y^2\theta_1\theta_2+iZ)\hat{{\underline{w}}}(\xi)]\\
&=(2\pi)^{-1/2}\int d{\xi}\,Y^{-1}e^{i\langle x|{\xi}\rangle}(1+iX\theta_1\theta_2)
(\hat{{\underline{v}}}(\xi)+iZ\hat{{\underline{w}}}(\xi)+Y^2\theta_1\theta_2\hat{{\underline{w}}}(\xi))\\
&=(2\pi)^{-1/2}\int d{\xi}\,e^{i\langle x|{\xi}\rangle}Y^{-1}(\hat{{\underline{v}}}(\xi)+iZ\hat{{\underline{w}}}(\xi))\\
&\qquad
+(2\pi)^{-1/2}\int d{\xi}\,e^{i\langle x|{\xi}\rangle}Y^{-1}[iX(\hat{{\underline{v}}}(\xi)+iZ\hat{{\underline{w}}}(\xi))+Y^2\hat{{\underline{w}}}(\xi)]\theta_1\theta_2.
\end{aligned}
$$

Since ${{\underline{w}}}=0$, we have
$$
v(t,q)=(2\pi)^{-1/2}\int d{\xi}\,e^{i\langle x|{\xi}\rangle}e^{it^2{\xi}/2}\sum_{\ell=0}^k\frac{2^{2\ell}k!}{(2\ell)!(k-\ell)!}t^{2\ell}(i{\xi})^{\ell}\hat{\varphi}({\xi})\bigg|_{x=q}.  \qquad\qed
$$

%\newpage
\section{An example of a system version of Egorov's theorem -- Bernardi's question}
It is well-known that Egorov's theorem
concerning the conjugation of $\Psi$DO(=pseudo-differential operator) with
FIOs(=Fourier integral operators) is a very powerfull tool for the study 
of $\Psi$DOs.

Using superanalysis, we extend 
that theorem to the $2\times2$ 
system of PDOs(=partial differential operators) or $\Psi$DOs.
As a by-product, we give a new geometrical interpretation of the similarity
transformations $e^{i{\mathbb H}}{\mathbb P}e^{-i{\mathbb H}}$ for any 
$2\times2$-matrices ${\mathbb P}$ and ${\mathbb H}={\mathbb H}^*$.

\subsection{Bernardi's question} 
Remarking that
$$
\begin{gathered}
(-\alpha \partial_x^2+\beta x^2)e^{-i\gamma x^2/2}
=[i\alpha\gamma+(\beta+\alpha\gamma^2)x^2]e^{-i\gamma x^2/2},\\
(c(x)\partial_x+\partial_xc(x))e^{-i\gamma x^2/2}
=({c'(x)}-2ic(x)\gamma x)e^{-i\gamma x^2/2},
\end{gathered}
$$
we have
\begin{equation}
\begin{aligned}
&\begin{pmatrix}
e^{i\gamma x^2/2}&0\\
0&e^{i\tilde\gamma x^2/2}
\end{pmatrix}
\begin{pmatrix}
-\alpha \partial_x^2+\beta x^2&2^{-1}(c(x)\partial_x+\partial_xc(x))\\
2^{-1}(d(x)\partial_x+\partial_xd(x))
&-\tilde{\alpha}\partial_x^2+\tilde{\beta}x^2
\end{pmatrix}\begin{pmatrix}
e^{-i\gamma x^2/2}&0\\
0&e^{-i\tilde\gamma x^2/2}
\end{pmatrix}\\
&\qquad\qquad
=\begin{pmatrix}
i\alpha\gamma+(\beta+{\alpha}\gamma^2)x^2
&2^{-1}({c'(x)}-2i\tilde\gamma xc(x))
e^{i(\gamma-\tilde\gamma)x^2/2}\\  
2^{-1}({d'(x)}-2i{\gamma} xd(x))
e^{-i(\gamma-\tilde\gamma)x^2/2}
&i\tilde\alpha\tilde\gamma
+(\tilde\beta+\tilde{\alpha}\tilde\gamma^2)x^2
\end{pmatrix}.
\end{aligned}
\label{b2001}
\end{equation}

In February 2001, Bernardi (as a chairman of a session where I gave a talk)
asked me whether it is possible to explain
\eqref{b2001} using superanalysis. 
Especially, why appear the terms $c'(x)$ and
$d'(x)$ in the off-diagonal part?

\subsection{An answer to Bernardi}
We re-interpret \eqref{b2001} as follows: For $u,v\in C^\infty(\euc)$,
\begin{equation}
\begin{aligned}
&\begin{pmatrix}
e^{i\gamma x^2/2}&0\\
0&e^{i\tilde\gamma x^2/2}
\end{pmatrix}
\begin{pmatrix}
\alpha D^2+\beta x^2&2^{-1}i(c(x)D+Dc(x))\\
2^{-1}i(d(x)D+Dd(x))&\tilde{\alpha} D^2+\tilde{\beta} x^2
\end{pmatrix}\begin{pmatrix}
e^{-i\gamma x^2/2}&0\\
0&e^{-i\tilde\gamma x^2/2}
\end{pmatrix}
\begin{pmatrix}
u\\v\end{pmatrix}\\
&\quad
=\begin{pmatrix}
(\beta+{\alpha}\gamma^2)x^2+i\alpha\gamma
+2i{\gamma}x\alpha\partial_x-\alpha\partial_x^2
&e^{i(\gamma-\tilde\gamma)x^2/2}
(2^{-1}{c'(x)}+c(x)\partial_x-i{\tilde\gamma}xc(x))\\  
e^{-i(\gamma-\tilde\gamma)x^2/2}
{2^{-1}d'(x)}+d(x)\partial_x-i{\gamma}xd(x)) 
&(\tilde\beta+\tilde{\alpha}\tilde\gamma^2)x^2
+i\tilde\alpha\tilde\gamma
+2i\tilde{\gamma}x\tilde\alpha\partial_x-\tilde\alpha\partial_x^2
\end{pmatrix}
\begin{pmatrix}
u\\v
\end{pmatrix}.
\end{aligned}
\label{b3}
\end{equation}
Since \eqref{b3} with $u=v=1$ gives \eqref{b2001}, we should explain the meaning
of \eqref{b3} instead of \eqref{b2001}.

Since we have
$$
\begin{gathered}
\big((\beta+{\alpha}\gamma^2)x^2+i\alpha\gamma
+2i{\gamma}x\alpha\partial_x-\alpha\partial_x^2\big)u(x)
=(2\pi)^{-1}\int_{\euc^2}d{\xi}dy\,
e^{i(x-y)\xi} {a}\left(\frac{x+y}{2},\xi\right)u(y),\\
e^{2^{-1}i(\gamma-\tilde\gamma)x^2}(c'/2+c\partial_x-i\tilde\gamma xc)u(x)
=(2\pi)^{-1}\int_{\euc^2}d{\xi}dy\,
e^{i(x-y)\xi} {c}\left(\frac{x+y}{2},\xi\right)u(y)
\end{gathered}
$$
with Weyl symbols
$$
\begin{gathered}
{a}(x,\xi)
=\alpha(\xi-{\gamma}x)^2+{\beta}x^2
=(\beta+\alpha\gamma^2)x^2-2\alpha\gamma x\xi+\alpha\xi^2,\\
{c}(x,\xi)
=ie^{2^{-1}i(\gamma-\tilde\gamma)x^2}c(x)\bigg(\xi
-\frac{\gamma+\tilde\gamma}{2}x\bigg),
\end{gathered}
$$
we get the Weyl symbol of the right-hand side of \eqref{b3}, given by
\begin{equation}
\begin{aligned}
&\begin{pmatrix}
\alpha(\xi-{\gamma}x)^2+{\beta}x^2
&ie^{i(\gamma-\tilde\gamma)x^2/2}c(x)(\xi-2^{-1}(\gamma+\tilde\gamma)x)\\   
ie^{-i(\gamma-\tilde\gamma)x^2/2}d(x)(\xi-2^{-1}(\gamma+\tilde\gamma)x)
&\tilde\alpha(\xi-{\tilde\gamma}x)^2+{\tilde\beta}x^2
\end{pmatrix}\\
&\qquad\sim
\frac{\alpha+\tilde\alpha}{2}\xi^2
-(\alpha\gamma+\tilde\alpha\tilde\gamma)x\xi
+\frac{(\beta+\alpha\gamma^2)+(\tilde\beta+\tilde\alpha\tilde\gamma^2)}
{2}x^2\\
&\qquad\qquad\qquad
-i\bigg[\frac{\alpha-\tilde\alpha}{2}\xi^2
-(\alpha\gamma-\tilde\alpha\tilde\gamma)x\xi
+\frac{(\beta+\alpha\gamma^2)-(\tilde\beta+\tilde\alpha\tilde\gamma^2)}{2}x^2
\bigg]\langle\theta|\pi\rangle\\
&\qquad\qquad\qquad
+ie^{-i(\gamma-\tilde\gamma)x^2/2}d(x)\bigg(\xi
-\frac{\gamma+\tilde\gamma}{2}x\bigg)\theta_1\theta_2
+ie^{i(\gamma-\tilde\gamma)x^2/2}c(x)\bigg(\xi
-\frac{\gamma+\tilde\gamma}{2}x\bigg)\pi_1\pi_2.
\end{aligned}
\label{b4}
\end{equation}

{\bf Superspace interpretation}: On the other hand, putting
$$
\sigma({\mathbb P})(x,\xi)=\begin{pmatrix}
\alpha\xi^2+\beta x^2&ic(x)\xi\\
id(x)\xi&\tilde\alpha\xi^2+\tilde\beta x^2
\end{pmatrix},\quad
\sigma({\mathbb H})(x)=\begin{pmatrix}
2^{-1}\gamma x^2&0\\
0&2^{-1}\tilde\gamma x^2
\end{pmatrix},
$$
we have
\begin{equation}
\begin{aligned}
\sigma(\sharp{\mathbb P}\flat)(x,\xi,\theta,\pi)
&=\sigma(\hat{\mathcal P})(x,\xi,\theta,\pi) 
={\mathcal P}(x,\xi,\theta,\pi)\\
&\!\!\!\!\!\!\!\!\!\!\!\!
=\frac{\alpha+\tilde\alpha}{2}\xi^2+\frac{\beta+\tilde\beta}{2}x^2
-i\bigg[\frac{\alpha-\tilde\alpha}{2}\xi^2+\frac{\beta-\tilde\beta}{2}x^2\bigg]
\langle\theta|\pi\rangle
+id(x)\xi\theta_1\theta_2+ic(x)\xi\pi_1\pi_2,\\
\sigma(\sharp{\mathbb H}\flat)(x,\theta,\pi)
&=\sigma(\hat{\mathcal H})(x,\theta,\pi)={\mathcal H}(x,\theta,\pi)
=\frac{(\gamma+\tilde\gamma)x^2}{4}
-i\frac{(\gamma-\tilde\gamma)x^2}{4}\langle\theta|\pi\rangle.
\end{aligned}
\label{BH}
\end{equation}
Therefore, we have
$$
\left\{
\begin{aligned}
&\dot x={\mathcal H}_\xi=0,\\
&\dot \xi=-{\mathcal H}_x=-\frac{\gamma+\tilde\gamma}{2} x
+i\frac{\gamma-\tilde\gamma}{2}x\langle\theta|\pi\rangle,\\
&\dot \theta_j=-{\mathcal H}_{\pi_j}
=-i\frac{\gamma-\tilde\gamma}{4}x^2\theta_j,\;(j=1,2),\\
&\dot\pi_j=-{\mathcal H}_{\theta_1}
=i\frac{\gamma-\tilde\gamma}{4}x^2\pi_j,\;(j=1,2),\\
\end{aligned}
\right.
\with
(x(0),\xi(0),\theta(0),\pi(0))
=(\unbx,\unbxi,\unbtheta,\unbpi)
$$
which yields the Hamilton flow corresponding to ${\mathcal H}$ as
$$
\begin{gathered}
{\mathcal C}(t)(\unbx,\unbxi,\unbtheta,\unbpi)=(x(t),\xi(t),\theta(t),\pi(t))
\with\\
x(t)=x(t,{\unbx,\unbxi,\unbtheta,\unbpi})=\unbx,\quad
\xi(t)=\xi(t,{\unbx,\unbxi,\unbtheta,\unbpi})=\unbxi-\frac{\gamma+\tilde\gamma}{2}{\unbx}t
+i\frac{\gamma-\tilde\gamma}{2}{\unbx}t\langle\unbtheta|\unbpi\rangle,\\
\theta_j(t)=\theta_j(t,{\unbx,\unbxi,\unbtheta,\unbpi})
=e^{-i(\gamma-\tilde\gamma)t\unbx^2/4}\unbtheta_j,\quad
\pi_j(t)=\pi(t,{\unbx,\unbxi,\unbtheta,\unbpi})
=e^{i(\gamma-\tilde\gamma)t\unbx^2/4}\unbpi_j \qquad
(j=1,2).
\end{gathered}
$$

Putting operators
$$
\hat{A}=-i\partial_{\unbx}-\frac{\gamma+\tilde\gamma}{2}{\unbx}t,\quad
\hat{B}=\frac{\gamma-\tilde\gamma}{2}{\unbx}t,\quad
\widehat{\sigma}_3=1-\unbtheta_1\frac{\partial}{\partial\unbtheta_1}
-\unbtheta_2\frac{\partial}{\partial\unbtheta_2},\quad
\widehat{\xi(t)}=\hat{A}-\hat{B}\widehat{\sigma}_3,
$$
with Weyl symbols
$$
\begin{gathered}
\sigma(\hat{A})(\unbx,\unbxi)=\xi-\frac{\gamma+\tilde\gamma}{2}{\unbx}t,\quad
\sigma(\hat{B})(\unbx,\unbxi)=\frac{\gamma-\tilde\gamma}{2}{\unbx}t,\quad
\sigma(\widehat{\sigma}_3)(\unbtheta,\unbpi)=-i\langle\unbtheta|\unbpi\rangle,\\
\sigma(\widehat{\xi(t)})(\unbx,\unbxi,\unbtheta,\unbpi)=\unbxi
-\frac{\gamma+\tilde\gamma}{2}{\unbx}t
+i\frac{\gamma-\tilde\gamma}{2}{\unbx}t\langle\unbtheta|\unbpi\rangle,
\quad \langle\theta(t)|\pi(t)\rangle=\langle\unbtheta|\unbpi\rangle,
\end{gathered}
$$
we have the following:
$$
\sigma(\hat{A}\hat{B}+\hat{B}\hat{A})({\unbx},\unbxi)=(\gamma-\tilde\gamma){\unbx}t
\bigg(\unbxi-\frac{\gamma+\tilde\gamma}{2}{\unbx}t\bigg),
\quad
\sigma(\widehat{\sigma}_3^2)(\unbtheta,\unbpi)=1.
$$

\begin{remark}
Let ${\mathfrak a}$, ${\mathfrak b}$ and ${\mathfrak c}$ be non-commutative
or commutative
``operators". For monomials $p_2(x,y)=xy$ and 
$p_3(x,y,z)=xyz$,
we define
$$
p_{2,\mathrm s}({\mathfrak a},{\mathfrak b})
=\begin{cases}
\frac{1}{2!}({\mathfrak a}{\mathfrak b}
+{\mathfrak b}{\mathfrak a})&\quad\mbox{if $[{\mathfrak a},{\mathfrak b}]\neq{0}$},\\
{\mathfrak a}{\mathfrak b}&\quad\mbox{if $[{\mathfrak a},{\mathfrak b}]=0$},
\end{cases}
$$
$$
p_{3,\mathrm s}({\mathfrak a},{\mathfrak b},{\mathfrak c})
=\begin{cases}
\frac{1}{3!}({\mathfrak a}{\mathfrak b}{\mathfrak c}
+{\mathfrak a}{\mathfrak c}{\mathfrak b}
+{\mathfrak b}{\mathfrak c}{\mathfrak a}
+{\mathfrak b}{\mathfrak a}{\mathfrak c}
+{\mathfrak c}{\mathfrak b}{\mathfrak a}
+{\mathfrak c}{\mathfrak a}{\mathfrak b})&\quad\mbox{if  ${\mathfrak a},{\mathfrak
b},{\mathfrak c}$ are non-commutative each other},\\
\frac{1}{2!}({\mathfrak a}{\mathfrak b}{\mathfrak c}
+{\mathfrak b}{\mathfrak c}{\mathfrak a})&\quad\mbox{if $[{\mathfrak
a},{\mathfrak b}]\neq{0}$, but $[{\mathfrak a},{\mathfrak c}]=0$ and $[{\mathfrak
b},{\mathfrak c}]=0$},\\ {\mathfrak a}{\mathfrak b}{\mathfrak c}&\quad
\mbox{if ${\mathfrak a},{\mathfrak b},{\mathfrak c}$ are commutative each other}.
\end{cases}
$$
\end{remark}

From these, we get
$$
\sigma\big(\widehat{\xi(t)}^2\big)({{\underline{x}}, {\underline{\xi}}, {\underline{\theta}},{\underline{\pi}}} )
={\unbxi}^2-(\gamma+\tilde\gamma){\unbx}t{\unbxi}
+\frac{\gamma^2+\tilde\gamma^2}{2}{\unbx}^2t^2
+i\bigg({\unbxi}-\frac{\gamma+\tilde\gamma}{2}{\unbx}t\bigg)
(\gamma-\tilde\gamma){\unbx}t\langle\unbtheta|\unbpi\rangle.
$$
In fact,
$$
\widehat{\xi(t)}^2=(\hat{A}-\hat{B}\widehat{\sigma}_3)(\hat{A}-\hat{B}\widehat{\sigma}_3)
=\hat{A}^2-(\hat{A}\hat{B}+\hat{B}\hat{A})\widehat{\sigma}_3
+\hat{B}^2\widehat{\sigma}_3^2,
$$
with
$$
\hat{A}^2+\hat{B}^2=-\partial_{\unbx}^2
+i\partial_{\unbx}\bigg(\frac{\gamma+\tilde\gamma}{2}{\unbx}t\bigg)
+\bigg(\frac{\gamma+\tilde\gamma}{2}{\unbx}t\bigg)i\partial_{\unbx}
+\frac{\gamma^2+\tilde\gamma^2}{2}{\unbx}^2t^2,
$$
and
$$
\sigma(\hat{A}^2+\hat{B}^2)
=\unbxi^2-i\unbxi(\gamma+\tilde\gamma){\unbx}t
+\frac{\gamma^2+\tilde\gamma^2}{2}{\unbx}^2t^2.
$$

Since $[\widehat{\xi(t)}^2,\widehat{\langle\theta(t)|\pi(t)\rangle}]=0$, we have
$$
\begin{aligned}
{{\sigma}}\big(\widehat{\xi(t)}^2\widehat{\langle\theta(t)|\pi(t)\rangle}\big)({{\underline{x}}, {\underline{\xi}}, {\underline{\theta}},{\underline{\pi}}} )
&=\sigma\big(\widehat{\xi(t)}^2\widehat{\langle\theta(t)|\pi(t)\rangle}\big)({{\underline{x}}, {\underline{\xi}}, {\underline{\theta}},{\underline{\pi}}} )\\
&=\bigg({\unbxi}^2-(\gamma+\tilde\gamma){\unbx}t{\unbxi}
+\frac{\gamma^2+\tilde\gamma^2}{2}{\unbx}^2t^2\bigg)
\langle\unbtheta|\unbpi\rangle\\
&\qquad
-i\bigg({\unbxi}-\frac{\gamma+\tilde\gamma}{2}{\unbx}t\bigg)
(\gamma-\tilde\gamma){\unbx}t
\bigg(\frac{1}{2}+2\unbtheta_1\unbtheta_2\unbpi_1\unbpi_2\bigg).
\end{aligned}
$$

Though $[\hat{\sigma}_3,\widehat{\theta_1\theta_2}]\neq{0}$ and 
$[\hat{\sigma}_3,\widehat{\pi_1\pi_2}]\neq{0}$, 
we have $[\widehat{\xi(t)},\widehat{\theta_1(t)\theta_2(t)}]=0$
and $[\widehat{\xi(t)},\widehat{\pi_1(t)\pi_2(t)}]=0$. 
Moreover, we get
$$
\begin{gathered}
\widehat{\xi(t)}\,\widehat{\theta_1(t)\theta_2(t)}(u_0+u_1\unbtheta_1\unbtheta_2)
=e^{-i(\gamma-\tilde\gamma){\unbx}^2t/2}
(-i\partial_{\unbx}-{\gamma}{\unbx}t)u_0\unbtheta_1\unbtheta_2,\\
\widehat{\xi(t)}\,\widehat{\pi_1(t)\pi_2(t)}(u_0+u_1\unbtheta_1\unbtheta_2)
=e^{i(\gamma-\tilde\gamma){\unbx}^2t/2}(-i\partial_{\unbx}-\tilde{\gamma}{\unbx}t)u_1.
\end{gathered}
$$
Therefore, we get
$$
\begin{aligned}
{{\sigma}}\big(\widehat{\xi(t)}\,\widehat{\theta_1(t)\theta_2(t)}\big)({{\underline{x}}, {\underline{\xi}}, {\underline{\theta}},{\underline{\pi}}} )
&=\sigma\big(\widehat{\xi(t)}\,\widehat{\theta_1(t)\theta_2(t)}\big)({{\underline{x}}, {\underline{\xi}}, {\underline{\theta}},{\underline{\pi}}} )\\
&=\bigg({\unbxi}-\frac{\gamma+\tilde\gamma}{2}{\unbx}t\bigg)
e^{-i(\gamma-\tilde\gamma)\unbx^2t/2}\unbtheta_1\unbtheta_2,\\
{{\sigma}}\big(\widehat{\xi(t)}\widehat{\pi_1(t)\pi_2(t)}\big)({{\underline{x}}, {\underline{\xi}}, {\underline{\theta}},{\underline{\pi}}} )
&=\sigma\big(\widehat{\xi(t)}\,\widehat{\pi_1(t)\pi_2(t)}\big)({{\underline{x}}, {\underline{\xi}}, {\underline{\theta}},{\underline{\pi}}} )\\
&=\bigg({\unbxi}-\frac{\gamma+\tilde\gamma}{2}{\unbx}t\bigg)
e^{i(\gamma-\tilde\gamma)\unbx^2t/2}\unbpi_1\unbpi_2.
\end{aligned}
$$

On the other hand, since $[{\widehat{d(x(t))}},\widehat{\xi(t)}]\neq{0}$ but
$[{\widehat{d(x(t))}},\widehat{\theta_1(t)\theta_2(t)}]=0$ and
$[\widehat{\xi(t)},\widehat{\theta_1(t)\theta_2(t)}]=0$, we have
$$
p_{3,{\mathrm{s}}}({\widehat{d(x(t))}},\widehat{\xi(t)},\widehat{\theta_1(t)\theta_2(t)})
=\frac{1}{2}({\widehat{d(x(t))}}\,\widehat{\xi(t)}\,\widehat{\theta_1(t)\theta_2(t)}+
\widehat{\xi(t)}\,\widehat{\theta_1(t)\theta_2(t)}\,{\widehat{d(x(t))}}),
$$
that is,
$$
p_{3,{\mathrm{s}}}({\widehat{d(x(t))}},\widehat{\xi(t)},\widehat{\theta_1(t)\theta_2(t)})
(u_0+u_1\unbtheta_1\unbtheta_2)
=\frac{1}{2}e^{-i(\gamma-\tilde\gamma){\unbx}^2t/2}
\{d({\unbx})(-i\partial_{\unbx}-{\gamma}{\unbx}t)+
(-i\partial_{\unbx}-{\gamma}{\unbx}t)d({\unbx})\}u_0\unbtheta_1\unbtheta_2.
$$
Analogous holds for
$p_{3,{\mathrm{s}}}({\widehat{c(x(t))}}\,\widehat{\xi(t)}\,\widehat{\pi_1(t)\pi_2(t)})$.

From these, we have
\begin{equation}
\begin{aligned}
{\mathcal P}&\big[{\mathcal C}(t)({{\underline{x}}, {\underline{\xi}}, {\underline{\theta}},{\underline{\pi}}} )\big]\\
&=\frac{\alpha+\tilde\alpha}{2}{{\sigma}}\big(\widehat{\xi(t)}^2\big)
+\frac{\beta+\tilde\beta}{2}{{\sigma}}\big(\widehat{x(t)}^2\big)\\
&\qquad
-i\frac{\alpha-\tilde\alpha}{2}
{{\sigma}}\big(\widehat{\xi(t)}^2\widehat{\langle\theta(t)|\pi(t)\rangle}\big)
-i\frac{\alpha-\tilde\alpha}{2}
{{\sigma}}(\widehat{x(t)}^2\widehat{\langle\theta(t)|\pi(t)\rangle})\\
&\qquad\qquad
+i{{\sigma}}
\big({\widehat{d(x(t))}}\,\widehat{\xi(t)}\,\widehat{\theta_1(t)\theta_2(t)}\big)
+i{{\sigma}}
\big({\widehat{c(x(t))}}\,\widehat{\xi(t)}\,\widehat{\pi_1(t)\pi_2(t)})({{\underline{x}}, {\underline{\xi}}, {\underline{\theta}},{\underline{\pi}}}\big)\\
&=\frac{\alpha+\tilde\alpha}{2}{\unbxi}^2+\frac{\beta+\tilde\beta}{2}{\unbx}^2
+\frac{\alpha\gamma^2t^2+\tilde\alpha\tilde\gamma^2t^2}{2}{\unbx}^2
-(\alpha{\gamma}t+\tilde\alpha{\tilde\gamma}t){\unbx}{\unbxi}\\
&\qquad
-i\bigg[\frac{\alpha-\tilde\alpha}{2}{\unbxi}^2
+\frac{\beta-\tilde\beta}{2}{\unbx}^2
-(\alpha{\gamma}t-\tilde\alpha{\tilde\gamma}t){\unbx}{\unbxi}
+\frac{\alpha\gamma^2t^2-\tilde\alpha\tilde\gamma^2t^2}{2}{\unbx}^2
\bigg]\langle\unbtheta|\unbpi\rangle\\
&\qquad
+id({\unbx})\bigg(\unbxi-\frac{\gamma+\tilde\gamma}{2}{\unbx}t\bigg)
e^{-i({\gamma}t-{\tilde\gamma}t){\unbx}^2/2}\unbtheta_1\unbtheta_2
+ic({\unbx})\bigg(\unbxi-\frac{{\gamma}t+{\tilde\gamma}t}{2}{\unbx}\bigg)
e^{i({\gamma}t-{\tilde\gamma}t){\unbx}^2/2}\unbpi_1\unbpi_2.
\end{aligned}
\label{b5}
\end{equation}
This equals to \eqref{b4} after replacing $\gamma\to \gamma t$ 
and $\tilde\gamma\to \tilde\gamma t$.

Therefore, denoting $\unbx$ simply by $x$, etc, we have proved
$$
\sigma({\sharp}e^{it{\mathbb H}}{\mathbb P}
e^{-it{\mathbb H}}{\flat})(x,\xi,\theta,\pi)
=\sigma(e^{it\hat{\mathcal H}}\hat{\mathcal P}
e^{-it\hat{\mathcal H}})(x,\xi,\theta,\pi)
={\mathcal P}\big[{\mathcal C}(t)(x,\xi,\theta,\pi)\big].  \qquad\qed
$$

\begin{remark}
In the above, we calculate the product of operators and find its symbol, rather directly.
In the near future, we need to give a product formula for operators as analogous to
``bosonic" case.
\end{remark}

%%%%%
%\newpage
\section{Functional Derivative Equations}
\subsection{Liouville equation}
I mentioned, at 7 or  8-th lecture, a function with countably infinite independent variables. There, I regard a function with an odd variable $\theta$ as a function with countably infinite independent variables 
$\{\theta_{\mathbf{J}}\in{\mathbb{C}}\}$
where $\theta=\sum_{{\mathbf{J}}\in{\mathcal{I}}}\theta_{\mathbf{J}}\sigma^{\mathbf{J}}$. This resembles to consider a functional 
%$F(g(\theta))\sim F(({\cdots}, g_1, g_0, g_1,{\cdots}))$ where $({\cdots}, g_1, g_0, g_1,{\cdots})$ are Fourier coefficients of $g(\theta)=\sum_n g_ne^{in\theta)\sim({\cdots}, g_1, g_0, g_1,{\cdots})$ for
%$\theta\in[0,2\pi]$, for example.

As is well-known, a non-linear system of ODEs may be regarded as a linear PDE and therefore ask what occurs when we have non-linear PDE on $\euc^d$ instead of ODE.

Typically, the solution of Hamilton equation relates to the solution of Liouville equation by the method of characteristics.
For $H(q,p)\in C^{\infty}(\euc^m:\euc)$, Hamilton equation is written down as
$$
{\begin{cases}
\dot{q}_j=H_{p_j}(q(t),p(t)),\\
\dot{p}_j=-H_{q_j}(q(t),p(t)),
\end{cases}}
\with
\binom{q(0)}{p(0)}
=\binom{\unbq}{\unbp},
$$
and the Liouville equation is
$$
\pdt u(t,q,p)=\{u,H(q,p)\}\with u(0,q,p)={\unbu}(q,p).
$$
Here, Poisson bracket is defined by
$$
\{f,g\}=\sum_{j=1}^m\bigg(\frac{\partial f}{\partial q_j}\frac{\partial f}{\partial p_j}
-\frac{\partial f}{\partial p_j}\frac{\partial f}{\partial q_j}\bigg).
$$
In general, Hamilton equation is a non-linear ODE and Liouville equation is a linear PDE.
Even non-linear, applying Galerkin method, PDE may be regarded as ODE with infinitely many components in certain function spaces.

If we take as the special initial data in Liouville equation, for example, ${\unbu}(q,p)$ may be a measure $\delta_{\unbq}(q)dq$ or $\delta_{\unbp}(p)dp$, respectively, we get the solution of Hamilton equation. Conversely, putting $u(t,{\unbq},{\unbp})={\unbu}(q(t,{\unbq},{\unbp}),p(t,{\unbq},{\unbp}))$ from the solution of Hamilton equation, we get the solution $u(t,{\unbq},{\unbp})dqdp$\footnote{here, we misuse measure $\delta_{\unbq}(q)dq$ and its density $\delta_{\unbq}(q)$ w.r.t. $dq$}.

\subsection{Hopf equation (H)}[E. Hopf~\cite{hop52}]
As Liouville equation corresponds to Hamilton equation, 
Hopf equation written by functional derivatives corresponds to Navier-Stokes equation.
More precisely,
\subsubsection{Navier-Stokes equation:}
Let a domain $\Omega$ in $\euc^m$ be given with smooth boundary $\partial\Omega$. Find a vector field$u(t,x)=\sum_{j=1}^m{u_j(t,x)}\frac{\partial}{\partial x_j}$ and a pressure $p(t,x)$ satisfying
$$
\begin{aligned}
&\frac{\partial}{\partial t} u(t,x)-\nu\Delta u(t,x)+(u{\cdot}\nabla)(t,x)+\nabla p=0,\\
&\dive u=0,\\
&u(0,x)=u_0(x),\quad u(t,x)\big|_{\partial\Omega}=0.
\end{aligned}
$$
Assuming this has a solution, we denote it $(T_tu_0)(x)=u(t,x)$.

For the sake of simplicity, we take a Riemannian manifold $M$ with metric $g_{ij}(x)dx^idx^j$.
Let ${\mathbf{L}}^2_{\sigma}(M)$ be a set of $L^2$-integrable solenoidal vector field on $M$,
and let ${\overset{\circ}{\Lambda}}{}_{\sigma}^1(M)$ be a solenoidal vector field with compact support on $M$.

Put ${\mathbf {H}}={\mathbf{L}}^2_{\sigma}(M,g)$ and $\tilde{\mathbf {H}}$ its dual.
Find  a functional $W(t,\pmb\eta)$ on $[0,{\infty})\times\tilde{\mathbf {H}}$, for
$(t,\eta)\in (0,{\infty})\times{\overset{\circ}{\Lambda}}{}_{\sigma}^1(M)$, it satisfies
\begin{equation}
\begin{aligned}
\pdt W(t,\eta )
&=\int_M\Big(
-i({\tilde {\mathcal T}}\eta)_{jk}(x)
\frac{\delta^2}{\delta{\eta_j(x)} \delta{\eta_k(x)}}{W(t,\eta)} \\ %\FDD({W(t,\eta)},{\eta_j(x)},{\eta_k(x)})\\
&\qquad\qquad\qquad+\nu(\Delta\eta)_j(x)\frac{\delta}{\delta{\eta_j (x)}}{W(t,\eta)} %\FD({W(t,\eta)},{\eta_j (x)})
+i{\eta_j(x)}{f^j(x,t)} W(t,\eta)\Big)d_g x ,
\end{aligned}
\label{H1}
\end{equation}
$$
\frac{1}{\sqrt {g(x)}}\pdxj \left({\sqrt {g(x)}} 
\frac{\delta}{\delta{\eta_j (x)}}{W(t,\eta)}\right) =0, %\FD({W(t,\eta )},{\eta_j (x)}) 
%\tag H.2
$$
$$ 
W(t,0)=1 \et W(0,\eta)= W_0 (\eta).
%\tag H.3
$$
and
$$
\begin{gathered}
W_0(0)=1,\quad 
{1\over {\sqrt {g(x)}}}{\pdxj}
\left(\sqrt{g(x)}\frac{\delta}{\delta{\eta_j (x)}}{W_0(\eta)}\right) = 0, \\
%\FD({W_0 (\eta)},{\eta_j (x)})\right) = 0, \\
f(x,t)=f^j(x,t)\pdxj\in L^2 (0,{\infty}: {\mathbf  V}^{-1} ).
\end{gathered}
%\tag H.4
$$

\def\FD(#1,#2){{{\delta {#1}}\over {\delta {#2}}}}

\subsubsection{Hopf-Foia{\c s} equation (HF)}[C. Foia{\c s}~\cite{foi73}]
Take ${\mathbf {H}}={\mathbf {L}}^2_{\sigma}(\Omega)$ as Hilbert space.
Find a family of Borel measures $\{\mu(t,\cdot)\}_{t\in(0,{\infty})} $ on ${\mathbf {H}} $, such that for a suitable class of  test functionals $ \Phi (t,u)$, it satisfies
\begin{equation}
\begin{aligned}
-\int_0^{\infty} & \int_{\mathbf {H}} 
{{\partial \Phi (t,u)}\over{\partial t}}\mu (t,du)dt
-\int_{{\mathbf {H}} }\Phi (0,u)\mu _0 (du) \\
&
=\int_0^{\infty}\int_{{\mathbf {H}}}\int_M
\left((\nabla_{u} u)^j(x)
-\nu(\Delta u)^j(x)- f^j (x,t)\right)
\FD({\Phi(t,u)},{u^j(x)})d_g x \,\mu (t,du) dt 
\end{aligned}
\label{HF}
\end{equation}
Here $\mu_0(\cdot)$ is a given Borel measure on ${\mathbf {H}} $.
This equation is obtained, for any $\omega\in{\mathcal B}({\mathbf {H}})$, putting
$\mu(t,\omega)=\mu_0(T_t^{-1}\omega)$, calculate
$$
\pdt \int_{{\mathbf {H}} }\Phi (t,T_tu)\mu_0 (du).
$$
Here, we assume $T_tu$ above is considered well-defined.

The solution of Hopf equation is obatained by putting $\Phi(t,u)=\rho(t)e^{i\langle u,\eta\rangle}$ into \eqref{HF}, and
$$
W(t,\eta)=\int_{\mathbf {H}}e^{i\langle u,\eta\rangle}\mu(t,du)
=\int_{\mathbf {H}}e^{i\langle T_tu,\eta\rangle}\mu_0(du)
\quad\text{with}\quad
\langle u,\eta\rangle=\int_M u^j(x)\eta_j(x)d_gx.
$$

\begin{center}
 
 =========== Mini Column 4 ========
 
 \end{center}
 
What are functional derivatives, which is abruptly mentioned above.
Let $\Omega$ be a domain in $\euc^m$. 
We consider a functional $W:X(\Omega)\ni \eta\to W(\eta)\in{\mathbb{C}}$ on suitable functional space $X(\Omega)$.
Taking a test function $\phi\in C_0^{\infty}(\Omega)\subset X(\Omega)$, if we have
$$
\frac{d}{dt} W(\eta+ t\phi)\big|_{t=0}
$$
then, we denote it as
$$
{}_{{\mathcal{D}}'}\langle w,\phi\rangle_{{\mathcal{D}}}
=\int_{\Omega} dx\,w(x){\phi(x)},\quad
w(x)=\frac{\delta W(\eta)}{\delta \eta(x)}\in {\mathcal{D}}'(\Omega)
$$
and call this functional derivative. We need higher order functional derivatives for $W$.
Formally we may put
$$
\frac{\delta^2 W(\eta)}{\delta \eta(x)\delta \eta(y)}\in {\mathcal{D}}'(\Omega\times\Omega)
$$
but what does this mean when $x=y$? See, Inoue~\cite{ino87}.

\begin{center}
 
 =========== End of Mini Column 4 ========
 
 \end{center}

$\bullet$ I feel the address of I.M. Gelfand~\cite{gel63} at ICM Amsterdam conference in 1954 suggests beautiful and important problems and I believe as he mentioned that we need to develop \underline{a very new theory}\\ 
\underline{of differential equations} to study quantum field theory or turbulence theory, for example, theory of functional derivative equations (FDE).
The configuration space where functional lives is a function space which is infinite-dimensional, therefore no suitable Lebesgue-like measure. This means it  is not yet possible to integrate functional freely, and no integration by parts, no Fourier transformation does exist. The tool which we are available now is Taylor expansion if it exists, therefore, only very algebraic treatise is possible. Concerning a simple model equation with removable $\infty$ by renormalization, see
Inoue~\cite{ino86}.

\subsubsection{FDE representing turbulence?} 
Though Hopf equation is related to the invariant measure w.r.t. the flow governed by Navier-Stokes equation,
I suspect that the equation related to turbulence will be Fokker-Planck type FDE derived from Navier-Stokes equation:

Find a measure $P(t,v)d_Fv$ satisfying below:
$$
\begin{aligned}
\pdt P(t,v) &=
\int_{\euc^3}d^3x
{\frac{\delta}{\delta v^i(x)}}
\Big\{\Big(v^j(x)\nabla_jv^i(x)+{\frac 1\rho}\nabla^i p(x)-\nu\Delta v^i(x)-f^i(x)\Big)
P(t,v)\Big\}\\
&\qquad\qquad\qquad
+{\frac{k_{\mathrm{B}}T\nu}{\rho}}
\int_{\euc^3}d^3x\Big(\nabla_j{\frac{\delta}{\delta v^i(x)}}\Big)^2P(t,v).
\end{aligned}
$$
Here
$$
p(x)={\frac\rho{4\pi}}\int_{\euc^3}d^3x'
{\frac{(\nabla'_iv^j(x'))(\nabla'_jv^i(x'))-\nabla'_if^i(x')}
{|x-x'|}}
$$
and  the functional derivatives are taken w.r.t. transversal velocity field
$$
v^j(x)=\int_{\euc^3}{\frac{d^3\xi}{(2\pi)^3}}
\Big(\delta^j_k-{\frac{\delta^j_\ell\xi_\ell\xi_k}{|\xi|^2}}\Big)v^k(x)e^{i\xi x}.
$$

\begin{problem}
Very recently, I make know the paper \cite{tro90} written by O.V. Troshkin, where he cited the ``result'' by W. Thomson (alias Lord Kelvin) such that W.T. obtained a wave equation for an incompressible fluid by averaging Euler's equation.
Troshkin claims that the formal analogy existing between waves of small disturbances of inviscous and incompressible turbulent medium and electromagnetic waves  is established.
Prove these facts mathematically using Reynolds equation by Foia{\c s}~\cite{foi73}.
\par
In the above, seemingly Thomson assumes the intrinsic fluctuation associated to Euler flow and averaging w.r.t. this.
On the other hand, we \cite{IF79} derive Navier-Stokes equation from Euler equation, by adding artificially white noise (extrinsic) fluctuation to each flow line of Euler equation. 
\end{problem}

\subsection{Equation for QED?}
As a functional derivative equation for QED=quantum electrodynamics), it might be the following forms a base?
$$
\left\{
\aligned
&\square %({\frac 1{c^2}}\partial_{x_0}^2-\Delta)
{\frac{\delta Z(\eta,{\bar\eta},J)}{\delta J_\mu(x)}}
=-i J_\mu(x){Z(\eta,\bar\eta,J)}
+ie\gamma^\mu 
{\frac{\delta^2 Z(\eta,\bar\eta,J)}{\delta \eta(x) \delta \bar\eta(x)}},\\
&(i\gamma^\mu {\vec{\partial}}_\mu -M)
{\frac{\delta Z(\eta,\bar\eta,J)}{\delta \bar\eta(x)}}
=-i\eta(x){Z(\eta,\bar\eta,J)}
+ie\gamma^\mu 
{\frac{\delta^2 Z(\eta,\bar\eta,J)}{\delta\bar\eta(x) \delta J_\mu(x)}},\\
&{\frac{\delta Z(\eta,\bar\eta,J)}{\delta\eta(x)}}
(i\gamma^\mu {\vec{{\partial}}}_\mu+M)
=i\bar\eta(x){Z(\eta,\bar\eta,J)}
-ie\gamma^\mu 
{\frac{\delta^2 Z(\eta,\bar\eta,J)}{\delta\eta(x) \delta J_\mu(x)}},\\
&\;Z(0,0,0) =1.
\endaligned
\right.
$$
\begin{remark}
This equation stems from functional method in QFT(Quantum Field Theory), more precisely, adjoining external forces to each component of Maxwell-Dirac equation, we get this equation.
In finite dimensional case, I know vaguely a story, for given non-linear ODE, adding fluctuating external force to it to have Langevin equation,
and solving that and taking average of solution w.r.t. fluctuation to get Fokker-Planck equation. In this story, if we replace this ODE with
the coupled Maxwell-Dirac equation, what facts do we get?
Considering like this, what type of classical property is inherited to the solution of quantum or statistical equation, and how the quantum or statistical effect is represented by ``classical quantity''?
\par 
 As the Feynman's path-integral representation gives directly quantum object from Lagrangian without solving Schr\''odinger equation, physicists write down the quantum quantity using path integral with Feynman measure and no use of FDE etc.
As is mentioned before explaining ``stationary phase method'', how to make rigorous their arguments?
\end{remark}

%%%%%
%\newpage
\section{Supersymmetric extension of the Riemannian metric $g_{jk}(q)dq^j dq^k$ on $\euc^d$}

In Witten's paper, he writes down ``classical object'' or rather  ``quantity before quantization'' corresponding to the deformed Laplace-Beltrami operator $d_{\phi}^*d_{\phi}+d_{\phi} d_{\phi}^*$ as if it is evident. Here the function $\phi$ is the Morse function on the manifold $M$. 

Mathematically it is not so clear what is the classical object\footnote{Semi-classical analysis is a study to get classical objects from the given quantum thing} for the given quantum operator. 
Therefore reversely, I try to give a prescription how one obtains the super symmetrically extended metric from the given Riemannian metric $g_{ij}(q)dq^idq^j$.
To make the situation simple, as a manifold, we take $\euc^d$\footnote{in this section, as the dimension of the configuration space, we use $d$ instead of $m$,} and as the given Lagrangian
\begin{equation*}
L(q,\dot{q})=
{\frac {1}{2}}
g_{ij}(q){\dot q}^i {\dot q}^j
+A_j(q){\dot q}^j
-V(q)\in C^\infty(T\eucd:\euc).
\end{equation*}
Using Legendre transformation, we associate a Hamiltonian
\begin{equation*}
H(q,p)= 
{\frac 12}g^{ij}(q) (p_i-A_i(q))( p_j-A_j(q)) + V(q)
\in C^\infty(T^*\eucd:\euc).
\end{equation*}
To such a Hamiltonian, via extending formally that Lagrangian,
we may associate a supersymmetric extension
\begin{equation*}
\begin{aligned}
\mathcal H (x,\xi,\theta,\pi) &= \frac 12 g^{ij}
\big(\xi_i-\frac {\sqrt{-1}}2 (g_{ik,l}-g_{il,k})\theta^k\pi^l-A_i\big)
\big(\xi_j-\frac {\sqrt{-1}}2 (g_{jm,n}-g_{jn,m})\theta^m\pi^n-A_j\big)\\
&\qquad
+\frac 12 R_{ikjl}
\theta^j \theta^l{\pi}^i {\pi}^k 
+{\frac12}g^{jk} W_{,j} W_{,k}
-W_{;ij}
\theta^i{\pi}^j 
\end{aligned}
\end{equation*}
which belongs to $\CSS({\fR^{2d|2d}}:\rev)$.
Here, the functions $g^{ij}=g^{ij}(x)$ of $x\in{\fR^{d|0}}$ etc., 
appeared above are Grassmann extensions of
the corresponding ones $g^{ij}=g^{ij}(q)$ of $q\in\eucd$ etc.

\subsection
{A prescription for a supersymmetric extension of a given $L(q,\dot{q})$}
We prepare two odd variables 
$\rho=(\rho_1,\rho_2)\in \rod^2$.
Instead of the path space $P$ considered in \S2,
we introduce another path space
${\tilde{\mathcal{P}}}_0 $
consisting of (super)fields 
${\Phi}={\Phi}(t,\rho)=({\Phi}^1(t,\rho), \dots ,{\Phi}^d(t,\rho)):(t,\rho)\in\euc\times\rod^2\to \rev^d$
given by
the following form:
\begin{equation} \label{SSE2.1}
\Phi ^j(t,\rho)=
x^j(t)+
{\sqrt{-1}}\rho_{\alpha}\epsilon_{\alpha\beta}\psi_{\beta}^j(t)+
\frac{\sqrt{-1}}{2}
\rho_{\alpha}\epsilon_{\alpha\beta}\rho_{\beta} F^j(t)
\end{equation}
where for a certain interval $I\subset\euc$
\begin{equation} %2.2
x^j(t)\in
C^{\infty}(I:\rev),\quad 
\psi_{\beta}^j(t)\in
C^{\infty}(I:\rod ),\quad
F^j(t)\in
C^{\infty}(I:\rev )
\end{equation}
%for $t\in I$,
with $\epsilon_{\alpha\beta}=- \epsilon_{\beta\alpha}$, 
$\epsilon_{12}=1$, $j=1,2,\cdots,d$ and $\alpha,\beta=1,2$.
\par
Introducing operators 
$ {\mathcal{D}}_{\alpha}^- $ as
\begin{equation} %2.3
{{\mathcal{D}}_{\alpha}^-} =
{\frac{\partial}{\partial\rho_\alpha}}-{\sqrt{-1}}{\rho}_\alpha{{\pdt}}
\for \alpha=1,2, 
\end{equation}
we put
\begin{equation} %2.4
\begin{aligned}
\tilde{\mathcal{L}}_0&=
\tilde{\mathcal{L}}_0 (\Phi,{\mathcal{D}}_{\alpha}^- \Phi)\\
&\defeq
-{\frac 14}
g_{jk}{\mathcal{D}}_{\alpha}^-\Phi^j %({\Phi})
\epsilon_{\alpha\beta}{\mathcal{D}}_{\beta}^-\Phi^k
+{\frac {\sqrt{-1}}4}g_{jk}
(\rho_\alpha A^j\epsilon_{\alpha\beta}{\mathcal{D}}_{\beta}^-\Phi^k
+{\mathcal{D}}_{\alpha}^-\Phi^j\epsilon_{\alpha\beta}\rho_\beta A^k)
-{\sqrt{-1}}W,
\end{aligned}
\end{equation}
where argument of $g_{ij}$, $A_j$ and $W$ is ${\Phi}$.
Here and what follows, for any smooth function $u$ on $\eucd$,
we extend it (called Grassmann continuation) on ${\fR}^{m|0}$ as
\begin{equation} %2.5
u({\Phi})
=u(x)+{\sqrt{-1}}u_{,k}(x)
\left(\rho_{\alpha}
\epsilon_{\alpha\beta}
\psi_{\beta}^k
+{\frac12}\rho_{\alpha}\epsilon_{\alpha\beta}\rho_{\beta} F^k\right)
-{\frac 12}u_{,kl}(x)\,
\rho_{\alpha}
\epsilon_{\alpha\beta}
\psi_{\beta}^k\,
\rho_{\beta}
\epsilon_{\beta\alpha}
\psi_{\alpha}^l
\end{equation}
for
\begin{equation}
 \Phi ^j= x^j +
{\sqrt{-1}}\rho_{\alpha}\epsilon_{\alpha\beta}\psi_{\beta}^j+
\frac {\sqrt{-1}}2
\rho_{\alpha}\epsilon_{\alpha\beta}\rho_{\beta} F^j 
\quad\text{where
$ x^i\in\rev ,\, \psi_{\alpha}^j \in\rod $ and 
$ F^k\in\rev$.}
\end{equation} 

{\bf Notations. }
We put
\begin{gather}
[jk]\defeq
\psi_{\alpha}^j\epsilon_{\alpha\beta}\psi_{\beta}^k=
\psi_1^i\psi_2^j- \psi_2^i\psi_1^j =
\psi_{\alpha}^k\epsilon_{\alpha\beta}\psi_{\beta}^j=[kj],\\
R_{ikjl}\defeq \frac 12 (
g_{ij,kl} +g_{kl,ij}
-g_{jk,il} -g_{il,jk})
+\Gamma^m_{ij} \Gamma^n_{kl} g_{mn}
-\Gamma^m_{il} \Gamma^n_{jk} g_{mn}\\
\intertext{and}
\nabla_j W=W_{,j}, \quad
\nabla^j W=g^{jk}W_{,k}, \quad
\nabla_j \nabla_k W
=\nabla_k \nabla_j W=W_{;jk} %W_{:jk}
=W_{,jk}-\Gamma^m_{jk}W_{,m}.\label{SSE2.19}
\end{gather}

{\bf Formulas. }
Following formulas are easily obtained using renumbering and symmetry
of indecies combining with the anticommutativity of 
$\{ \psi _{\alpha}^i \}$'s.

\begin{equation} \label{SSE2.8}
(g_{ij,kl}
+g_{mn}
\Gamma^m_{ij}\Gamma^n_{kl})
[ij][kl]
=\frac 23
R_{ikjl}[ij][kl],
\end{equation}
\begin{equation}\label{SSE2.20}
\begin{aligned}
g_{ij,kl}[ij][kl] 
& =g_{ij,kl}(\psi_1^i \psi_2^j \psi_1^k \psi_2^l
+\psi_1^i \psi_2^j \psi_1^l \psi_2^k
+\psi_1^j \psi_2^i \psi_1^k \psi_2^l
+\psi_1^j \psi_2^i \psi_1^l \psi_2^k )\\
& =4 g_{ij,kl} \psi_1^i \psi_2^j \psi_1^k \psi_2^l
= g_{kl,ij}[ij][kl], 
\end{aligned}
\end{equation}
\begin{equation}\label{SSE2.21}
\begin{aligned}
g_{il,kj}[ij][kl] 
& =g_{il,kj}(\psi_1^i \psi_2^j \psi_1^k \psi_2^l
+\psi_1^i \psi_2^j \psi_1^l \psi_2^k
+\psi_1^j \psi_2^i \psi_1^k \psi_2^l
+\psi_1^j \psi_2^i \psi_1^l \psi_2^k )\\
& =2 g_{il,kj}\psi_1^i \psi_2^j \psi_1^k \psi_2^l
= g_{kj,il}[ij][kl]
\end{aligned}
\end{equation}
\begin{align}
g_{il,kj}\psi_1^i \psi_2^j \psi_1^k \psi_2^l
& =-g_{ij,kl} \psi_1^i \psi_2^j \psi_1^k \psi_2^l,\label{SSE2.22}\\
\Gamma^m_{ij} \Gamma^n_{kl} g_{mn}[ij][kl]
& =4\Gamma^m_{ij} \Gamma^n_{kl} g_{mn} 
\psi_1^i \psi_2^j \psi_1^k \psi_2^l,\label{SSE2.23}\\
\Gamma^m_{il} \Gamma^n_{kj} g_{mn}[ij][kl]
& =-2\Gamma^m_{ij} \Gamma^n_{kl} g_{mn} 
\psi_1^i \psi_2^j \psi_1^k \psi_2^l.\label{SSE2.24} 
\end{align}
\par
Simple but lengthy calculations yield that
\begin{equation}\label{SSE2.6}
\begin{aligned} 
{\mathcal{L}}^{\prime}_0 (x,\dot x, & \psi_{\alpha},\dot\psi_{\alpha},F) 
\defeq
\int_{\fR^{0|2}}d{\rho}\,\tilde{\mathcal{L}}_0
(\Phi(t,\rho), {\mathcal{D}}_{\alpha}^-\Phi(t,\rho))  \\
&={\frac 12}
g_{jk}(\dot x^j\dot x^k + F^j F^k + {\sqrt{-1}}
\psi_{\alpha}^j \dot\psi_{\alpha}^k)
+{\frac 18}g_{ij,kl}[ij][kl]\\
&\qquad 
+A_j\dot x^j-{\frac {\sqrt{-1}}2}A_{j,k}\psi_{\alpha}^j \psi_{\alpha}^k
-{\frac {\sqrt{-1}}2}
\Gamma^k_{ij}[ij]g_{kl}F^l
+{\frac {\sqrt{-1}}2}g_{ij}\psi_{\alpha}^i\Gamma^j_{kl}
\dot x^k\psi_{\alpha}^l\\
&\qquad\qquad
+W_{,k} F^k -
{\frac {\sqrt{-1}}2}W_{;kl}[kl].
\end{aligned}
\end{equation}

Assuming the auxiliary field $F=(F^1,\dots,F^d)$ satisfies
\begin{equation} \label{SSE2.7}
{\frac{\delta{\mathcal{L}}^{\prime}_0}{\delta F^k}}=0, \quad i.e. \quad
F^k= \frac {\sqrt{-1}}2 
\Gamma^k_{ij}[ij]
- g^{kl}W_{,l},
\end{equation}
we get
\begin{equation}\label{SSE2.18}
\begin{aligned}
\mathcal{L}_0 (x,\dot x,\psi_{\alpha},\dot\psi_{\alpha})
= {\frac 12} &
g_{jk}\, \dot x^j \dot x^k
+\frac {\sqrt{-1}}2 g_{jk}\psi_{\alpha}^j
{\frac D{dt}} \psi_{\alpha}^k 
+\frac 1{12} R_{ikjl}[ij][kl]\\
& +A_j\dot x^j-{\frac {\sqrt{-1}}2}A_{j,k}\psi_{\alpha}^j \psi_{\alpha}^k
-{\frac12}g^{jk}W_{,j}W_{,k}
-{\frac {\sqrt{-1}}2}\nabla_j W_{;jk}[jk],
\end{aligned}
\end{equation}
where
\begin{equation}
{\frac D{dt}} \psi_{\alpha}^k \defeq
\dot\psi_{\alpha}^k+\Gamma^k_{pl}\dot x^p \psi_{\alpha}^l.
\end{equation}
 
\begin{remark}
(a) Above derivation of \eqref{SSE2.18} is essentially due to
Davis et al.\cite{D-M-P-V},
though in their calculations the coefficient 1/8 in \eqref{SSE2.6}
is replaced by 1. Moreover, they didn't mention the necessity
of using the Grassmann algebra with infinite number of generators,
which is necessary to define odd derivatives uniquely.
\newline
(b) To eliminate the auxiliary fields $ F^i $, we assume that the `equation of
motion' described in \eqref{SSE2.7} holds.
Though there is a work, for example Cooper and Freedman \cite{CF},
which asserts that $F^i$ is calculated out after integrating
the partition function expressed by the Feynman measure,
it seems curious to use the ``quantum argument'' when we are discussing
the ``classical objects''. 
(In any way, there does not exist 
the `Fubini theorem' with respect to the `Feynman measure'.)
\end{remark}

\subsection{A prescription for a supersymmetric extension of Hamiltonian $H(q,p)$} 
We restart from the Lagrangian $\mathcal{L}_0$ \eqref{SSE2.18} ignoring
the procedures itself.
As the variables $\{ \psi_{\alpha}^i \}$
are assumed to be ``real" and anticommutative, we define from them
the ``complex" odd variables as follows:
\begin{gather}
\psi^j={\frac 1{\sqrt 2}}(\psi_1^j+ {\sqrt{-1}}\psi_2^j), \quad
\bar\psi^j={\frac 1{\sqrt 2}}(\psi_1^j- {\sqrt{-1}}\psi_2^j),\\
\intertext{that is}
\psi_1^j
={\frac 1{\sqrt 2}} 
(\psi^j+ \bar\psi^j), \quad
\psi_2^j
={\frac 1{{\sqrt 2}{\sqrt{-1}}}}
(\psi^j-\bar\psi^j).
\end{gather}
Then, clearly we have
\begin{equation}
\begin{gathered}
\{{\bar\psi}^i,{\bar\psi}^j \}=
\{{\psi}^i,{\psi}^j \}=
\{{\bar\psi}^i,{\psi}^j \}=
\{{\psi}^i,{\bar\psi}^j \}=0\\
[ij]
={\sqrt{-1}}(\psi^i{\bar\psi}^j-{\bar\psi}^i\psi^j).
\end{gathered}
\end{equation}
By the same calculation as before,
\begin{equation} \label{SSE2.33}
\begin{aligned}
R_{ikjl}[ij][kl]
&=-6(g_{ij,kl}+
\Gamma^m_{ij} \Gamma^n_{kl} g_{mn})
{\psi}^i\bar \psi^j {\psi}^k \bar\psi^l\\
&=6(g_{ik,jl}+
\Gamma^m_{ik} \Gamma^n_{jl} g_{mn})
{\psi}^i \bar\psi^j {\psi}^k \bar\psi^l.
\end{aligned}
\end{equation}
So, we get
\begin{equation}\label{SSE2.34}
\begin{aligned} 
{\mathcal{L}} (x,\dot x,\psi,\dot\psi,
\bar\psi,\dot{\bar\psi})
&=
{\frac 12}
g_{jk}\, \dot x^j \dot x^k
+ \frac {\sqrt{-1}}2 g_{jk}
(\psi^j {\frac D{dt}} \bar\psi^k 
+\bar\psi^j {\frac D{dt}}\psi^k)
-\frac 14 R_{ijkl}
{\psi}^i \bar\psi^j {\psi}^k \bar\psi^l \\
&\qquad
+A_j\dot x^j-{\frac {\sqrt{-1}}2}B_{jk}{\psi}^j\bar\psi^k
+{\frac 12}A_jA^j\\
&\qquad\qquad
-{\frac12} \nabla^j W \nabla_j W
+{\frac 12} \nabla_i \nabla_j W 
({\psi}^i\bar\psi^j - \bar\psi^i{\psi}^j).
\end{aligned}
\end{equation}
\par
Introducing new variables by
\begin{equation}
\left\{
\begin{aligned}
\xi_i & =\frac {\delta {\mathcal{L}}}
{\delta {\dot x^i}}
=g_{ij}(x)\dot x^j+
\frac {\sqrt{-1}}2 g_{ij,k}(x)
(\bar\psi^j\psi^k+ \psi^j\bar\psi^k)+A_i(x),\\
\phi_i & =\frac {\delta {\mathcal{L}}}
{\delta {\dot \psi^i}}=-
\frac {\sqrt{-1}}2 g_{ij}(x) \bar\psi^j,\\
\bar\phi_i & =\frac {\delta {\mathcal{L}}}
{\delta {\dot{\bar\psi^i}}}=
-\frac {\sqrt{-1}}2 g_{ij}(x) \psi^j,
\end{aligned}
\right.
\end{equation}
we get
\begin{equation}
\begin{aligned}
\mathcal{H} 
& =\mathcal{H} (x,\xi,\psi,\bar\psi)
\defeq \dot x\xi+
\dot\psi\phi+ \dot{\bar\psi}{\bar\phi} -\mathcal{L}\\
& = \frac 12 g^{ij}
\big(\xi _i - \frac {\sqrt{-1}}2 g_{ik,l}(\bar\psi^k\psi^l+ \psi^k\bar\psi^l)
-A_i\big)
\big(\xi _j - \frac {\sqrt{-1}}2 g_{jm,n}(\bar\psi^m\psi^n+ \psi^m\bar\psi^n)
-A_j\big)\\
& \qquad -\frac 12 R_{ikjl}
{\bar\psi}^i \psi^j {\bar\psi}^k \psi^l
+{\frac12}g^{jk} W_{,j} W_{,k}
+{\frac 12}W_{;ij}
({\bar\psi}^i\psi^j - \psi^i{\bar\psi}^j ).
\end{aligned}
\end{equation}
Now, rewriting the variables
$ \psi^i,\ \bar\psi^i $ as
$\theta^i, \, \pi^i $, we get finally
\begin{equation} \label{SSE2.27}
\begin{aligned}
\mathcal{H} &=\mathcal{H} (x,\xi,\theta,\pi)\\
&= \frac 12 g^{ij}
\big(\xi_i - \frac {\sqrt{-1}}2 (g_{ik,l}-g_{il,k})\theta^k\pi^l-A_i\big)
\big(\xi_j - \frac {\sqrt{-1}}2 (g_{jm,n}-g_{jn,m})\theta^m\pi^n-A_j\big)\\
&\qquad
+\frac 12 R_{ikjl}
\theta^j \theta^l{\pi}^i {\pi}^k 
+{\frac12}g^{jk} W_{,j} W_{,k}
-W_{;ij}
\theta^i{\pi}^j,
\end{aligned}
\end{equation}
which is thought as the supersymmetric extension of $H(q,p)$
when $V={\frac12}g^{jk} W_{,j} W_{,k}$.

\subsection{Supersymmetry and supercharges}
Preparing a pair of ``real" Grassmann parameters 
$ \varepsilon_{\alpha} \in \rod $
for $\alpha=1,2$, 
we introduce a one parameter group of transformations 
$ T_s $ ($s\in\euc$)
from
$(t,\rho_1,\rho_2)\in{\mathfrak{R}}^{1|2}$ to 
$(t',\rho'_1,\rho'_2)\in{\mathfrak{R}}^{1|2}$
defined by
\begin{equation} \label{SSE2.25}
\left\{
\aligned
t' &  =t-is(\varepsilon_1\rho_2-\varepsilon_2\rho_1),\\
\rho_1^{\prime} & = \rho_1- s \varepsilon_1,\\
\rho_2^{\prime} & = \rho_2- s \varepsilon_2
\endaligned
\right.
\end{equation}
and also two operators
\begin{equation} \label{SSE2.26}
{\mathcal{D}}_{\alpha}^+={\frac {\partial}{\partial\rho_{\alpha}}}
+i\rho_{\alpha}{\frac {\partial}{\partial t}} 
\quad \text{for} \quad \alpha=1,2.
\end{equation}
Clearly, the infinitesimal generator of transformations above
is given by
\begin{equation} %2.27
{\frac {\partial}{\partial s}} 
v(T_s(t,\rho_1,\rho_2)) {\big |}_{s=0} =
-(\varepsilon_1 {\mathcal{D}}_2^+ - \varepsilon_2 {\mathcal{D}}_1^+)
v(t,\rho_1,\rho_2)
\end{equation}
for any smooth function
$v(t,\rho_1,\rho_2)$
from ${\mathfrak{R}}^{1|2}$ to $\rev$.
Here, we remark that
$$
-(\varepsilon_1 {\mathcal{D}}_1^+ + \varepsilon_2 {\mathcal{D}}_2^+)
v(t,\rho_1,\rho_2)
= \delta t
{\frac {\partial v}{\partial t}}
+\delta\rho_1
{\frac {\partial v}{\partial\rho_1}}
+\delta\rho_2
{\frac {\partial v}{\partial\rho_2}}
$$
with
$$
\delta t= -i(\varepsilon_1\rho_1+ \varepsilon_2\rho_2),\quad
\delta\rho_1 = - \varepsilon_1,\quad
\delta\rho_2 = - \varepsilon_2.
$$
\par
Moreover,
$v(t,\rho_1,\rho_2)$ is called supersymmetric if
\begin{equation} %2.28
\delta v(t,\rho_1,\rho_2) =
-(\varepsilon_1 {\mathcal{D}}_1+ \varepsilon_2 {\mathcal{D}}_2)
v(t,\rho).
\end{equation}
Above relation implies the following:
If $\Phi^j(t,\rho_1,\rho_2)$ given in \eqref{SSE2.1} is supersymmetric, we have
\begin{equation}\label{SSE2.29}
\begin{aligned}
\delta \Phi ^j(t,\rho_1,\rho_2) & \equiv
\delta x^j(t)+
i\rho_{\alpha}\epsilon_{\alpha\beta}\,\delta\psi_{\beta}^j(t)+
\frac i2
\rho_{\alpha}\epsilon_{\alpha\beta}\rho_{\beta}\delta F^j(t)\\
& = -(\varepsilon_1 {\mathcal{D}}_1+ \varepsilon_2 {\mathcal{D}}_2)
\Phi ^j(t,\rho_1,\rho_2)
\end{aligned}
\end{equation}
Since $\fR$ is infinite dimensional,
if $\omega $ satisfies $\delta\rho_1 \omega =0$ and 
$\rho_1 \omega =0$ then $\omega=0$. 
This yields that
\begin{equation} \label{SSE2.30}
\left\{
\begin{aligned}
\delta x^j & = -i 
\varepsilon _{\alpha}
\epsilon_{\alpha \beta}
\psi_{\beta}^j,\\
\delta\psi_{\alpha}^j & = -\varepsilon _{\alpha} F^j
-\epsilon_{\alpha \beta} \varepsilon _{\beta} \dot x^j ,\\
\delta F^j & = i \varepsilon _{\alpha} \dot\psi_{\alpha}^j .
\end{aligned}
\right.
\end{equation}
Using the relation \eqref{SSE2.16}, we get
\begin{equation} \label{SSE2.31}
\left\{
\begin{aligned}
\delta x^j & = -i 
\varepsilon _{\alpha}
\epsilon_{\alpha \beta}
\psi_{\beta}^j,\\
\delta\psi_{\alpha}^j & = 
-\varepsilon _{\alpha} 
(\frac i2 \Gamma^j_{kl}[kl] - \nabla^j W)
-\epsilon_{\alpha \beta} \varepsilon _{\beta} \dot x^j .
\end{aligned}
\right.
\end{equation}
\par
From this, we have the following quantities, called supercharges,
\begin{equation} \label{SSE2.32}
{\mathcal{Q}}_{\alpha}=\psi_{\alpha}^i g_{ij} \dot x^j
-\epsilon_{\alpha\beta}
\psi_{\beta}^i \nabla_i W
\end{equation}
which is conserved by the flow defined by the above Lagrangian.

On the other hand, the following supersymmetric Lagrangian is introduced by physicist:
\begin{align*}
\mathcal{L}_0 (x;\dot{x},\psi_{\alpha};\dot\psi_{\alpha})
= {\frac{1}{2}}
g_{jk}\, \dot{x}^j \dot {x}^k
+ \frac{\sqrt{-1}}{2} g_{jk}
\bar\psi^j \gamma^0
& {\frac D{dt}} \psi^k 
+\frac {1}{12} R_{ikjl}
\bar\psi^i \psi^j \bar\psi^k \psi^l\\
&- {\frac{1}{2}}
\nabla^j W \nabla_j W
- {\frac{1}{2}}
\nabla_i \nabla_j W [\bar\psi^i,\psi^j].
\end{align*}
Here,
$$
\psi^j=\begin{pmatrix}
 \psi_1^j\\ \psi_2^j \end{pmatrix} ,\quad
\bar\psi^j={\sqrt{-1}}(\psi_2^j, -\psi_1^j)
={}^t{\begin{pmatrix} \psi_1^j\\ \psi_2^j \end{pmatrix} }^* \gamma^0,\quad
\gamma^0=\begin{pmatrix} 0 & - {\sqrt{-1}}\\ {\sqrt{-1}} & 0 \end{pmatrix} ,
$$
L. Alvarez-Gaum\'e \cite{AG83} used above with $W=0$ on a general manifold $M$.

%%%%%
%\newpage
\section[Hamilton flow for Weyl equation with variable coefficients]{Hamilton flow for Weyl equation with external electro-magnetic field}
In this section, we consider the Weyl equation with external electro-magnetic field \eqref{eW-1.1} %(0.7) in Chapter 7%\eqref{fw+mi} 
with its symbol below: % \eqref{Weyl-2symbol}: 
\begin{equation}
{\mathcal{H}}(t,x,\xi,\theta,\pi)=\sum_{j=1}^3c\sigma_j(\theta,\pi)\big(\xi_j-\frac{e}{c} A_j(t,x)\big)+eA_0(t,x).
\label{Weyl-2symbol}
\end{equation}
Corresponding Hamilton equation is, for $j=1,2,3$, $l=1,2$,
 \begin{equation}
\left\{
\begin{aligned}
&\dt x_j=\frac{\partial{\mathcal{H}}(t,x,\xi,\theta,\pi)}{\partial \xi_j},\quad
\dt \xi_j=-\frac{\partial{\mathcal{H}}(t,x,\xi,\theta,\pi)}{\partial x_j},\\
&\dt \theta_l=-\frac{\partial{\mathcal{H}}(t,x,\xi,\theta,\pi)}{\partial \pi_l},\quad
\dt \pi_l=-\frac{\partial{\mathcal{H}}(t,x,\xi,\theta,\pi)}{\partial \theta_l}.
\end{aligned}
\right.
\label{fwmi1.9}
\end{equation}
We take this as a simple example for the necessity of the countablely infinite Grassmann generators.

\begin{proposition}\label{prop:existence}
Assume $(A_0(t,q), A_1(t,q), A_2(t,q), A_3(t,q))\in C^{\infty}(\euc\times\euc^3:\euc)$ in (0.7) in Chapter 7. %\eqref{fw+mi}.
Then, for any initial data $(x(0),\xi(0),\theta(0),\pi(0))=({\unbx,\unbxi,\unbtheta,\unbpi})
\in{\fR}^{6|4}\cong{\mathcal{T}}^*{\fR}^{3|2}$, \eqref{fwmi1.9} has the unique, global in time, solution $(x(t),\xi(t),\theta(t),\pi(t))$.
\end{proposition}
\begin{remark}
We require only smoothness for $\{A_j(t,q)\}_{j=0}^3$ without strict conditions on the behavior when $|q|\to\infty$.
\end{remark}

Odd variables part of \eqref{fwmi1.9} are rewritten as 
\begin{equation}
\dt
\begin{pmatrix}
\theta_1\\
\theta_2\\
\pi_1\\
\pi_2
\end{pmatrix}
=ic\hbar^{-1}{\mathbb X}(t)
\begin{pmatrix}
\theta_1\\
\theta_2\\
\pi_1\\
\pi_2
\end{pmatrix}
\with
\begin{pmatrix}
\theta_1({\unbt})\\
\theta_2({\unbt})\\
\pi_1({\unbt})\\
\pi_2({\unbt})
\end{pmatrix}
=\begin{pmatrix}
\unbtheta_1\\
\unbtheta_2\\
\unbpi_1\\
\unbpi_2
\end{pmatrix}.
\label{rt3.1}
\end{equation}
Here, we put
\begin{equation}
\eta_j(t)=\xi_j(t)-\frac{e}{c}A_j(t,x(t)),
\label{rteta}
\end{equation}
and
\begin{equation}
{\mathbb X}(t)
=\begin{pmatrix}
-\eta_3(t){\mathbb{I}}_2&{\hbar}^{-1}(\eta_1(t)-i\eta_2(t)){\pmb{\sigma}}_2\\
\hbar(\eta_1(t)+i\eta_2(t)){\pmb{\sigma}}_2&\eta_3(t){\mathbb{I}}_2
\end{pmatrix}
\begin{comment}
\begin{aligned}
&{\mathbb X}(t)\\
&=
\begin{pmatrix}
-\eta_3(t)&0&0&{i}\hbar^{-1}(\eta_1(t)-i\eta_2(t))\\
0&-\eta_3(t)&-{i}\hbar^{-1}(\eta_1(t)-i\eta_2(t))&0\\
0&{i}\hbar(\eta_1(t)+i\eta_2(t))&\eta_3(t)&0\\
-{i}\hbar(\eta_1(t)+i\eta_2(t))&0&0&\eta_3(t)
\end{pmatrix}.
\end{aligned}
\end{comment}
\label{rt3.2}
\end{equation}

Moreover, for $\sigma_j(t)=\sigma_j(\theta(t),\pi(t))$,
\begin{equation}
{\mathbb Y}(t)=\begin{pmatrix}
0& -\eta_3(t) &\eta_2(t)\\
\eta_3(t)&0&-\eta_1(t)\\
-\eta_2(t)&\eta_1(t)&0
\end{pmatrix}.
\label{rt3.4}
\end{equation}
by simple calculation
\begin{equation}
\frac{d}{dt} %\dt
\begin{pmatrix}
\sigma_1\\
\sigma_2\\
\sigma_3
\end{pmatrix}
=2c\hbar^{-1}{\mathbb Y}(t)
\begin{pmatrix}
\sigma_1\\
\sigma_2\\
\sigma_3
\end{pmatrix}
\with
\begin{pmatrix}
\sigma_1({\unbt})\\
\sigma_2({\unbt})\\
\sigma_3({\unbt})
\end{pmatrix}
=\begin{pmatrix}
\unbsigma_1\\
\unbsigma_2\\
\unbsigma_3
\end{pmatrix}
=\begin{pmatrix}
\unbtheta_1\unbtheta_2+\hbar^{-2}\unbpi_1\unbpi_2\\
i(\unbtheta_1\unbtheta_2-\hbar^{-2}\unbpi_1\unbpi_2)\\
-i\hbar^{-1}(\unbtheta_1\unbpi_1+\unbtheta_2\unbpi_2)
\end{pmatrix}.
\label{rt3.3}
\end{equation}

Now we begin our proof.  We decompose dependent variables $(x,\xi,\theta,\pi)$ by degree:
\begin{equation}
x_j(t)=\sum_{\ell=0}^\infty x_j^{[2\ell]}(t), \quad 
\xi_j(t)=\sum_{\ell=0}^\infty \xi_j^{[2\ell]}(t),\quad
\theta_k(t)=\sum_{\ell=0}^\infty \theta_k^{[2\ell+1]}(t), \quad 
\pi_k(t)=\sum_{\ell=0}^\infty \pi_k^{[2\ell+1]}(t).
\label{rt3.5}
\end{equation}

For given $m=0,1,2,\cdots$,
\begin{equation}
\left\{
\begin{aligned}
&\dt x_j^{[2m]}=c\sigma_j^{[2m]}\where \sigma_j^{[0]}=0,\\
&\dt \xi_j^{[2m]}={e}\sum_{\ell=1}^m\sum_{k=1}^3 \sigma_k^{[2\ell]}
\frac{\partial A_k^{[2m-2\ell]}}{\partial x_j}
-{e}\frac{\partial {A}_0^{[2m]}}{\partial x_j}
\end{aligned}\right.
\with
\begin{pmatrix}
x^{[2m]}({\unbt})\\
\xi^{[2m]}({\unbt})
\end{pmatrix}
=
\begin{pmatrix}
\unbx^{[2m]}\\
\unbxi^{[2m]}
\end{pmatrix},
\label{rt3.6}
\end{equation}
\begin{equation}
\dt
\begin{pmatrix}
\theta_1^{[2m+1]}\\
\theta_2^{[2m+1]}\\
\pi_1^{[2m+1]}\\
\pi_2^{[2m+1]}\\
\end{pmatrix}
=ic\hbar^{-1}\sum_{\ell=0}^m{\mathbb X}^{[2\ell]}(t)
\begin{pmatrix}
\theta_1^{[2m+1-2\ell]}\\
\theta_2^{[2m+1-2\ell]}\\
\pi_1^{[2m+1-2\ell]}\\
\pi_2^{[2m+1-2\ell]}\\
\end{pmatrix}
\with
\begin{pmatrix}
\theta_1^{[2m+1]}({\unbt})\\
\theta_2^{[2m+1]}({\unbt})\\
\pi_1^{[2m+1]}({\unbt})\\
\pi_2^{[2m+1]}({\unbt})\\
\end{pmatrix}
=\begin{pmatrix}
\unbtheta_1^{[2m+1]}\\
\unbtheta_2^{[2m+1]}\\
\unbpi_1^{[2m+1]}\\
\unbpi_2^{[2m+1]}\\
\end{pmatrix},
\label{rt3.7}
\end{equation}
and
\begin{equation}
\dt
\begin{pmatrix}
\sigma_1^{[2m]}\\
\sigma_2^{[2m]}\\
\sigma_3^{[2m]}
\end{pmatrix}
=\sum_{\ell=0}^{m-1}2c\hbar^{-1}{\mathbb Y}^{[2\ell]}(t)
\begin{pmatrix}
\sigma_1^{[2m-2\ell]}\\
\sigma_2^{[2m-2\ell]}\\
\sigma_3^{[2m-2\ell]}
\end{pmatrix}
\with
\begin{pmatrix}
\sigma_1^{[2m]}({\unbt})\\
\sigma_2^{[2m]}({\unbt})\\
\sigma_3^{[2m]}({\unbt})
\end{pmatrix}
=\begin{pmatrix}
\unbsigma_1^{[2m]}\\
\unbsigma_2^{[2m]}\\
\unbsigma_3^{[2m]}
\end{pmatrix}.
\label{rt3.8}
\end{equation}
Here, ${\mathbb X}^{[2\ell]}(t)$, ${\mathbb Y}^{[2\ell]}(t)$, 
$\sigma^{[2\ell]}(t)$ are degree $2\ell$ part of${\mathbb X}(t)$, 
${\mathbb Y}(t)$, $\sigma(t)$, respectively.
$$
\begin{gathered}
\eta_k^{[2\ell]}(t)=\xi_k^{[2\ell]}(t)-\frac{e}{c}A_k^{[2\ell]}(t,x),\\
A_k^{[2\ell]}(t,x)=\sum
_{\scriptstyle{|\alpha|\le2\ell}\atop\scriptstyle{\ell_1+\ell_2+\ell_3=\ell}}
%{\begin{Sb}|\alpha|\le2\ell\\
%\ell_1+\ell_2+\ell_3=\ell\end{Sb}}
\frac1{\alpha!}\partial_q^\alpha A_k(t,x^{[0]})\cdot
(x_{1}^{\alpha_1})^{[2\ell_1]}
(x_{2}^{\alpha_2})^{[2\ell_2]}
(x_{3}^{\alpha_3})^{[2\ell_3]},\\
\frac{\partial{A}_0^{[2\ell]}(t,x)}{\partial x_j}=\sum
_{\scriptstyle{|\alpha|\le2\ell}\atop\scriptstyle{\ell_1+\ell_2+\ell_3=\ell}}
%{\begin{Sb}|\alpha|\le2\ell\\
%\ell_1+\ell_2+\ell_3=\ell\end{Sb}}
\frac1{\alpha!}\partial_q^\alpha \partial_{q_j}{A}_0(t,x^{[0]})\cdot
(x_{1}^{\alpha_1})^{[2\ell_1]}
(x_{2}^{\alpha_2})^{[2\ell_2]}
(x_{3}^{\alpha_3})^{[2\ell_3]}
\end{gathered}
$$
And
$$
\left\{
\begin{aligned}
\sigma_1^{[2m]}&=\sum_{\ell=0}^{m-1}
\Big(\theta_1^{[2\ell+1]}\theta_2^{[2m-2\ell-1]}
+\hbar^{-2}\pi_1^{[2\ell+1]}\pi_2^{[2m-2\ell-1]}\Big),\\
\sigma_2^{[2m]}&=i\sum_{\ell=0}^{m-1}
\Big(\theta_1^{[2\ell+1]}\theta_2^{[2m-2\ell-1]}
-\hbar^{-2}\pi_1^{[2\ell+1]}\pi_2^{[2m-2\ell-1]}\Big),\\
\sigma_3^{[2m]}&=-i\hbar^{-1}\sum_{\ell=0}^{m-1}
\Big(\theta_1^{[2\ell+1]}\pi_1^{[2m-2\ell-1]}
+\theta_2^{[2\ell+1]}\pi_2^{[2m-2\ell-1]}\Big).
\end{aligned}
\right.
$$
\par
[0] Putting $m=0$ in \eqref{rt3.6}, for $j=1,2,3$,
\begin{equation*}
\dt x_j^{[0]}(t)=0\et 
\dt\xi_j^{[0]}(t)=-{e}\frac{\partial{A}_0^{[0]}(t,x^{[0]})}{\partial x_j}
=-{e}\partial_{q_j}{A}_0^{[0]}(t,x^{[0]}).
%\label{rt3.9}
\end{equation*}
Therefore, for any $t\in \euc$ and $j=1,2,3$,
\begin{equation*}
x_j^{[0]}(t)=\unbx_j^{[0]}\et
\xi_j^{[0]}(t)=\unbxi_j^{[0]}
-{e}\int_{\unbt}^t ds\,\partial_{q_j}{A}_0^{[0]}(s,\unbx^{[0]}).
%\label{rt3.10}
\end{equation*}
\par
[1] Putting these result into \eqref{rt3.7} with $m=0$
\begin{equation}
\dt
\begin{pmatrix}
\theta_1^{[1]}\\
\theta_2^{[1]}\\
\pi_1^{[1]}\\
\pi_2^{[1]}
\end{pmatrix}
=ic\hbar^{-1}{\mathbb X}^{[0]}(t)
\begin{pmatrix}
\theta_1^{[1]}\\
\theta_2^{[1]}\\
\pi_1^{[1]}\\
\pi_2^{[1]}
\end{pmatrix}
\with
\begin{pmatrix}
\theta_1^{[1]}({\unbt})\\
\theta_2^{[1]}({\unbt})\\
\pi_1^{[1]}({\unbt})\\
\pi_2^{[1]}({\unbt})
\end{pmatrix}
=\begin{pmatrix}
\unbtheta_1^{[1]}\\
\unbtheta_2^{[1]}\\
\unbpi_1^{[1]}\\
\unbpi_2^{[1]}
\end{pmatrix}.
\label{rt3.11}
\end{equation}
Here, ${\mathbb X}^{[0]}(t)$ is $4\times 4$-matrix whose components are complex valued, depending on 
$$
(t,{\unbt},\unbx^{[0]},\unbxi^{[0]},
\partial_q^\beta {A}_0,\partial_q^\alpha A\,;|\alpha|=0,|\beta|\le 1).
$$
More precisely,  components of ${\mathbb X}^{[0]}(t)$ are given as
$$
\eta_j^{[0]}(t)=\xi_j^{[0]}(t)-\frac{e}{c}A_j(t,\unbx^{[0]})
=\unbxi_j^{[0]}-{e}\int_{\unbt}^t ds\,\partial_{q_j}{A}_0^{[0]}(s,\unbx^{[0]})
-\frac{e}{c}A_j^{[0]}(t,\unbx^{[0]}).
$$
ODE \eqref{rt3.11} has smooth coefficients w.r.t. $t$ with value in $({\fR}^{0|1})^4$, which has a unique global in time solution depending on as follows:
Putting $A=(A_1,A_2,A_3)$, we have
\begin{equation}
\left\{
\begin{aligned}
&\theta_j^{[1]}(t)=\theta_j^{[1]}(t,\unbx^{[0]},\unbxi^{[0]},
\unbtheta^{[1]},\unbpi^{[1]},
\partial_q^\beta {A}_0,\partial_q^\alpha A\,;|\alpha|=0,|\beta|\le 1),\;
\text{linear w.r.t. $\unbtheta^{[1]},\unbpi^{[1]}$},\\
&\pi_j^{[1]}(t)=\pi_j^{[1]}(t,\unbx^{[0]},\unbxi^{[0]},
\unbtheta^{[1]},\unbpi^{[1]},
\partial_q^\beta {A}_0,\partial_q^\alpha A\,;|\alpha|=0,|\beta|\le 1),\;
\text{linear w.r.t. $\unbtheta^{[1]},\unbpi^{[1]}$}.
\end{aligned}
\right.\label{rt3.12}
\end{equation}
\par
[2] Putting $m=1$ in \eqref{rt3.6},
\begin{equation*}
\left\{
\begin{aligned}
&\dt
x_j^{[2]}=c\sigma_j^{[2]},\\
&\dt
\xi_j^{[2]}={e}\sum_{k=1}^3
\sigma_k^{[2]}\frac{\partial A_k^{[0]}}{\partial x_j}
-{e}\frac{\partial {A}_0^{[2]}}{\partial x_j}
\end{aligned}
\right.
\with
\begin{pmatrix}
x^{[2]}({\unbt})=\unbx^{[2]}\\
\xi^{[2]}({\unbt})=\unbxi^{[2]}
\end{pmatrix}.
%\label{rt3.13}
\end{equation*}
From $m=1$ in \eqref{rt3.8} and \eqref{rt3.12}, for $j=1,2,3$,
$$
\left\{
\begin{aligned}
\sigma_1^{[2]}&=\theta_1^{[1]}\theta_2^{[1]}+\hbar^{-2}\pi_1^{[1]}\pi_2^{[1]},\\
\sigma_2^{[2]}&=i(\theta_1^{[1]}\theta_2^{[1]}-\hbar^{-2}\pi_1^{[1]}\pi_2^{[1]}),\\
\sigma_3^{[2]}&=-i\hbar^{-1}(\theta_1^{[1]}\pi_1^{[1]}+\theta_2^{[1]}\pi_2^{[1]})
\end{aligned}
\right.
\et
\left\{
\begin{aligned}
{A}_0^{[2]}(x)&=\sum_{k=1}^3\partial_{q_k}{A}_0(x^{[0]})x_{k}^{[2]},\\
\frac{\partial {A}_0^{[2]}}{\partial x_j}&=
\sum_{k=1}^3\partial_{q_kq_j}{A}_0(x^{[0]})x_{k}^{[2]}.
\end{aligned}
\right.
$$
Therefore, for $j=1,2,3$,
\begin{equation*}
\left\{
\begin{aligned}
&x_j^{[2]}(t)=x_j^{[2]}(t,\unbx^{[2\ell]},\unbxi^{[0]},
\unbtheta^{[1]},\unbpi^{[1]},
\partial_q^\beta {A}_0,\partial_q^\alpha A\,;0\le\ell\le 1,|\alpha|=0,|\beta|\le 1),\\
&\xi_j^{[2]}(t)=\xi_j^{[2]}(t,\unbx^{[2\ell]},\unbxi^{[2\ell]},
\unbtheta^{[1]},\unbpi^{[1]},
\partial_q^\beta {A}_0,\partial_q^\alpha A\,;0\le\ell\le 1,|\alpha|\le 1,|\beta|\le 2).
\end{aligned}
\right.
%\label{rt3.14}
\end{equation*}
\par
[3] Putting $m=1$ in \eqref{rt3.7},
\begin{equation*}
\dt
\begin{pmatrix}
\theta_1^{[3]}\\
\theta_2^{[3]}\\
\pi_1^{[3]}\\
\pi_2^{[3]}
\end{pmatrix}
=ic\hbar^{-1}{\mathbb X}^{[0]}(t)
\begin{pmatrix}
\theta_1^{[3]}\\
\theta_2^{[3]}\\
\pi_1^{[3]}\\
\pi_2^{[3]}
\end{pmatrix}
+ic\hbar^{-1}{\mathbb X}^{[2]}(t)
\begin{pmatrix}
\theta_1^{[1]}\\
\theta_2^{[1]}\\
\pi_1^{[1]}\\
\pi_2^{[1]}
\end{pmatrix}
\with
\begin{pmatrix}
\theta_1^{[3]}({\unbt})\\
\theta_2^{[3]}({\unbt})\\
\pi_1^{[3]}({\unbt})\\
\pi_2^{[3]}({\unbt})
\end{pmatrix}
=\begin{pmatrix}
\unbtheta_1^{[3]}\\
\unbtheta_2^{[3]}\\
\unbpi_1^{[3]}\\
\unbpi_2^{[3]}
\end{pmatrix}.
%\label{rt3.15}
\end{equation*}
Here, $4\times 4$-matrix ${\mathbb X}^{[2]}(t)$ has components, valued in $\cev$ depending on 
$$
(t,\unbx^{[2\ell]},\unbxi^{[2\ell]},
\unbtheta^{[1]},\unbpi^{[1]},
\partial_q^\beta {A}_0,\partial_q^\alpha A\,;
0\le\ell\le 1,|\alpha|\le 1,|\beta|\le 2).$$

From these, for $k=1,2$,
\begin{equation*}
\left\{
\begin{aligned}
&\theta_k^{[3]}(t)=\theta_k^{[3]}(t,\unbx^{[2\ell]},\unbxi^{[2\ell]},
\unbtheta^{[2\ell+1]},\unbpi^{[2\ell+1]},
\partial_q^\beta{A}_0,\partial_q^\alpha{A}\,;
0\le\ell\le 1,|\alpha|\le 1, |\beta|\le 2),\\
&\pi_k^{[3]}(t)=\pi_k^{[3]}(t,\unbx^{[2\ell]},\unbxi^{[2\ell]},
\unbtheta^{[2\ell+1]},\unbpi^{[2\ell+1]},
\partial_q^\beta{A}_0,\partial_q^\alpha{A}\,;
0\le\ell\le 1,|\alpha|\le 1, |\beta|\le 2).
\end{aligned}
\right.
%\label{rt3.16}
\end{equation*}
\par
[4] Repeating this procedure, we get
\begin{equation*}
\left\{
\begin{aligned}
&x^{[2m]}(t)=x^{[2m]}(t,\unbx^{[2\ell]},\unbxi^{[2\ell]},
\unbtheta^{[2\ell-1]},\unbpi^{[2\ell-1]},
\partial_q^\beta{A}_0, \partial_q^\alpha A\,;
0\le\ell\le m,|\alpha|\le m-1,|\beta|\le m),\\
&\xi^{[2m]}(t)=\xi^{[2m]}(t,\unbx^{[2\ell]},\unbxi^{[2\ell]},
\unbtheta^{[2\ell-1]},\unbpi^{[2\ell-1]},
\partial_q^\beta{A}_0, \partial_q^\alpha A\,;
0\le\ell\le m,|\alpha|\le m,|\beta|\le m+1),\\
&\theta^{[2m+1]}(t)=\theta^{[2m+1]}(t,\unbx^{[2\ell]},\unbxi^{[2\ell]},
\unbtheta^{[2\ell+1]},\unbpi^{[2\ell+1]},
\partial_q^\beta{A}_0, \partial_q^\alpha A\,;
0\le\ell\le m,|\alpha|\le m,|\beta|\le m+1),\\
&\pi^{[2m+1]}(t)=\pi^{[2m+1]}(t,\unbx^{[2\ell]},\unbxi^{[2\ell]},
\unbtheta^{[2\ell+1]},\unbpi^{[2\ell+1]},
\partial_q^\beta{A}_0, \partial_q^\alpha A\,;
0\le\ell\le m,|\alpha|\le m,|\beta|\le m+1).
\end{aligned}
\right. %\label{rt3.17}
\end{equation*}
These prove the existence, moreover, since for each degree, the solution of \eqref{rt3.6} with \eqref{rt3.7} is unique, so follows the uniqueness of the solution of \eqref{fwmi1.9}.$\qquad\qed$

Moreover, we get easily
\begin{corollary}
If $(x(t),\xi(t),\theta(t),\pi(t))\in C^1(\euc:{\mathcal{T}}^*{\mathfrak {R}}^{3|2})$ is a solution of \eqref{fwmi1.9}, then\begin{equation}
\dt {\mathcal{H}}(t,x(t),\xi(t),\theta(t),\pi(t))
=\frac{\partial{\mathcal{H}}}{\partial t}(t,x(t),\xi(t),\theta(t),\pi(t)).
\label{rt3.18}
\end{equation}
\end{corollary}

Putting
$$
B_{jk}(t,x)=\frac{\partial A_k(t,x)}{\partial x_j}
-\frac{\partial A_j(t,x)}{\partial x_k},
$$
and rewriting
\begin{equation}
\left\{
\begin{aligned}
&\dt x_j=c\sigma_j(\theta,\pi),\\
&\dt \eta_j
=\sum_{k=1}^3{e}\sigma_k(\theta,\pi)B_{jk}(t,x)
-{e}\frac{\partial {A}_0(t,x)}{\partial x_j},
\end{aligned}
\right.
\label{rt2.6-11}
\end{equation}
we have
\begin{corollary}
For $\{A_j(t,q)\}_{j=0}^3\in C^\infty(\euc\times\euc^3:\euc)$, putting the initial data
$$
({\tilde x}({\unbt}),
{\tilde \eta}({\unbt}),
{\tilde \theta}({\unbt}),
{\tilde \pi}({\unbt}))=
(\unbx,\unbeta,\unbtheta,\unbpi) \where
\unbeta_j=\unbxi_j-\frac{e}{c}A_j({\unbt},\unbx)
$$
for \eqref{rt2.6-11} with \eqref{rt3.1}, there exists a unique solution
$({\tilde x}(t),{\tilde \eta}(t),
{\tilde \theta}(t),{\tilde \pi}(t))\in C^1(\euc:{\mathcal{T}}^*{\mathfrak{R}}^{3|2})$.
These are related with $(x(t),\xi(t),\theta(t),\pi(t))$ as
\begin{equation*}
\left\{
\begin{aligned}
&x_j(t,{\unbt}\,;{\unbx,\unbxi,\unbtheta,\unbpi})
={\tilde x}_j(t,{\unbt}\,;
\unbx,\unbxi-\frac{e}{c}A({\unbt},\unbx),\unbtheta,\unbpi),\\
&\xi_j(t,{\unbt}\,;{\unbx,\unbxi,\unbtheta,\unbpi})
={\tilde\eta}_j(t,{\unbt}\,;
\unbx,\unbxi-\frac{e}{c}A({\unbt},\unbx),\unbtheta,\unbpi)
+\frac{e}{c}A_j(t,{\tilde x}(t,{\unbt}\,;
\unbx,\unbxi-\frac{e}{c}A({\unbt},\unbx),\unbtheta,\unbpi)),\\
&\theta_k(t,{\unbt}\,;{\unbx,\unbxi,\unbtheta,\unbpi})
={\tilde\theta}_k(t,{\unbt}\,;
\unbx,\unbxi-\frac{e}{c}A({\unbt},\unbx),\unbtheta,\unbpi),\\
&\pi_k(t,{\unbt}\,;{\unbx,\unbxi,\unbtheta,\unbpi})
={\tilde\pi}_k(t,{\unbt}\,;
\unbx,\unbxi-\frac{e}{c}A({\unbt},\unbx),\unbtheta,\unbpi).
\end{aligned}\right.
\end{equation*}
\end{corollary}

\begin{remark}
The solution of the Hamilton flow corresponding to free Weyl equation in Proposition~\ref{prop:existence} is solved explicitly.
Obtaining such an explicit solution is not necessarily happened frequently, in general we have only its existence abstractly.
Fortunately, because of the countable degree stems from the countable Grassmann generators, we get rather easily the existence proof. But it is rather complicated to have the estimates w.r.t. the initial data.
\end{remark}

%\newpage

\end{document}